%% file: TR-Experimental_Control_Complexity_FV_BV_PV-oneside-without-comments.tex
\documentclass[a4paper,10pt,twoside]{scrartcl}
\usepackage[utf8x]{inputenc}

\usepackage{tabularx}
\usepackage{multirow}
\usepackage{eepic,epic}
\usepackage{epsfig}
\usepackage{graphicx}
\usepackage{color}
\usepackage{mathptmx}
\usepackage{amssymb}
\usepackage{amsmath}
\usepackage{pifont}
\usepackage{nicefrac}
\usepackage{caption}
\usepackage{booktabs}
\usepackage{subfig}

\newcommand{\EP}[3]{
\begin{center}
{\small 
\begin{tabularx}{0.95\columnwidth}{ll}
\toprule
\multicolumn{2}{c}{\sc{#1}} \\
\midrule
{\bf Given:}& \parbox[t]{0.8\columnwidth}{#2\vspace*{1mm}} \\
{\bf Question:}& \parbox[t]{0.8\columnwidth}{#3\vspace*{.5mm}} \\ 
\bottomrule
\end{tabularx}
}
\end{center}
}

\makeatletter%
\newcommand{\doublespacing}{\let\CS=
\@currsize\renewcommand{\baselinestretch}{1.75}\tiny\CS}
\newcommand{\extradoublespacing}{\let\CS=
\@currsize\renewcommand{\baselinestretch}{1.9}\tiny\CS}
\newcommand{\draftspacing}{\let\CS=
\@currsize\renewcommand{\baselinestretch}{2.0}\tiny\CS}
\newcommand{\hugedraftspacing}{\let\CS=
\@currsize\renewcommand{\baselinestretch}{2.4}\tiny\CS}
\makeatother%

\newcommand{\OMIT}[1]{} %

\newcommand\qedblob{\ding{113}}
\def\literalqed{{\ \nolinebreak\hfill\mbox{\qedblob\quad}}}

\newcommand{\seq}{\subseteq}

\newcommand{\maj}[1]{\mathit{maj}(#1)}

\newcommand{\scoresublevel}[3]{\mathit{score}_{#1}^{#2}(#3)}

\newcommand{\np}{\mbox{\rm NP}}

\newcommand{\vs}{\mathcal{E}}

\newenvironment{desctight}
  {\begin{list}{}{\setlength\labelwidth{0pt}%
        \setlength{\itemsep}{0.5pt}%
        \setlength{\parsep}{0pt}%
        \setlength\itemindent{-\leftmargin}%
        }}
    {\end{list}}

\usepackage{hyperref}

\begin{document}
\title{Control Complexity in Bucklin, Fallback, and Plurality
  Voting: An Experimental Approach\thanks{This work was supported in
    part by DFG grants RO~\mbox{1202/15-1} and RO~\mbox{1202/12-1}
    (within the EUROCORES programme LogICCC of the ESF), SFF
    grant ``Cooperative Normsetting'' of HHU D{\"u}sseldorf, and a
    DAAD grant for a PPP project in the PROCOPE programme. Author URLs: 
\mbox{\tt{}ccc.cs.uni-duesseldorf.de/\mbox{\tiny$\sim\,$}rothe},
\mbox{\tt{}ccc.cs.uni-duesseldorf.de/\mbox{\tiny$\sim\,$}schend}. }}

\author{
J\"{o}rg Rothe
\ and \ 
Lena Schend
\\
 Institut f\"{u}r Informatik\\
Heinrich-Heine-Universit\"{a}t D\"{u}sseldorf \\
40225 D\"{u}sseldorf, Germany 
}
\date{\today}
\maketitle

\begin{abstract}
  Walsh~\cite{wal:c:empirical-study-of-manipulability-of-STV,wal:c:phase-transition-in-manipulating-veto}, Davies et
  al.~\cite{dav-kat-nar-wal:c:empirical-study-borda-manipulation,dav-kat-nar-wal:c:complexity-and-algorithms-for-borda},
  and Narodytska et
  al.~\cite{nar-wal-xia:c:manipulation-of-nanson-and-baldwin}
  studied various voting systems empirically and showed that they can
  often be manipulated effectively, despite their manipulation
  problems being $\np$-hard.  Such an experimental approach is sorely
  missing for $\np$-hard control problems, where control refers to
  attempts to tamper with the outcome of elections by
  adding/deleting/partitioning either voters or candidates.
  We experimentally tackle $\np$-hard
  control problems for
  Bucklin and fallback voting.
  Among natural voting systems with
  efficient winner determination, fallback voting is currently known to
  display the broadest resistance to control in terms of
  $\np$-hardness, and Bucklin voting has been shown to behave almost
  as well in terms of control resistance~\cite{erd-rot:c:fallback-voting,erd-pir-rot:c:voter-partition-in-bucklin-and-fallback-voting,erd-fel-pir-rot:t-v1:control-in-bucklin-and-fallback-voting}.
  We also investigate control resistance experimentally
  for plurality voting, one of the first voting systems analyzed 	
  with respect to electoral
  control~\cite{bar-tov-tri:j:control,hem-hem-rot:j:destructive-control}.

  Our findings indicate that $\np$-hard control problems can often be
  solved effectively in practice.  Moreover, our experiments allow a more
  fine-grained analysis and comparison---across various control scenarios,
  vote distribution models, and voting systems---than merely stating
  $\np$-hardness for all these control problems.
\end{abstract}

\newpage
\tableofcontents

\newpage
\section{Introduction and Motivation}
\label{sec:introduction}
Electoral
control~\cite{bar-tov-tri:j:control,hem-hem-rot:j:destructive-control}
refers to attempts to tamper with the outcome of elections by
adding/deleting/partitioning either the voters or the candidates.  To
protect elections against such control attempts and other ways of
manipulation (see, e.g., the
surveys~\cite{fal-hem-hem:j:cacm-survey,fal-pro:j:manipulation}), much
work has been done recently to show that the attacker's task can be
computationally hard: Certain voting systems are resistant to
manipulation~\cite{fal-hem-hem:j:cacm-survey,fal-pro:j:manipulation,con-san-lan:j:when-hard-to-manipulate}
or
control~\cite{bar-tov-tri:j:control,hem-hem-rot:j:destructive-control,erd-rot:c:fallback-voting,erd-pir-rot:c:voter-partition-in-bucklin-and-fallback-voting}
in certain scenarios.  However, most of this work is concerned with
$\np$-hardness results, which is a worst-case measure of complexity
and leaves open the possibility that many elections can still be
manipulated or controlled in a reasonable amount of time.

To avoid this disadvantage, manipulation and control problems have
also been tackled from different angles.  For example, from a
theoretical perspective, Zuckerman et
al.~\cite{zuc-pro-ros:j:coalitional-manipulation} proposed
approximation algorithms for $\np$-hard manipulation problems and
Faliszewski et al.~\cite{fal-hem-hem-rot:j:single-peaked-preferences}
showed that restricting to single-peaked electorates may strip
manipulation and control problems off their $\np$-hardness shields.
From an experimental perspective, in a series of papers Walsh et
al.~\cite{wal:c:empirical-study-of-manipulability-of-STV,wal:c:phase-transition-in-manipulating-veto,dav-kat-nar-wal:c:empirical-study-borda-manipulation,dav-kat-nar-wal:c:complexity-and-algorithms-for-borda}
(see
also~\cite{col-tea:c:complexity-of-manipulating-elections})
studied various voting systems empirically, such as single
transferable vote (STV), veto, and Borda, and showed that they can
often be manipulated effectively, even though their manipulation
problems are $\np$-hard.  Such an experimental approach is sourly
missing for $\np$-hard control problems to date.

This paper is the first attempt to tackle $\np$-hard control problems
via an experimental analysis.  Among natural voting systems with
efficient winner determination, the system currently known to display
the broadest resistance ($\np$-hardness) to control is fallback
voting, proposed by Brams and
Sanver~\cite{bra-san:j:preference-approval-voting} via combining
approval with Bucklin voting.  Erd\'{e}lyi et
al.~\cite{erd-rot:c:fallback-voting,erd-pir-rot:c:voter-partition-in-bucklin-and-fallback-voting}
showed that fallback voting is resistant to $20$ out of the $22$
standard types of control and that Bucklin voting behaves almost as
good. 
Shortly after
these results with all proofs were made public in a technical report
dated March 11, 2011,
\cite{erd-fel-pir-rot:t-v1:control-in-bucklin-and-fallback-voting},
Menton~\cite{men:t:normalized-range-voting-broadly-resists-control}
reported analogous results for normalized range voting (the version of
his technical report that establishes a matching number of resistances
is dated April 25, 2011).

We empirically investigate six voter control scenarios for 
Bucklin and fallback voting and two
for plurality voting. Furthermore we investigate twelve candidate 
control scenarios for all three voting
systems\footnote{Note that there are overall eight voter 
control scenarios but we only analyze those where for the corresponding 
control problem no deterministic polynomial-time algorithm is known. 
Furthermore we do not analyze two types 
of control by adding candidates, namely the case where the number of 
candidates that can be added is not limited, so that we investigate 
$18$ of the $22$ known types of electoral control.}, i.e., while Walsh et
al.~\cite{wal:c:empirical-study-of-manipulability-of-STV,wal:c:phase-transition-in-manipulating-veto,dav-kat-nar-wal:c:empirical-study-borda-manipulation,dav-kat-nar-wal:c:complexity-and-algorithms-for-borda}
focused on \emph{constructive} manipulation problems only (where the
aim is to make a candidate win), we study both \emph{constructive} and
\emph{destructive} control problems (the latter aiming at preventing
some candidate's victory).  When generating random elections in our
experiments, we consider two probability distributions: the
\emph{Impartial Culture model} (where votes are distributed uniformly
and are drawn independently) and the \emph{Two Mainstreams model},
introduced here to model two mainstreams in society by adapting the
\emph{Polya Eggenberger urn model}~\cite{ber:j:paradox-urn-model}.

After introducing the investigated voting systems and types of
electoral controls in Section~\ref{sec:prel}, we present the
experimental setting and implemented algorithms in
Sections~\ref{sec:exp} and~\ref{sec:alg}.  Section~\ref{sec:results}
summarizes some of our findings and observations for particular
control scenarios.  We conclude by providing a brief discussion of the
main findings of our experiments in Section~\ref{sec:conclusions},
which allow a more fine-grained analysis and comparison---across
various control scenarios, vote distribution models, and voting
systems---than merely stating $\np$-hardness for all these problems.
A comprehensive presentation of all results can be found in the
appendix.

\section{Preliminaries}
\label{sec:prel}

\subsection{Elections and Voting Systems}
\label{sec:prel-elections}

An \emph{election} is a pair $(C,V)$ consisting of a finite candidate
set $C=\{c_1,c_2,\ldots,c_m\}$ and a finite list of voters
$V=(v_1,v_2,\ldots,v_n)$ expressing their preferences over the
candidates in~$C$.  How the votes are represented depends on the
voting system used.  A \emph{voting system} $\vs$ determines how the
voters' ballots are cast and who has won a given election $(C,V)$,
where the set $W\seq C$ of winners may be empty or have one or more
elements.  We call an election with votes cast according to a voting
system $\vs$ an \emph{$\vs$ election}.
Here we focus on the systems Bucklin voting, fallback voting, and plurality voting.

\emph{Bucklin voting} is a preference-based voting system named after
James W. Bucklin~\cite{hoa-hal:b:proportional-representation}.
``Preference-based'' means that the voters' ballots are (strict)
linear orders over all candidates in~$C$.
For example, if $C=\{c_1,c_2,c_3\}$ and a vote $v$ is given by
$c_2\,c_1\,c_3$, then this voter $v$ strictly prefers $c_2$ to $c_1$
and $c_1$ to~$c_3$.

Let $(C,V)$ be a given Bucklin election.  The \emph{level $i$ score of
  a candidate $c \in C$} ($\scoresublevel{(C,V)}{i}{c}$, for short) is
the number of voters in $V$ ranking $c$ among their top $i$ positions.
Letting the \emph{strict majority threshold} of a list $V$ of votes be
$\maj{V}=\left\lfloor \nicefrac{\|V\|}{2}\right\rfloor +1$, the
\emph{Bucklin score of $c \in C$} is defined to be the smallest $i$
such that $\scoresublevel{(C,V)}{i}{c}\geq \maj{V}$.  Every candidate
with the smallest Bucklin score (say $\ell$) and the highest level
$\ell$ score is a \emph{level $\ell$ Bucklin winner} (\emph{BV
  winner}, for short).  Note that there always exists a Bucklin
winner, level $1$ Bucklin winners are always unique, but on levels
$\ell \geq 2$ there can be more than one BV winner.

\emph{Fallback voting} is a hybrid voting system introduced by Brams
and Sanver~\cite{bra-san:j:preference-approval-voting}.  It combines
Bucklin voting with \emph{approval
  voting}~\cite{bra-fis:b:approval-voting}.  In a fallback election,
each voter determines those candidates he or she approves of and
provides a linear order of the approved candidates.  So for
$C=\{c_1,c_2,c_3\}$ a vote in a fallback election could be of the form
$c_3\,c_1$ meaning that this voter approves of $c_1$ and $c_3$,
strictly preferring $c_3$ to~$c_1$, and disapproves of~$c_2$.

Winners are determined as follows in fallback voting: Given a fallback
election $(C,V)$, the notions of \emph{level $i$ score} of a candidate
$c \in C$ and \emph{level $i$ fallback winner} are defined analogously
as in Bucklin voting.  If there is a level $\ell$ fallback winner with
$\ell \leq \|C\|$, then he or she is the \emph{fallback winner in
  $(C,V)$}.  Otherwise (i.e., if no fallback winner exists in
$(C,V)$), every candidate with a highest \emph{approval score} (which
is the number of voters approving of this candidate) is a
\emph{fallback winner in $(C,V)$}.  The second case can occur in
fallback elections, since the voters can prevent the candidates from
gaining points, and so it is possible that no candidate reaches or
exceeds the strict majority threshold on any level.

Note that Bucklin elections can be seen as fallback elections where
each voter approves of all candidates. So Bucklin voting is a special
case of fallback voting.

In \emph{plurality voting}, the most preferred candidate in each vote
gains one point, and the candidates with the most points are the
\emph{plurality winners}.  Note that there always exists at least one
plurality winner.  This voting rule is preference-based as well, even
though the ranking of the candidates after the top candidate is
irrelevant.

\subsection{Electoral Control and Control Complexity}
\label{sec:prel-control}

\emph{Electoral control} is a way to tamper with the outcome of an
election by changing the structure of the election
itself~\cite{bar-tov-tri:j:control,hem-hem-rot:j:destructive-control}.
These structural changes include adding, deleting, and partitioning
voters or candidates.  In the model of electoral control these changes
are exerted by an external actor, the ``chair,'' having full knowledge
of the voters' preferences. For a detailed discussion of why and where
this assumption is appropriate when investigating control complexity,
see \cite{hem-hem-rot:j:destructive-control}.  Bartholdi et
al.~\cite{bar-tov-tri:j:control} introduced
the notion of \emph{constructive control} where the chair's goal is to
make a distinguished candidate end up winning alone the resulting
election.  The case where the chair's control action aims at
preventing a given candidate from being a unique winner is called
\emph{destructive control} and has been introduced by Hemaspaandra et
al.~\cite{hem-hem-rot:j:destructive-control}.

To study the complexity of control in different scenarios, a decision
problem is defined for each type of electoral control.  

\EP{$\vs$-Constructive Control by Deleting Voters ($\vs$-CCDV)}
{An $\vs$ election $(C,V)$, a distinguished candidate $c\in C$, and a
  positive integer $k\leq \|V\|$.}
{Is there a subset $V'\seq V$ with $\|V'\|\leq k$ such that $c$ is the
  unique $\vs$ winner of election $(C,V-V')$?}

\EP{$\vs$-Constructive Control by Adding Voters ($\vs$-CCAV)}
{An $\vs$ election $(C,V\cup V')$, where $V\cap V'=\emptyset$ and $V$
  is the list of registered voters and $V'$ is the list of
  unregistered voters, a distinguished candidate $c\in C$, and a
  positive integer $k\leq \|V'\|$.}
{Is there a subset $V''\seq V'$ with $\|V''\|\leq k$ such that $c$ is
  the unique $\vs$ winner of election $(C,V\cup V'')$?}

Constructive control by partition of voters is modeled via a two-stage
election, where in the first stage $V$ is partitioned into $V_1$ and
$V_2$ and the winners (subject to the tie-handling rule used, see
below) of subelections $(C,V_1)$ and $(C,V_2)$ run against each other
in the final stage, with respect to~$V$.  Hemaspaandra et
al.~\cite{hem-hem-rot:j:destructive-control} introduced the
tie-handling rules ``Ties Promote'' (TP) in which all subelection
winners participate in the runoff, and ``Ties Eliminate'' (TE) in
which only a unique winner from either subelection can move on to the
runoff (if there is more than one winner, none of them moves on).

\EP{$\vs$-Constructive Control by Partition of Voters ($\vs$-CCPV)}
{An $\vs$ election $(C,V)$ and a distinguished candidate $c\in C$.}
{Is there a partition $(V_1,V_2)$ of $V$ such that $c$ is the unique
  $\vs$ winner of election $(W_1\cup W_2,V)$, where $W_i$, $i\in\{1,2\}$, is
  the set of $\vs$ winners of subelection $(C,V_i)$ surviving the
  tie-handling rule?}

Depending on the tie-handling rule used, we obtain the problems
\textsc{$\vs$-CCPV-TP} and \textsc{$\vs$-CCPV-TE}.  For the
destructive cases, we simply
ask whether it is possible to \emph{prevent} the distinguished
candidate from being a unique winner, yielding the destructive control
problems \textsc{$\vs$-DCDV},
\textsc{$\vs$-DCAV}\textsc{$\vs$-DCPV-TP}, and \textsc{$\vs$-DCPV-TE}.
  Each of the four
problems just defined models ``two-district gerrymandering.''

\EP{$\vs$-Constructive Control by Deleting Candidates ($\vs$-CCDC)}
{An $\vs$ election $(C,V)$ and a distinguished candidate $c\in C$.}
{ Does there exist a subset $C' \seq C$ such that
  $\|C'\| \leq k$ and $c$ is the unique $\vs$ winner of election $(C-C',V)$}

\EP{$\vs$-Constructive Control by Adding Candidates ($\vs$-CCAC)}
{An $\vs$ election $(C \cup D, V)$, $C \cap D = \emptyset$,
  a distinguished candidate $c \in C$, and a nonnegative integer~$k$.
  ($C$ is the set of originally qualified candidates and $D$ is the
  set of spoiler candidates that may be added.)}
{Does there exist a subset $D' \seq D$ such that $\| D' \|
  \leq k$ and $c$ is the unique $\vs$ winner of election $(C \cup D', V)$?}

\EP{$\vs$-Constructive Control by Partition of Candidates ($\vs$-CCPC)}
{An $\vs$ election $(C,V)$ and a distinguished candidate $c\in C$.}
{ Is it possible to partition $C$ into $C_1$ and $C_2$
  such that $c$ is the unique $\vs$ winner of election $(W_1 \cup C_2,V)$, 
where $W_1$ is the set of $\vs$ winners of subelection $(C_1,V)$}

\EP{$\vs$-Constructive Control by Runoff-Partition of Candidates ($\vs$-CCroPC)}
{An $\vs$ election $(C,V)$ and a distinguished candidate $c\in C$.}
{Is it possible to partition $C$ into $C_1$ and $C_2$
  such that $c$ is the unique $\vs$ winner of election $(W_1 \cup W_2,V)$, 
where~$W_i$, $i \in \{1,2\}$,
  is the set of $\vs$ winners of subelection $(C_i,V)$?}

Summing up, we now have defined twelve candidate control problems and thus
a total of $20$ control problems. Note that the classic standard control 
scenarios include a version of control by adding candidates where the number 
of candidates that may be added is not bound by a constant given in the 
instance. With that we have fourteen candidate control problems and $22$
different types of control but we do not consider these two cases in our 
experimental analysis.

Let $\mathfrak{C}$ be a type of electoral control. Using the notions
defined by Bartholdi et al.~\cite{bar-tov-tri:j:control} (see
also~\cite{hem-hem-rot:j:destructive-control}), we say a voting system
$\vs$ is \emph{immune to $\mathfrak{C}$} if the chair never succeeds
in exerting control of type~$\mathfrak{C}$.  If $\vs$ is not immune to
$\mathfrak{C}$, it is \emph{susceptible to~$\mathfrak{C}$}.  If $\vs$
is susceptible to a control type~$\mathfrak{C}$, we say it is
\emph{vulnerable to $\mathfrak{C}$} if the corresponding decision
problem is decidable in deterministic polynomial time, and we say it
is \emph{resistant to $\mathfrak{C}$} if the corresponding decision
problem is $\np$-hard.

\subsection{Control Complexity in Bucklin, Fallback, and Plurality Voting}

Plurality voting is one of the first voting systems for which the
complexity of constructive control~\cite{bar-tov-tri:j:control} and
destructive control~\cite{hem-hem-rot:j:destructive-control} has been
studied in the above scenarios.
Control in fallback voting and Bucklin voting has been
previously studied by Erd\'{e}lyi et
al.~\cite{erd-rot:c:fallback-voting,erd-pir-rot:c:voter-partition-in-bucklin-and-fallback-voting}
with respect to classical complexity and also with respect to
parameterized complexity~\cite{erd-fel:c:fallback-voting}.  In terms of
$\np$-hardness, among natural systems with polynomial-time winner
determination fallback voting has the most resistances to control
(namely, $20$ out of the $22$ standard control types) and Bucklin
voting behaves similarly well---just one case is open
(Bucklin-DCPV-TP).  Table~\ref{tab:res-fv-bv-pv} gives an overview of
known complexity results for electoral control in these three systems.

\begin{table}
\small
\centering
\begin{tabular}{@{}l c ccc c ccc c ccc@{}}
\toprule
 & & \multicolumn{3}{c}{\text{Fallback Voting}}
 & & \multicolumn{3}{c}{\text{Bucklin Voting}}
 & & \multicolumn{3}{c}{\text{Plurality Voting}}\\
\cmidrule(r){3-5}\cmidrule(l){3-5}\cmidrule(r){7-9}\cmidrule(l){7-9}\cmidrule(r){11-13}\cmidrule(l){11-13}
\text{Control by} & & \text{Constr.} 	& & \text{Destr.}
		  & & \text{Constr.} 	& & \text{Destr.}
                  & & \text{Constr.} 	& & \text{Destr.} \\
\cmidrule(r){1-1}\cmidrule(l){1-1}\cmidrule(r){3-3}\cmidrule(l){3-3}\cmidrule(r){5-5}\cmidrule(l){5-5}\cmidrule(r){7-7}\cmidrule(l){7-7}\cmidrule(r){9-9}\cmidrule(l){9-9}\cmidrule(r){11-11}\cmidrule(l){11-11}\cmidrule(r){13-13}\cmidrule(l){13-13}
\text{Adding Candidates}       & ~~~ &     R & ~~ &     R & ~~~ &     R & ~~ &     R & ~~~ &     R & ~~ &     R\\
\text{Deleting Candidates}     & &     R & &     R & &     R & &     R & &    R & &     R\\
\multirow{2}{*}{\text{Partition of Candidates}}
                           & & TE: R & & TE: R & & TE: R & & TE: R & & TE: R & & TE: R \\
                           & & TP: R & & TP: R & & TP: R & & TP: R & & TP: R & & TP: R\\
\multirow{2}{*}{\text{Runoff-Partition of Candidates}}
                           & & TE: R & & TE: R & & TE: R & & TE: R & & TE: R & & TE: R \\
                           & & TP: R & & TP: R & & TP: R & & TP: R & & TP: R & & TP: R\\
\text{Adding Voters}       & ~~~ &     R & ~~ &     V & ~~~ &     R & ~~ &     V & ~~~ &     V & ~~ &     V\\
\text{Deleting Voters}     & &     R & &     V & &     R & &     V & &    V & &     V\\
\multirow{2}{*}{\text{Partition of Voters}}
                           & & TE: R & & TE: R & & TE: R & & TE: R & & TE: V & & TE: V \\
                           & & TP: R & & TP: R & & TP: R & & TP: S & & TP: R & & TP: R\\
\bottomrule
\end{tabular}
\caption{\label{tab:res-fv-bv-pv} 
  Known results on the control complexity of fallback, Bucklin,
  and plurality voting.
  Key: R means resistance, V means vulnerability,
  and S means susceptibility.
}
\end{table}

\section{Experimental Setting}
\label{sec:exp}

In this section we describe the experimental setting.
As stated in Section~\ref{sec:prel-control}, the instances of control
by adding and deleting both candidates and voters contain a parameter
$k$ bounding the number of candidates/voters that can be deleted or
added.  In our experiments, we confine ourselves to the case of
$k=\left\lfloor \nicefrac{n}{3}\right\rfloor$, where $n$ is the number
of voters.  Since every yes-instance for a given $k$ is also a
yes-instance for $k'\geq k$, the number of yes-instances found in our
experiments are a lower bound for the number of yes-instances when
more voters can be deleted or added.

We randomly generated elections $(C,V)$ with $m=\|C\|$ and $n=\|V\|$
for all combinations of $n,m\in\{4,8,16,32,64,128\}$.  Each
combination of $n$ and $m$ is one data point for which we evaluated
$500$ of these elections, trying to determine for each given election
whether or not control is possible.  How the elections have been
generated and how the algorithms are designed will be described below.

Before we specify the different distribution models underlying our
election generation, we explain how random votes can be cast in
Bucklin and fallback voting and how many different votes exist in both
voting systems.

Assuming that the generated election has $m$ candidates, in Bucklin
voting a random vote can be obtained by generating a random
permutation over the $m$ different candidates.  Clearly, the overall
number of different votes in Bucklin elections is~$m!$.
 
In fallback voting random votes can be generated as follows: In a
first step, a preference (i.e., linear order) $p$ over all $m$
candidates is drawn randomly under a certain distribution (see below)
from all $m!$ possible preferences.  In a second step, the number of
approved candidates, say $\ell$, is drawn from the possible numbers
$\{0,1,2,\ldots,m\}$.  The preference $p$ and $\ell$ are drawn
independently.  Then, the generated vote consists of the first $\ell$
candidates in~$p$.  Generalizing this, we know that there can be
$\sum_{\ell=0}^{m}\binom{m}{\ell}l!$ different votes in fallback
elections with $m$ candidates.  With this in mind, we now specify the
two distribution models we will be working with.

In the \emph{Impartial Culture model} (\emph{IC model}) we assume
uniformly distributed votes and draw each vote independently out of
all possible preferences.

In the second model, which we call the \emph{Two Mainstreams model}
(\emph{TM model}), we adapt the \emph{Polya Eggenberger urn model}
(\emph{PE model}, see \cite{ber:j:paradox-urn-model}) that has been
used by Walsh~\cite{wal:c:empirical-study-of-manipulability-of-STV} in
the following way: We draw two votes out of an urn containing all
possible, say $t$, votes (with $t=m!$ or
$t=\sum_{\ell=0}^{m}\binom{m}{\ell}\ell!$, depending on the voting
system).  Each of these votes can be interpreted as a representative
of one ``main stream'' in society (e.g., liberal and conservative).
Then each vote is put back into the urn with $k$ additional votes of
the same form.  Out of this urn the votes for the election are drawn
randomly with replacement.  So we have that each voter's preference is
with probability $\nicefrac{1}{3}$ from the first mainstream, with
probability $\nicefrac{1}{3}$ from the second mainstream, and with
probability $\nicefrac{1}{3}$ it is a different preference.
The main difference to the above-mentioned Polya Eggenberger urn model
is that the voters do not influence each other. We do have correlated
votes in the sense that with a certain probability voters vote like
other voters but there are no direct dependencies between the voters,
whereas in the PE model the preference of the first voter influences
the preference of the second voter, who in turn influences the
preference of the third voter, and so on.  Of course, an investigation
of control in elections generated in this model could be interesting
as well but we postpone this to future work.

Note that for control by adding voters, a second list of votes has to
be generated, namely the ballots of the unregistered voters that may
be added. In our setting, the list of unregistered voters is of the
same size as the list of registered voters and both lists are
generated with the same underlying distribution model.

\section{A High-Level Description of the Algorithms}
\label{sec:alg}

All algorithms for the different types of control share the same
essential method of testing various subsets, and they differ only in
the type of preordering and internal testing.

Before actually searching for a successful sublist of voters or subset
of candidates, the algorithms check conditions that would indicate
that the given instance is a no-instance.  If the tested conditions do
not hold the candidates or voters are preordered to ensure that the
algorithm tries promising subsets or sublists first. Depending on the
control type at hand some of the following conditions are tested.

\begin{desctight}
\item[Condition 1:] The distinguished candidate is positioned on the
  last place in every vote, or is positioned on the last place or is
  disapproved by every voter if $(C,V)$ is a fallback election.
\item[Condition 2:] For each $k'\leq k$, determine the smallest $i$
  and $j$ such that
\begin{eqnarray*}
\scoresublevel{(C,V)}{i}{c'}
\geq \left\lfloor \nicefrac{(\|V\|-k')}{2}\right\rfloor+1+k' 
& \mbox{ and } &
\scoresublevel{(C,V)}{j}{c}
\geq \left\lfloor \nicefrac{(\|V\|-k')}{2}\right\rfloor+1
\label{eq:cond2}
\end{eqnarray*}
hold for $c'\in C-\{c\}$.  It holds that $i\leq j-1$ for all $k'\leq k$.
 
 \item[Condition 3:] For each $k'\leq k$ determine the smallest $i$
   and $j$ such that
\begin{eqnarray*}
\scoresublevel{(C,V)}{i}{c'}
  \geq  \left\lfloor (\|V\|+k')/2\right\rfloor+1 & \mbox{ and} &
\scoresublevel{(C,V)}{j}{c}
  \geq  \left\lfloor (\|V\|+k')/2\right\rfloor+1-k'
\end{eqnarray*}
hold for $c'\in C-\{c\}$. It holds that $i\leq j-1$ for all $k'\leq
k$.
\item[Condition 4:] In the given election, the winner has a strict
  majority on the first level.

\end{desctight}

Condition~1 is tested for every constructive control type except the
partition of candidates cases (with and without runoff).

In the following sections we will describe the different algorithms
for the different types of electoral controls and the implemented
preorderings.

\subsection{Algorithms for Voter Control}

We begin with the algorihtms for the voter control cases. For both
constructive control by adding and deleting voters, Condition~1 is
tested.  Note that for the adding voter cases this condition has to
hold for both voter lists, the registered voters and the unregistered
voters.

For constructive control by deleting voters, Condition~2 is
additionally tested.  If Condition~2 holds, $c$ cannot be made a
unique winner by deleting at most $k$ voters because even if all $k$
voters would harm the strongest rival $c'$ of $c$ the most and $c$ not
at all, the rival would still reach a strict majority on a lower level
than~$c$.

For constructive control by adding voters, Condition~1 and 3 are tested. 
If Condition~3 holds for the given election and the given
distinguished candidate $c$, then even if all added voters helped only
$c$ on the lowest level, there would still be at least one other
candidate reaching a strict majority on a level lower than $c$.

For the voter-partition cases, we have Condition~4 indicating that
control is not possible for both the constructive and destructive
case, namely that in the given election there is a unique winner on
the first level. It is easy to see that for every possible partition
$(V_1,V_2)$ of $V$ a level~$1$ winner is also a level~$1$ winner in at
least one of the subelections. Since level~$1$ winners are always
unique, independent of the tie-handling model, this candidate always
participates in the runoff and will therefore always be the unique
level~$1$ winner of the resulting two-stage election. So no
distinguished candidate can ever be made the unique winner by
partitioning the voters.
So the algorithms for destructive and constructive control by
partition of voters first check Condition~4 where the latter checks
Condition~1 as well.

After having excluded these trivial cases, the different algorithms search for
a successful sublist of~$V$ after having ordered the voters. 

\begin{figure}[h]
\centering
\includegraphics[scale=0.3]{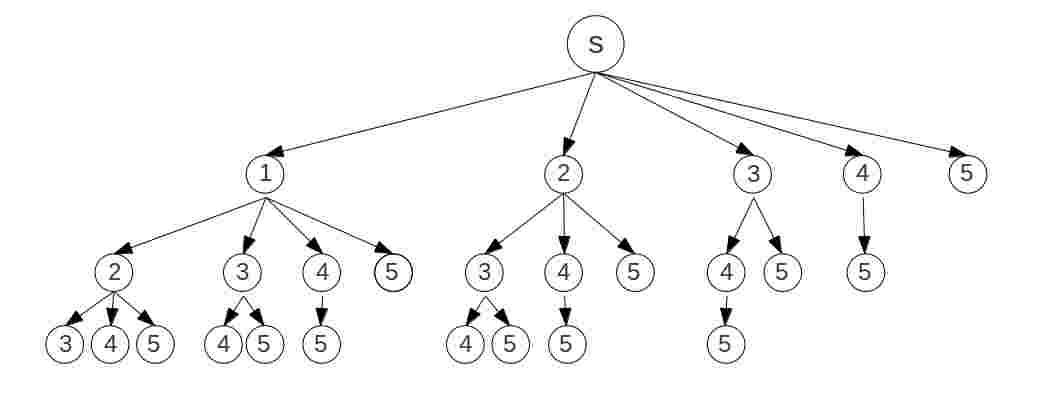}
\caption{Tree for $n=5$ voters where up to $k=3$ voters may be
  deleted.  A node $i$ corresponds to voter $v_i$ after the
  preordering.}
\label{fig:del-tree}
\end{figure}

We will describe this procedure for constructive control by deleting voters
where the voters are ordered ascending for~$c$.  
That is, after the preordering $v_1$ is a
voter positioning $c$ worst and $v_n$ is a voter positioning $c$ best
among all voters.  In fallback voting, the ``worst position'' for a
candidate is to be not approved at all. The algorithm now
starts with deleting those votes $c$ benefits least of.  It follows
the procedure of a depth-first search on a tree of height $k$ that is
structured as shown in Figure~\ref{fig:del-tree}.

In each node, it is tested whether deleting the voters on the path is
a successful control action. So on the path $s\rightarrow 1
\rightarrow 2 \rightarrow 3$ the algorithm tests the sublists $(v_1),
(v_1,v_2),(v_1,v_2,v_3)$ and then tracks back testing the sublists
$(v_1,v_2,v_4),(v_1,v_2,v_5),(v_1,v_3),(v_1,v_3,v_4)$, and so on.  The
branches on the left side are visited first and due to the preordering
of the votes, these are the votes $c$ benefits least of.
 
For the adding-voters case, the voters in the list of unregistered
voters are ordered in a descending order for the distinguished
candidate and the algorithm proceeds similar to the algorithm for the
deleting-voters case, trying to find a successful sublist for the
control.  With this preordering, the algorithm first tests those
voters the distinguished candidate can benefit most from when these are
added to the voter list.

For the partition cases the algorithm considers every possible sublist
of the voter list up to size $k=\left\lfloor \nicefrac{n}{2}
\right\rfloor$ as $V_1$, sets $V_2=V-V_1$, and tests whether this is a
successful control action or not. For the constructive cases the
voters are preordered descendingly with respect to the distinguished
candidate whereas for the destructive control cases no preordering is
implemented.

\subsection{Algorithms for Candidate Control}

The algorithms for the candidate control scenarios test Condition~1 in
the constructive cases except where the candidates are partitioned.
For the destructive cases, on the other hand, Condition~4 is always
tested. Note that for the adding candidates cases both conditions must
hold in the election over both the registered and the unregistered
candidates.

After testing for trivial instances the algorithms make use of the
same approach of testing systematically preordered candidate subsets
to find a successful control action. Here, the candidates are also
ordered with respect to the distinguished candidate, where a
descending order means that the first candidate has the most voters
positioning him or her before the distinguished candidate and the last
candidate has the fewest voters doing that. An ascending order is
defined analogously.  Again, in the adding candidates case, the votes
over all candidates (including the unregistered) are considered.

The descending ordering is used for finding control actions for the
constructive case of deleting candidates and the destructive cases of
adding and partitioning candidates with and without runoff.  In the
costructive case of the deleting candidates scenario we want to make
the distinguished candidate the winner, so the algorithm tries to
delete those candidates whose deletion moves the distinguished
candidate forward in as many votes as possible.  On the contrary the
algorithm for destructive control by adding candidates tries to
prevent the distinguished candidate from winning, so candidates are
added that move the distinguished candidate back in as many votes as
possible.  The algorithm for both destructive cases of partition of
candidates assignes the distinguished candidate to the subset $C_1$
and tries to prevent him or her from moving to the final election. So
the other participants in $C_1$ are chosen from the remaining
candidates after having ordered them descending with respect to the
distinguished candidate.

With analog arguments it is obvious that an ascending order of the
candidates with respect to the distinguished candidate is used for the
destructive case of deleting candidates, the destructive case of
adding candidates, and the constructive case of partition of
candidates.  Note that the algorithm for constructive control by
partition of candidates positions the distinguished candidate in $C_2$
and assigns those other candidates to $C_2$ that are positioned behind
the distinguished candidate in as many votes as possible since those
are direct rivals for the distinguished candidate in the final
election.\footnote{Recall that the candidates in $C_2$ participate
  directly in the final round whereas those candidates in $C_1$ have
  to compete against each other in a pre-round election.}

\subsection*{}

 Obviously,
in the worst-case, the algorithms check all possible subsets of size
at most $k$, so they have a worst-case running time of
$\sum_{\ell=1}^{k}\binom{n}{\ell}$.  
To handle the worst-case scenarios, a time limit of ten minutes has been implemented
such that the algorithms stop when exceeding this limit, indicating by
the output that the search process is aborted unsuccessfully.  Setting
the time limit higher can only increase the number of yes-instances,
so again, the results obtained in our experiments give a lower bound
for the number of yes-instances in the generated elections.
In our experiments we implemented the same timeout value for all 
investigated types of control. As our results in Section~\ref{sec:results} 
will show, the different control types react differently to this 
constant timeout threshold, 
so tuning of the timeout-parameter would be an interesting issue for
further experiments. Also, varying the timeout value with respect to the 
election size at hand might be an interesting approach. 

The algorithms and data-generation programs are implemented in  
\emph{Octave~3.2} and the experiments were run on a 2,67 GHz
 Core-I5 750 with 8GB RAM. 

\section{Summary of Experimental Results}
\label{sec:results}

Tables~\ref{tab:overview-exp-res-fv-bv} and~\ref{tab:overview-exp-res-pv}
summarize our experimental results
on control in Bucklin, fallback, and plurality voting.  We investigated 
the three
voting systems only for those control types they are not known to be
vulnerable to, which is indicated by an R- or an S-entry in
Table~\ref{tab:res-fv-bv-pv}. That is, destructive control by
adding and by deleting voters (\text{DCAV} and \text{DCDV}) are
omitted in Tables~\ref{tab:overview-exp-res-fv-bv} and~\ref{tab:overview-exp-res-pv}. 
Also, since our
algorithms use the parameter $k$ bounding the number of candidates to
be added, constructive and destructive control by adding an unlimited
number of candidates (\text{CCAUC} and \text{DCAUC}) are not
considered either.  For each combination of any of the remaining $18$
control types, any of the two voting systems Bucklin and fallback voting,
and any of the two distribution models (IC and TM), we tested a total of
$18,000 = 36 \cdot 500$ elections, varying over the $36$ data points with
different values for $m$ and~$n$, as explained above.  This gives a
total of $1,296,000 = 18 \cdot 4 \cdot 18,000$ generated and tested
elections.
For plurality voting we investigated fourteen types of electoral control 
leading to $504,000 = 14 \cdot 2 \cdot 18,0000 $ generated and tested elections.

\begin{table}
 \centering
\small
\begin{tabular}{@{}l c rrcrrcrr c rrcrrcrr @{}}
\toprule
& & \multicolumn{8}{c}{\text{Fallback Voting}} 	
& &  \multicolumn{8}{c}{\text{Bucklin Voting}}  \\
\cmidrule(rl){3-10}\cmidrule(l){12-19}
&& \multicolumn{2}{c}{\text{min}} && \multicolumn{2}{c}{\text{max}} && 
\multicolumn{2}{c}{\text{to}}
&& \multicolumn{2}{c}{\text{min}} && \multicolumn{2}{c}{\text{max}} && 
\multicolumn{2}{c}{\text{to}}\\
\cmidrule(rl){3-4}\cmidrule(rl){6-7}\cmidrule(rl){9-10}\cmidrule(rl){12-13}\cmidrule(rl){15-16}\cmidrule(l){18-19}
&& \text{IC} & \text{TM} && \text{IC} & \text{TM} && \text{IC} & \text{TM}
&& \text{IC} & \text{TM} && \text{IC} & \text{TM} && \text{IC} & \text{TM} \\
\cmidrule(rl){3-3}\cmidrule(rl){4-4}\cmidrule(rl){6-6}\cmidrule(rl){7-7}\cmidrule(rl){9-9}\cmidrule(rl){10-10}\cmidrule(rl){12-12}\cmidrule(rl){13-13}\cmidrule(rl){15-15}\cmidrule(rl){16-16}\cmidrule(rl){18-18}\cmidrule(l){19-19}
\textsc{CCAC} && 1 & \text{0} && 11 & \text{7} && 51 & \text{50} &
	& 0 & \text{0} && 23 & \text{11} && 50 & \text{49}  \\
\textsc{DCAC} && 53 & \text{39} && 92 & \text{71} && 11 & \text{14} &
	& 71 & \text{42} && 99 & \text{77} && 6 & \text{12} \\
\textsc{CCDC} && 13 & \text{15} && 33 & \text{36} && 37 & \text{37} &
	& 13 & \text{17} && 58 & \text{45} && 34 & \text{37} \\
\textsc{DCDC} && 8 & \text{12} && 78& \text{63} && 15 & \text{22} &
	& 48 & \text{25} && 99 & \text{77} && 7 & \text{18} \\
\textsc{CCPC-TE} && 0 & \text{0} && 19 & \text{18} && 62 & \text{64} &
	& 1 & \text{0} && 57 & \text{37} && 57 & \text{62} \\
\bottomrule 
\end{tabular}
\captionsetup{font=small}
\caption{\label{tab:overview-exp-res-fv-bv}
Overview of experimental results on control in Bucklin and fallback voting.
Key: The ``min'' and ``max'' columns give the minimal and maximal
percentage of yes-instances observed in all tested instances for the given control type, 
including those elections where timeouts occured;
``to'' gives the percentage of timeouts that occurred for the total of
18,000 elections tested in this control case.
}
\end{table}

\begin{table}
 \centering
\small
\begin{tabular}{@{}l c rrcrrcrr @{}}
\toprule
& & \multicolumn{8}{c}{\text{Plurality Voting}} 	
  \\
\cmidrule(l){3-10}
&& \multicolumn{2}{c}{\text{min}} && \multicolumn{2}{c}{\text{max}} && 
\multicolumn{2}{c}{\text{to}}
\\
\cmidrule(rl){3-4}\cmidrule(rl){6-7}\cmidrule(l){9-10}
&& \text{IC} & \text{TM} && \text{IC} & \text{TM} && \text{IC} & \text{TM}
 \\
\cmidrule(rl){3-3}\cmidrule(rl){4-4}\cmidrule(rl){6-6}\cmidrule(rl){7-7}\cmidrule(rl){9-9}\cmidrule(l){10-10} 
\textsc{CCAC} && 0 & \text{0} && 20 & \text{3} && 50 & \text{34} 
	 \\
\textsc{DCAC} && 70 & \text{47} && 99 & \text{60} && 7 & \text{25} 
	 \\
\textsc{CCDC} && 5 & \text{22} && 66 & \text{40} && 37 & \text{35} 
 \\
\textsc{DCDC} && 8 & \text{12} && 78& \text{63} && 15 & \text{22}  
\\
\textsc{CCPC-TE} && 0 & \text{0} && 60 & \text{21} && 58 & \text{65}  
\\
\textsc{DCPC-TE} && 1 & \text{2} && 100 & \text{59} && 22 & \text{41} 
 \\
\textsc{CCPC-TP} &&0 & \text{0} && 64 & \text{24} && 58 & \text{65}  
 \\
\textsc{DCPC-TP} && 1 & \text{3} && 100 & \text{59} && 22 & \text{44}  
 \\
\textsc{CCRPC-TE} && 0 & \text{0} && 65 & \text{19} && 50 & \text{63} 
 \\
\textsc{DCRPC-TE} && 25 & \text{14} && 100 & \text{61} && 12 & \text{37} 
 \\
\textsc{CCRPC-TP} && 0 & \text{0} && 65 & \text{21} && 50 & \text{61}  
 \\
\textsc{DCRPC-TP} && 23 & \text{14} && 100 & \text{61} && 13 & \text{35} 
  \\
\textsc{CCPV-TP} && 0 & \text{0} && 54 & \text{13} && 24 & \text{23}  
 \\
\textsc{DCPV-TP} && 55 & \text{15} && 100 & \text{59} && 4 & \text{35} 
  \\
\bottomrule 
\end{tabular}
\captionsetup{font=small}
\caption{\label{tab:overview-exp-res-pv}
Overview of experimental results on control in plurality voting.
Key: The ``min'' and ``max'' columns give the minimal and maximal
percentage of yes-instances observed in all tested instances for the given control type, 
including those elections where timeouts occurred;
``to'' gives the percentage of timeouts that occurred for the total of
18,000 elections tested in this control case.
}
\end{table}

Tables~\ref{tab:overview-exp-res-fv-bv} and~\ref{tab:overview-exp-res-pv}
give an overview of the percentage
of timeouts for each such combination of control type/voting
system/distribution model, and also the minimal and maximal
percentage of yes-instances observed.  We do not discuss the results
for all these cases in detail here.  Rather, we will focus on
adding/deleting candidates/voters and partition of voters to very
briefly discuss some observations from our experiments, to exemplify
some of the numbers in Tables~\ref{tab:overview-exp-res-fv-bv} and~\ref{tab:overview-exp-res-pv}. 
For those cases that we discuss in detail, we provide plots giving the 
percentage of yes-instances, timeouts, and average computational costs 
for all different election sizes that were tested.
Note that
a comprehensive presentation of all results 
containing the above information 
for all cases can be found in the appendix.

\subsubsection*{Constructive Control by Deleting and by Adding Voters:}
We here briefly discuss some results for control by deleting voters
only, since those for control by adding voters are very similar, in
both Bucklin and fallback voting. 

\begin{figure}[ht]
\centering
\subfloat[Percentage of yes-instances.]{\label{fig:ccdv-bv-d0-m-res}
\includegraphics[scale=0.27]{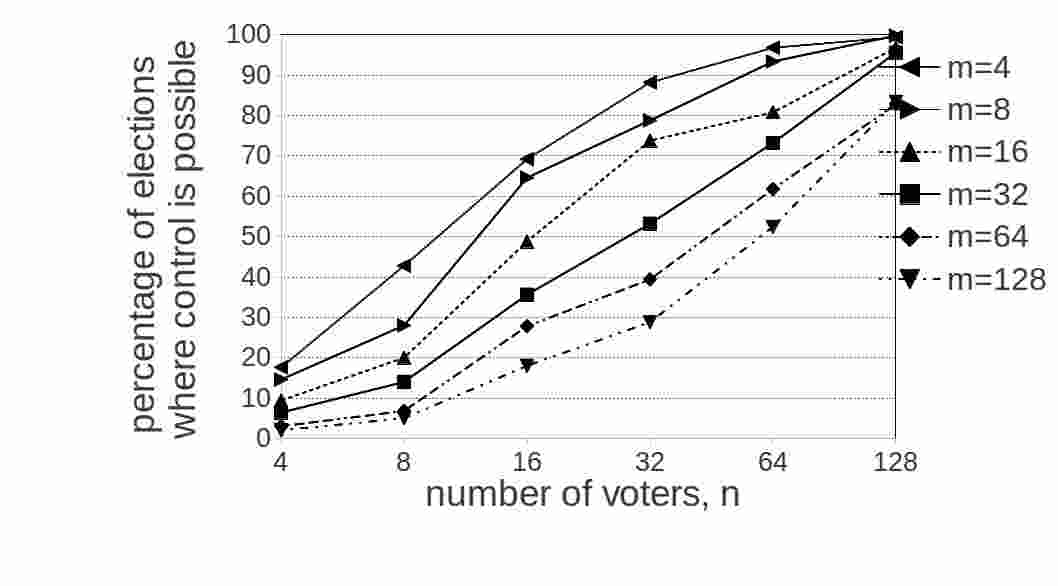}
}
\subfloat[Percentage of timeouts.]{
\label{fig:ccdv-bv-d0-m-tos}
\scriptsize
\raisebox{3.0cm}{
\begin{tabular}{@{}l cccccc@{}}
 \toprule
 \textbf{m\textbackslash n} &  $4$ & $8$ & $16$ & $32$ & $64$ & $128$\\
 \cmidrule(r){1-1} \cmidrule(rl){2-2} \cmidrule(rl){3-3} 
 \cmidrule(rl){4-4} \cmidrule(rl){5-5} \cmidrule(rl){6-6} 
 \cmidrule(l){7-7} 
  $4$   & $0$ 	& $0$ 	& $0$ 	& $10$ 	& $3$ 	& $1$  \\
  $8$   & $0$ 	& $0$ 	& $0$ 	& $20$ 	& $7$ 	& $0$  \\
  $16$  & $0$ 	& $0$ 	& $0$ 	& $23$ 	& $19$ 	& $13$  \\
  $32$  & $0$ 	& $0$ 	& $0$ 	& $39$	& $27$ 	& $4$  \\
  $64$  & $0$ 	& $0$ 	& $1$ 	& $49$	& $38$ 	& $17$  \\
  $128$ & $0$ 	& $0$ 	& $31$ 	& $53$	& $47$	& $17$  \\
 \bottomrule
 \end{tabular}}
 }
 
  \subfloat[Average time the algorithm needs to give a definite output,
 instances where timeouts occur are excluded.]{\label{fig:ccdv-bv-d0-m-comp-cost}
\includegraphics[scale=0.27]{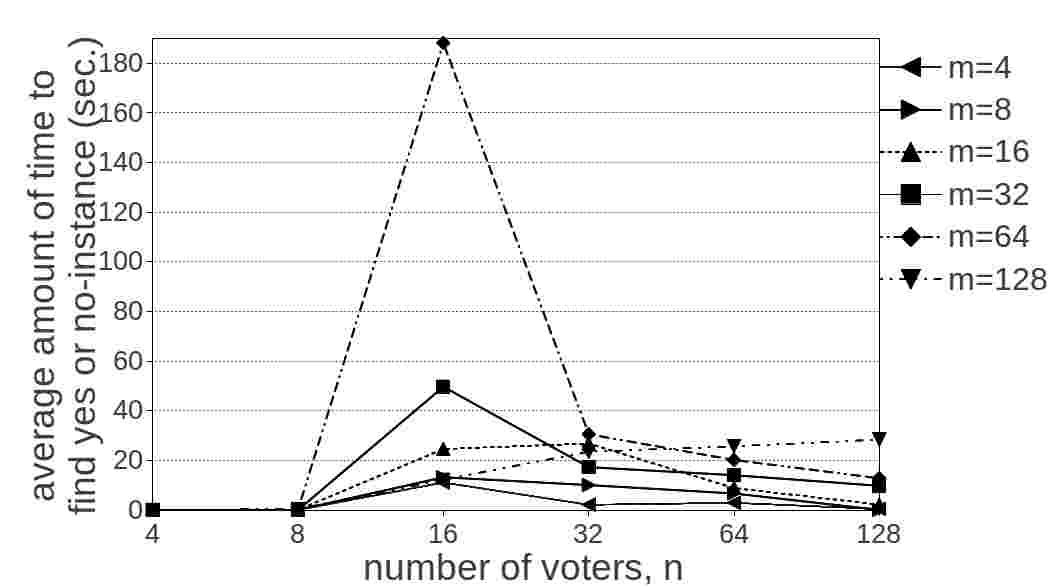}
}
\caption{Bucklin voting in the IC model for
\textsc{CCDV}.
}
\label{fig:ccdv-bv-d0-m}
\end{figure}

In the IC model, increasing the
number of candidates decreases the number of yes-instances in the
generated 
Bucklin
elections
as can be seen in Figure~\ref{fig:ccdv-bv-d0-m} showing the results 
for control by deleting voters for 
Bucklin voting  in the IC model.  On the other hand, the number of
yes-instances increases with the number of voters growing.  In the Two
Mainstreams model, the same correlations can be observed but here,
again, the total number and percentage of yes-instances is smaller
than in the IC model.

Fallback voting behaves very similarly, so for both distributions and
both voting systems increasing the number of candidates makes
successful actions of control by deleting voters less likely.

In both voting systems and in both distribution models, timeouts occur
whenever the number of voters exceeds~$32$. If the number of candidates
is~$128$, we have timeouts already with $16$ voters.
This can also be seen in the development of the computational costs shown 
in Figure~\ref{fig:ccdv-bv-d0-m-comp-cost} after the peak for $n=16$. 
For bigger electorates the average computational costs drop since the number 
of timeouts increases with the number of no-instances diminishing. 
\subsubsection*{Control By Partition of Voters:}

As
mentioned in Section~\ref{sec:prel}, control by 
partition of voters comes in four problem variants, where each case
must be investigated separately.

For constructive control by partition of voters in model TP we made the following 
observations:
Similarly to 
control by deleting or by adding voters,
 the number of 
controllable elections increases with the number of voters increasing.
This was observed for all three voting systems investigated.
We have seen that in at most $13\%$ 
of the tested plurality elections in 
the TM model a successful 
control action can be found.  Note that no timeouts occur for up to $32$
candidates, so more than $87\%$ of the elections tested are
demonstrably not controllable in these cases. 
For both distribution models, plurality elections produce fewer 
timeouts than the corresponding fallback or Bucklin elections.  This
suggests that the control problem for plurality voting is
easier to solve on average than for fallback or Bucklin voting.
Using the 
tie-handling model TE instead of TP, in both Bucklin and fallback 
voting an increase of yes-instances in the constructive cases is evident. 
By contrast, in the destructive counterparts no significant difference
can be observed with respect to the tie-handling rule used.

\begin{figure}[ht]
\centering
\subfloat[Percentage of yes-instances.]{\label{fig:dpvtp-fv-d0-m-res}
\includegraphics[scale=0.27]{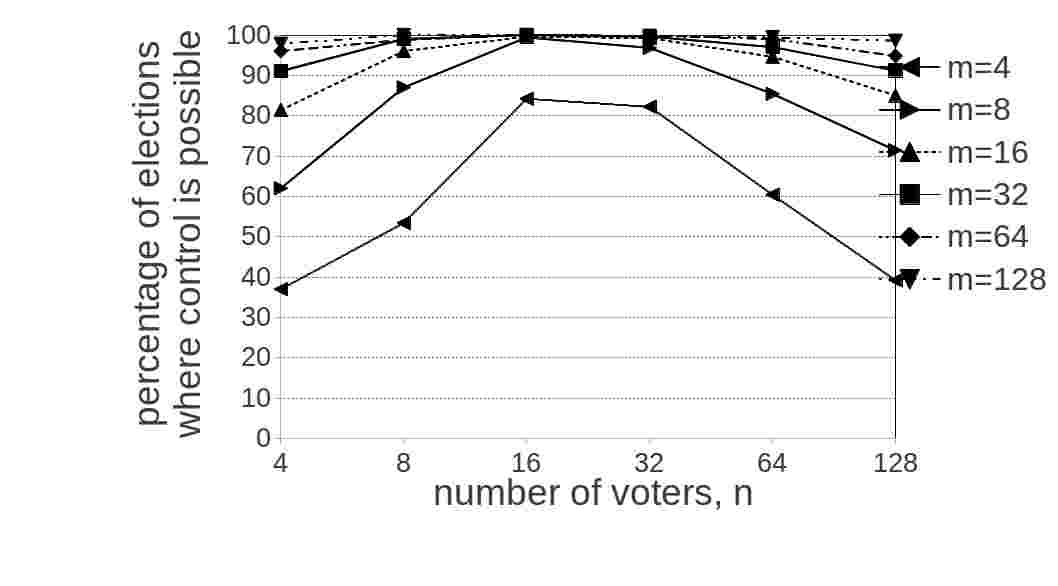}
}
\subfloat[Percentage of timeouts.]{
\label{fig:dpvtp-fv-d0-m-tos}
\scriptsize
\raisebox{3.0cm}{
\begin{tabular}{@{}l cccccc@{}}
 \toprule
 \textbf{m\textbackslash n} &  $4$ & $8$ & $16$ & $32$ & $64$ & $128$\\
 \cmidrule(r){1-1} \cmidrule(rl){2-2} \cmidrule(rl){3-3} 
 \cmidrule(rl){4-4} \cmidrule(rl){5-5} \cmidrule(rl){6-6} 
 \cmidrule(l){7-7} 
  $4$   & $0$ 	& $0$ 	& $0$ 	& $18$ 	& $40$ 	& $61$  \\
  $8$   & $0$ 	& $0$ 	& $0$ 	& $3$ 	& $15$ 	& $29$  \\
  $16$  & $0$ 	& $0$ 	& $0$ 	& $1$ 	& $5$ 	& $15$  \\
  $32$  & $0$ 	& $0$ 	& $0$ 	& $0$	& $3$ 	& $9$  \\
  $64$  & $0$ 	& $0$ 	& $0$ 	& $0$	& $1$ 	& $5$  \\
  $128$ & $0$ 	& $0$ 	& $0$ 	& $0$	& $0$	& $1$  \\
 \bottomrule
 \end{tabular}}
 }
 
  \subfloat[Average time the algorithm needs to give a definite output,
 instances where timeouts occur are excluded.]{\label{fig:dcpv-tp-fv-d0-m-comp-cost}
\includegraphics[scale=0.27]{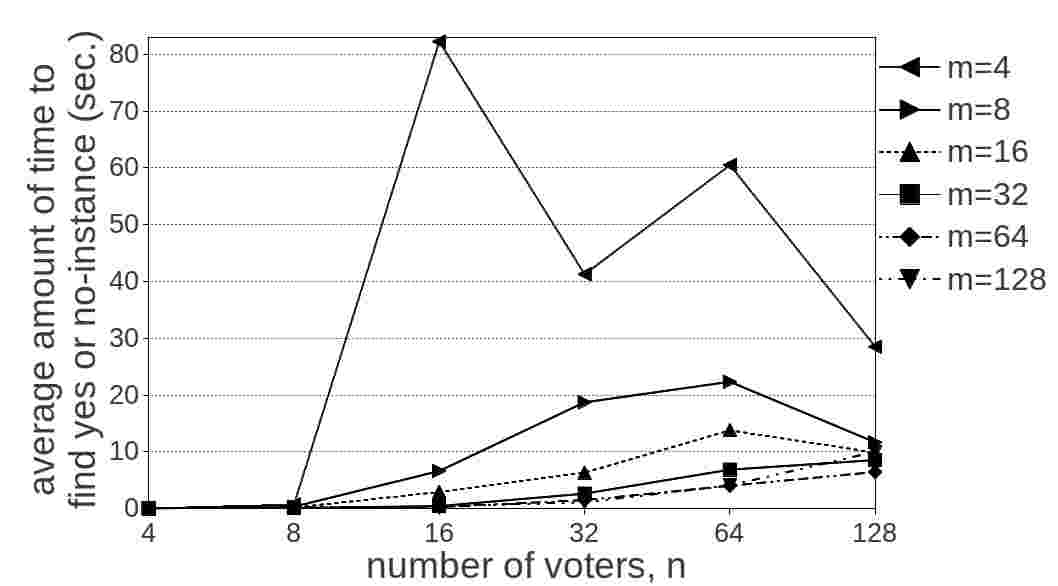}
}
\caption{Fallback voting in the IC model for
\textsc{DCPV-TP}.
}
\label{fig:dpvtp-fv-d0-m}
\end{figure}

The most striking results are those obtained for the destructive
cases.  Here we have that, for all three voting systems (and both
tie-handling models for fallback and Bucklin voting) in the TM model,
 the average number of
controllable elections is very high; and in the IC model, control is
almost always possible, see Figure~\ref{fig:dpvtp-fv-d0-m}. 

 In light of the fact that for these cases the
resistance proofs of Erd\'{e}lyi et
al.~\cite{erd-rot:c:fallback-voting,erd-pir-rot:c:voter-partition-in-bucklin-and-fallback-voting}
for fallback and Bucklin voting tend to be the most involved ones 
(yielding the most complex instances
for showing $\np$-hardness), these results might be surprising at
first glance.  However, one explanation
for the observed results can be found in exactly this fact: The
elections constructed in these reductions have a very complex structure
which seems to be unlikely to occur in randomly generated elections
(at least in elections generated under the distribution models
discussed in this paper). 
Another explanation is that the problems used to reduce from
in~\cite{erd-rot:c:fallback-voting,erd-pir-rot:c:voter-partition-in-bucklin-and-fallback-voting}
tend to be easy to solve for small input sizes, but
due to the complexity of the
reduction, the resulting elections have many voters/candidates
compared to the elections generated for the conducted experiments.

In the destructive cases, the number of timeouts is for all three voting
systems
the lowest among all control types investigated. In Bucklin elections
with uniformly distributed votes and for destructive control by
partition of voters in model TP, for only $3.32\%$ of the elections no
decision can be made within the time limit. 
As can be seen in the table, timeouts begin to occur for those elections 
where the number of voters exceeds~$16$.  But, again, we have to emphasize that 
these values are very low compared to other types of control. 
This explains the plateaus all graphs show.  On the one hand,
increasing the number of voters 
increases the number of yes-instances.  But on the other 
hand, for more than $16$ voters timeouts begin to diminish the 
fraction of observed yes-instances.
Also, the average running time of the algorithm for those instances 
where the time limit is not exceeded is rather low, compared to other types of 
control, see Figure~\ref{fig:dcpv-tp-fv-d0-m-comp-cost}.
The highest computational costs occur for those election sizes 
where the most no-instances were observed.
As expected, in the
corresponding constructive cases the number of timeouts is
significantly higher and so are the average computational costs.

\subsubsection*{Control By Adding Candidates:}

\begin{figure}[ht]
\centering
\subfloat[Percentage of yes-instances.]{\label{fig:ccac-fv-d21-n-res}
\includegraphics[scale=0.27]{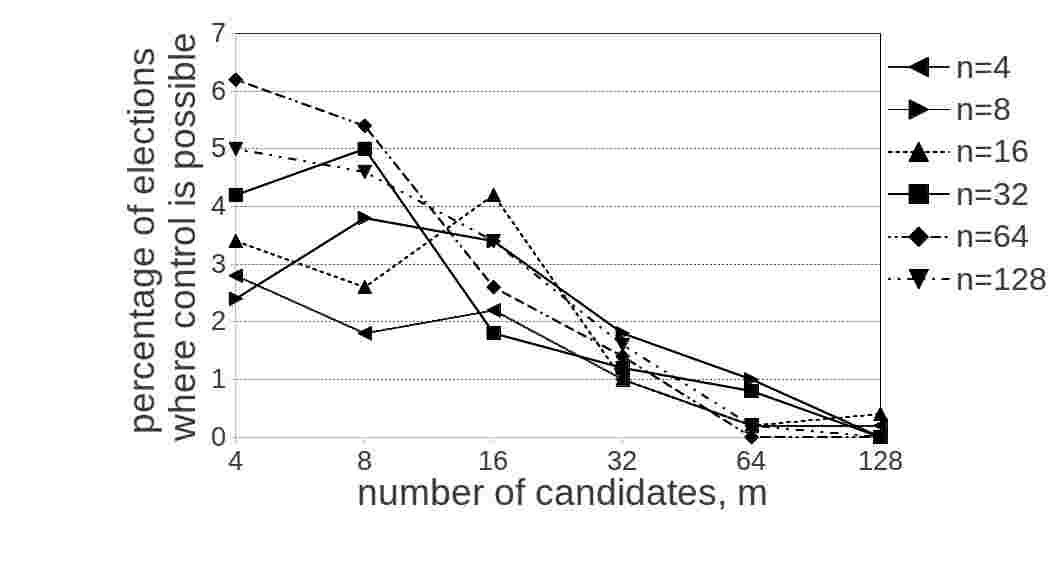}
}
\subfloat[Percentage of timeouts.]{
\label{fig:ccac-fv-d21-n-tos}
\scriptsize
\raisebox{3.0cm}{
\begin{tabular}{@{}l cccccc@{}}
 \toprule
 \textbf{n\textbackslash m} &  $4$ & $8$ & $16$ & $32$ & $64$ & $128$\\
 \cmidrule(r){1-1} \cmidrule(rl){2-2} \cmidrule(rl){3-3} 
 \cmidrule(rl){4-4} \cmidrule(rl){5-5} \cmidrule(rl){6-6} 
 \cmidrule(l){7-7} 
  $4$   & $0$ 	& $0$ 	& $0$ 	& $83$ 	& $81$ 	& $81$  \\
  $8$   & $0$ 	& $0$ 	& $0$ 	& $93$ 	& $92$ 	& $93$  \\
  $16$  & $0$ 	& $0$ 	& $0$ 	& $98$ 	& $99$ 	& $98$  \\
  $32$  & $0$ 	& $0$ 	& $0$	& $99$	& $99$ 	& $100$  \\
  $64$  & $0$ 	& $0$ 	& $0$	& $99$	& $100$ & $100$  \\
  $128$ & $0$ 	& $97$ 	& $98$	& $100$	& $100$	& $100$  \\
 \bottomrule
 \end{tabular}}
 }
 
 \subfloat[Average time the algorithm needs to give a definite output,
 instances where timeouts occur are excluded.]{\label{fig:ccac-fv-d21-n-comp-cost}
\includegraphics[scale=0.27]{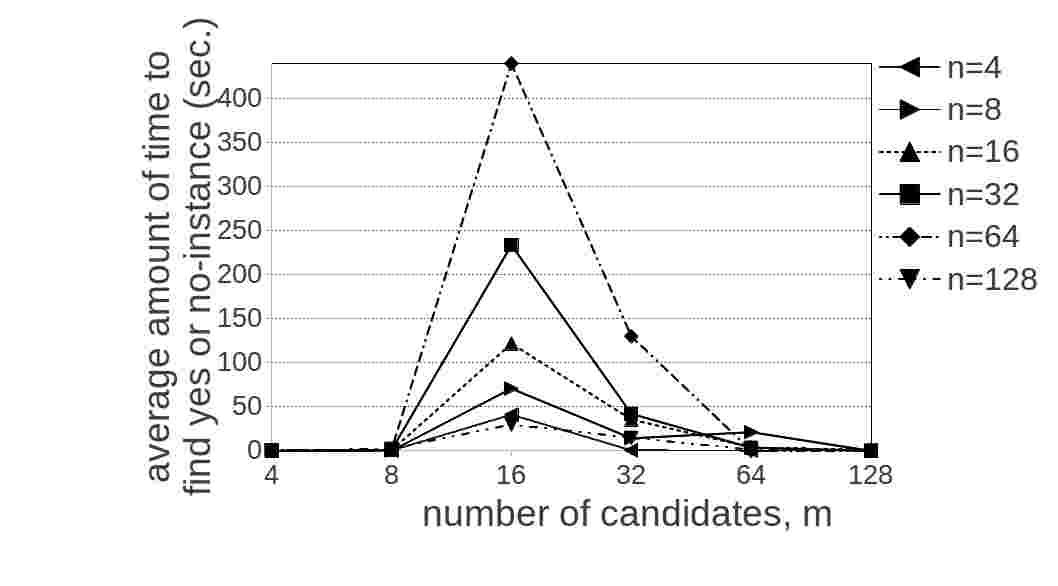}
}
\caption{Fallback voting in the TM model for
\textsc{CCAC}.
}
\label{fig:ccac-fv-d21-n}
\end{figure}

So far, the results for constructive control by adding 
candidates show the highest number of timeouts. 
For those election sizes where no timeouts occur (i.e.,
where the determination of yes- or no-instances is successful), we
have that not many elections can be
controlled successfully in either of the two voting systems.
In Figure~\ref{fig:ccac-fv-d21-n}, we see the
results for fallback voting in the TM model, exemplifying in 
Figure~\ref{fig:ccac-fv-d21-n-res} the low
number of yes-instances for this type of control.  For example,
in the ``max'' column in Table~\ref{tab:overview-exp-res-fv-bv},
the highest percentage of controllable fallback elections is $11\%$ in
the IC model and only $7\%$ in the TM model.  
Plurality voting shows similar results as fallback voting with at most 
$20\%$ yes-instances in the IC model and less than 
$4\%$ in the TM model.
Figure~\ref{fig:ccac-fv-d21-n-tos} gives the detailed occurrence of 
timeouts for the different election sizes and Figure~\ref{fig:ccac-fv-d21-n-comp-cost}
shows the average time needed to determine whether a given 
fallback election generated under the TM model can be controlled by adding 
candidates or not. Remember that in the latter figure the average values do not 
consider those elections where the algorithm exceeded the time
limit of 600 seconds.
Together with the timeout table we can see in Figure~\ref{fig:ccac-fv-d21-n-res} 
that in elections with up to $16$ candidates the number of non-controllable 
elections is very high and increases with the number of candidates increasing.
When more than $16$ candidates participate in an election the number of 
no-instances diminishes as drastically as the timeout rate grows.
Looking at the computational costs in Figure~\ref{fig:ccac-fv-d21-n-comp-cost} 
we can see this in the peaks for $m=16$. Since by design our algorithm needs 
generally more time to determine that an instance is a no-instance than finding a 
yes-instance, the high number of no-instances for $16$ candidates 
inflates the average computing time. 
Knowing that the average computing time for finding yes-instances is not 
particularly high for this type of control (see Figure~\ref{fig:ccac-fv-d21-cp-cost}
in the 
appendix) we might conjecture 
that for the bigger election 
sizes the instances where no distinction 
could be made by our algorithm are no-instances rather than yes-instances.
This indicates that this control type presumably has the lowest overall 
number of yes-instances.  Thus, even for small election sizes,
this type of control seems to be hard to exert succesfully.

Turning now to the destructive variant of control by adding candidates,
for Bucklin elections generated with the IC model,
$71\%$ is a lower bound for the number of controllable elections. 
For up to $16$ candidates, a successful control action can be found in
nearly all elections.  The results for the TM model reconfirm the observation 
made before, namely that the tendencies in both models are similar, but 
with at most $77\%$ and at least $42\%$ of controllable elections the overall 
numbers are again lower than in the IC model.
The latter results hold for fallback elections generated in the TM model 
as well, whereas in the IC model at least $53\%$ and no more
than $92\%$ of the fallback elections are controllable. 
In the tested plurality elections generated with the IC model, similarly
to Bucklin voting, more than $70\%$ and nearly up to $100\%$ are 
controllable. In the TM model, roughly between $50\%$ and $60\%$ of
yes-instances are found for those election sizes where no timeouts occur, so 
between $40\%$ and $50\%$ of these plurality elections are not controllable.
In this control scenario, for about $46\%$ of the elections no
definite output is given in the constructive case, whereas in only
about $8\%$ of the elections timeouts occur in the destructive case.

\section{Discussion and Conclusions}
\label{sec:conclusions}

We have empirically studied the complexity of $\np$-hard
control problems for plurality, fallback, and Bucklin voting in the 
most important of the common control scenarios.  This is the first
such study for control problems in voting and complements the
corresponding results
\cite{wal:c:empirical-study-of-manipulability-of-STV,wal:c:phase-transition-in-manipulating-veto,dav-kat-nar-wal:c:empirical-study-borda-manipulation,dav-kat-nar-wal:c:complexity-and-algorithms-for-borda,nar-wal-xia:c:manipulation-of-nanson-and-baldwin,col-tea:c:complexity-of-manipulating-elections}
for manipulating elections.
In general, our findings indicate that control can often be exerted
effectively in practice, despite the $\np$-hardness of the
corresponding problems.  Our experiments also allow a more
fine-grained analysis and comparison---across various control
scenarios, vote distribution models, and voting systems---than merely
stating $\np$-hardness for all these problems.
Tables~\ref{tab:overview-exp-res-fv-bv} and~\ref{tab:overview-exp-res-pv} give an overview of our experimental
results.  A detailed analysis and discussion follows here, and a
comprehensive presentation of all experimental data and results can be
found in the appendix.

\subsubsection*{IC versus TM:}
Comparing the results for the different distribution models,
we see that in every voting system for all control types studied
(except fallback voting in constructive control by deleting candidates) 
the overall number of yes-instances is higher in the IC than in the
TM model.  
 This may result from the fact that 
in elections with uniformly distributed votes, all candidates 
are likely to be approximately equally preferred by the voters. So 
both constructive and destructive control actions are easier to find. 
This also explains the observation that the IC model mostly produces fewer 
timeouts.
\subsubsection*{Constructive versus Destructive Control:}
For all investigated types of control where both constructive 
and destructive control was investigated, we found that the 
destructive control types are experimentally much easier than their
constructive counterparts, culminating in almost $100\%$ of controllable 
elections for certain control types in the IC model.
These findings confirm---and strengthen---the theoretical insight of
Hemaspaandra
et al.~\cite{hem-hem-rot:j:destructive-control} that the destructive
control problems disjunctively truth-table-reduce to their
constructive counterparts and thus are never harder to solve, up to a
polynomial factor (see also the corresponding observation of Conitzer
et al.~\cite{con-san-lan:j:when-hard-to-manipulate} regarding
manipulation):
In fact, destructive control tends to be even much easier than 
constructive control.

\subsubsection*{Comparison Across Voting Systems:}
For constructive control, we have seen that 
fallback and Bucklin voting show similar tendencies and numbers of 
yes-instances, especially regarding voter control.  We also observed
that their constructive voter control 
problems are in general harder to solve than those for plurality voting.
In all three voting systems, constructive control by adding candidates
seems to be one of the hardest control problems 
showing the smallest numbers of yes-instances in the TM model. 
Only for the constructive partition-of-candidates cases higher 
numbers of timeouts were observed.
\subsubsection*{Adding Candidates versus Deleting Candidates:}
Comparing control by adding candidates to 
control by deleting candidates in the constructive case
we observed that the number of 
yes-instances for control by deleting candidates is 
significantly higher.
These findings are perhaps not overly surprising,
since in the voting systems considered here adding candidates 
to an election can only worsen the position of the designated 
candidate in the votes.  That is, constructive control can  
be exerted successfully only if by adding candidates rivals of the 
designated candidate lose enough points so as to get
defeated by him or her.  This, in turn, can happen only if the 
designated candidate was already a highly preferred candidate in the 
original election.

\subsubsection*{Voter Control versus Candidate Control:}
For fallback and Bucklin voting, we can also compare 
constructive candidate and voter control directly. 
In both voting systems and both distribution models, 
the number of yes-instances 
for constructive control by adding voters is around four times 
higher than the number of yes-instances in the corresponding 
candidate control type, which confirms the argument above, saying that 
adding candidates cannot push the designated candidate directly.
Constructive control by deleting voters can be successfully exerted 
more frequently when votes are less correlated, 
whereas the proportion of successful control actions for deleting candidates 
is about the same for both considered distribution models. 
The observed differences between these types of voter and candidate 
control may result from the fact 
that adding or deleting candidates only shifts the position 
of the designated candidate, which may not influence the outcome 
of the election as directly as increasing or decreasing the 
candidates' scores by adding or deleting voters does.  This explains why
voter control can be tackled more easily than candidate control
by greedy approaches such as ours. 
\subsubsection*{Concluding Remarks:}
Just as
Walsh~\cite{wal:c:phase-transition-in-manipulating-veto,wal:c:empirical-study-of-manipulability-of-STV}
observes for manipulation in the veto rule and in STV, for all types
of control investigated in our experiments, the curves do not show the
typical phase transition known for ``really hard'' computational
problems such as the satisfiability problem
(see~\cite{gen-wal:c:phase-trans-real-problems,che-kan-tay:c:really-hard-problems}
for a detailed discussion of this issue).

These observations raise the question of how other distribution
models
influence
the outcome of such experiments.
Furthermore, the algorithms
implemented could be improved in terms of considering a higher number
of elections per data point, increasing the election sizes, or
allowing a higher number of voters or candidates to be deleted or added in the
corresponding control scenarios.  Besides this,
other voting systems can be analyzed since only their winner determination
has to be implemented 
in addition to a few minor
adjustments 
such as trivial-case checks for the investigated
 control scenarios tailored to the voting system 
at hand. 

\subsubsection*{Acknowledgments:} 
We thank Toby Walsh for interesting discussions,
Volker Aurich for giving us access to his computer lab, and 
Guido K\"onigstein for his help in setting up our experiments.

\bibliographystyle{alpha}

 \appendix

\include{experiments_appendix_tr_oneside_without_comments}

\end{document}

%% file: experiments_appendix_tr_oneside_without_comments.tex
\section{Fallback Voting}

\subsection{Constructive Control by Adding Candidates}
\begin{center}

\begin{figure}[ht]
\centering
	\includegraphics[scale=0.3]{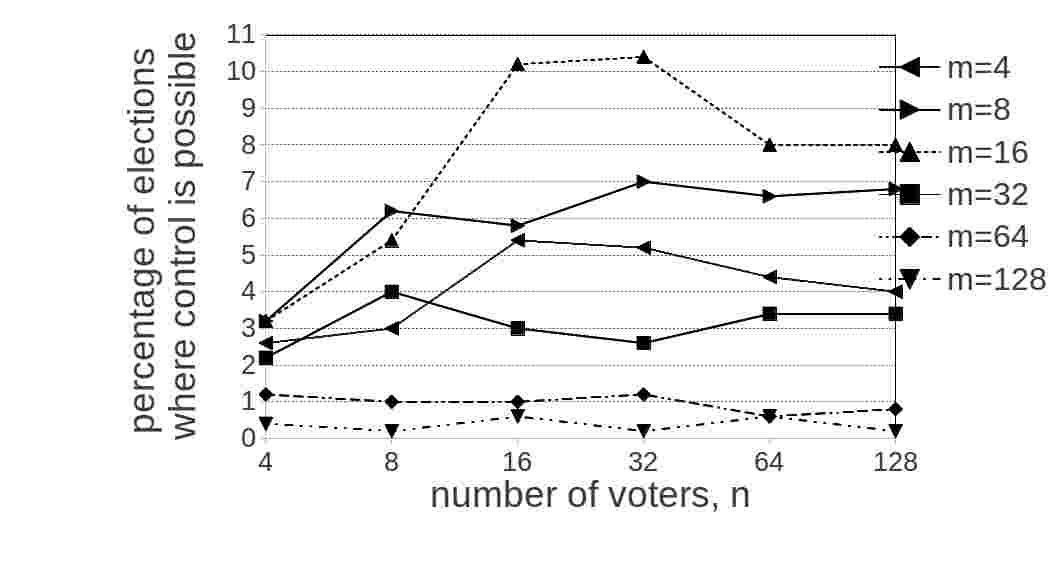}
		\caption{Results for fallback voting in the IC model for 
constructive control by adding candidates. Number of candidates is fixed. }
\end{figure}
\begin{tabular}{||c||c|c|c|c|c|c||c|c|c|c|c|c||}
\hline
				& \multicolumn{6}{c||}{$m=4$} 				& \multicolumn{6}{|c||}{$m=8$}\\\hline
$n$ 		& $4$ 	& $8$ 	& $16$ 	& $32$ 	& $64$ 	& $128$ & $4$ 	& $8$ 	& $16$ 	& $32$ 	& $64$ 	& $128$\\\hline
$\#$ cp 	& $13$ 	& $15$ 	& $270$ & $26$  & $22$ 	& $20$  & $16$ 	& $31$ 	& $29$ 	& $35$ 	& $33$ 	& $34$ \\\hline
$\#$ ci 	& $487$ & $485$ & $473$ & $474$ & $478$ & $480$ & $484$ & $469$ & $471$ & $465$ & $467$ & $466$ \\\hline
$\#$ to 	& $0$ 	& $0$ 	& $0$ 	& $0$  	& $0$ 	& $0$  	& $0$ 	& $0$ 	& $0$ 	& $0$ 	& $0$ 	& $0$ \\\hline
				& \multicolumn{6}{c||}{$m=16$} 				& \multicolumn{6}{|c||}{$m=32$}\\\hline
$n$ 		& $4$ 	& $8$ 	& $16$ 	& $32$ 	& $64$ 	& $128$ & $4$ 	& $8$ 	& $16$ 	& $32$ 	& $64$ 	& $128$\\\hline
$\#$ cp 	& $16$ 	& $27$ 	& $51$ 	& $52$  & $40$ 	& $40$  & $11$ 	& $20$ 	& $15$ 	& $13$ 	& $17$ 	& $17$ \\\hline
$\#$ ci 	& $484$ & $473$ & $449$ & $448$	& $460$ & $0$  	& $36$ 	& $3$ 	& $0$ 	& $0$ 	& $0$ 	& $0$ \\\hline
$\#$ to 	& $0$ 	& $0$ 	& $0$ 	& $0$  	& $0$ 	& $460$ & $453$ & $477$ & $485$ & $487$ & $483$	& $483$ \\\hline
				& \multicolumn{6}{c||}{$m=64$} 				& \multicolumn{6}{|c||}{$m=128$}\\\hline
$n$ 		& $4$ 	& $8$ 	& $16$ 	& $32$ 	& $64$ 	& $128$ & $4$ 	& $8$ 	& $16$ 	& $32$ 	& $64$ 	& $128$\\\hline
$\#$ cp 	& $6$ 	& $5$ 	& $5$ 	& $6$  	& $3$ 	& $4$  	& $2$ 	& $1$ 	& $3$ 	& $1$ 	& $3$ 	& $1$ \\\hline
$\#$ ci 	& $28$ 	& $0$ 	& $0$ 	& $0$  	& $0$ 	& $0$  	& $26$ 	& $3$ 	& $0$ 	& $0$ 	& $0$ 	& $0$ \\\hline
$\#$ to 	& $466$ & $495$ & $495$ & $494$ & $497$ & $496$ & $472$ & $497$ & $497$ & $499$ & $497$ & $499$ \\\hline
\end{tabular}

\newpage
\begin{figure}[ht]
\centering
	\includegraphics[scale=0.3]{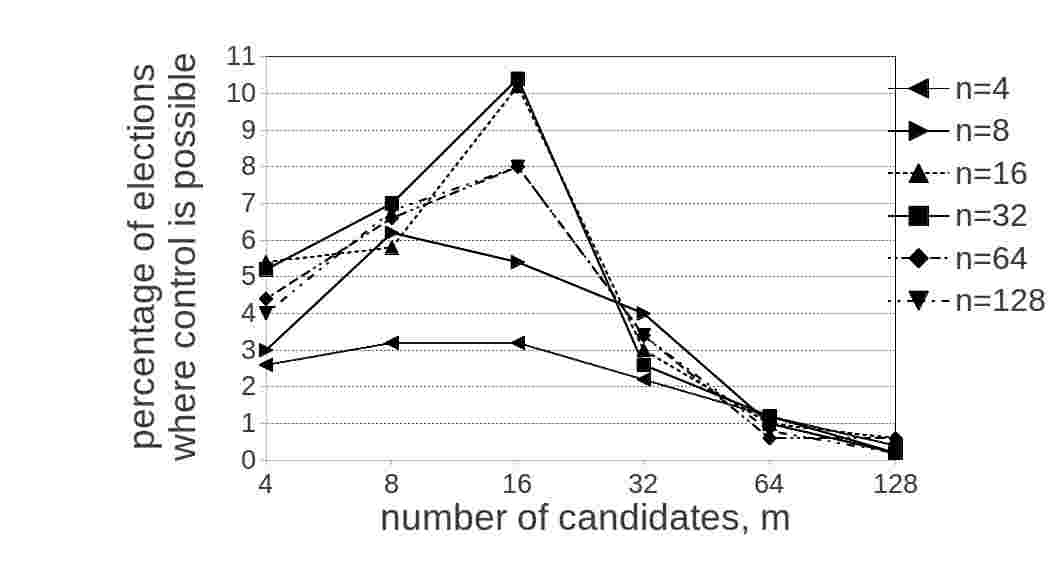}
		\caption{Results for fallback voting in the IC model for 
constructive control by adding candidates. Number of voters is fixed. }
\end{figure}
\begin{tabular}{||c||c|c|c|c|c|c||c|c|c|c|c|c||}
\hline
		& \multicolumn{6}{c||}{$n=4$} 						& \multicolumn{6}{|c||}{$n=8$}\\\hline
$m$ 		& $4$ 	& $8$ 	& $16$ 	& $32$ 	& $64$ 	& $128$ & $4$ 	& $8$ 	& $16$ 	& $32$ 	& $64$ 	& $128$\\\hline
$\#$ cp 	& $13$ 	& $16$ 	& $16$ 	& $11$  & $6$ 	& $2$  	& $15$ 	& $31$ 	& $27$ 	& $20$ 	& $5$ 	& $1$ \\\hline
$\#$ ci 	& $487$ & $484$ & $484$ & $36$  & $28$ 	& $26$  & $485$ & $469$ & $473$ & $3$ 	& $0$ 	& $3$ \\\hline
$\#$ to 	& $0$ 	& $0$ 	& $0$ 	& $453$ & $466$ & $472$ & $0$ 	& $0$ 	& $0$ 	& $477$ & $495$ & $496$ \\\hline
				& \multicolumn{6}{c||}{$n=16$} 				& \multicolumn{6}{|c||}{$n=32$}\\\hline
$m$ 		& $4$ 	& $8$ 	& $16$ 	& $32$ 	& $64$ 	& $128$ & $4$ 	& $8$ 	& $16$ 	& $32$ 	& $64$ 	& $128$\\\hline
$\#$ cp 	& $27$ 	& $29$ 	& $51$ 	& $15$  & $5$ 	& $3$  	& $26$ 	& $35$ 	& $52$ 	& $13$ 	& $6$ 	& $1$ \\\hline
$\#$ ci 	& $473$ & $471$ & $449$ & $0$  	& $0$ 	& $0$  	& $474$ & $465$ & $448$ & $0$ 	& $0$ 	& $0$ \\\hline
$\#$ to 	& $0$ 	& $0$ 	& $0$ 	& $485$ & $495$ & $497$ & $0$ 	& $0$ 	& $0$ 	& $487$ & $494$ & $499$ \\\hline
				& \multicolumn{6}{c||}{$n=64$} 				& \multicolumn{6}{|c||}{$n=128$}\\\hline
$m$ 		& $4$ 	& $8$ 	& $16$ 	& $32$ 	& $64$ 	& $128$ & $4$ 	& $8$ 	& $16$ 	& $32$ 	& $64$ 	& $128$\\\hline
$\#$ cp 	& $22$ 	& $33$ 	& $40$ 	& $17$  & $3$ 	& $3$  	& $20$ 	& $34$ 	& $40$ 	& $17$ 	& $4$ 	& $1$ \\\hline
$\#$ ci 	& $478$ & $467$ & $460$ & $0$  	& $0$ 	& $0$  	& $480$ & $466$ & $0$ 	& $0$ 	& $0$ 	& $0$ \\\hline
$\#$ to 	& $0$ 	& $0$ 	& $0$ 	& $483$ & $497$ & $497$ & $0$ 	& $0$ 	& $460$ & $483$ & $496$ & $499$ \\\hline
\end{tabular}

\newpage
\begin{figure}[ht]
\centering
	\includegraphics[scale=0.3]{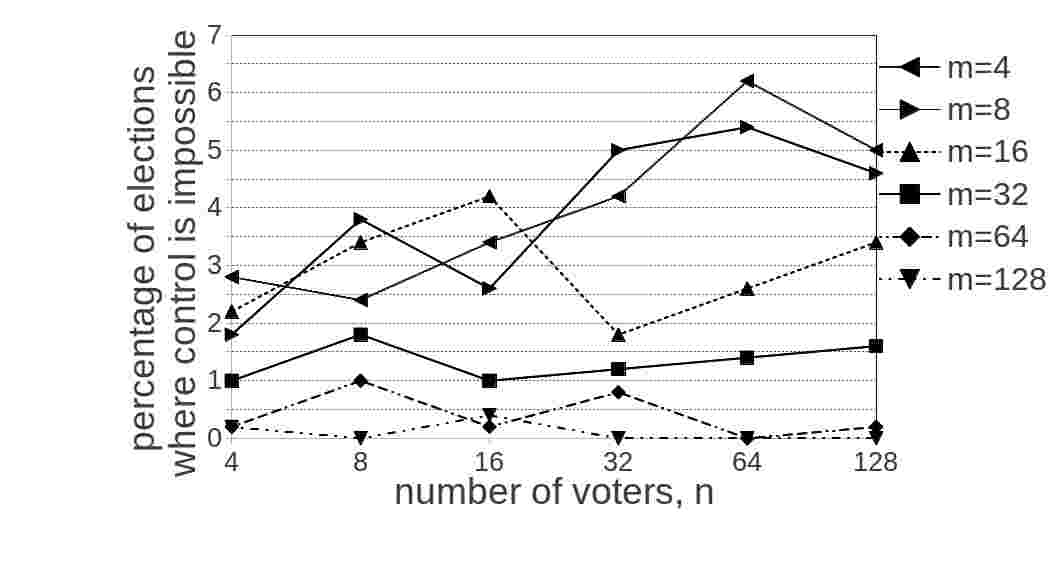}
		\caption{Results for fallback voting in the TM model for 
constructive control by adding candidates. Number of candidates is fixed. }
\end{figure}
\begin{tabular}{||c||c|c|c|c|c|c||c|c|c|c|c|c||}
\hline
				& \multicolumn{6}{c||}{$m=4$} 				& \multicolumn{6}{|c||}{$m=8$}\\\hline
$n$ 		& $4$ 	& $8$ 	& $16$ 	& $32$ 	& $64$ 	& $128$ & $4$ 	& $8$ 	& $16$ 	& $32$ 	& $64$ 	& $128$\\\hline
$\#$ cp 	& $14$ 	& $12$ 	& $17$ 	& $21$  & $31$ 	& $25$  & $9$ 	& $19$ 	& $13$ 	& $25$ 	& $27$ 	& $23$ \\\hline
$\#$ ci 	& $486$ & $488$ & $483$ & $479$ & $469$ & $475$ & $491$ & $481$ & $487$ & $475$ & $473$ & $477$ \\\hline
$\#$ to 	& $0$ 	& $0$ 	& $0$ 	& $0$  	& $0$ 	& $0$  	& $0$ 	& $0$ 	& $0$ 	& $0$ 	& $0$ 	& $0$ \\\hline
				& \multicolumn{6}{c||}{$m=16$} 				& \multicolumn{6}{|c||}{$m=32$}\\\hline
$n$ 		& $4$ 	& $8$ 	& $16$ 	& $32$ 	& $64$ 	& $128$ & $4$ 	& $8$ 	& $16$ 	& $32$ 	& $64$ 	& $128$\\\hline
$\#$ cp 	& $11$ 	& $17$ 	& $21$ 	& $9$  	& $13$ 	& $17$  & $5$ 	& $9$ 	& $5$ 	& $6$ 	& $7$ 	& $8$ \\\hline
$\#$ ci 	& $489$ & $483$ & $479$ & $491$	& $487$ & $0$  	& $82$ 	& $26$ 	& $6$ 	& $0$ 	& $0$ 	& $0$ \\\hline
$\#$ to 	& $0$ 	& $0$ 	& $0$ 	& $0$  	& $0$ 	& $483$ & $413$ & $465$ & $489$ & $494$ & $493$ & $492$ \\\hline
				& \multicolumn{6}{c||}{$m=64$} 				& \multicolumn{6}{|c||}{$m=128$}\\\hline
$n$ 		& $4$ 	& $8$ 	& $16$ 	& $32$ 	& $64$ 	& $128$ & $4$ 	& $8$ 	& $16$ 	& $32$ 	& $64$ 	& $128$\\\hline
$\#$ cp 	& $1$ 	& $5$ 	& $1$ 	& $4$  	& $0$ 	& $1$  	& $1$ 	& $0$ 	& $2$ 	& $0$ 	& $0$ 	& $0$ \\\hline
$\#$ ci 	& $92$ 	& $33$ 	& $5$ 	& $0$  	& $0$ 	& $0$  	& $95$ 	& $36$ 	& $10$ 	& $0$ 	& $0$ 	& $0$ \\\hline
$\#$ to 	& $407$ & $462$ & $494$ & $496$ & $500$ & $499$	& $404$ & $464$ & $488$ & $500$ & $500$	& $500$ \\\hline
\end{tabular}
\newpage
\begin{figure}[ht]
\centering
	\includegraphics[scale=0.3]{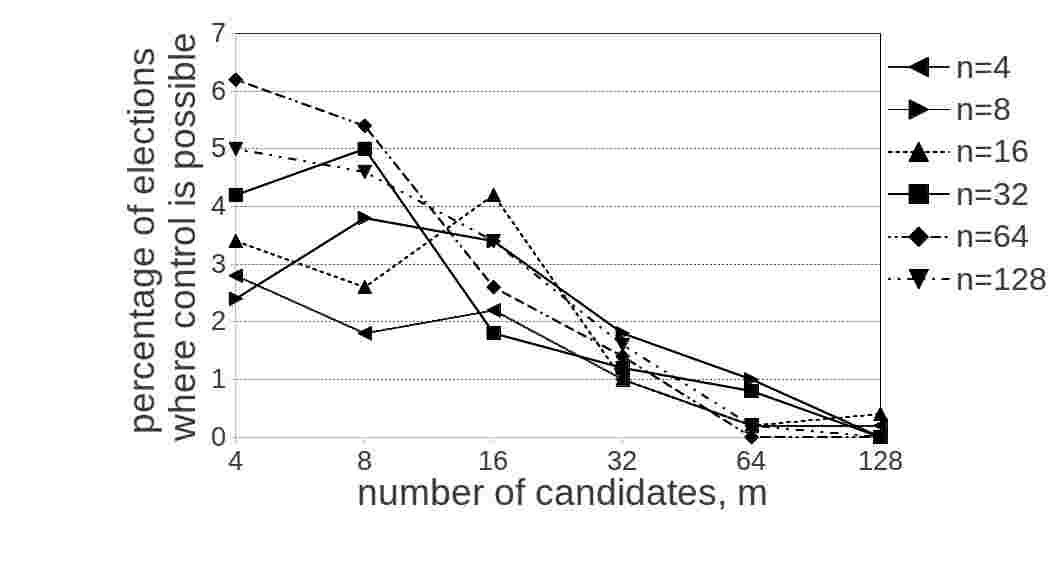}
		\caption{Results for fallback voting in the TM model for 
constructive control by adding candidates. Number of voters is fixed. }
\end{figure}
\begin{tabular}{||c||c|c|c|c|c|c||c|c|c|c|c|c||}
\hline
		& \multicolumn{6}{c||}{$n=4$} 			& \multicolumn{6}{|c||}{$n=8$}\\\hline
$m$ 		& $4$ 	& $8$ 	& $16$ 	& $32$ 	& $64$ 	& $128$ & $4$ 	& $8$ 	& $16$ 	& $32$ 	& $64$ 	& $128$\\\hline
$\#$ cp 	& $14$ 	& $9$ 	& $11$ 	& $5$  	& $1$ 	& $1$  	& $12$ 	& $19$ 	& $17$ 	& $9$ 	& $5$ 	& $0$ \\\hline
$\#$ ci 	& $486$ & $491$ & $489$ & $82$  & $92$ 	& $95$  & $488$ & $481$ & $483$ & $26$ 	& $33$ 	& $36$ \\\hline
$\#$ to 	& $0$ 	& $0$ 	& $0$ 	& $413$ & $407$ & $404$ & $0$ 	& $0$ 	& $0$ 	& $465$ & $462$ & $464$ \\\hline
		& \multicolumn{6}{c||}{$n=16$} 			& \multicolumn{6}{|c||}{$n=32$}\\\hline
$m$ 		& $4$ 	& $8$ 	& $16$ 	& $32$ 	& $64$ 	& $128$ & $4$ 	& $8$ 	& $16$ 	& $32$ 	& $64$ 	& $128$\\\hline
$\#$ cp 	& $17$ 	& $13$ 	& $21$ 	& $5$  	& $1$ 	& $2$  	& $21$ 	& $25$ 	& $9$ 	& $6$ 	& $4$ 	& $0$ \\\hline
$\#$ ci 	& $483$ & $487$ & $479$ & $6$  	& $5$ 	& $10$  & $479$ & $475$ & $491$ & $0$ 	& $0$ 	& $0$ \\\hline
$\#$ to 	& $0$ 	& $0$ 	& $0$ 	& $489$ & $494$ & $488$ & $0$ 	& $0$ 	& $0$ 	& $494$ & $496$ & $500$ \\\hline
		& \multicolumn{6}{c||}{$n=64$} 			& \multicolumn{6}{|c||}{$n=128$}\\\hline
$m$ 		& $4$ 	& $8$ 	& $16$ 	& $32$ 	& $64$ 	& $128$ & $4$ 	& $8$ 	& $16$ 	& $32$ 	& $64$ 	& $128$\\\hline
$\#$ cp 	& $31$ 	& $27$ 	& $13$ 	& $7$  	& $0$ 	& $0$  	& $25$ 	& $23$ 	& $17$ 	& $8$ 	& $1$ 	& $0$ \\\hline
$\#$ ci 	& $469$ & $473$ & $487$ & $0$  	& $0$ 	& $0$  	& $475$ & $477$ & $0$ 	& $0$ 	& $0$ 	& $0$ \\\hline
$\#$ to 	& $0$ 	& $0$ 	& $0$ 	& $493$ & $500$ & $500$ & $0$ 	& $0$ 	& $483$ & $492$ & $499$ & $500$ \\\hline
\end{tabular}
\end{center}

\clearpage
\subsubsection{Computation Costs}
\begin{figure}[ht]
\centering
	\includegraphics[scale=0.3]{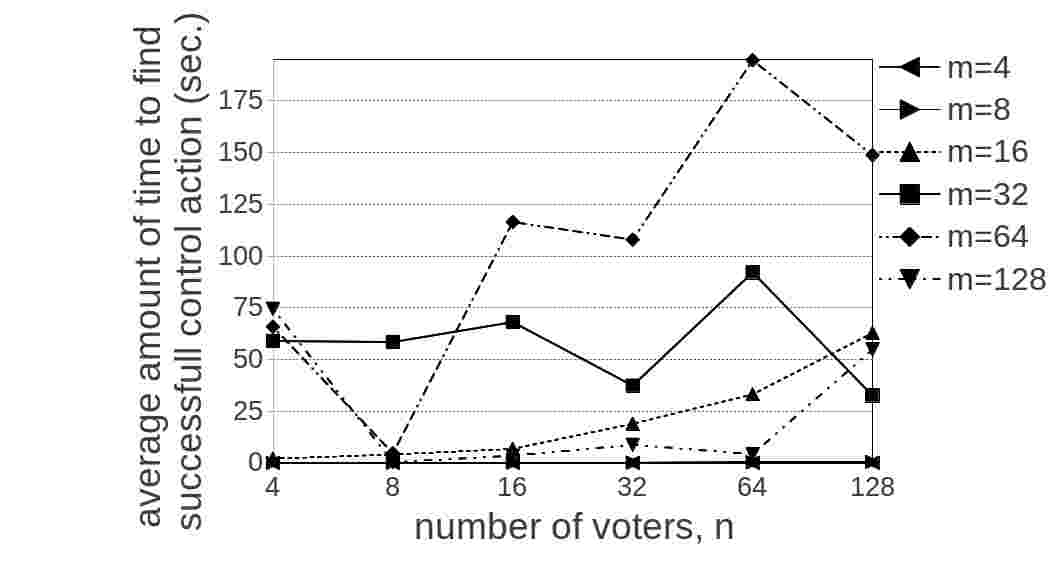}
	\caption{Average time the algorithm needs to find a successful control action for 
	constructive control by adding candidates
	in fallback elections in the IC model. The maximum is $194,7$ seconds.}
\end{figure}
\begin{figure}[ht]
\centering
	\includegraphics[scale=0.3]{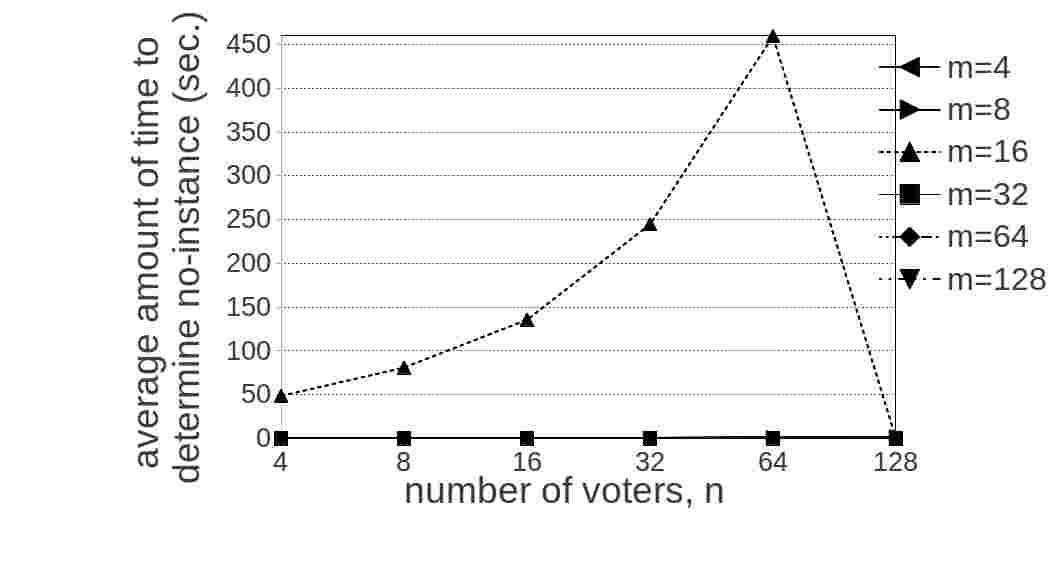}
	\caption{Average time the algorithm needs to determine no-instance of 
		constructive control by adding candidates
	in fallback elections in the IC model. The maximum is $459,54$ seconds.}
\end{figure}
\begin{figure}[ht]
\centering
	\includegraphics[scale=0.3]{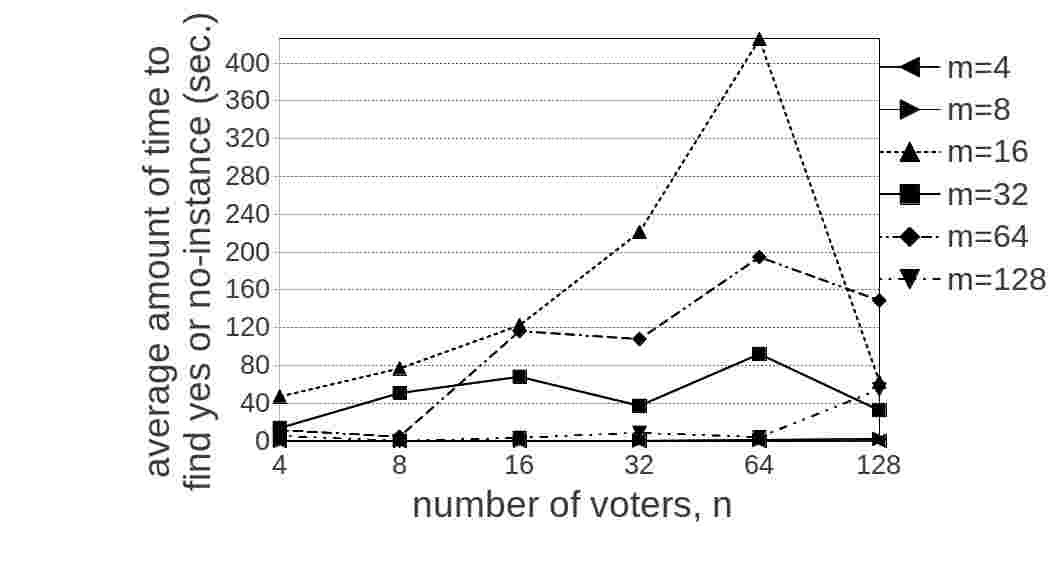}
	\caption{Average time the algorithm needs to give a definite output for 
	constructive control by adding candidates
	in fallback elections in the IC model. The maximum is $425,43$ seconds.}
\end{figure}

\begin{figure}[ht]
\centering
	\includegraphics[scale=0.3]{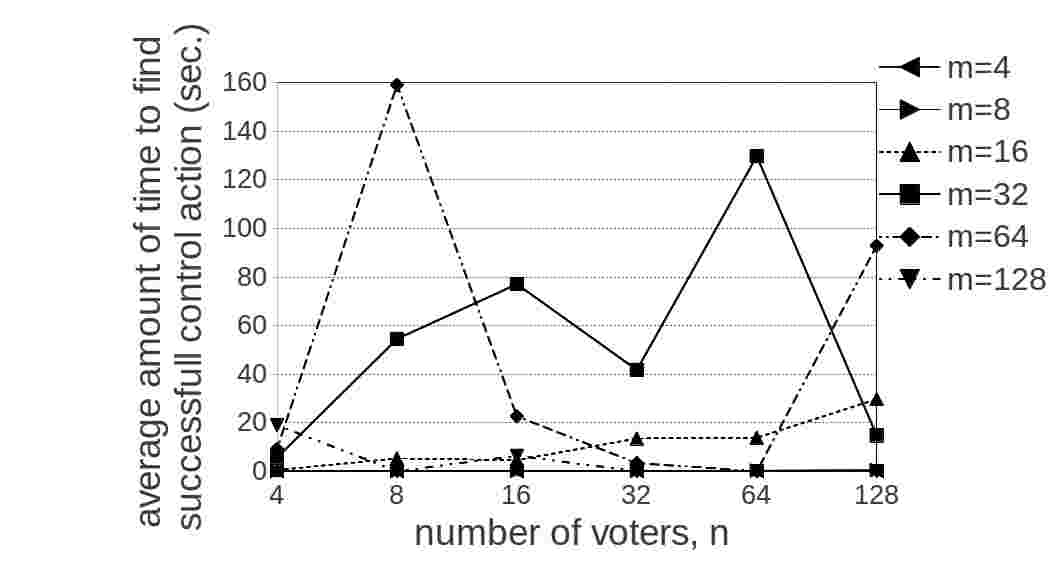}
	\caption{Average time the algorithm needs to find a successful control action for 
	constructive control by adding candidates
	in fallback elections in the TM model. The maximum is $159,22$ seconds.}
	\label{fig:ccac-fv-d21-cp-cost}
\end{figure}

\begin{figure}[ht]
\centering
	\includegraphics[scale=0.3]{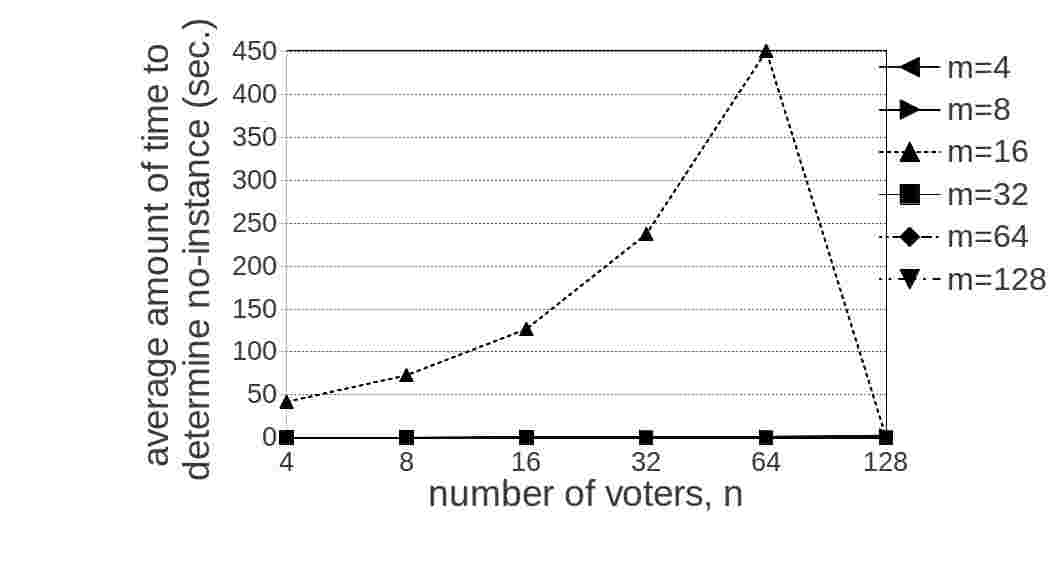}
	\caption{Average time the algorithm needs to determine no-instance of 
		constructive control by adding candidates
	in fallback elections in the TM model. The maximum is $451,05$ seconds.}
\end{figure}
\begin{figure}[ht]
\centering
	\includegraphics[scale=0.3]{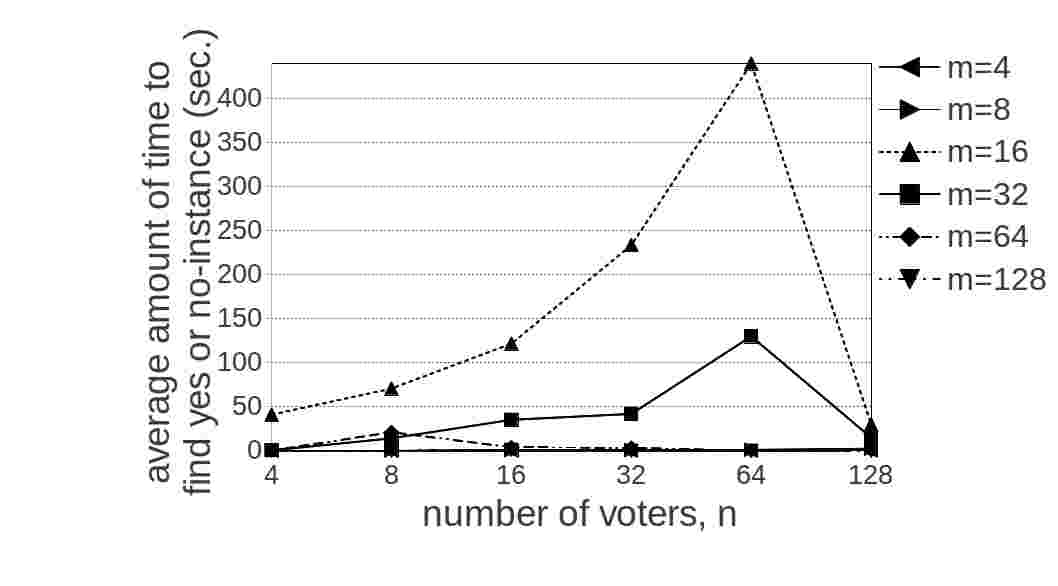}
	\caption{Average time the algorithm needs to give a definite output for 
	constructive control by adding candidates
	in fallback elections in the TM model. The maximum is $439,68$ seconds.}
\end{figure}
\clearpage
\subsection{Destructive Control by Adding Candidates}
\begin{center}

\begin{figure}[ht]
\centering
	\includegraphics[scale=0.3]{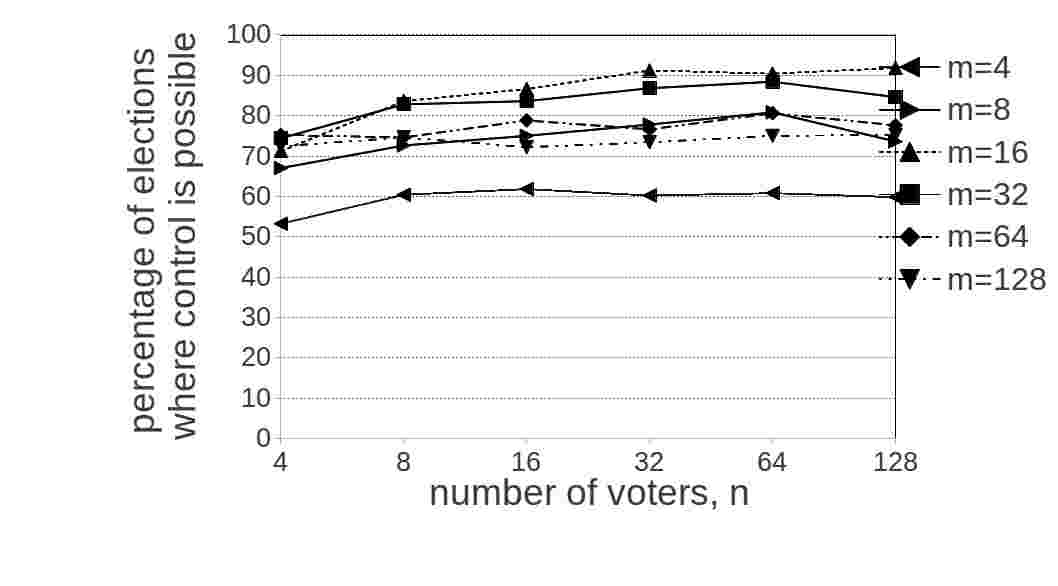}
		\caption{Results for fallback voting in the IC model for 
destructive control by adding candidates. Number of candidates is fixed. }
\end{figure}

\begin{tabular}{||c||c|c|c|c|c|c||c|c|c|c|c|c||}
\hline
				& \multicolumn{6}{c||}{$m=4$} 				& \multicolumn{6}{|c||}{$m=8$}\\\hline
$n$ 		& $4$ 	& $8$ 	& $16$ 	& $32$ 	& $64$ 	& $128$ & $4$ 	& $8$ 	& $16$ 	& $32$ 	& $64$ 	& $128$\\\hline
$\#$ cp 	& $266$ & $302$ & $309$ & $301$ & $304$ & $299$ & $335$ & $363$ & $375$ & $389$ & $404$ & $368$ \\\hline
$\#$ ci 	& $234$ & $198$ & $191$ & $199$ & $196$ & $201$ & $165$ & $137$ & $125$ & $111$ & $96$ 	& $132$ \\\hline
$\#$ to 	& $0$ 	& $0$ 	& $0$ 	& $0$  	& $0$ 	& $0$  	& $0$ 	& $0$ 	& $0$ 	& $0$ 	& $0$ 	& $0$ \\\hline
				& \multicolumn{6}{c||}{$m=16$} 				& \multicolumn{6}{|c||}{$m=32$}\\\hline
$n$ 		& $4$ 	& $8$ 	& $16$ 	& $32$ 	& $64$ 	& $128$ & $4$ 	& $8$ 	& $16$ 	& $32$ 	& $64$ 	& $128$\\\hline
$\#$ cp 	& $356$ & $418$ & $433$ & $456$ & $452$ & $459$ & $372$ & $414$ & $418$ & $434$ & $442$ & $423$ \\\hline
$\#$ ci 	& $144$ & $82$ 	& $67$ 	& $44$  & $48$ 	& $0$  	& $0$ 	& $0$ 	& $0$ 	& $0$ 	& $0$ 	& $0$ \\\hline
$\#$ to 	& $0$ 	& $0$ 	& $0$ 	& $0$  	& $0$ 	& $41$  & $128$ & $86$ 	& $82$ 	& $66$ 	& $58$ 	& $77$ \\\hline
				& \multicolumn{6}{c||}{$m=64$} 				& \multicolumn{6}{|c||}{$m=128$}\\\hline
$n$ 		& $4$ 	& $8$ 	& $16$ 	& $32$ 	& $64$ 	& $128$ & $4$ 	& $8$ 	& $16$ 	& $32$ 	& $64$ 	& $128$\\\hline
$\#$ cp 	& $376$ & $373$ & $394$ & $383$ & $403$ & $388$ & $362$ & $373$ & $361$ & $367$ & $375$ & $376$ \\\hline
$\#$ ci 	& $0$ 	& $0$ 	& $0$ 	& $0$  	& $0$ 	& $0$  	& $0$ 	& $0$ 	& $0$ 	& $0$ 	& $0$ 	& $0$ \\\hline
$\#$ to 	& $124$ & $127$ & $106$ & $117$ & $97$ 	& $112$ & $138$ & $127$ & $139$ & $133$ & $125$ & $124$ \\\hline
\end{tabular}

\newpage
\begin{figure}[ht]
\centering
	\includegraphics[scale=0.3]{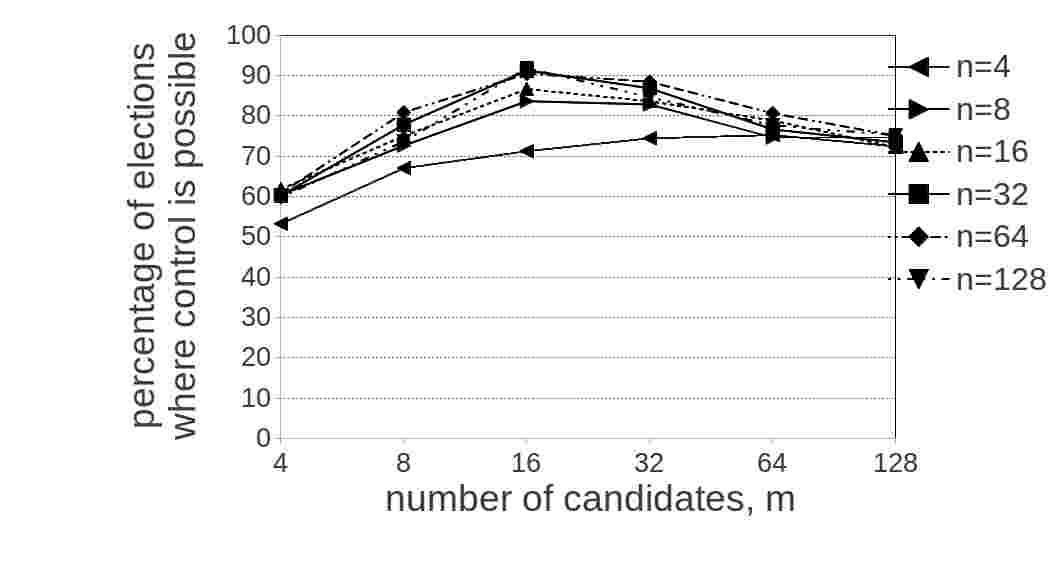}
		\caption{Results for fallback voting in the IC model for 
destructive control by adding candidates. Number of voters is fixed. }
\end{figure}

\begin{tabular}{||c||c|c|c|c|c|c||c|c|c|c|c|c||}
\hline
		& \multicolumn{6}{c||}{$n=4$} 						& \multicolumn{6}{|c||}{$n=8$}\\\hline
$m$ 		& $4$ 	& $8$ 	& $16$ 	& $32$ 	& $64$ 	& $128$ & $4$ 	& $8$ 	& $16$ 	& $32$ 	& $64$ 	& $128$\\\hline
$\#$ cp 	& $266$ & $335$ & $356$ & $372$ & $376$ & $362$ & $302$ & $363$ & $418$ & $414$ & $373$ & $373$ \\\hline
$\#$ ci 	& $234$ & $165$ & $144$ & $0$  	& $0$ 	& $0$  	& $198$ & $137$ & $82$ 	& $0$ 	& $0$ 	& $0$ \\\hline
$\#$ to 	& $0$ 	& $0$ 	& $0$ 	& $128$ & $124$ & $138$ & $0$ 	& $0$ 	& $0$ 	& $86$ 	& $127$ & $127$ \\\hline
				& \multicolumn{6}{c||}{$n=16$} 				& \multicolumn{6}{|c||}{$n=32$}\\\hline
$m$ 		& $4$ 	& $8$ 	& $16$ 	& $32$ 	& $64$ 	& $128$ & $4$ 	& $8$ 	& $16$ 	& $32$ 	& $64$ 	& $128$\\\hline
$\#$ cp 	& $309$ & $375$ & $433$ & $418$ & $394$ & $361$ & $301$ & $389$ & $456$ & $434$ & $383$ & $367$ \\\hline
$\#$ ci 	& $191$ & $125$ & $67$ 	& $0$  	& $0$ 	& $0$  	& $199$ & $111$ & $44$ 	& $0$ 	& $0$ 	& $0$ \\\hline
$\#$ to 	& $0$ 	& $0$ 	& $0$ 	& $82$  & $106$ & $139$ & $0$ 	& $0$ 	& $0$ 	& $66$ 	& $117$ & $133$ \\\hline
				& \multicolumn{6}{c||}{$n=64$} 				& \multicolumn{6}{|c||}{$n=128$}\\\hline
$m$ 		& $4$ 	& $8$ 	& $16$ 	& $32$ 	& $64$ 	& $128$ & $4$ 	& $8$ 	& $16$ 	& $32$ 	& $64$ 	& $128$\\\hline
$\#$ cp 	& $304$ & $404$ & $452$ & $442$ & $403$ & $375$ & $299$ & $368$ & $459$ & $423$ & $388$ & $376$ \\\hline
$\#$ ci 	& $196$ & $96$ 	& $48$ 	& $0$  	& $0$ 	& $0$  	& $201$ & $132$ & $0$ 	& $0$ 	& $0$ 	& $0$ \\\hline
$\#$ to 	& $0$ 	& $0$ 	& $0$ 	& $58$  & $97$ 	& $125$ & $0$ 	& $0$ 	& $41$ 	& $77$ 	& $112$ & $124$ \\\hline
\end{tabular}
\newpage
\begin{figure}[ht]
\centering
	\includegraphics[scale=0.3]{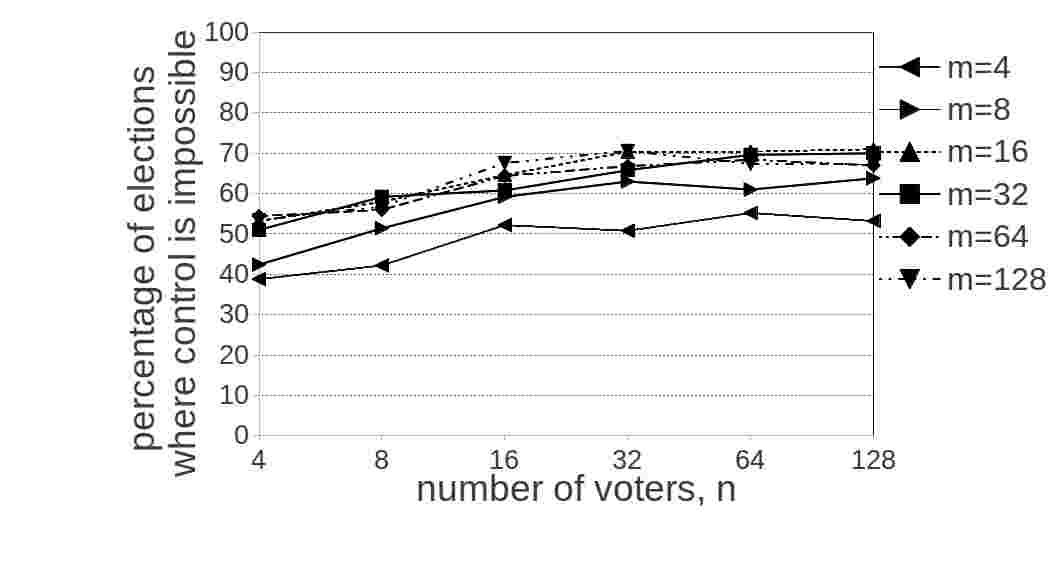}
		\caption{Results for fallback voting in the TM model for 
destructive control by adding candidates. Number of candidates is fixed. }
\end{figure}
\begin{tabular}{||c||c|c|c|c|c|c||c|c|c|c|c|c||}
\hline
		& \multicolumn{6}{c||}{$m=4$} 			& \multicolumn{6}{|c||}{$m=8$}\\\hline
$n$ 		& $4$ 	& $8$ 	& $16$ 	& $32$ 	& $64$ 	& $128$ & $4$ 	& $8$ 	& $16$ 	& $32$ 	& $64$ 	& $128$\\\hline
$\#$ cp 	& $194$ & $211$ & $261$ & $254$ & $276$ & $266$ & $212$ & $257$ & $296$ & $315$ & $305$ & $319$ \\\hline
$\#$ ci 	& $306$ & $289$ & $239$ & $246$ & $224$ & $234$ & $288$ & $243$ & $204$ & $185$ & $195$ & $181$ \\\hline
$\#$ to 	& $0$ 	& $0$ 	& $0$ 	& $0$  	& $0$ 	& $0$  	& $0$ 	& $0$ 	& $0$ 	& $0$ 	& $0$ 	& $0$ \\\hline
		& \multicolumn{6}{c||}{$m=16$} 			& \multicolumn{6}{|c||}{$m=32$}\\\hline
$n$ 		& $4$ 	& $8$ 	& $16$ 	& $32$ 	& $64$ 	& $128$ & $4$ 	& $8$ 	& $16$ 	& $32$ 	& $64$ 	& $128$\\\hline
$\#$ cp 	& $266$ & $290$ & $323$ & $352$ & $352$ & $355$ & $255$ & $296$ & $304$ & $329$ & $348$ & $350$ \\\hline
$\#$ ci 	& $234$ & $210$ & $177$ & $148$ & $148$ & $13$  & $123$ & $99$ 	& $62$ 	& $27$ 	& $5$ 	& $7$ \\\hline
$\#$ to 	& $0$ 	& $0$ 	& $0$ 	& $0$  	& $0$ 	& $132$ & $122$ & $105$ & $134$	& $144$ & $147$ & $143$ \\\hline
		& \multicolumn{6}{c||}{$m=64$} 			& \multicolumn{6}{|c||}{$m=128$}\\\hline
$n$ 		& $4$ 	& $8$ 	& $16$ 	& $32$ 	& $64$ 	& $128$ & $4$ 	& $8$ 	& $16$ 	& $32$ 	& $64$ 	& $128$\\\hline
$\#$ cp 	& $272$ & $280$ & $322$ & $334$ & $342$ & $335$ & $267$ & $284$ & $338$ & $353$ & $337$ & $336$ \\\hline
$\#$ ci 	& $118$ & $108$ & $58$ 	& $19$  & $4$ 	& $6$  	& $135$ & $105$ & $58$ 	& $20$ 	& $6$ 	& $0$ \\\hline
$\#$ to 	& $110$ & $112$ & $120$ & $147$ & $154$ & $159$ & $98$ 	& $111$ & $104$ & $127$ & $157$ & $164$ \\\hline
\end{tabular}
\newpage
\begin{figure}[ht]
\centering
	\includegraphics[scale=0.3]{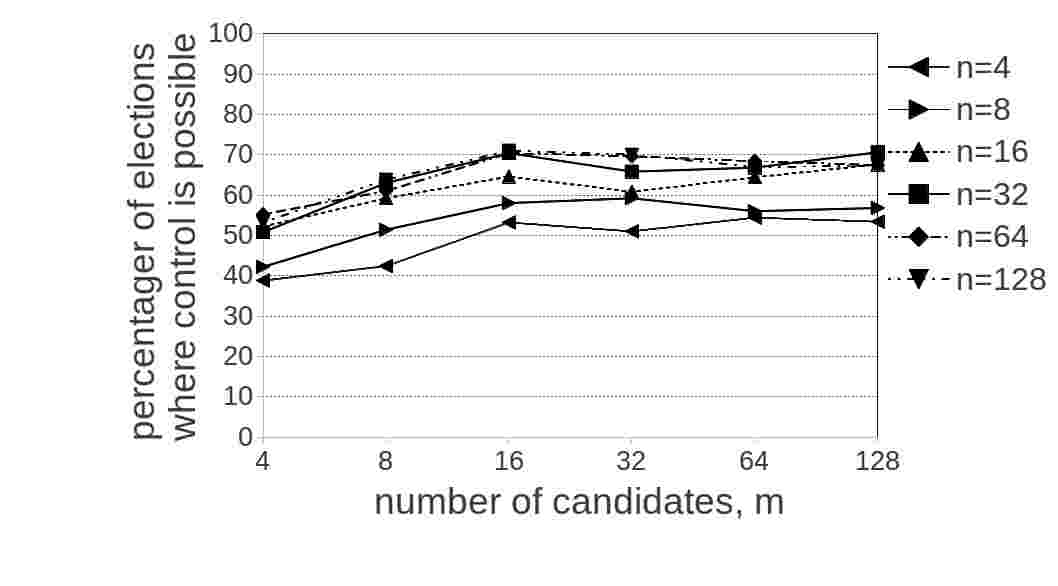}
		\caption{Results for fallback voting in the TM model for 
destructive control by adding candidates. Number of voters is fixed. }
\end{figure}
\begin{tabular}{||c||c|c|c|c|c|c||c|c|c|c|c|c||}
\hline
& \multicolumn{6}{c||}{$n=4$} 								& \multicolumn{6}{|c||}{$n=8$}\\\hline
$m$ 		& $4$ 	& $8$ 	& $16$ 	& $32$ 	& $64$ 	& $128$ & $4$ 	& $8$ 	& $16$ 	& $32$ 	& $64$ 	& $128$\\\hline
$\#$ cp 	& $194$ & $212$ & $266$ & $255$ & $272$ & $267$ & $211$ & $257$ & $290$ & $296$ & $280$ & $284$ \\\hline
$\#$ ci 	& $306$ & $288$ & $234$ & $123$ & $118$ & $135$ & $289$ & $243$ & $210$ & $99$ 	& $108$ & $105$ \\\hline
$\#$ to 	& $0$ 	& $0$ 	& $0$ 	& $122$ & $110$ & $98$  & $0$ 	& $0$ 	& $0$ 	& $105$ & $112$ & $111$ \\\hline
				& \multicolumn{6}{c||}{$n=16$} 				& \multicolumn{6}{|c||}{$n=32$}\\\hline
$m$ 		& $4$ 	& $8$ 	& $16$ 	& $32$ 	& $64$ 	& $128$ & $4$ 	& $8$ 	& $16$ 	& $32$ 	& $64$ 	& $128$\\\hline
$\#$ cp 	& $261$ & $296$ & $323$ & $304$ & $322$ & $338$ & $254$ & $315$ & $352$ & $329$ & $334$ & $353$ \\\hline
$\#$ ci 	& $239$ & $204$ & $177$ & $62$  & $58$ 	& $58$  & $246$ & $185$ & $148$ & $27$ 	& $19$ 	& $20$ \\\hline
$\#$ to 	& $0$ 	& $0$ 	& $0$ 	& $134$ & $120$ & $104$ & $0$ 	& $0$ 	& $0$ 	& $144$ & $147$ & $127$ \\\hline
				& \multicolumn{6}{c||}{$n=64$} 				& \multicolumn{6}{|c||}{$n=128$}\\\hline
$m$ 		& $4$ 	& $8$ 	& $16$ 	& $32$ 	& $64$ 	& $128$ & $4$ 	& $8$ 	& $16$ 	& $32$ 	& $64$ 	& $128$\\\hline
$\#$ cp 	& $276$ & $305$ & $352$ & $348$ & $342$ & $337$ & $266$ & $319$ & $355$ & $350$ & $335$ & $336$ \\\hline
$\#$ ci 	& $224$ & $195$ & $148$ & $5$  	& $4$ 	& $6$  	& $234$ & $181$ & $13$ 	& $7$ 	& $6$ 	& $0$ \\\hline
$\#$ to 	& $0$ 	& $0$ 	& $0$ 	& $147$ & $154$ & $157$ & $0$ 	& $0$ 	& $132$ & $143$ & $159$ & $164$ \\\hline
\end{tabular}
\end{center}

\clearpage
\subsubsection{Computational Costs}
\begin{figure}[ht]
\centering
	\includegraphics[scale=0.3]{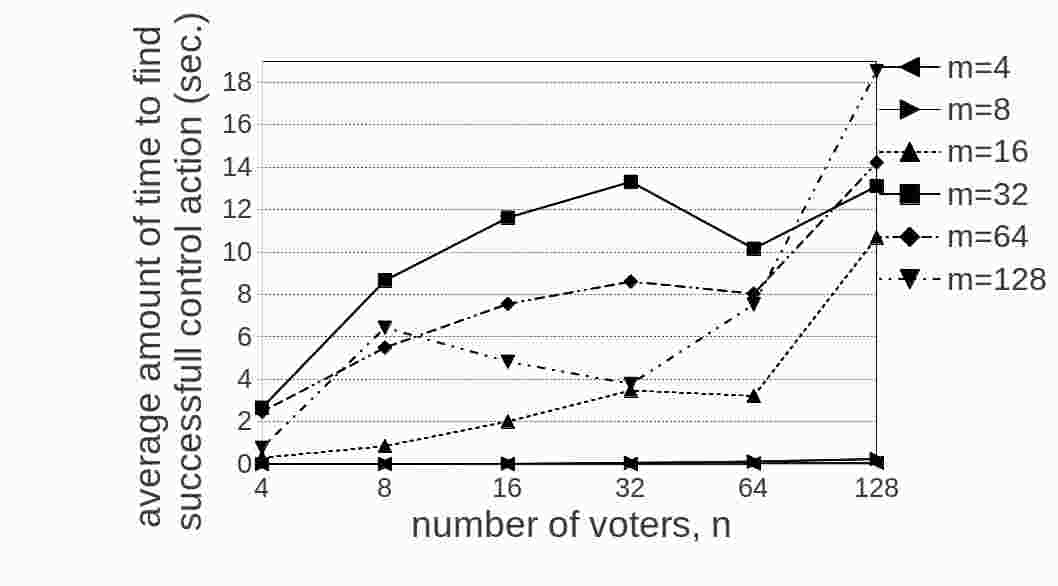}
	\caption{Average time the algorithm needs to find a successful control action for 
	destructive control by adding candidates
	in fallback elections in the IC model. The maximum is $18,53$ seconds.}
\end{figure}
\begin{figure}[ht]
\centering
	\includegraphics[scale=0.3]{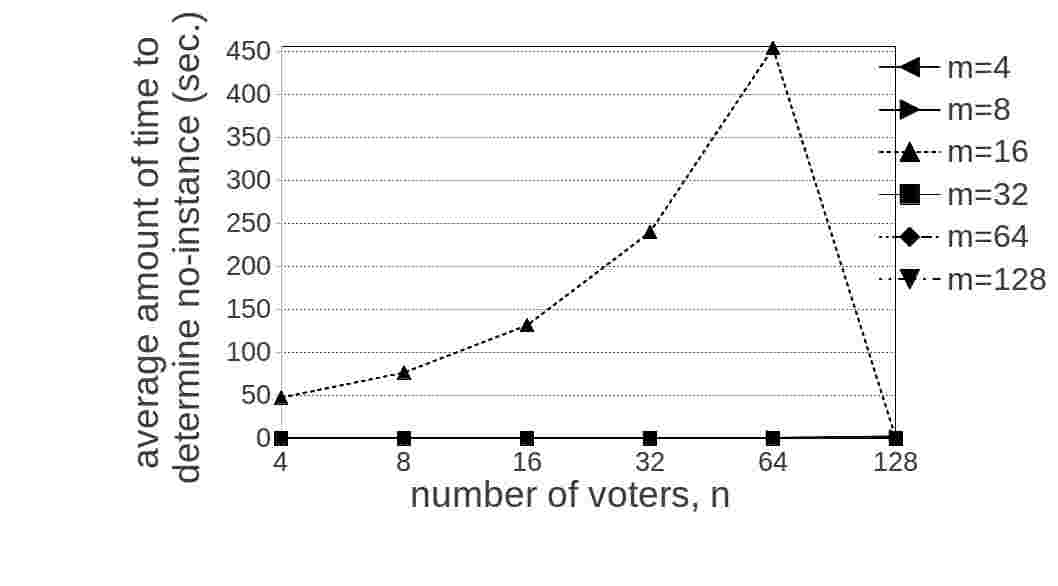}
	\caption{Average time the algorithm needs to determine no-instance of 
		destructive control by adding candidates
	in fallback elections in the IC model. The maximum is $453,8$ seconds.}
\end{figure}
\begin{figure}[ht]
\centering
	\includegraphics[scale=0.3]{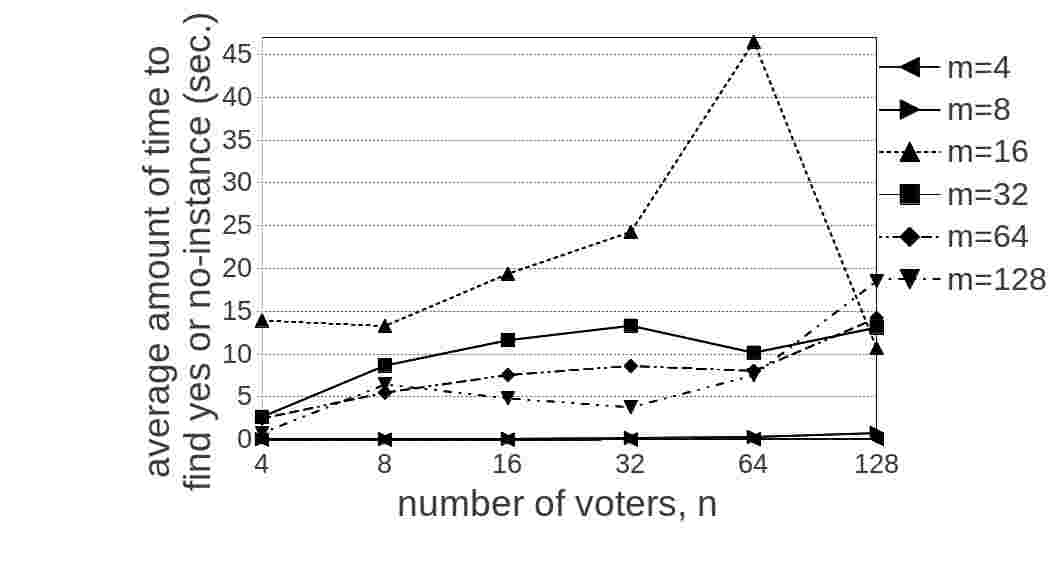}
	\caption{Average time the algorithm needs to give a definite output for 
	destructive control by adding candidates
	in fallback elections in the IC model. The maximum is $46,48$ seconds.}
\end{figure}
\begin{figure}[ht]
\centering
	\includegraphics[scale=0.3]{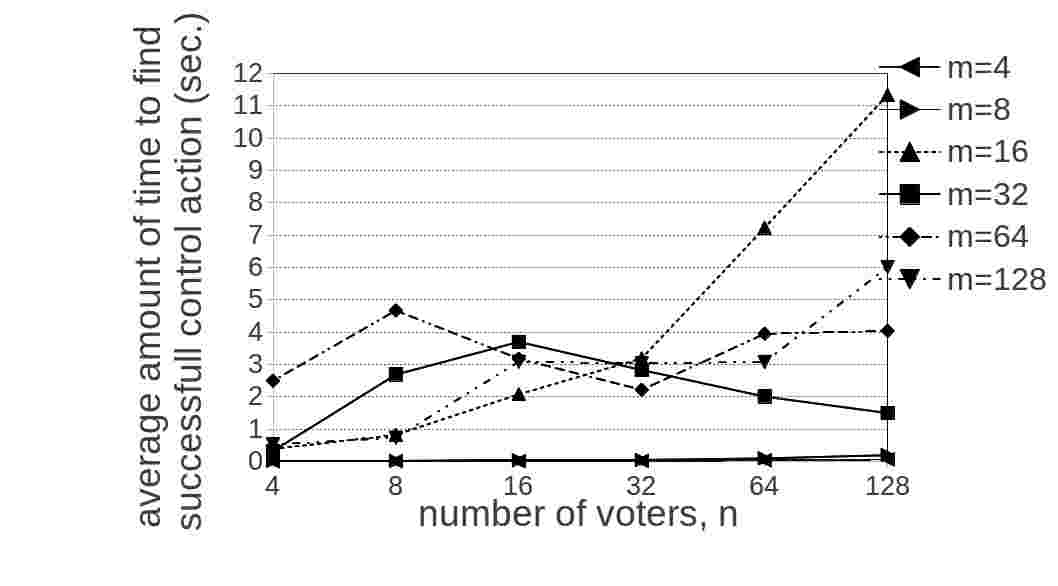}
	\caption{Average time the algorithm needs to find a successful control action for 
	destructive control by adding candidates
	in fallback elections in the TM model. The maximum is $11,33$ seconds.}
\end{figure}
\begin{figure}[ht]
\centering
	\includegraphics[scale=0.3]{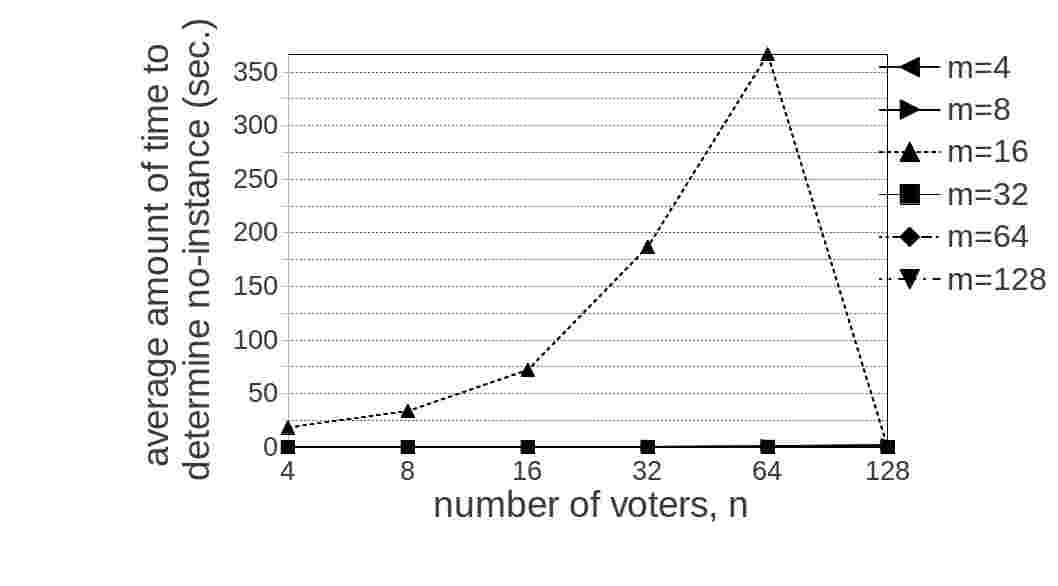}
	\caption{Average time the algorithm needs to determine no-instance of 
		destructive control by adding candidates
	in fallback elections in the TM model. The maximum is $366,67$ seconds.}
\end{figure}
\begin{figure}[ht]
\centering
	\includegraphics[scale=0.3]{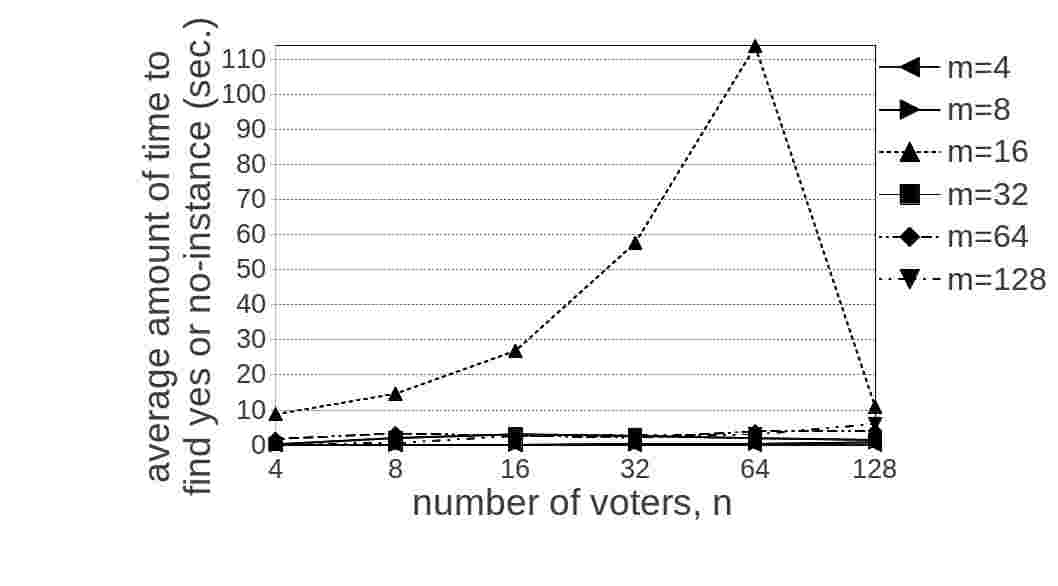}
	\caption{Average time the algorithm needs to give a definite output for 
	destructive control by adding candidates
	in fallback elections in the TM model. The maximum is $113,63$ seconds.}
\end{figure}
\clearpage
\subsection{Constructive Control by Deleting Candidates}
\begin{center}
\begin{figure}[ht]
\centering
	\includegraphics[scale=0.3]{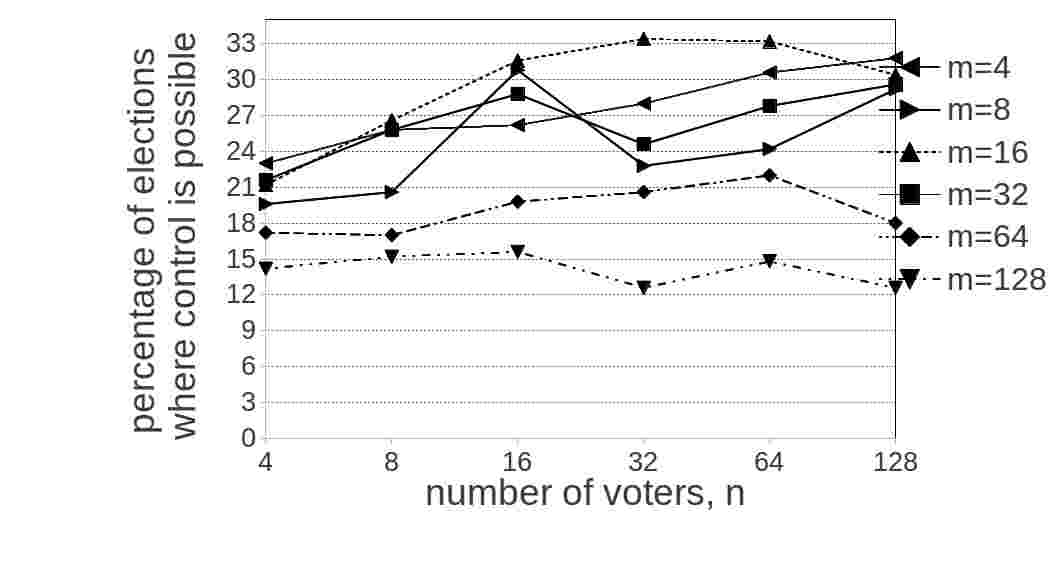}
		\caption{Results for fallback voting in the IC model for 
constructive control by deleting candidates. Number of candidates is fixed. }
\end{figure}

\begin{tabular}{||c||c|c|c|c|c|c||c|c|c|c|c|c||}
\hline
				& \multicolumn{6}{c||}{$m=4$} 				& \multicolumn{6}{|c||}{$m=8$}\\\hline
$n$ 		& $4$ 	& $8$ 	& $16$ 	& $32$ 	& $64$ 	& $128$ & $4$ 	& $8$ 	& $16$ 	& $32$ 	& $64$ 	& $128$\\\hline
$\#$ cp 	& $115$	& $129$	& $131$	& $140$ & $153$	& $159$ & $98$ 	& $103$	& $154$	& $114$	& $121$	& $146$ \\\hline
$\#$ ci 	& $385$	& $371$	& $369$	& $360$ & $347$	& $341$	& $402$	& $397$	& $346$	& $386$	& $379$ & $354$ \\\hline
$\#$ to 	& $0$ 	& $0$ 	& $0$ 	& $0$  	& $0$ 	& $0$  	& $0$ 	& $0$ 	& $0$ 	& $0$ 	& $0$ 	& $0$ \\\hline
				& \multicolumn{6}{c||}{$m=16$} 				& \multicolumn{6}{|c||}{$m=32$}\\\hline
$n$ 		& $4$ 	& $8$ 	& $16$ 	& $32$ 	& $64$ 	& $128$ & $4$ 	& $8$ 	& $16$ 	& $32$ 	& $64$ 	& $128$\\\hline
$\#$ cp 	& $106$	& $133$	& $158$ & $167$	& $166$	& $152$ & $109$ & $129$ & $144$	& $123$	& $139$ & $148$ \\\hline
$\#$ ci 	& $394$	& $367$	& $342$	& $333$	& $334$ & $348$	& $33$ 	& $2$ 	& $0$ 	& $0$ 	& $0$ 	& $0$ \\\hline
$\#$ to 	& $0$ 	& $0$ 	& $0$ 	& $0$  	& $0$ 	& $0$  	& $358$ & $369$ & $356$ & $377$ & $361$	& $352$ \\\hline
				& \multicolumn{6}{c||}{$m=64$} 				& \multicolumn{6}{|c||}{$m=128$}\\\hline
$n$ 		& $4$ 	& $8$ 	& $16$ 	& $32$ 	& $64$ 	& $128$ & $4$ 	& $8$ 	& $16$ 	& $32$ 	& $64$ 	& $128$\\\hline
$\#$ cp 	& $86$ 	& $85$ 	& $99$ 	& $103$ & $110$	& $90$  & $71$ 	& $76$ 	& $78$ 	& $63$ 	& $74$ 	& $63$ \\\hline
$\#$ ci 	& $29$ 	& $1$ 	& $0$ 	& $0$  	& $0$ 	& $0$  	& $0$ 	& $28$ 	& $30$ 	& $23$ 	& $0$ 	& $0$ \\\hline
$\#$ to 	& $385$ & $414$ & $401$ & $397$ & $390$ & $410$ & $429$	& $396$	& $392$ & $414$ & $426$ & $437$ \\\hline
\end{tabular}

\newpage
\begin{figure}[ht]
\centering
	\includegraphics[scale=0.3]{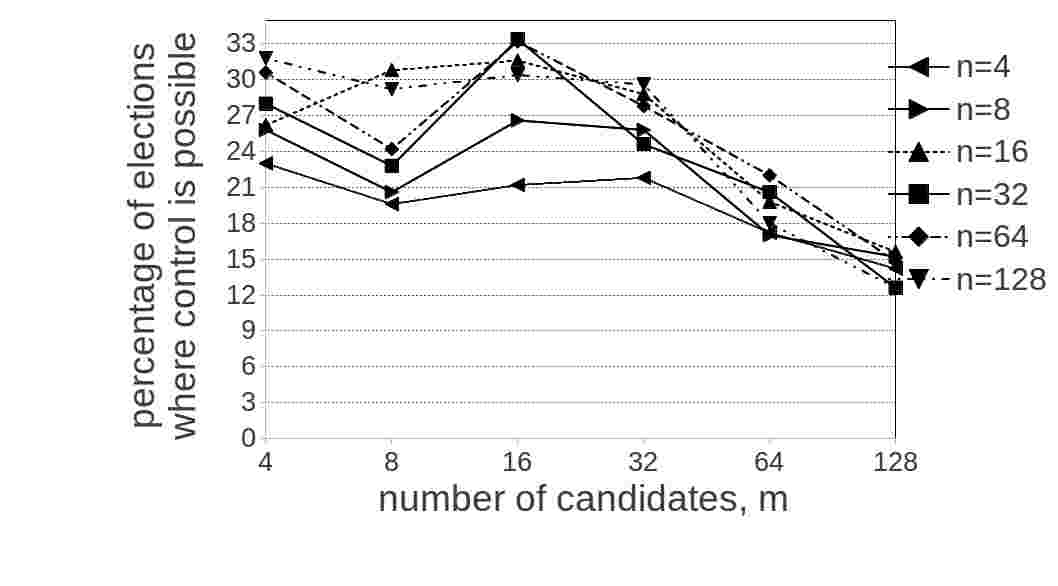}
		\caption{Results for fallback voting in the IC model for 
constructive control by deleting candidates. Number of voters is fixed. }
\end{figure}

\begin{tabular}{||c||c|c|c|c|c|c||c|c|c|c|c|c||}
\hline
		& \multicolumn{6}{c||}{$n=4$} 						& \multicolumn{6}{|c||}{$n=8$}\\\hline
$m$ 		& $4$ 	& $8$ 	& $16$ 	& $32$ 	& $64$ 	& $128$ & $4$ 	& $8$ 	& $16$ 	& $32$ 	& $64$ 	& $128$\\\hline
$\#$ cp 	& $115$	& $98$ 	& $106$	& $109$	& $86$ 	& $71$ 	& $129$ & $103$	& $133$	& $129$ & $85$ 	& $76$ \\\hline
$\#$ ci 	& $385$	& $402$	& $394$	& $33$  & $29$ 	& $0$  	& $371$	& $397$	& $367$	& $2$ 	& $1$ 	& $28$ \\\hline
$\#$ to 	& $0$ 	& $0$ 	& $0$ 	& $358$ & $385$	& $429$	& $0$ 	& $0$ 	& $0$ 	& $369$ & $414$ & $396$ \\\hline
				& \multicolumn{6}{c||}{$n=16$} 				& \multicolumn{6}{|c||}{$n=32$}\\\hline
$m$ 		& $4$ 	& $8$ 	& $16$ 	& $32$ 	& $64$ 	& $128$ & $4$ 	& $8$ 	& $16$ 	& $32$ 	& $64$ 	& $128$\\\hline
$\#$ cp 	& $131$	& $154$	& $158$	& $144$ & $99$ 	& $78$  & $140$	& $114$	& $167$	& $123$	& $103$	& $63$ \\\hline
$\#$ ci 	& $369$	& $346$	& $342$	& $0$  	& $0$ 	& $30$  & $360$	& $386$	& $333$	& $0$ 	& $0$ 	& $23$ \\\hline
$\#$ to 	& $0$ 	& $0$ 	& $0$ 	& $356$	& $401$ & $392$	& $0$ 	& $0$ 	& $0$ 	& $377$	& $397$	& $414$ \\\hline
				& \multicolumn{6}{c||}{$n=64$} 				& \multicolumn{6}{|c||}{$n=128$}\\\hline
$m$ 		& $4$ 	& $8$ 	& $16$ 	& $32$ 	& $64$ 	& $128$ & $4$ 	& $8$ 	& $16$ 	& $32$ 	& $64$ 	& $128$\\\hline
$\#$ cp 	& $153$	& $121$	& $166$ & $139$	& $110$ & $74$ 	& $159$	& $146$	& $152$	& $148$ & $90$ 	& $63$ \\\hline
$\#$ ci 	& $347$	& $379$	& $334$ & $0$  	& $0$ 	& $0$  	& $341$	& $354$	& $348$	& $0$ 	& $0$ 	& $0$ \\\hline
$\#$ to 	& $0$ 	& $0$ 	& $0$ 	& $361$ & $390$ & $426$	& $0$ 	& $0$ 	& $0$ 	& $352$	& $410$	& $437$ \\\hline
\end{tabular}

\newpage
\begin{figure}[ht]
\centering
	\includegraphics[scale=0.3]{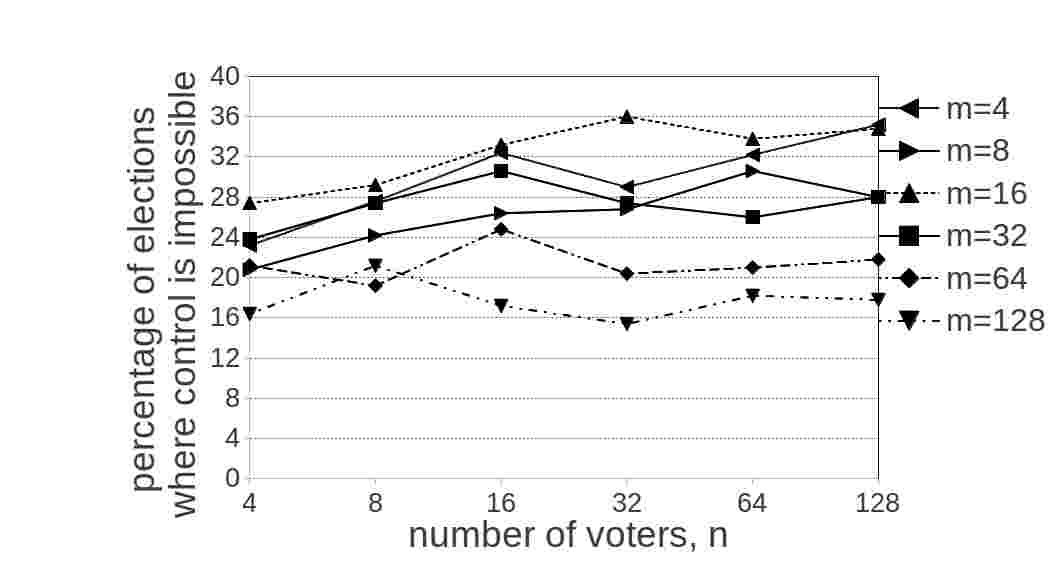}
		\caption{Results for fallback voting in the TM model for 
constructive control by deleting candidates. Number of candidates is fixed. }
\end{figure}

\begin{tabular}{||c||c|c|c|c|c|c||c|c|c|c|c|c||}
\hline
				& \multicolumn{6}{c||}{$m=4$} 				& \multicolumn{6}{|c||}{$m=8$}\\\hline
$n$ 		& $4$ 	& $8$ 	& $16$ 	& $32$ 	& $64$ 	& $128$ & $4$ 	& $8$ 	& $16$ 	& $32$ 	& $64$ 	& $128$\\\hline
$\#$ cp 	& $116$ & $138$	& $162$	& $145$	& $161$	& $176$	& $104$	& $132$	& $134$	& $134$ & $153$	& $140$ \\\hline
$\#$ ci 	& $384$ & $362$	& $338$	& $355$	& $339$	& $324$	& $396$	& $368$	& $366$	& $347$	& $347$	& $360$ \\\hline
$\#$ to 	& $0$ 	& $0$ 	& $0$ 	& $0$  	& $0$ 	& $0$  	& $0$ 	& $0$ 	& $0$ 	& $0$ 	& $0$ 	& $0$ \\\hline
				& \multicolumn{6}{c||}{$m=16$} 				& \multicolumn{6}{|c||}{$m=32$}\\\hline
$n$ 		& $4$ 	& $8$ 	& $16$ 	& $32$ 	& $64$ 	& $128$ & $4$ 	& $8$ 	& $16$ 	& $32$ 	& $64$ 	& $128$\\\hline
$\#$ cp 	& $137$	& $146$	& $166$	& $180$	& $169$	& $174$	& $119$	& $137$ & $153$	& $137$ & $130$	& $140$ \\\hline
$\#$ ci 	& $363$ & $354$	& $334$	& $320$	& $331$	& $326$	& $66$ 	& $29$ 	& $0$ 	& $0$ 	& $0$ 	& $0$ \\\hline
$\#$ to 	& $0$ 	& $0$ 	& $0$ 	& $0$  	& $0$ 	& $0$  	& $315$	& $334$	& $347$	& $363$	& $370$	& $360$ \\\hline
				& \multicolumn{6}{c||}{$m=64$} 				& \multicolumn{6}{|c||}{$m=128$}\\\hline
$n$ 		& $4$ 	& $8$ 	& $16$ 	& $32$ 	& $64$ 	& $128$ & $4$ 	& $8$ 	& $16$ 	& $32$ 	& $64$ 	& $128$\\\hline
$\#$ cp 	& $106$	& $96$ 	& $124$ & $102$ & $105$	& $109$ & $82$ 	& $106$ & $86$ 	& $77$ 	& $91$ 	& $89$ \\\hline
$\#$ ci 	& $76$ 	& $33$ 	& $10$ 	& $0$  	& $0$ 	& $0$  	& $77$ 	& $0$ 	& $0$ 	& $0$ 	& $0$ 	& $0$ \\\hline
$\#$ to 	& $318$ & $371$	& $366$ & $398$	& $395$	& $391$	& $341$	& $394$	& $414$	& $423$	& $409$ & $411$ \\\hline
\end{tabular}

\newpage
\begin{figure}[ht]
\centering
	\includegraphics[scale=0.3]{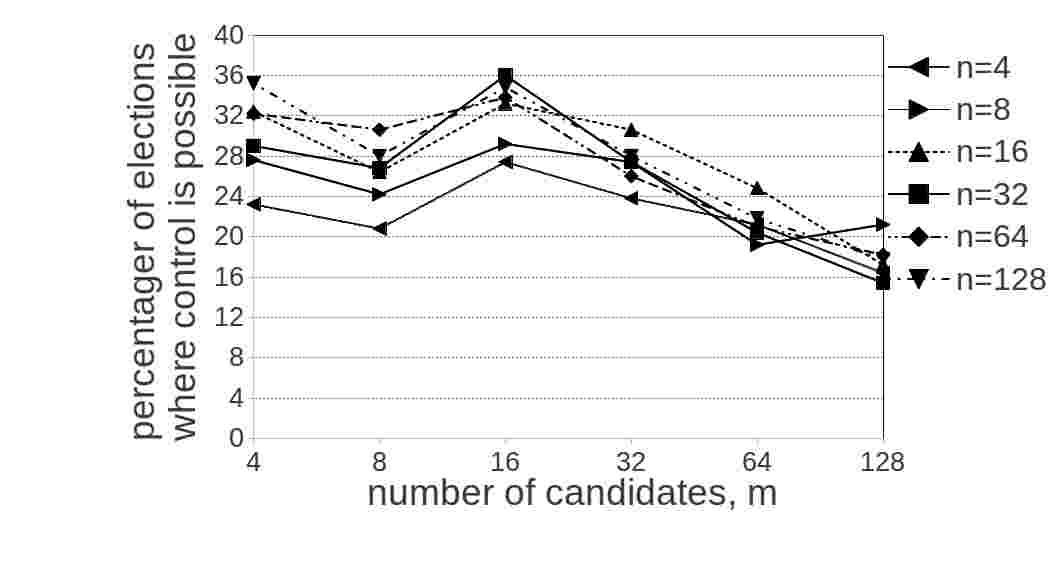}
		\caption{Results for fallback voting in the TM model for 
constructive control by deleting candidates. Number of voters is fixed. }
\end{figure}

\begin{tabular}{||c||c|c|c|c|c|c||c|c|c|c|c|c||}
\hline
& \multicolumn{6}{c||}{$n=4$} 								& \multicolumn{6}{|c||}{$n=8$}\\\hline
$m$ 		& $4$ 	& $8$ 	& $16$ 	& $32$ 	& $64$ 	& $128$ & $4$ 	& $8$ 	& $16$ 	& $32$ 	& $64$ 	& $128$\\\hline
$\#$ cp 	& $116$	& $104$	& $137$	& $119$	& $106$ & $82$  & $138$	& $121$	& $146$	& $137$	& $96$ 	& $106$ \\\hline
$\#$ ci 	& $384$	& $396$	& $363$	& $66$  & $76$ 	& $77$  & $362$	& $379$	& $354$	& $29$ 	& $33$ 	& $0$ \\\hline
$\#$ to 	& $0$ 	& $0$ 	& $0$ 	& $315$ & $318$ & $341$ & $0$ 	& $0$ 	& $0$ 	& $334$ & $371$ & $394$ \\\hline
				& \multicolumn{6}{c||}{$n=16$} 				& \multicolumn{6}{|c||}{$n=32$}\\\hline
$m$ 		& $4$ 	& $8$ 	& $16$ 	& $32$ 	& $64$ 	& $128$ & $4$ 	& $8$ 	& $16$ 	& $32$ 	& $64$ 	& $128$\\\hline
$\#$ cp 	& $162$	& $132$	& $166$	& $153$	& $124$	& $86$ 	& $145$	& $134$	& $180$	& $137$	& $102$	& $77$ \\\hline
$\#$ ci 	& $338$	& $368$	& $334$	& $0$  	& $0$ 	& $0$  	& $355$	& $366$	& $320$ & $0$ 	& $0$ 	& $0$ \\\hline
$\#$ to 	& $0$ 	& $0$ 	& $0$ 	& $347$	& $366$	& $414$ & $0$ 	& $0$ 	& $0$ 	& $363$	& $398$	& $423$ \\\hline
				& \multicolumn{6}{c||}{$n=64$} 				& \multicolumn{6}{|c||}{$n=128$}\\\hline
$m$ 		& $4$ 	& $8$ 	& $16$ 	& $32$ 	& $64$ 	& $128$ & $4$ 	& $8$ 	& $16$ 	& $32$ 	& $64$ 	& $128$\\\hline
$\#$ cp 	& $161$	& $153$	& $169$	& $130$	& $105$	& $91$ 	& $176$	& $140$	& $174$	& $140$ & $109$ & $89$ \\\hline
$\#$ ci 	& $339$	& $347$	& $331$	& $0$  	& $0$ 	& $0$  	& $324$	& $360$	& $326$	& $0$ 	& $0$ 	& $0$ \\\hline
$\#$ to 	& $0$ 	& $0$ 	& $0$ 	& $370$ & $395$ & $409$	& $0$ 	& $0$ 	& $0$ 	& $360$ & $391$ & $411$ \\\hline
\end{tabular}
\end{center}

\clearpage
\subsubsection{Computation Costs}
\begin{figure}[ht]
\centering
	\includegraphics[scale=0.3]{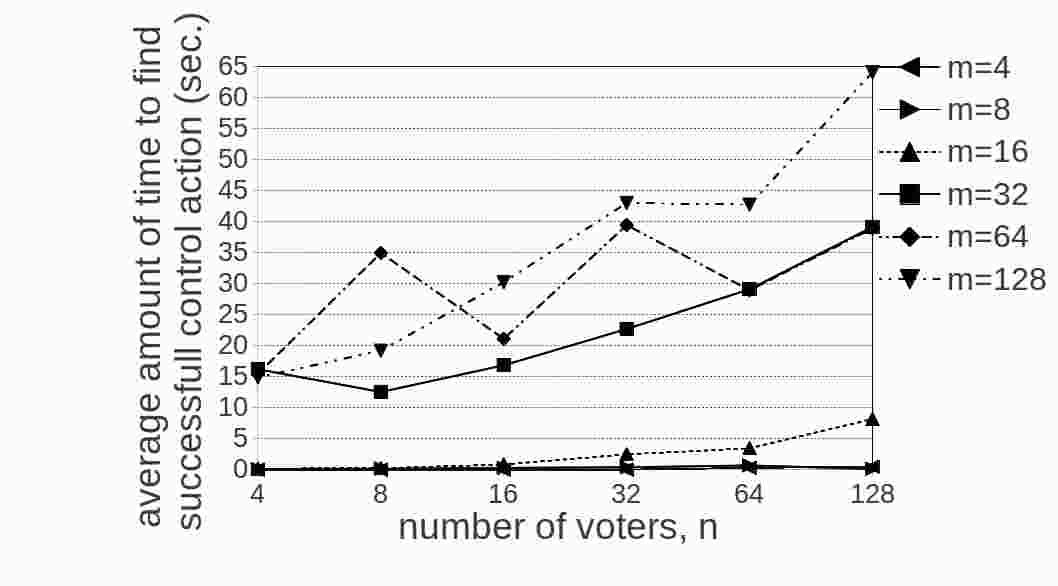}
	\caption{Average time the algorithm needs to find a successful control action for 
	constructive control by deleting candidates
	in fallback elections in the IC model. The maximum is $64,05$ seconds.}
\end{figure}
\begin{figure}[ht]
\centering
	\includegraphics[scale=0.3]{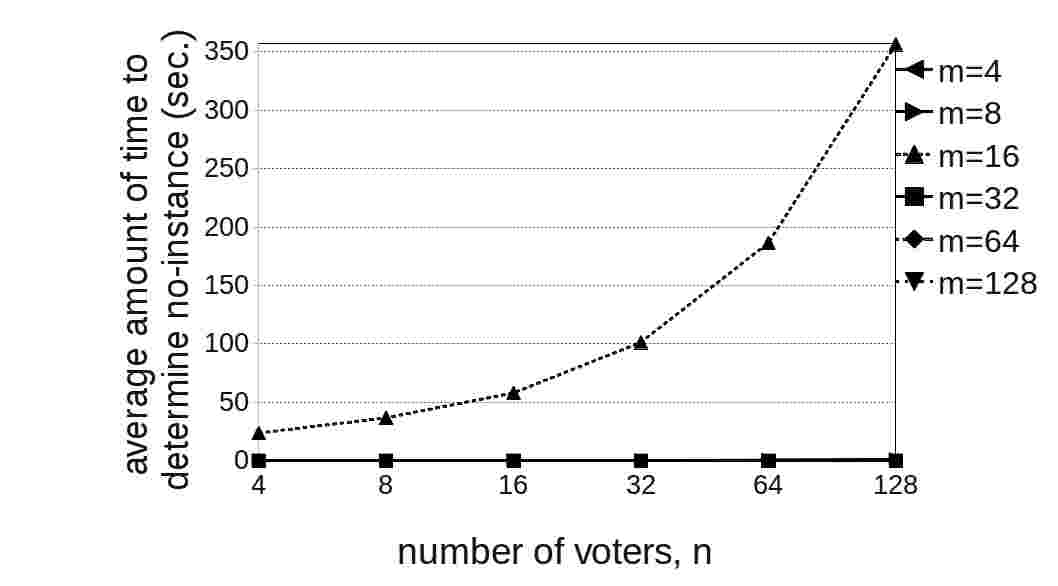}
	\caption{Average time the algorithm needs to determine no-instance of 
		constructive control by deleting candidates
	in fallback elections in the IC model. The maximum is $356,18$ seconds.}
\end{figure}

\begin{figure}[ht]
\centering
	\includegraphics[scale=0.3]{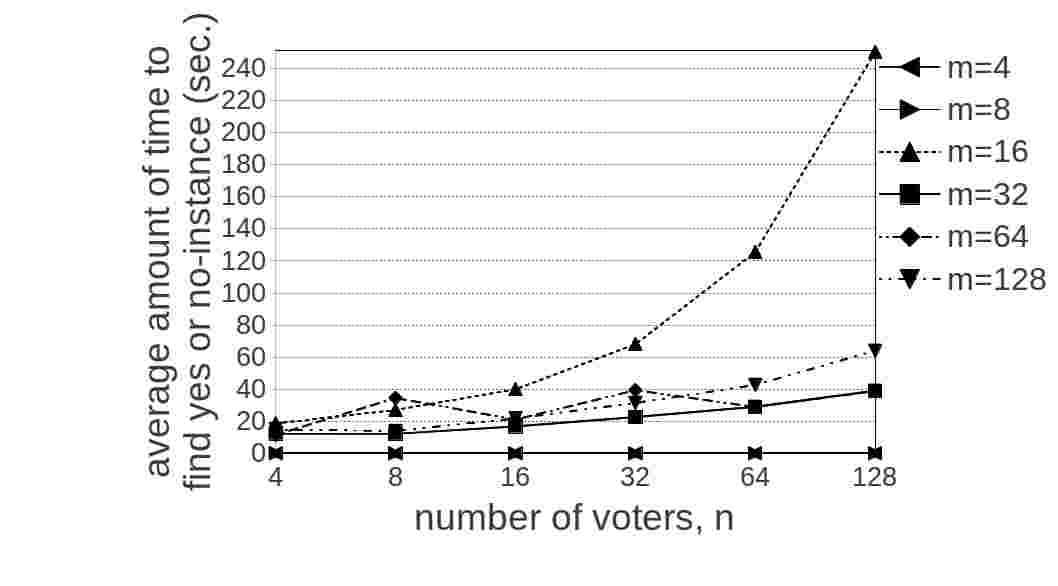}
	\caption{Average time the algorithm needs to give a definite output for 
	constructive control by deleting candidates
	in fallback elections in the IC model. The maximum is $250,39$ seconds.}
\end{figure}
\begin{figure}[ht]
\centering
	\includegraphics[scale=0.3]{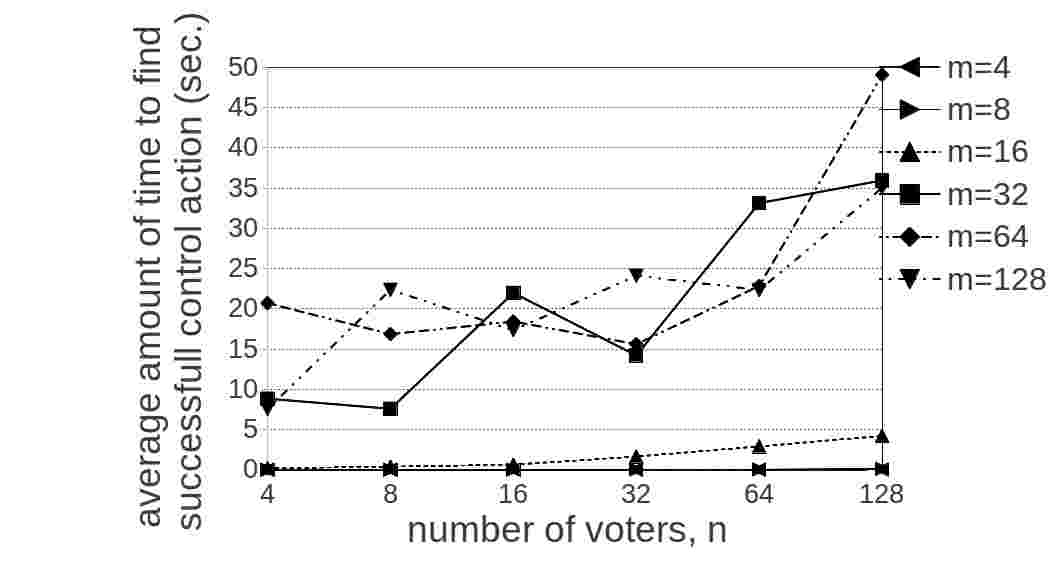}
	\caption{Average time the algorithm needs to find a successful control action for 
	constructive control by deleting candidates
	in fallback elections in the TM model. The maximum is $49,05$ seconds.}
\end{figure}
\begin{figure}[ht]
\centering
	\includegraphics[scale=0.3]{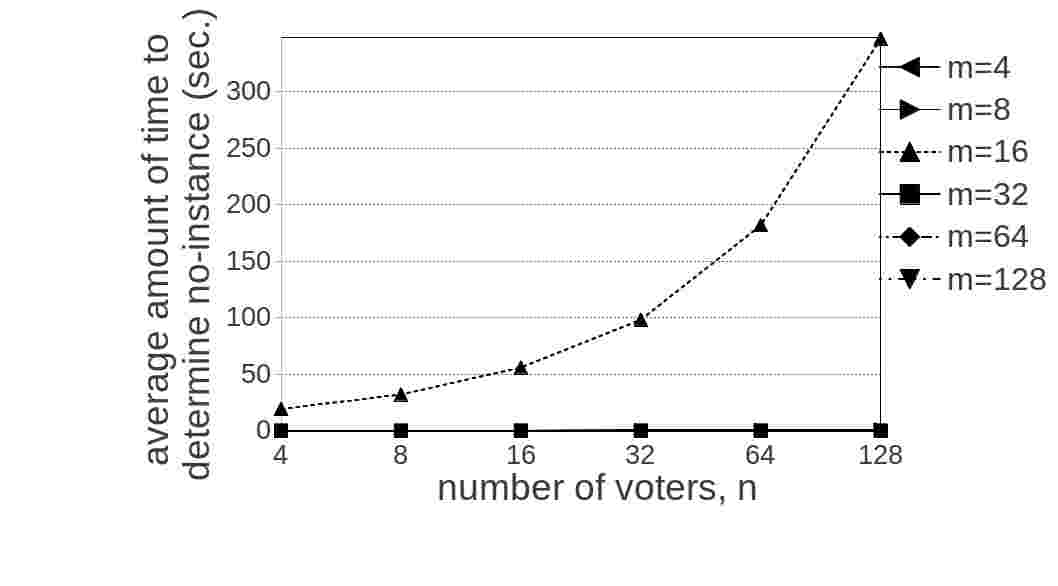}
	\caption{Average time the algorithm needs to determine no-instance of 
		constructive control by deleting candidates
	in fallback elections in the TM model. The maximum is $347,06$ seconds.}
\end{figure}
\begin{figure}[ht]
\centering
	\includegraphics[scale=0.3]{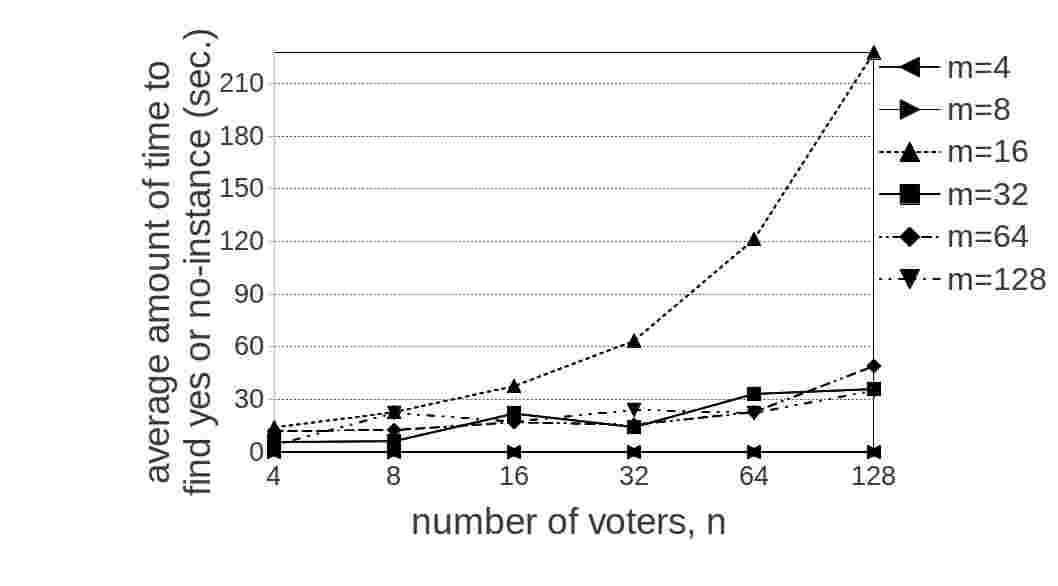}
	\caption{Average time the algorithm needs to give a definite output for 
	constructive control by deleting candidates
	in fallback elections in the TM model. The maximum is $227,76$ seconds.}
\end{figure}

\clearpage
\subsection{Destructive Control by Deleting Candidates}
\begin{center}
\begin{figure}[ht]
\centering
	\includegraphics[scale=0.3]{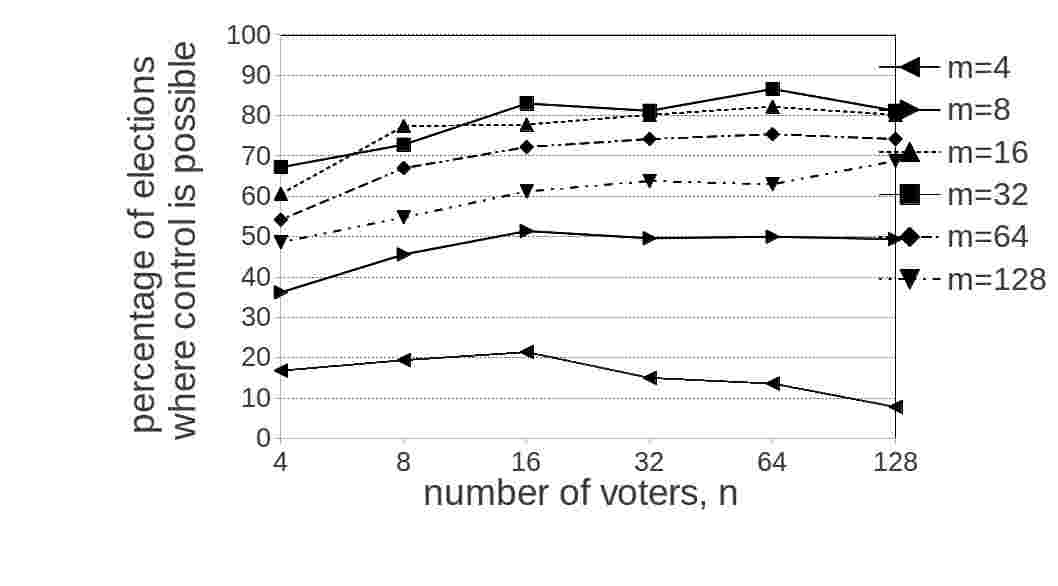}
		\caption{Results for fallback voting in the IC model for 
destructive control by deleting candidates. Number of candidates is fixed. }
\end{figure}

\begin{tabular}{||c||c|c|c|c|c|c||c|c|c|c|c|c||}
\hline
				& \multicolumn{6}{c||}{$m=4$} 				& \multicolumn{6}{|c||}{$m=8$}\\\hline
$n$ 		& $4$ 	& $8$ 	& $16$ 	& $32$ 	& $64$ 	& $128$ & $4$ 	& $8$ 	& $16$ 	& $32$ 	& $64$ 	& $128$\\\hline
$\#$ cp 	& $84$ 	& $97$ 	& $107$	& $75$  & $68$ 	& $39$ 	& $181$ & $228$	& $257$ & $248$	& $250$ & $247$ \\\hline
$\#$ ci 	& $416$	& $403$	& $393$	& $425$ & $432$	& $461$	& $319$	& $272$ & $243$	& $252$ & $250$	& $253$ \\\hline
$\#$ to 	& $0$ 	& $0$ 	& $0$ 	& $0$  	& $0$ 	& $0$  	& $0$ 	& $0$ 	& $0$ 	& $0$ 	& $0$ 	& $0$ \\\hline
				& \multicolumn{6}{c||}{$m=16$} 				& \multicolumn{6}{|c||}{$m=32$}\\\hline
$n$ 		& $4$ 	& $8$ 	& $16$ 	& $32$ 	& $64$ 	& $128$ & $4$ 	& $8$ 	& $16$ 	& $32$ 	& $64$ 	& $128$\\\hline
$\#$ cp 	& $303$ & $387$ & $389$ & $99$ 	& $411$ & $401$	& $336$ & $364$	& $415$ & $406$ & $433$ & $406$ \\\hline
$\#$ ci 	& $197$	& $113$ & $111$	& $99$  & $89$ 	& $99$  & $1$ 	& $0$ 	& $0$ 	& $0$ 	& $0$ 	& $0$ \\\hline
$\#$ to 	& $0$ 	& $0$ 	& $0$ 	& $0$  	& $0$ 	& $0$  	& $163$ & $136$ & $85$ 	& $94$ 	& $67$ 	& $94$ \\\hline
				& \multicolumn{6}{c||}{$m=64$} 				& \multicolumn{6}{|c||}{$m=128$}\\\hline
$n$ 		& $4$ 	& $8$ 	& $16$ 	& $32$ 	& $64$ 	& $128$ & $4$ 	& $8$ 	& $16$ 	& $32$ 	& $64$ 	& $128$\\\hline
$\#$ cp 	& $271$ & $335$ & $361$ & $371$ & $377$ & $371$ & $243$ & $274$ & $306$ & $319$	& $315$ & $344$ \\\hline
$\#$ ci 	& $0$ 	& $0$ 	& $0$ 	& $0$  	& $0$ 	& $0$  	& $0$ 	& $0$ 	& $0$ 	& $0$ 	& $0$ 	& $0$ \\\hline
$\#$ to 	& $229$ & $165$ & $139$ & $129$ & $123$ & $129$ & $257$ & $226$ & $194$ & $181$ & $185$ & $156$ \\\hline
\end{tabular}

\newpage
\begin{figure}[ht]
\centering
	\includegraphics[scale=0.3]{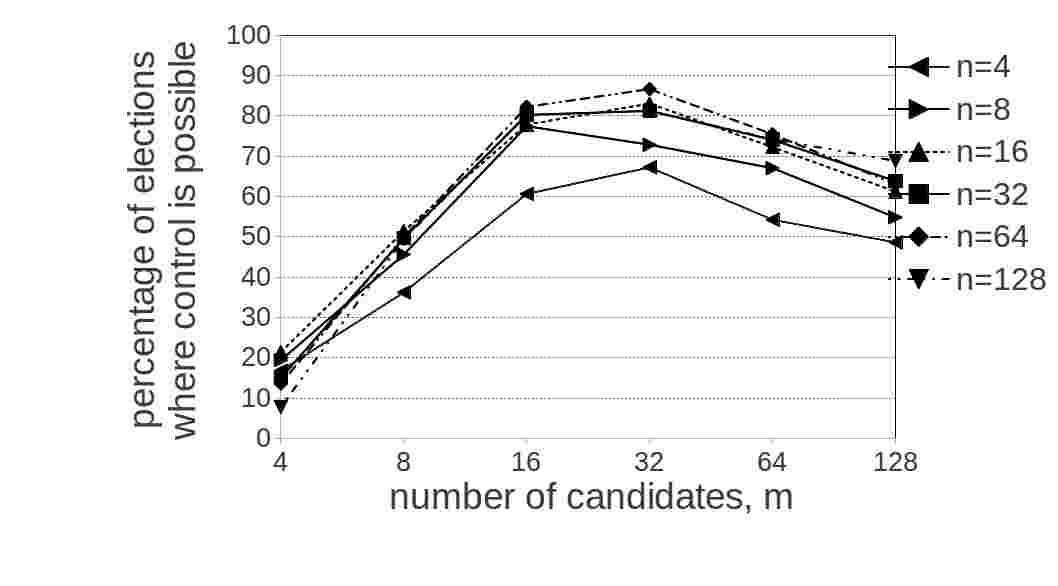}
		\caption{Results for fallback voting in the IC model for 
destructive control by deleting candidates. Number of voters is fixed. }
\end{figure}

\begin{tabular}{||c||c|c|c|c|c|c||c|c|c|c|c|c||}
\hline
		& \multicolumn{6}{c||}{$n=4$} 						& \multicolumn{6}{|c||}{$n=8$}\\\hline
$m$ 		& $4$ 	& $8$ 	& $16$ 	& $32$ 	& $64$ 	& $128$ & $4$ 	& $8$ 	& $16$ 	& $32$ 	& $64$ 	& $128$\\\hline
$\#$ cp 	& $84$ 	& $181$ & $303$ & $336$ & $271$ & $243$ & $97$ 	& $228$ & $387$ & $364$ & $335$ & $274$ \\\hline
$\#$ ci 	& $416$	& $319$ & $197$ & $1$  	& $0$ 	& $0$  	& $403$	& $272$ & $113$ & $0$ 	& $0$ 	& $0$ \\\hline
$\#$ to 	& $0$ 	& $0$ 	& $0$ 	& $163$ & $229$ & $257$ & $0$ 	& $0$ 	& $0$ 	& $136$ & $165$ & $226$ \\\hline
				& \multicolumn{6}{c||}{$n=16$} 				& \multicolumn{6}{|c||}{$n=32$}\\\hline
$m$ 		& $4$ 	& $8$ 	& $16$ 	& $32$ 	& $64$ 	& $128$ & $4$ 	& $8$ 	& $16$ 	& $32$ 	& $64$ 	& $128$\\\hline
$\#$ cp 	& $107$ & $257$ & $389$ & $415$ & $361$ & $306$ & $75$ 	& $248$	& $401$ & $405$ & $371$ & $319$ \\\hline
$\#$ ci 	& $393$	& $243$ & $111$ & $0$  	& $0$ 	& $0$  	& $425$ & $252$ & $99$ 	& $0$ 	& $0$ 	& $0$ \\\hline
$\#$ to 	& $0$ 	& $0$ 	& $0$ 	& $85$  & $139$ & $194$ & $0$ 	& $0$ 	& $0$ 	& $94$ 	& $19$ 	& $181$ \\\hline
				& \multicolumn{6}{c||}{$n=64$} 				& \multicolumn{6}{|c||}{$n=128$}\\\hline
$m$ 		& $4$ 	& $8$ 	& $16$ 	& $32$ 	& $64$ 	& $128$ & $4$ 	& $8$ 	& $16$ 	& $32$ 	& $64$ 	& $128$\\\hline
$\#$ cp 	& $68$ 	& $250$ & $411$ & $433$ & $377$ & $315$ & $39$ 	& $247$ & $401$ & $406$ & $371$ & $344$ \\\hline
$\#$ ci 	& $432$	& $250$ & $89$ 	& $0$  	& $0$ 	& $0$  	& $461$ & $253$ & $99$ 	& $0$ 	& $0$ 	& $0$ \\\hline
$\#$ to 	& $0$ 	& $0$ 	& $0$ 	& $67$  & $123$ & $185$ & $0$ 	& $0$ 	& $0$ 	& $94$ 	& $129$ & $156$ \\\hline
\end{tabular}
\newpage
\begin{figure}[ht]
\centering
	\includegraphics[scale=0.3]{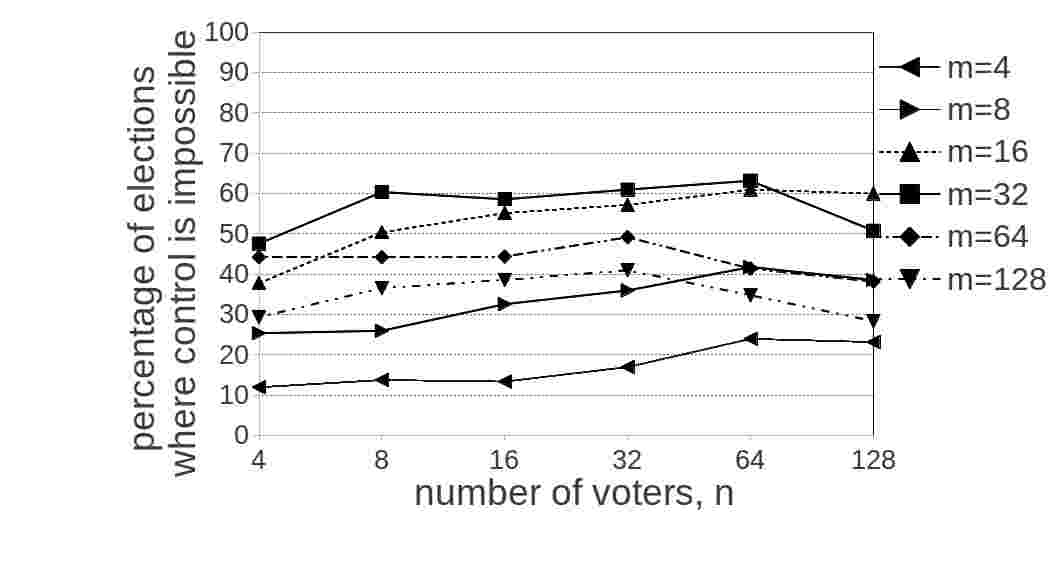}
		\caption{Results for fallback voting in the TM model for 
destructive control by deleting candidates. Number of candidates is fixed. }
\end{figure}

\begin{tabular}{||c||c|c|c|c|c|c||c|c|c|c|c|c||}
\hline
				& \multicolumn{6}{c||}{$m=4$} 				& \multicolumn{6}{|c||}{$m=8$}\\\hline
$n$ 		& $4$ 	& $8$ 	& $16$ 	& $32$ 	& $64$ 	& $128$ & $4$ 	& $8$ 	& $16$ 	& $32$ 	& $64$ 	& $128$\\\hline
$\#$ cp 	& $60$ 	& $69$ 	& $67$ 	& $85$ 	& $120$ & $116$ & $127$ & $130$ & $163$ & $180$ & $209$ & $193$ \\\hline
$\#$ ci 	& $440$	& $431$ & $433$ & $415$ & $380$ & $384$ & $373$ & $370$ & $337$ & $320$ & $291$ & $307$ \\\hline
$\#$ to 	& $0$ 	& $0$ 	& $0$ 	& $0$  	& $0$ 	& $0$  	& $0$ 	& $0$ 	& $0$ 	& $0$ 	& $0$ 	& $0$ \\\hline
				& \multicolumn{6}{c||}{$m=16$} 				& \multicolumn{6}{|c||}{$m=32$}\\\hline
$n$ 		& $4$ 	& $8$ 	& $16$ 	& $32$ 	& $64$ 	& $128$ & $4$ 	& $8$ 	& $16$ 	& $32$ 	& $64$ 	& $128$\\\hline
$\#$ cp 	& $189$ & $252$ & $276$ & $286$ & $305$ & $300$ & $238$ & $302$ & $293$ & $305$ & $316$	& $254$ \\\hline
$\#$ ci 	& $311$	& $248$ & $224$ & $214$ & $195$ & $200$ & $153$ & $104$ & $81$ 	& $34$ 	& $21$ 	& $15$ \\\hline
$\#$ to 	& $0$ 	& $0$ 	& $0$ 	& $0$  	& $0$ 	& $0$  	& $109$ & $94$ 	& $126$ & $161$ & $163$ & $231$ \\\hline
				& \multicolumn{6}{c||}{$m=64$} 				& \multicolumn{6}{|c||}{$m=128$}\\\hline
$n$ 		& $4$ 	& $8$ 	& $16$ 	& $32$ 	& $64$ 	& $128$ & $4$ 	& $8$ 	& $16$ 	& $32$ 	& $64$ 	& $128$\\\hline
$\#$ cp 	& $221$ & $221$ & $222$ & $26$ 	& $207$ & $191$ & $147$ & $183$ & $193$ & $205$ & $174$ & $142$ \\\hline
$\#$ ci 	& $110$ & $96$ 	& $69$ 	& $23$  & $8$ 	& $6$  	& $123$ & $101$ & $66$ 	& $18$ 	& $7$ 	& $7$ \\\hline
$\#$ to 	& $169$	& $183$ & $209$ & $231$ & $285$ & $303$	& $230$	& $216$ & $241$ & $277$ & $319$ & $351$ \\\hline
\end{tabular}
\newpage
\begin{figure}[ht]
\centering
	\includegraphics[scale=0.3]{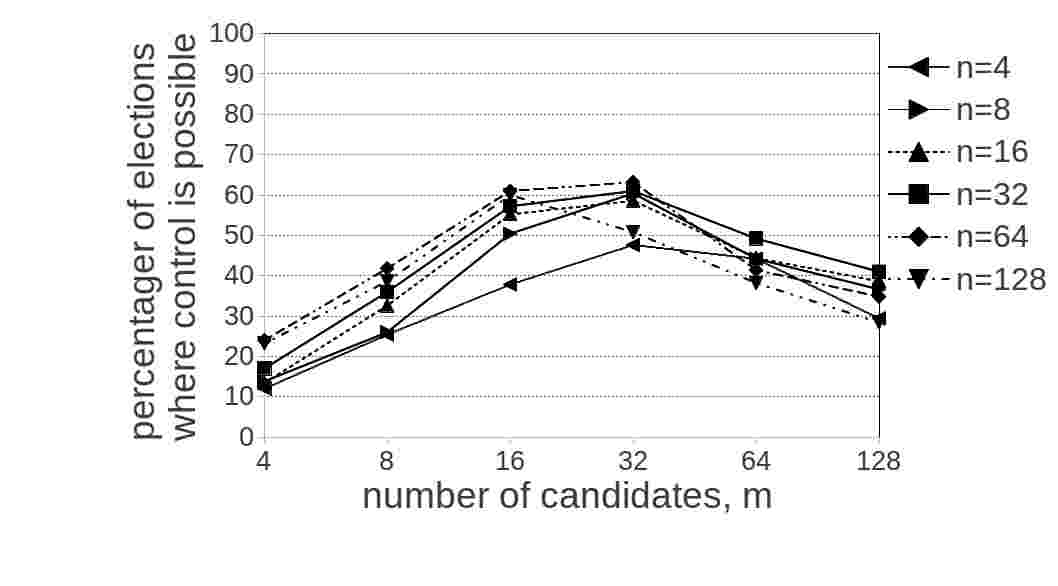}
		\caption{Results for fallback voting in the TM model for 
destructive control by deleting candidates. Number of voters is fixed. }
\end{figure}

\begin{tabular}{||c||c|c|c|c|c|c||c|c|c|c|c|c||}
\hline
& \multicolumn{6}{c||}{$n=4$} 								& \multicolumn{6}{|c||}{$n=8$}\\\hline
$m$ 		& $4$ 	& $8$ 	& $16$ 	& $32$ 	& $64$ 	& $128$ & $4$ 	& $8$ 	& $16$ 	& $32$ 	& $64$ 	& $128$\\\hline
$\#$ cp 	& $60$ 	& $127$ & $189$ & $238$ & $221$ & $147$ & $69$ 	& $130$ & $252$ & $302$ & $221$ & $183$ \\\hline
$\#$ ci 	& $440$	& $373$ & $311$ & $153$ & $110$ & $123$ & $431$ & $370$ & $248$ & $104$ & $96$ 	& $101$ \\\hline
$\#$ to 	& $0$ 	& $0$ 	& $0$ 	& $109$ & $169$ & $230$ & $0$ 	& $0$ 	& $0$ 	& $94$ 	& $183$ & $216$ \\\hline
				& \multicolumn{6}{c||}{$n=16$} 				& \multicolumn{6}{|c||}{$n=32$}\\\hline
$m$ 		& $4$ 	& $8$ 	& $16$ 	& $32$ 	& $64$ 	& $128$ & $4$ 	& $8$ 	& $16$ 	& $32$ 	& $64$ 	& $128$\\\hline
$\#$ cp 	& $67$ 	& $163$ & $276$ & $293$ & $222$ & $193$ & $85$ 	& $180$ & $286$ & $305$ & $246$ & $205$ \\\hline
$\#$ ci 	& $433$ & $337$ & $224$ & $81$  & $69$ 	& $66$  & $415$	& $320$ & $214$ & $34$ 	& $23$ 	& $18$ \\\hline
$\#$ to 	& $0$ 	& $0$ 	& $0$ 	& $126$ & $209$ & $241$ & $0$ 	& $0$ 	& $0$ 	& $161$ & $231$ & $277$ \\\hline
				& \multicolumn{6}{c||}{$n=64$} 				& \multicolumn{6}{|c||}{$n=128$}\\\hline
$m$ 		& $4$ 	& $8$ 	& $16$ 	& $32$ 	& $64$ 	& $128$ & $4$ 	& $8$ 	& $16$ 	& $32$ 	& $64$ 	& $128$\\\hline
$\#$ cp 	& $120$ & $209$ & $305$ & $316$ & $207$ & $174$ & $116$ & $193$	& $300$ & $254$ & $191$ & $142$ \\\hline
$\#$ ci 	& $380$	& $291$ & $195$ & $21$  & $8$ 	& $7$  	& $384$ & $307$ & $200$ & $15$ 	& $6$ 	& $0$ \\\hline
$\#$ to 	& $0$ 	& $0$ 	& $0$ 	& $163$ & $285$ & $319$ & $0$ 	& $0$ 	& $0$ 	& $231$ & $303$	& $351$ \\\hline
\end{tabular}
\end{center}

\clearpage
\subsubsection{Computational Costs}
\begin{figure}[ht]
\centering
	\includegraphics[scale=0.3]{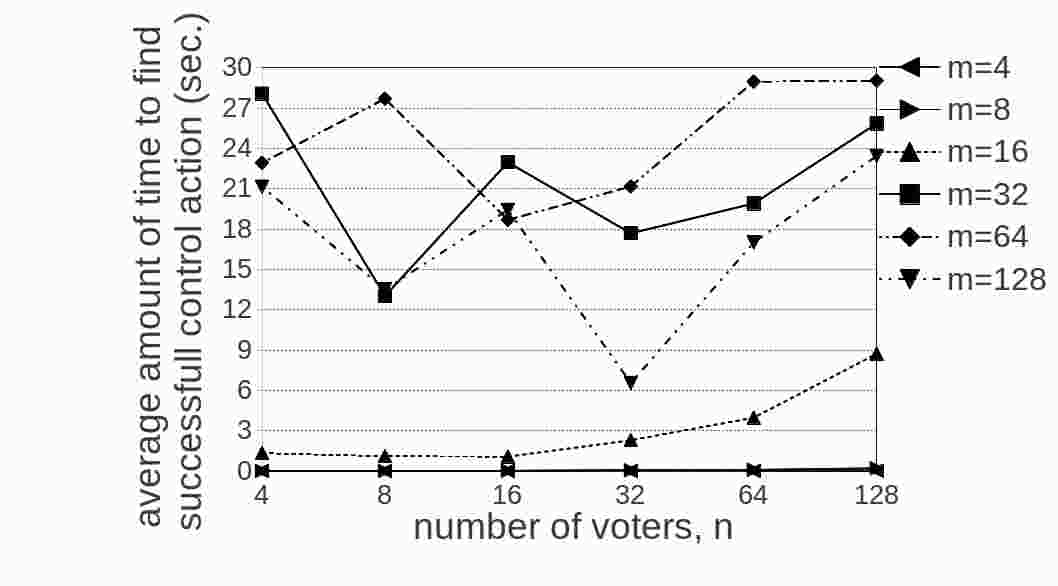}
	\caption{Average time the algorithm needs to find a successful control action for 
	destructive control by deleting candidates
	in fallback elections in the IC model. The maximum is $29,04$ seconds.}
\end{figure}
\begin{figure}[ht]
\centering
	\includegraphics[scale=0.3]{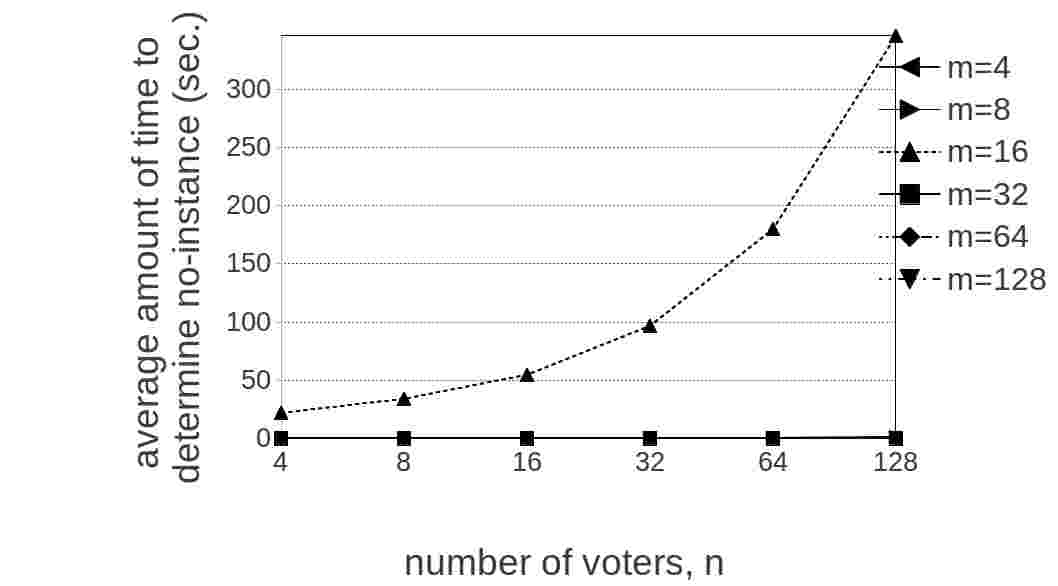}
	\caption{Average time the algorithm needs to determine no-instance of 
		destructive control by deleting candidates
	in fallback elections in the IC model. The maximum is $345,94$ seconds.}
\end{figure}

\begin{figure}[ht]
\centering
	\includegraphics[scale=0.3]{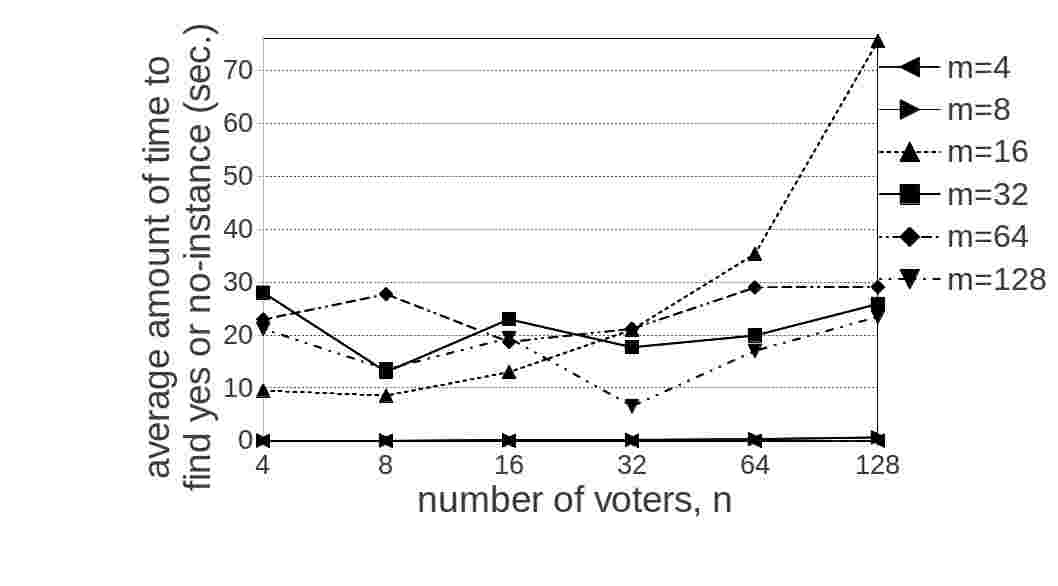}
	\caption{Average time the algorithm needs to give a definite output for 
	destructive control by deleting candidates
	in fallback elections in the IC model. The maximum is $75,5$ seconds.}
\end{figure}

\begin{figure}[ht]
\centering
	\includegraphics[scale=0.3]{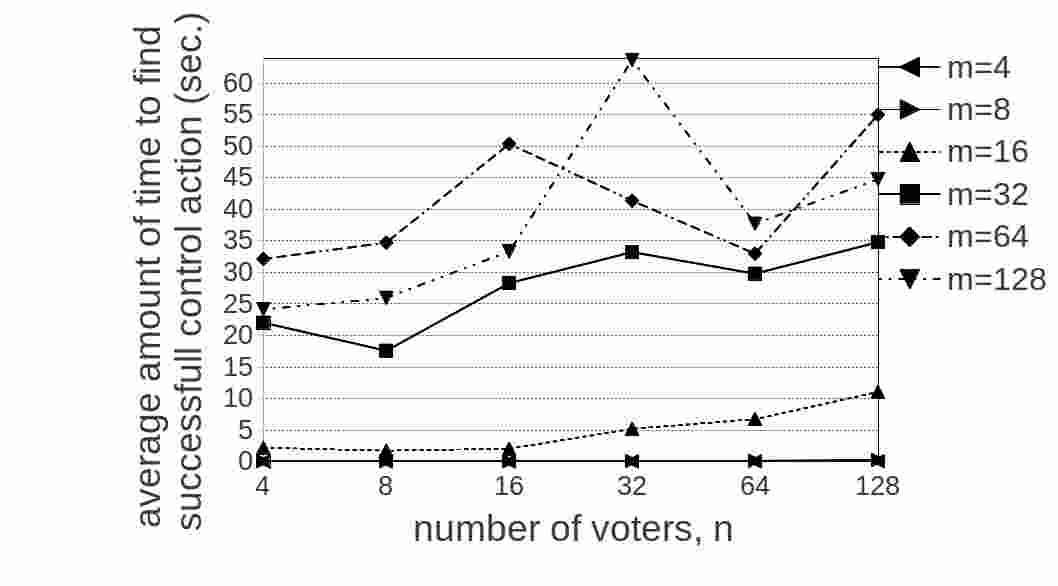}
	\caption{Average time the algorithm needs to find a successful control action for 
	destructive control by deleting candidates
	in fallback elections in the TM model. The maximum is $63,71$ seconds.}
\end{figure}
\begin{figure}[ht]
\centering
	\includegraphics[scale=0.3]{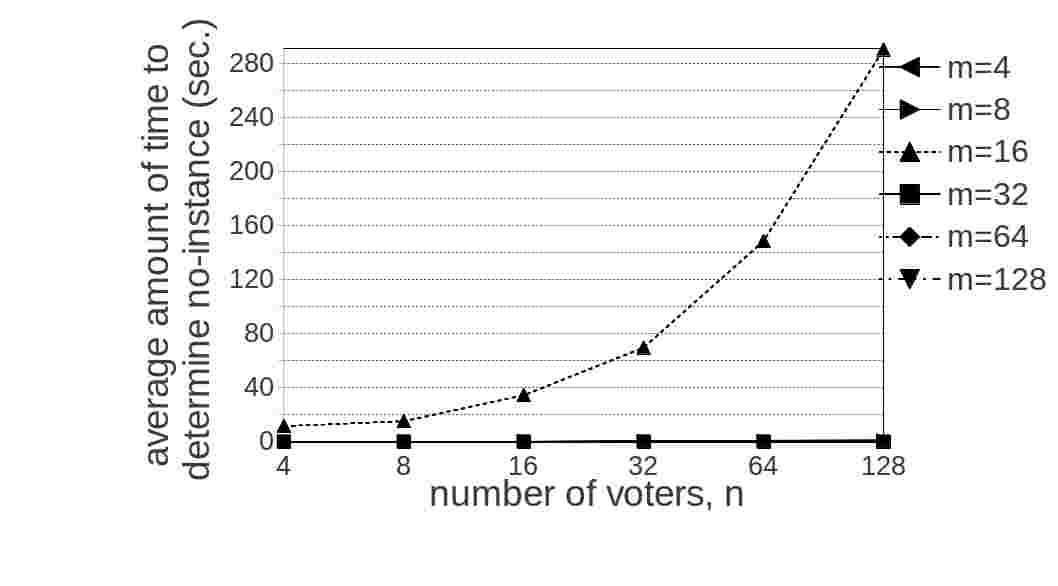}
	\caption{Average time the algorithm needs to determine no-instance of 
		destructive control by deleting candidates
	in fallback elections in the TM model. The maximum is $290,26$ seconds.}
\end{figure}

\begin{figure}[ht]
\centering
	\includegraphics[scale=0.3]{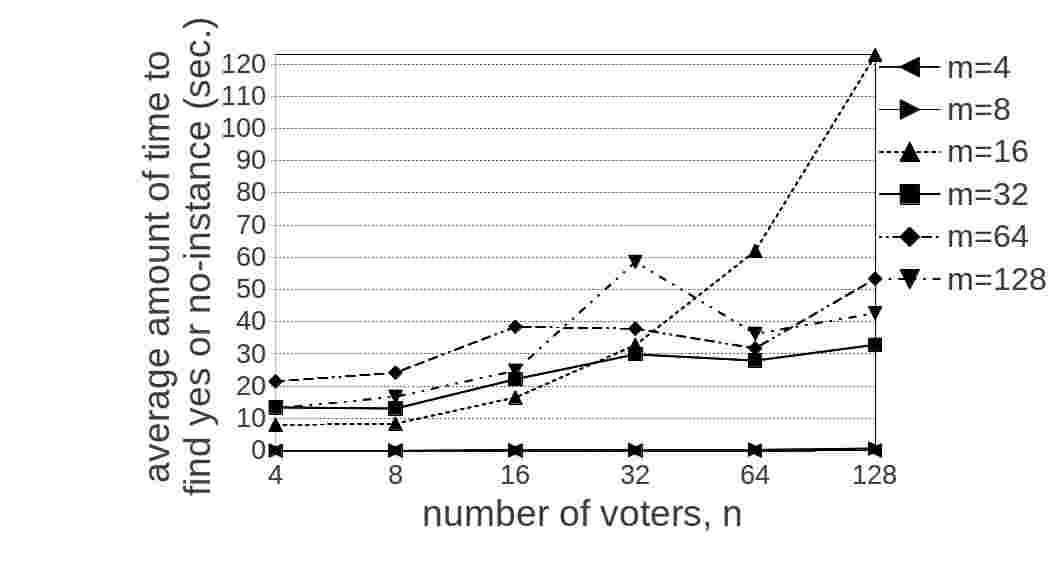}
	\caption{Average time the algorithm needs to give a definite output for 
	destructive control by deleting candidates
	in fallback elections in the TM model. The maximum is $122,72$ seconds.}
\end{figure}
\clearpage
\subsection{Constructive Control by Partition of Candidates in Model TE}
\begin{center}

\begin{figure}[ht]
\centering
	\includegraphics[scale=0.3]{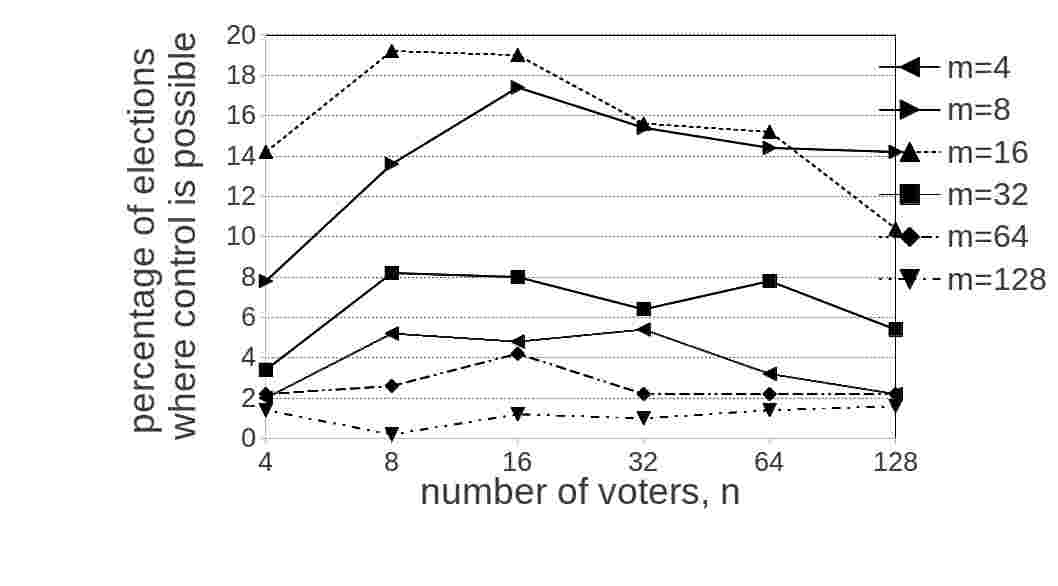}
		\caption{Results for fallback voting in the IC model for 
constructive control by partition of candidates in model TE. Number of candidates is fixed. }
\end{figure}
\begin{tabular}{||c||c|c|c|c|c|c||c|c|c|c|c|c||}
\hline

		& \multicolumn{6}{c||}{$m=4$} 			& \multicolumn{6}{|c||}{$m=8$}\\\hline
$n$ 		& $4$ 	& $8$ 	& $16$ 	& $32$ 	& $64$ 	& $128$ & $4$ 	& $8$ 	& $16$ 	& $32$ 	& $64$ 	& $128$\\\hline
$\#$ cp 	& $10$ 	& $26$ 	& $24$ 	& $27$  & $16$ 	& $11$  & $39$ 	& $68$ 	& $87$ 	& $77$ 	& $72$ 	& $71$ \\\hline
$\#$ ci 	& $490$ & $474$ & $476$ & $473$ & $484$ & $489$ & $461$ & $432$ & $413$ & $423$ & $428$ & $429$ \\\hline
$\#$ to 	& $0$ 	& $0$ 	& $0$ 	& $0$  	& $0$ 	& $0$  	& $0$ 	& $0$ 	& $0$ 	& $0$ 	& $0$ 	& $0$ \\\hline
		& \multicolumn{6}{c||}{$m=16$} 			& \multicolumn{6}{|c||}{$m=32$}\\\hline
$n$ 		& $4$ 	& $8$ 	& $16$ 	& $32$ 	& $64$ 	& $128$ & $4$ 	& $8$ 	& $16$ 	& $32$ 	& $64$ 	& $128$\\\hline
$\#$ cp 	& $71$ 	& $96$ 	& $95$ 	& $78$  & $76$ 	& $52$ 	& $17$ 	& $41$ 	& $40$ 	& $32$ 	& $39$ 	& $27$ \\\hline
$\#$ ci 	& $0$ 	& $0$ 	& $0$ 	& $0$  	& $0$ 	& $0$  	& $0$ 	& $0$ 	& $0$ 	& $0$ 	& $0$ 	& $0$ \\\hline
$\#$ to 	& $429$ & $404$ & $405$ & $422$ & $424$ & $448$	& $483$ & $459$ & $460$ & $468$ & $461$ & $473$ \\\hline
		& \multicolumn{6}{c||}{$m=64$} 			& \multicolumn{6}{|c||}{$m=128$}\\\hline
$n$ 		& $4$ 	& $8$ 	& $16$ 	& $32$ 	& $64$ 	& $128$ & $4$ 	& $8$ 	& $16$ 	& $32$ 	& $64$ 	& $128$\\\hline
$\#$ cp 	& $11$ 	& $13$ 	& $21$ 	& $11$  & $11$ 	& $11$  & $7$ 	& $1$ 	& $6$ 	& $5$ 	& $7$ 	& $8$ \\\hline
$\#$ ci 	& $0$ 	& $0$ 	& $0$ 	& $0$  	& $0$ 	& $0$  	& $0$ 	& $0$ 	& $0$ 	& $0$ 	& $0$ 	& $0$ \\\hline
$\#$ to 	& $489$ & $487$	& $479$ & $489$ & $489$	& $489$	& $493$ & $499$ & $494$ & $495$ & $493$ & $492$ \\\hline
\end{tabular}
\newpage
\begin{figure}[ht]
\centering
	\includegraphics[scale=0.3]{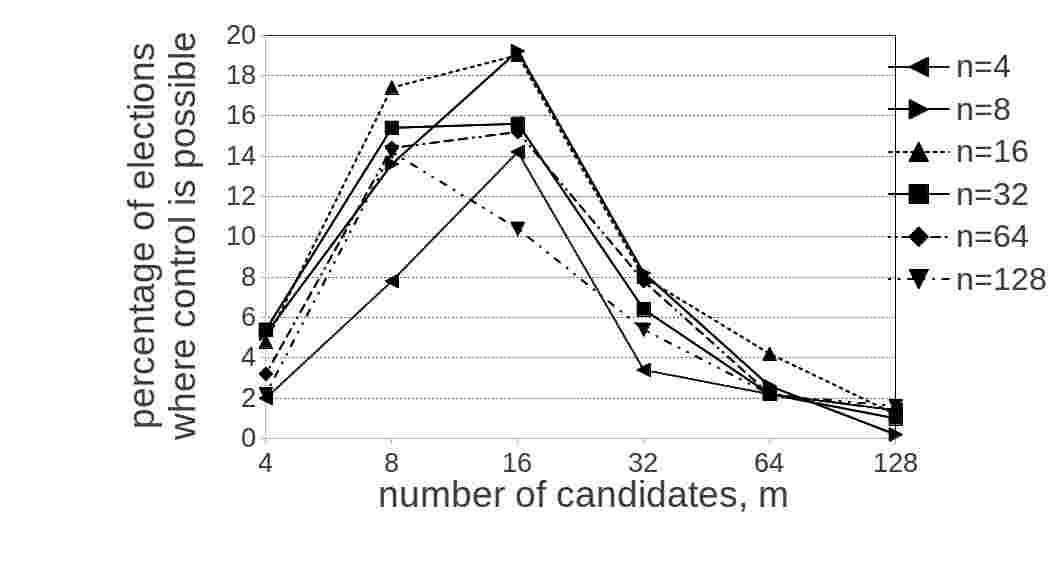}
		\caption{Results for fallback voting in the IC model for 
constructive control by partition of candidates in model TE. Number of voters is fixed. }
\end{figure}
\begin{tabular}{||c||c|c|c|c|c|c||c|c|c|c|c|c||}
\hline

		& \multicolumn{6}{c||}{$n=4$} 			& \multicolumn{6}{|c||}{$n=8$}\\\hline
$m$ 		& $4$ 	& $8$ 	& $16$ 	& $32$ 	& $64$ 	& $128$ & $4$ 	& $8$ 	& $16$ 	& $32$ 	& $64$ 	& $128$\\\hline
$\#$ cp 	& $10$ 	& $39$ 	& $71$ 	& $17$  & $11$ 	& $7$  	& $26$ 	& $68$ 	& $96$ 	& $41$ 	& $13$ 	& $1$ \\\hline
$\#$ ci 	& $490$ & $461$ & $0$ 	& $0$  	& $0$ 	& $0$  	& $474$ & $432$	& $0$ 	& $0$ 	& $0$ 	& $0$ \\\hline
$\#$ to 	& $0$ 	& $0$ 	& $429$ & $483$ & $489$ & $493$	& $0$ 	& $0$ 	& $404$ & $459$ & $487$ & $499$ \\\hline
		& \multicolumn{6}{c||}{$n=16$} 			& \multicolumn{6}{|c||}{$n=32$}\\\hline
$m$ 		& $4$ 	& $8$ 	& $16$ 	& $32$ 	& $64$ 	& $128$ & $4$ 	& $8$ 	& $16$ 	& $32$ 	& $64$ 	& $128$\\\hline
$\#$ cp 	& $24$ 	& $87$ 	& $95$ 	& $40$  & $21$ 	& $6$  	& $27$ 	& $77$ 	& $78$ 	& $32$ 	& $11$ 	& $5$ \\\hline
$\#$ ci 	& $476$ & $413$ & $0$ 	& $0$  	& $0$ 	& $0$  	& $473$ & $423$ & $0$ 	& $0$ 	& $0$ 	& $0$ \\\hline
$\#$ to 	& $0$ 	& $0$ 	& $405$ & $460$ & $479$ & $494$ & $0$ 	& $0$ 	& $422$ & $468$ & $489$ & $495$ \\\hline
		& \multicolumn{6}{c||}{$n=64$} 			& \multicolumn{6}{|c||}{$n=128$}\\\hline
$m$ 		& $4$ 	& $8$ 	& $16$ 	& $32$ 	& $64$ 	& $128$ & $4$ 	& $8$ 	& $16$ 	& $32$ 	& $64$ 	& $128$\\\hline
$\#$ cp 	& $16$ 	& $72$ 	& $76$ 	& $39$  & $11$ 	& $7$  	& $11$ 	& $71$ 	& $52$ 	& $27$ 	& $11$ 	& $8$ \\\hline
$\#$ ci 	& $484$	& $428$	& $0$ 	& $0$  	& $0$ 	& $0$  	& $489$ & $429$ & $0$ 	& $0$ 	& $0$ 	& $0$ \\\hline
$\#$ to 	& $0$ 	& $0$ 	& $424$ & $461$ & $489$ & $493$ & $0$ 	& $0$ 	& $448$ & $473$ & $489$ & $492$ \\\hline
\end{tabular}
\newpage
\begin{figure}[ht]
\centering
	\includegraphics[scale=0.3]{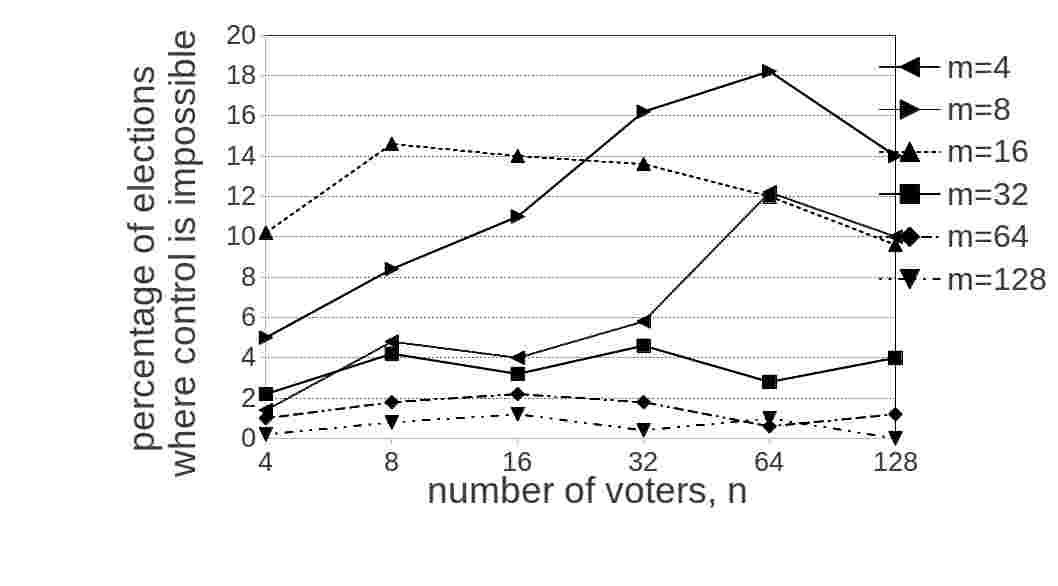}
		\caption{Results for fallback voting in the TM model for 
constructive control by partition of candidates in model TE. Number of candidates is fixed. }
\end{figure}
\begin{tabular}{||c||c|c|c|c|c|c||c|c|c|c|c|c||}
\hline

				& \multicolumn{6}{c||}{$m=4$} 				& \multicolumn{6}{|c||}{$m=8$}\\\hline
$n$ 		& $4$ 	& $8$ 	& $16$ 	& $32$ 	& $64$ 	& $128$ & $4$ 	& $8$ 	& $16$ 	& $32$ 	& $64$ 	& $128$\\\hline
$\#$ cp 	& $7$ 	& $24$ 	& $20$ 	& $29$  & $61$ 	& $50$  & $25$ 	& $42$ 	& $55$ 	& $81$ 	& $91$ 	& $70$ \\\hline
$\#$ ci 	& $493$ & $476$ & $480$ & $471$ & $439$ & $450$ & $475$ & $458$ & $445$ & $419$ & $409$ & $430$ \\\hline
$\#$ to 	& $0$ 	& $0$ 	& $0$ 	& $0$  	& $0$ 	& $0$  	& $0$ 	& $0$ 	& $0$ 	& $0$ 	& $0$ 	& $0$ \\\hline
				& \multicolumn{6}{c||}{$m=16$} 				& \multicolumn{6}{|c||}{$m=32$}\\\hline
$n$ 		& $4$ 	& $8$ 	& $16$ 	& $32$ 	& $64$ 	& $128$ & $4$ 	& $8$ 	& $16$ 	& $32$ 	& $64$ 	& $128$\\\hline
$\#$ cp 	& $51$ 	& $73$ 	& $70$ 	& $68$  & $60$ 	& $48$  & $11$ 	& $21$ 	& $16$ 	& $23$ 	& $14$ 	& $20$ \\\hline
$\#$ ci 	& $0$ 	& $0$ 	& $0$ 	& $0$  	& $0$ 	& $0$  	& $0$ 	& $0$ 	& $0$ 	& $0$ 	& $0$ 	& $0$ \\\hline
$\#$ to 	& $449$ & $427$ & $430$ & $432$ & $440$ & $452$ & $489$ & $479$ & $484$ & $477$ & $486$ & $480$ \\\hline
				& \multicolumn{6}{c||}{$m=64$} 				& \multicolumn{6}{|c||}{$m=128$}\\\hline
$n$ 		& $4$ 	& $8$ 	& $16$ 	& $32$ 	& $64$ 	& $128$ & $4$ 	& $8$ 	& $16$ 	& $32$ 	& $64$ 	& $128$\\\hline
$\#$ cp 	& $5$ 	& $9$ 	& $11$ 	& $9$  	& $3$ 	& $6$  	& $1$ 	& $4$ 	& $6$ 	& $2$ 	& $5$ 	& $0$ \\\hline
$\#$ ci 	& $0$ 	& $0$ 	& $0$ 	& $0$  	& $0$ 	& $0$  	& $0$ 	& $0$ 	& $0$ 	& $0$ 	& $0$ 	& $0$ \\\hline
$\#$ to 	& $495$ & $491$ & $489$ & $491$ & $497$ & $494$ & $499$ & $496$ & $494$ & $498$ & $495$ & $500$ \\\hline
\end{tabular}
\newpage
\begin{figure}[ht]
\centering
	\includegraphics[scale=0.3]{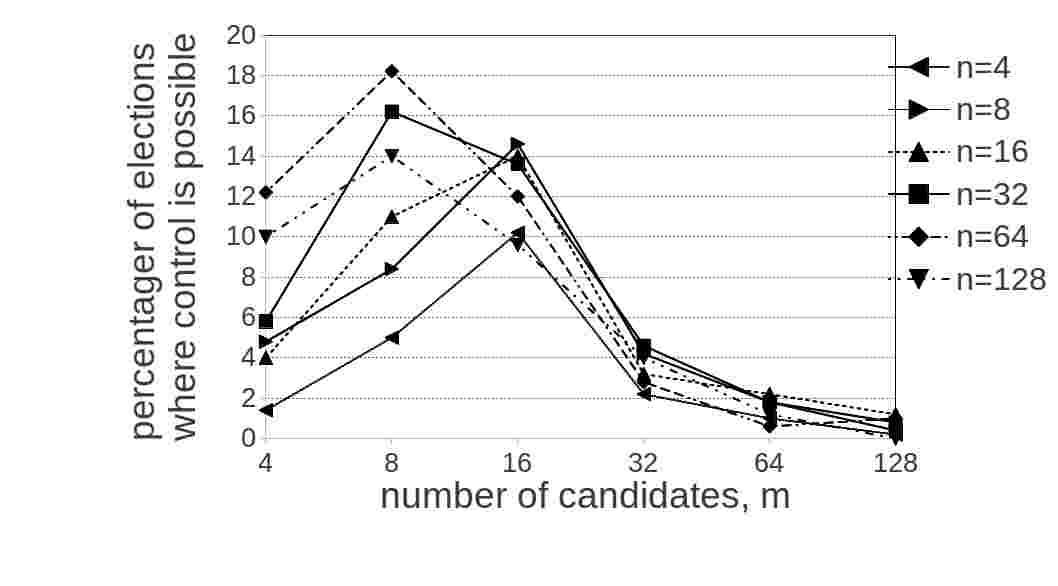}
		\caption{Results for fallback voting in the TM model for 
constructive control by partition of candidates in model TE. Number of voters is fixed. }
\end{figure}
\begin{tabular}{||c||c|c|c|c|c|c||c|c|c|c|c|c||}
\hline

		& \multicolumn{6}{c||}{$n=4$} 			& \multicolumn{6}{|c||}{$n=8$}\\\hline
$m$ 		& $4$ 	& $8$ 	& $16$ 	& $32$ 	& $64$ 	& $128$ & $4$ 	& $8$ 	& $16$ 	& $32$ 	& $64$ 	& $128$\\\hline
$\#$ cp 	& $7$ 	& $25$ 	& $51$ 	& $11$  & $5$ 	& $1$  	& $24$ 	& $42$ 	& $73$ 	& $21$ 	& $9$ 	& $4$ \\\hline
$\#$ ci 	& $493$ & $475$ & $0$ 	& $0$  	& $0$ 	& $0$  	& $476$ & $458$ & $0$ 	& $0$ 	& $0$ 	& $0$ \\\hline
$\#$ to 	& $0$ 	& $0$ 	& $449$ & $489$ & $495$ & $499$ & $0$ 	& $0$ 	& $427$ & $479$ & $491$ & $496$ \\\hline
		& \multicolumn{6}{c||}{$n=16$} 			& \multicolumn{6}{|c||}{$n=32$}\\\hline
$m$ 		& $4$ 	& $8$ 	& $16$ 	& $32$ 	& $64$ 	& $128$ & $4$ 	& $8$ 	& $16$ 	& $32$ 	& $64$ 	& $128$\\\hline
$\#$ cp 	& $20$ 	& $55$ 	& $70$ 	& $16$  & $11$ 	& $6$  	& $29$ 	& $81$ 	& $68$ 	& $23$ 	& $9$ 	& $2$ \\\hline
$\#$ ci 	& $480$ & $455$ & $0$ 	& $0$  	& $0$ 	& $0$  	& $471$ & $419$ & $0$ 	& $0$ 	& $0$ 	& $0$ \\\hline
$\#$ to 	& $0$ 	& $0$ 	& $430$ & $484$ & $489$ & $494$ & $0$ 	& $0$ 	& $432$ & $477$ & $491$ & $498$ \\\hline
		& \multicolumn{6}{c||}{$n=64$} 			& \multicolumn{6}{|c||}{$n=128$}\\\hline
$m$ 		& $4$ 	& $8$ 	& $16$ 	& $32$ 	& $64$ 	& $128$ & $4$ 	& $8$ 	& $16$ 	& $32$ 	& $64$ 	& $128$\\\hline
$\#$ cp 	& $61$ 	& $91$ 	& $60$ 	& $14$  & $3$ 	& $5$  	& $5$ 	& $70$ 	& $48$ 	& $20$ 	& $6$ 	& $0$ \\\hline
$\#$ ci 	& $439$ & $409$ & $0$ 	& $0$  	& $0$ 	& $0$  	& $495$ & $430$ & $0$ 	& $0$ 	& $0$ 	& $0$ \\\hline
$\#$ to 	& $0$ 	& $0$ 	& $440$	& $486$ & $497$ & $495$ & $0$ 	& $0$ 	& $452$ & $480$ & $494$ & $500$ \\\hline
\end{tabular}
\end{center}
\clearpage
\subsubsection{Computational Costs}
\begin{figure}[ht]
\centering
	\includegraphics[scale=0.3]{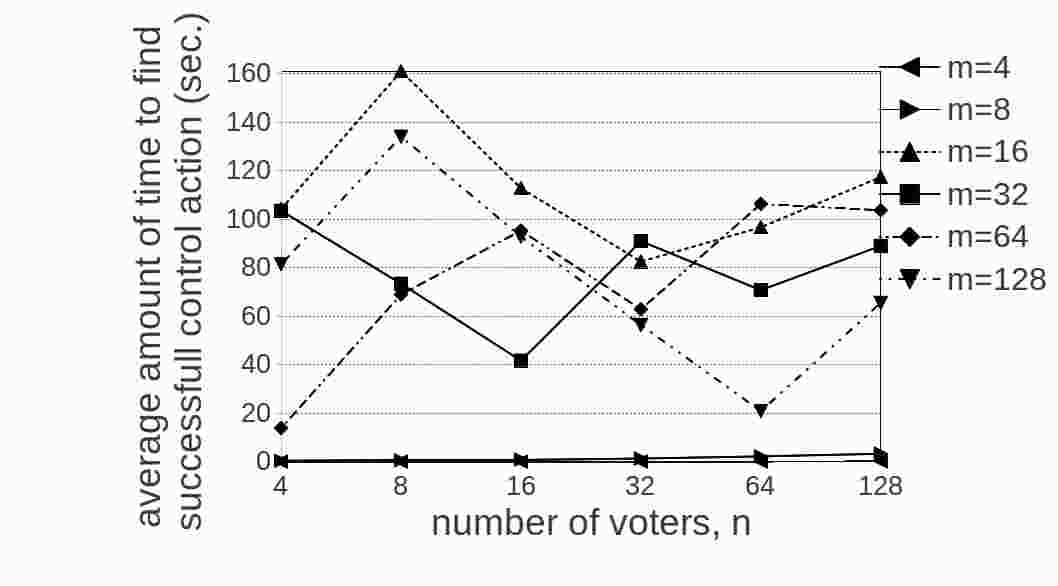}
	\caption{Average time the algorithm needs to find a successful control action for 
	constructive control by partition of candidates in model TE
	in fallback elections in the IC model. The maximum is $160,89$ seconds.}
\end{figure}
\begin{figure}[ht]
\centering
	\includegraphics[scale=0.3]{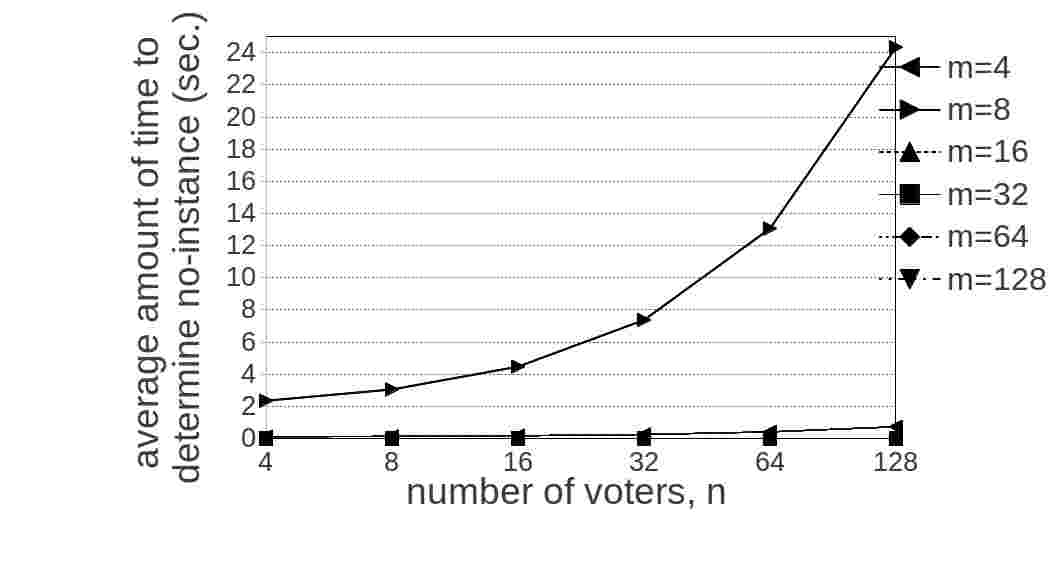}
	\caption{Average time the algorithm needs to determine no-instance of 
		constructive control by partition of candidates in model TE
	in fallback elections in the IC model. The maximum is $24,34$ seconds.}
\end{figure}
\begin{figure}[ht]
\centering
	\includegraphics[scale=0.3]{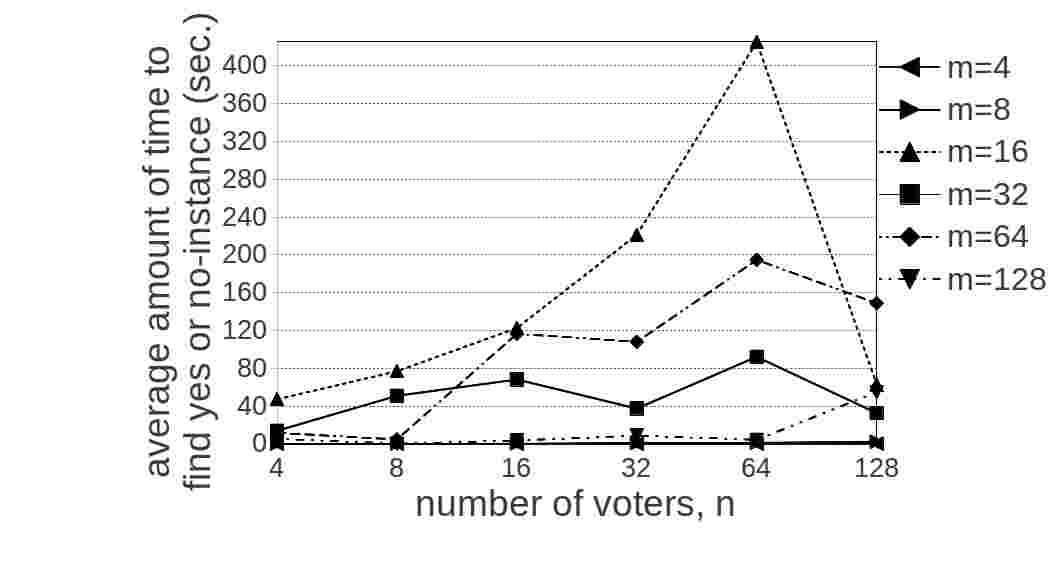}
	\caption{Average time the algorithm needs to give a definite output for 
	constructive control by partition of candidates in model TE
	in fallback elections in the IC model. The maximum is $425,43$ seconds.}
\end{figure}

\begin{figure}[ht]
\centering
	\includegraphics[scale=0.3]{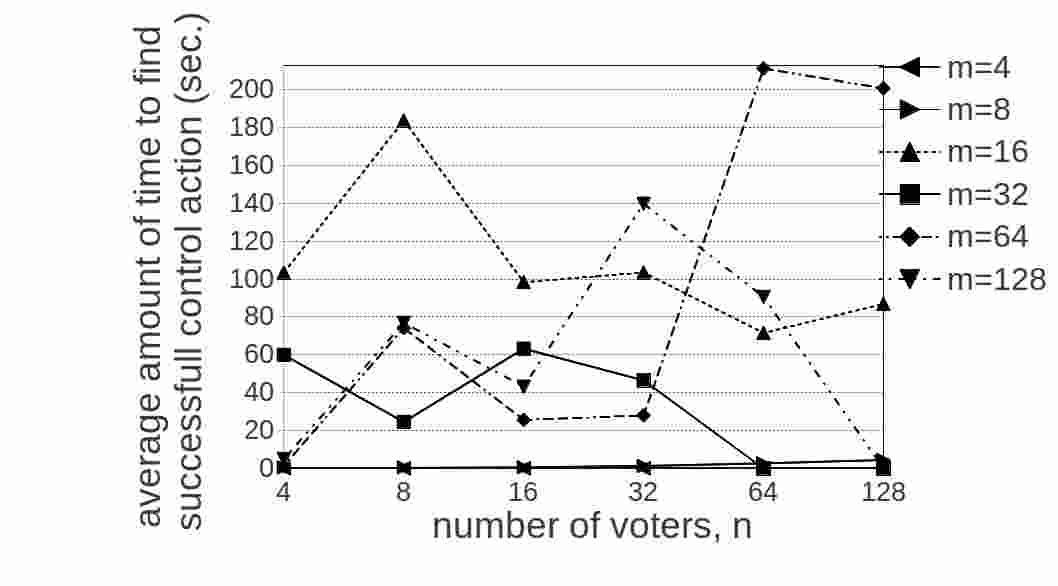}
	\caption{Average time the algorithm needs to find a successful control action for 
	constructive control by partition of candidates in model TE
	in fallback elections in the TM model. The maximum is $211,22$ seconds.}
\end{figure}

\begin{figure}[ht]
\centering
	\includegraphics[scale=0.3]{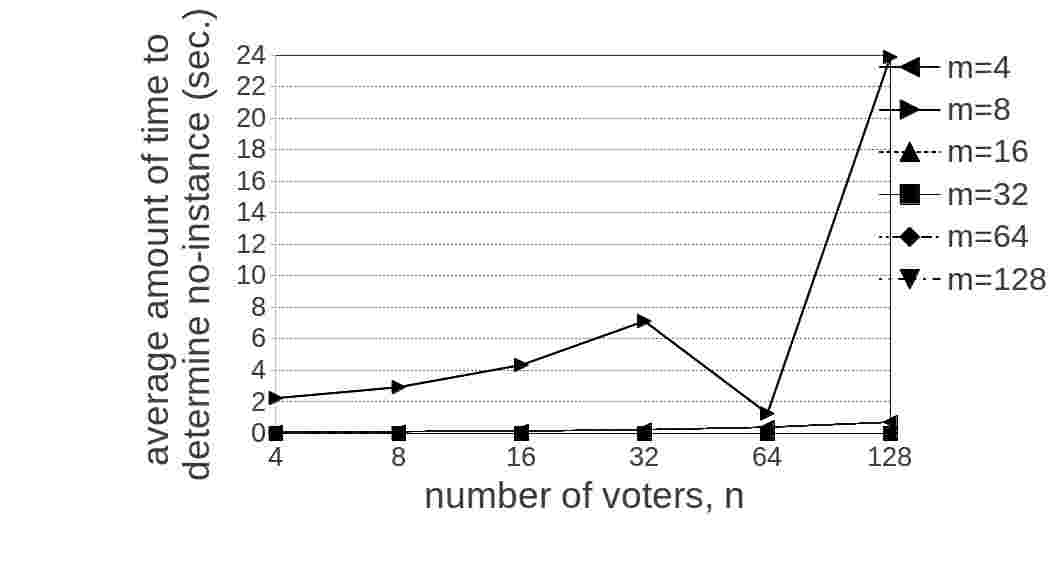}
	\caption{Average time the algorithm needs to determine no-instance of 
		constructive control by partition of candidates in model TE
	in fallback elections in the TM model. The maximum is $23,89$ seconds.}
\end{figure}

\begin{figure}[ht]
\centering
	\includegraphics[scale=0.3]{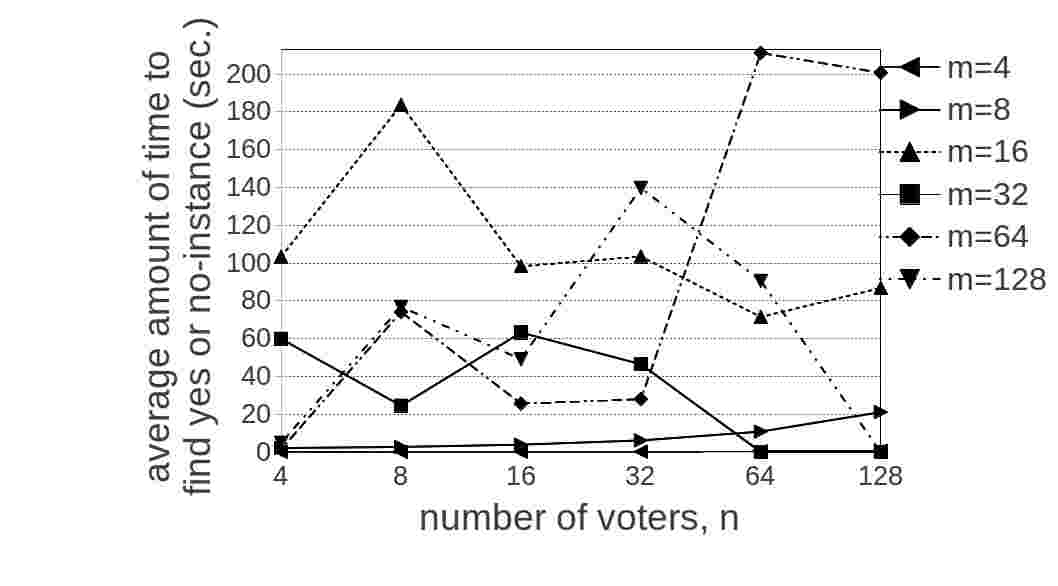}
	\caption{Average time the algorithm needs to give a definite output for 
	constructive control by partition of candidates in model TE
	in fallback elections in the TM model. The maximum is $211,22$ seconds.}
\end{figure}
\clearpage
\subsection{Destructive Control by Partition of Candidates in Model TE}
\begin{center}
\begin{figure}[ht]
\centering
	\includegraphics[scale=0.3]{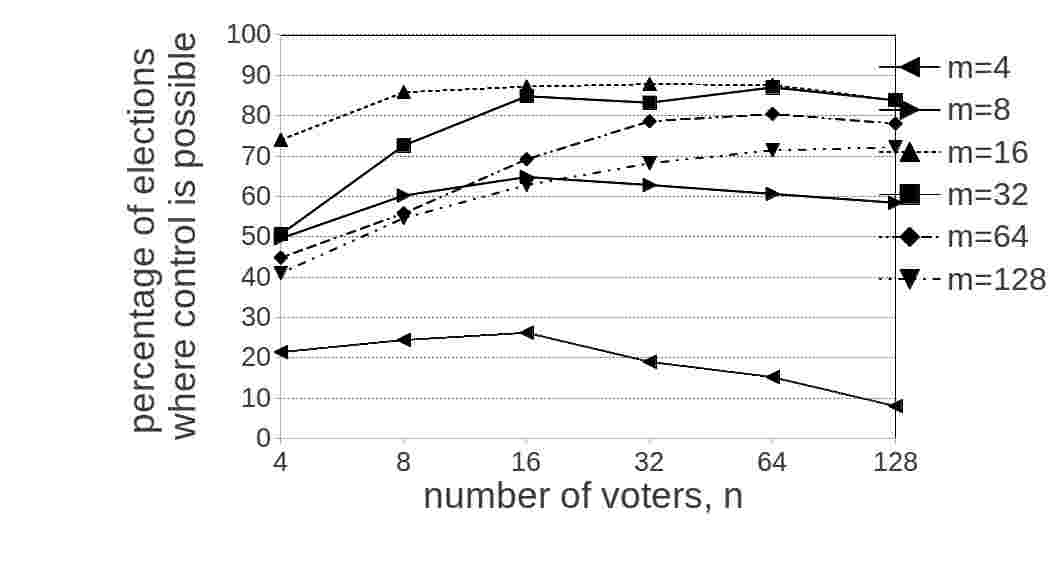}
		\caption{Results for fallback voting in the IC model for 
destructive control by partition of candidates in model TE. Number of candidates is fixed. }
\end{figure}

\begin{tabular}{||c||c|c|c|c|c|c||c|c|c|c|c|c||}
\hline

		& \multicolumn{6}{c||}{$m=4$} 			& \multicolumn{6}{|c||}{$m=8$}\\\hline
$n$ 		& $4$ 	& $8$ 	& $16$ 	& $32$ 	& $64$ 	& $128$ & $4$ 	& $8$ 	& $16$ 	& $32$ 	& $64$ 	& $128$\\\hline
$\#$ cp 	& $107$ & $122$ & $131$ & $95$  & $76$ 	& $40$  & $248$ & $301$	& $324$ & $314$ & $303$ & $292$ \\\hline
$\#$ ci 	& $393$ & $378$ & $369$ & $405$ & $424$ & $460$ & $252$ & $199$	& $176$ & $186$	& $197$ & $208$ \\\hline
$\#$ to 	& $0$ 	& $0$ 	& $0$ 	& $0$  	& $0$ 	& $0$  	& $0$ 	& $0$ 	& $0$ 	& $0$ 	& $0$ 	& $0$ \\\hline
		& \multicolumn{6}{c||}{$m=16$} 			& \multicolumn{6}{|c||}{$m=32$}\\\hline
$n$ 		& $4$ 	& $8$ 	& $16$ 	& $32$ 	& $64$ 	& $128$ & $4$ 	& $8$ 	& $16$ 	& $32$ 	& $64$ 	& $128$\\\hline
$\#$ cp 	& $370$ & $429$ & $436$ & $439$ & $438$ & $419$ & $253$ & $363$ & $424$	& $416$ & $435$ & $419$ \\\hline
$\#$ ci 	& $9$ 	& $0$ 	& $0$ 	& $0$  	& $0$ 	& $0$  	& $1$ 	& $0$ 	& $0$ 	& $0$ 	& $0$ 	& $0$ \\\hline
$\#$ to 	& $121$ & $71$ 	& $64$ 	& $61$  & $62$ 	& $81$  & $246$ & $137$ & $76$ 	& $84$ 	& $65$ 	& $81$ \\\hline
		& \multicolumn{6}{c||}{$m=64$} 			& \multicolumn{6}{|c||}{$m=128$}\\\hline
$n$ 		& $4$ 	& $8$ 	& $16$ 	& $32$ 	& $64$ 	& $128$ & $4$ 	& $8$ 	& $16$ 	& $32$ 	& $64$ 	& $128$\\\hline
$\#$ cp 	& $224$ & $279$ & $346$ & $393$ & $402$ & $390$ & $205$ & $273$ & $314$ & $341$ & $357$ & $361$ \\\hline
$\#$ ci 	& $0$ 	& $0$ 	& $0$ 	& $0$  	& $0$ 	& $0$  	& $0$ 	& $0$ 	& $0$ 	& $0$ 	& $0$ 	& $0$ \\\hline
$\#$ to 	& $276$ & $221$ & $154$ & $107$ & $98$ 	& $110$ & $295$ & $227$ & $186$ & $159$ & $143$ & $139$ \\\hline
\end{tabular}
\newpage
\begin{figure}[ht]
\centering
	\includegraphics[scale=0.3]{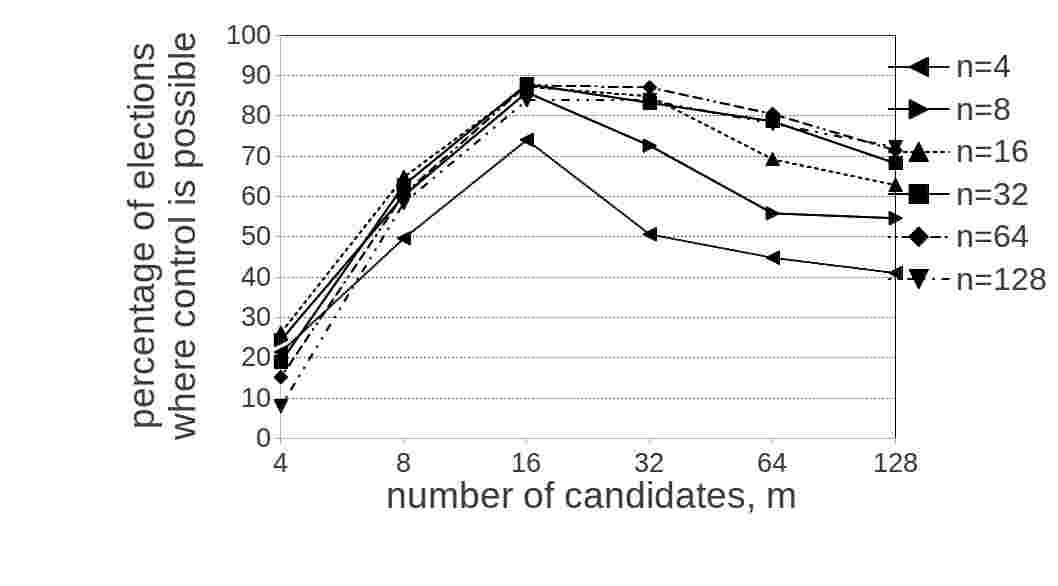}
		\caption{Results for fallback voting in the IC model for 
destructive control by partition of candidates in model TE. Number of voters is fixed. }
\end{figure}

\begin{tabular}{||c||c|c|c|c|c|c||c|c|c|c|c|c||}
\hline

		& \multicolumn{6}{c||}{$n=4$} 						& \multicolumn{6}{|c||}{$n=8$}\\\hline
$m$ 		& $4$ 	& $8$ 	& $16$ 	& $32$ 	& $64$ 	& $128$ & $4$ 	& $8$ 	& $16$ 	& $32$ 	& $64$ 	& $128$\\\hline
$\#$ cp 	& $107$ & $248$ & $370$ & $253$ & $224$ & $205$ & $122$ & $301$ & $429$ & $363$ & $279$ & $273$ \\\hline
$\#$ ci 	& $393$ & $252$ & $9$ 	& $1$  	& $0$ 	& $0$  	& $378$ & $199$ & $0$ 	& $0$ 	& $0$ 	& $0$ \\\hline
$\#$ to 	& $0$ 	& $0$ 	& $121$ & $246$ & $276$ & $295$ & $0$ 	& $0$ 	& $71$ 	& $137$ & $221$ & $227$ \\\hline
				& \multicolumn{6}{c||}{$n=16$} 				& \multicolumn{6}{|c||}{$n=32$}\\\hline
$m$ 		& $4$ 	& $8$ 	& $16$ 	& $32$ 	& $64$ 	& $128$ & $4$ 	& $8$ 	& $16$ 	& $32$ 	& $64$ 	& $128$\\\hline
$\#$ cp 	& $131$ & $324$ & $436$ & $424$ & $346$ & $314$ & $95$ 	& $314$ & $439$ & $416$ & $393$ & $341$ \\\hline
$\#$ ci 	& $369$ & $176$ & $0$ 	& $0$  	& $0$ 	& $0$  	& $405$ & $186$ & $0$ 	& $0$ 	& $0$ 	& $0$ \\\hline
$\#$ to 	& $0$ 	& $0$ 	& $64$ 	& $76$  & $154$ & $186$ & $0$ 	& $0$ 	& $61$ 	& $84$ 	& $107$ & $159$ \\\hline
				& \multicolumn{6}{c||}{$n=64$} 				& \multicolumn{6}{|c||}{$n=128$}\\\hline
$m$ 		& $4$ 	& $8$ 	& $16$ 	& $32$ 	& $64$ 	& $128$ & $4$ 	& $8$ 	& $16$ 	& $32$ 	& $64$ 	& $128$\\\hline
$\#$ cp 	& $76$ 	& $303$ & $438$ & $435$ & $402$ & $357$ & $40$ 	& $292$ & $419$ & $419$ & $390$ & $361$ \\\hline
$\#$ ci 	& $424$ & $197$ & $0$ 	& $0$  	& $0$ 	& $0$  	& $460$ & $208$ & $0$ 	& $0$ 	& $0$ 	& $0$ \\\hline
$\#$ to 	& $0$ 	& $0$ 	& $62$ 	& $65$  & $98$ 	& $143$ & $0$ 	& $0$ 	& $81$ 	& $81$ 	& $110$ & $139$ \\\hline
\end{tabular}
\newpage
\begin{figure}[ht]
\centering
	\includegraphics[scale=0.3]{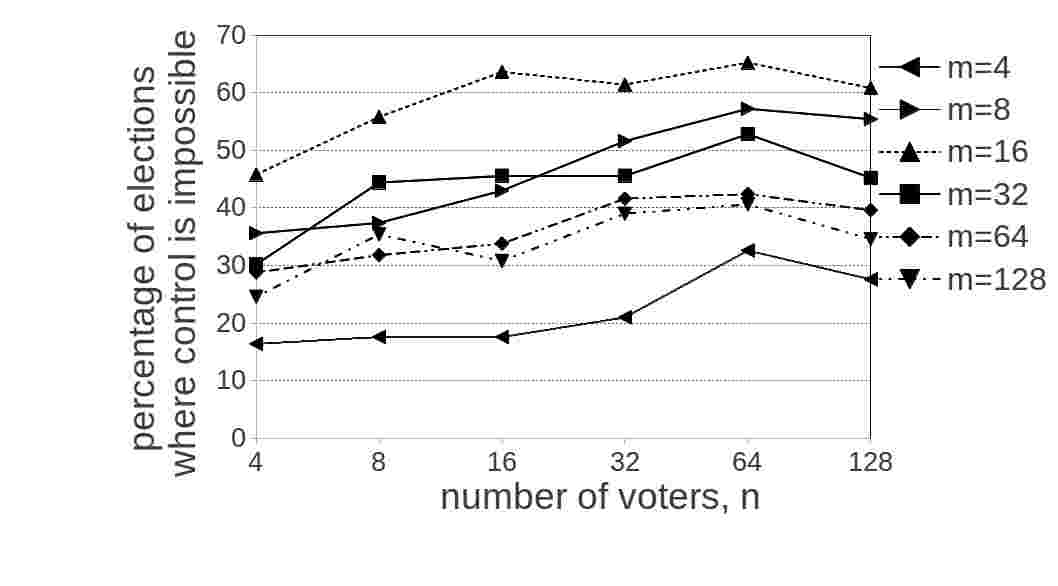}
		\caption{Results for fallback voting in the TM model for 
destructive control by partition of candidates in model TE. Number of candidates is fixed. }
\end{figure}
\begin{tabular}{||c||c|c|c|c|c|c||c|c|c|c|c|c||}
\hline

				& \multicolumn{6}{c||}{$m=4$} 				& \multicolumn{6}{|c||}{$m=8$}\\\hline
$n$ 		& $4$ 	& $8$ 	& $16$ 	& $32$ 	& $64$ 	& $128$ & $4$ 	& $8$ 	& $16$ 	& $32$ 	& $64$ 	& $128$\\\hline
$\#$ cp 	& $82$ 	& $88$ 	& $88$ 	& $105$ & $163$ & $138$ & $178$ & $187$ & $215$ & $258$ & $286$ & $277$ \\\hline
$\#$ ci 	& $418$ & $412$ & $412$ & $395$ & $337$ & $362$ & $322$ & $313$ & $285$ & $242$ & $214$ & $223$ \\\hline
$\#$ to 	& $0$ 	& $0$ 	& $0$ 	& $0$  	& $0$ 	& $0$  	& $0$ 	& $0$ 	& $0$ 	& $0$ 	& $0$ 	& $0$ \\\hline
				& \multicolumn{6}{c||}{$m=16$} 				& \multicolumn{6}{|c||}{$m=32$}\\\hline
$n$ 		& $4$ 	& $8$ 	& $16$ 	& $32$ 	& $64$ 	& $128$ & $4$ 	& $8$ 	& $16$ 	& $32$ 	& $64$ 	& $128$\\\hline
$\#$ cp 	& $229$ & $279$ & $318$ & $307$ & $326$ & $304$ & $151$ & $222$ & $228$ & $228$ & $264$ & $226$ \\\hline
$\#$ ci 	& $159$ & $138$ & $83$ 	& $56$  & $30$ 	& $29$  & $153$ & $104$ & $81$ 	& $34$ 	& $21$ 	& $15$ \\\hline
$\#$ to 	& $112$ & $83$ 	& $99$ 	& $137$ & $144$ & $167$ & $196$ & $174$ & $191$ & $238$ & $215$ & $259$ \\\hline
				& \multicolumn{6}{c||}{$m=64$} 				& \multicolumn{6}{|c||}{$m=128$}\\\hline
$n$ 		& $4$ 	& $8$ 	& $16$ 	& $32$ 	& $64$ 	& $128$ & $4$ 	& $8$ 	& $16$ 	& $32$ 	& $64$ 	& $128$\\\hline
$\#$ cp 	& $144$ & $159$ & $169$ & $208$ & $212$	& $198$ & $123$ & $177$ & $154$ & $195$ & $203$ & $173$ \\\hline
$\#$ ci 	& $110$ & $96$ 	& $69$ 	& $23$  & $8$ 	& $6$  	& $123$ & $101$ & $60$ 	& $18$ 	& $7$ 	& $7$ \\\hline
$\#$ to 	& $246$ & $245$ & $262$ & $269$ & $280$ & $296$ & $254$ & $222$ & $242$ & $287$ & $290$ & $320$ \\\hline
\end{tabular}
\newpage
\begin{figure}[ht]
\centering
	\includegraphics[scale=0.3]{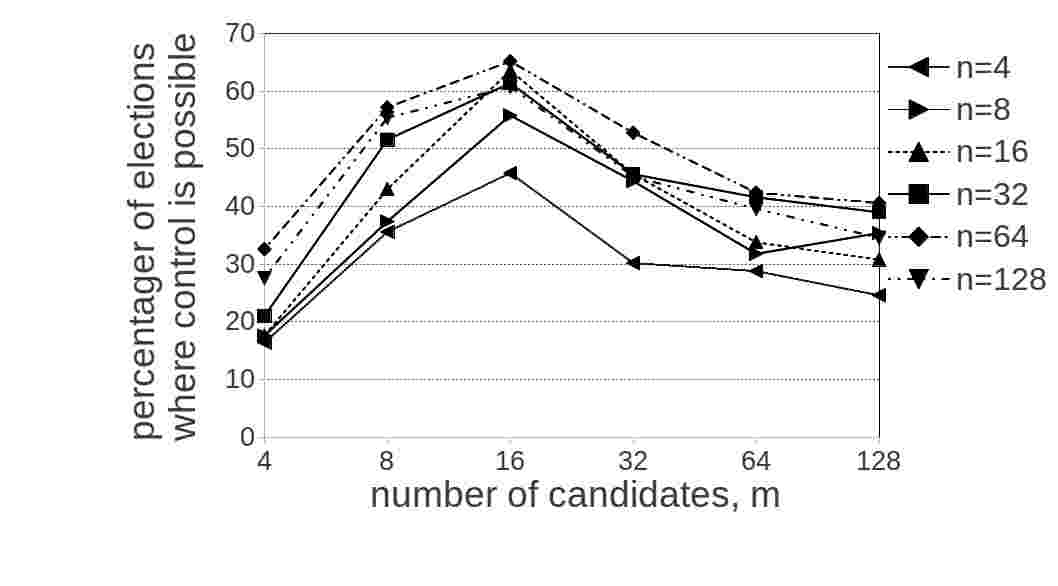}
		\caption{Results for fallback voting in the TM model for 
destructive control by partition of candidates in model TE. Number of voters is fixed. }
\end{figure}
\begin{tabular}{||c||c|c|c|c|c|c||c|c|c|c|c|c||}
\hline

		& \multicolumn{6}{c||}{$n=4$} 			& \multicolumn{6}{|c||}{$n=8$}\\\hline
$m$ 		& $4$ 	& $8$ 	& $16$ 	& $32$ 	& $64$ 	& $128$ & $4$ 	& $8$ 	& $16$ 	& $32$ 	& $64$ 	& $128$\\\hline
$\#$ cp 	& $82$ 	& $178$ & $229$ & $151$ & $144$ & $123$ & $88$ 	& $187$ & $279$ & $222$ & $159$ & $177$ \\\hline
$\#$ ci 	& $418$ & $322$ & $159$ & $153$ & $110$ & $123$ & $412$ & $313$ & $138$ & $104$ & $96$ 	& $101$ \\\hline
$\#$ to 	& $0$ 	& $0$ 	& $112$ & $196$ & $246$ & $254$ & $0$ 	& $0$ 	& $83$ 	& $174$ & $245$ & $222$ \\\hline
		& \multicolumn{6}{c||}{$n=16$} 			& \multicolumn{6}{|c||}{$n=32$}\\\hline
$m$ 		& $4$ 	& $8$ 	& $16$ 	& $32$ 	& $64$ 	& $128$ & $4$ 	& $8$ 	& $16$ 	& $32$ 	& $64$ 	& $128$\\\hline
$\#$ cp 	& $88$ 	& $215$ & $318$ & $228$ & $169$ & $154$ & $105$ & $258$ & $307$ & $228$ & $208$ & $195$ \\\hline
$\#$ ci 	& $412$ & $285$ & $83$ 	& $81$  & $69$ 	& $60$  & $395$ & $242$ & $56$ 	& $34$ 	& $23$ 	& $18$ \\\hline
$\#$ to 	& $0$ 	& $0$ 	& $99$ 	& $191$ & $262$ & $242$ & $0$ 	& $0$ 	& $137$ & $238$ & $269$ & $287$ \\\hline
		& \multicolumn{6}{c||}{$n=64$} 			& \multicolumn{6}{|c||}{$n=128$}\\\hline
$m$ 		& $4$ 	& $8$ 	& $16$ 	& $32$ 	& $64$ 	& $128$ & $4$ 	& $8$ 	& $16$ 	& $32$ 	& $64$ 	& $128$\\\hline
$\#$ cp 	& $163$ & $286$ & $326$ & $264$ & $212$ & $203$ & $138$ & $277$ & $304$ & $226$ & $198$ & $173$ \\\hline
$\#$ ci 	& $337$ & $214$ & $30$ 	& $21$  & $8$ 	& $7$  	& $362$ & $223$ & $29$ 	& $15$ 	& $6$ 	& $7$ \\\hline
$\#$ to 	& $0$ 	& $0$ 	& $144$ & $215$ & $280$ & $290$ & $0$ 	& $0$ 	& $167$ & $259$ & $296$ & $320$ \\\hline
\end{tabular}

\end{center}
\clearpage
\subsubsection{Computational Costs}
\begin{figure}[ht]
\centering
	\includegraphics[scale=0.3]{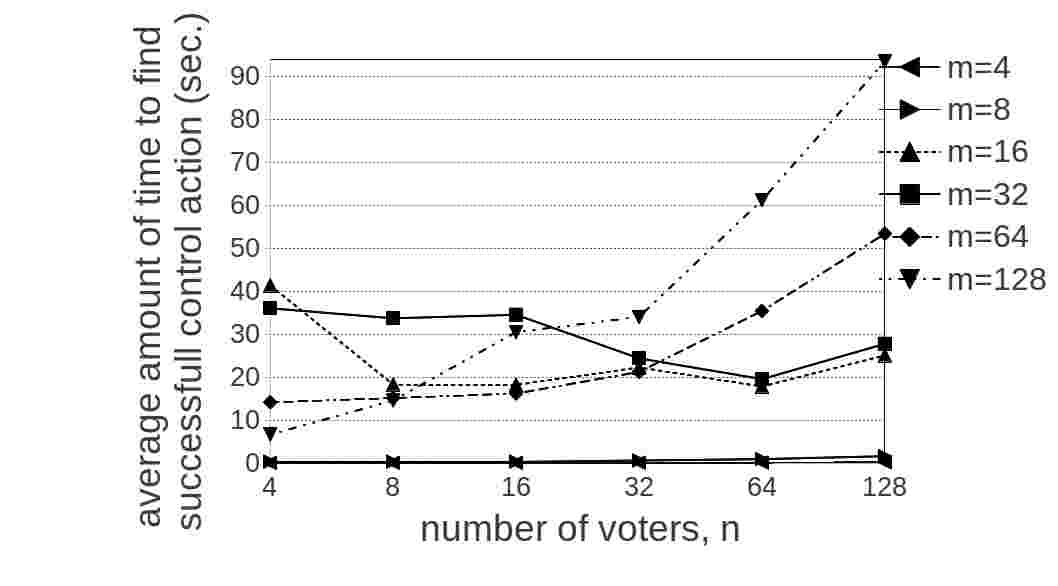}
	\caption{Average time the algorithm needs to find a successful control action for 
	destructive control by partition of candidates in model TE
	in fallback elections in the IC model. The maximum is $93,6$ seconds.}
\end{figure}
\begin{figure}[ht]
\centering
	\includegraphics[scale=0.3]{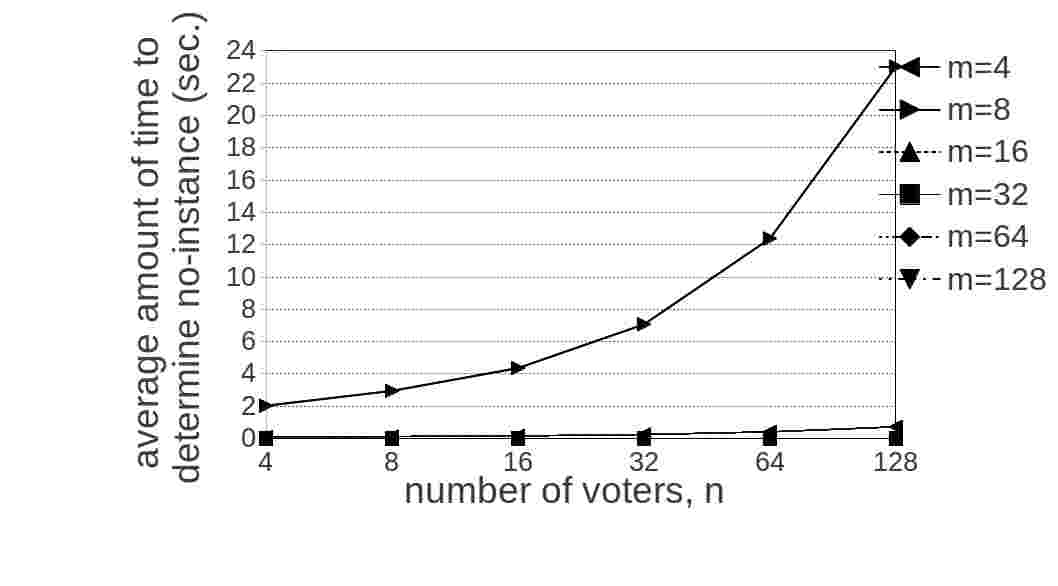}
	\caption{Average time the algorithm needs to determine no-instance of 
		destructive control by partition of candidates in model TE
	in fallback elections in the IC model. The maximum is $23,03$ seconds.}
\end{figure}
\begin{figure}[ht]
\centering
	\includegraphics[scale=0.3]{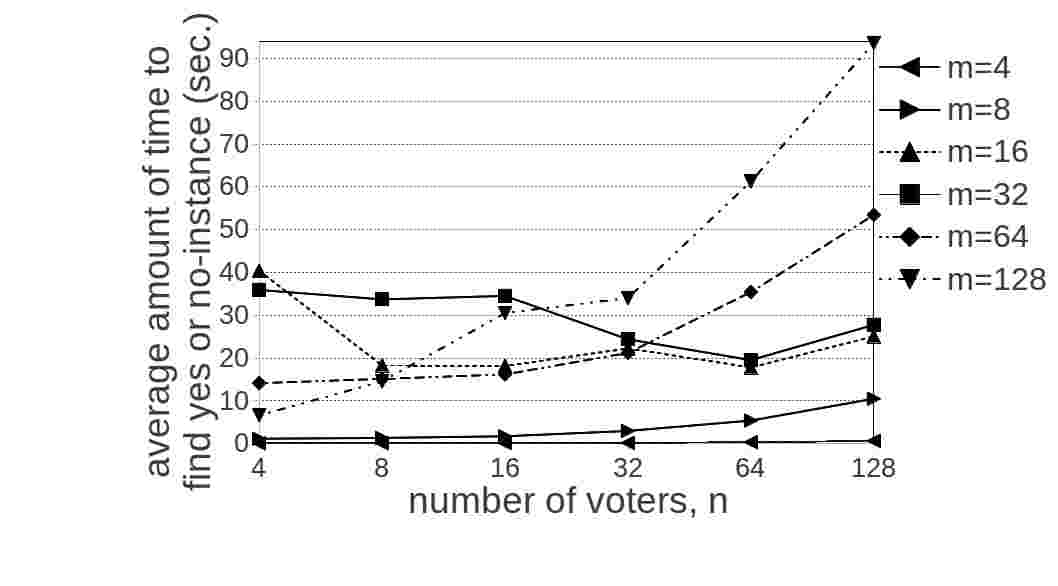}
	\caption{Average time the algorithm needs to give a definite output for 
	destructive control by partition of candidates in model TE
	in fallback elections in the IC model. The maximum is $93,6$ seconds.}
\end{figure}

\begin{figure}[ht]
\centering
	\includegraphics[scale=0.3]{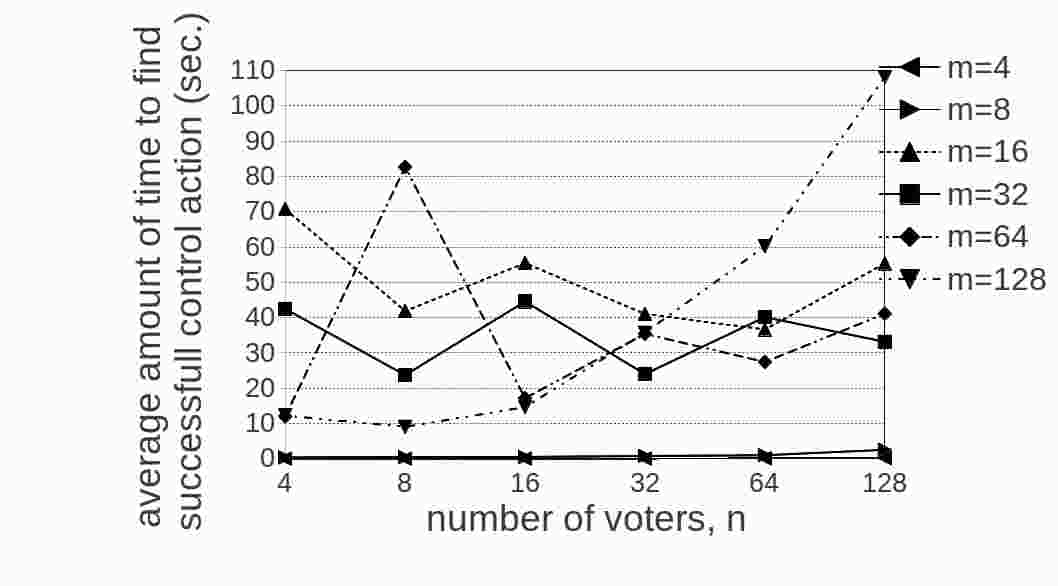}
	\caption{Average time the algorithm needs to find a successful control action for 
	destructive control by partition of candidates in model TE
	in fallback elections in the TM model. The maximum is $108,15$ seconds.}
\end{figure}

\begin{figure}[ht]
\centering
	\includegraphics[scale=0.3]{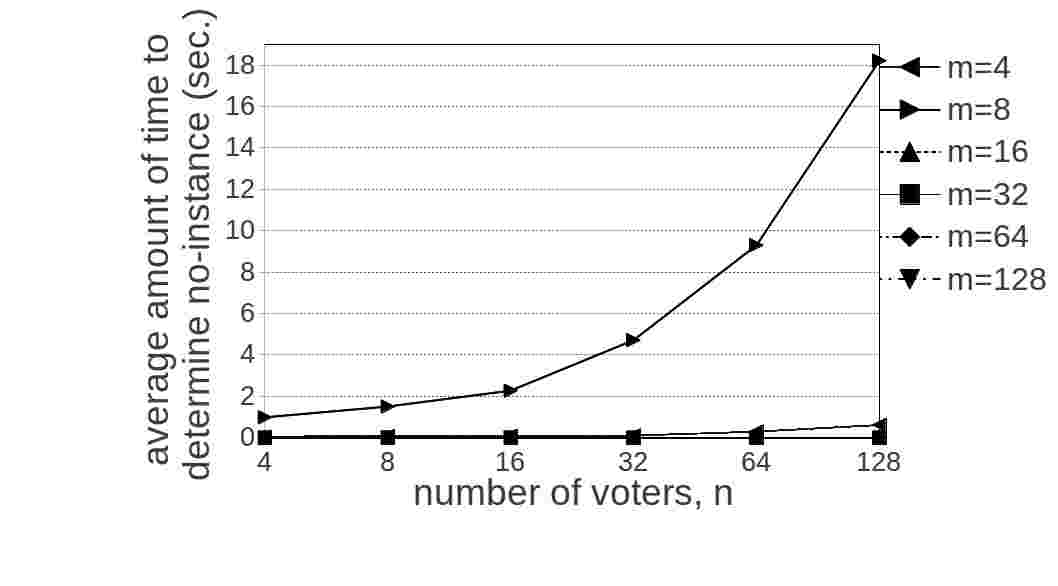}
	\caption{Average time the algorithm needs to determine no-instance of 
		destructive control by partition of candidates in model TE
	in fallback elections in the TM model. The maximum is $18,22$ seconds.}
\end{figure}

\begin{figure}[ht]
\centering
	\includegraphics[scale=0.3]{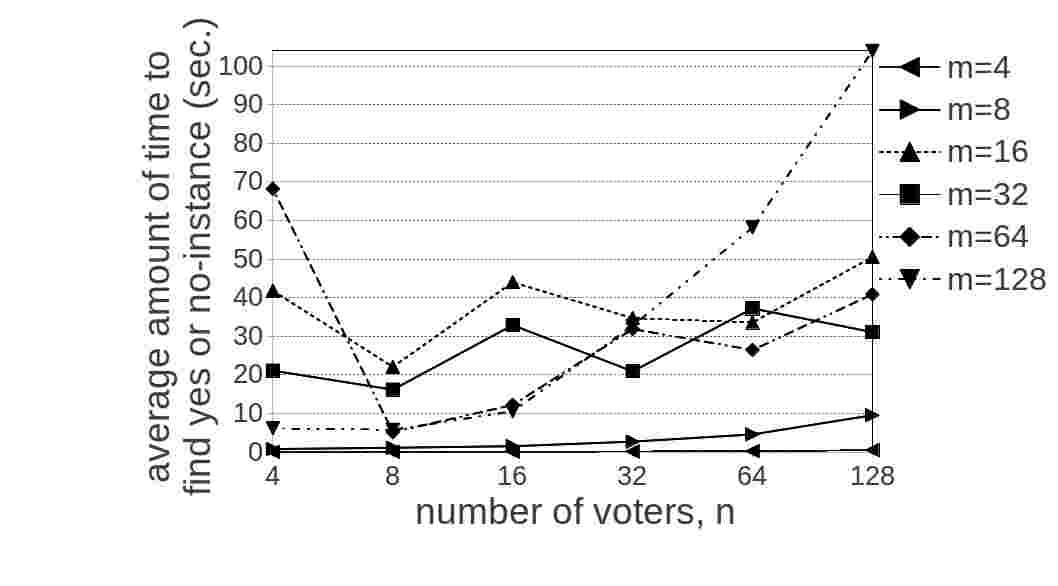}
	\caption{Average time the algorithm needs to give a definite output for 
	destructive control by partition of candidates in model TE
	in fallback elections in the TM model. The maximum is $103,95$ seconds.}
\end{figure}

\clearpage
\subsection{Constructive Control by Partition of Candidates in Model TP}
\begin{center}

\begin{figure}[ht]
\centering
	\includegraphics[scale=0.3]{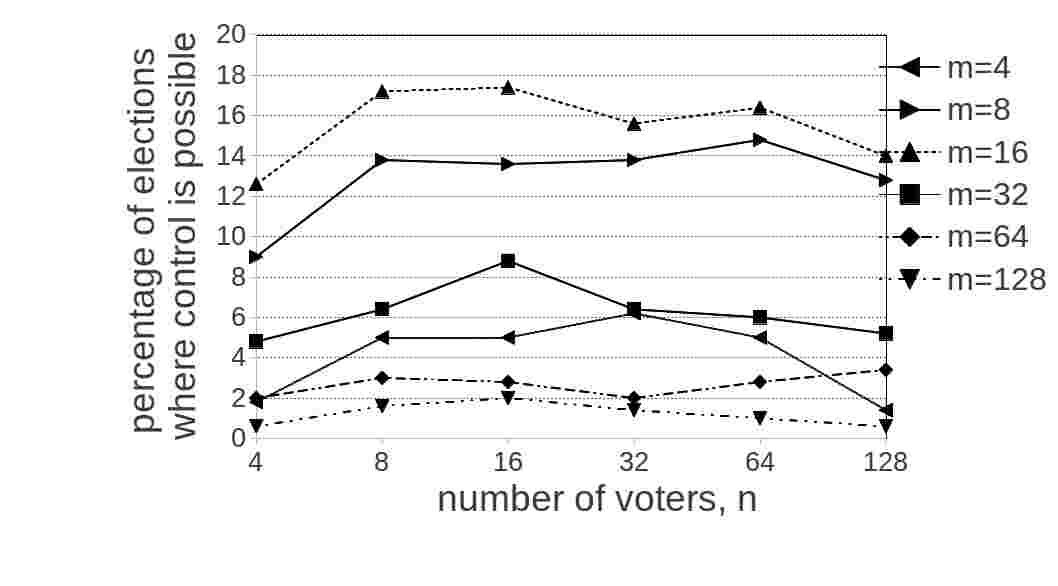}
		\caption{Results for fallback voting in the IC model for 
constructive control by partition of candidates in model TP. Number of candidates is fixed. }
\end{figure}

\begin{tabular}{||c||c|c|c|c|c|c||c|c|c|c|c|c||}
\hline

		& \multicolumn{6}{c||}{$m=4$} 			& \multicolumn{6}{|c||}{$m=8$}\\\hline
$n$ 		& $4$ 	& $8$ 	& $16$ 	& $32$ 	& $64$ 	& $128$ & $4$ 	& $8$ 	& $16$ 	& $32$ 	& $64$ 	& $128$\\\hline
$\#$ cp 	& $9$ 	& $25$ 	& $25$ 	& $31$  & $25$ 	& $7$  	& $45$ 	& $69$ 	& $68$ 	& $69$ 	& $74$ 	& $64$ \\\hline
$\#$ ci 	& $491$ & $475$ & $475$ & $469$ & $475$ & $493$ & $455$ & $431$ & $432$ & $431$ & $426$ & $436$ \\\hline
$\#$ to 	& $0$ 	& $0$ 	& $0$ 	& $0$  	& $0$ 	& $0$  	& $0$ 	& $0$ 	& $0$ 	& $0$ 	& $0$ 	& $0$ \\\hline
		& \multicolumn{6}{c||}{$m=16$} 			& \multicolumn{6}{|c||}{$m=32$}\\\hline
$n$ 		& $4$ 	& $8$ 	& $16$ 	& $32$ 	& $64$ 	& $128$ & $4$ 	& $8$ 	& $16$ 	& $32$ 	& $64$ 	& $128$\\\hline
$\#$ cp 	& $63$ 	& $86$ 	& $87$ 	& $78$  & $82$ 	& $70$  & $24$ 	& $32$ 	& $44$ 	& $32$ 	& $30$ 	& $26$ \\\hline
$\#$ ci 	& $0$ 	& $0$ 	& $0$ 	& $0$  	& $0$ 	& $0$  	& $0$ 	& $0$ 	& $0$ 	& $0$ 	& $0$ 	& $0$ \\\hline
$\#$ to 	& $437$ & $414$ & $413$ & $422$ & $418$ & $430$ & $476$ & $468$ & $456$ & $468$ & $470$ & $474$ \\\hline
		& \multicolumn{6}{c||}{$m=64$} 			& \multicolumn{6}{|c||}{$m=128$}\\\hline
$n$ 		& $4$ 	& $8$ 	& $16$ 	& $32$ 	& $64$ 	& $128$ & $4$ 	& $8$ 	& $16$ 	& $32$ 	& $64$ 	& $128$\\\hline
$\#$ cp 	& $10$ 	& $15$ 	& $14$ 	& $10$  & $14$ 	& $17$  & $3$ 	& $8$ 	& $10$ 	& $7$ 	& $5$ 	& $3$ \\\hline
$\#$ ci 	& $0$ 	& $0$ 	& $0$ 	& $0$  	& $0$ 	& $0$  	& $0$ 	& $0$ 	& $0$ 	& $0$ 	& $0$ 	& $0$ \\\hline
$\#$ to 	& $490$ & $485$ & $486$ & $490$ & $486$ & $483$ & $497$ & $492$ & $490$ & $493$ & $495$ & $497$ \\\hline
\end{tabular}
\newpage
\begin{figure}[ht]
\centering
	\includegraphics[scale=0.3]{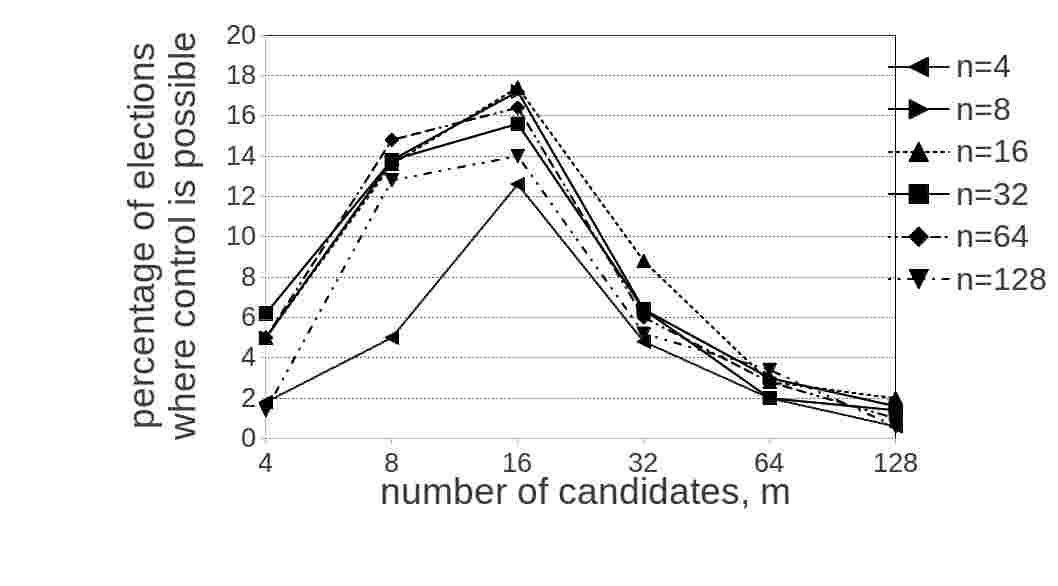}
		\caption{Results for fallback voting in the IC model for 
constructive control by partition of candidates in model TP. Number of voters is fixed. }
\end{figure}

\begin{tabular}{||c||c|c|c|c|c|c||c|c|c|c|c|c||}
\hline

		& \multicolumn{6}{c||}{$n=4$} 			& \multicolumn{6}{|c||}{$n=8$}\\\hline
$m$ 		& $4$ 	& $8$ 	& $16$ 	& $32$ 	& $64$ 	& $128$ & $4$ 	& $8$ 	& $16$ 	& $32$ 	& $64$ 	& $128$\\\hline
$\#$ cp 	& $9$ 	& $45$ 	& $63$ 	& $24$  & $10$ 	& $3$  	& $25$ 	& $69$ 	& $86$ 	& $32$ 	& $15$ 	& $8$ \\\hline
$\#$ ci 	& $491$ & $455$ & $0$ 	& $0$  	& $0$ 	& $0$  	& $475$ & $431$ & $0$ 	& $0$ 	& $0$ 	& $0$ \\\hline
$\#$ to 	& $0$ 	& $0$ 	& $437$ & $476$ & $490$ & $497$ & $0$ 	& $0$ 	& $414$ & $468$ & $485$ & $492$ \\\hline
		& \multicolumn{6}{c||}{$n=16$} 			& \multicolumn{6}{|c||}{$n=32$}\\\hline
$m$ 		& $4$ 	& $8$ 	& $16$ 	& $32$ 	& $64$ 	& $128$ & $4$ 	& $8$ 	& $16$ 	& $32$ 	& $64$ 	& $128$\\\hline
$\#$ cp 	& $25$ 	& $68$ 	& $87$ 	& $44$  & $14$ 	& $10$  & $31$ 	& $69$ 	& $78$ 	& $32$ 	& $10$ 	& $7$ \\\hline
$\#$ ci 	& $475$ & $432$ & $0$ 	& $0$  	& $0$ 	& $0$  	& $469$ & $431$ & $0$ 	& $0$ 	& $0$ 	& $0$ \\\hline
$\#$ to 	& $0$ 	& $0$ 	& $413$ & $456$ & $486$ & $490$ & $0$ 	& $0$ 	& $422$ & $468$ & $490$ & $493$ \\\hline
		& \multicolumn{6}{c||}{$n=64$} 			& \multicolumn{6}{|c||}{$n=128$}\\\hline
$m$ 		& $4$ 	& $8$ 	& $16$ 	& $32$ 	& $64$ 	& $128$ & $4$ 	& $8$ 	& $16$ 	& $32$ 	& $64$ 	& $128$\\\hline
$\#$ cp 	& $25$ 	& $74$ 	& $82$ 	& $30$  & $14$ 	& $5$  	& $7$ 	& $64$ 	& $7$ 	& $26$ 	& $17$ 	& $3$ \\\hline
$\#$ ci 	& $475$ & $426$ & $0$ 	& $0$  	& $0$ 	& $0$  	& $493$ & $436$ & $0$ 	& $0$ 	& $0$ 	& $0$ \\\hline
$\#$ to 	& $0$ 	& $0$ 	& $418$ & $470$ & $486$ & $495$ & $0$ 	& $0$ 	& $493$ & $474$ & $483$ & $497$ \\\hline
\end{tabular}
\newpage
\begin{figure}[ht]
\centering
	\includegraphics[scale=0.3]{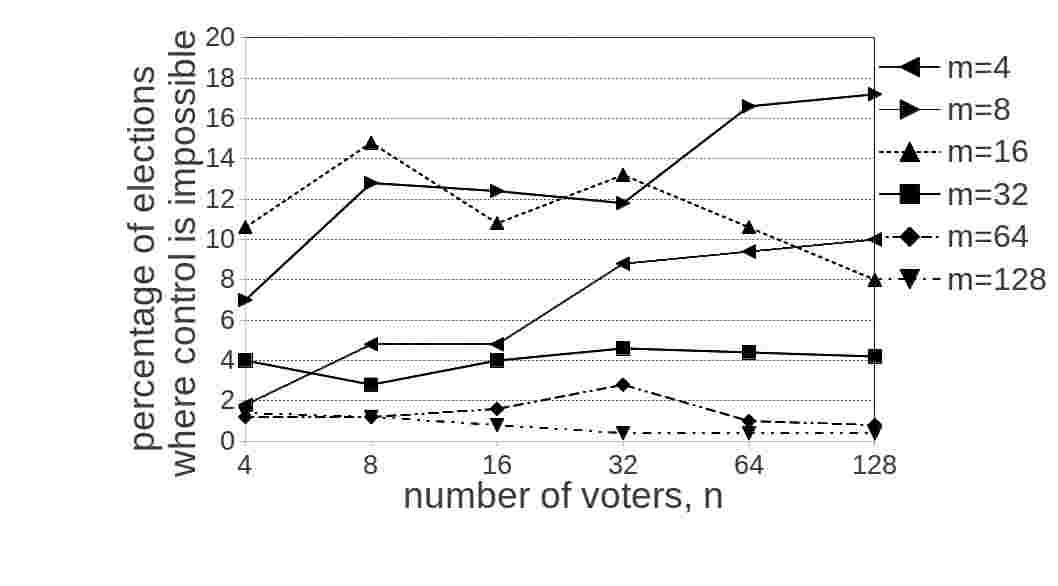}
		\caption{Results for fallback voting in the TM model for 
constructive control by partition of candidates in model TP. Number of candidates is fixed. }
\end{figure}

\begin{tabular}{||c||c|c|c|c|c|c||c|c|c|c|c|c||}
\hline

		& \multicolumn{6}{c||}{$m=4$} 			& \multicolumn{6}{|c||}{$m=8$}\\\hline
$n$ 		& $4$ 	& $8$ 	& $16$ 	& $32$ 	& $64$ 	& $128$ & $4$ 	& $8$ 	& $16$ 	& $32$ 	& $64$ 	& $128$\\\hline
$\#$ cp 	& $9$ 	& $24$ 	& $24$ 	& $44$  & $47$ 	& $50$  & $35$ 	& $64$ 	& $62$ 	& $59$ 	& $83$ 	& $86$ \\\hline
$\#$ ci 	& $491$ & $476$ & $476$ & $456$ & $453$ & $450$ & $465$ & $436$ & $438$ & $441$	& $417$ & $414$ \\\hline
$\#$ to 	& $0$ 	& $0$ 	& $0$ 	& $0$  	& $0$ 	& $0$  	& $0$ 	& $0$ 	& $0$ 	& $0$ 	& $0$ 	& $0$ \\\hline
		& \multicolumn{6}{c||}{$m=16$} 			& \multicolumn{6}{|c||}{$m=32$}\\\hline
$n$ 		& $4$ 	& $8$ 	& $16$ 	& $32$ 	& $64$ 	& $128$ & $4$ 	& $8$ 	& $16$ 	& $32$ 	& $64$ 	& $128$\\\hline
$\#$ cp 	& $53$ 	& $74$ 	& $54$ 	& $66$  & $53$ 	& $40$  & $20$ 	& $14$ 	& $20$ 	& $23$ 	& $22$ 	& $21$ \\\hline
$\#$ ci 	& $0$ 	& $0$ 	& $0$ 	& $0$  	& $0$ 	& $0$  	& $0$ 	& $0$ 	& $0$ 	& $0$ 	& $0$ 	& $0$ \\\hline
$\#$ to 	& $447$ & $426$ & $446$ & $434$ & $447$	& $460$ & $480$ & $486$ & $480$ & $477$ & $478$ & $479$ \\\hline
		& \multicolumn{6}{c||}{$m=64$} 			& \multicolumn{6}{|c||}{$m=128$}\\\hline
$n$ 		& $4$ 	& $8$ 	& $16$ 	& $32$ 	& $64$ 	& $128$ & $4$ 	& $8$ 	& $16$ 	& $32$ 	& $64$ 	& $128$\\\hline
$\#$ cp 	& $6$ 	& $6$ 	& $8$ 	& $14$  & $5$ 	& $4$  	& $7$ 	& $6$ 	& $4$ 	& $2$ 	& $2$ 	& $2$ \\\hline
$\#$ ci 	& $0$ 	& $0$ 	& $0$ 	& $0$  	& $0$ 	& $0$  	& $0$ 	& $0$ 	& $0$ 	& $0$ 	& $0$ 	& $0$ \\\hline
$\#$ to 	& $494$	& $494$ & $492$ & $486$ & $495$ & $496$	& $493$ & $494$ & $496$ & $498$ & $498$ & $498$ \\\hline
\end{tabular}
\newpage
\begin{figure}[ht]
\centering
	\includegraphics[scale=0.3]{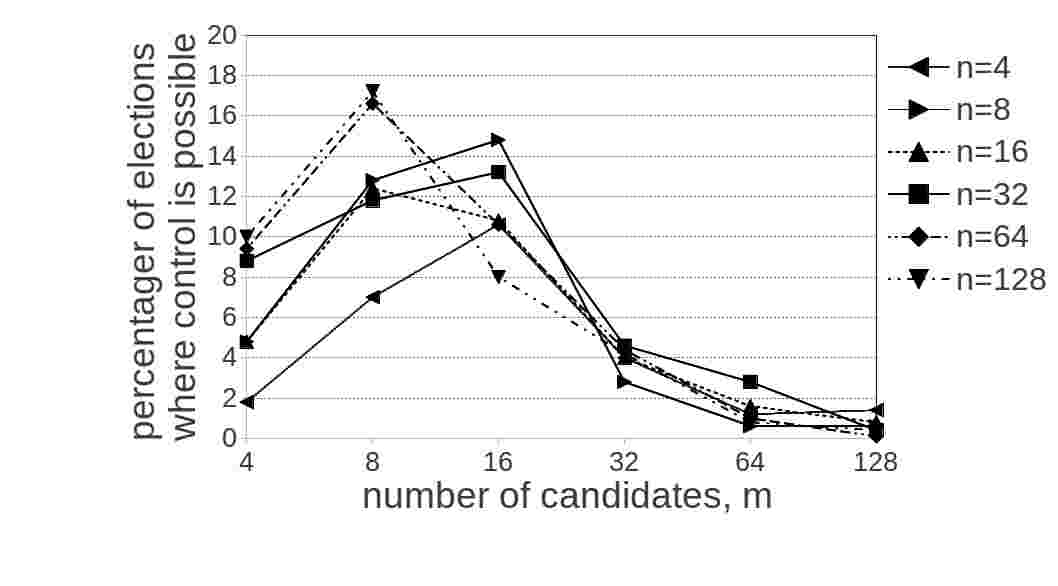}
		\caption{Results for fallback voting in the TM model for 
constructive control by partition of candidates in model TP. Number of voters is fixed. }
\end{figure}

\begin{tabular}{||c||c|c|c|c|c|c||c|c|c|c|c|c||}
\hline

		& \multicolumn{6}{c||}{$n=4$} 			& \multicolumn{6}{|c||}{$n=8$}\\\hline
$m$ 		& $4$ 	& $8$ 	& $16$ 	& $32$ 	& $64$ 	& $128$ & $4$ 	& $8$ 	& $16$ 	& $32$ 	& $64$ 	& $128$\\\hline
$\#$ cp 	& $9$ 	& $35$ 	& $53$ 	& $20$  & $6$ 	& $7$  	& $24$ 	& $64$ 	& $74$ 	& $14$ 	& $6$ 	& $6$ \\\hline
$\#$ ci 	& $491$	& $465$ & $0$ 	& $0$  	& $0$ 	& $0$  	& $476$ & $436$ & $0$ 	& $0$ 	& $0$ 	& $0$ \\\hline
$\#$ to 	& $0$ 	& $0$ 	& $447$ & $480$ & $494$ & $493$ & $0$ 	& $0$ 	& $426$	& $486$ & $494$ & $494$ \\\hline
		& \multicolumn{6}{c||}{$n=16$} 			& \multicolumn{6}{|c||}{$n=32$}\\\hline
$m$ 		& $4$ 	& $8$ 	& $16$ 	& $32$ 	& $64$ 	& $128$ & $4$ 	& $8$ 	& $16$ 	& $32$ 	& $64$ 	& $128$\\\hline
$\#$ cp 	& $24$ 	& $62$ 	& $54$ 	& $20$ 	& $8$ 	& $4$  	& $44$ 	& $59$ 	& $66$ 	& $23$ 	& $14$ 	& $2$ \\\hline
$\#$ ci 	& $476$ & $438$ & $0$ 	& $0$  	& $0$ 	& $0$  	& $456$ & $441$ & $0$ 	& $0$ 	& $0$ 	& $0$ \\\hline
$\#$ to 	& $0$ 	& $0$ 	& $446$ & $480$	& $492$ & $496$ & $0$ 	& $0$ 	& $434$ & $477$ & $486$ & $498$ \\\hline
		& \multicolumn{6}{c||}{$n=64$} 			& \multicolumn{6}{|c||}{$n=128$}\\\hline
$m$ 		& $4$ 	& $8$ 	& $16$ 	& $32$ 	& $64$ 	& $128$ & $4$ 	& $8$ 	& $16$ 	& $32$ 	& $64$ 	& $128$\\\hline
$\#$ cp 	& $47$ 	& $83$ 	& $53$ 	& $22$ 	& $5$ 	& $2$  	& $50$ 	& $86$ 	& $40$ 	& $21$ 	& $4$ 	& $2$ \\\hline
$\#$ ci 	& $453$ & $417$ & $0$ 	& $0$  	& $0$ 	& $0$  	& $450$	& $414$ & $0$ 	& $0$ 	& $0$ 	& $0$ \\\hline
$\#$ to 	& $0$ 	& $0$ 	& $447$ & $478$ & $495$ & $498$ & $0$ 	& $0$ 	& $460$ & $479$	& $496$ & $498$ \\\hline
\end{tabular}
\end{center}
\clearpage
\subsubsection{Computational Costs}
\begin{figure}[ht]
\centering
	\includegraphics[scale=0.3]{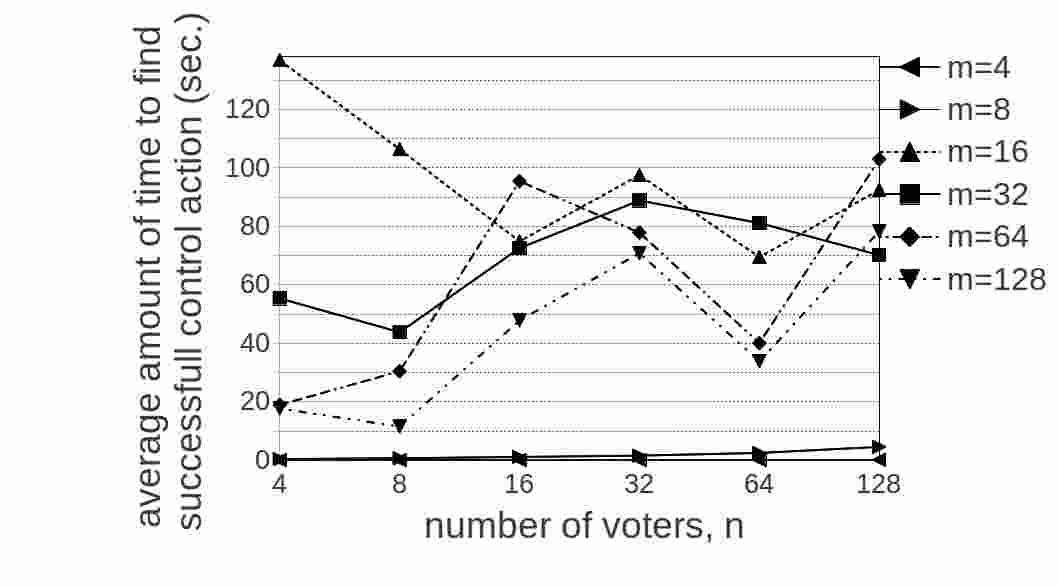}
	\caption{Average time the algorithm needs to find a successful control action for 
	constructive control by partition of candidates in model TP
	in fallback elections in the IC model. The maximum is $136,83$ seconds.}
\end{figure}

\begin{figure}[ht]
\centering
	\includegraphics[scale=0.3]{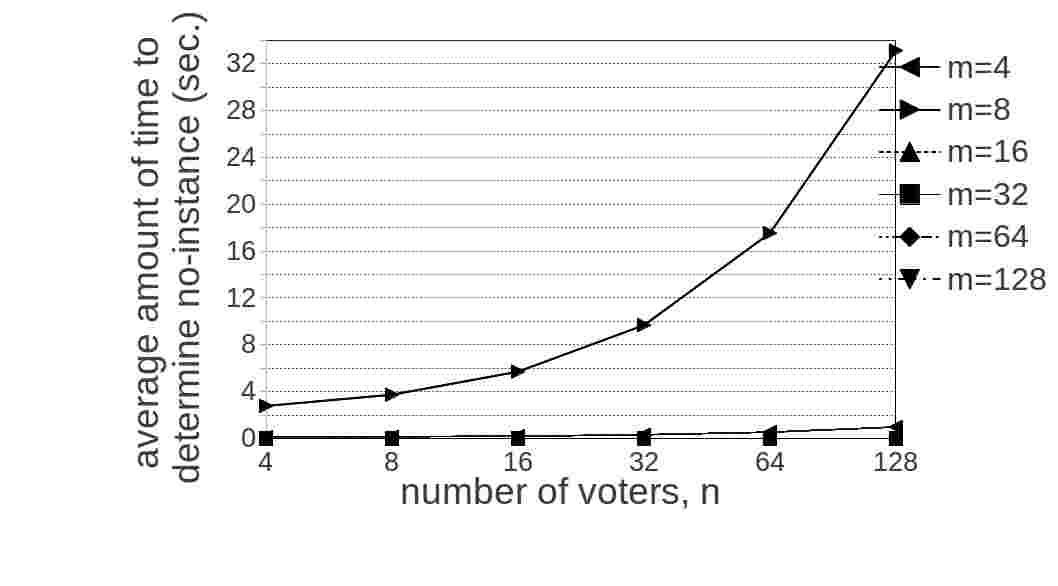}
	\caption{Average time the algorithm needs to determine no-instance of 
		constructive control by partition of candidates in model TP
	in fallback elections in the IC model. The maximum is $33,13$ seconds.}
\end{figure}

\begin{figure}[ht]
\centering
	\includegraphics[scale=0.3]{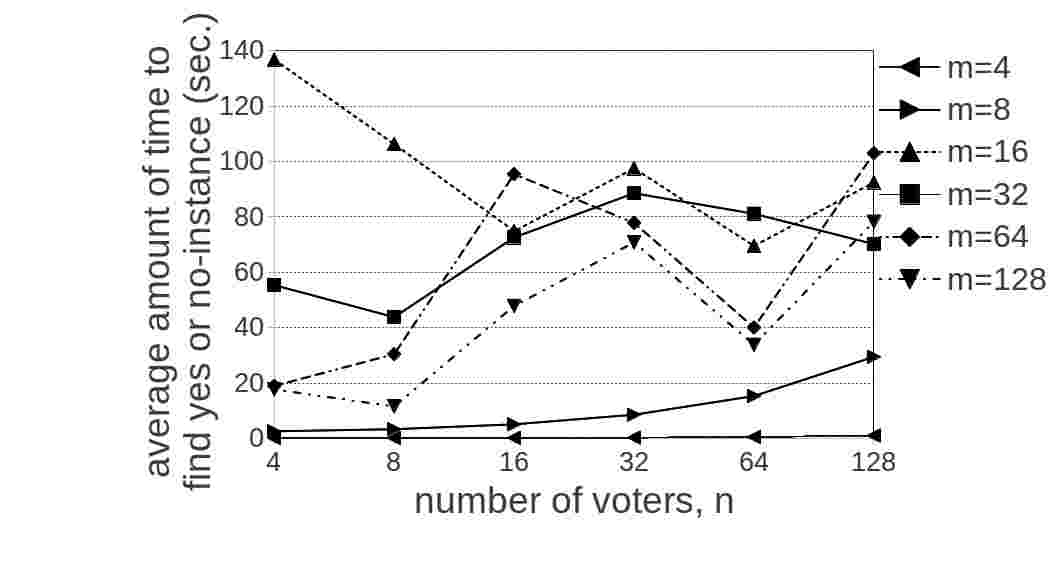}
	\caption{Average time the algorithm needs to give a definite output for 
	constructive control by partition of candidates in model TP
	in fallback elections in the IC model. The maximum is $136,83$ seconds.}
\end{figure}

\begin{figure}[ht]
\centering
	\includegraphics[scale=0.3]{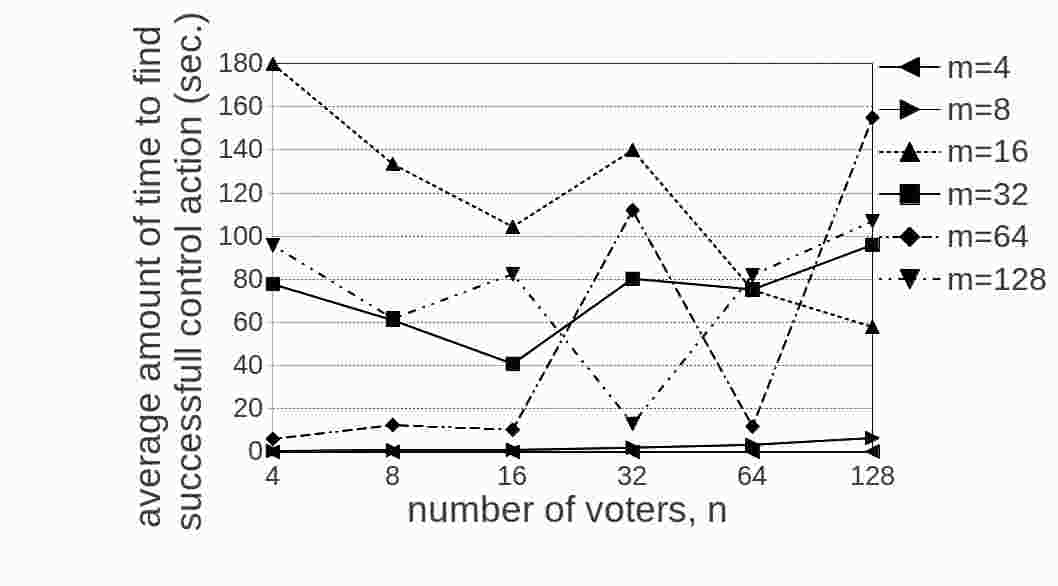}
	\caption{Average time the algorithm needs to find a successful control action for 
	constructive control by partition of candidates in model TP
	in fallback elections in the TM model. The maximum is $179,79$ seconds.}
\end{figure}

\begin{figure}[ht]
\centering
	\includegraphics[scale=0.3]{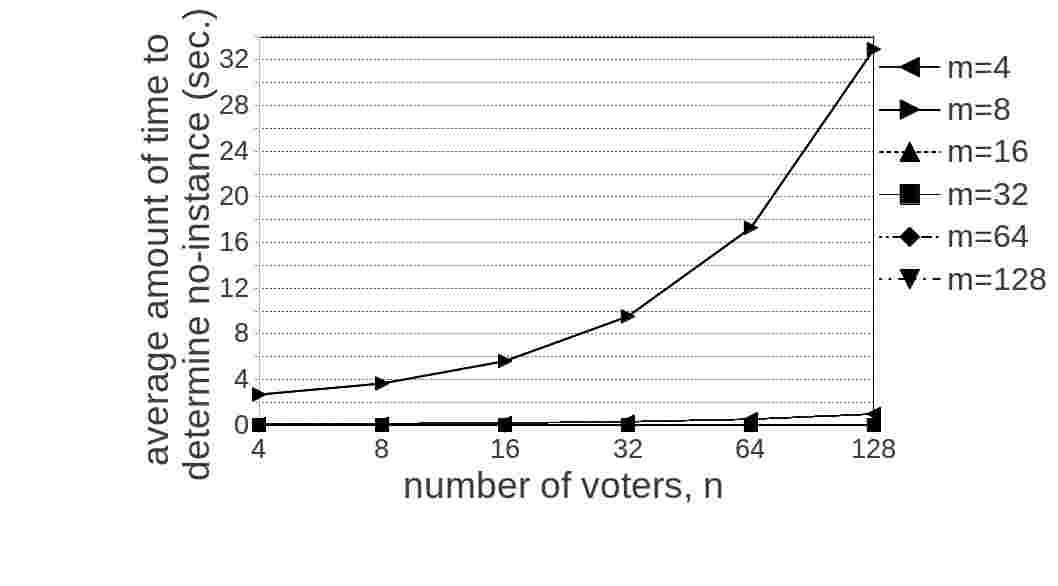}
	\caption{Average time the algorithm needs to determine no-instance of 
		constructive control by partition of candidates in model TP
	in fallback elections in the TM model. The maximum is $32,92$ seconds.}
\end{figure}

\begin{figure}[ht]
\centering
	\includegraphics[scale=0.3]{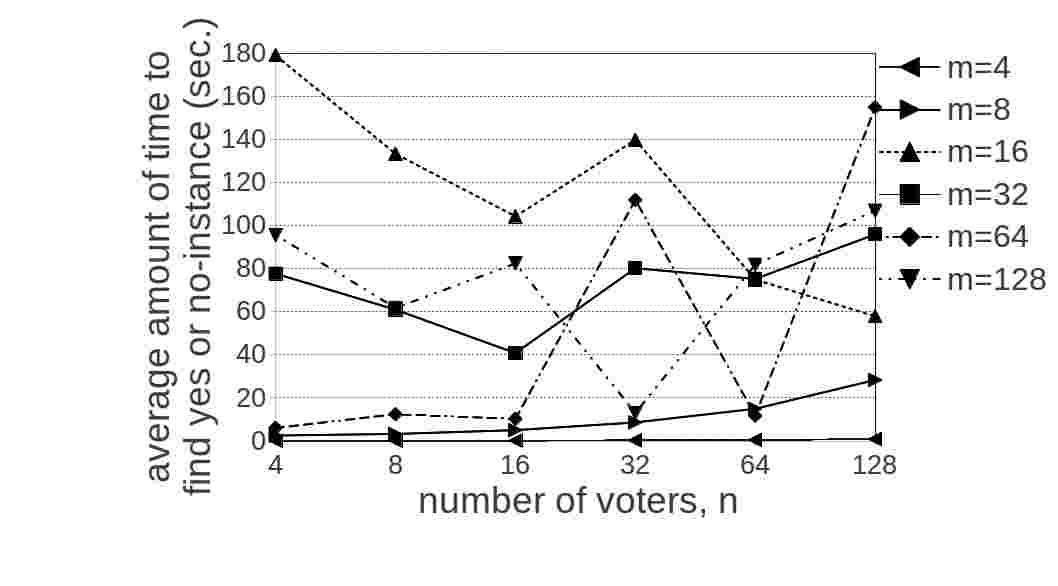}
	\caption{Average time the algorithm needs to give a definite output for 
	constructive control by partition of candidates in model TP
	in fallback elections in the TM model. The maximum is $179,19$ seconds.}
\end{figure}

\clearpage
\subsection{Destructive Control by Partition of Candidates in Model TP}
\begin{center}
\begin{figure}[ht]
\centering
	\includegraphics[scale=0.3]{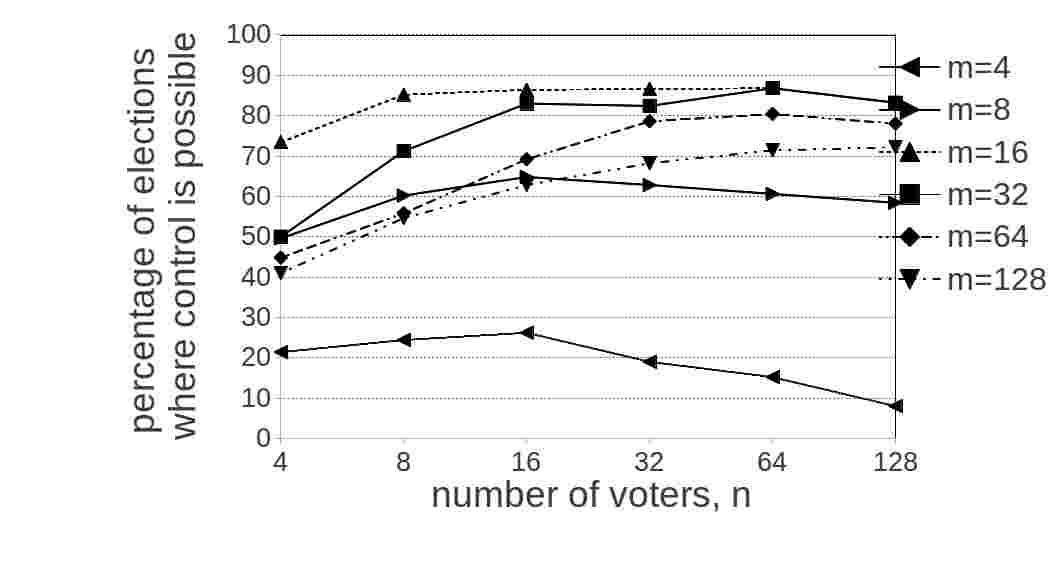}
		\caption{Results for fallback voting in the IC model for 
destructive control by partition of candidates in model TP. Number of candidates is fixed. }
\end{figure}
\begin{tabular}{||c||c|c|c|c|c|c||c|c|c|c|c|c||}
\hline

		& \multicolumn{6}{c||}{$m=4$} 			& \multicolumn{6}{|c||}{$m=8$}\\\hline
$n$ 		& $4$ 	& $8$ 	& $16$ 	& $32$ 	& $64$ 	& $128$ & $4$ 	& $8$ 	& $16$ 	& $32$ 	& $64$ 	& $128$\\\hline
$\#$ cp 	& $107$ & $122$ & $131$	& $95$  & $76$ 	& $40$  & $248$ & $301$ & $324$ & $314$ & $303$ & $292$ \\\hline
$\#$ ci 	& $393$ & $378$ & $369$ & $405$ & $424$ & $460$ & $252$ & $199$ & $176$ & $186$ & $197$ & $208$ \\\hline
$\#$ to 	& $0$ 	& $0$ 	& $0$ 	& $0$  	& $0$ 	& $0$  	& $0$ 	& $0$ 	& $0$ 	& $0$ 	& $0$ 	& $0$ \\\hline
		& \multicolumn{6}{c||}{$m=16$} 			& \multicolumn{6}{|c||}{$m=32$}\\\hline
$n$ 		& $4$ 	& $8$ 	& $16$ 	& $32$ 	& $64$ 	& $128$ & $4$ 	& $8$ 	& $16$ 	& $32$ 	& $64$ 	& $128$\\\hline
$\#$ cp 	& $367$ & $426$ & $432$ & $433$ & $434$ & $416$ & $250$ & $356$ & $415$ & $412$ & $434$ & $416$ \\\hline
$\#$ ci 	& $9$ 	& $0$ 	& $0$ 	& $0$  	& $0$ 	& $0$  	& $1$ 	& $0$ 	& $0$ 	& $0$ 	& $0$ 	& $0$ \\\hline
$\#$ to 	& $124$ & $74$ 	& $68$ 	& $67$  & $66$ 	& $84$  & $249$ & $144$ & $85$ 	& $88$ 	& $66$ 	& $84$ \\\hline
		& \multicolumn{6}{c||}{$m=64$} 			& \multicolumn{6}{|c||}{$m=128$}\\\hline
$n$ 		& $4$ 	& $8$ 	& $16$ 	& $32$ 	& $64$ 	& $128$ & $4$ 	& $8$ 	& $16$ 	& $32$ 	& $64$ 	& $128$\\\hline
$\#$ cp 	& $224$	& $279$ & $346$ & $393$ & $402$ & $390$ & $205$ & $273$	& $314$ & $341$ & $357$ & $361$ \\\hline
$\#$ ci 	& $0$ 	& $0$ 	& $0$ 	& $0$  	& $0$ 	& $0$  	& $0$ 	& $0$ 	& $0$ 	& $0$ 	& $0$ 	& $0$ \\\hline
$\#$ to 	& $276$ & $221$ & $154$ & $107$ & $98$ 	& $110$ & $295$ & $227$ & $186$ & $159$ & $143$ & $139$ \\\hline
\end{tabular}
\newpage
\begin{figure}[ht]
\centering
	\includegraphics[scale=0.3]{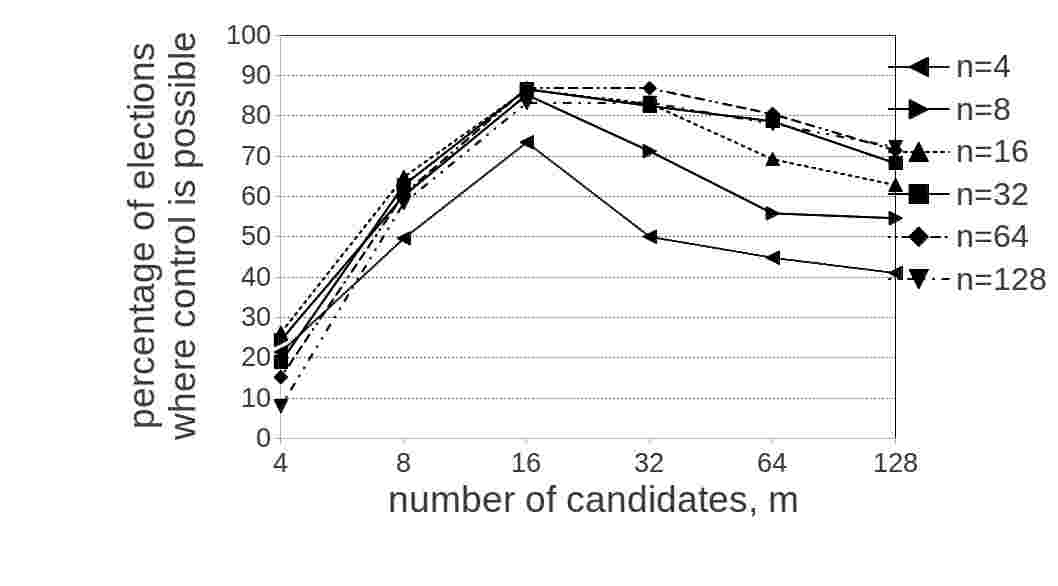}
		\caption{Results for fallback voting in the IC model for 
destructive control by partition of candidates in model TP. Number of voters is fixed. }
\end{figure}

%
\newpage
\begin{figure}[ht]
\centering
	\includegraphics[scale=0.3]{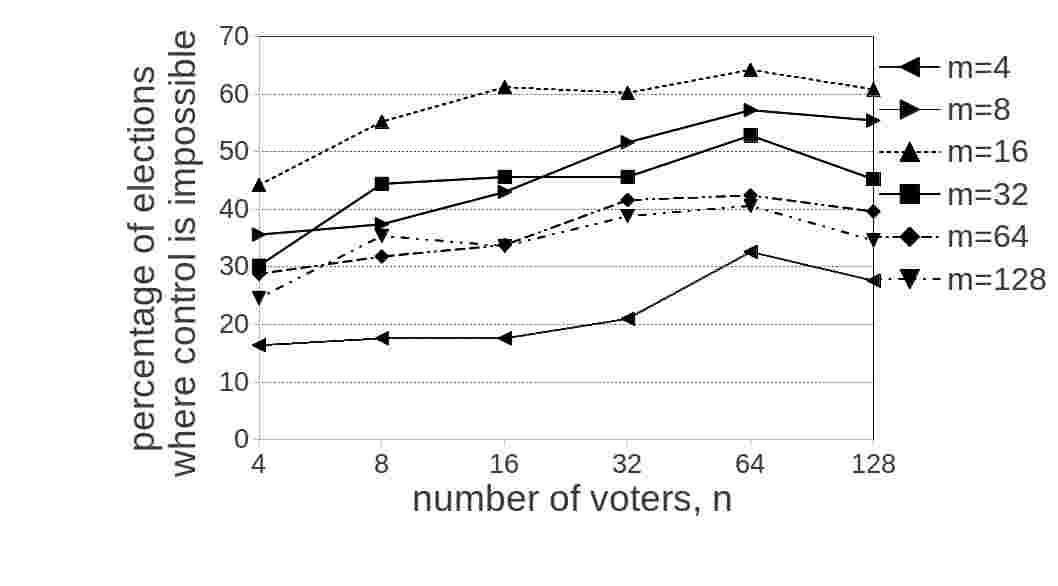}
		\caption{Results for fallback voting in the TM model for 
destructive control by partition of candidates in model TP. Number of candidates is fixed. }
\end{figure}

%
\newpage
\begin{figure}[ht]
\centering
	\includegraphics[scale=0.3]{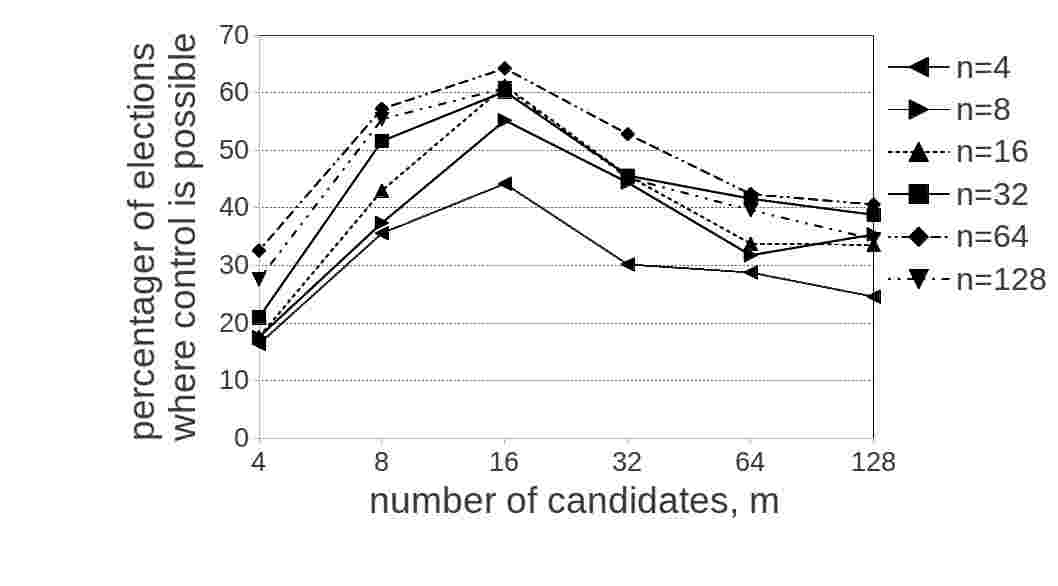}
		\caption{Results for fallback voting in the TM model for 
destructive control by partition of candidates in model TP. Number of voters is fixed. }
\end{figure}

%
\end{center}
\clearpage
\subsubsection{Computational Costs}
\begin{figure}[ht]
\centering
	\includegraphics[scale=0.3]{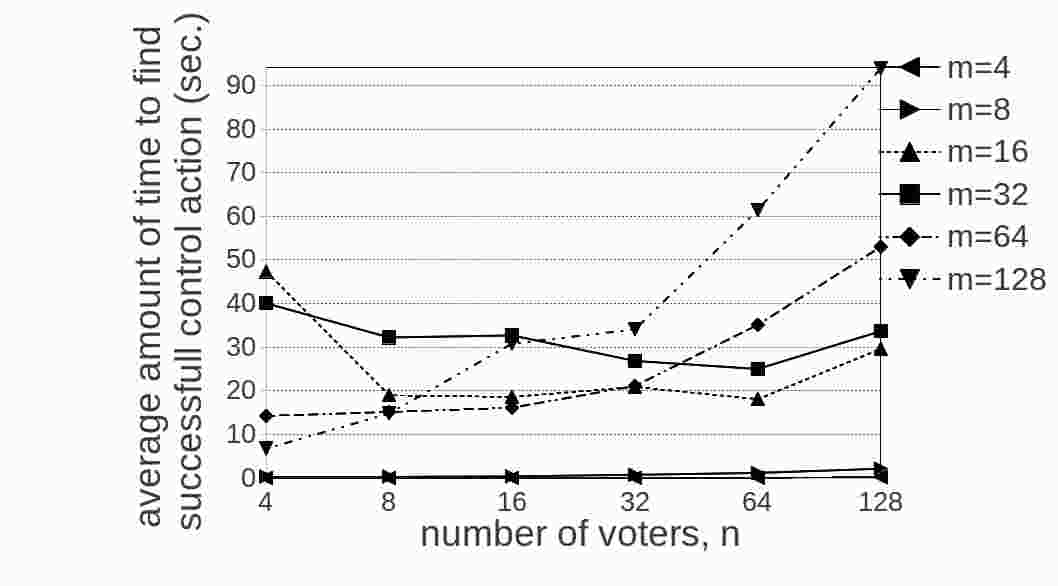}
	\caption{Average time the algorithm needs to find a successful control action for 
	destructive control by partition of candidates in model TP
	in fallback elections in the IC model. The maximum is $93,92$ seconds.}
\end{figure}

\begin{figure}[ht]
\centering
	\includegraphics[scale=0.3]{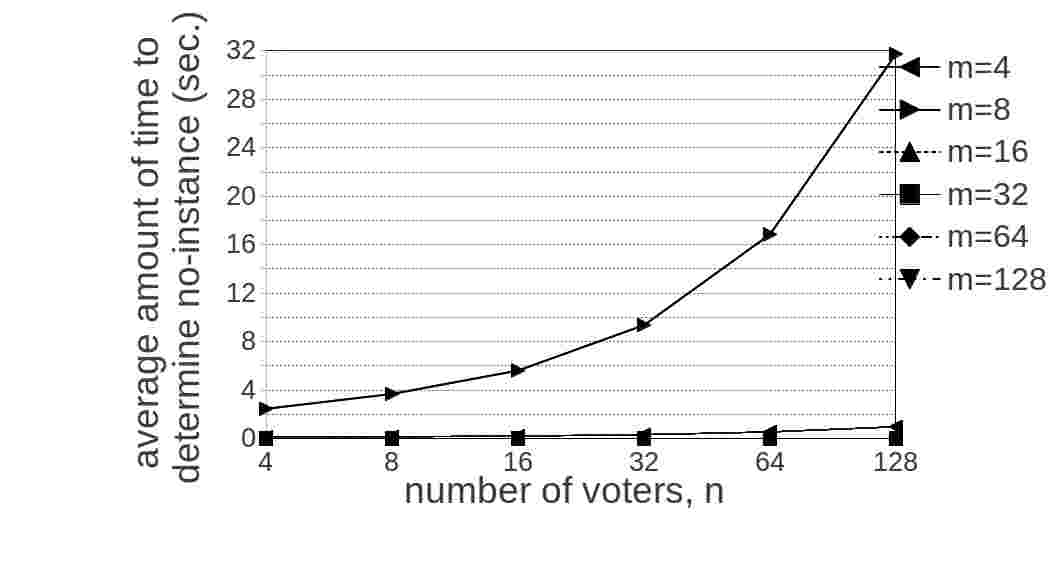}
	\caption{Average time the algorithm needs to determine no-instance of 
		destructive control by partition of candidates in model TP
	in fallback elections in the IC model. The maximum is $31,75$ seconds.}
\end{figure}

\begin{figure}[ht]
\centering
	\includegraphics[scale=0.3]{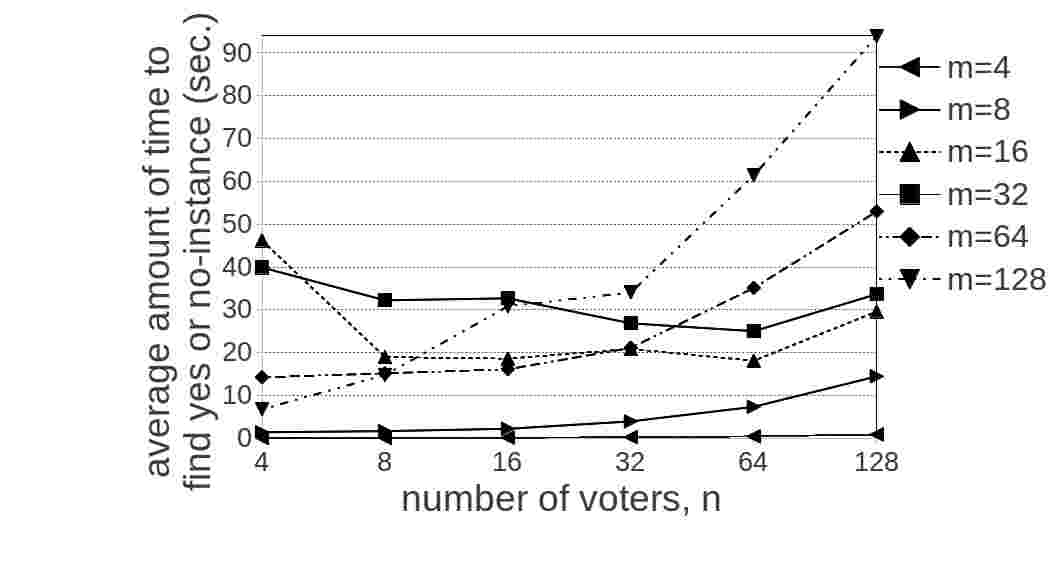}
	\caption{Average time the algorithm needs to give a definite output for 
	destructive control by partition of candidates in model TP
	in fallback elections in the IC model. The maximum is $93,92$ seconds.}
\end{figure}

\begin{figure}[ht]
\centering
	\includegraphics[scale=0.3]{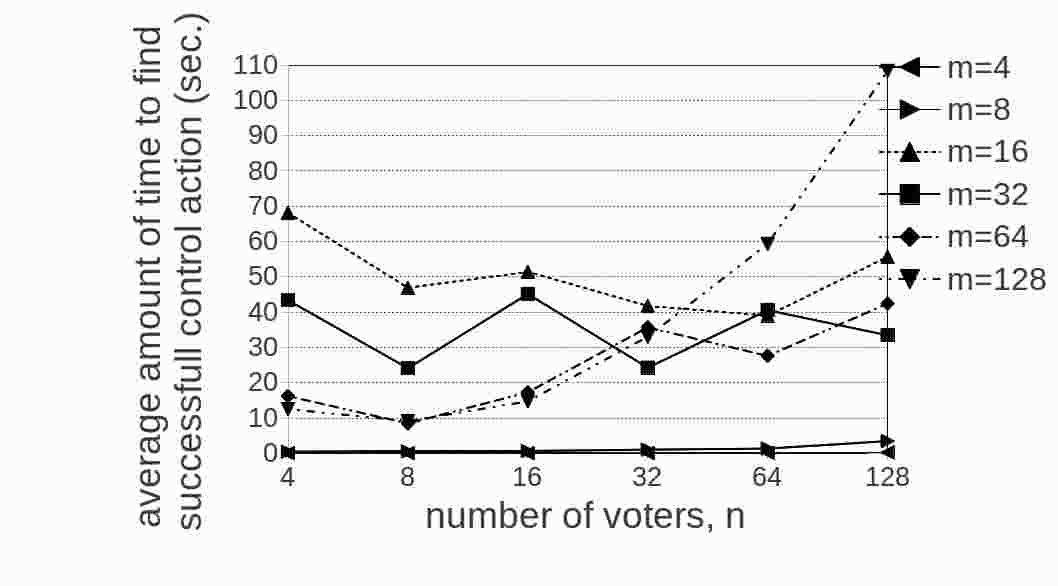}
	\caption{Average time the algorithm needs to find a successful control action for 
	destructive control by partition of candidates in model TP
	in fallback elections in the TM model. The maximum is $108,35$ seconds.}
\end{figure}

\begin{figure}[ht]
\centering
	\includegraphics[scale=0.3]{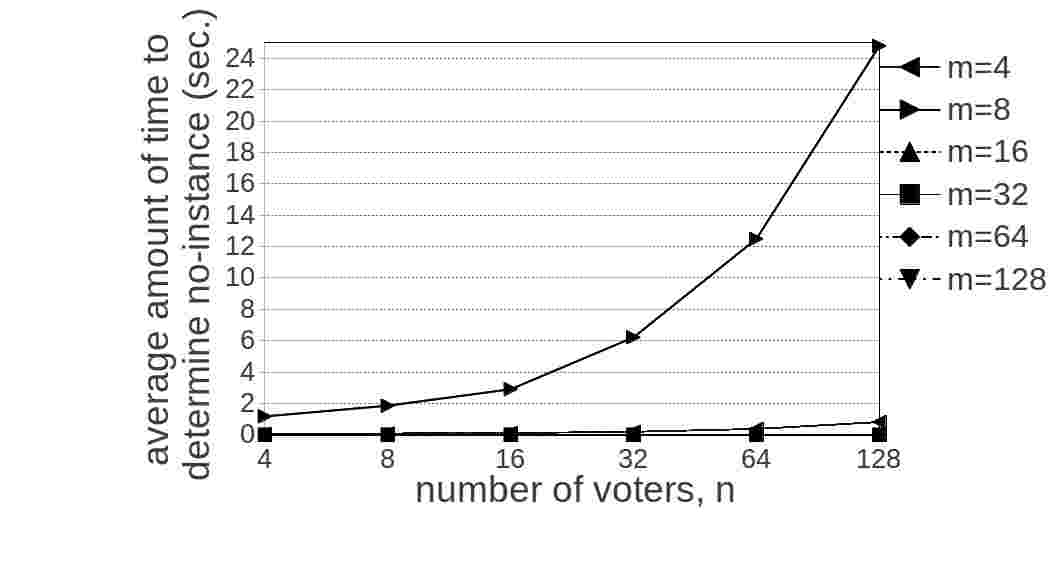}
	\caption{Average time the algorithm needs to determine no-instance of 
		destructive control by partition of candidates in model TP
	in fallback elections in the TM model. The maximum is $24,8$ seconds.}
\end{figure}

\begin{figure}[ht]
\centering
	\includegraphics[scale=0.3]{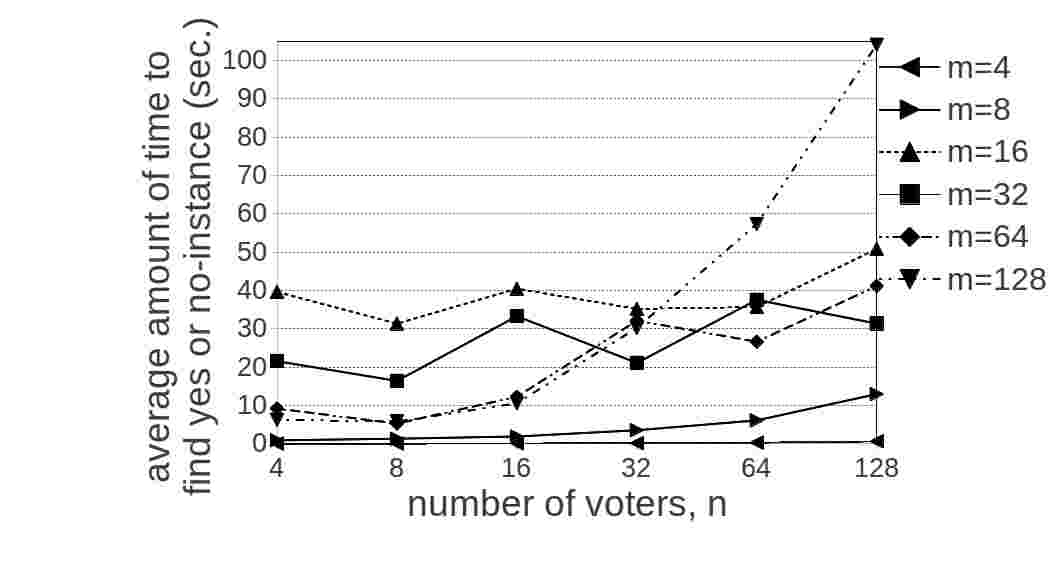}
	\caption{Average time the algorithm needs to give a definite output for 
	destructive control by partition of candidates in model TP
	in fallback elections in the TM model. The maximum is $104,12$ seconds.}
\end{figure}

\clearpage
\subsection{Constructive Control by Runoff Partition of Candidates in Model TE}
\begin{center}

\begin{figure}[ht]
\centering
	\includegraphics[scale=0.3]{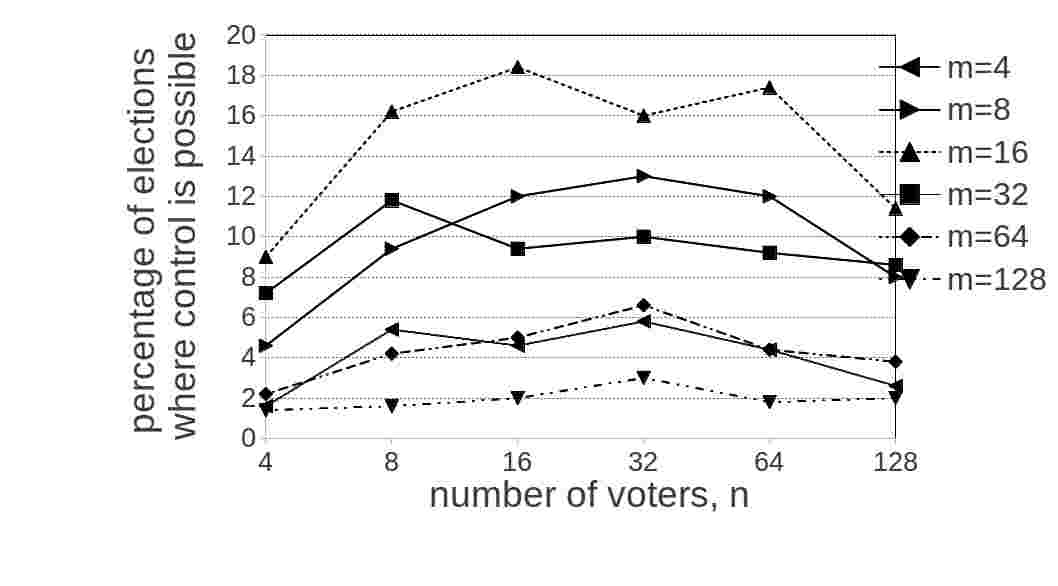}
		\caption{Results for fallback voting in the IC model for 
constructive control by runoff-partition  of candidates in model TE. Number of candidates is fixed. }
\end{figure}

\newpage
\begin{figure}[ht]
\centering
	\includegraphics[scale=0.3]{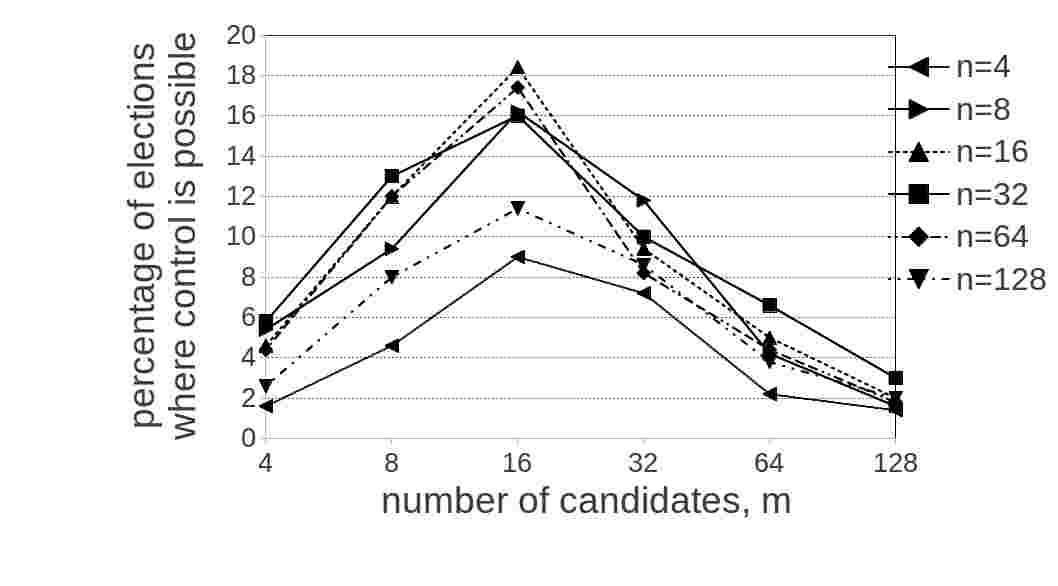}
		\caption{Results for fallback voting in the IC model for 
constructive control by runoff-partition  of candidates in model TE. Number of voters is fixed. }
\end{figure}

%
\newpage
\begin{figure}[ht]
\centering
	\includegraphics[scale=0.3]{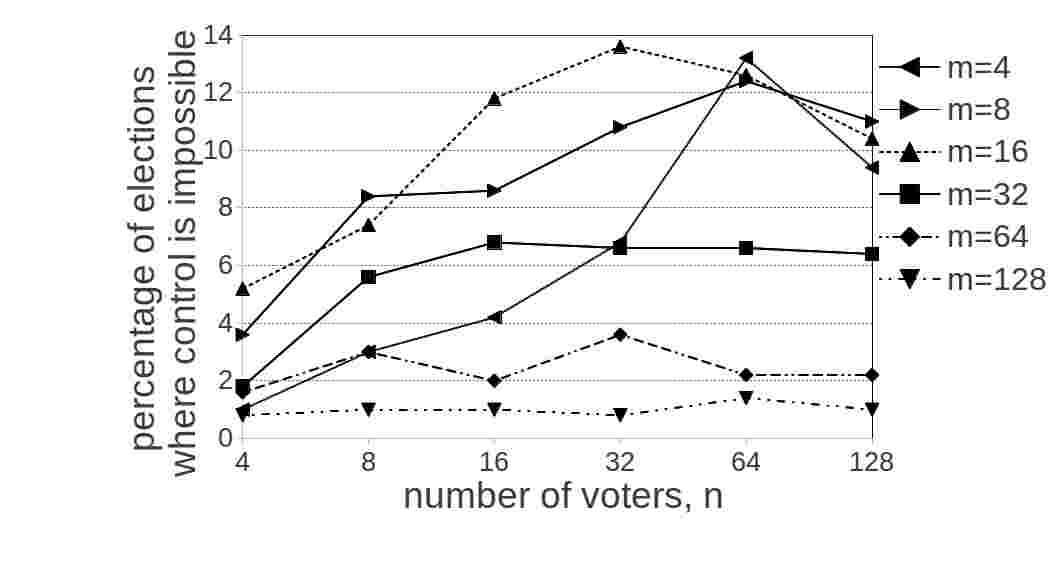}
		\caption{Results for fallback voting in the TM model for 
constructive control by runoff-partition  of candidates in model TE. Number of candidates is fixed. }
\end{figure}

%
\newpage
\begin{figure}[ht]
\centering
	\includegraphics[scale=0.3]{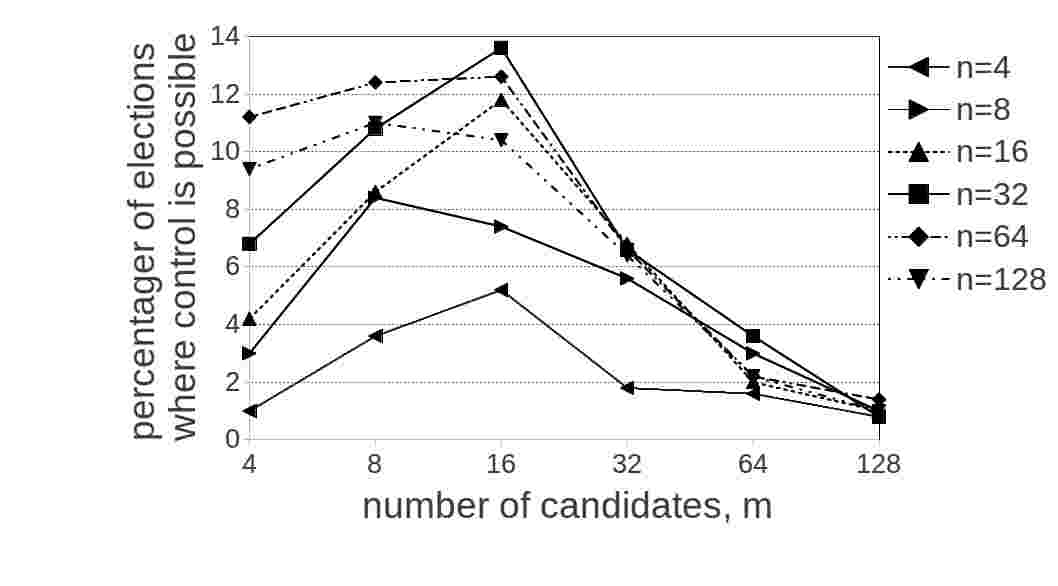}
		\caption{Results for fallback voting in the TM model for 
constructive control by runoff-partition  of candidates in model TE. Number of voters is fixed. }
\end{figure}

%
\end{center}
\clearpage
\subsubsection{Computational Costs}
\begin{figure}[ht]
\centering
	\includegraphics[scale=0.3]{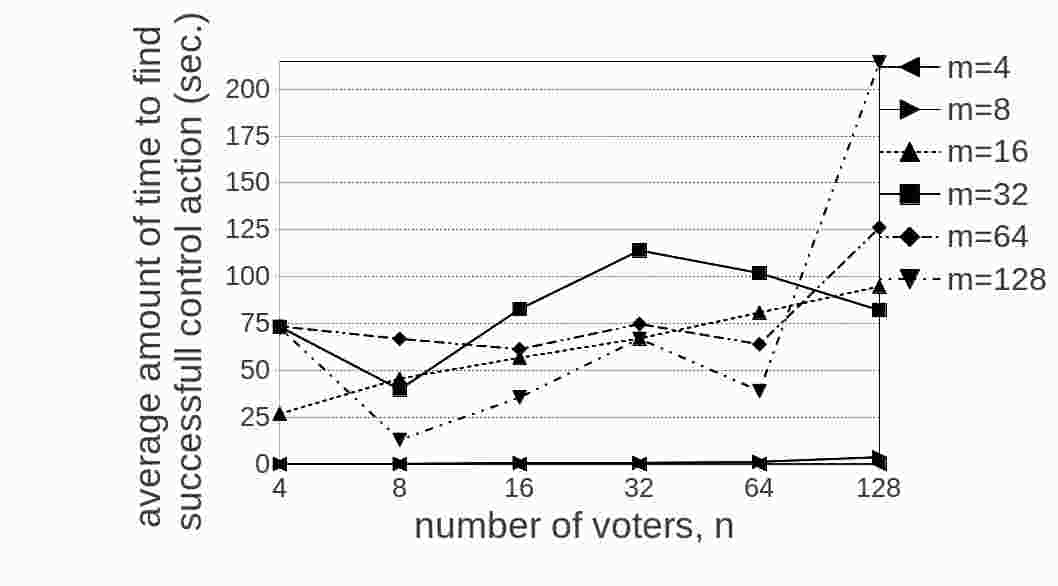}
	\caption{Average time the algorithm needs to find a successful control action for 
	constructive control by runoff-partition  of candidates in model TE
	in fallback elections in the IC model. The maximum is $214,19$ seconds.}
\end{figure}

\begin{figure}[ht]
\centering
	\includegraphics[scale=0.3]{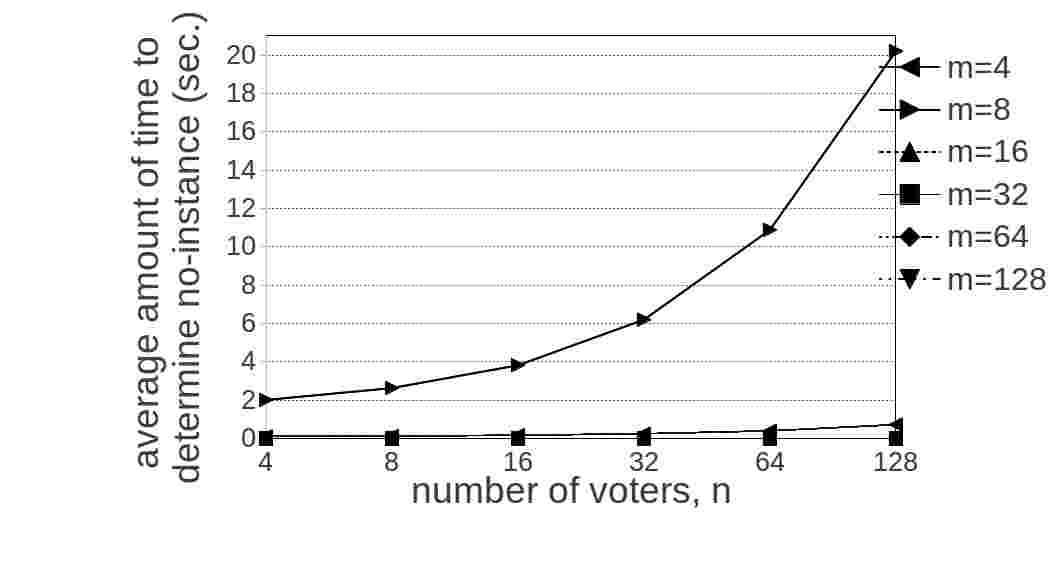}
	\caption{Average time the algorithm needs to determine no-instance of 
		constructive control by runoff-partition  of candidates in model TE
	in fallback elections in the IC model. The maximum is $20,2$ seconds.}
\end{figure}
\begin{figure}[ht]
\centering
	\includegraphics[scale=0.3]{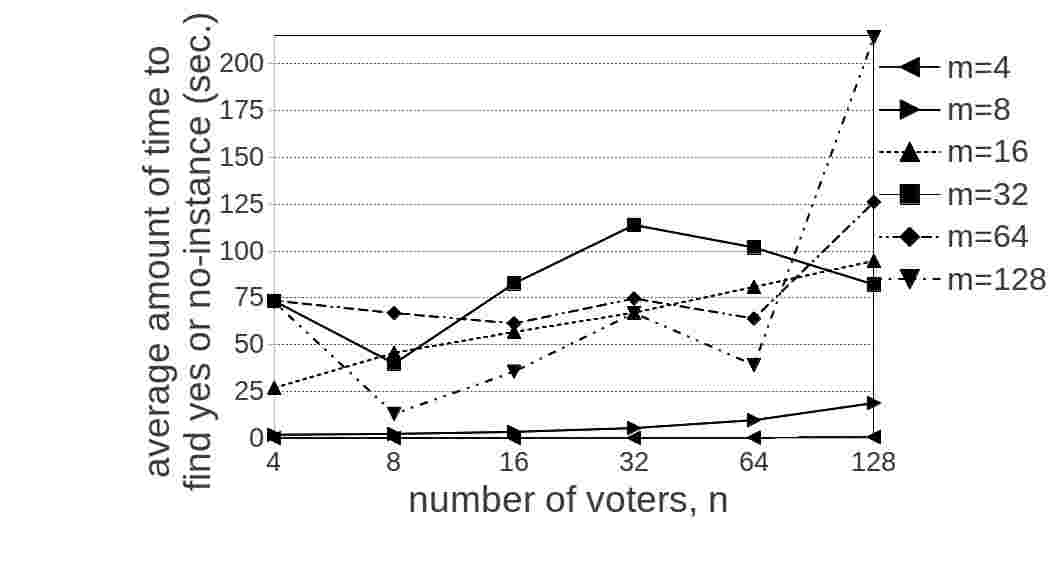}
	\caption{Average time the algorithm needs to give a definite output for 
	constructive control by runoff-partition  of candidates in model TE
	in fallback elections in the IC model. The maximum is $214,19$ seconds.}
\end{figure}

\begin{figure}[ht]
\centering
	\includegraphics[scale=0.3]{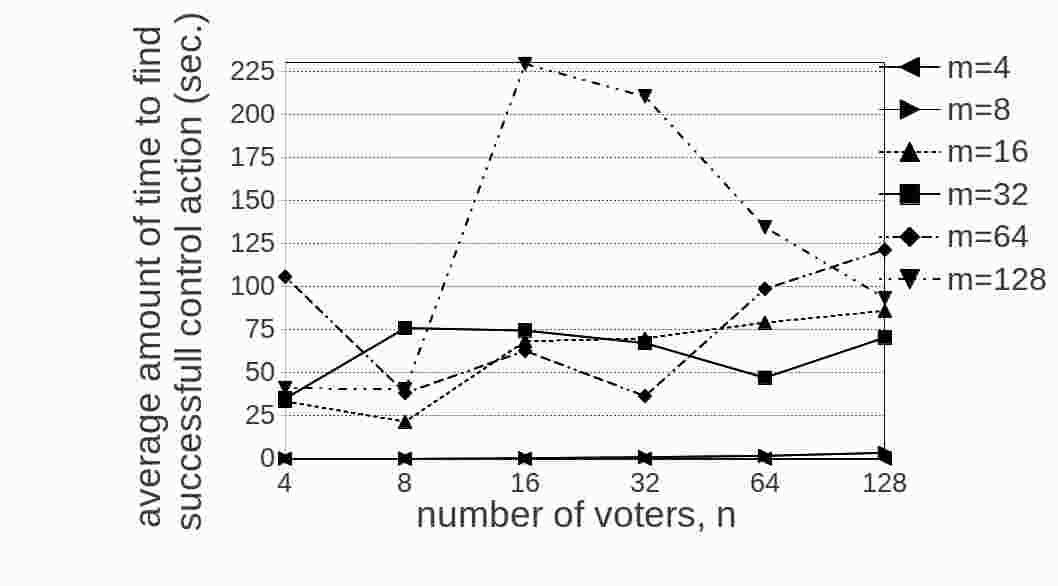}
	\caption{Average time the algorithm needs to find a successful control action for 
	constructive control by runoff-partition  of candidates in model TE
	in fallback elections in the TM model. The maximum is $229,26$ seconds.}
\end{figure}
\begin{figure}[ht]
\centering
	\includegraphics[scale=0.3]{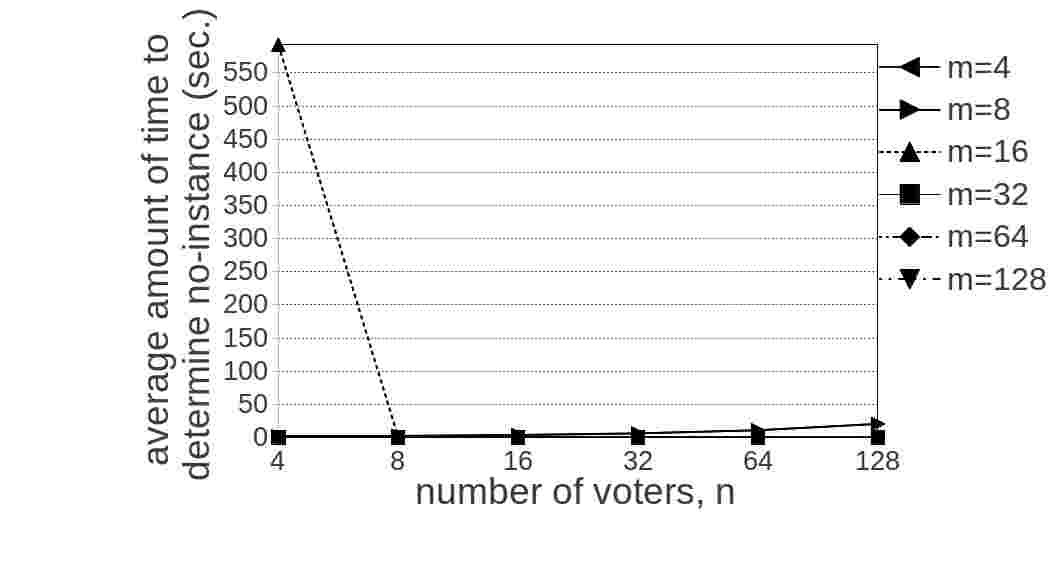}
	\caption{Average time the algorithm needs to determine no-instance of 
		constructive control by runoff-partition  of candidates in model TE
	in fallback elections in the TM model. The maximum is $592,3$ seconds.}
\end{figure}
\begin{figure}[ht]
\centering
	\includegraphics[scale=0.3]{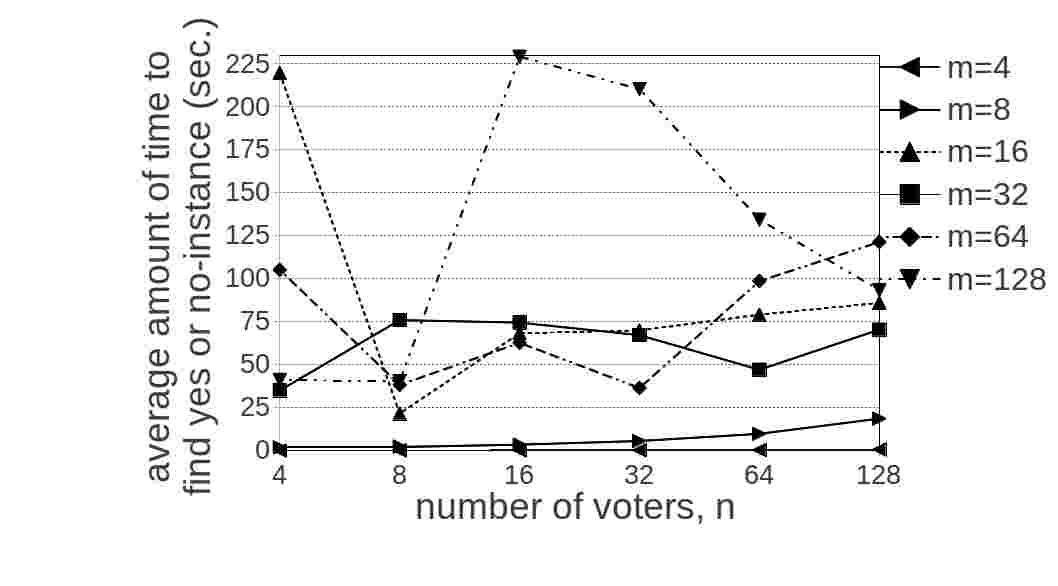}
	\caption{Average time the algorithm needs to give a definite output for 
	constructive control by runoff-partition  of candidates in model TE
	in fallback elections in the TM model. The maximum is $229,26$ seconds.}
\end{figure}

\clearpage
\subsection{Destructive Control by Runoff Partition of Candidates in Model TE}
\begin{center}
\begin{figure}[ht]
\centering
	\includegraphics[scale=0.3]{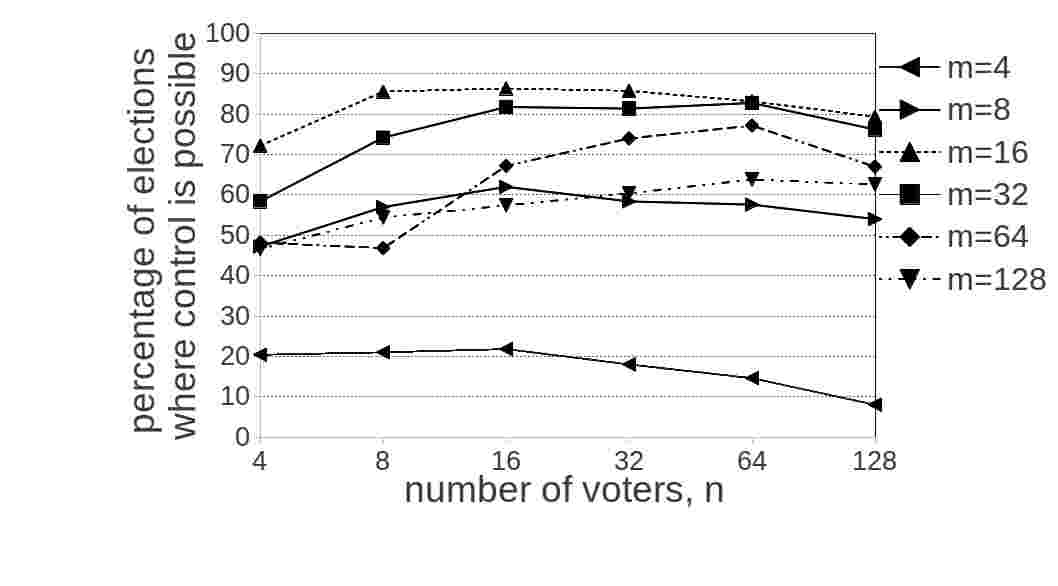}
		\caption{Results for fallback voting in the IC model for 
destructive control by runoff-partition  of candidates in model TE. Number of candidates is fixed. }
\end{figure}

\newpage
\begin{figure}[ht]
\centering
	\includegraphics[scale=0.3]{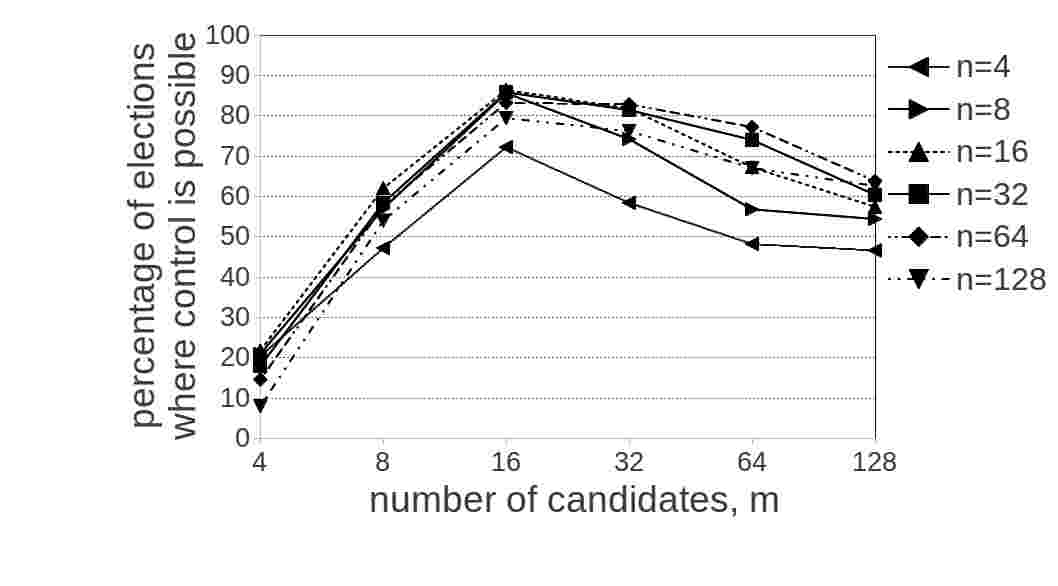}
		\caption{Results for fallback voting in the IC model for 
destructive control by runoff-partition  of candidates in model TE. Number of voters is fixed. }
\end{figure}
%
\newpage
\begin{figure}[ht]
\centering
	\includegraphics[scale=0.3]{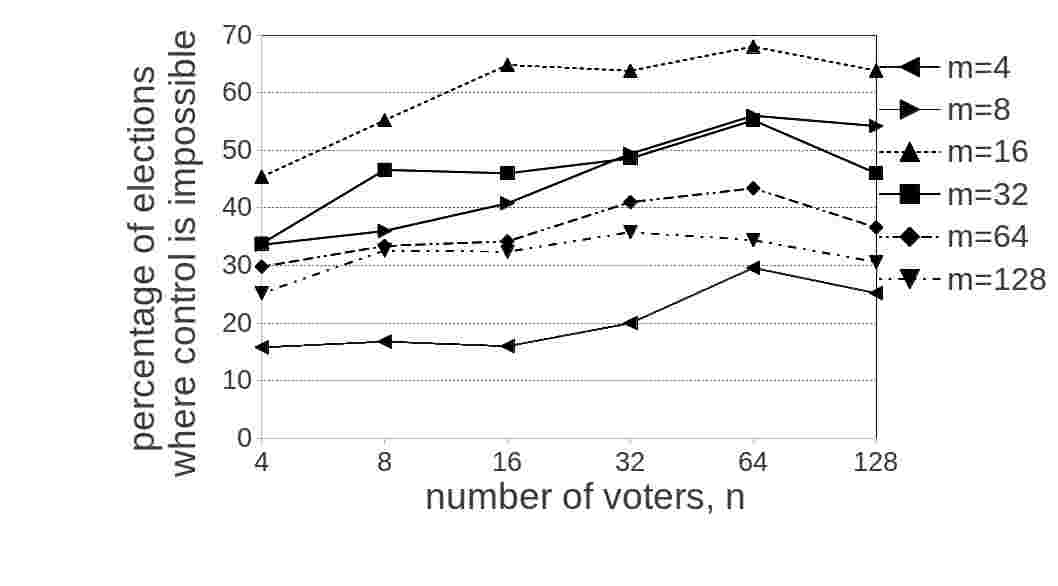}
		\caption{Results for fallback voting in the TM model for 
destructive control by runoff-partition  of candidates in model TE. Number of candidates is fixed. }
\end{figure}
%

\newpage
\begin{figure}[ht]
\centering
	\includegraphics[scale=0.3]{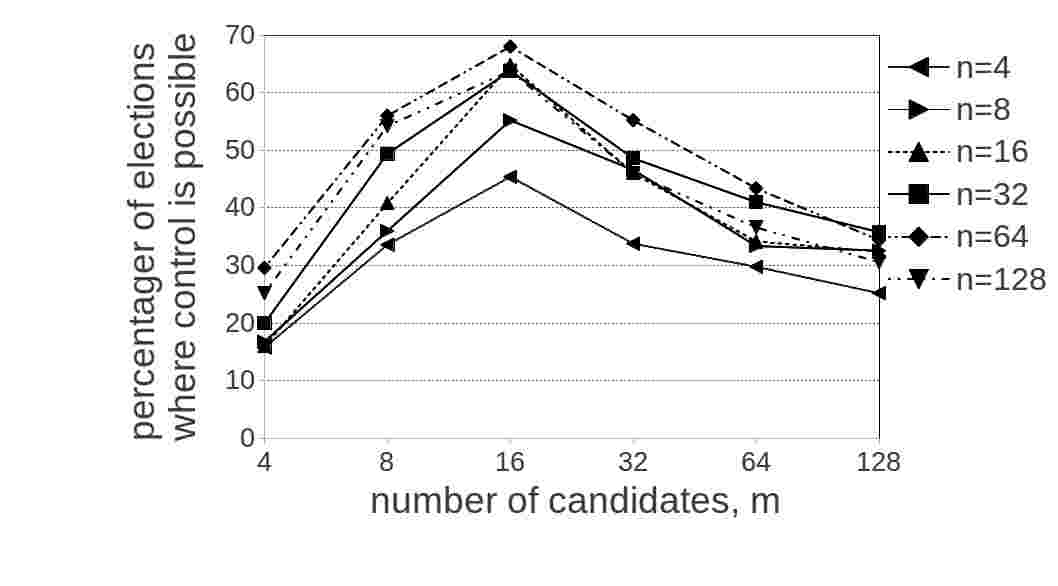}
		\caption{Results for fallback voting in the TM model for 
destructive control by runoff-partition  of candidates in model TE. Number of voters is fixed. }
\end{figure}

%
\end{center}
\clearpage
\subsubsection{Computational Costs}
\begin{figure}[ht]
\centering
	\includegraphics[scale=0.3]{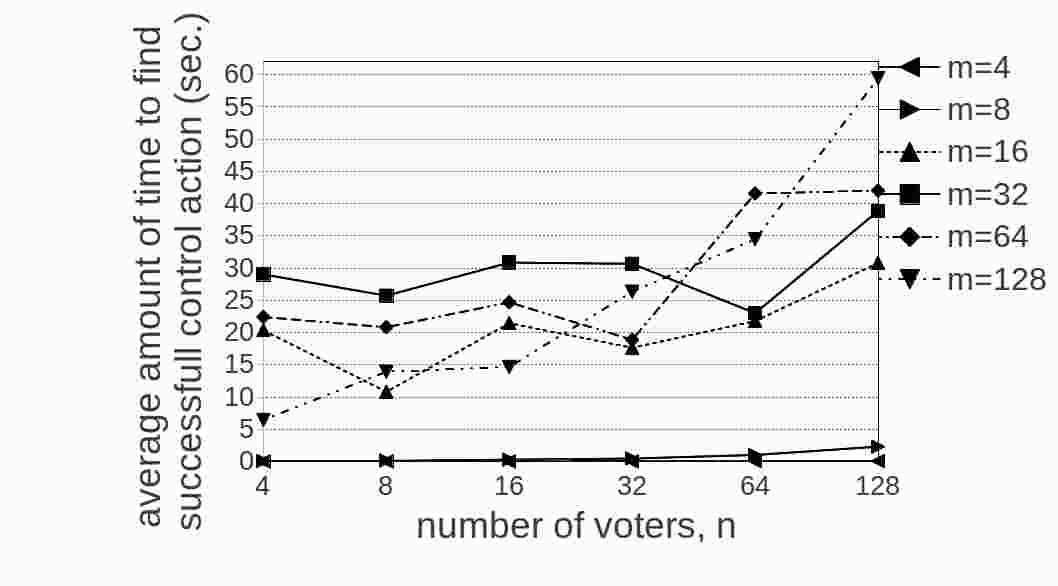}
	\caption{Average time the algorithm needs to find a successful control action for 
	destructive control by runoff-partition  of candidates in model TE
	in fallback elections in the IC model. The maximum is $59,44$ seconds.}
\end{figure}
\begin{figure}[ht]
\centering
	\includegraphics[scale=0.3]{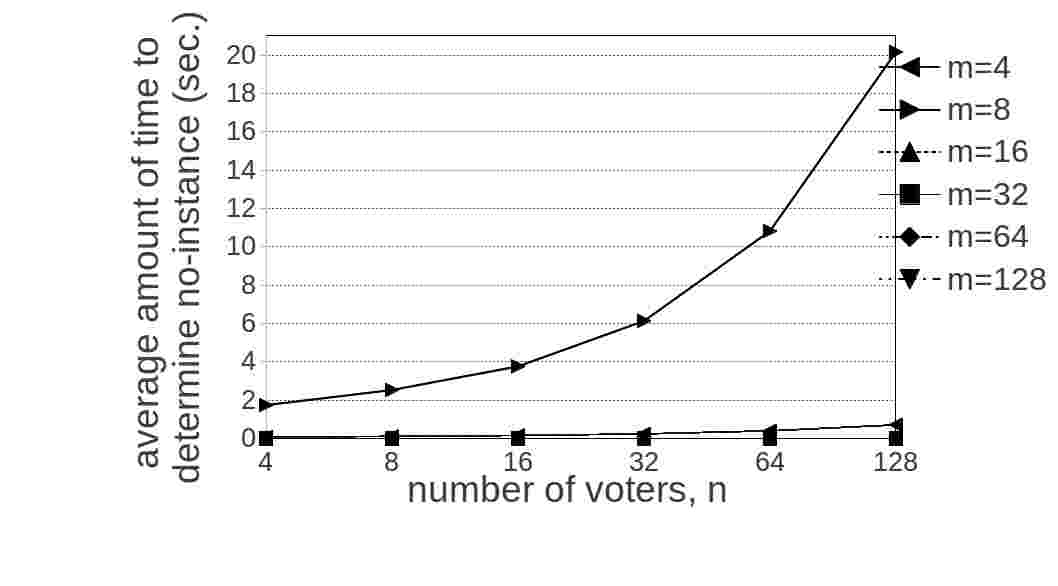}
	\caption{Average time the algorithm needs to determine no-instance of 
		destructive control by runoff-partition  of candidates in model TE
	in fallback elections in the IC model. The maximum is $20,16$ seconds.}
\end{figure}
\begin{figure}[ht]
\centering
	\includegraphics[scale=0.3]{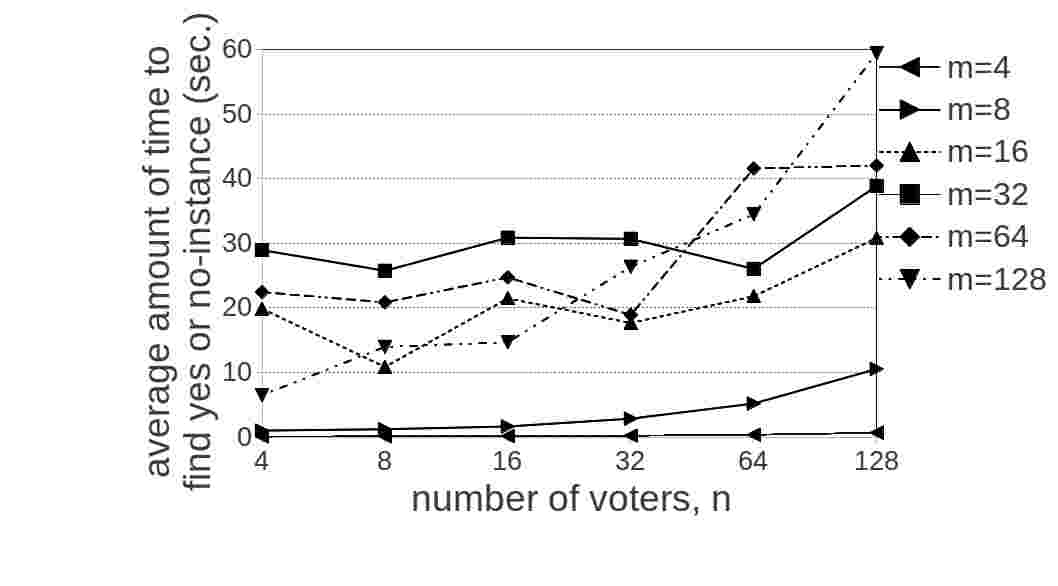}
	\caption{Average time the algorithm needs to give a definite output for 
	destructive control by runoff-partition  of candidates in model TE
	in fallback elections in the IC model. The maximum is $59,44$ seconds.}
\end{figure}
\begin{figure}[ht]
\centering
	\includegraphics[scale=0.3]{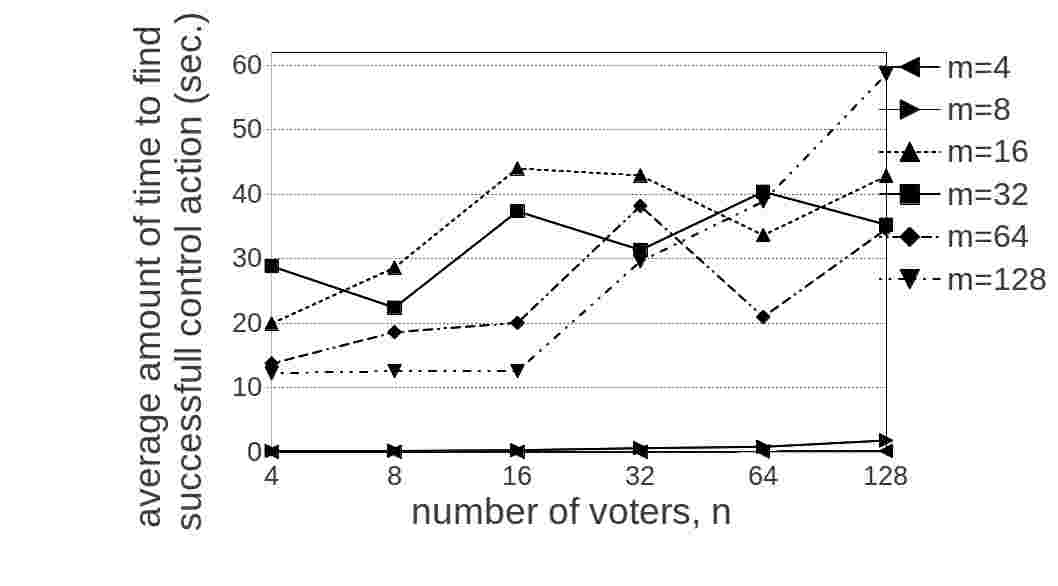}
	\caption{Average time the algorithm needs to find a successful control action for 
	destructive control by runoff-partition  of candidates in model TE
	in fallback elections in the TM model. The maximum is $58,6$ seconds.}
\end{figure}
\begin{figure}[ht]
\centering
	\includegraphics[scale=0.3]{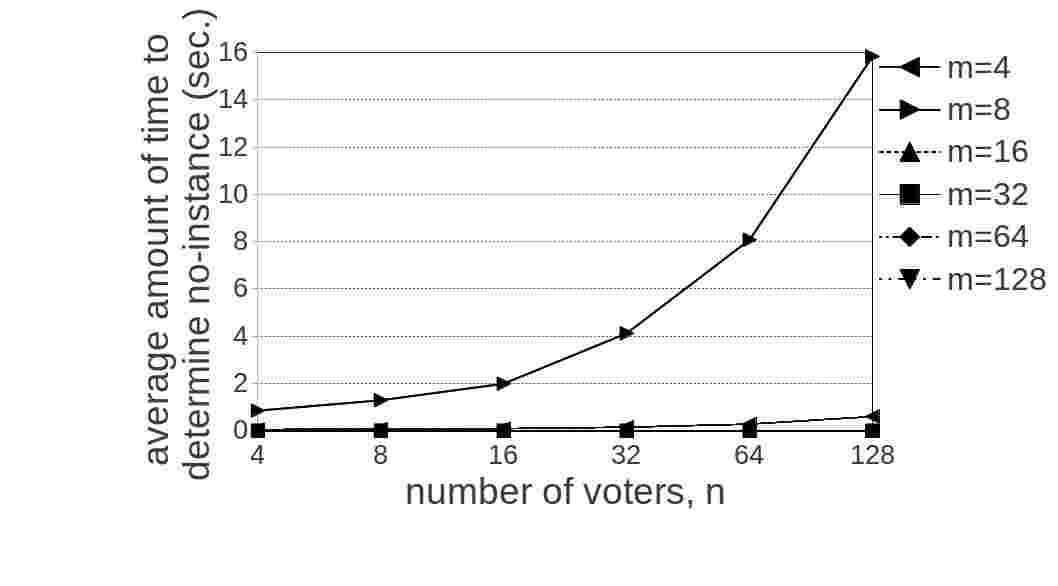}
	\caption{Average time the algorithm needs to determine no-instance of 
		destructive control by runoff-partition  of candidates in model TE
	in fallback elections in the TM model. The maximum is $15,84$ seconds.}
\end{figure}
\begin{figure}[ht]
\centering
	\includegraphics[scale=0.3]{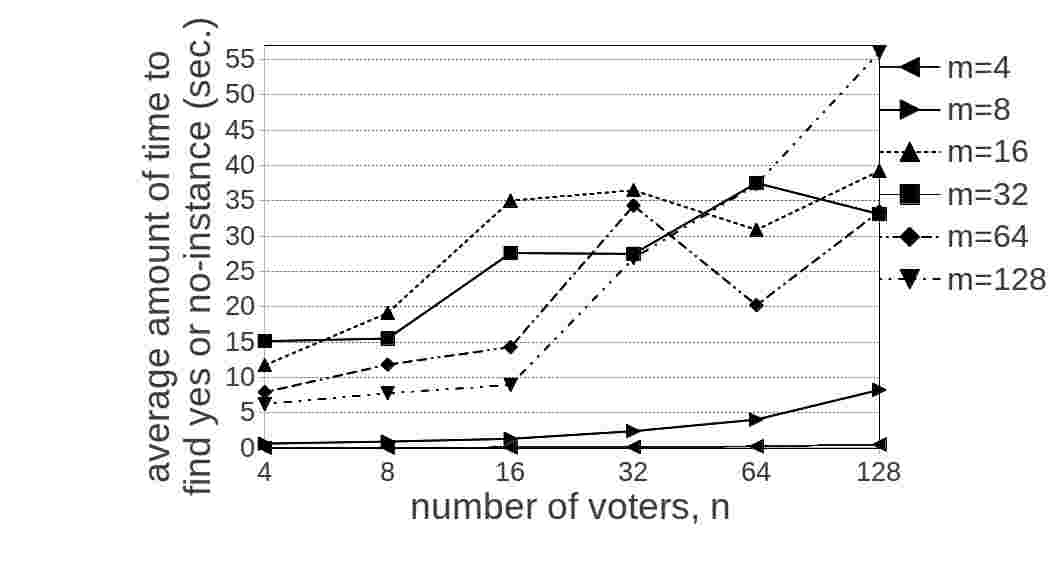}
	\caption{Average time the algorithm needs to give a definite output for 
	destructive control by runoff-partition  of candidates in model TE
	in fallback elections in the TM model. The maximum is $56,03$ seconds.}
\end{figure}

\clearpage
\subsection{Constructive Control by Runoff Partition of Candidates in Model TP}
\begin{center}
\begin{figure}[ht]
\centering
	\includegraphics[scale=0.3]{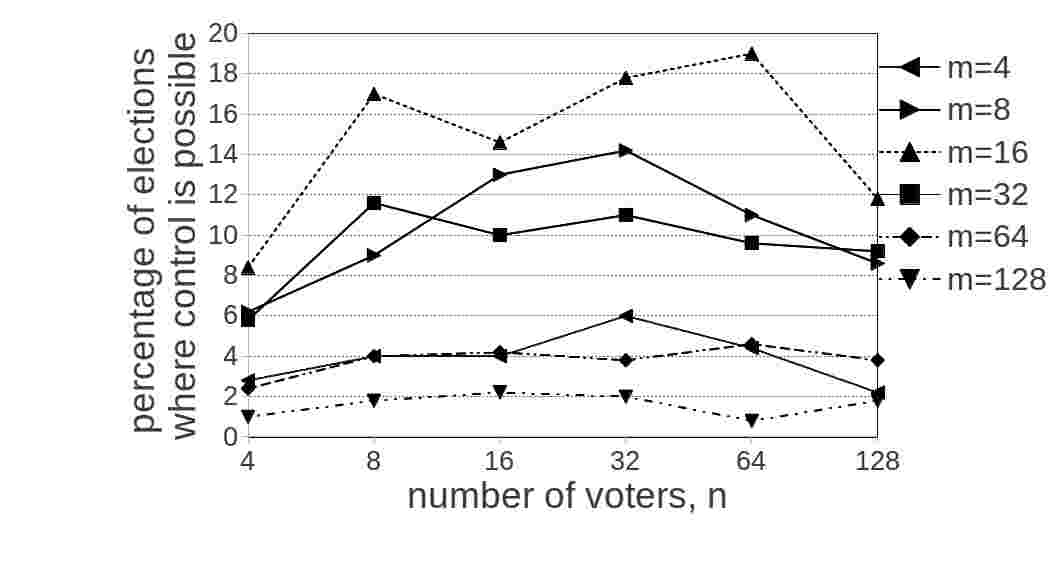}
		\caption{Results for fallback voting in the IC model for 
constructive control by runoff-partition  of candidates in model TP. Number of candidates is fixed. }
\end{figure}

\newpage
\begin{figure}[ht]
\centering
	\includegraphics[scale=0.3]{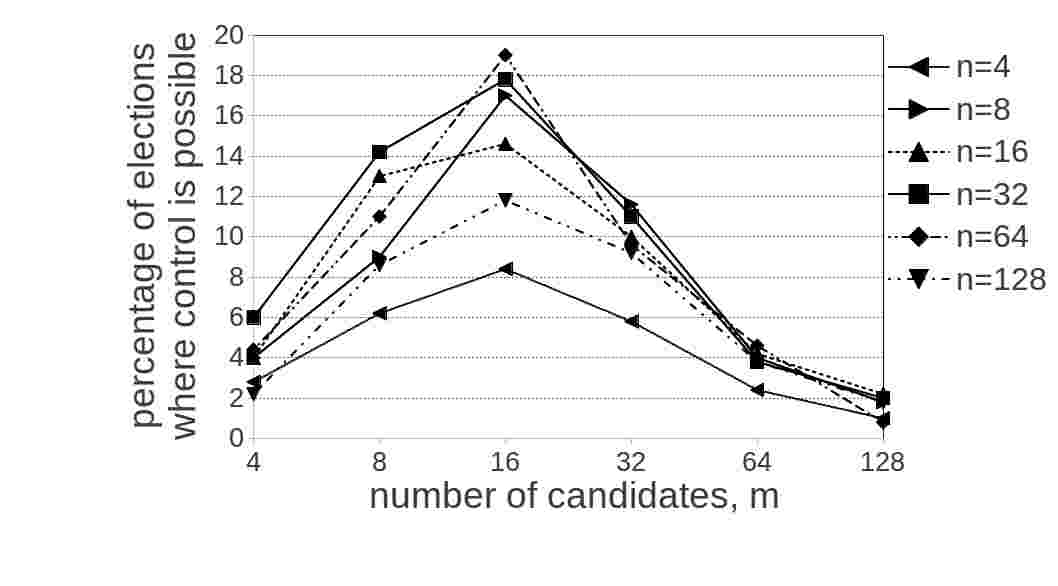}
		\caption{Results for fallback voting in the IC model for 
constructive control by runoff-partition  of candidates in model TP. Number of voters is fixed. }
\end{figure}
%
\newpage
\begin{figure}[ht]
\centering
	\includegraphics[scale=0.3]{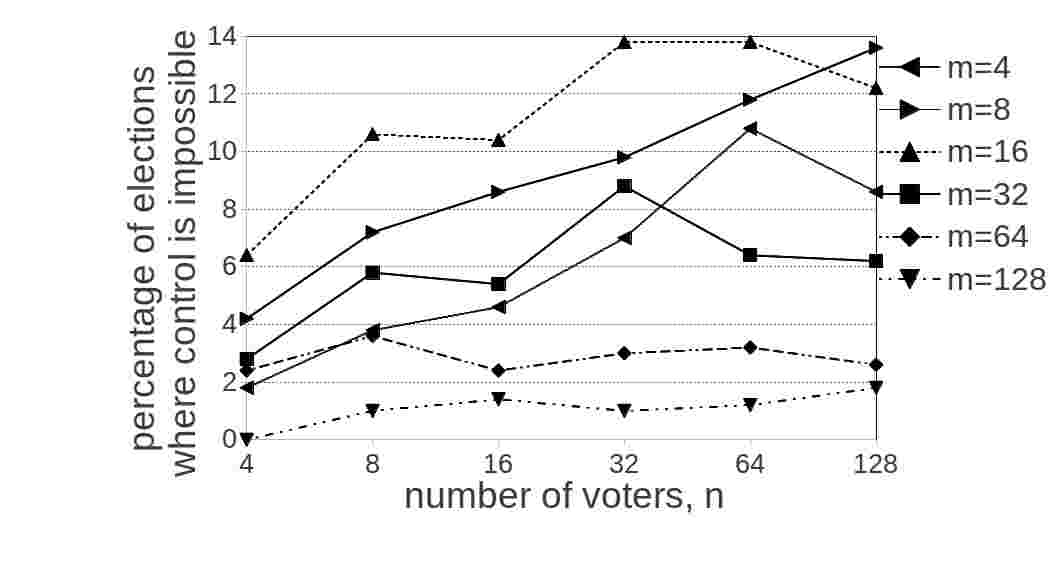}
		\caption{Results for fallback voting in the TM model for 
constructive control by runoff-partition  of candidates in model TP. Number of candidates is fixed. }
\end{figure}
%
\newpage
\begin{figure}[ht]
\centering
	\includegraphics[scale=0.3]{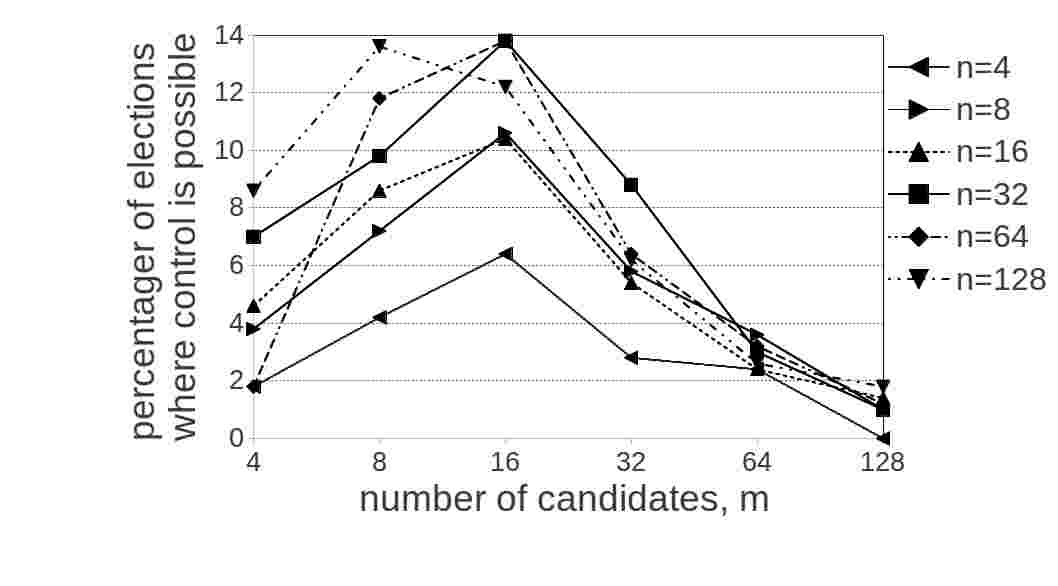}
		\caption{Results for fallback voting in the TM model for 
constructive control by runoff-partition  of candidates in model TP. Number of voters is fixed. }
\end{figure}
%
\end{center}

\clearpage
\subsubsection{Computational Costs}
\begin{figure}[ht]
\centering
	\includegraphics[scale=0.3]{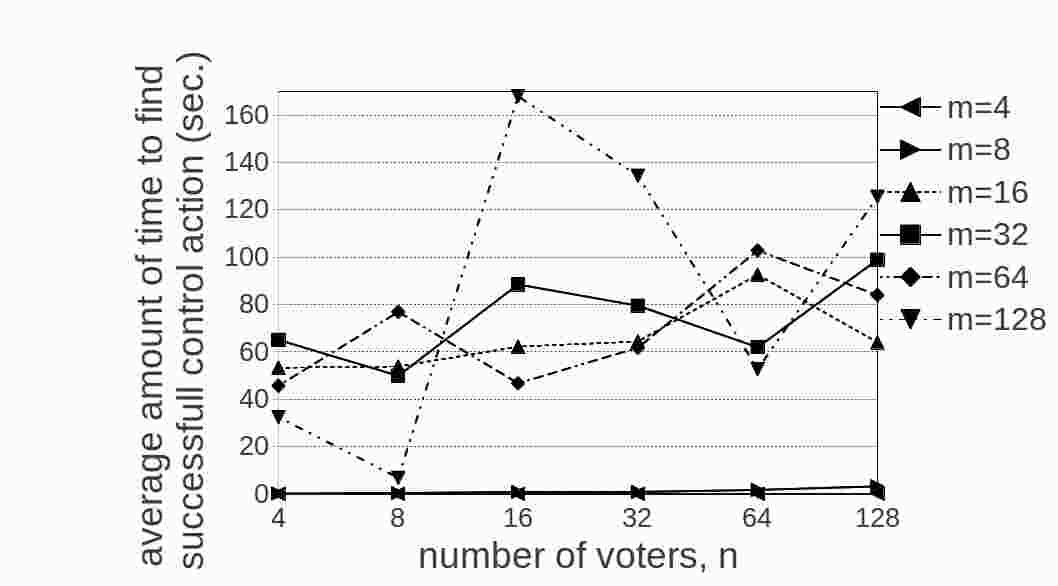}
	\caption{Average time the algorithm needs to find a successful control action for 
	constructive control by runoff-partition  of candidates in model TP
	in fallback elections in the IC model. The maximum is $168,02$ seconds.}
\end{figure}
\begin{figure}[ht]
\centering
	\includegraphics[scale=0.3]{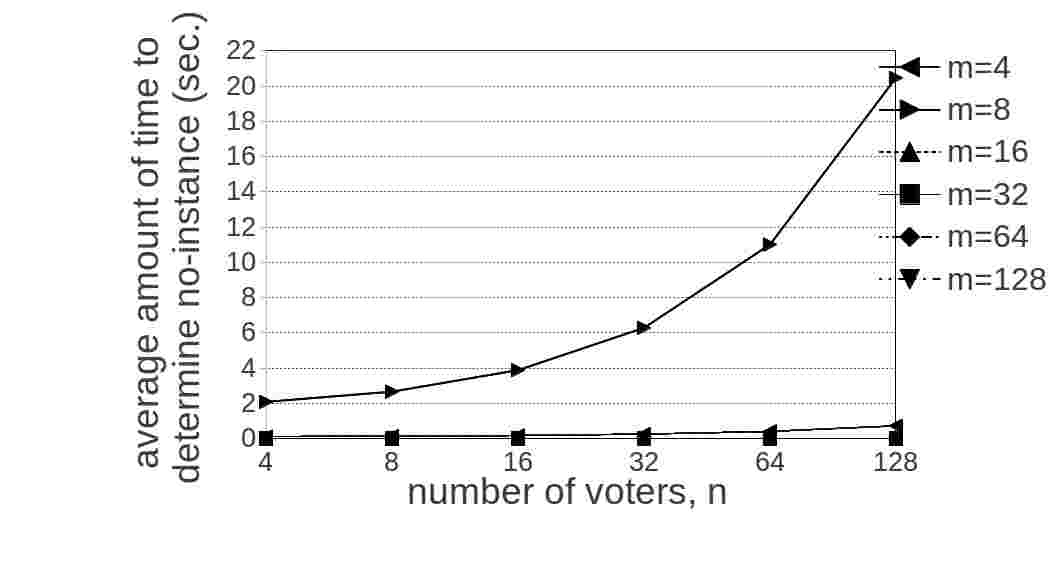}
	\caption{Average time the algorithm needs to determine no-instance of 
		constructive control by runoff-partition  of candidates in model TP
	in fallback elections in the IC model. The maximum is $20,47$ seconds.}
\end{figure}
\begin{figure}[ht]
\centering
	\includegraphics[scale=0.3]{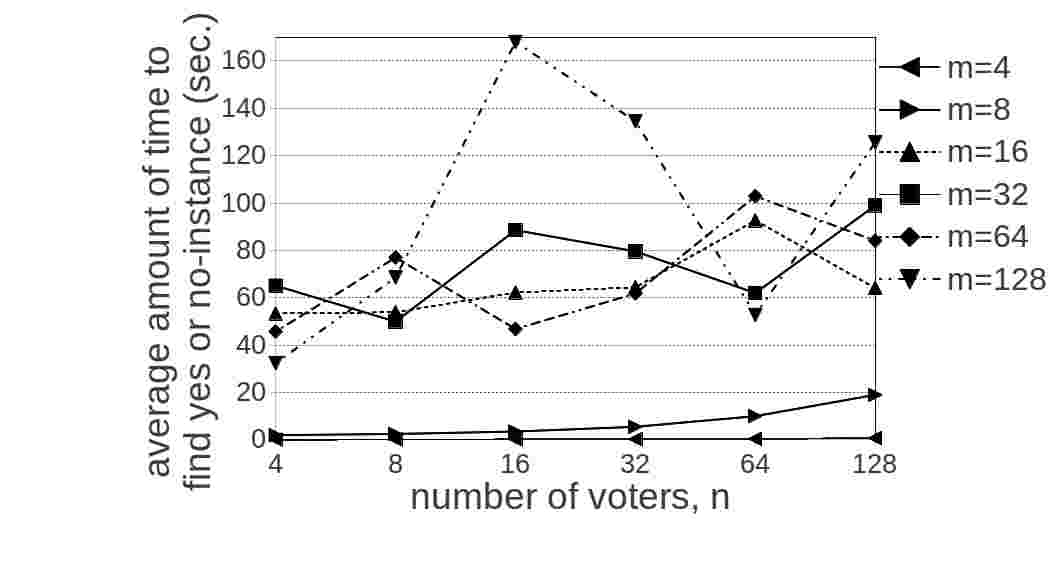}
	\caption{Average time the algorithm needs to give a definite output for 
	constructive control by runoff-partition  of candidates in model TP
	in fallback elections in the IC model. The maximum is $168,02$ seconds.}
\end{figure}
\begin{figure}[ht]
\centering
	\includegraphics[scale=0.3]{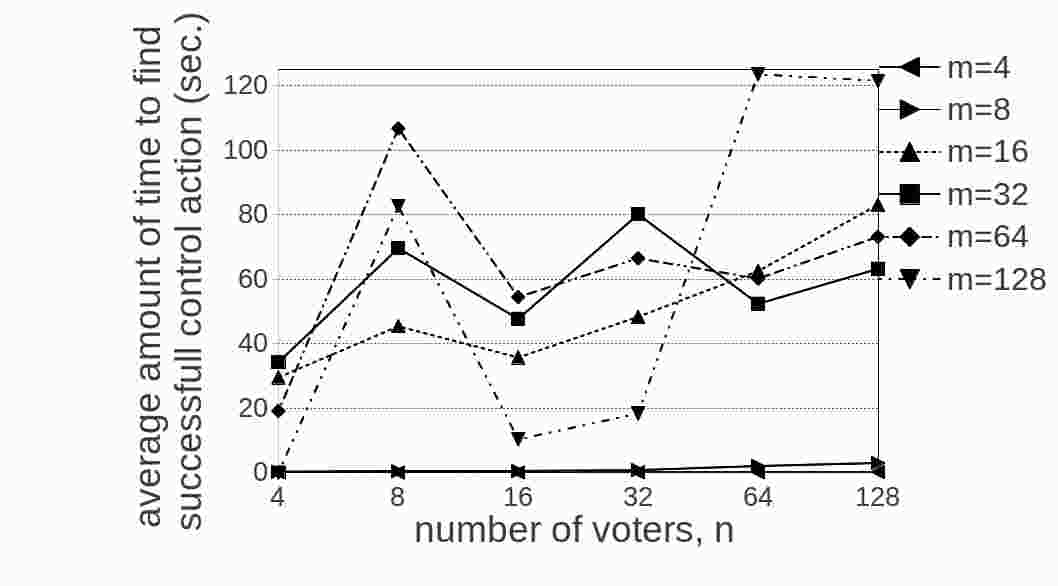}
	\caption{Average time the algorithm needs to find a successful control action for 
	constructive control by runoff-partition  of candidates in model TP
	in fallback elections in the TM model. The maximum is $123,59$ seconds.}
\end{figure}
\begin{figure}[ht]
\centering
	\includegraphics[scale=0.3]{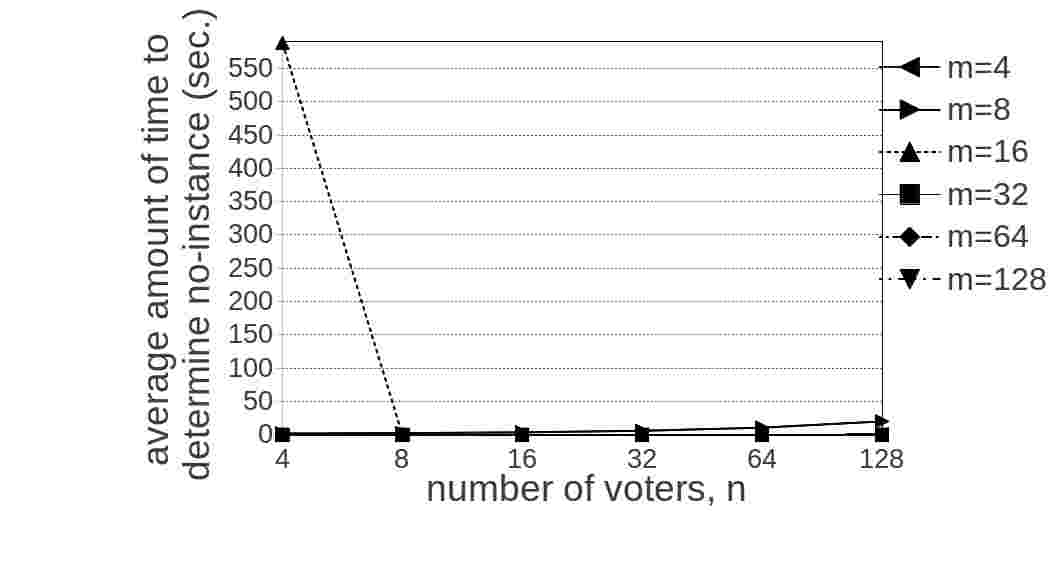}
	\caption{Average time the algorithm needs to determine no-instance of 
		constructive control by runoff-partition  of candidates in model TP
	in fallback elections in the TM model. The maximum is $587,95$ seconds.}
\end{figure}
\begin{figure}[ht]
\centering
	\includegraphics[scale=0.3]{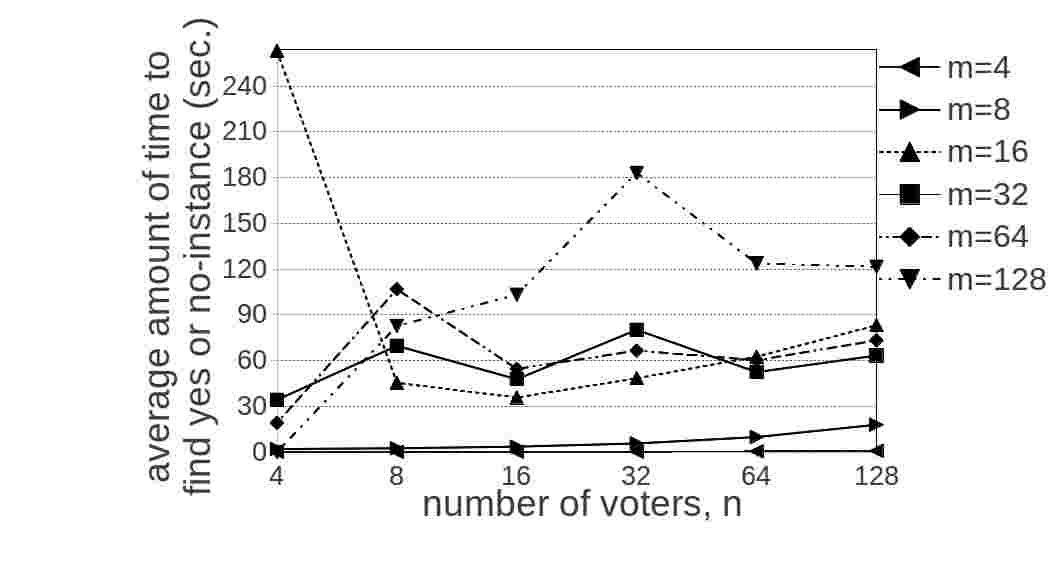}
	\caption{Average time the algorithm needs to give a definite output for 
	constructive control by runoff-partition  of candidates in model TP
	in fallback elections in the TM model. The maximum is $263,04$ seconds.}
\end{figure}

\clearpage
\subsection{Destructive Control by Runoff Partition of Candidates in Model TP}
\begin{center}
\begin{figure}[ht]
\centering
	\includegraphics[scale=0.3]{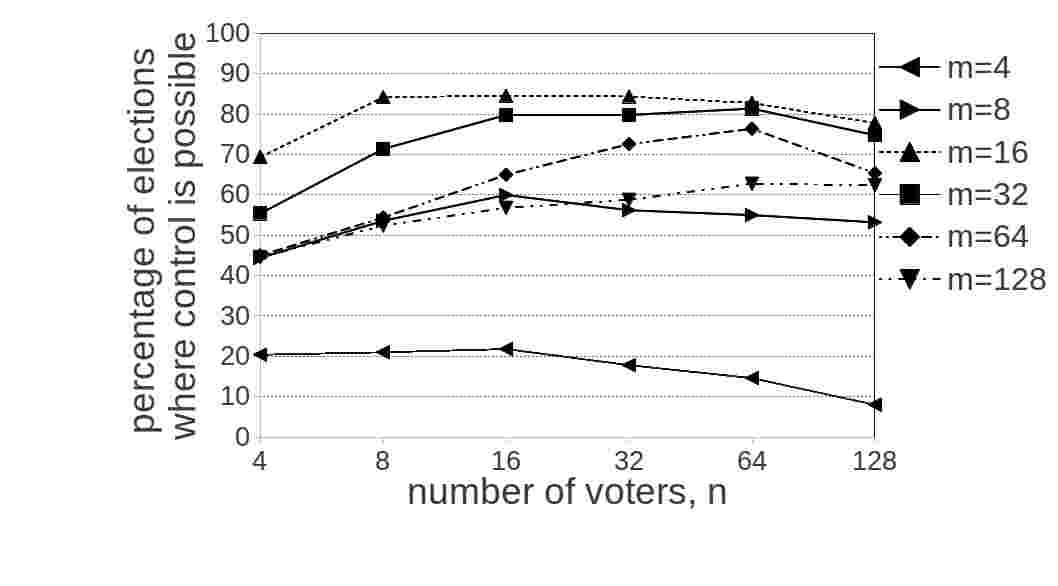}
		\caption{Results for fallback voting in the IC model for 
destructive control by runoff-partition  of candidates in model TP. Number of candidates is fixed. }
\end{figure}

\newpage
\begin{figure}[ht]
\centering
	\includegraphics[scale=0.3]{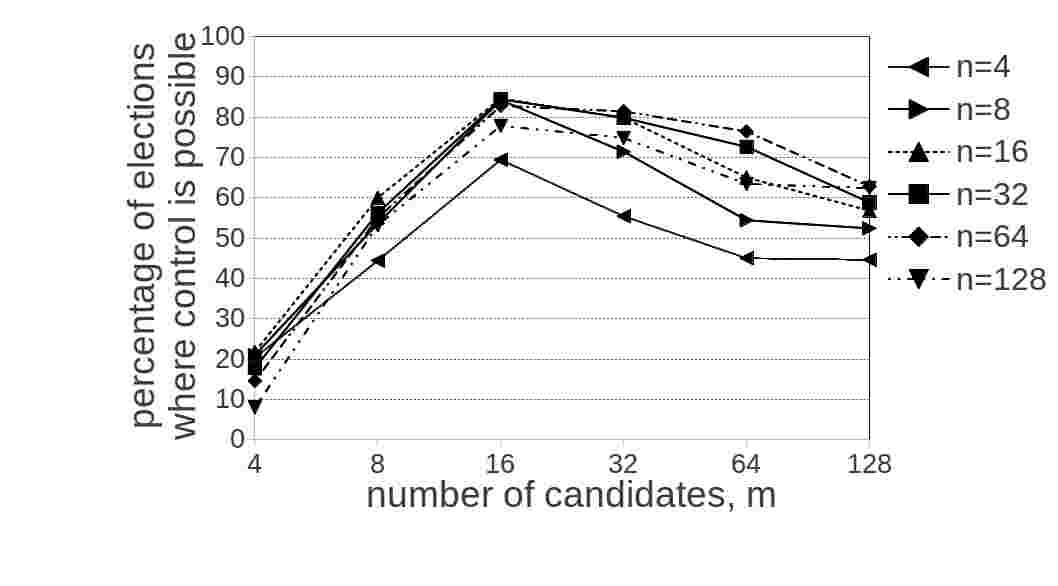}
		\caption{Results for fallback voting in the IC model for 
destructive control by runoff-partition  of candidates in model TP. Number of voters is fixed. }
\end{figure}
%
\newpage
\begin{figure}[ht]
\centering
	\includegraphics[scale=0.3]{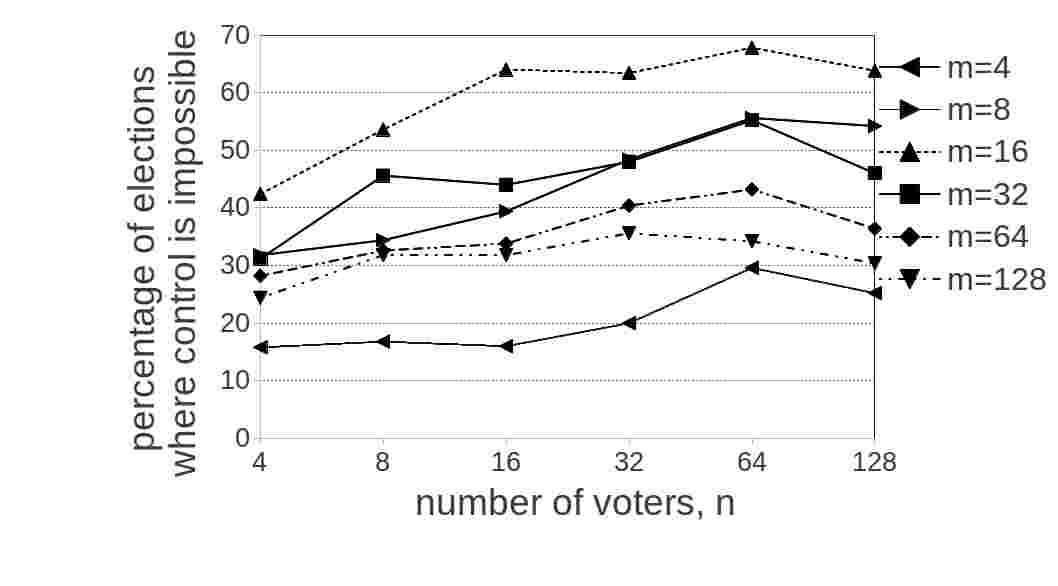}
		\caption{Results for fallback voting in the TM model for 
destructive control by runoff-partition  of candidates in model TP. Number of candidates is fixed. }
\end{figure}
%
\newpage
\begin{figure}[ht]
\centering
	\includegraphics[scale=0.3]{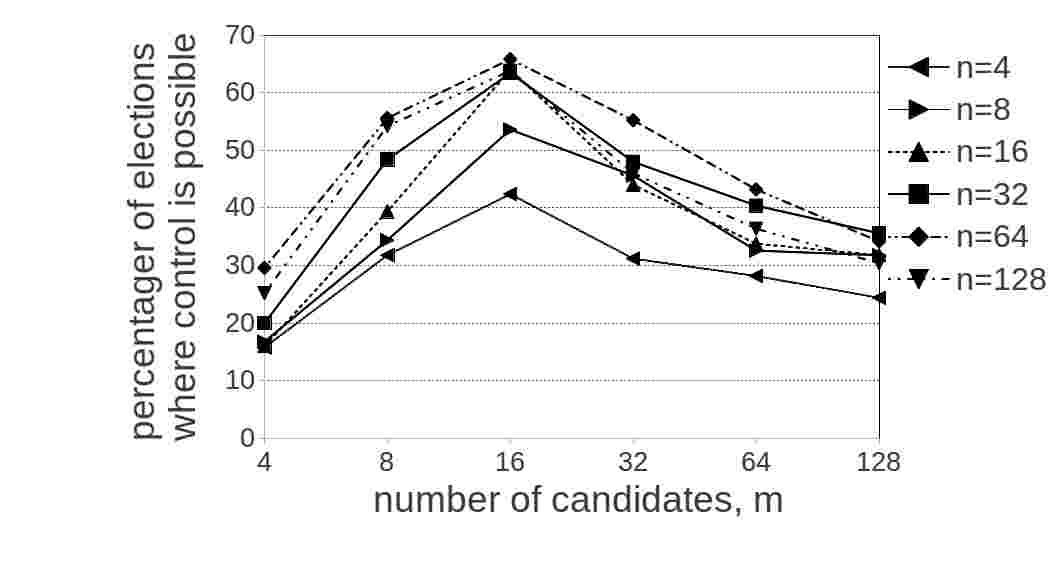}
		\caption{Results for fallback voting in the TM model for 
destructive control by runoff-partition  of candidates in model TP. Number of voters is fixed. }
\end{figure}
%
\end{center}

\clearpage
\subsubsection{Computational Costs}
\begin{figure}[ht]
\centering
	\includegraphics[scale=0.3]{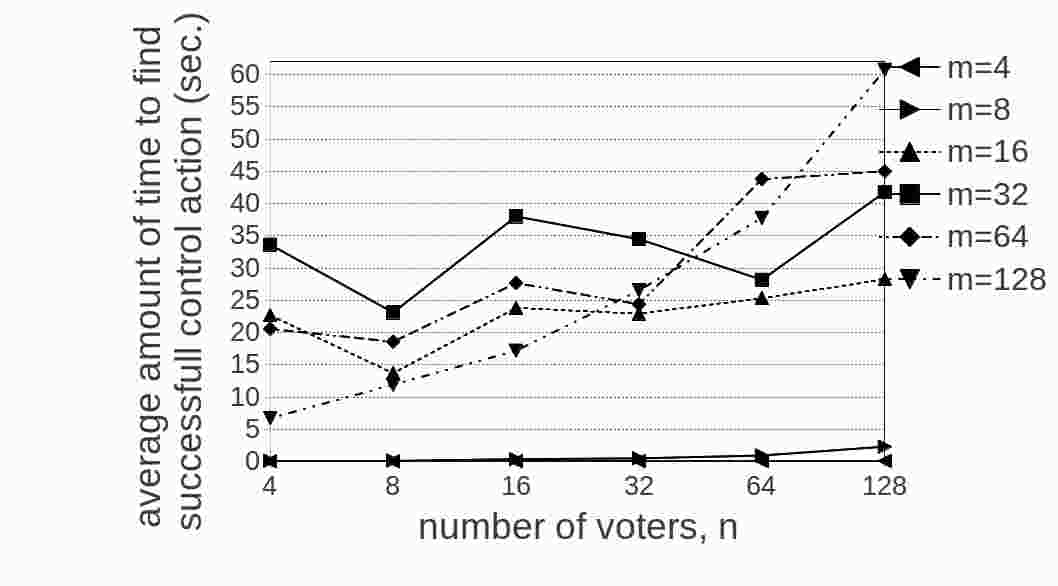}
	\caption{Average time the algorithm needs to find a successful control action for 
	destructive control by runoff-partition  of candidates in model TP
	in fallback elections in the IC model. The maximum is $60,76$ seconds.}
\end{figure}
\begin{figure}[ht]
\centering
	\includegraphics[scale=0.3]{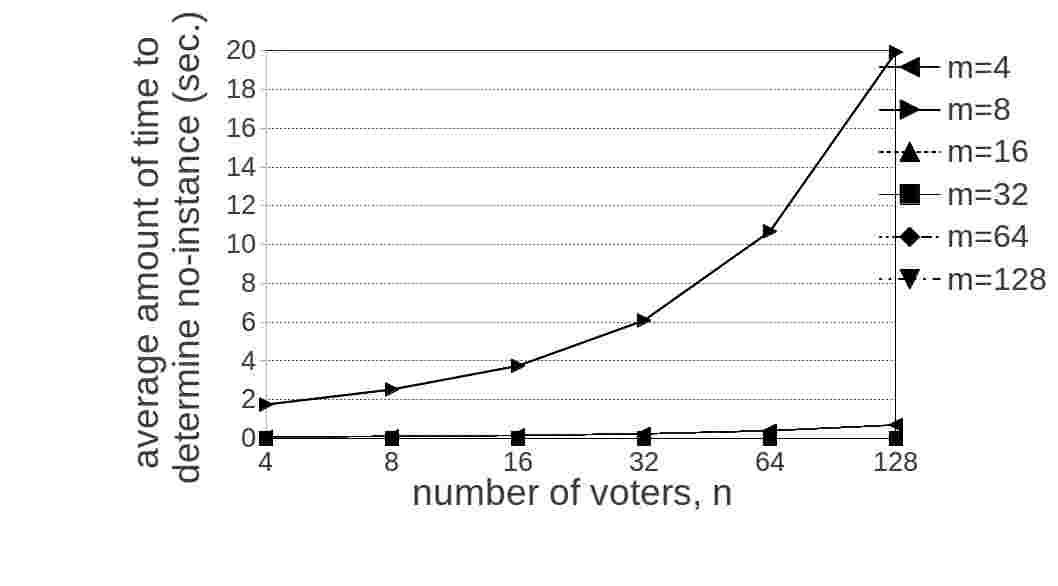}
	\caption{Average time the algorithm needs to determine no-instance of 
		destructive control by runoff-partition  of candidates in model TP
	in fallback elections in the IC model. The maximum is $19,94$ seconds.}
\end{figure}
\begin{figure}[ht]
\centering
	\includegraphics[scale=0.3]{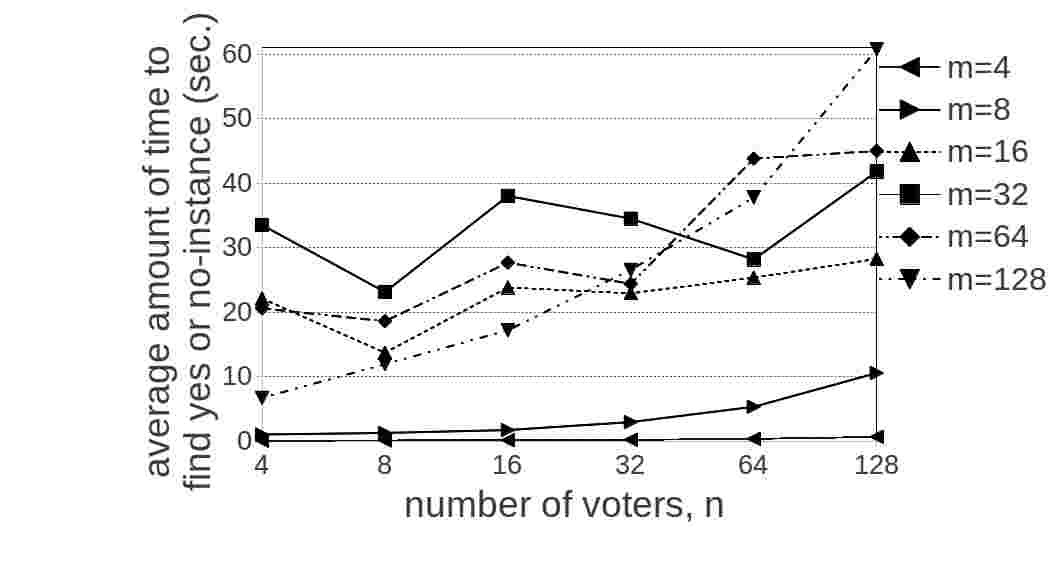}
	\caption{Average time the algorithm needs to give a definite output for 
	destructive control by runoff-partition  of candidates in model TP
	in fallback elections in the IC model. The maximum is $60,76$ seconds.}
\end{figure}
\begin{figure}[ht]
\centering
	\includegraphics[scale=0.3]{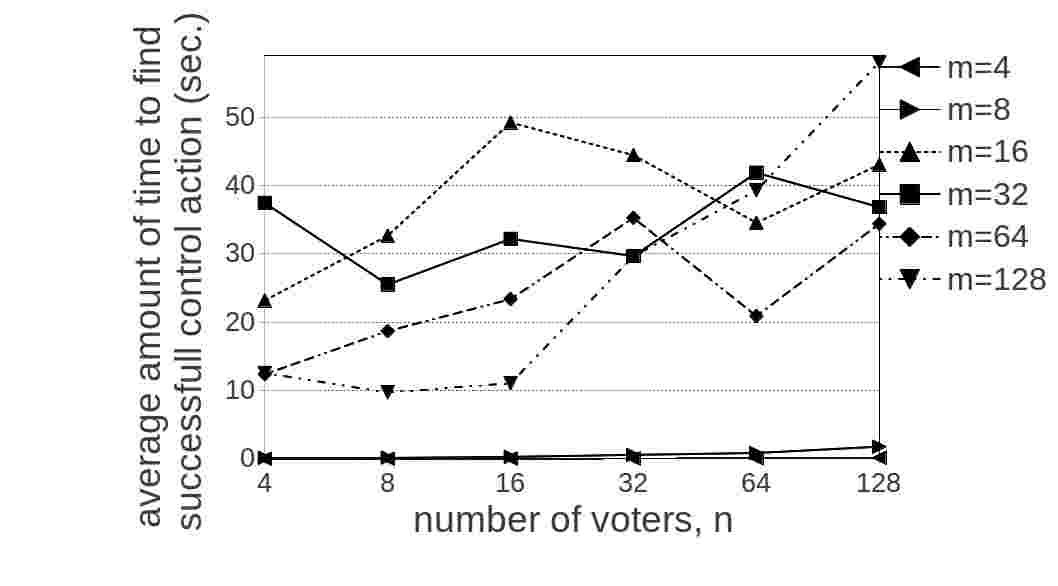}
	\caption{Average time the algorithm needs to find a successful control action for 
	destructive control by runoff-partition  of candidates in model TP
	in fallback elections in the TM model. The maximum is $58,04$ seconds.}
\end{figure}
\begin{figure}[ht]
\centering
	\includegraphics[scale=0.3]{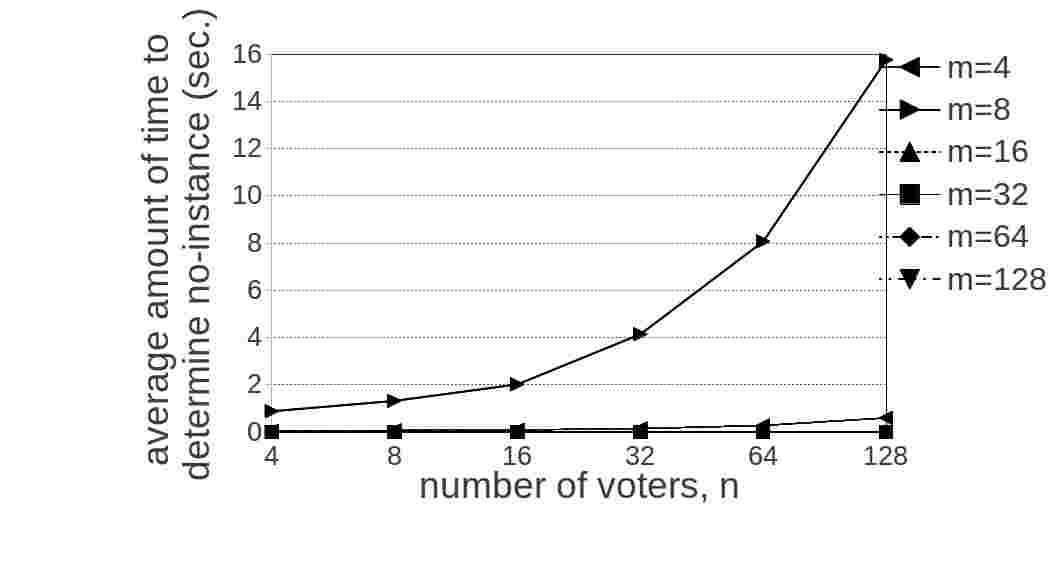}
	\caption{Average time the algorithm needs to determine no-instance of 
		destructive control by runoff-partition  of candidates in model TP
	in fallback elections in the TM model. The maximum is $15,75$ seconds.}
\end{figure}
\begin{figure}[ht]
\centering
	\includegraphics[scale=0.3]{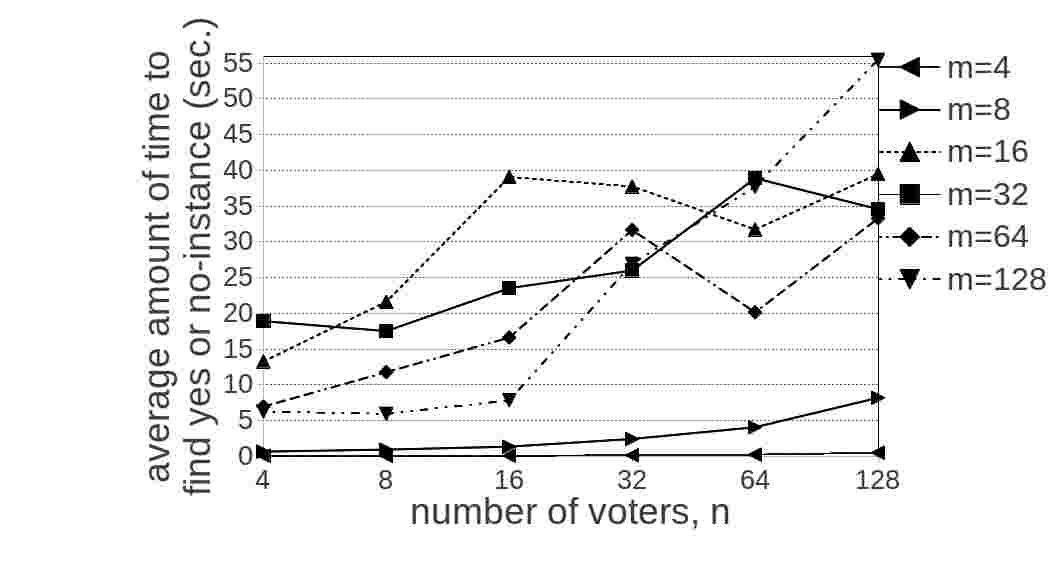}
	\caption{Average time the algorithm needs to give a definite output for 
	destructive control by runoff-partition  of candidates in model TP
	in fallback elections in the TM model. The maximum is $55,48$ seconds.}
\end{figure}

\clearpage
\subsection{Constructive Control by Adding Voters}
\begin{center}
\begin{figure}[ht]
\centering
	\includegraphics[scale=0.3]{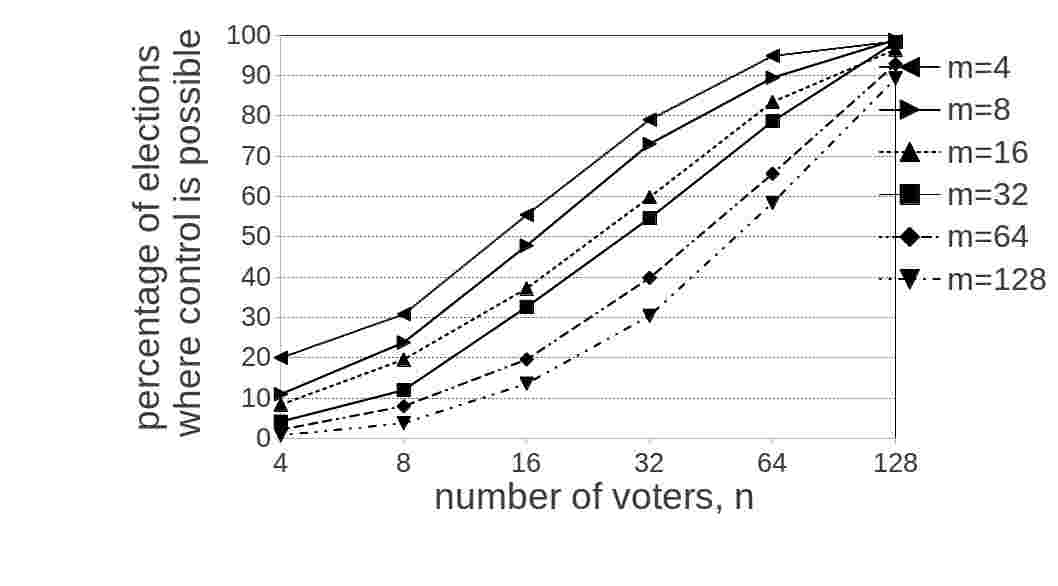}
	\caption{Results for fallback voting in the IC model for 
constructive control by adding voters. Number of candidates is fixed. }
\end{figure}


\newpage
\begin{figure}[ht]
\centering
	\includegraphics[scale=0.3]{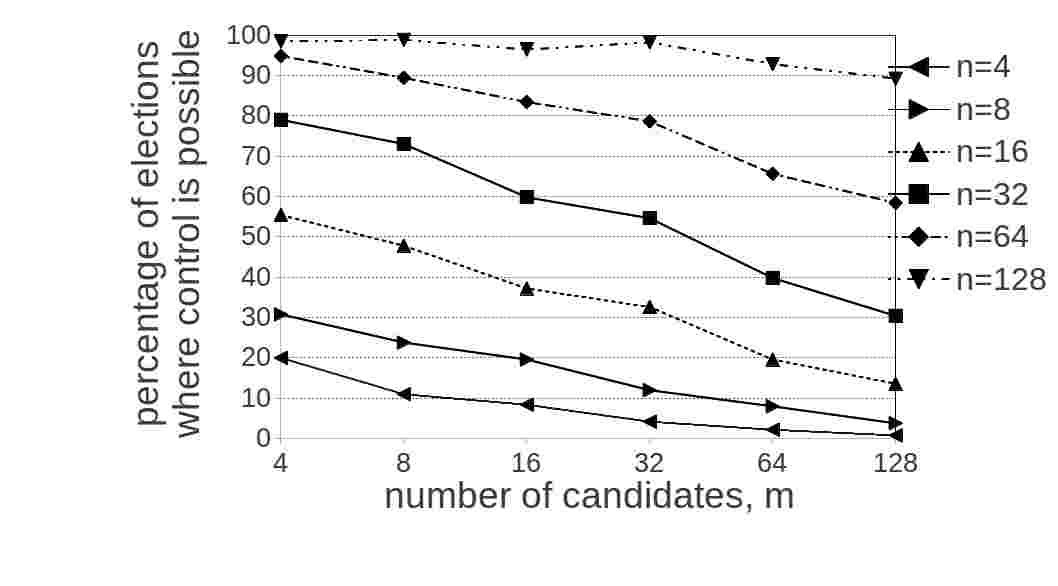}
	\caption{Results for fallback voting in the IC model for 
constructive control by adding voters. Number of voters is fixed. }
\end{figure}

%

\newpage
\begin{figure}[ht]
\centering
	\includegraphics[scale=0.3]{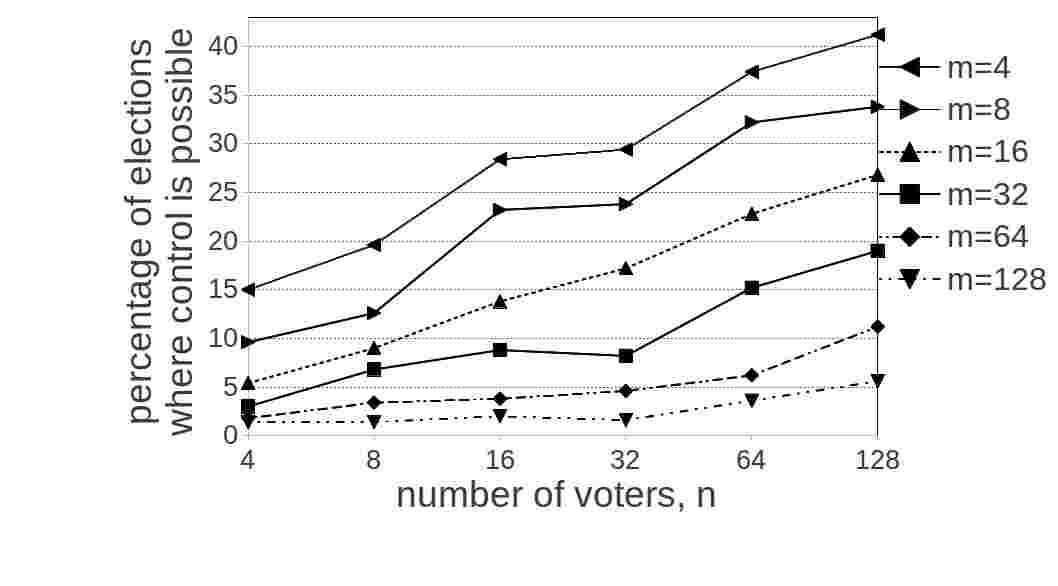}
	\caption{Results for fallback voting in the TM model for 
constructive control by adding voters. Number of candidates is fixed. }
\end{figure}

%

\newpage
\begin{figure}[ht]
\centering
	\includegraphics[scale=0.3]{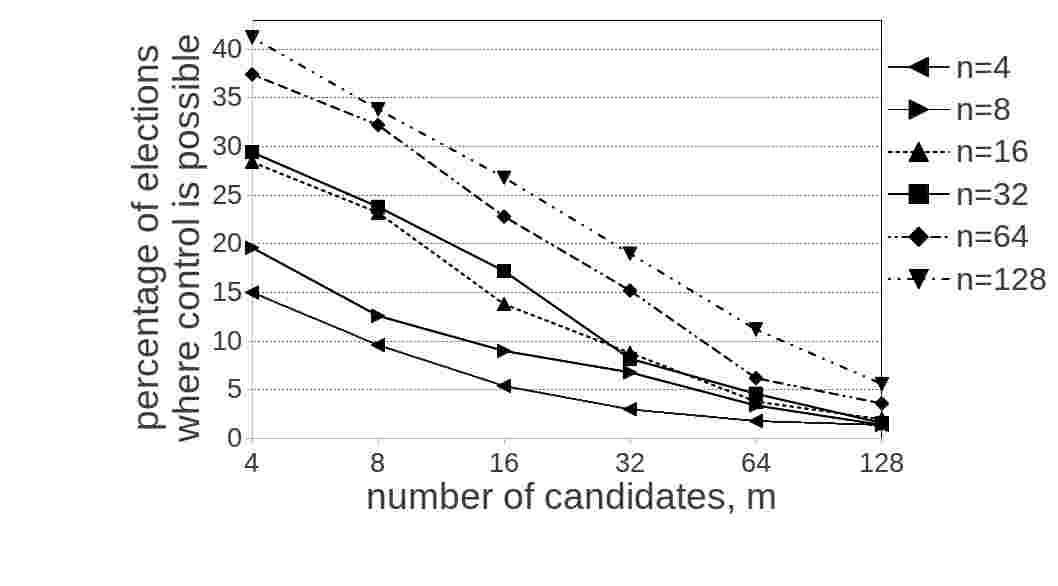}
	\caption{Results for fallback voting in the TM model for 
constructive control by adding voters. Number of voters is fixed. }
\end{figure}

%
\end{center}

\clearpage
\subsubsection{Computational Costs}
\begin{figure}[ht]
\centering
	\includegraphics[scale=0.3]{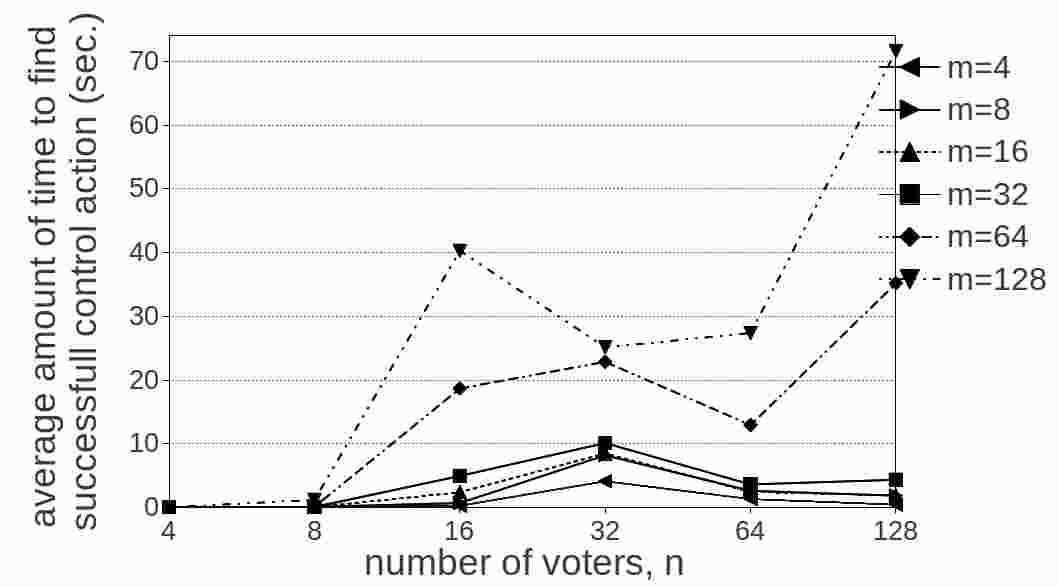}
	\caption{Average time the algorithm needs to find a successful control action for 
	constructive control by adding voters
	in fallback elections in the IC model. The maximum is $71,62$ seconds.}
\end{figure}

\begin{figure}[ht]
\centering
	\includegraphics[scale=0.3]{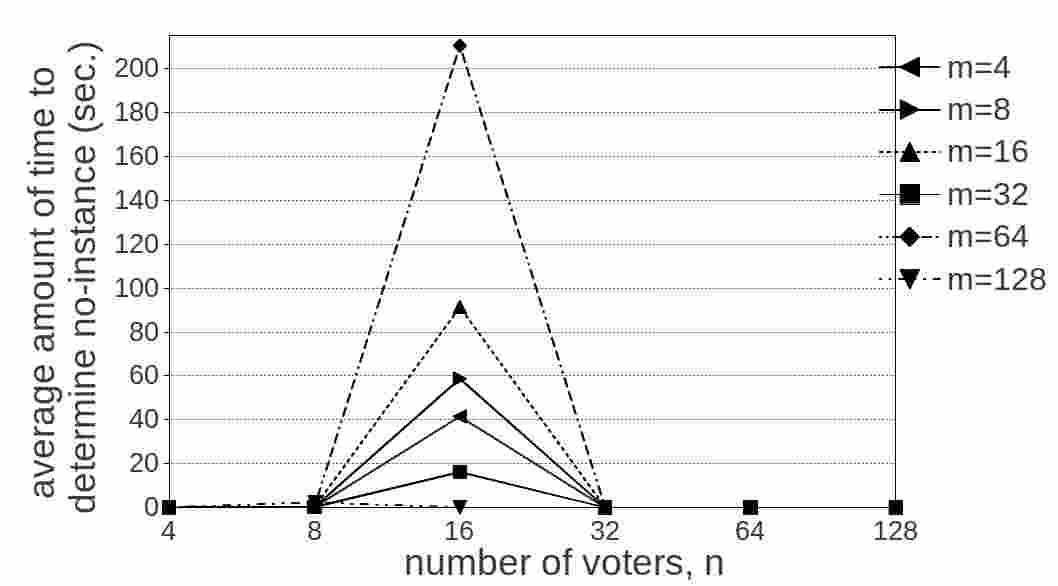}
	\caption{Average time the algorithm needs to determine no-instance of 
		constructive control by adding voters
	in fallback elections in the IC model. The maximum is $210,51$ seconds.}
\end{figure}

\begin{figure}[ht]
\centering
	\includegraphics[scale=0.3]{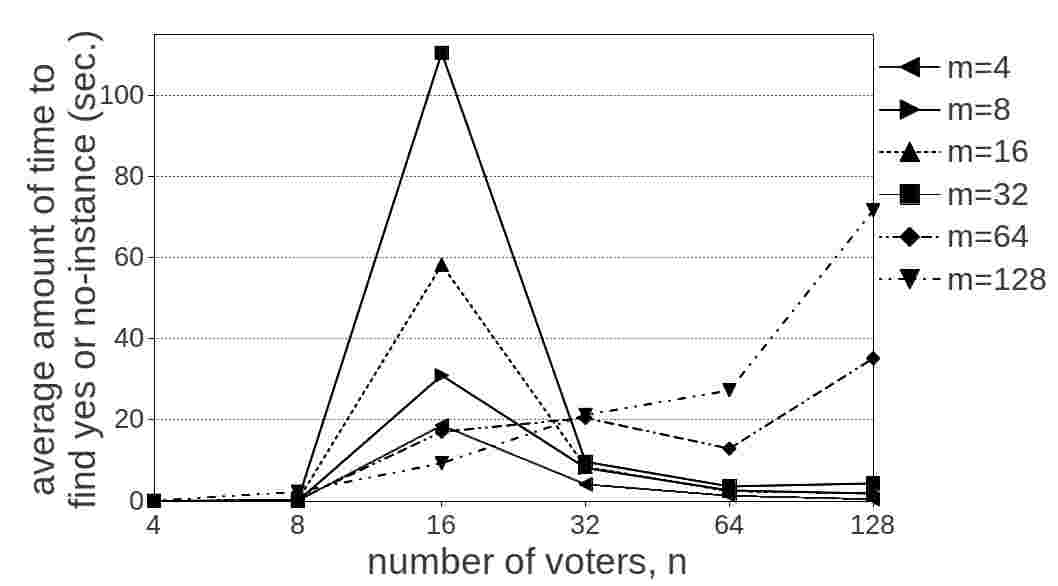}
	\caption{Average time the algorithm needs to give a definite output for 
	constructive control by adding voters
	in fallback elections in the IC model. The maximum is $110,45$ seconds.}
\end{figure}

\begin{figure}[ht]
\centering
	\includegraphics[scale=0.3]{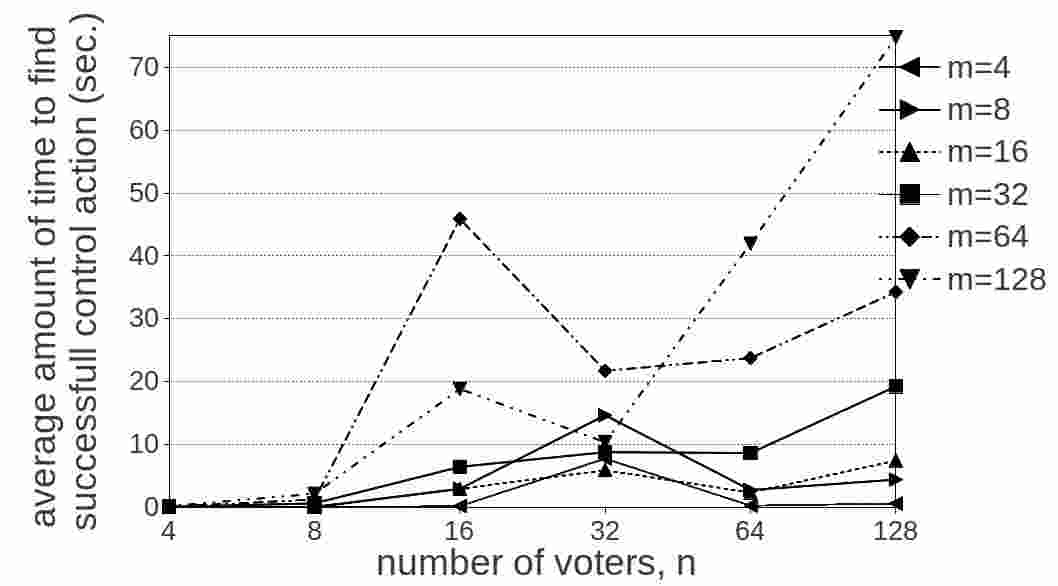}
	\caption{Average time the algorithm needs to find a successful control action for 
	constructive control by adding voters
	in fallback elections in the TM model. The maximum is $74,8$ seconds.}
\end{figure}

\begin{figure}[ht]
\centering
	\includegraphics[scale=0.3]{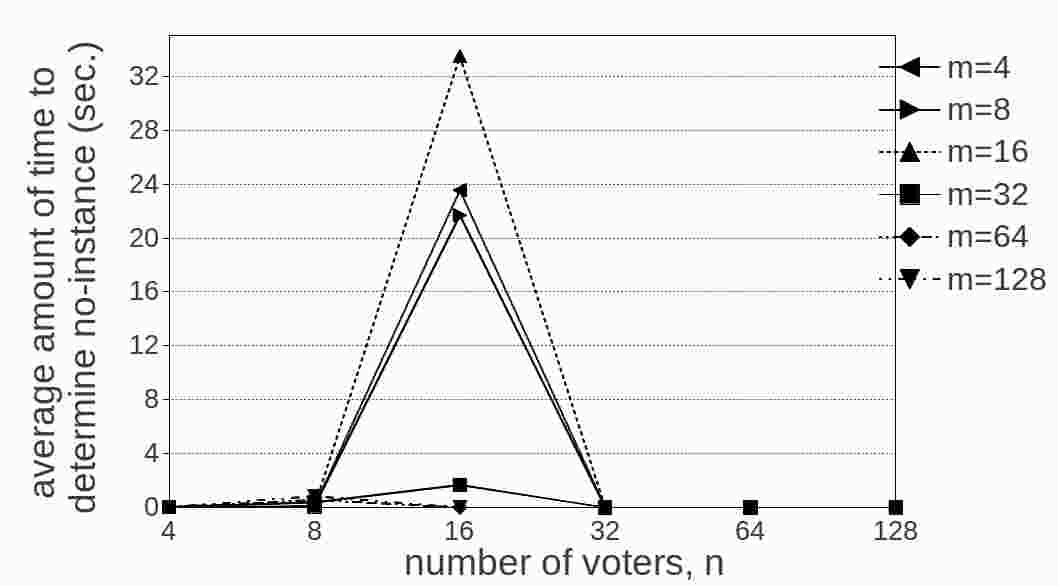}
	\caption{Average time the algorithm needs to determine no-instance of 
		constructive control by adding voters
	in fallback elections in the TM model. The maximum is $233,48$ seconds.}
\end{figure}

\begin{figure}[ht]
\centering
	\includegraphics[scale=0.3]{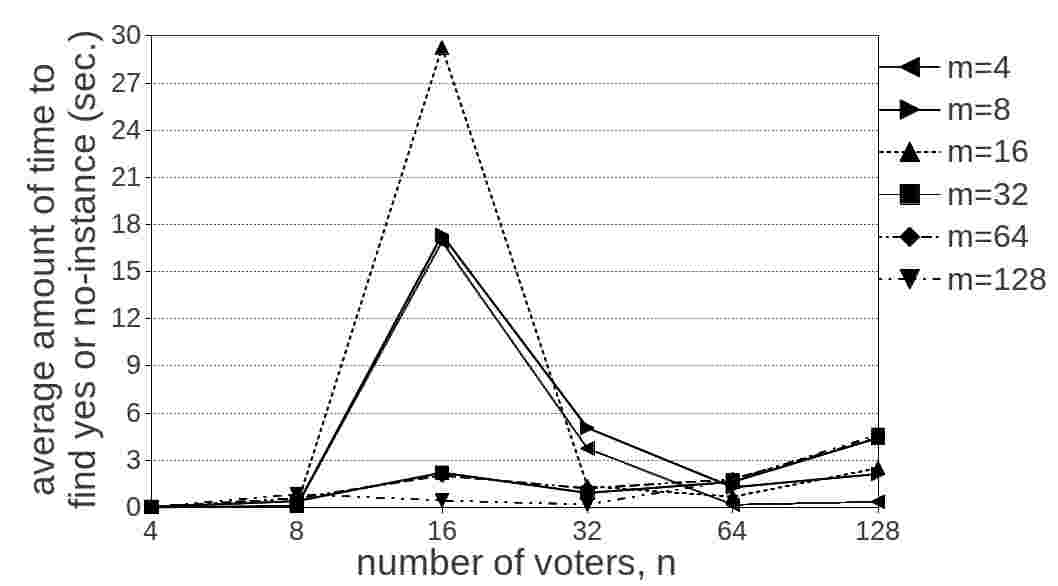}
	\caption{Average time the algorithm needs to give a definite output for 
	constructive control by adding voters
	in fallback elections in the TM model. The maximum is $29,26$ seconds.}
\end{figure}

\clearpage
\subsection{Constructive Control by Deleting Voters}
\begin{center}
\begin{figure}[ht]
\centering
	\includegraphics[scale=0.3]{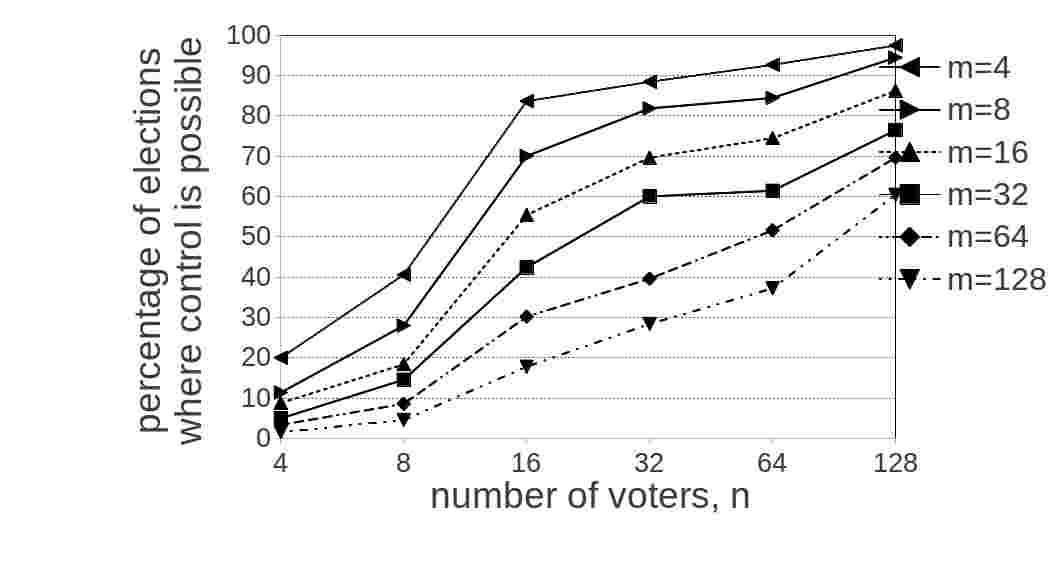}
		\caption{Results for fallback voting in the IC model for 
constructive control by deleting voters. Number of candidates is fixed. }
\end{figure}


\newpage
\begin{figure}[ht]
\centering
	\includegraphics[scale=0.3]{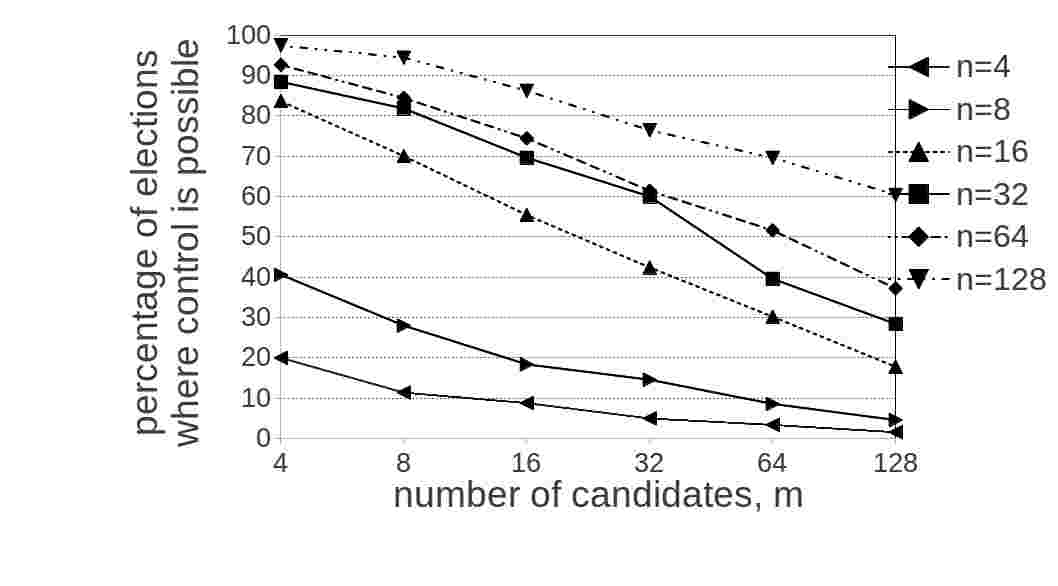}
		\caption{Results for fallback voting in the IC model for 
constructive control by deleting voters. Number of voters is fixed. }
\end{figure}

%

\newpage
\begin{figure}[ht]
\centering
	\includegraphics[scale=0.3]{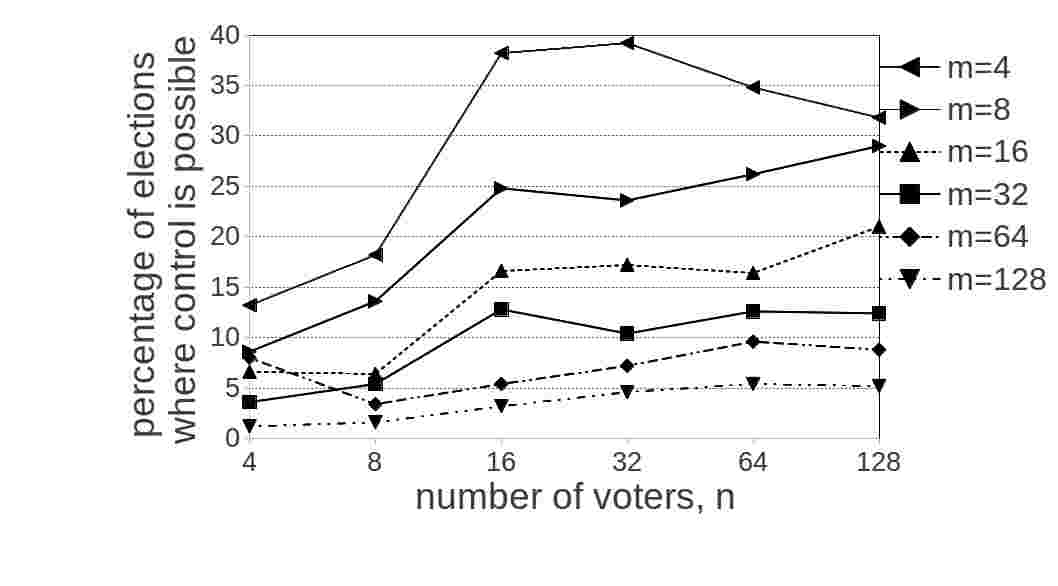}
		\caption{Results for fallback voting in the TM model for 
constructive control by deleting voters. Number of candidates is fixed. }
\end{figure}

%

\newpage
\begin{figure}[ht]
\centering
	\includegraphics[scale=0.3]{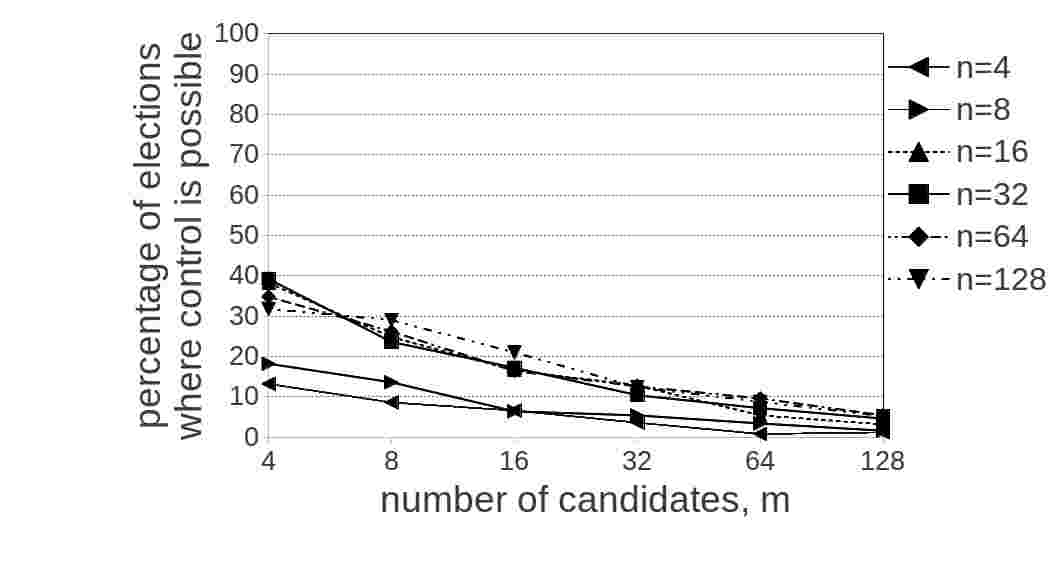}
		\caption{Results for fallback voting in the TM model for 
constructive control by deleting voters. Number of voters is fixed. }
\end{figure}

%
\end{center}

\clearpage
\subsubsection{Computational Costs}
\begin{figure}[ht]
\centering
	\includegraphics[scale=0.3]{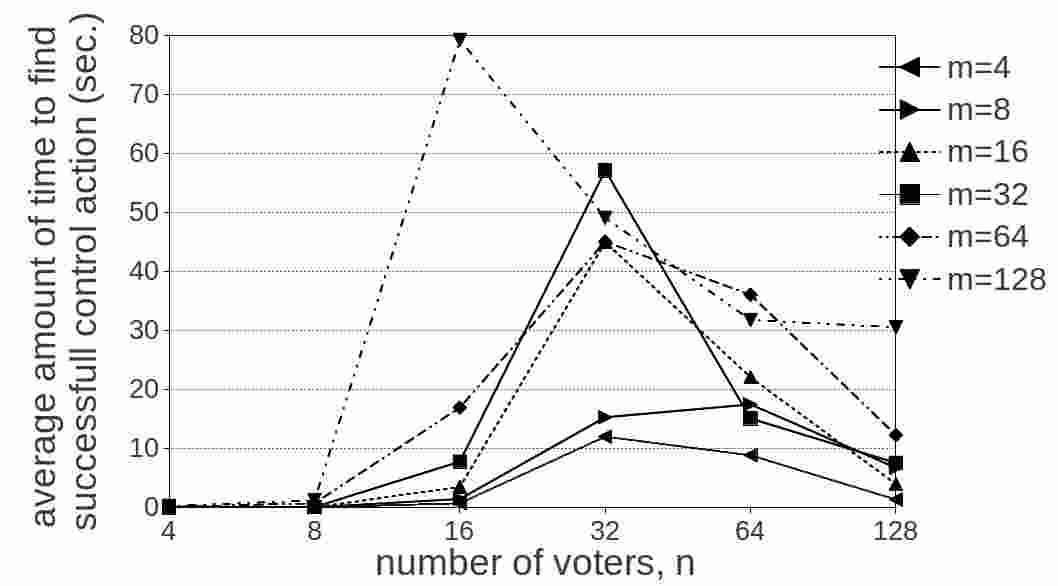}
	\caption{Average time the algorithm needs to find a successful control action for 
	constructive control by deleting voters
	in fallback elections in the IC model. The maximum is $79,32$ seconds.}
\end{figure}

\begin{figure}[ht]
\centering
	\includegraphics[scale=0.3]{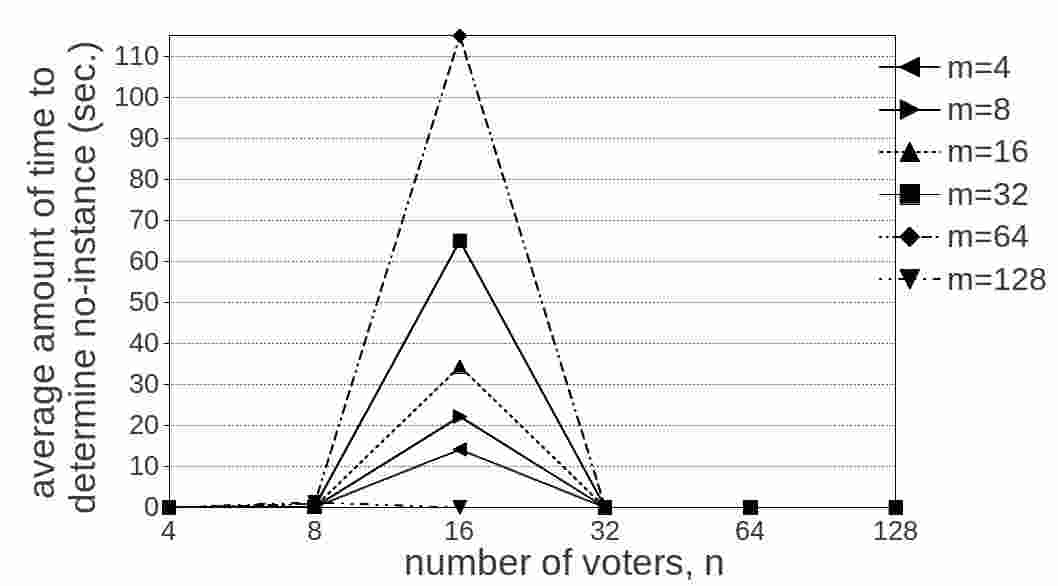}
	\caption{Average time the algorithm needs to determine no-instance of 
		constructive control by deleting voters
	in fallback elections in the IC model. The maximum is $114,99$ seconds.}
\end{figure}

\begin{figure}[ht]
\centering
	\includegraphics[scale=0.3]{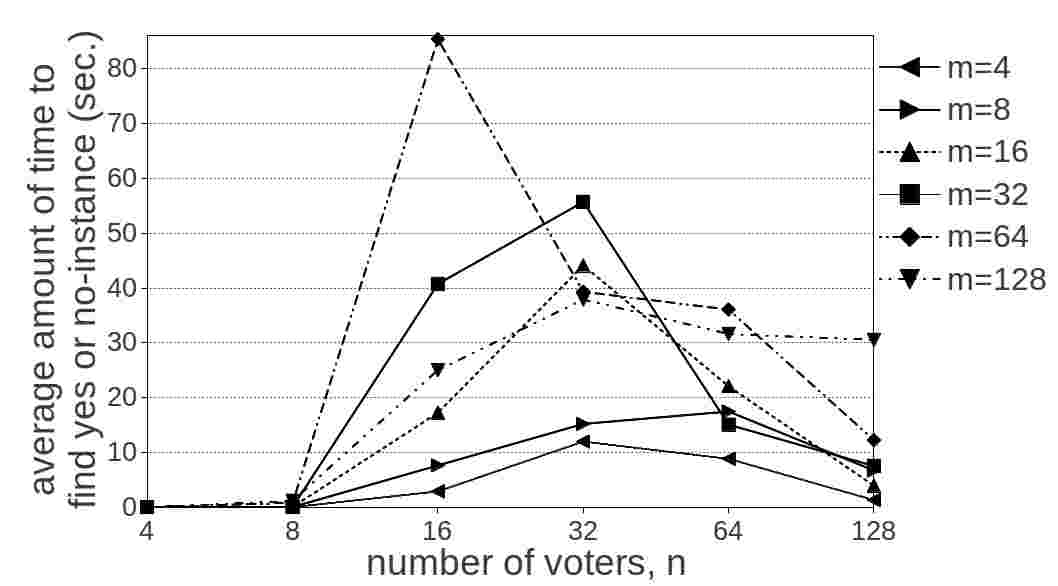}
	\caption{Average time the algorithm needs to give a definite output for 
	constructive control by deleting voters
	in fallback elections in the IC model. The maximum is $88,57$ seconds.}
\end{figure}

\begin{figure}[ht]
\centering
	\includegraphics[scale=0.3]{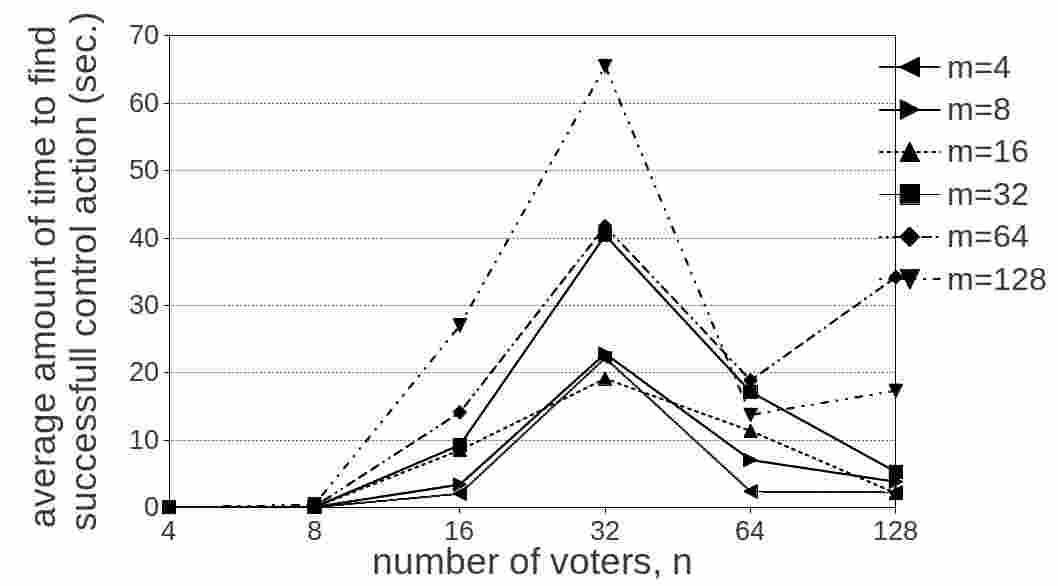}
	\caption{Average time the algorithm needs to find a successful control action for 
	constructive control by deleting voters
	in fallback elections in the TM model. The maximum is $65,48$ seconds.}
\end{figure}

\begin{figure}[ht]
\centering
	\includegraphics[scale=0.3]{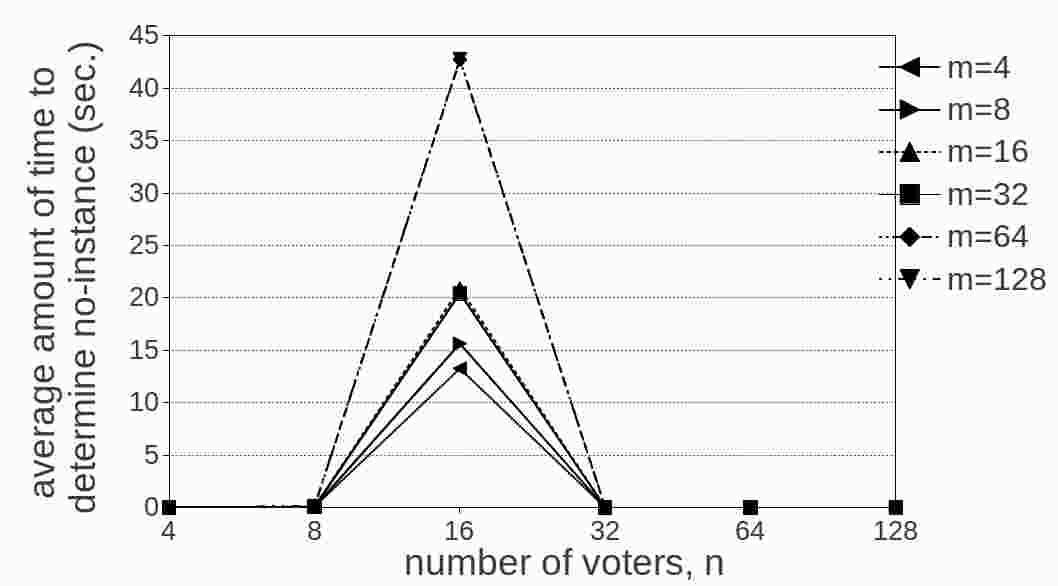}
	\caption{Average time the algorithm needs to determine no-instance of 
		constructive control by deleting voters
	in fallback elections in the TM model. The maximum is $42,8$ seconds.}
\end{figure}

\begin{figure}[ht]
\centering
	\includegraphics[scale=0.3]{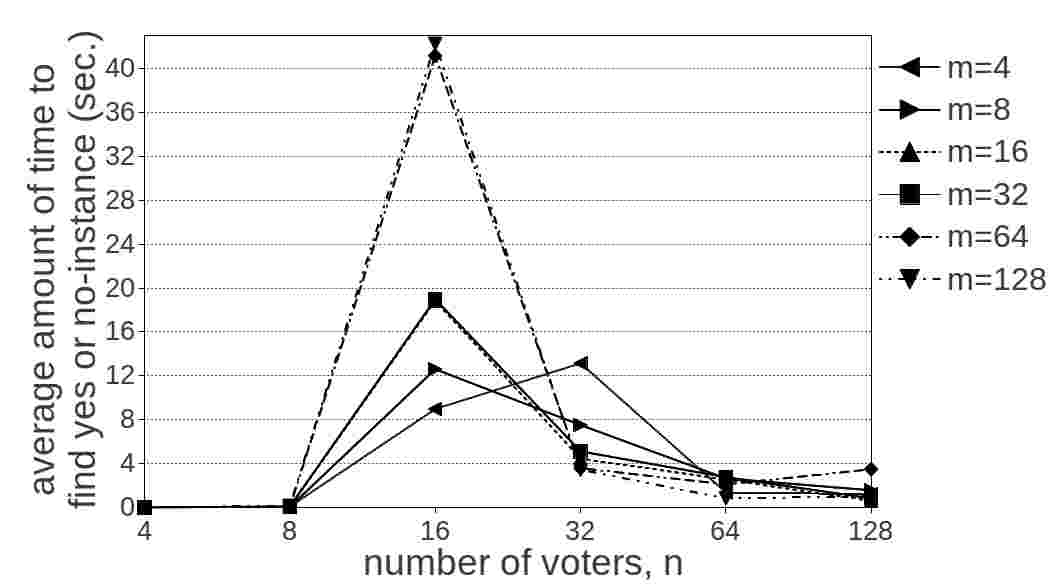}
	\caption{Average time the algorithm needs to give a definite output for 
	constructive control by deleting voters
	in fallback elections in the TM model. The maximum is $42,28$ seconds.}
\end{figure}

\clearpage
\subsection{Constructive Control by Partition of Voters in Model TE}
\begin{center}
\begin{figure}[ht]
\centering
	\includegraphics[scale=0.3]{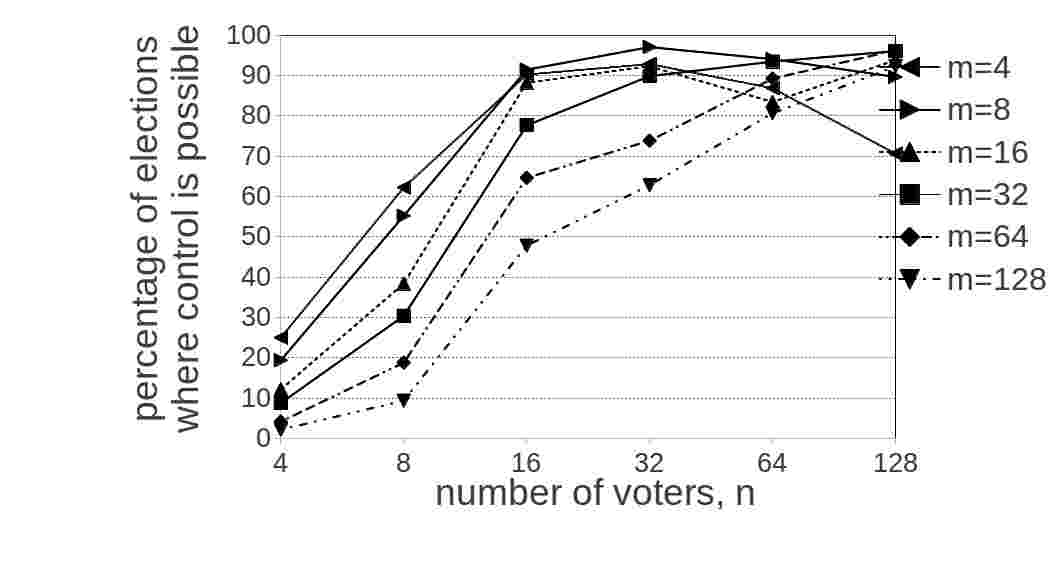}
	\caption{Results for fallback voting in the IC model for 
constructive control by partition of voters in model TE. Number of candidates is fixed. }
\end{figure}


\newpage
\begin{figure}[ht]
\centering
	\includegraphics[scale=0.3]{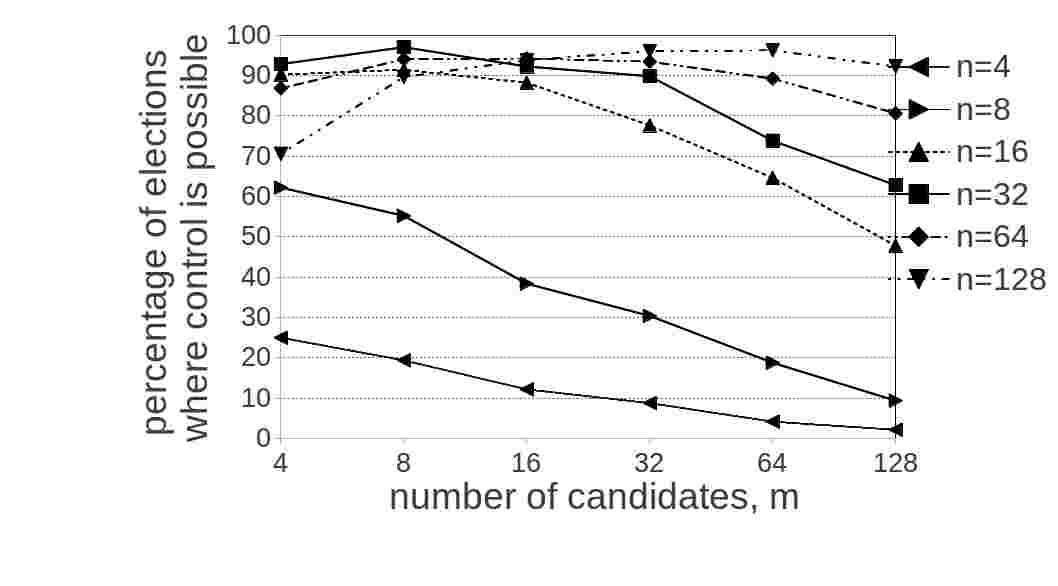}
	\caption{Results for fallback voting in the IC model for 
constructive control by partition of voters in model TE.  Number of voters is fixed. }
\end{figure}

%

\newpage
\begin{figure}[ht]
\centering
	\includegraphics[scale=0.3]{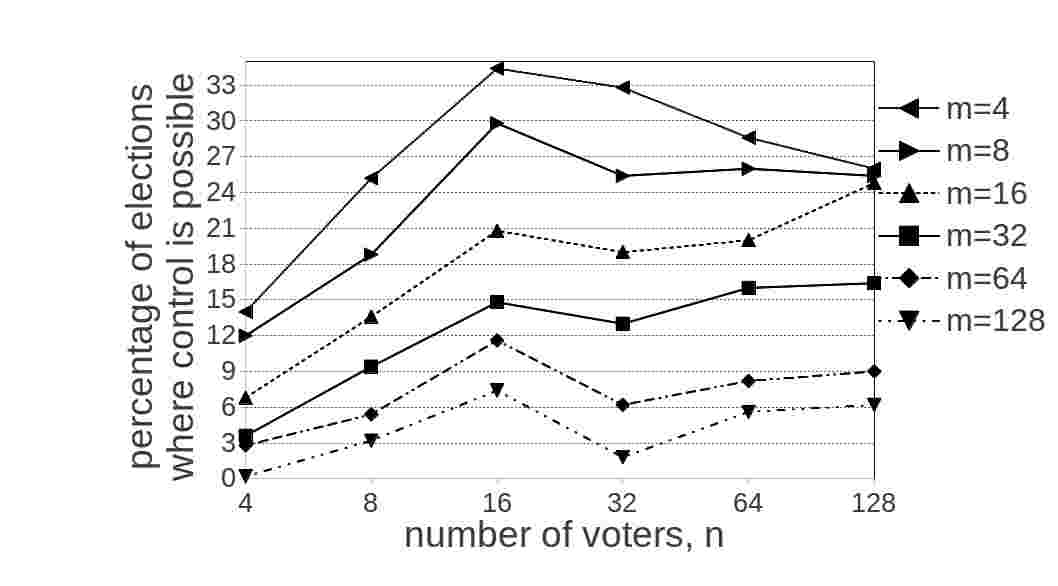}
	\caption{Results for fallback voting in the TM model for 
constructive control by partition of voters in model TE.  Number of candidates is fixed. }
\end{figure}

%
%

\newpage
\begin{figure}[ht]
\centering
	\includegraphics[scale=0.3]{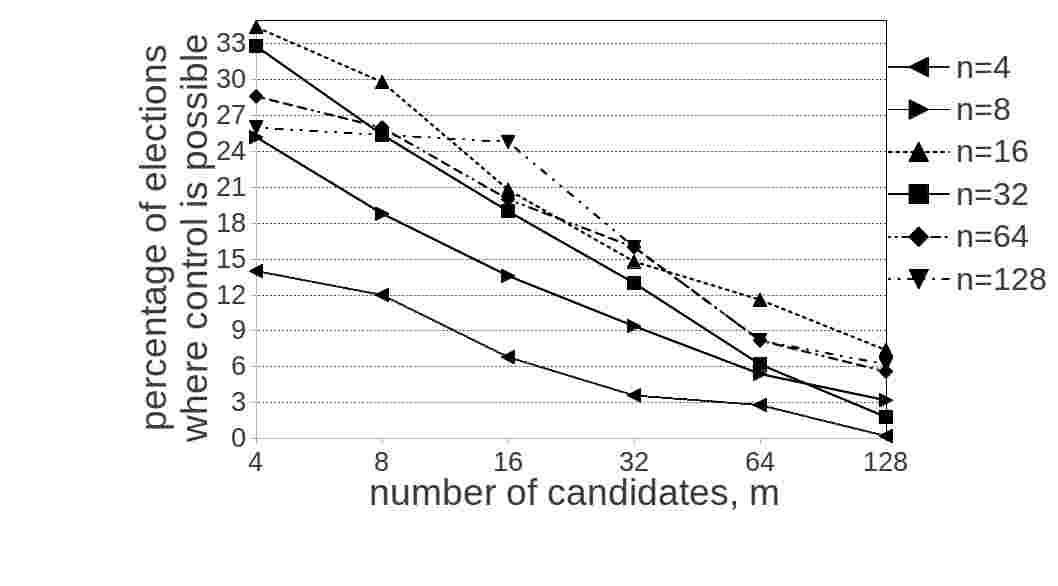}
	\caption{Results for fallback voting in the TM model for 
constructive control by partition of voters in model TE.  Number of voters is fixed. }
\end{figure}

%
\end{center}

\clearpage
\subsubsection{Computational Costs}
\begin{figure}[ht]
\centering
	\includegraphics[scale=0.3]{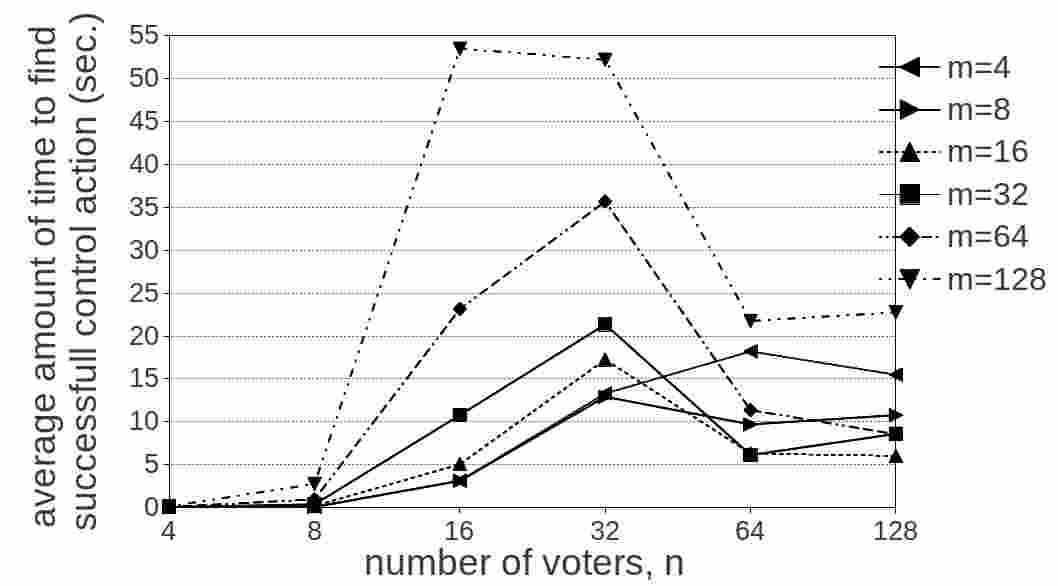}
	\caption{Average time the algorithm needs to find a successful control action for 
	constructive control by partition of voters in model TE
	in fallback elections in the IC model. The maximum is $53,51$ seconds.}
\end{figure}

\begin{figure}[ht]
\centering
	\includegraphics[scale=0.3]{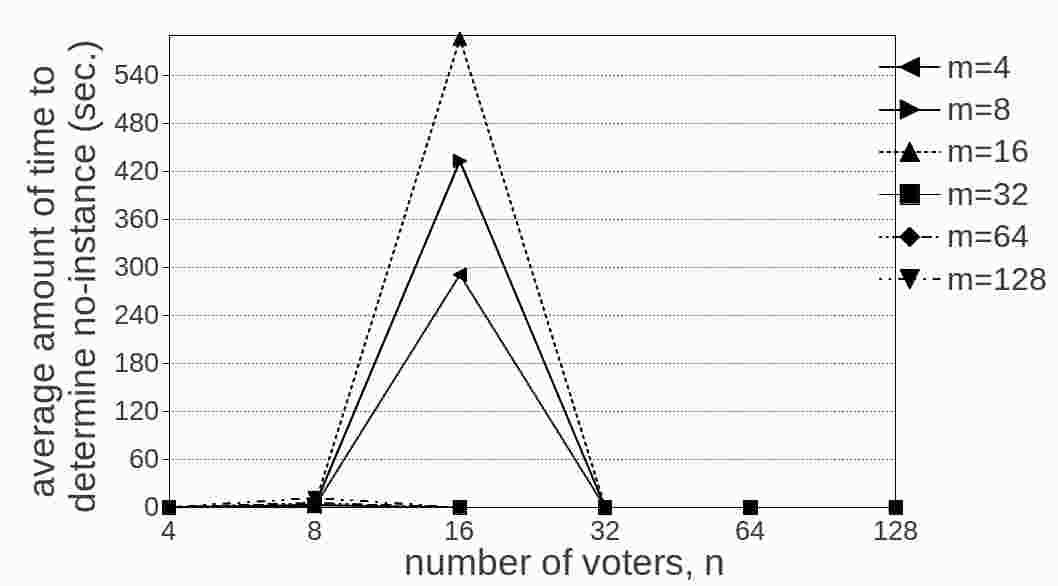}
	\caption{Average time the algorithm needs to determine no-instance of 
		constructive control by partition of voters in model TE
	in fallback elections in the IC model. The maximum is $586,56$ seconds.}
\end{figure}

\begin{figure}[ht]
\centering
	\includegraphics[scale=0.3]{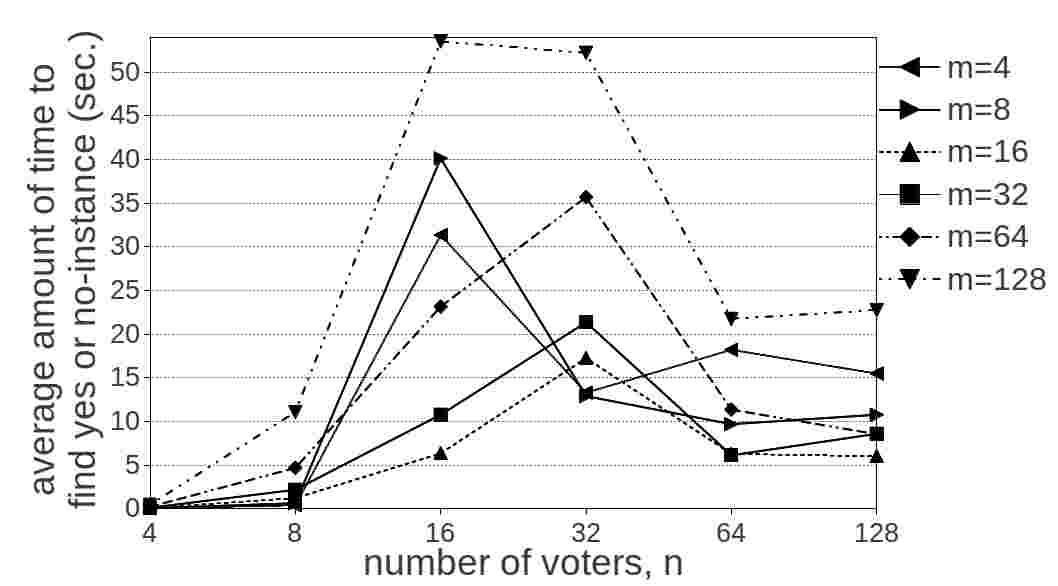}
	\caption{Average time the algorithm needs to give a definite output for 
	constructive control by adding voters
	in fallback elections in the IC model. The maximum is $53,51$ seconds.}
\end{figure}

\begin{figure}[ht]
\centering
	\includegraphics[scale=0.3]{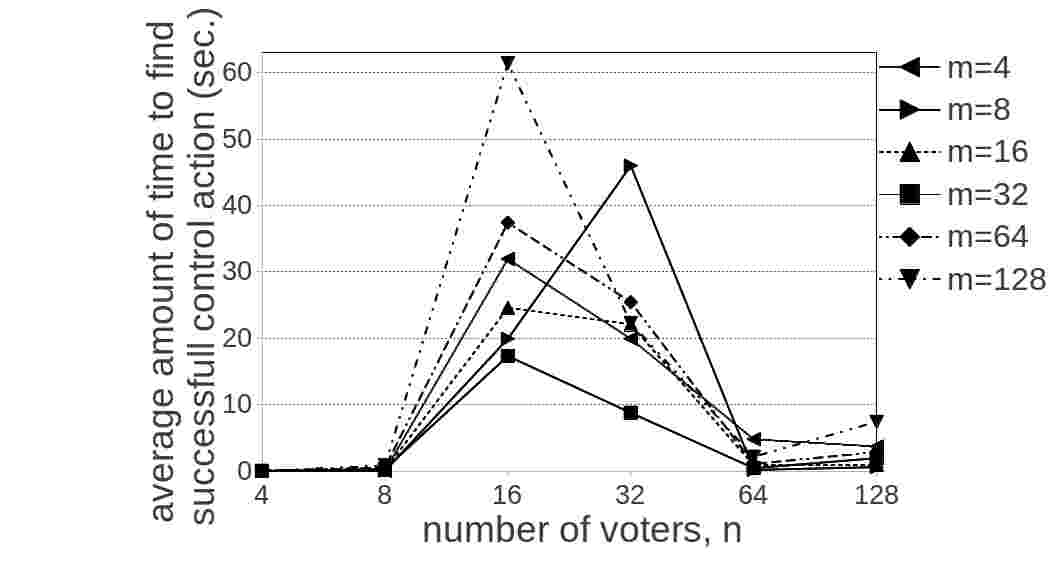}
	\caption{Average time the algorithm needs to find a successful control action for 
	constructive control by partition of voters in model TE
	in fallback elections in the TM model. The maximum is $61,41$ seconds.}
\end{figure}

\begin{figure}[ht]
\centering
	\includegraphics[scale=0.3]{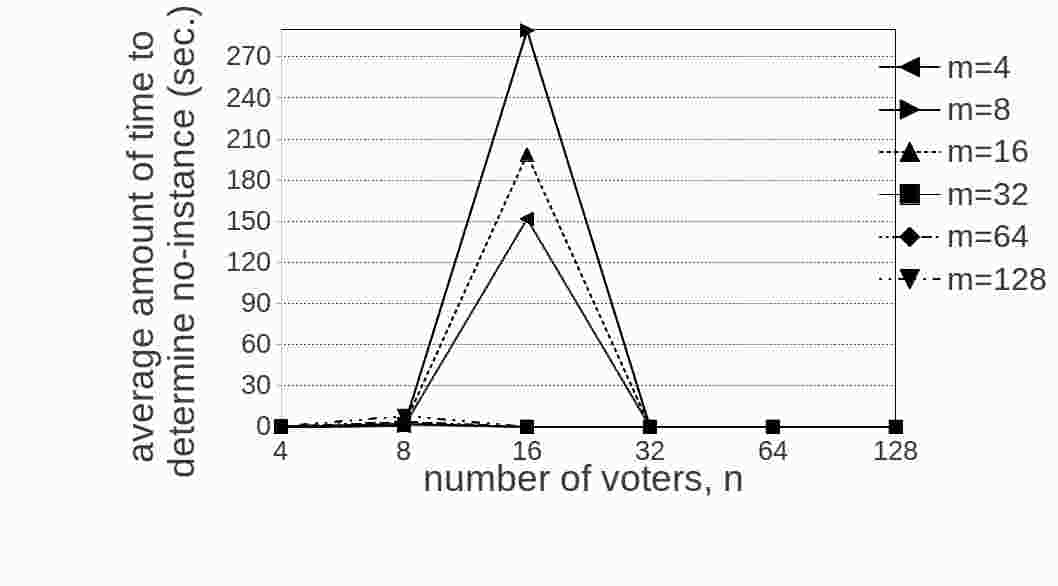}
	\caption{Average time the algorithm needs to determine no-instance of 
		constructive control by partition of voters in model TE
	in fallback elections in the TM model. The maximum is $289,18$ seconds.}
\end{figure}

\begin{figure}[ht]
\centering
	\includegraphics[scale=0.3]{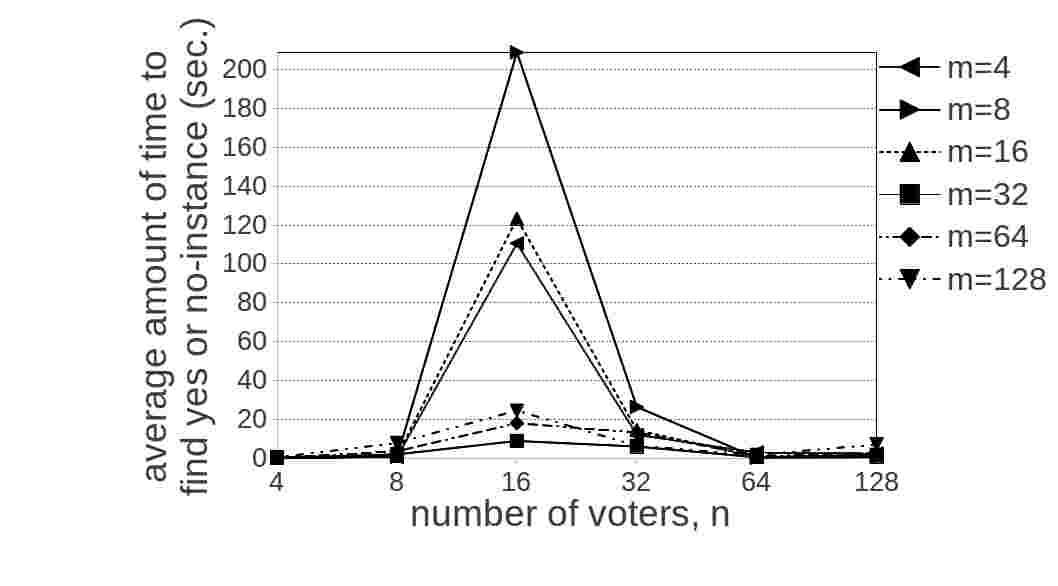}
	\caption{Average time the algorithm needs to give a definite output for 
	constructive control by partition of voters in model TE
	in fallback elections in the TM model. The maximum is $208,95$ seconds.}
\end{figure}

\clearpage
\subsection{Destructive Control by Partition of Voters in Model TE}
\begin{center}
\begin{figure}[ht]
\centering
	\includegraphics[scale=0.3]{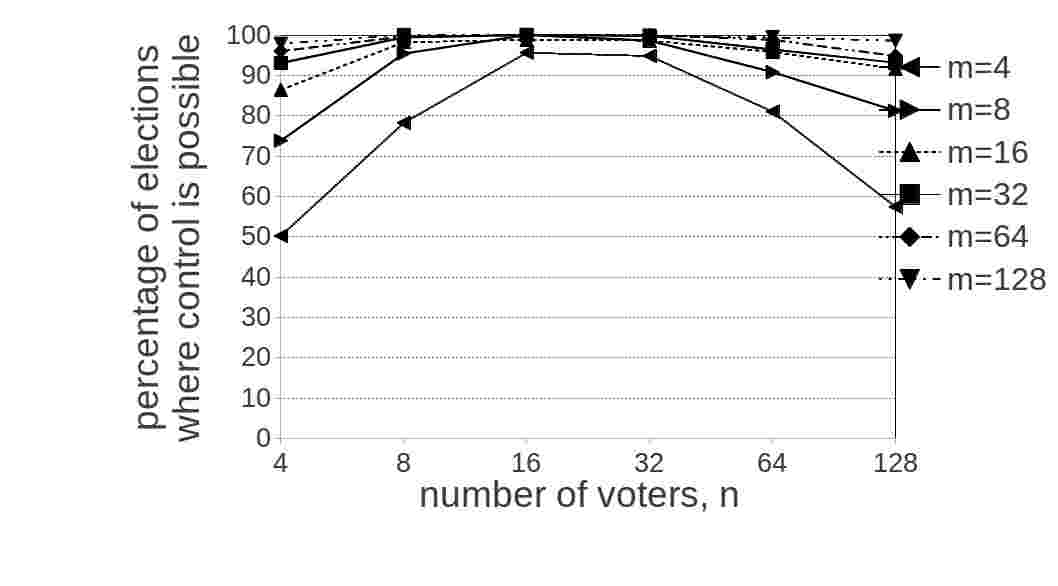}
	\caption{Results for fallback voting in the IC model for 
destructive control by partition of voters in model TE.  Number of candidates is fixed. }
\end{figure}


\newpage
\begin{figure}[ht]
\centering
	\includegraphics[scale=0.3]{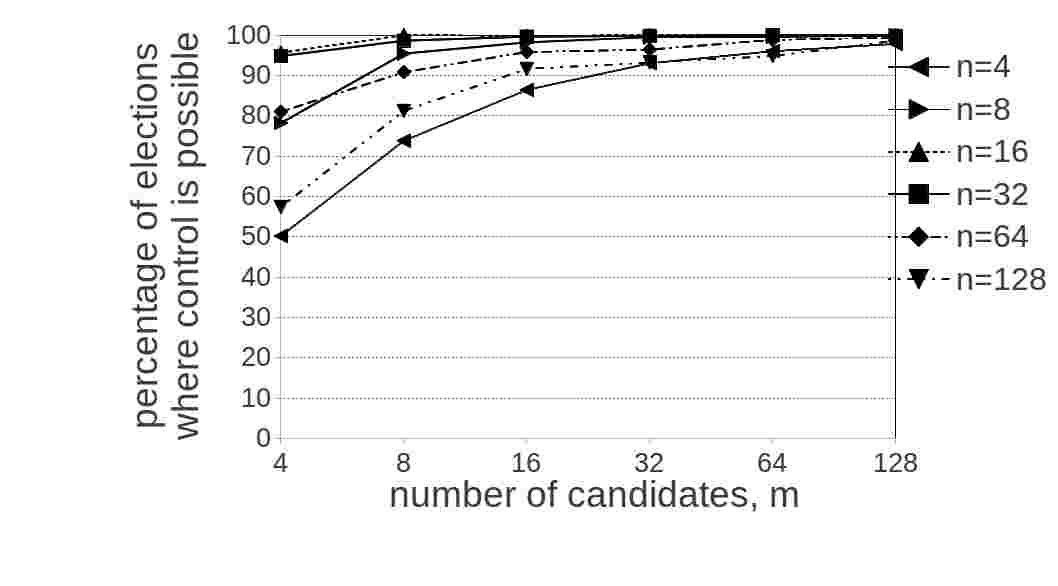}
	\caption{Results for fallback voting in the IC model for 
destructive control by partition of voters in model TE.  Number of voters is fixed.}
\end{figure}

%

\newpage
\begin{figure}[ht]
\centering
	\includegraphics[scale=0.3]{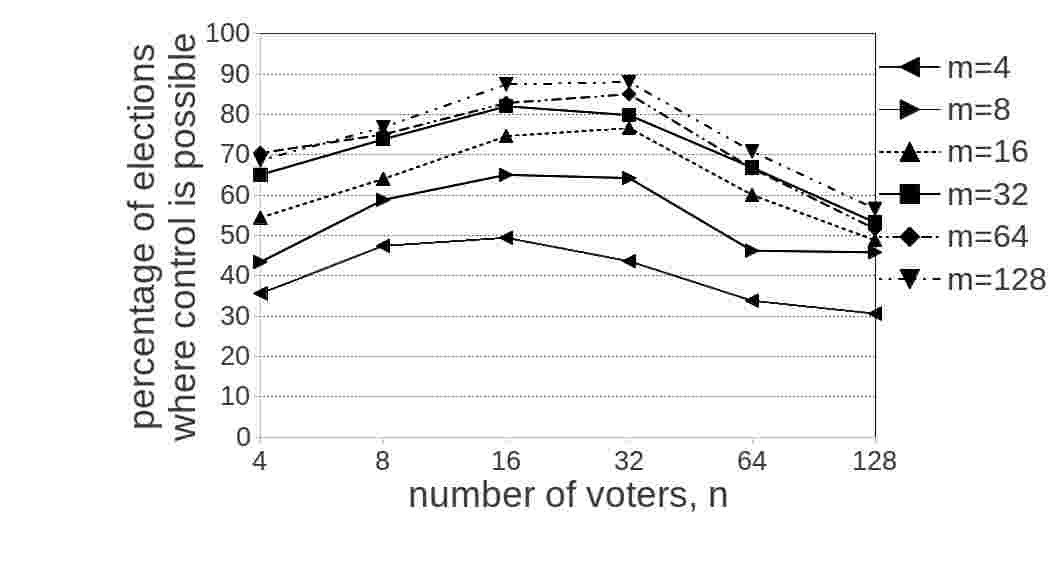}
	\caption{Results for fallback voting in the TM model for 
destructive control by partition of voters in model TE.  Number of candidates is fixed.}
\end{figure}

%
%

\newpage
\begin{figure}[ht]
\centering
	\includegraphics[scale=0.3]{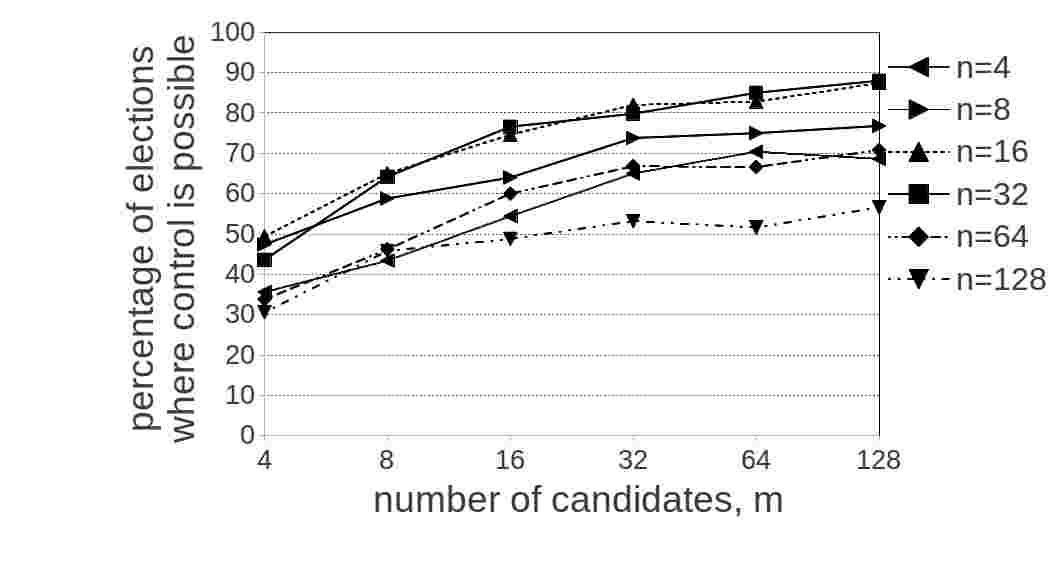}
	\caption{Results for fallback voting in the TM model for 
destructive control by partition of voters in model TE.  Number of voters is fixed.}
\end{figure}

%
\end{center}

\clearpage
\subsubsection{Computational Costs}
\begin{figure}[ht]
\centering
	\includegraphics[scale=0.3]{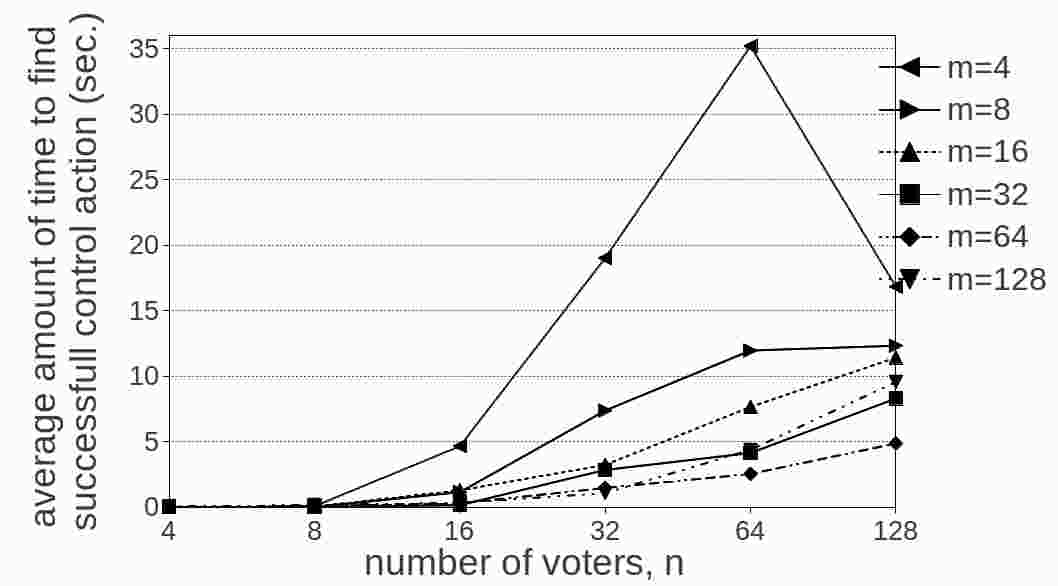}
	\caption{Average time the algorithm needs to find a successful control action for 
	destructive control by partition of voters in model TE
	in fallback elections in the IC model. The maximum is $35,22$ seconds.}
\end{figure}

\begin{figure}[ht]
\centering
	\includegraphics[scale=0.3]{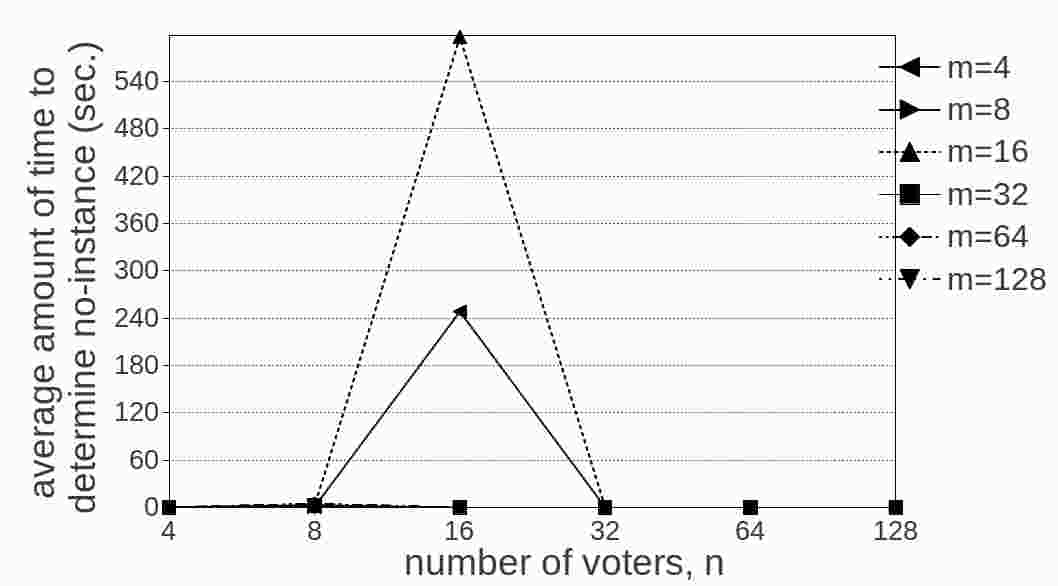}
	\caption{Average time the algorithm needs to determine no-instance of 
		destructive control by partition of voters in model TE
	in fallback elections in the IC model. The maximum is $596,35$ seconds.}
\end{figure}

\begin{figure}[ht]
\centering
	\includegraphics[scale=0.3]{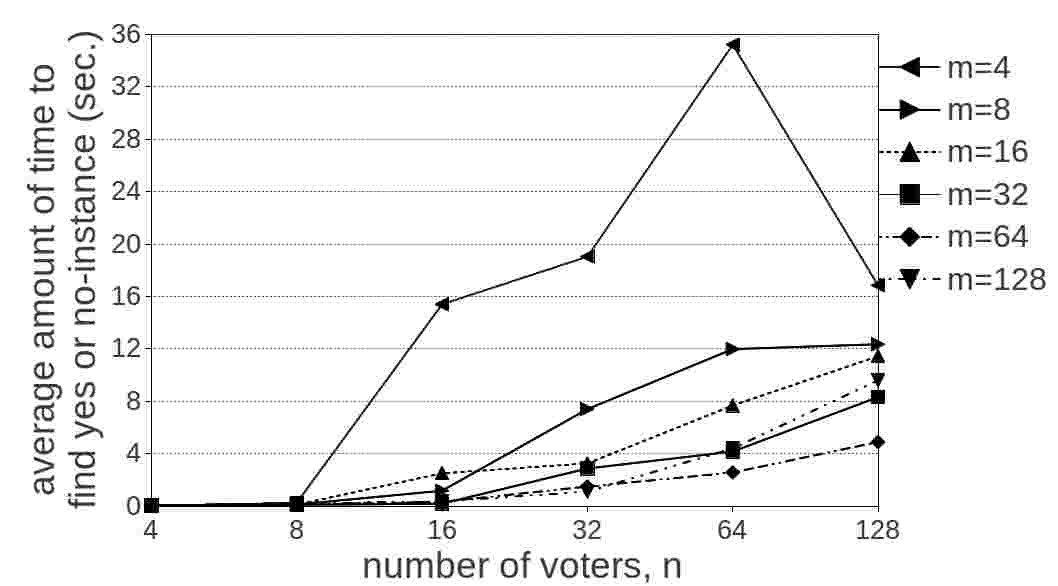}
	\caption{Average time the algorithm needs to give a definite output for 
	destructive control by partition of voters in model TE
	in fallback elections in the IC model. The maximum is $35,22$ seconds.}
\end{figure}

\begin{figure}[ht]
\centering
	\includegraphics[scale=0.3]{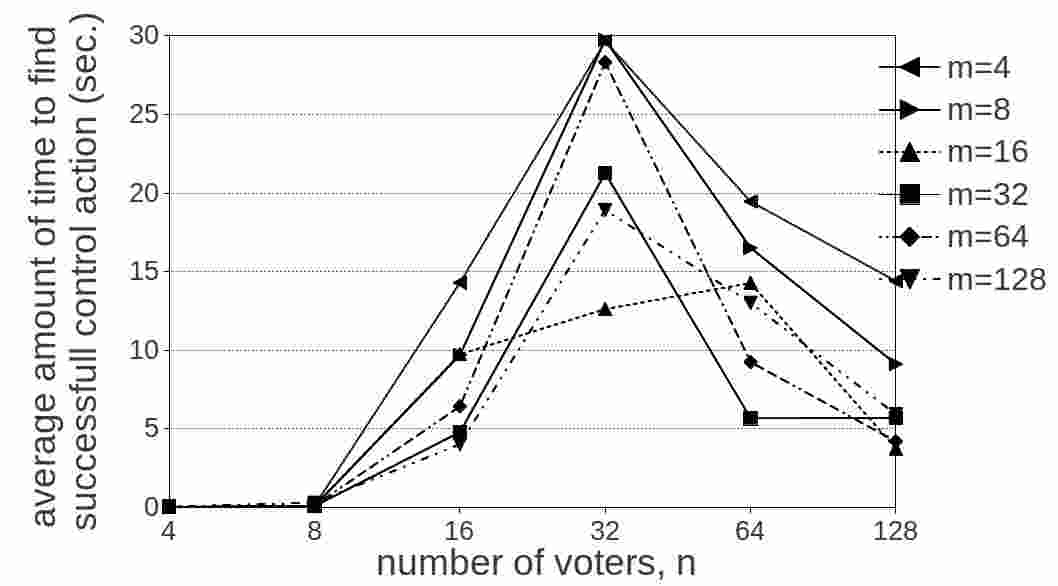}
	\caption{Average time the algorithm needs to find a successful control action for 
	destructive control by partition of voters in model TE
	in fallback elections in the TM model. The maximum is $29,74$ seconds.}
\end{figure}

\begin{figure}[ht]
\centering
	\includegraphics[scale=0.3]{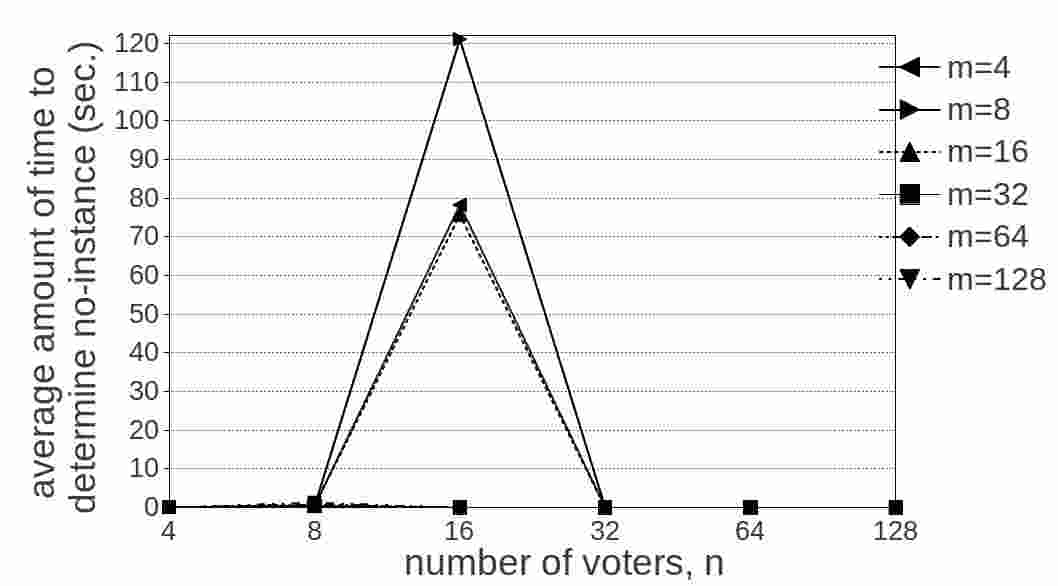}
	\caption{Average time the algorithm needs to determine no-instance of 
		destructive control by partition of voters in model TE
	in fallback elections in the TM model. The maximum is $121,13$ seconds.}
\end{figure}

\begin{figure}[ht]
\centering
	\includegraphics[scale=0.3]{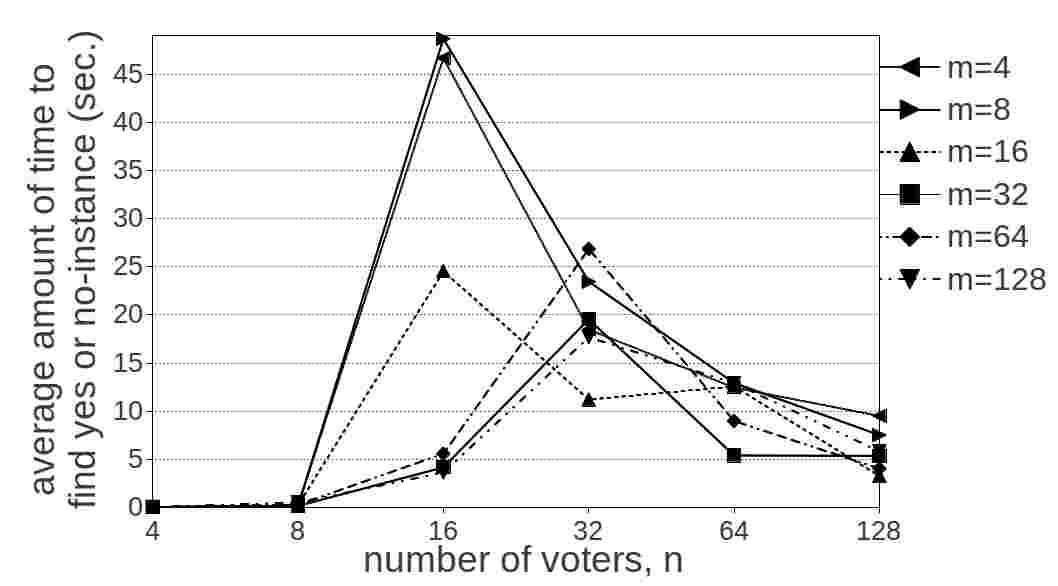}
	\caption{Average time the algorithm needs to give a definite output for 
	destructive control by partition of voters in model TE
	in fallback elections in the TM model. The maximum is $48,7$ seconds.48,7 seconds.}
\end{figure}

\clearpage
\subsection{Constructive Control by Partition of Voters in Model TP}
\begin{center}
\begin{figure}[ht]
\centering
	\includegraphics[scale=0.3]{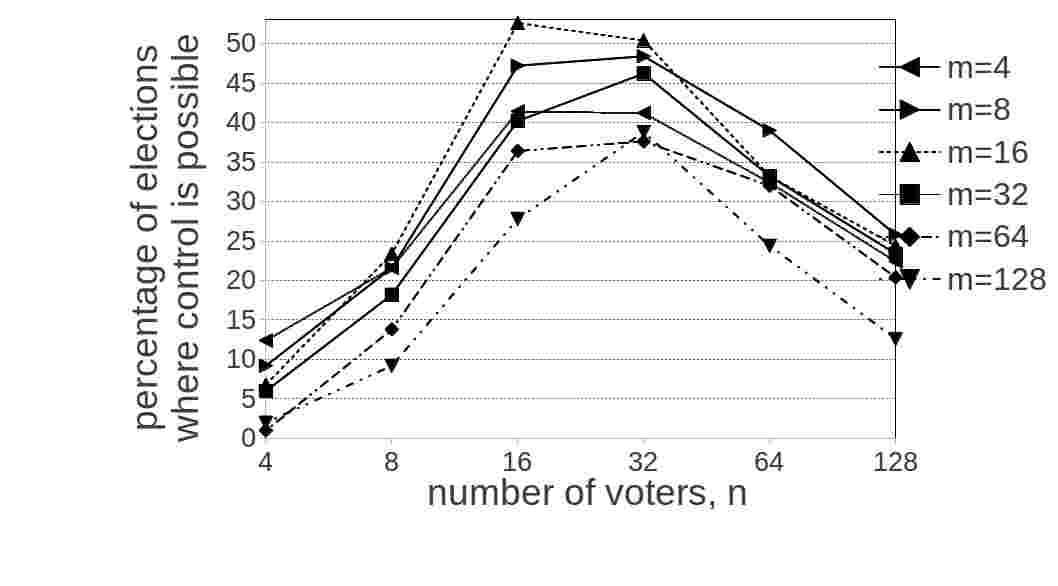}
	\caption{Results for fallback voting in the IC model for 
constructive control by partition of voters in model TP.  Number of candidates is fixed.}
\end{figure}


\newpage
\begin{figure}[ht]
\centering
	\includegraphics[scale=0.3]{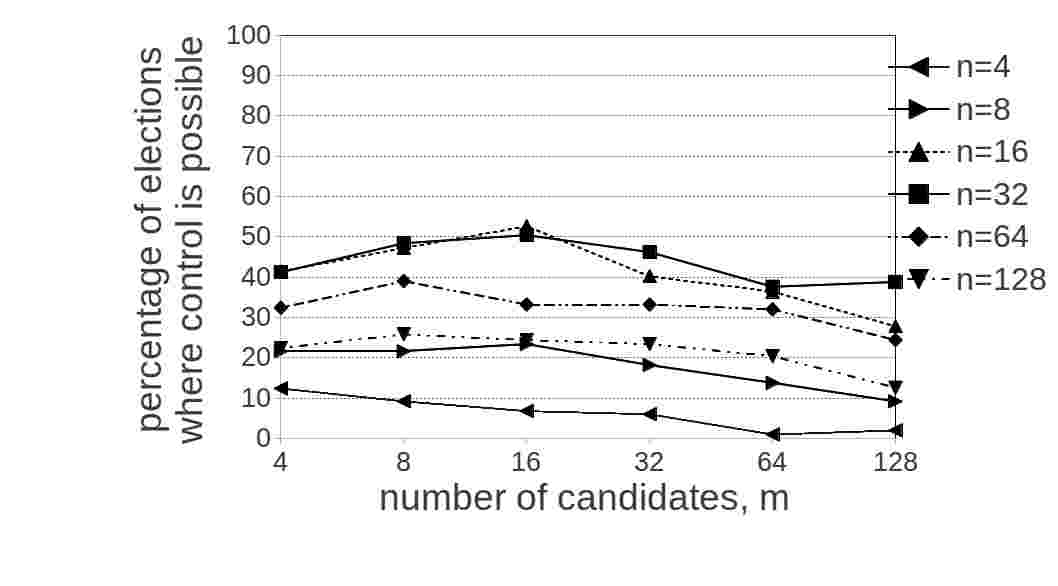}
	\caption{Results for fallback voting in the IC model for 
constructive control by partition of voters in model TP.  Number of voters is fixed.}
\end{figure}

%

\newpage
\begin{figure}[ht]
\centering
	\includegraphics[scale=0.3]{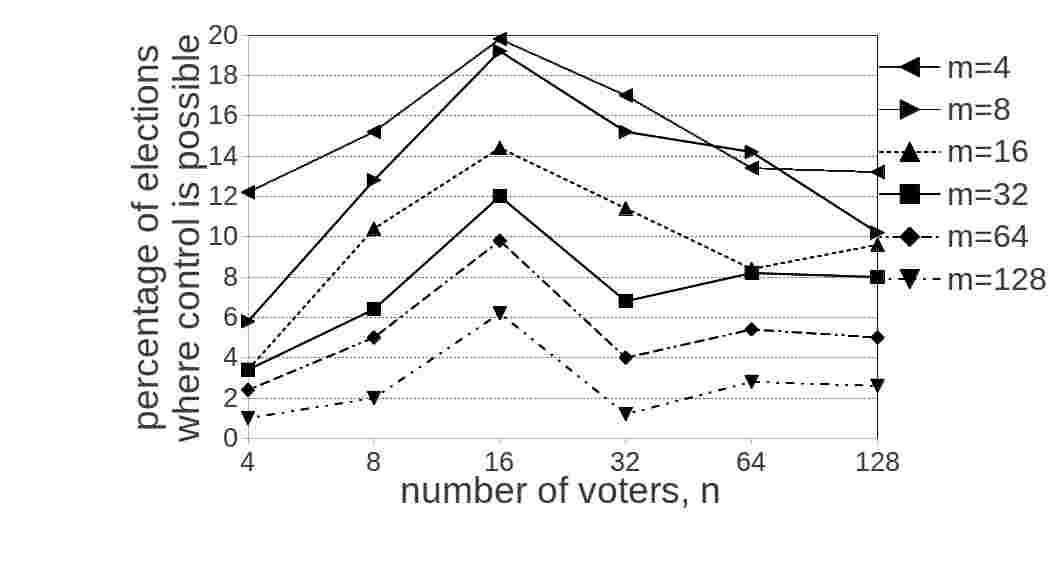}
	\caption{Results for fallback voting in the TM model for 
constructive control by partition of voters in model TP.  Number of candidates is fixed.}
\end{figure}

%
%

\newpage
\begin{figure}[ht]
\centering
	\includegraphics[scale=0.3]{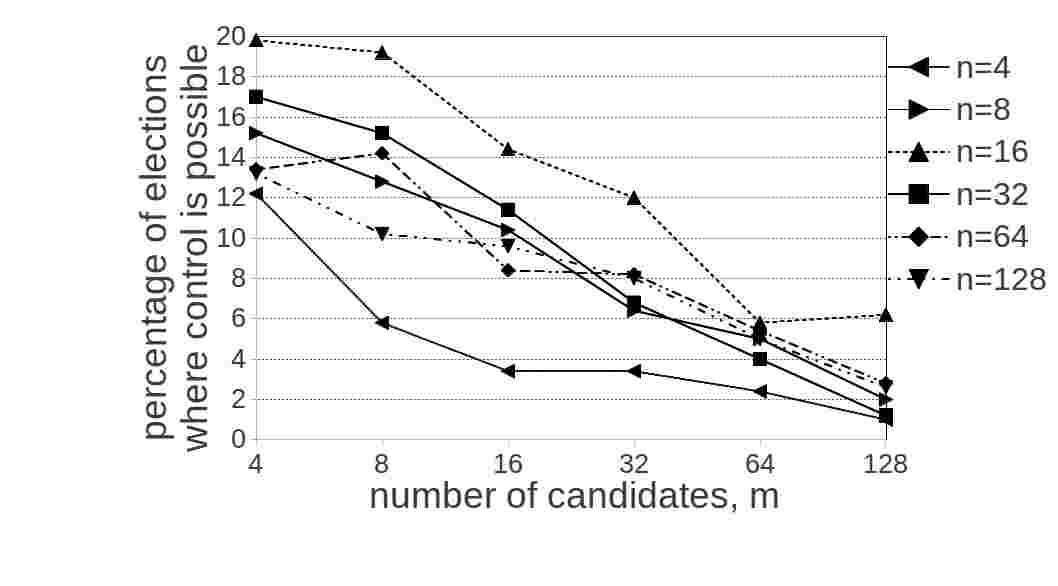}
	\caption{Results for fallback voting in the TM model for 
constructive control by partition of voters in model TP.  Number of voters is fixed.}
\end{figure}

%
\end{center}

\clearpage
\subsubsection{Computational Costs}
\begin{figure}[ht]
\centering
	\includegraphics[scale=0.3]{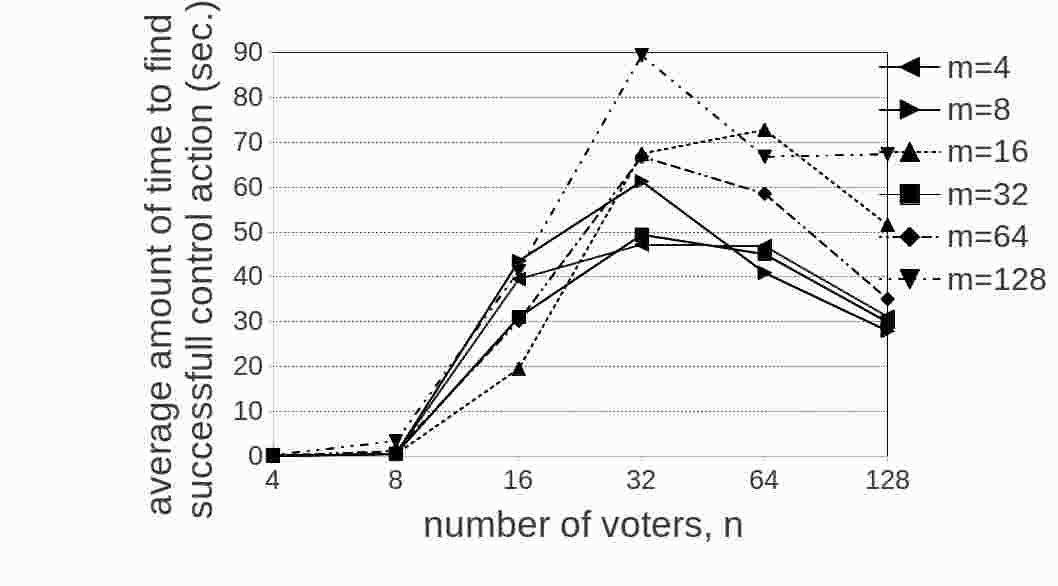}
	\caption{Average time the algorithm needs to find a successful control action for 
	constructive control by partition of voters in model TP
	in fallback elections in the IC model. The maximum is $89,37$ seconds.}
\end{figure}

\begin{figure}[ht]
\centering
	\includegraphics[scale=0.3]{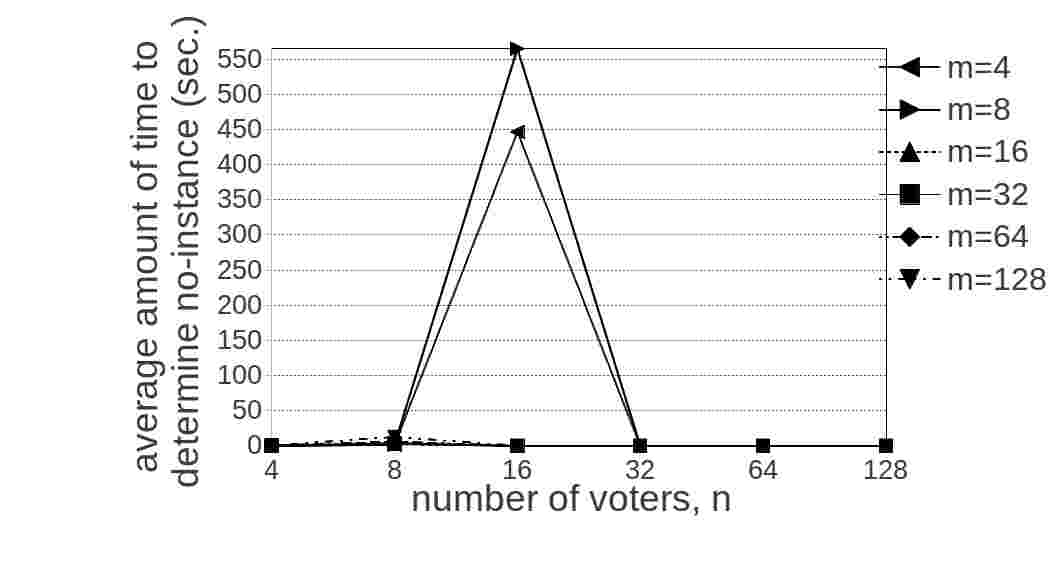}
	\caption{Average time the algorithm needs to determine no-instance of 
		constructive control by partition of voters in model TP
	in fallback elections in the IC model. The maximum is $564,65$ seconds.}
\end{figure}

\begin{figure}[ht]
\centering
	\includegraphics[scale=0.3]{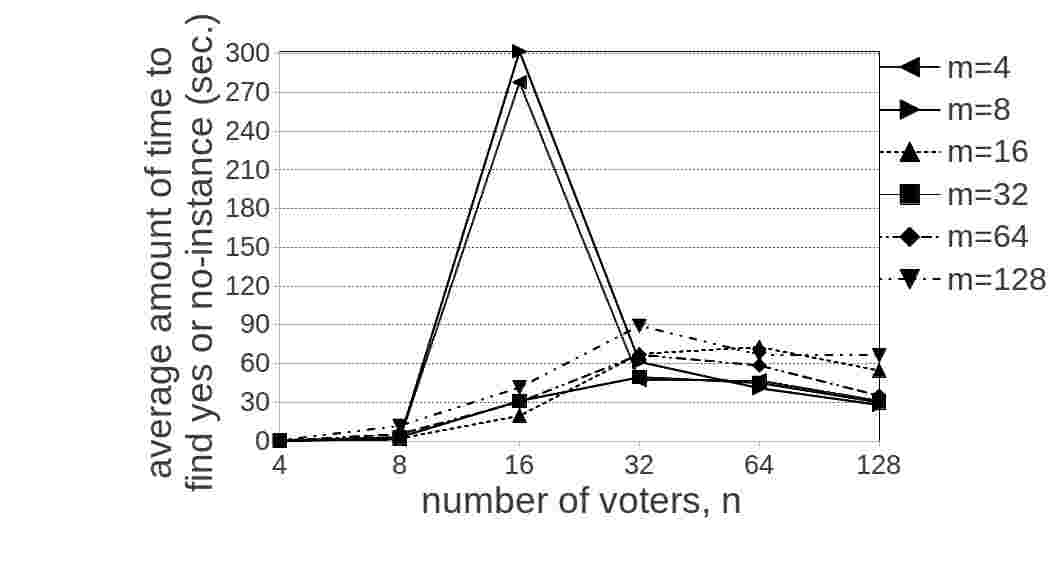}
	\caption{Average time the algorithm needs to give a definite output for 
	constructive control by partition of voters in model TP
	in fallback elections in the IC model. The maximum is $301,87$ seconds.}
\end{figure}

\begin{figure}[ht]
\centering
	\includegraphics[scale=0.3]{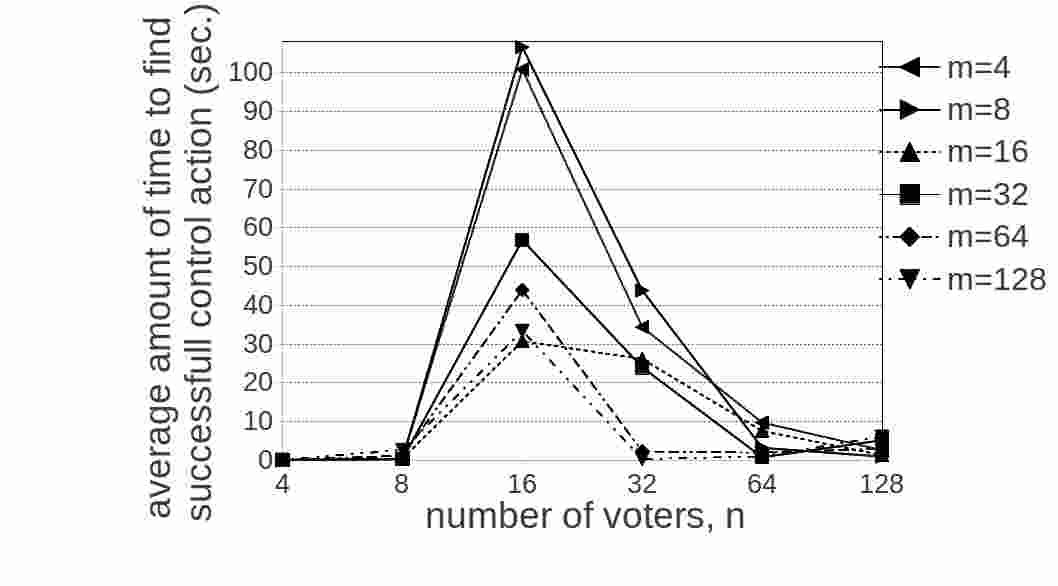}
	\caption{Average time the algorithm needs to find a successful control action for 
	constructive control by partition of voters in model TP
	in fallback elections in the TM model. The maximum is $106,58$ seconds.}
\end{figure}

\begin{figure}[ht]
\centering
	\includegraphics[scale=0.3]{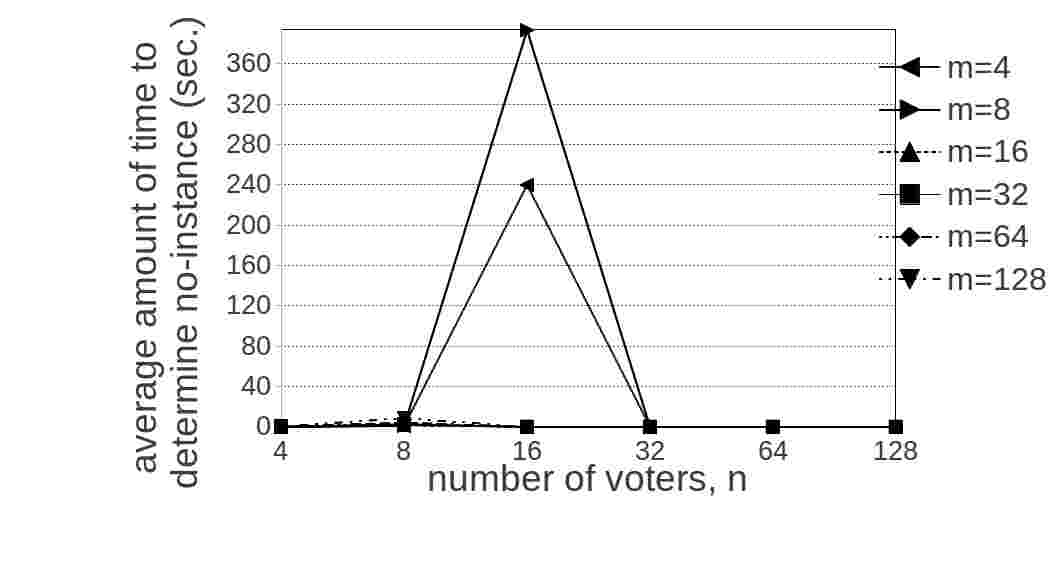}
	\caption{Average time the algorithm needs to determine no-instance of 
		constructive control by partition of voters in model TP
	in fallback elections in the TM model. The maximum is $393,18$ seconds.}
\end{figure}

\begin{figure}[ht]
\centering
	\includegraphics[scale=0.3]{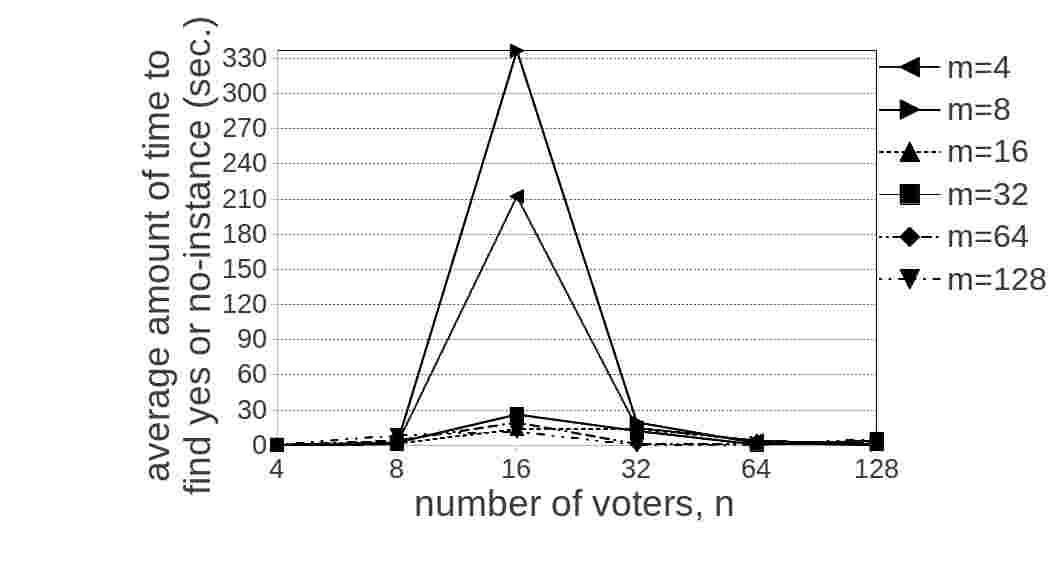}
	\caption{Average time the algorithm needs to give a definite output for 
	constructive control by partition of voters in model TP
	in fallback elections in the TM model. The maximum is $336,56$ seconds.}
\end{figure}

\clearpage
\subsection{Destructive Control by Partition of Voters in Model TP}
\begin{center}
\begin{figure}[ht]
\centering
	\includegraphics[scale=0.3]{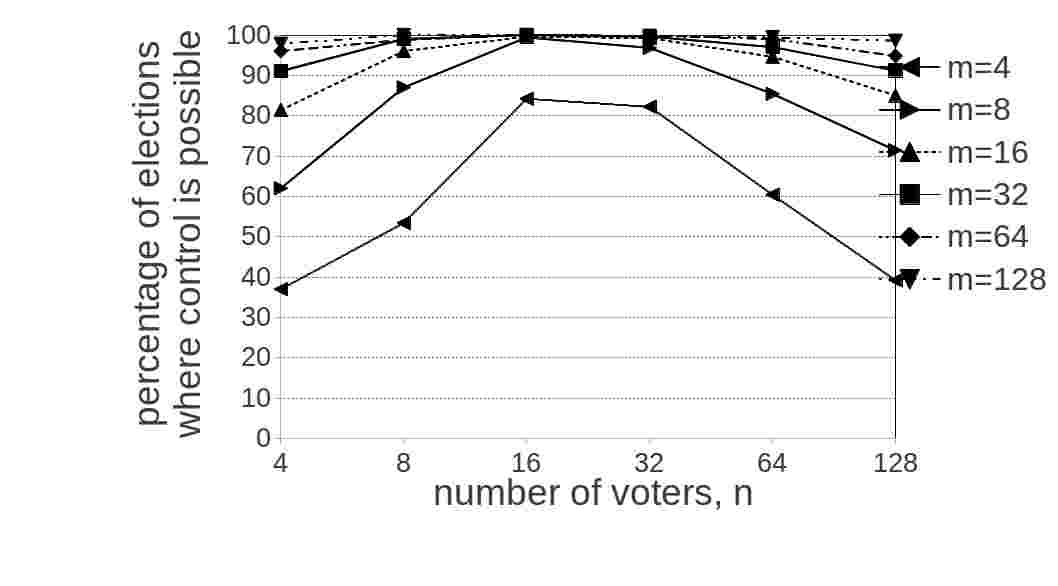}
	\caption{Results for fallback voting in the IC model for 
destructive control by partition of voters in model TP.  Number of candidates is fixed.}
\end{figure}


\newpage
\begin{figure}[ht]
\centering
	\includegraphics[scale=0.3]{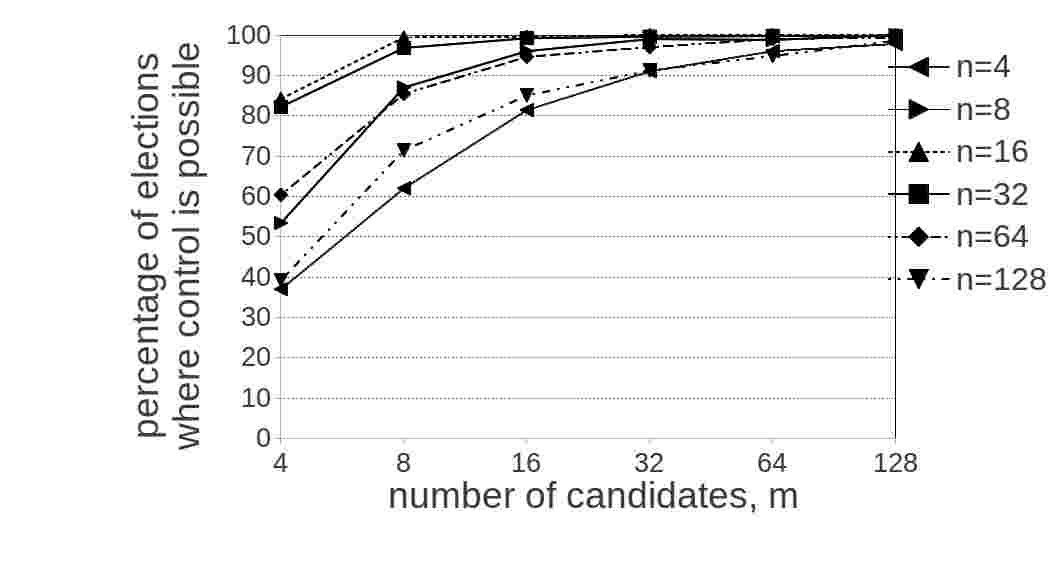}
	\caption{Results for fallback voting in the IC model for 
destructive control by partition of voters in model TP.  Number of voters is fixed.}
\end{figure}

%

\newpage
\begin{figure}[ht]
\centering
	\includegraphics[scale=0.3]{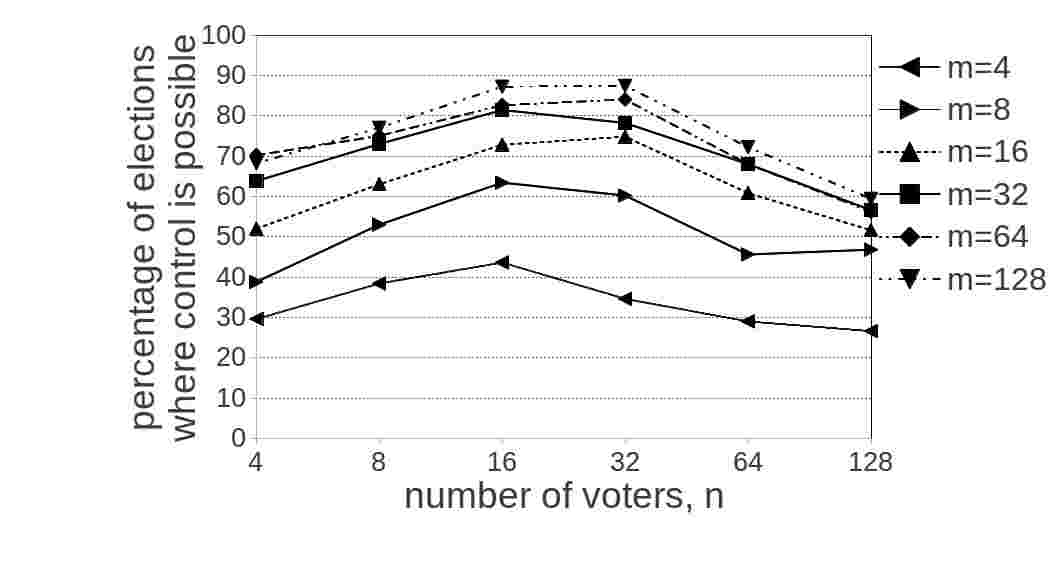}
	\caption{Results for fallback voting in the TM model for 
destructive control by partition of voters in model TP.  Number of candidates is fixed.}
\end{figure}

%
%
\newpage
\begin{figure}[ht]
\centering
	\includegraphics[scale=0.3]{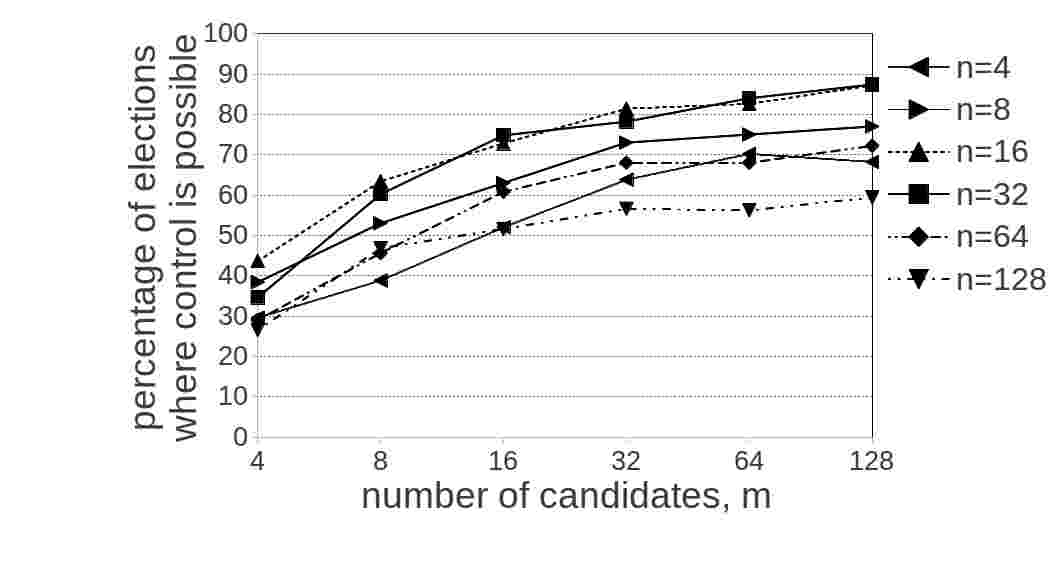}
	\caption{Results for fallback voting in the TM model for 
destructive control by partition of voters in model TP.  Number of voters is fixed.}
\end{figure}

%
\end{center}

\clearpage
\subsubsection{Computational Costs}
\begin{figure}[ht]
\centering
	\includegraphics[scale=0.3]{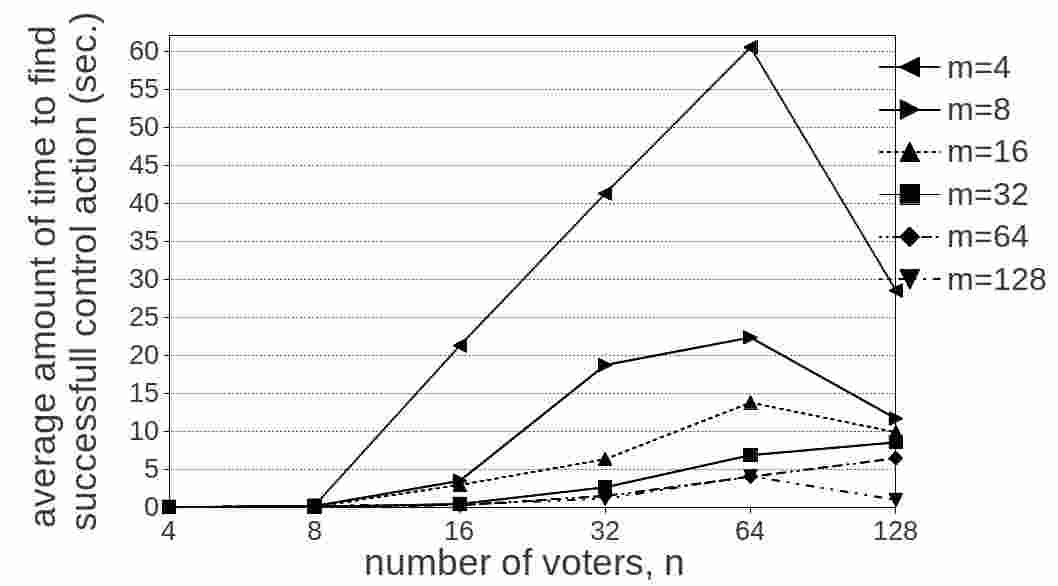}
	\caption{Average time the algorithm needs to find a successful control action for 
	destructive control by partition of voters in model TP
	in fallback elections in the IC model. The maximum is $60,49$ seconds.}
\end{figure}

\begin{figure}[ht]
\centering
	\includegraphics[scale=0.3]{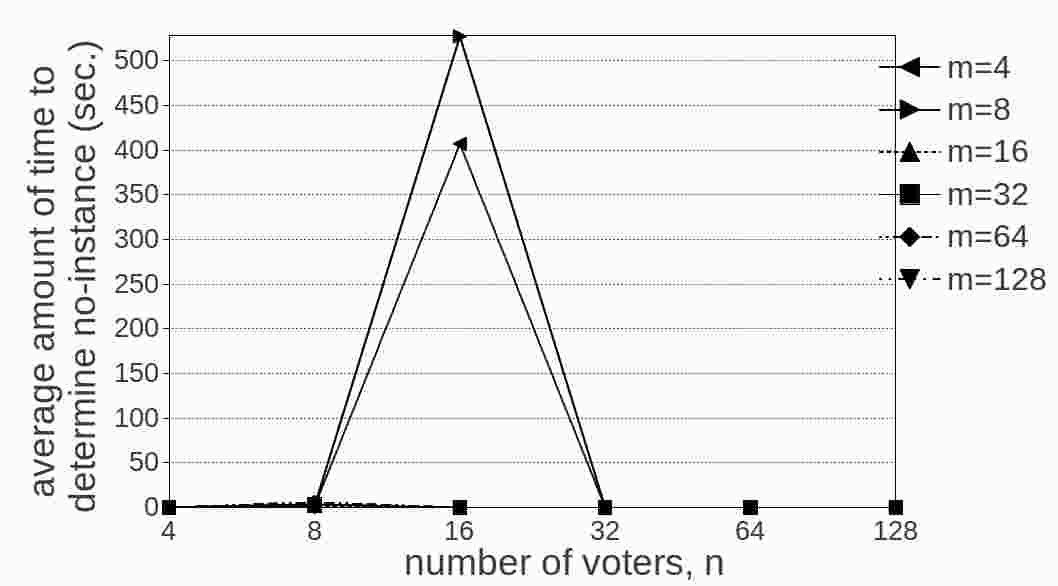}
	\caption{Average time the algorithm needs to determine no-instance of 
		destructive control by partition of voters in model TP
	in fallback elections in the IC model. The maximum is $527,4$ seconds.}
\end{figure}

\begin{figure}[ht]
\centering
	\includegraphics[scale=0.3]{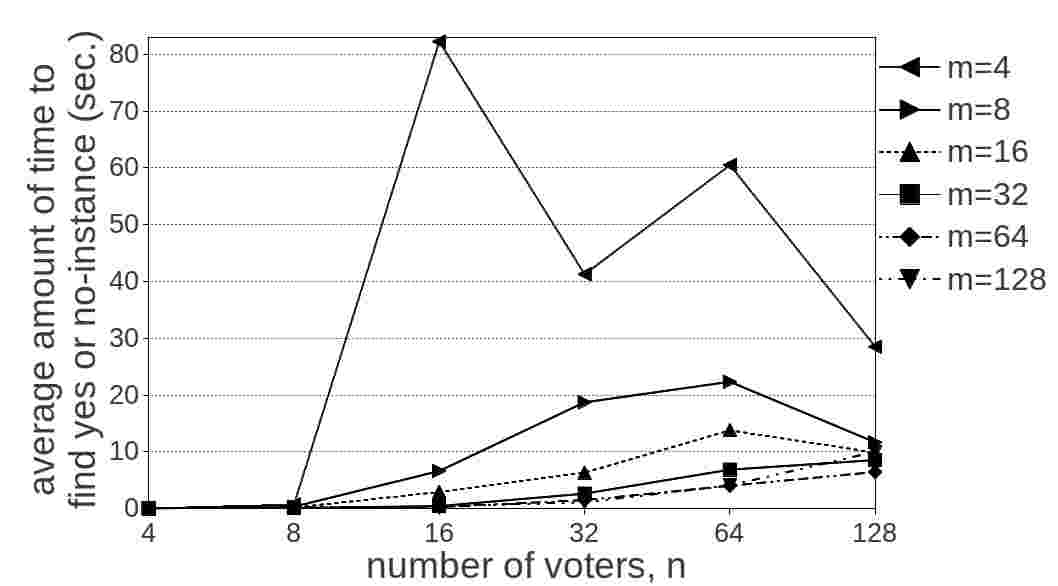}
	\caption{Average time the algorithm needs to give a definite output for 
	destructive control by partition of voters in model TP
	in fallback elections in the IC model. The maximum is $82,22$ seconds.}
\end{figure}

\begin{figure}[ht]
\centering
	\includegraphics[scale=0.3]{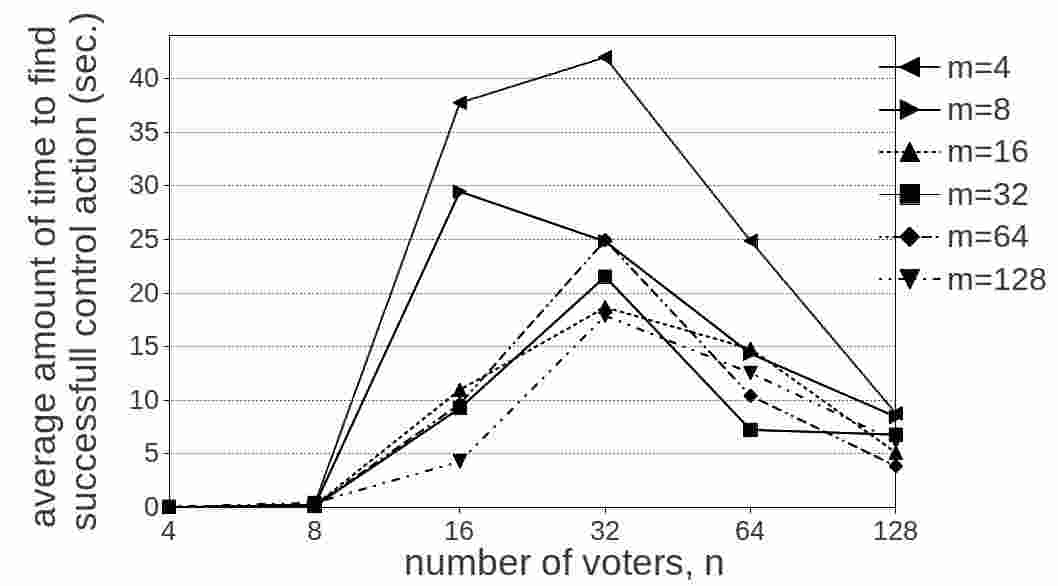}
	\caption{Average time the algorithm needs to find a successful control action for 
	destructive control by partition of voters in model TP
	in fallback elections in the TM model. The maximum is $42$ seconds.}
\end{figure}

\begin{figure}[ht]
\centering
	\includegraphics[scale=0.3]{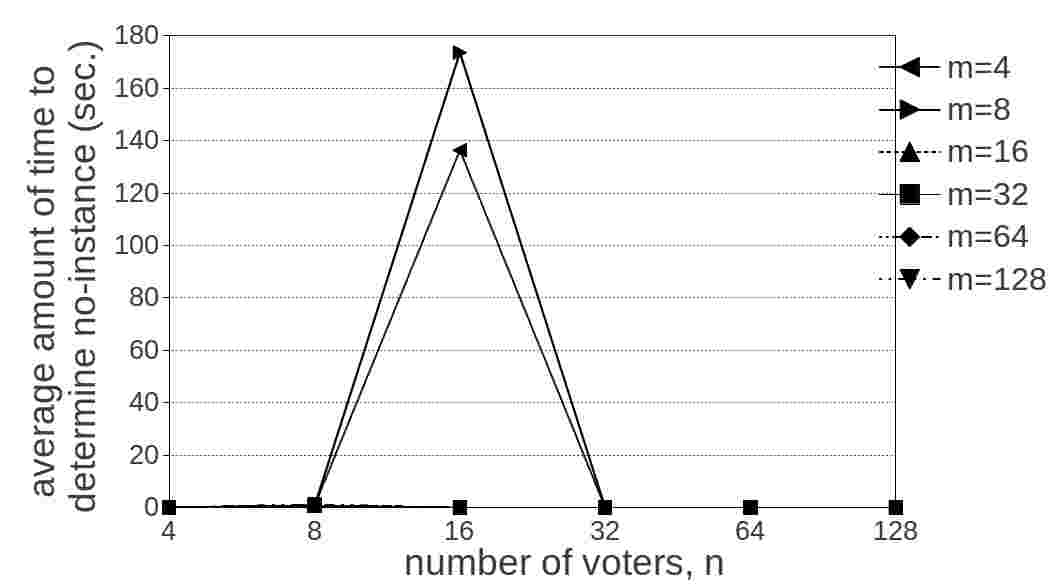}
	\caption{Average time the algorithm needs to determine no-instance of 
		destructive control by partition of voters in model TP
	in fallback elections in the TM model. The maximum is $173,62$ seconds.}
\end{figure}

\begin{figure}[ht]
\centering
	\includegraphics[scale=0.3]{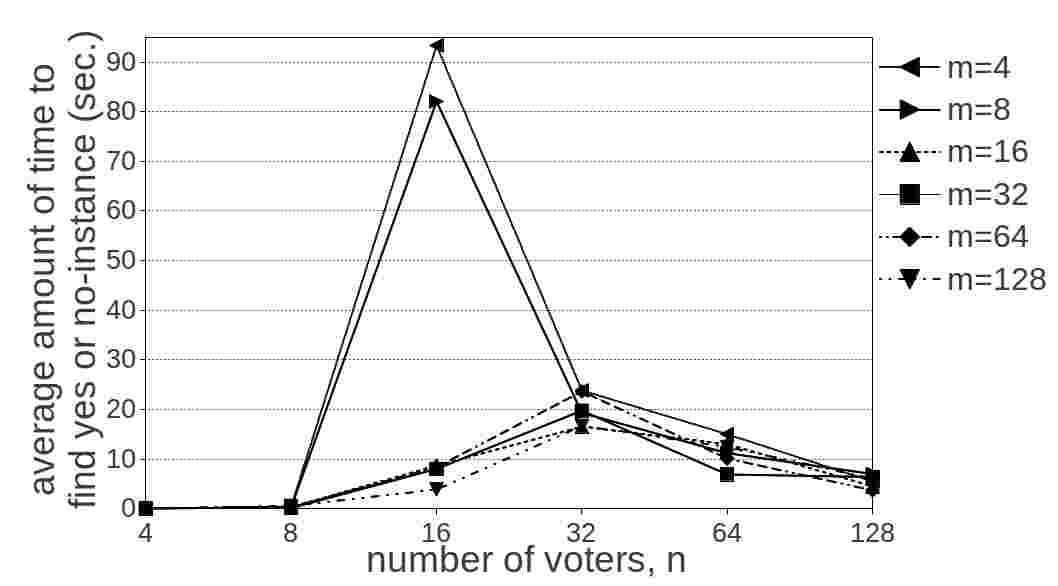}
	\caption{Average time the algorithm needs to give a definite output for 
	destructive control by partition of voters in model TP
	in fallback elections in the TM model. The maximum is $93,34$ seconds.}
\end{figure}

\clearpage
\section{Bucklin Voting}

\subsection{Constructive Control by Adding Candidates}
\begin{center}
\begin{figure}[ht]
\centering
	\includegraphics[scale=0.3]{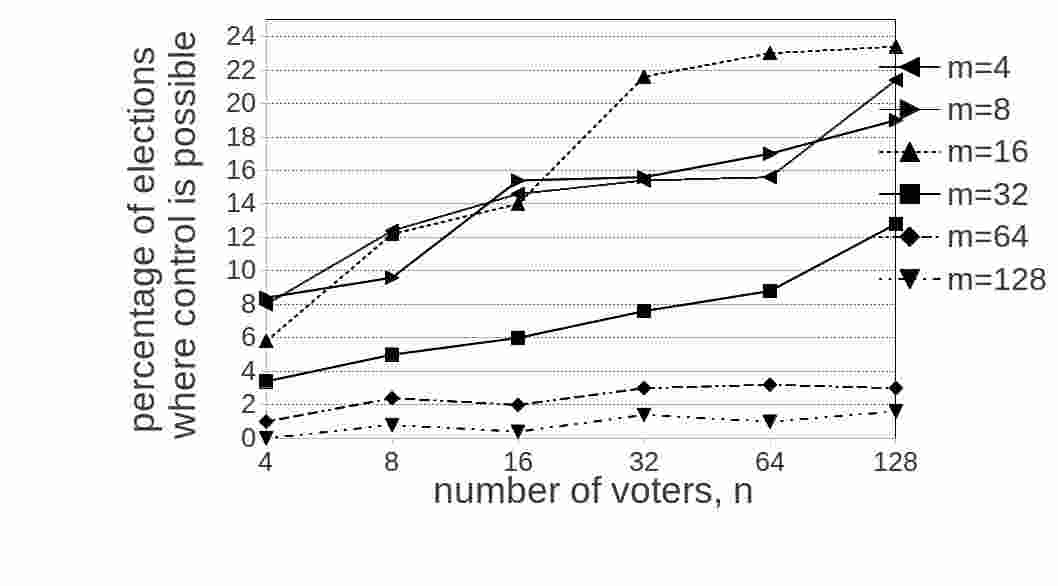}
		\caption{Results for Bucklin voting in the IC model for 
constructive control by adding candidates. Number of candidates is fixed. }
\end{figure}

\newpage
\begin{figure}[ht]
\centering
	\includegraphics[scale=0.3]{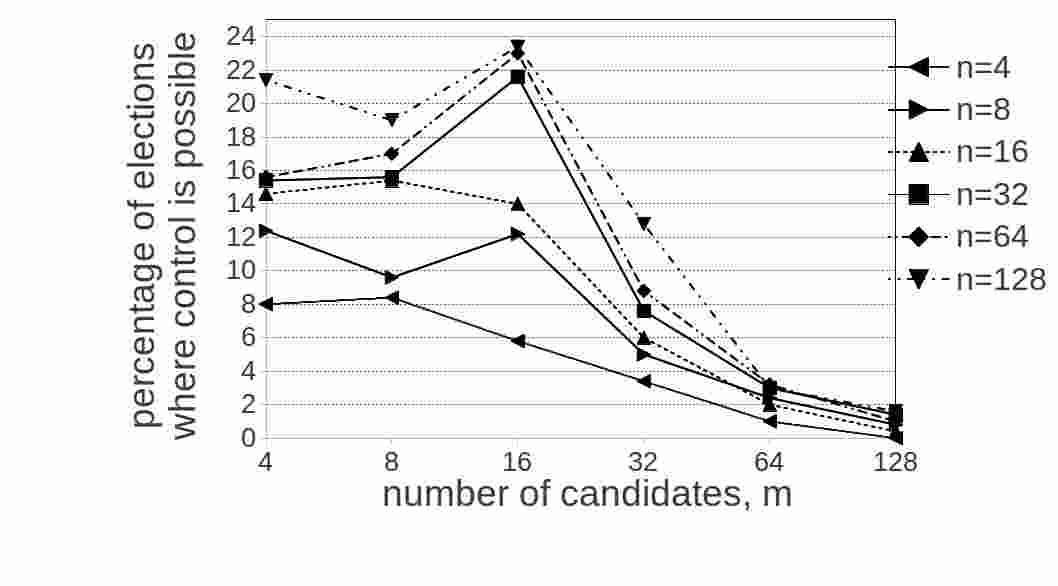}
		\caption{Results for Bucklin voting in the IC model for 
constructive control by adding candidates. Number of voters is fixed. }
\end{figure}
%
\newpage
\begin{figure}[ht]
\centering
	\includegraphics[scale=0.3]{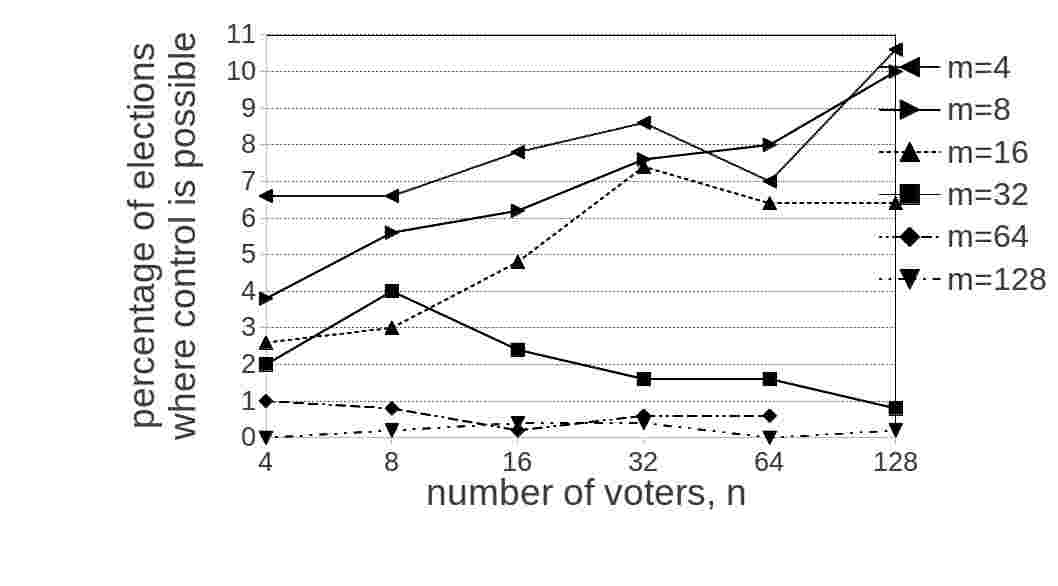}
	\caption{Results for Bucklin voting in the TM model for 
constructive control by adding candidates. Number of candidates is fixed. }
\end{figure}
%
\newpage
\begin{figure}[ht]
\centering
	\includegraphics[scale=0.3]{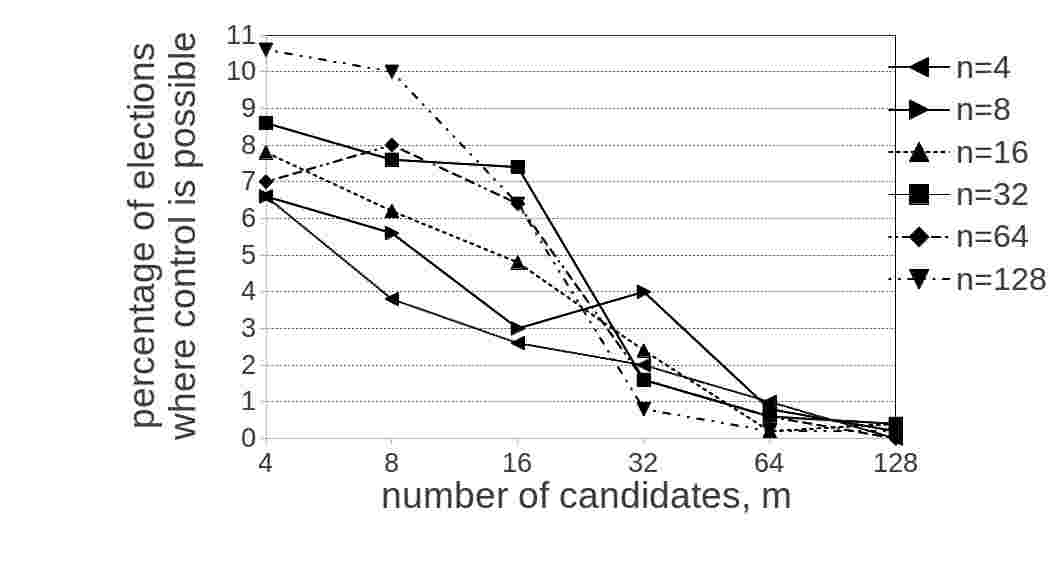}
	\caption{Results for Bucklin voting in the TM model for 
constructive control by adding candidates. Number of voters is fixed. }
\end{figure}
%
\end{center}

\clearpage
\subsubsection{Computational Costs}
\begin{figure}[ht]
\centering
	\includegraphics[scale=0.3]{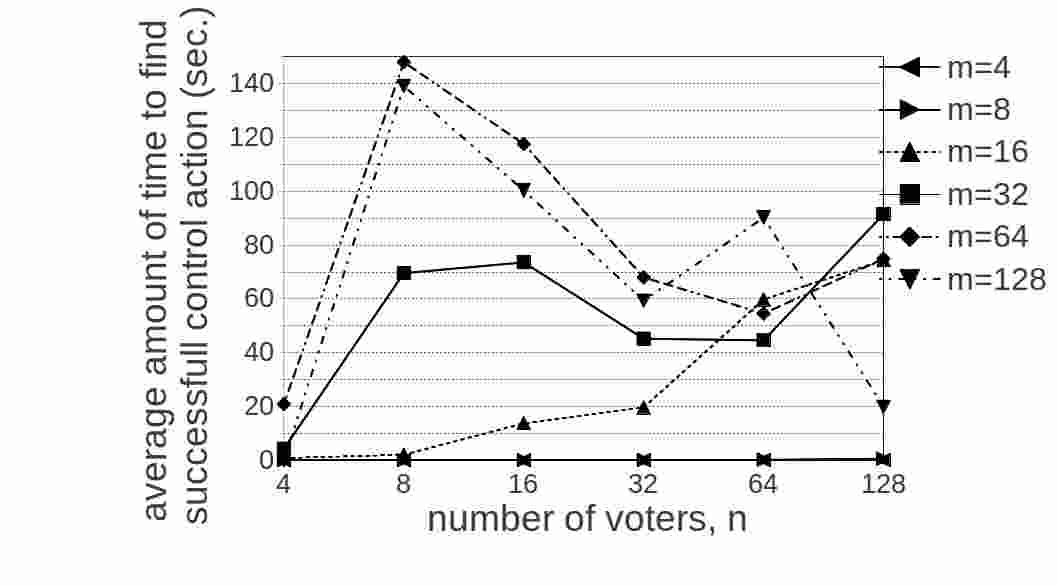}
	\caption{Average time the algorithm needs to find a successful control action for 
	constructive control by adding candidates
	in Bucklin elections in the IC model. The maximum is $148$ seconds.}
\end{figure}
\begin{figure}[ht]
\centering
	\includegraphics[scale=0.3]{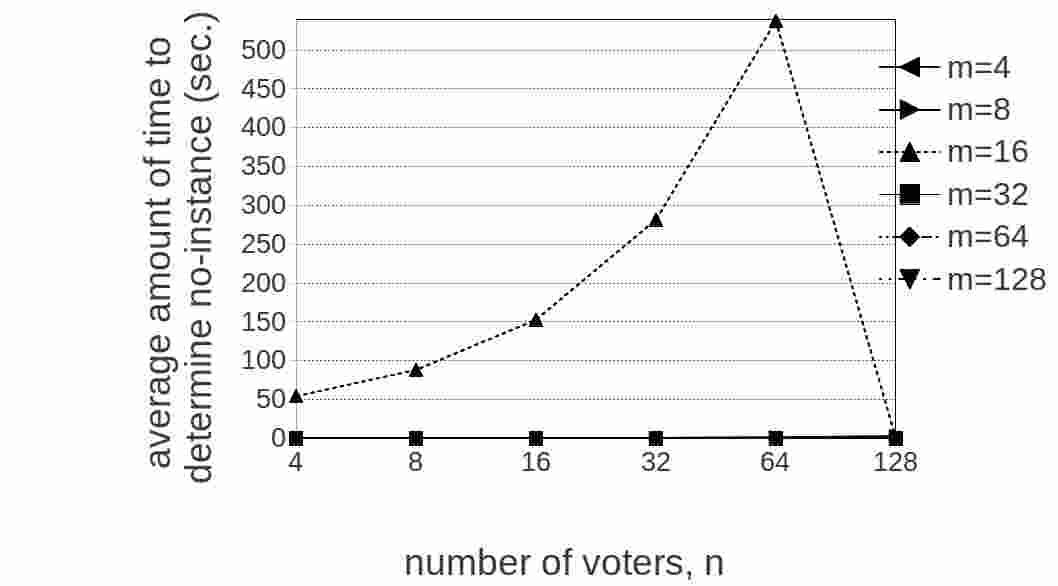}
	\caption{Average time the algorithm needs to determine no-instance of 
		constructive control by adding candidates
	in Bucklin elections in the IC model. The maximum is $537,35$ seconds.}
\end{figure}
\begin{figure}[ht]
\centering
	\includegraphics[scale=0.3]{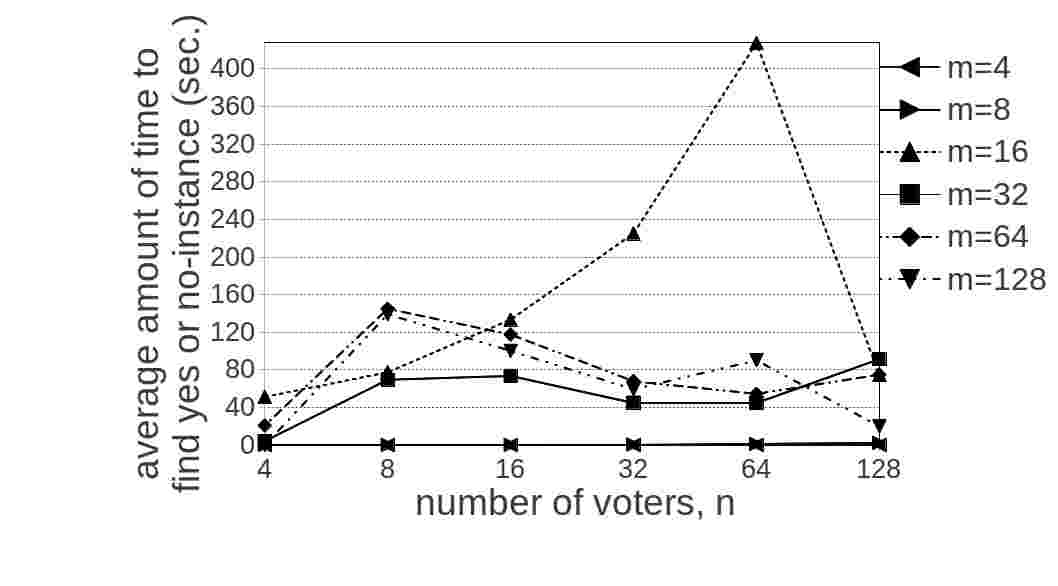}
	\caption{Average time the algorithm needs to give a definite output for 
	constructive control by adding candidates
	in Bucklin elections in the IC model. The maximum is $427,51$ seconds.}
\end{figure}
\begin{figure}[ht]
\centering
	\includegraphics[scale=0.3]{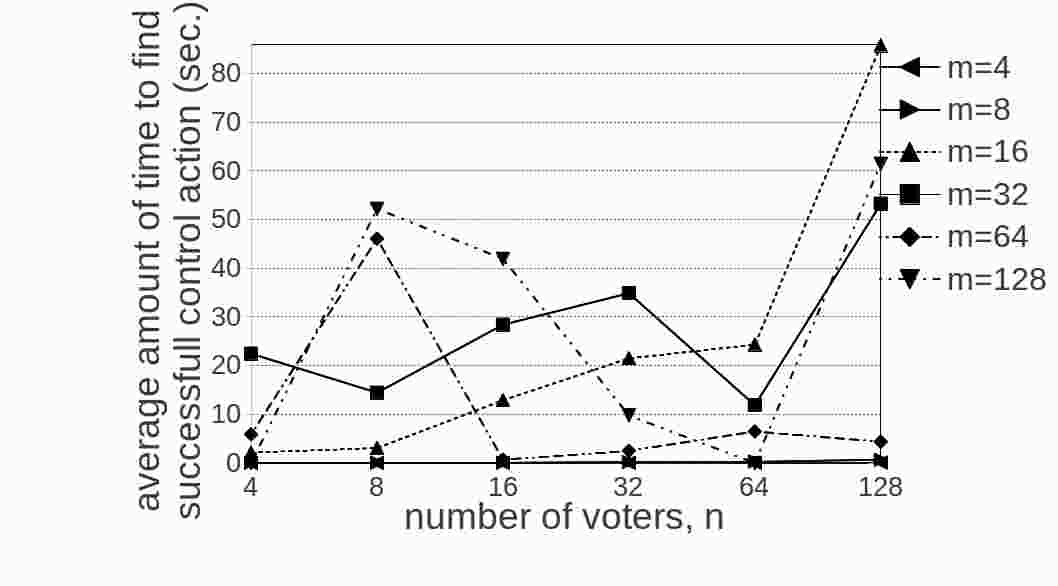}
	\caption{Average time the algorithm needs to find a successful control action for 
	constructive control by adding candidates
	in Bucklin elections in the TM model. The maximum is $85,81$ seconds.}
\end{figure}
\begin{figure}[ht]
\centering
	\includegraphics[scale=0.3]{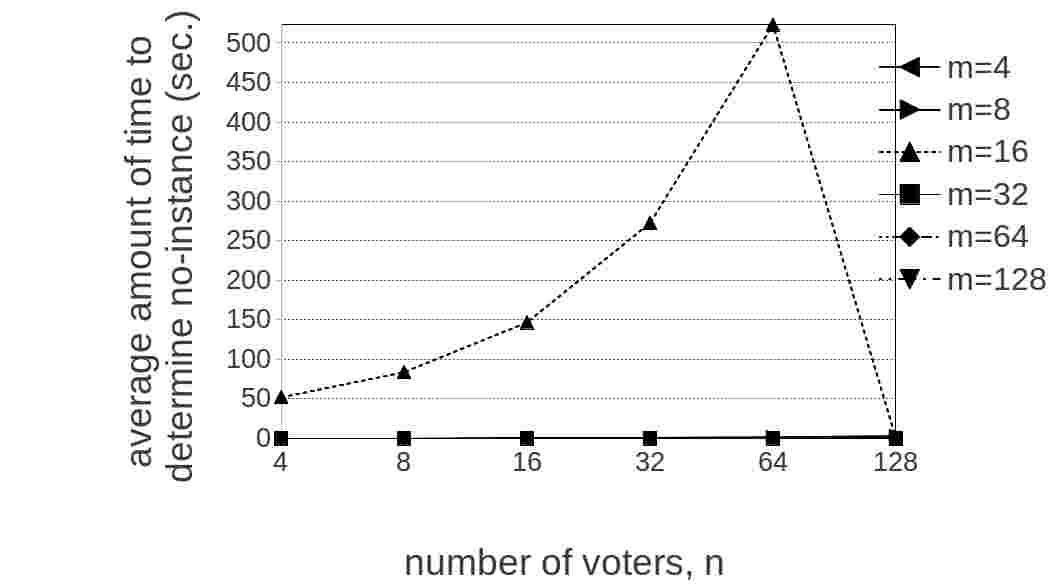}
	\caption{Average time the algorithm needs to determine no-instance of 
		constructive control by adding candidates
	in Bucklin elections in the TM model. The maximum is $523,05$ seconds.}
\end{figure}
\begin{figure}[ht]
\centering
	\includegraphics[scale=0.3]{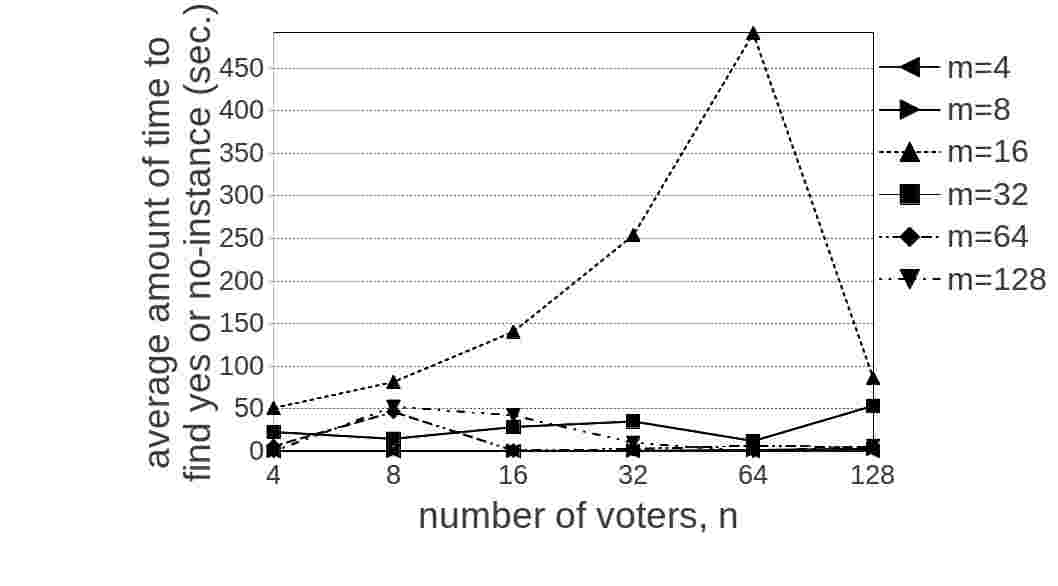}
	\caption{Average time the algorithm needs to give a definite output for 
	constructive control by adding candidates
	in Bucklin elections in the TM model. The maximum is $491,13$ seconds.}
\end{figure}

\clearpage
\subsection{Destructive Control by Adding Candidates}
\begin{center}
\begin{figure}[ht]
\centering
	\includegraphics[scale=0.3]{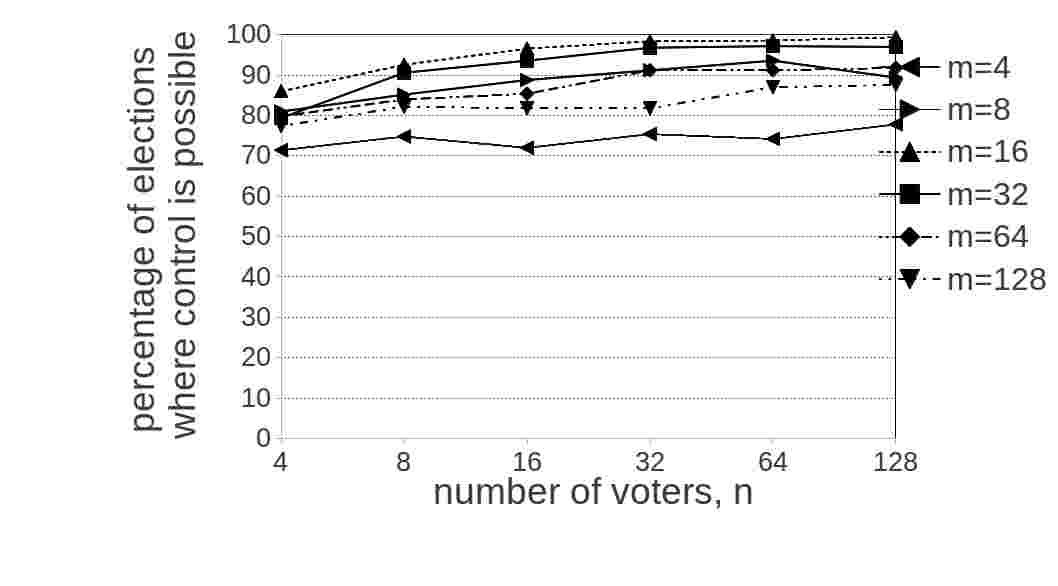}
		\caption{Results for Bucklin voting in the IC model for 
destructive control by adding candidates. Number of candidates is fixed. }
\end{figure}

\newpage
\begin{figure}[ht]
\centering
	\includegraphics[scale=0.3]{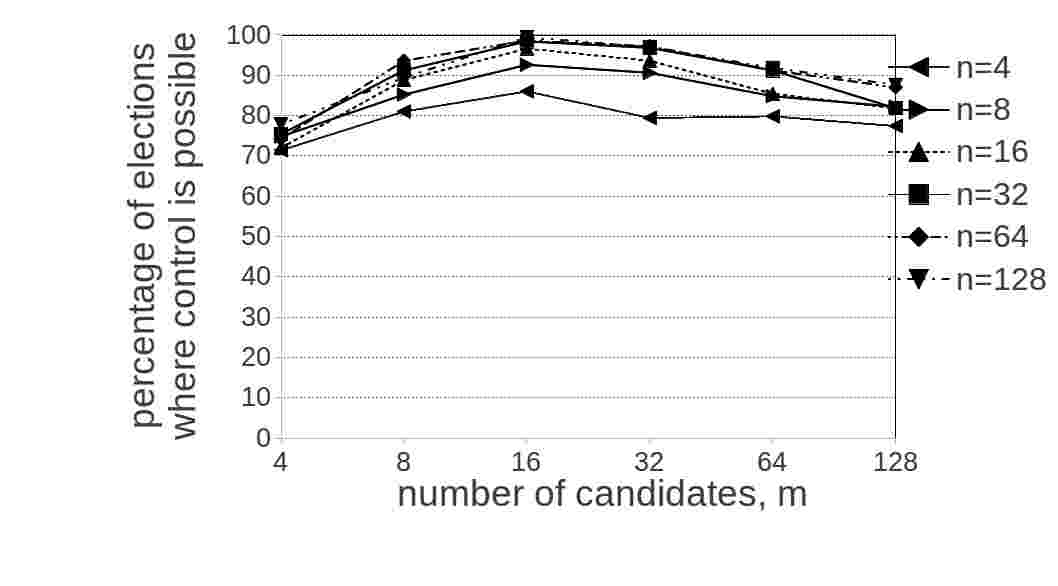}
		\caption{Results for Bucklin voting in the IC model for 
destructive control by adding candidates. Number of voters is fixed. }
\end{figure}
%
\newpage
\begin{figure}[ht]
\centering
	\includegraphics[scale=0.3]{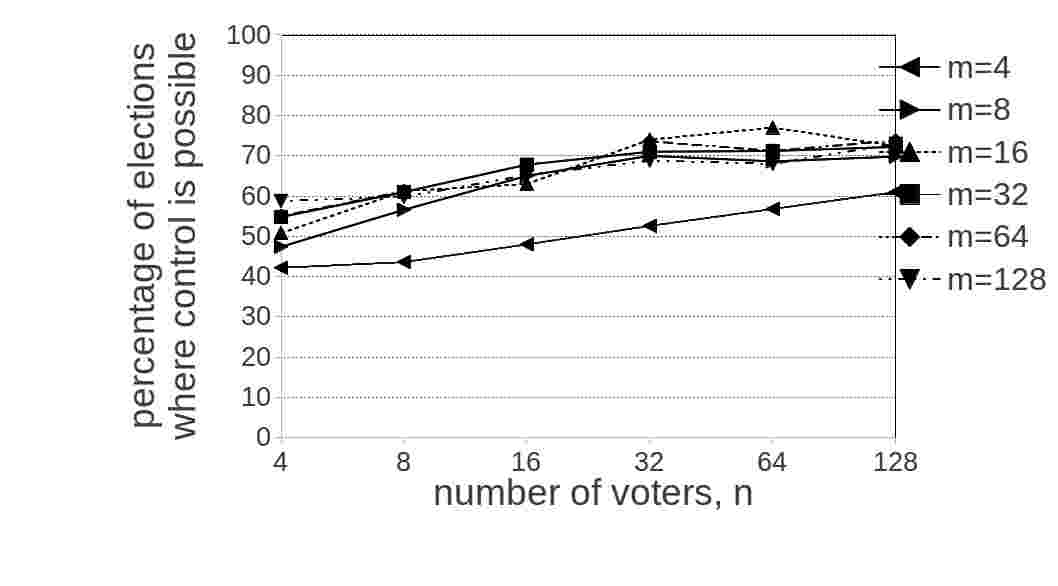}
	\caption{Results for Bucklin voting in the TM model for 
destructive control by adding candidates. Number of candidates is fixed. }
\end{figure}
%
\newpage
\begin{figure}[ht]
\centering
	\includegraphics[scale=0.3]{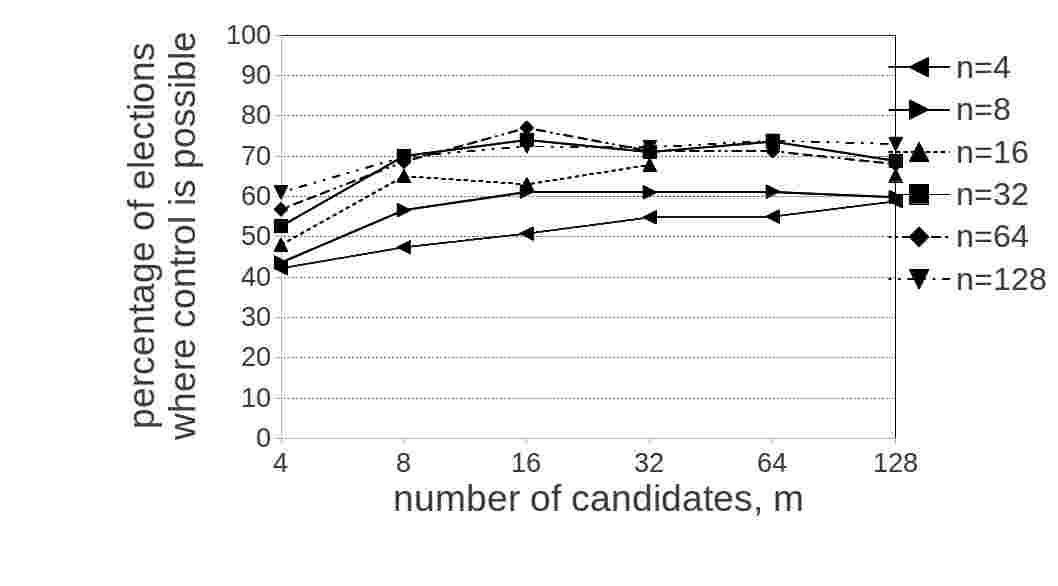}
	\caption{Results for Bucklin voting in the TM model for 
destructive control by adding candidates. Number of voters is fixed. }
\end{figure}

%
\end{center}

\clearpage
\subsubsection{Computational Costs}
\begin{figure}[ht]
\centering
	\includegraphics[scale=0.3]{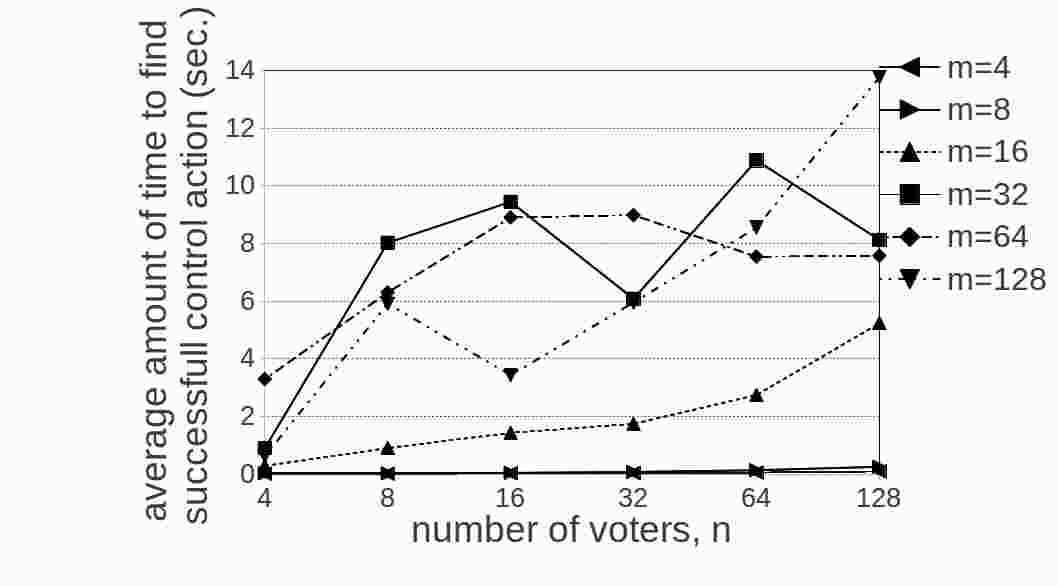}
	\caption{Average time the algorithm needs to find a successful control action for 
	destructive control by adding candidates
	in Bucklin elections in the IC model. The maximum is $13,77$ seconds.}
\end{figure}
\begin{figure}[ht]
\centering
	\includegraphics[scale=0.3]{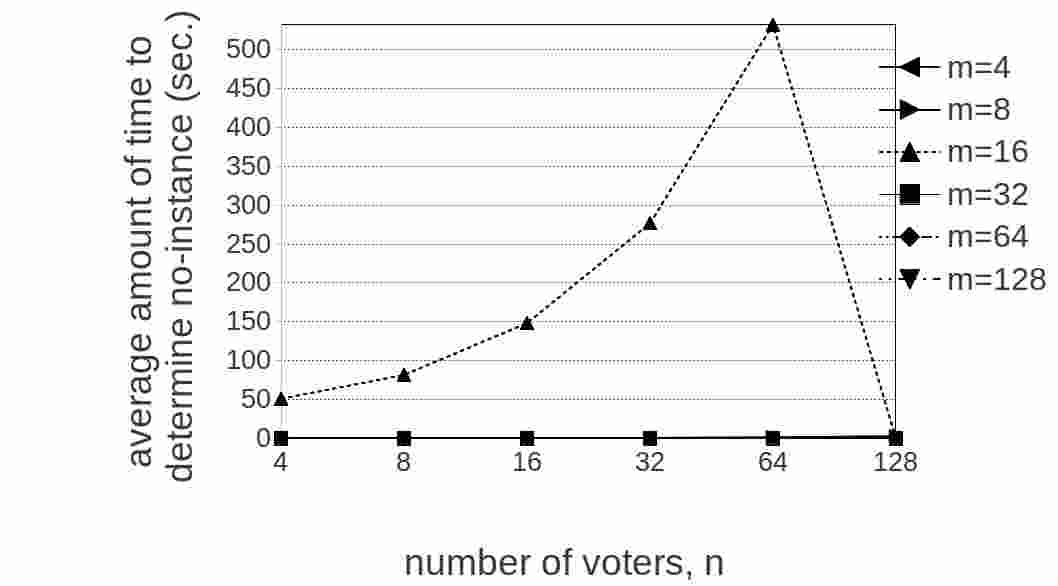}
	\caption{Average time the algorithm needs to determine no-instance of 
		destructive control by adding candidates
	in Bucklin elections in the IC model. The maximum is $532,06$ seconds.}
\end{figure}
\begin{figure}[ht]
\centering
	\includegraphics[scale=0.3]{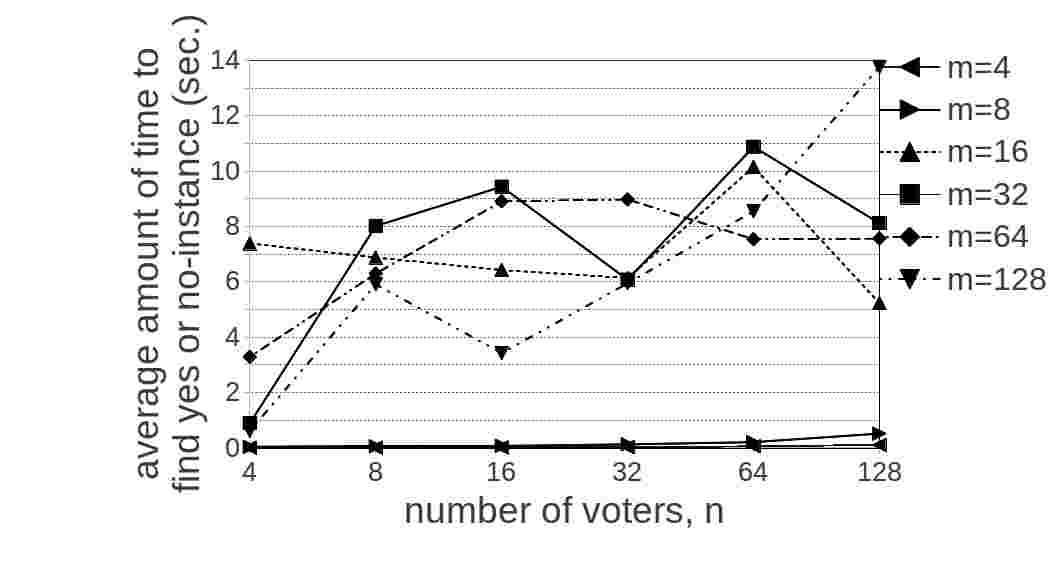}
	\caption{Average time the algorithm needs to give a definite output for 
	destructive control by adding candidates
	in Bucklin elections in the IC model. The maximum is $13,77$ seconds.}
\end{figure}
\begin{figure}[ht]
\centering
	\includegraphics[scale=0.3]{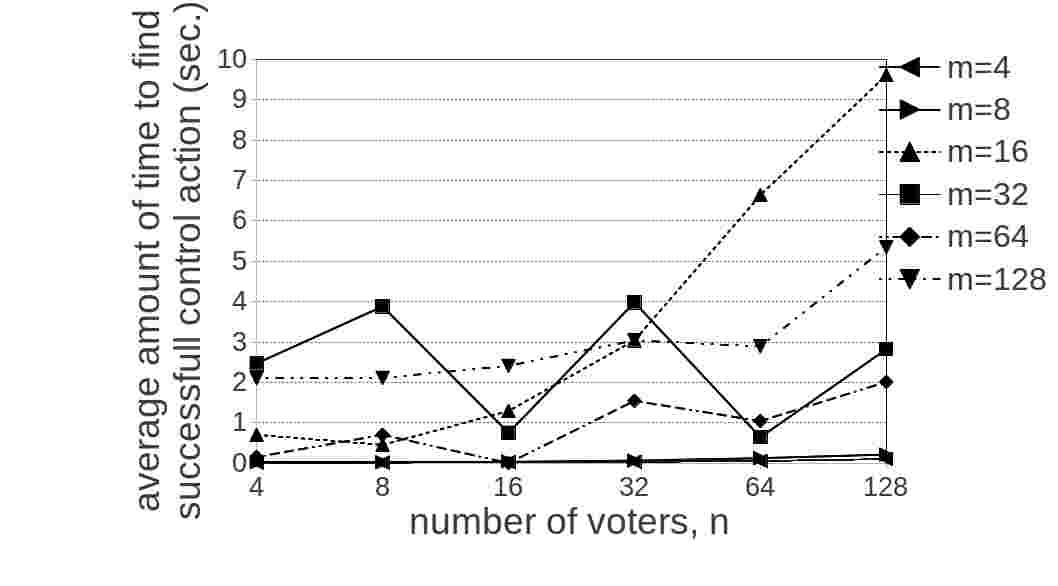}
	\caption{Average time the algorithm needs to find a successful control action for 
	destructive control by adding candidates
	in Bucklin elections in the TM model. The maximum is $9,61$ seconds.}
\end{figure}
\begin{figure}[ht]
\centering
	\includegraphics[scale=0.3]{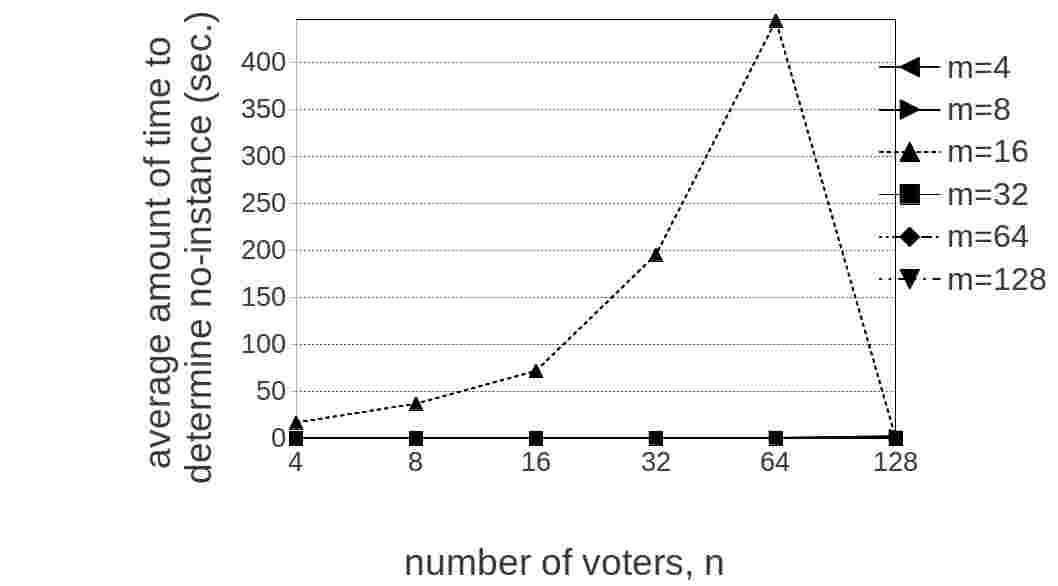}
	\caption{Average time the algorithm needs to determine no-instance of 
		destructive control by adding candidates
	in Bucklin elections in the TM model. The maximum is $444,22$ seconds.}
\end{figure}
\begin{figure}[ht]
\centering
	\includegraphics[scale=0.3]{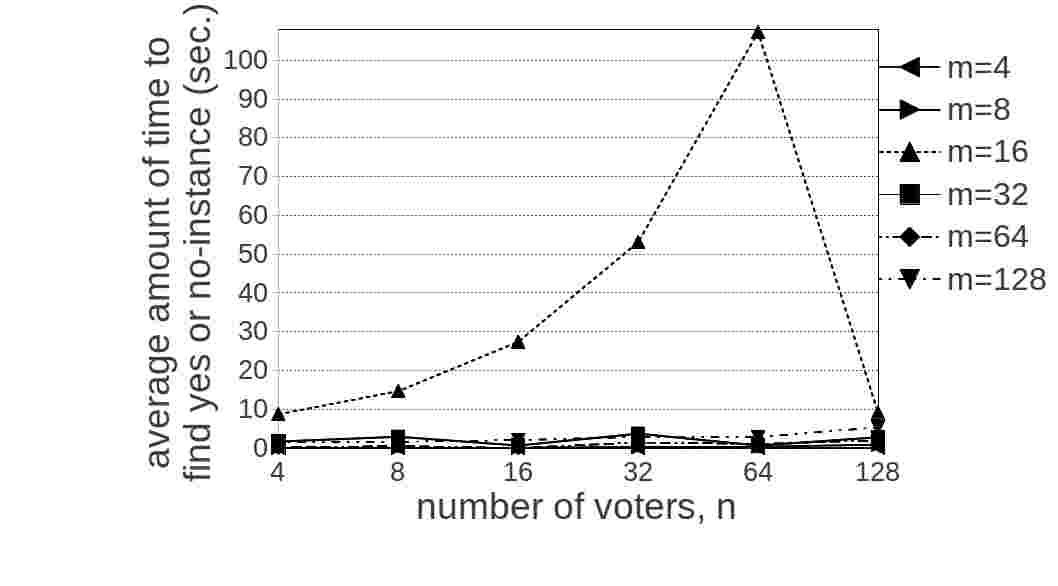}
	\caption{Average time the algorithm needs to give a definite output for 
	destructive control by adding candidates
	in Bucklin elections in the TM model. The maximum is $107,28$ seconds.}
\end{figure}
\clearpage
\subsection{Constructive Control by Deleting Candidates}

\begin{center}
\begin{figure}[ht]
\centering
	\includegraphics[scale=0.3]{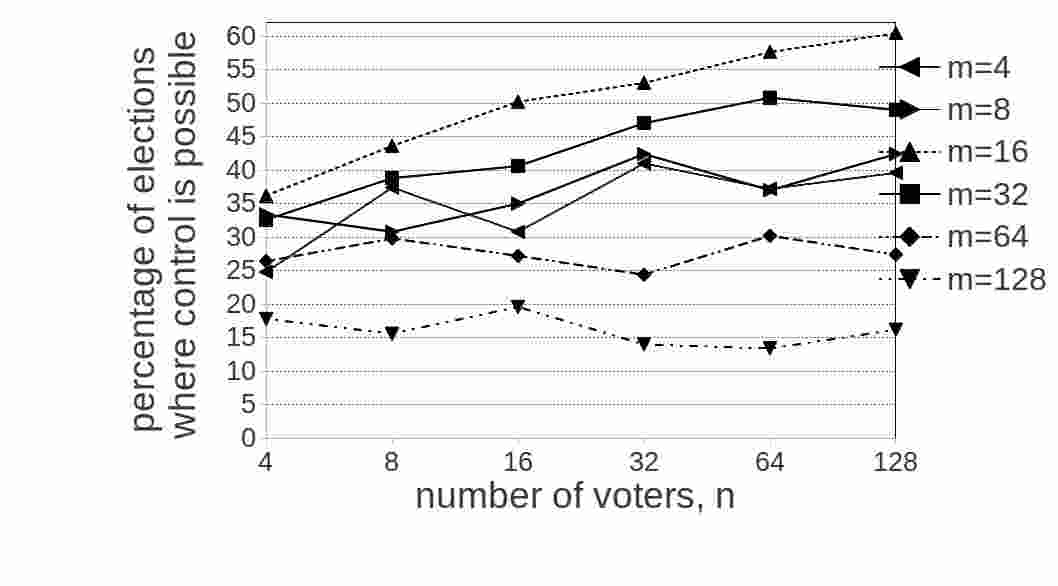}
		\caption{Results for Bucklin voting in the IC model for 
constructive control by deleting candidates. Number of candidates is fixed. }
\end{figure}


\newpage
\begin{figure}[ht]
\centering
	\includegraphics[scale=0.3]{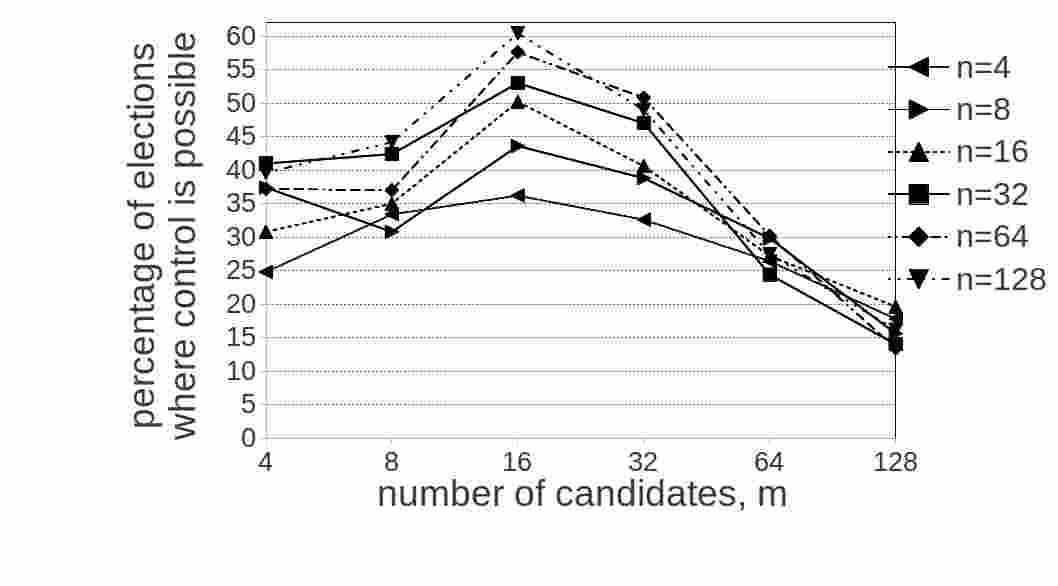}
		\caption{Results for Bucklin voting in the IC model for 
constructive control by deleting candidates. Number of voters is fixed. }
\end{figure}

%

\newpage
\begin{figure}[ht]
\centering
	\includegraphics[scale=0.3]{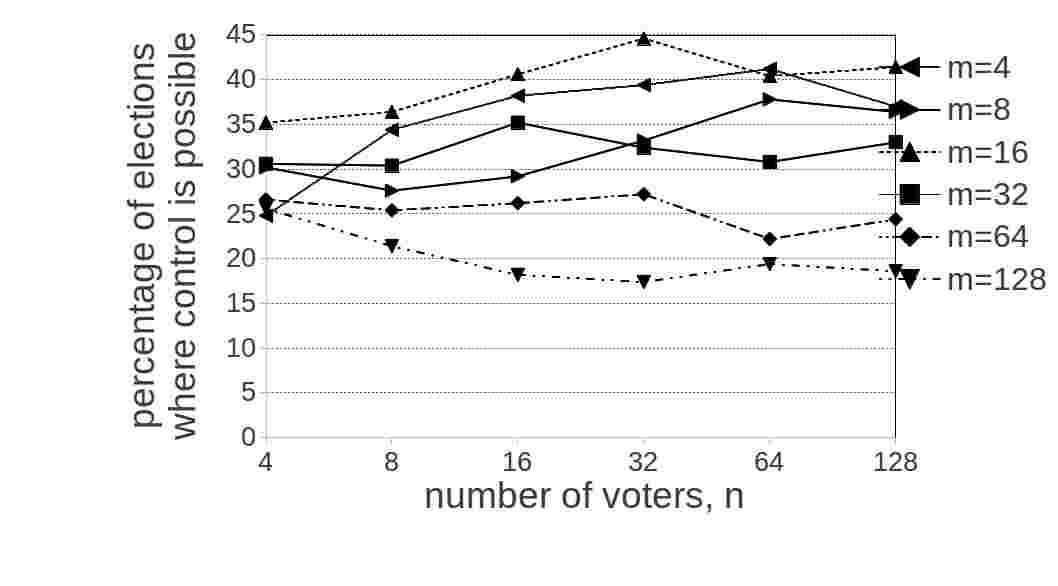}
	\caption{Results for Bucklin voting in the TM model for 
constructive control by deleting candidates. Number of candidates is fixed. }
\end{figure}

%

\newpage
\begin{figure}[ht]
\centering
	\includegraphics[scale=0.3]{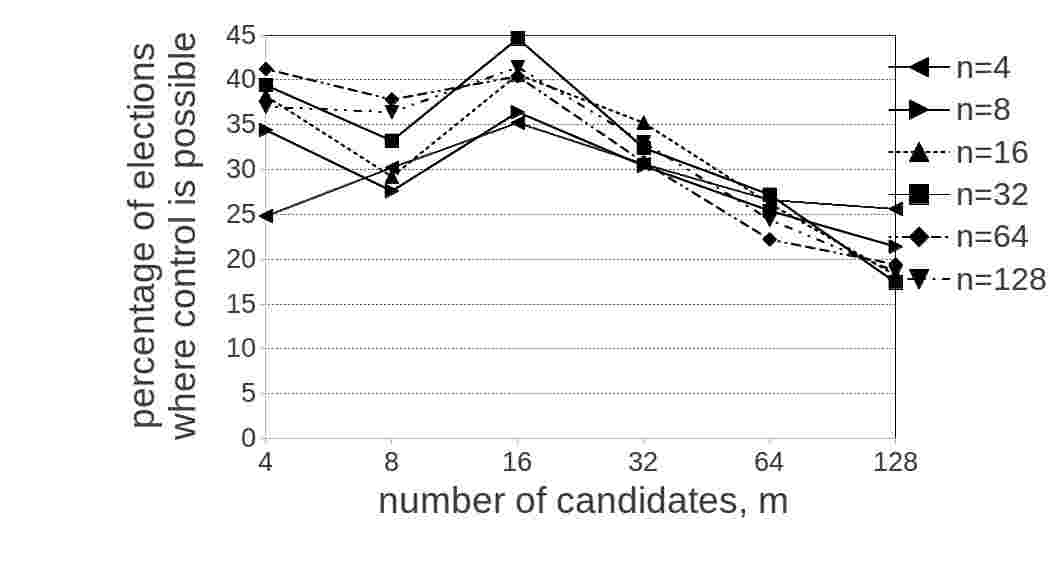}
	\caption{Results for Bucklin voting in the TM model for 
constructive control by deleting candidates. Number of voters is fixed. }
\end{figure}

%
\end{center}

\clearpage
\subsubsection{Computational Costs}
\begin{figure}[ht]
\centering
	\includegraphics[scale=0.3]{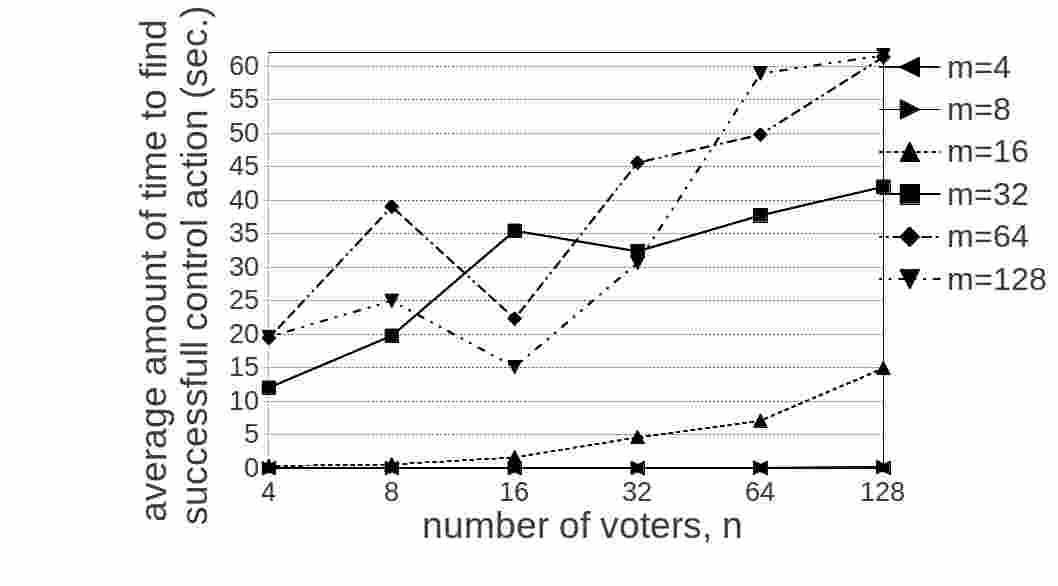}
	\caption{Average time the algorithm needs to find a successful control action for 
	constructive control by deleting candidates
	in Bucklin elections in the IC model. The maximum is $61,64$ seconds.}
\end{figure}
\begin{figure}[ht]
\centering
	\includegraphics[scale=0.3]{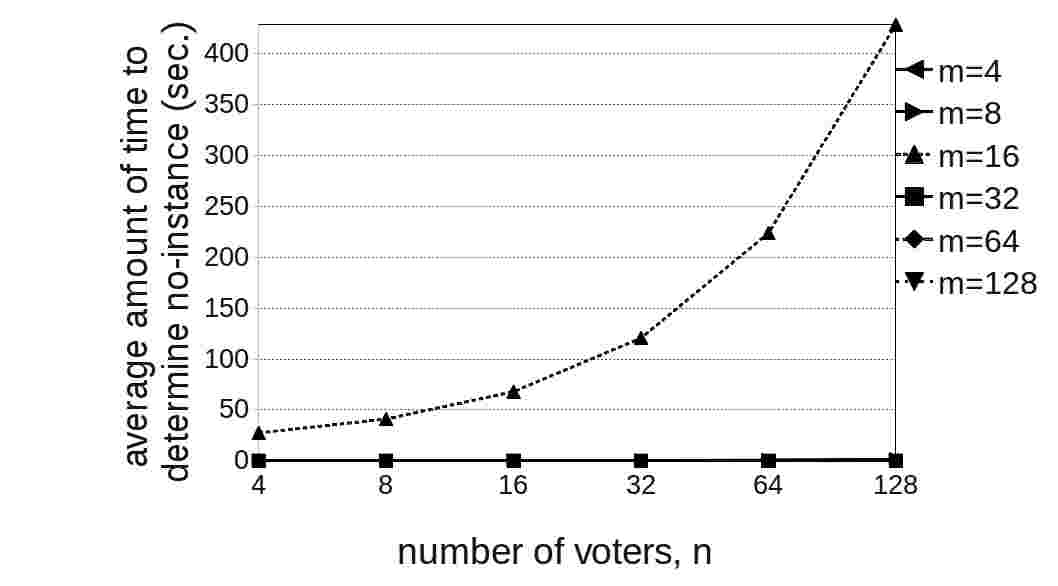}
	\caption{Average time the algorithm needs to determine no-instance of 
		constructive control by deleting candidates
	in Bucklin elections in the IC model. The maximum is $428,06$ seconds.}
\end{figure}
\begin{figure}[ht]
\centering
	\includegraphics[scale=0.3]{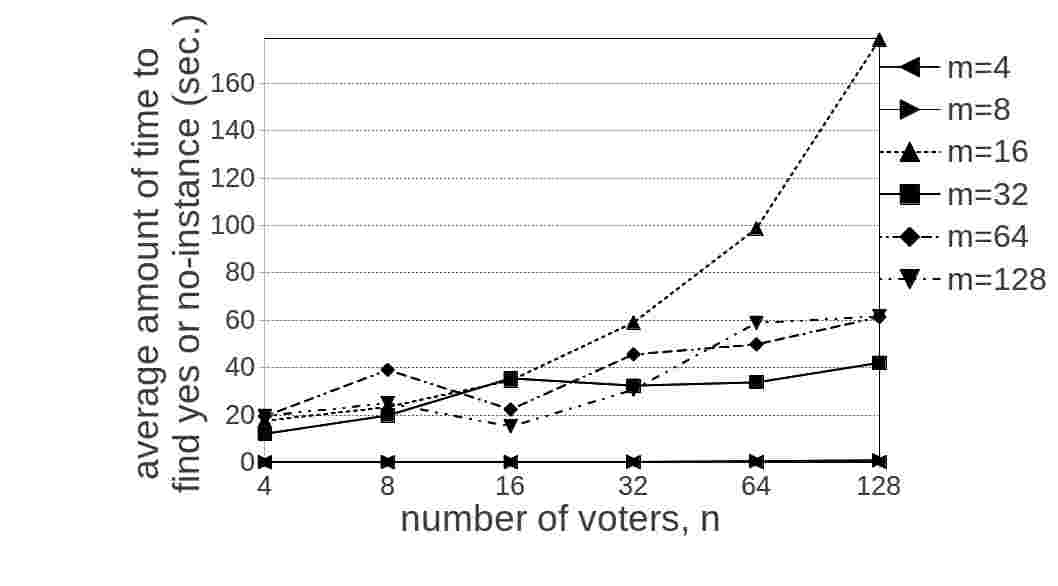}
	\caption{Average time the algorithm needs to give a definite output for 
	constructive control by deleting candidates
	in Bucklin elections in the IC model. The maximum is $178,54$ seconds.}
\end{figure}

\begin{figure}[ht]
\centering
	\includegraphics[scale=0.3]{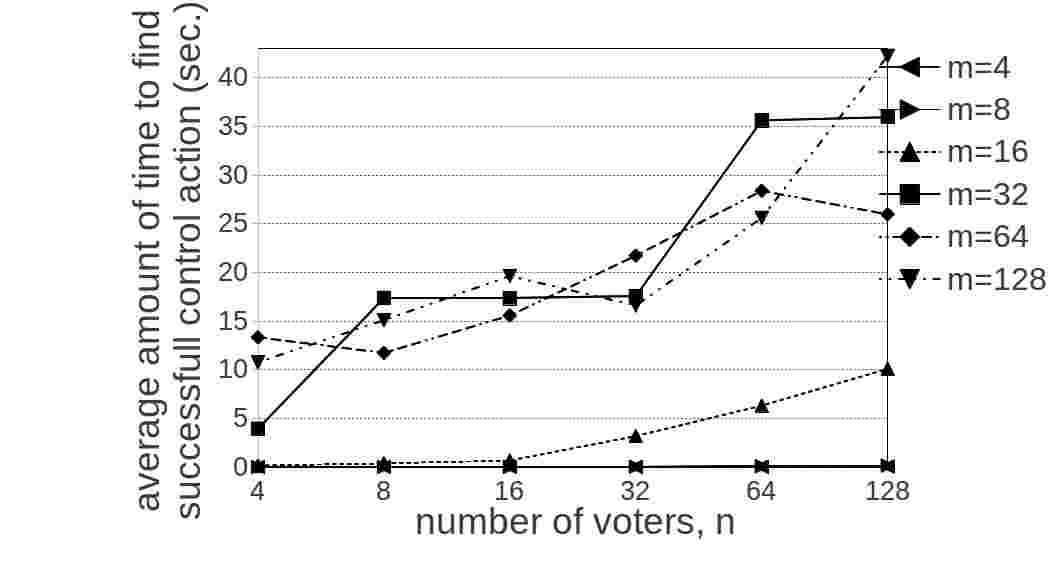}
	\caption{Average time the algorithm needs to find a successful control action for 
	constructive control by deleting candidates
	in Bucklin elections in the TM model. The maximum is $42,25$ seconds.}
\end{figure}
\begin{figure}[ht]
\centering
	\includegraphics[scale=0.3]{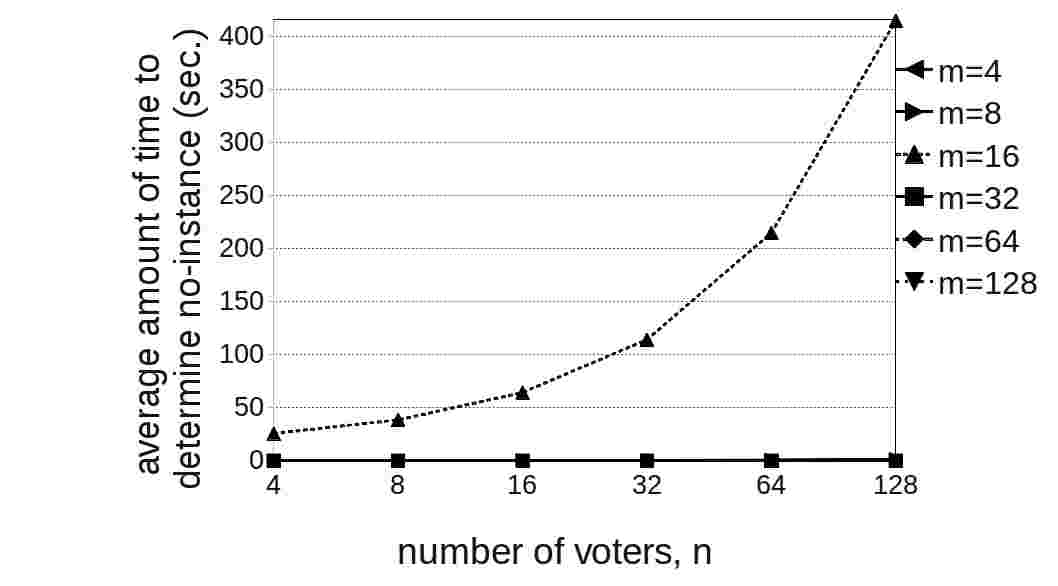}
	\caption{Average time the algorithm needs to determine no-instance of 
		constructive control by deleting candidates
	in Bucklin elections in the TM model. The maximum is $415,02$ seconds.}
\end{figure}
\begin{figure}[ht]
\centering
	\includegraphics[scale=0.3]{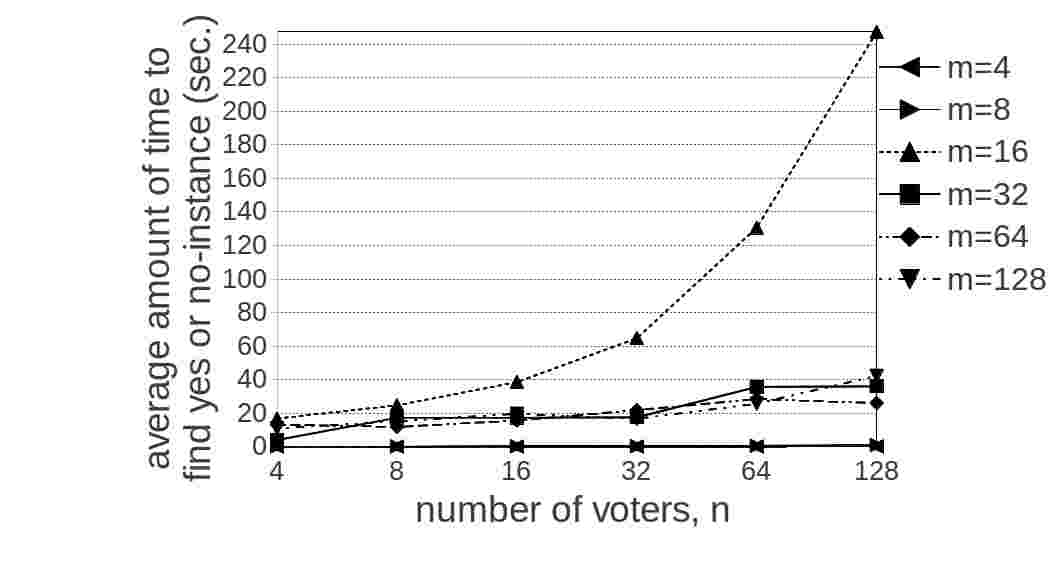}
	\caption{Average time the algorithm needs to give a definite output for 
	constructive control by deleting candidates
	in Bucklin elections in the TM model. The maximum is $247,38$ seconds.}
\end{figure}
\clearpage
\subsection{Destructive Control by Deleting Candidates}
\begin{center}
\begin{figure}[ht]
\centering
	\includegraphics[scale=0.3]{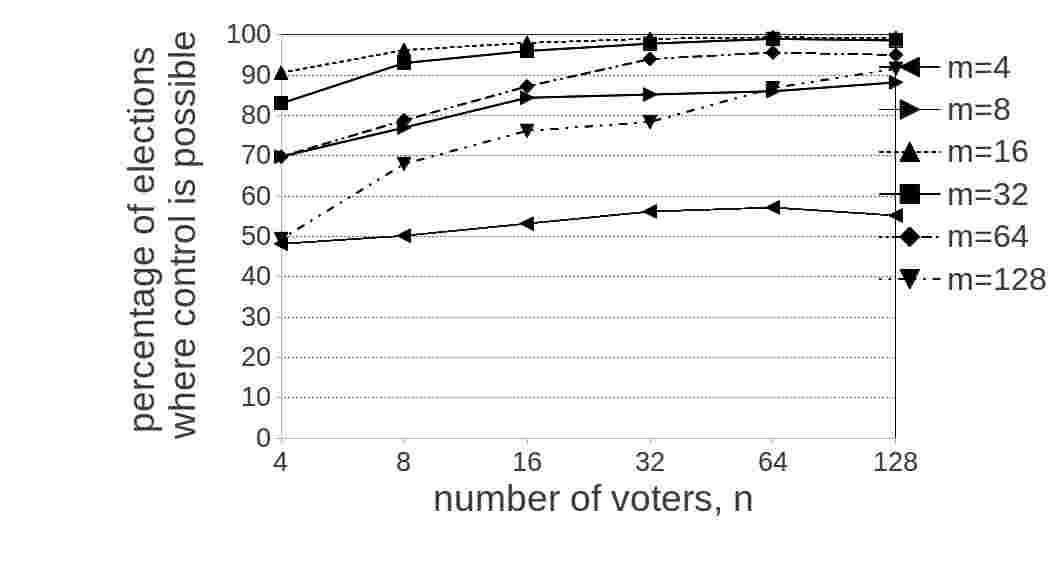}
		\caption{Results for Bucklin voting in the IC model for 
destructive control by deleting candidates. Number of candidates is fixed. }
\end{figure}


\newpage
\begin{figure}[ht]
\centering
	\includegraphics[scale=0.3]{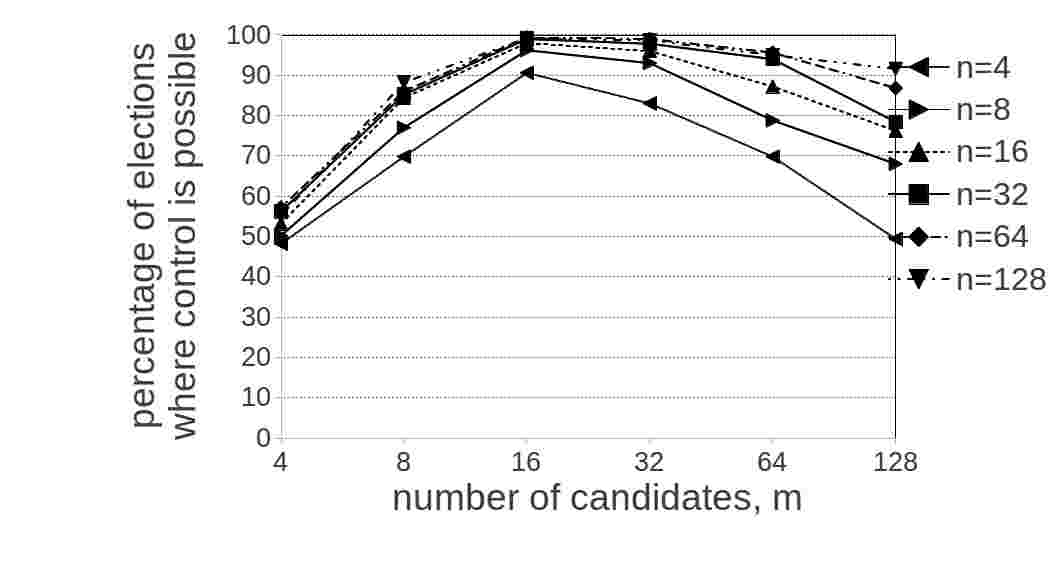}
		\caption{Results for Bucklin voting in the IC model for 
destructive control by deleting candidates. Number of voters is fixed. }
\end{figure}

%
\newpage
\begin{figure}[ht]
\centering
	\includegraphics[scale=0.3]{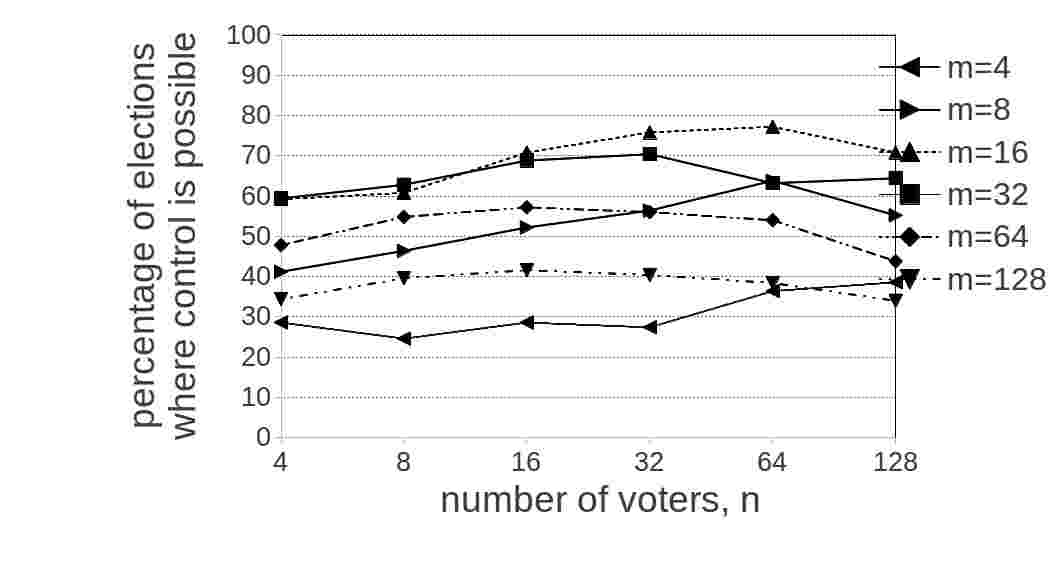}
	\caption{Results for Bucklin voting in the TM model for 
destructive control by deleting candidates. Number of candidates is fixed. }
\end{figure}

%
\newpage
\begin{figure}[ht]
\centering
	\includegraphics[scale=0.3]{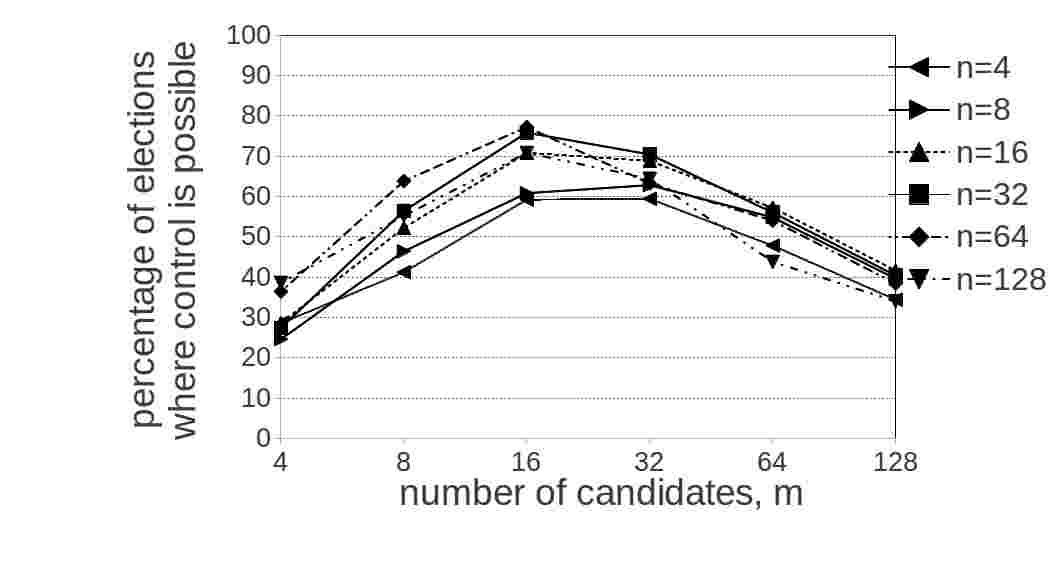}
	\caption{Results for Bucklin voting in the TM model for 
destructive control by deleting candidates. Number of voters is fixed. }
\end{figure}

%
\end{center}

\clearpage
\subsubsection{Computational Costs}
\begin{figure}[ht]
\centering
	\includegraphics[scale=0.3]{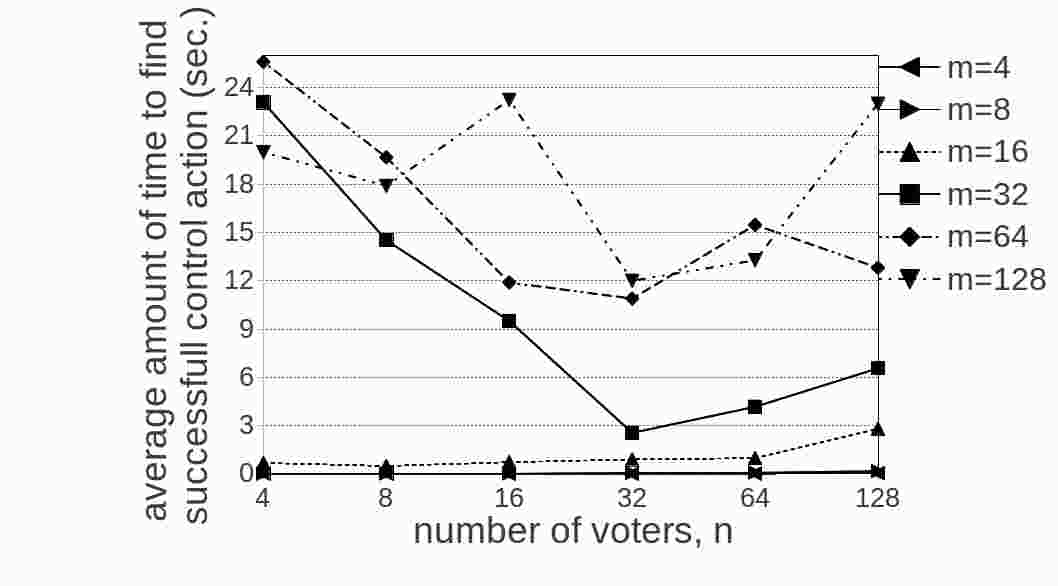}
	\caption{Average time the algorithm needs to find a successful control action for 
	destructive control by deleting candidates
	in Bucklin elections in the IC model. The maximum is $25,59$ seconds.}
\end{figure}
\begin{figure}[ht]
\centering
	\includegraphics[scale=0.3]{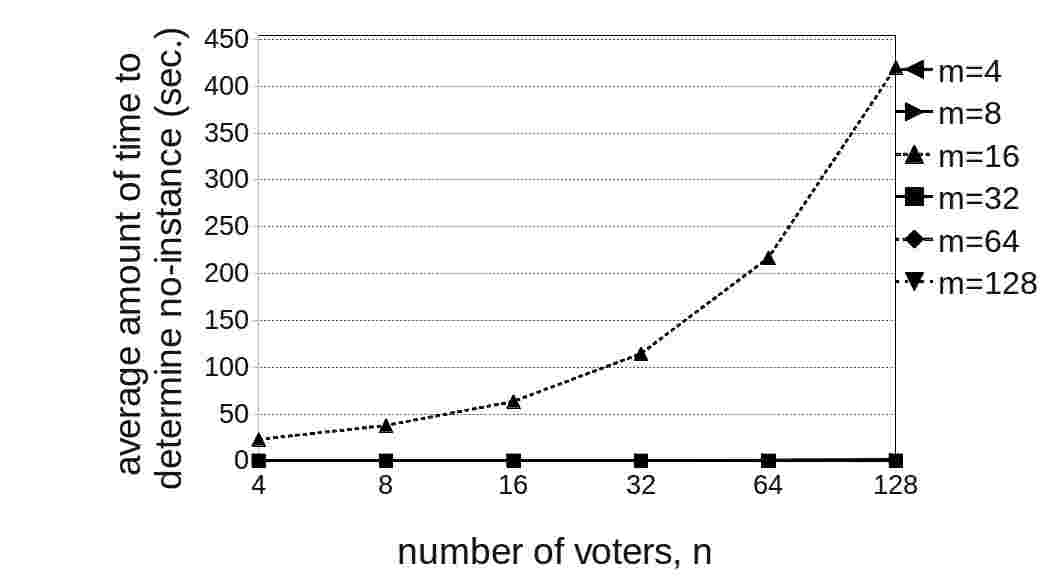}
	\caption{Average time the algorithm needs to determine no-instance of 
		destructive control by deleting candidates
	in Bucklin elections in the IC model. The maximum is $420,05$   seconds.}
\end{figure}
\begin{figure}[ht]
\centering
	\includegraphics[scale=0.3]{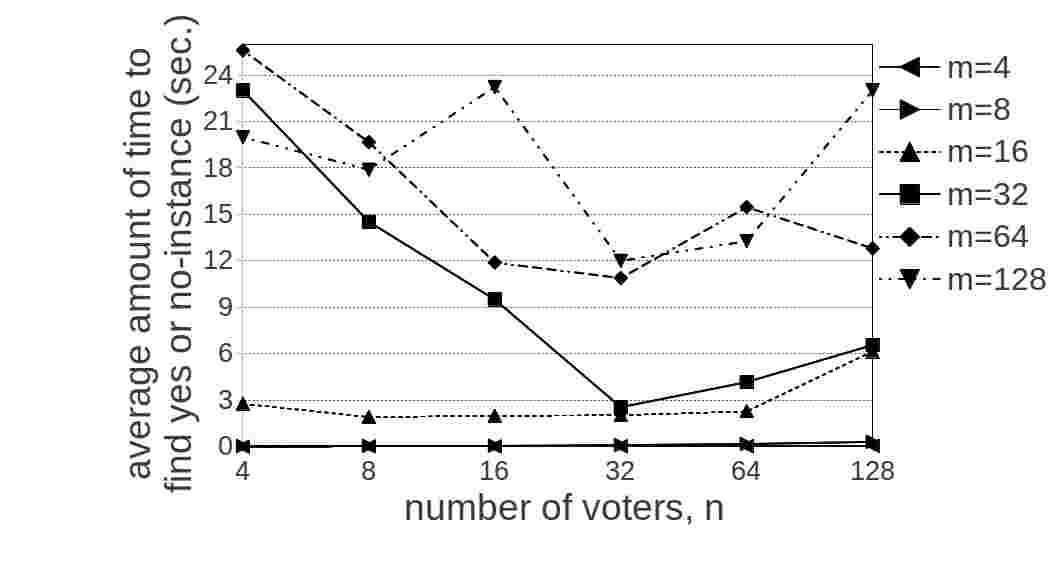}
	\caption{Average time the algorithm needs to give a definite output for 
	destructive control by deleting  candidates
	in Bucklin elections in the IC model. The maximum is $25,59$ seconds.}
\end{figure}
\begin{figure}[ht]
\centering
	\includegraphics[scale=0.3]{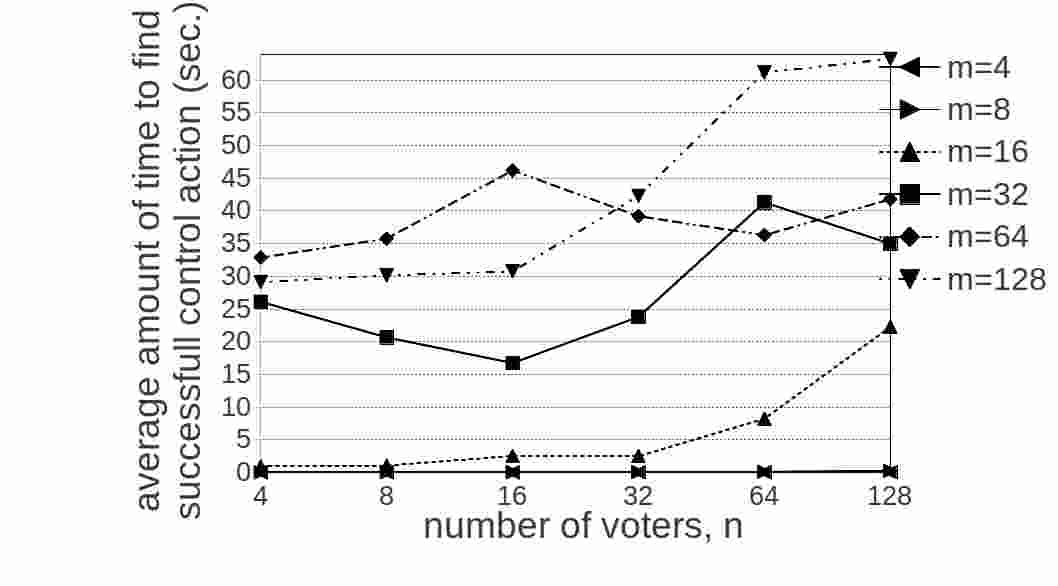}
	\caption{Average time the algorithm needs to find a successful control action for 
	destructive control by deleting candidates
	in Bucklin elections in the TM model. The maximum is $63,26$ seconds.}
\end{figure}
\begin{figure}[ht]
\centering
	\includegraphics[scale=0.3]{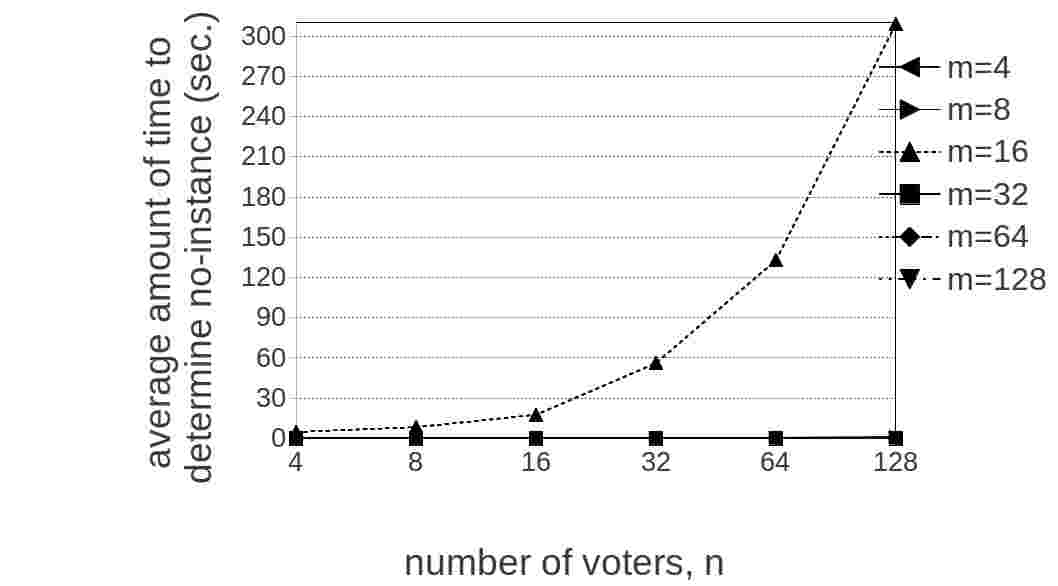}
	\caption{Average time the algorithm needs to determine no-instance of 
		destructive control by deleting candidates
	in Bucklin elections in the TM model. The maximum is $309,1$ seconds.}
\end{figure}
\begin{figure}[ht]
\centering
	\includegraphics[scale=0.3]{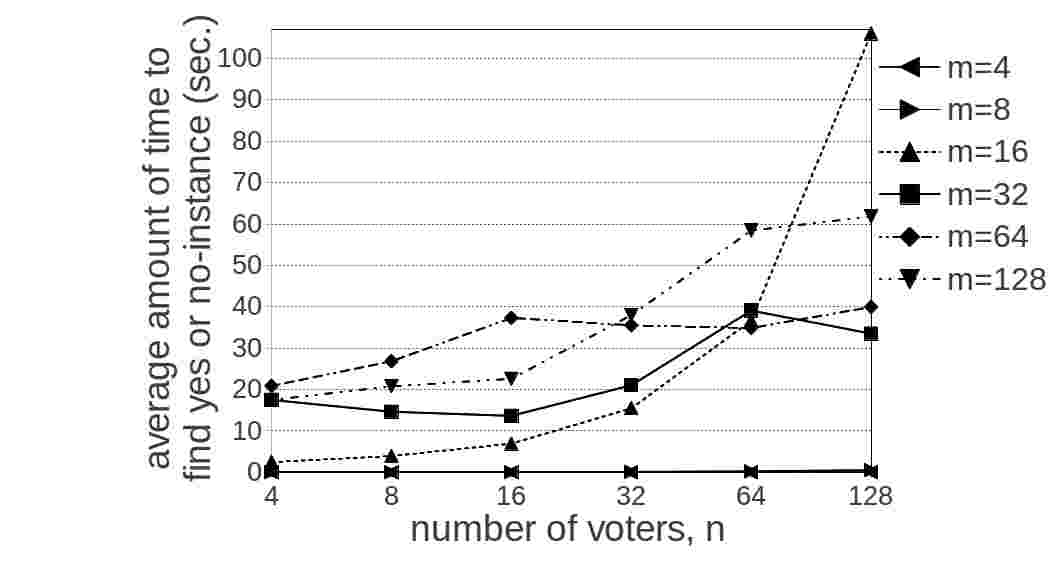}
	\caption{Average time the algorithm needs to give a definite output for 
	destructive control by deleting candidates
	in Bucklin elections in the TM model. The maximum is $106,02$ seconds.}
\end{figure}

\clearpage
\subsection{Constructive Control by Partition of Candidates in Model TE}
\begin{center}
\begin{figure}[ht]
\centering
	\includegraphics[scale=0.3]{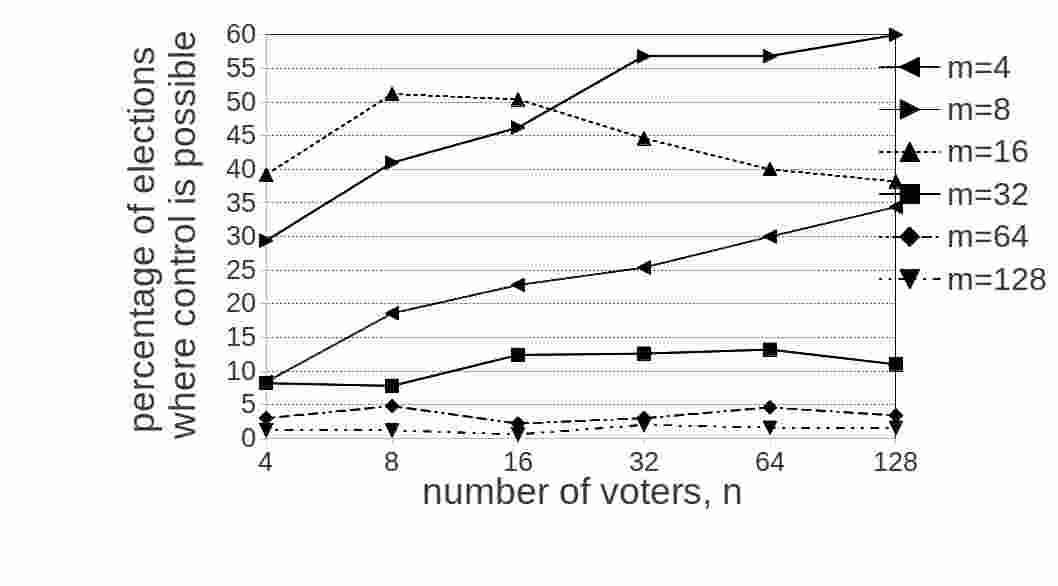}
		\caption{Results for Bucklin voting in the IC model for 
constructive control by partition of candidates in model TE. Number of voters is fixed. }
\end{figure}

\newpage
\begin{figure}[ht]
\centering
	\includegraphics[scale=0.3]{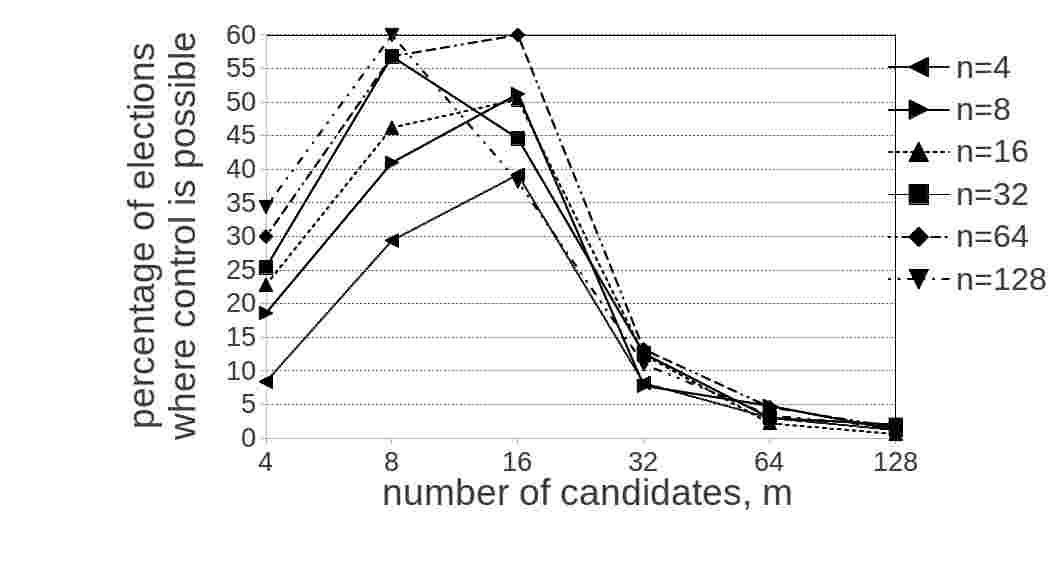}
		\caption{Results for Bucklin voting in the IC model for 
constructive control by partition of candidates in model TE. Number of voters is fixed. }
\end{figure}

%
\newpage
\begin{figure}[ht]
\centering
	\includegraphics[scale=0.3]{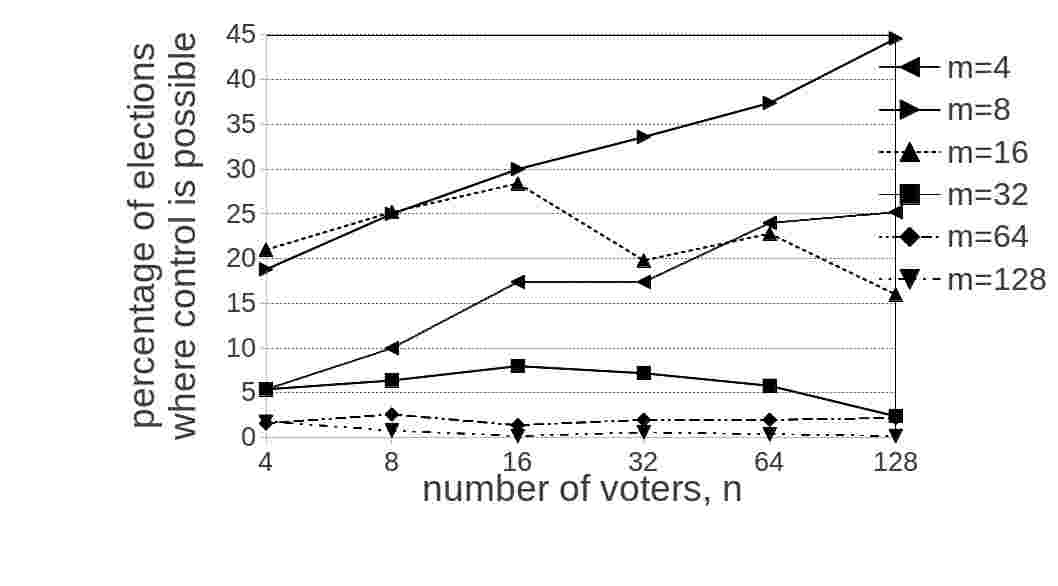}
	\caption{Results for Bucklin voting in the TM model for 
constructive control by partition of candidates in model TE. Number of candidates is fixed. }
\end{figure}

%
\newpage
\begin{figure}[ht]
\centering
	\includegraphics[scale=0.3]{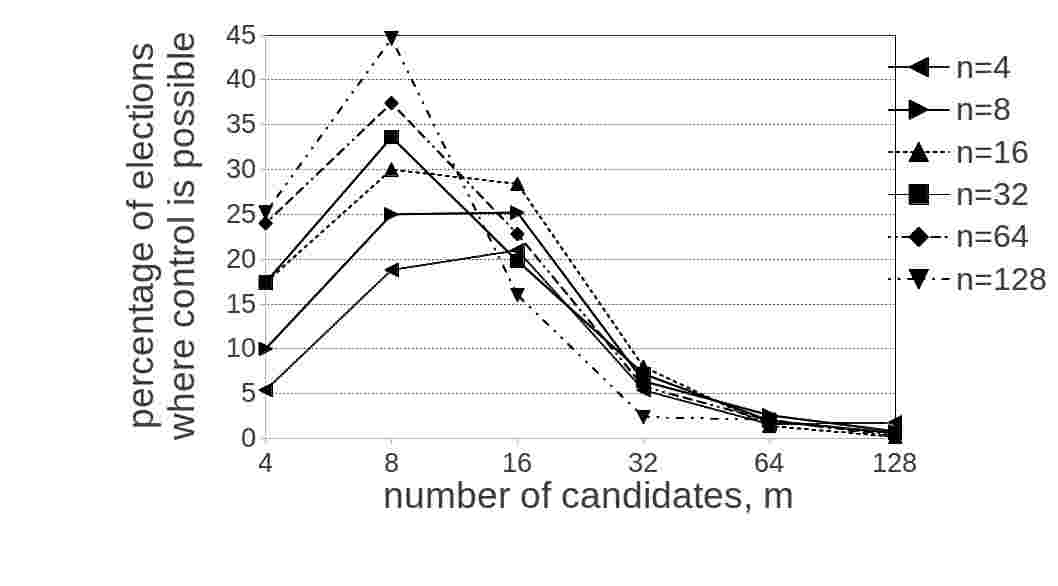}
	\caption{Results for Bucklin voting in the TM model for 
constructive control by partition of candidates in model TE. Number of voters is fixed. }
\end{figure}

%
\end{center}
\clearpage
\subsubsection{Computational Costs}
\begin{figure}[ht]
\centering
	\includegraphics[scale=0.3]{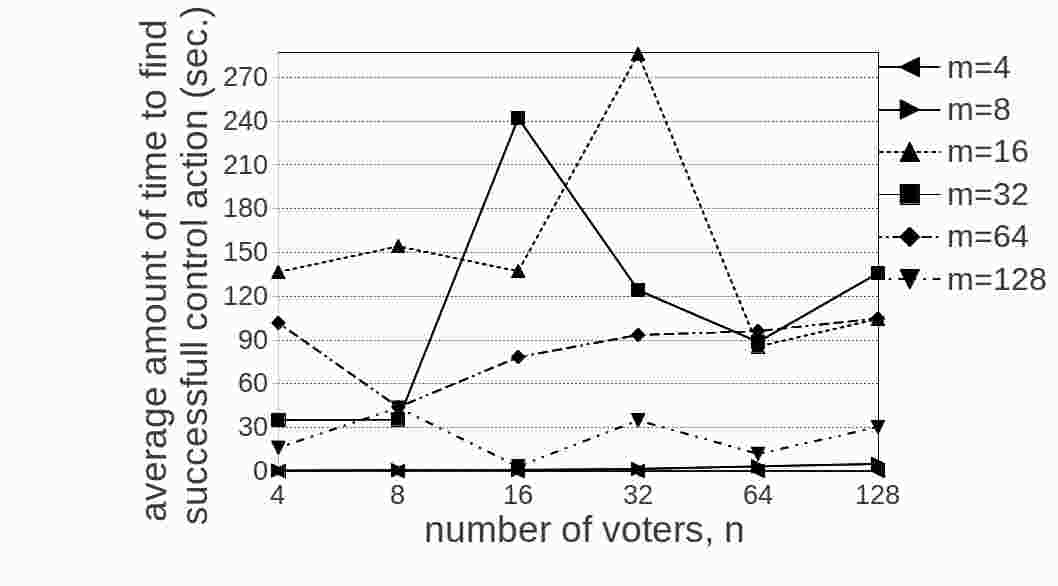}
	\caption{Average time the algorithm needs to find a successful control action for 
	constructive control by partition of candidates in model TE
	in Bucklin elections in the IC model. The maximum is $286,05$ seconds.}
\end{figure}
\begin{figure}[ht]
\centering
	\includegraphics[scale=0.3]{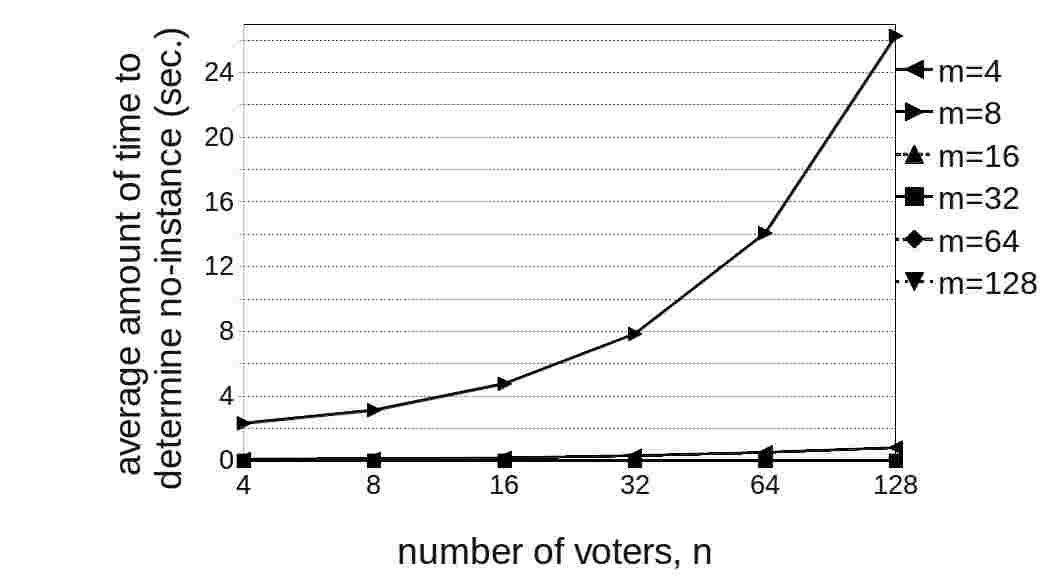}
	\caption{Average time the algorithm needs to determine no-instance of 
		constructive control by partition of candidates in model TE
	in Bucklin elections in the IC model. The maximum is $26,26$ seconds.}
\end{figure}
\begin{figure}[ht]
\centering
	\includegraphics[scale=0.3]{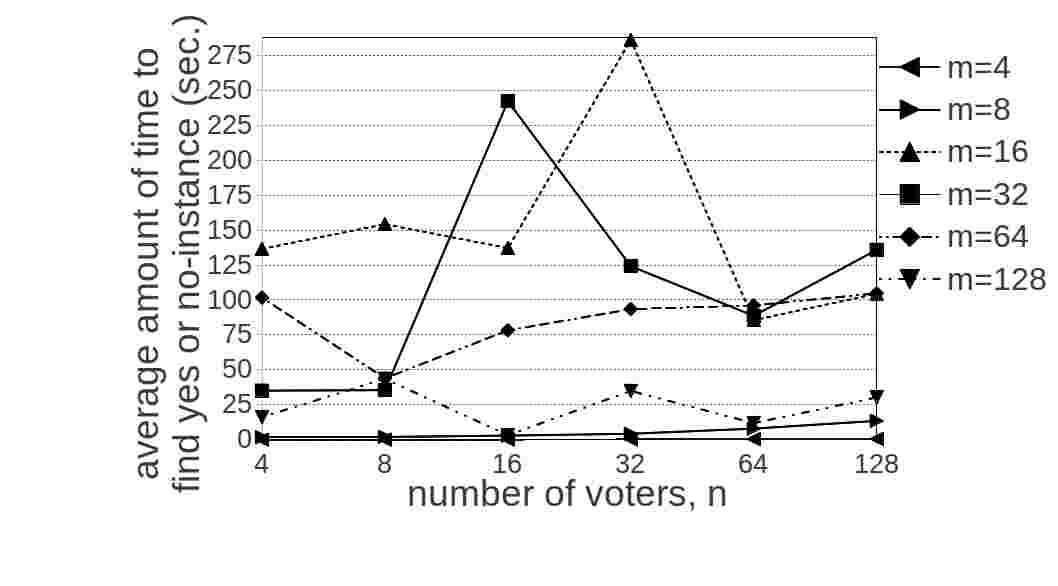}
	\caption{Average time the algorithm needs to give a definite output for 
	constructive control by partition of candidates in model TE
	in Bucklin elections in the IC model. The maximum is $286,05$ seconds.}
\end{figure}
\begin{figure}[ht]
\centering
	\includegraphics[scale=0.3]{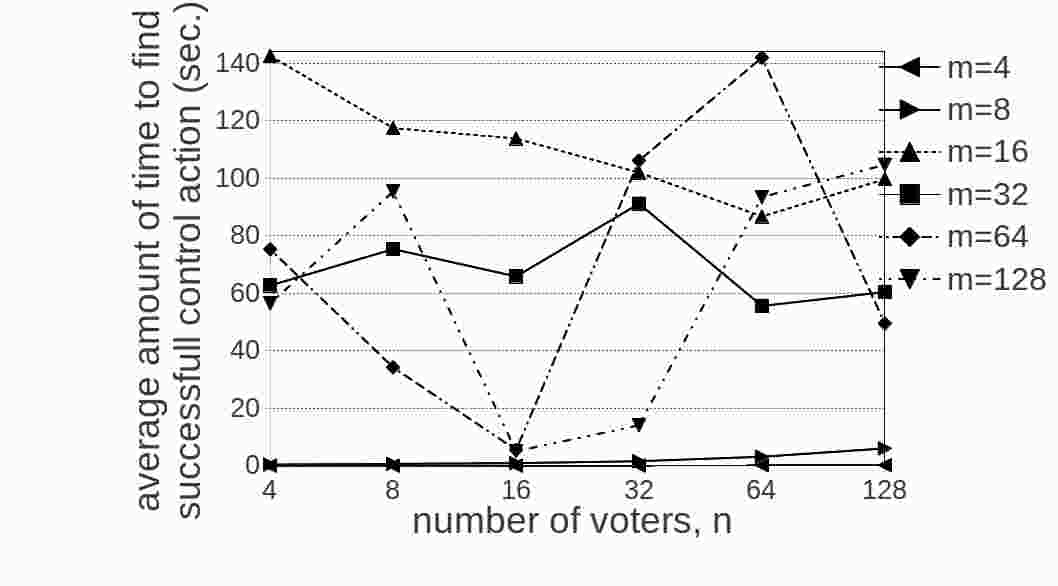}
	\caption{Average time the algorithm needs to find a successful control action for 
	constructive control by partition of candidates in model TE
	in Bucklin elections in the TM model. The maximum is $142,58$ seconds.}
\end{figure}
\begin{figure}[ht]
\centering
	\includegraphics[scale=0.3]{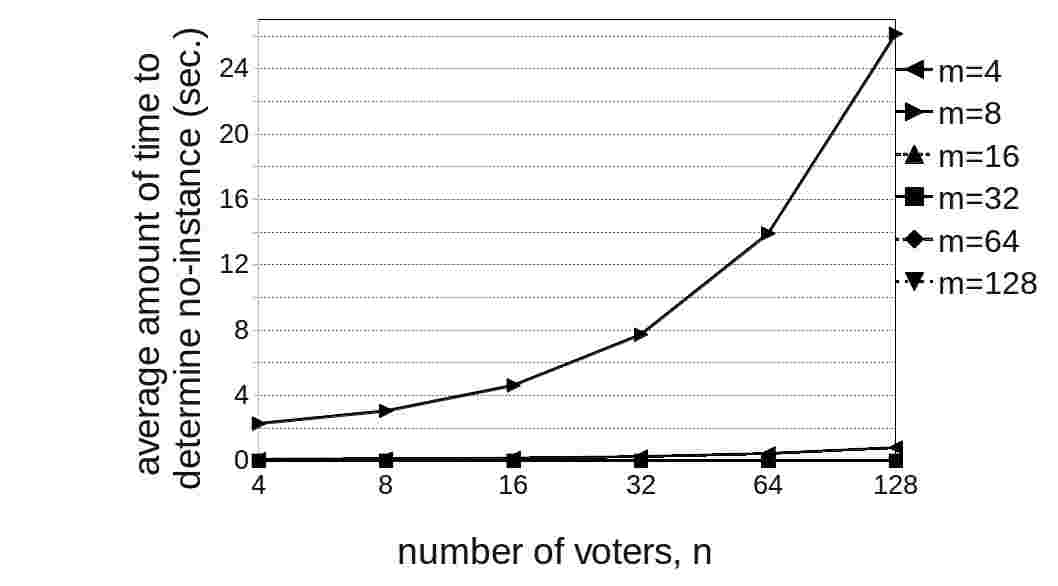}
	\caption{Average time the algorithm needs to determine no-instance of 
		constructive control by partition of candidates in model TE
	in Bucklin elections in the TM model. The maximum is $26,14$ seconds.}
\end{figure}

\begin{figure}[ht]
\centering
	\includegraphics[scale=0.3]{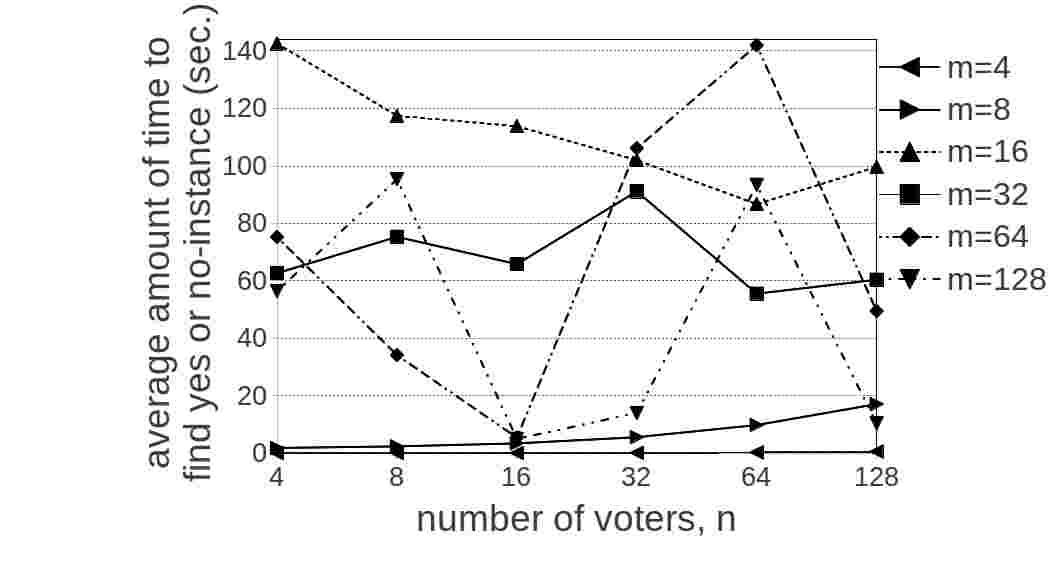}
	\caption{Average time the algorithm needs to give a definite output for 
	constructive control by partition of candidates in model TE
	in Bucklin elections in the TM model. The maximum is $142,58$ seconds.}
\end{figure}

\clearpage
\subsection{Destructive Control by Partition of Candidates in Model TE}
\begin{center}
\begin{figure}[ht]
\centering
	\includegraphics[scale=0.3]{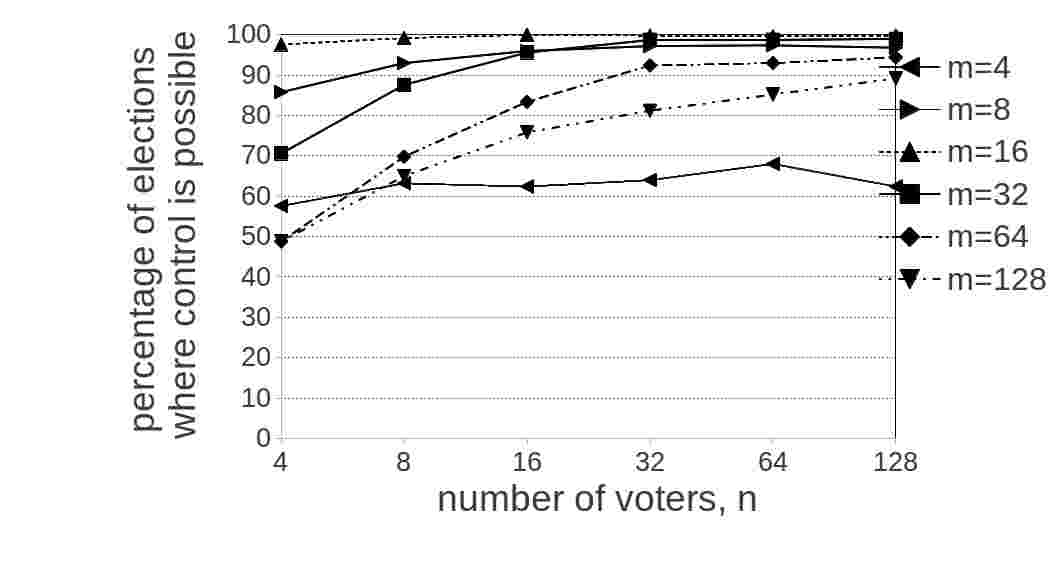}
		\caption{Results for Bucklin voting in the IC model for 
destructive control by partition of candidates in model TE. Number of candidates is fixed. }
\end{figure}


\newpage
\begin{figure}[ht]
\centering
	\includegraphics[scale=0.3]{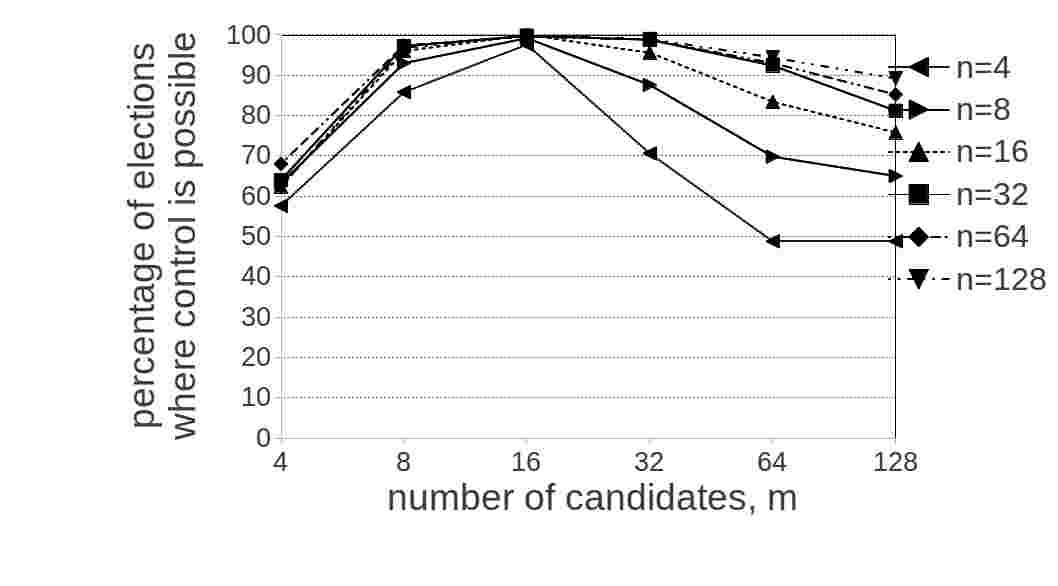}
		\caption{Results for Bucklin voting in the IC model for 
destructive control by partition of candidates in model TE. Number of voters is fixed. }
\end{figure}

%

\newpage
\begin{figure}[ht]
\centering
	\includegraphics[scale=0.3]{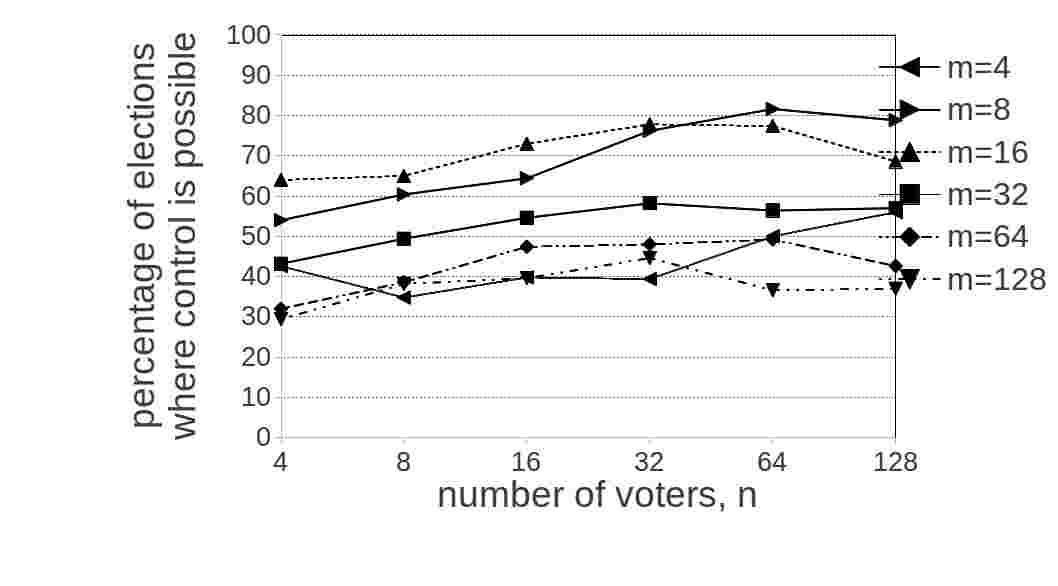}
	\caption{Results for Bucklin voting in the TM model for 
destructive control by partition of candidates in model TE. Number of candidates is fixed. }
\end{figure}
%
\newpage
\begin{figure}[ht]
\centering
	\includegraphics[scale=0.3]{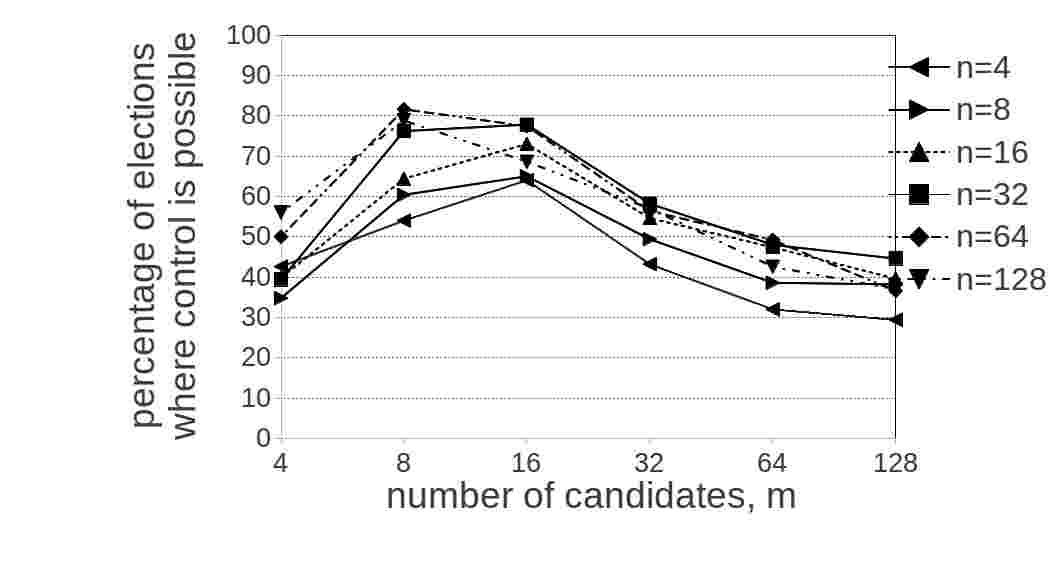}
	\caption{Results for Bucklin voting in the TM model for 
destructive control by partition of candidates in model TE. Number of voters is fixed. }
\end{figure}
%
\end{center} 

\clearpage
\subsubsection{Computational Costs}
\begin{figure}[ht]
\centering
	\includegraphics[scale=0.3]{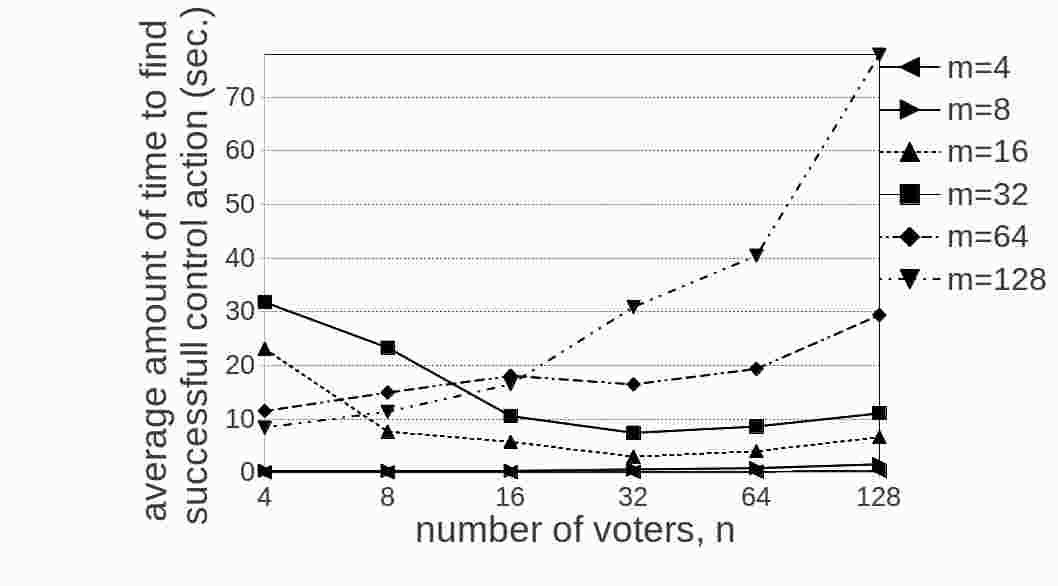}
	\caption{Average time the algorithm needs to find a successful control action for 
	destructive control by partition of candidates in model TE
	in Bucklin elections in the IC model. The maximum is $36,79$ seconds.}
\end{figure}
\begin{figure}[ht]
\centering
	\includegraphics[scale=0.3]{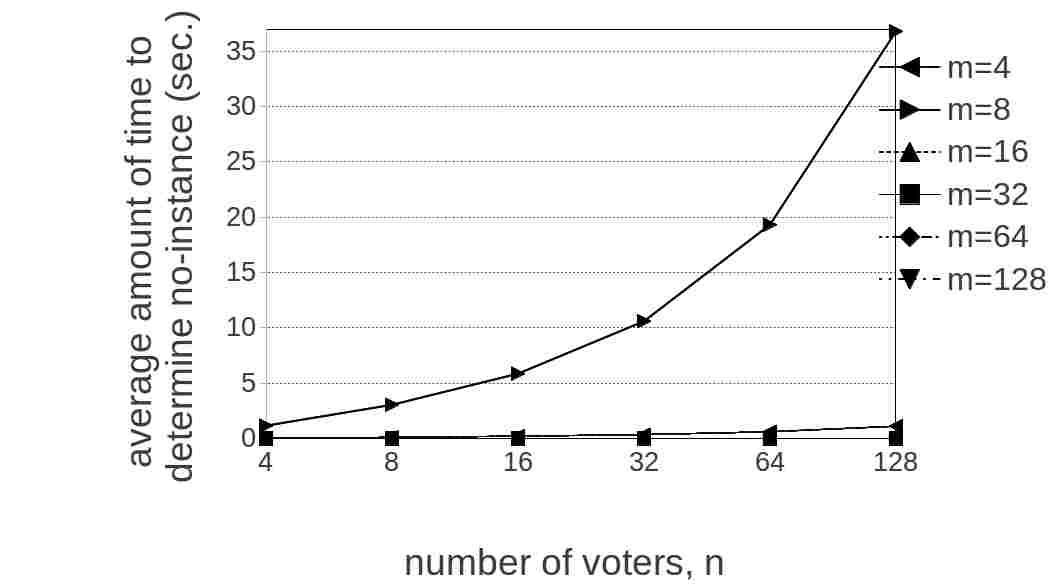}
	\caption{Average time the algorithm needs to determine no-instance of 
		destructive control by partition of candidates in model TE
	in Bucklin elections in the IC model. The maximum is $77,9$ seconds.}
\end{figure}
\begin{figure}[ht]
\centering
	\includegraphics[scale=0.3]{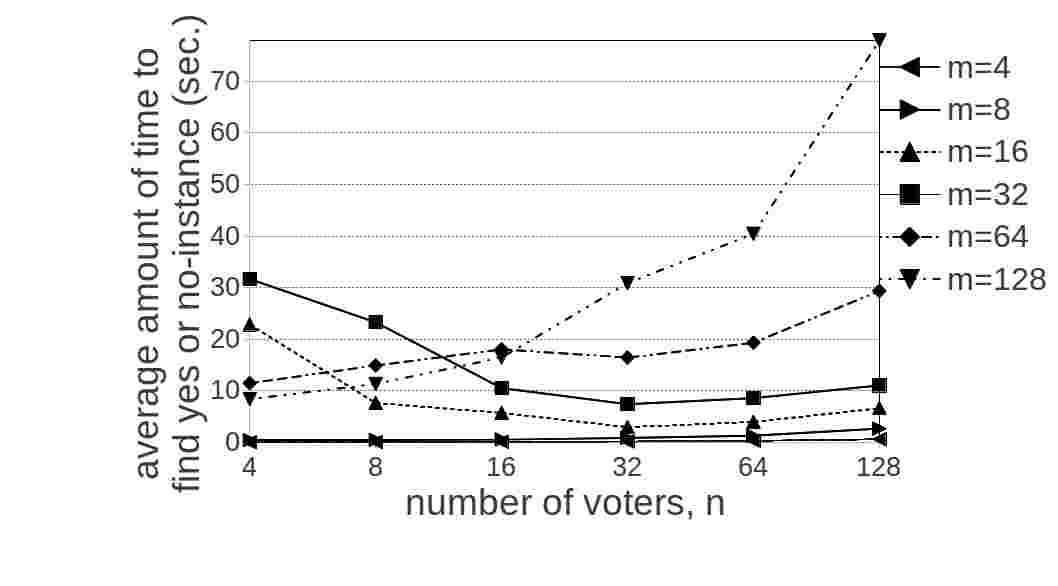}
	\caption{Average time the algorithm needs to give a definite output for 
	destructive control by partition of candidates in model TE
	in Bucklin elections in the IC model. The maximum is $77,9$ seconds.}
\end{figure}

\begin{figure}[ht]
\centering
	\includegraphics[scale=0.3]{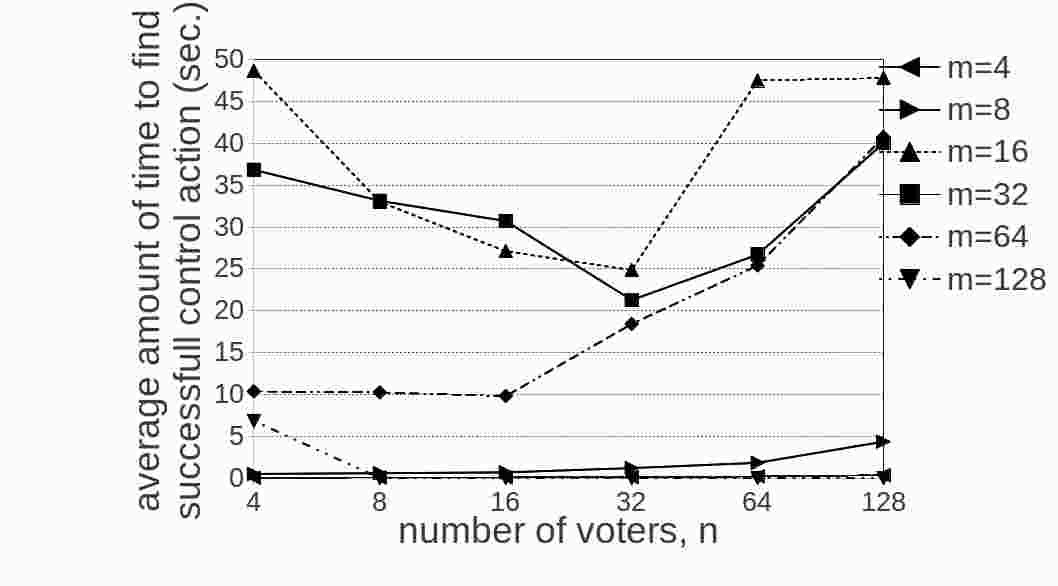}
	\caption{Average time the algorithm needs to find a successful control action for 
	destructive control by partition of candidates in model TE
	in Bucklin elections in the TM model. The maximum is $48,67$ seconds.}
\end{figure}
\begin{figure}[ht]
\centering
	\includegraphics[scale=0.3]{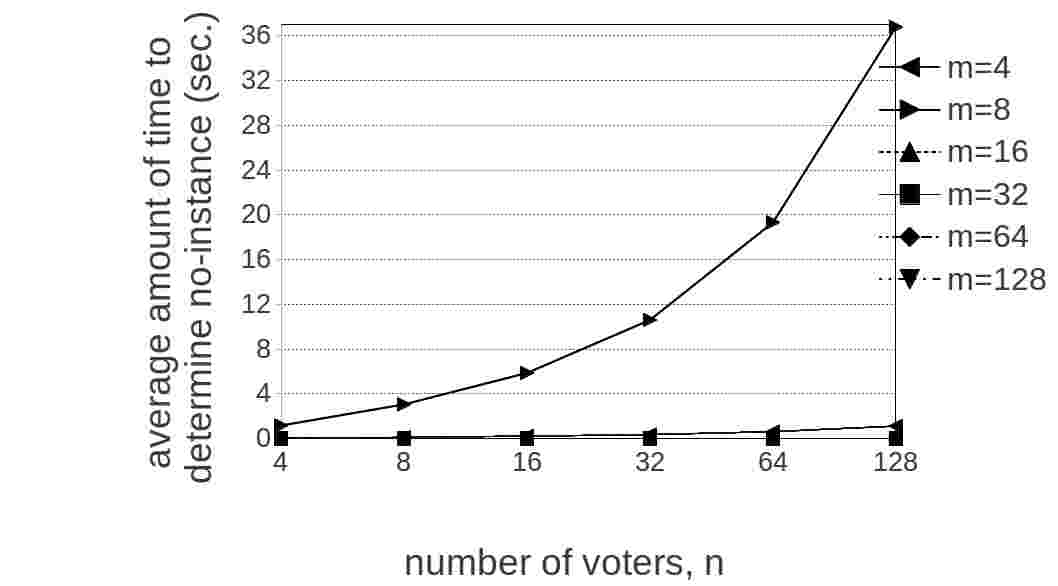}
	\caption{Average time the algorithm needs to determine no-instance of 
		destructive control by partition of candidates in model TE
	in Bucklin elections in the TM model. The maximum is $36,79$ seconds.}
\end{figure}
\begin{figure}[ht]
\centering
	\includegraphics[scale=0.3]{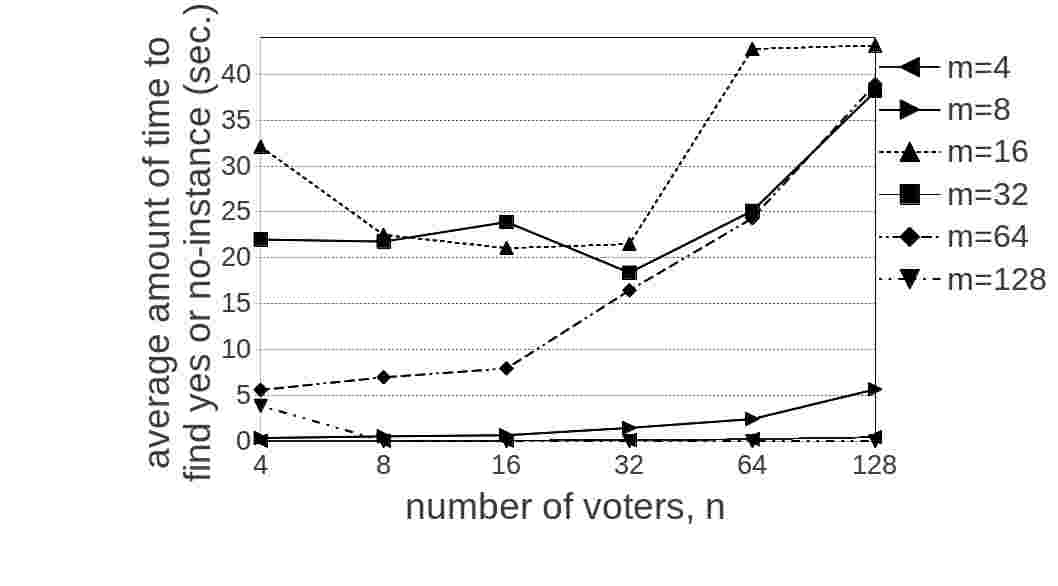}
	\caption{Average time the algorithm needs to give a definite output for 
	destructive control by partition of candidates in model TE
	in Bucklin elections in the TM model. The maximum is $43,14$ seconds.}
\end{figure}

\clearpage
\subsection{Constructive Control by Partition of Candidates in Model TP}
\begin{center}
\begin{figure}[ht]
\centering
	\includegraphics[scale=0.3]{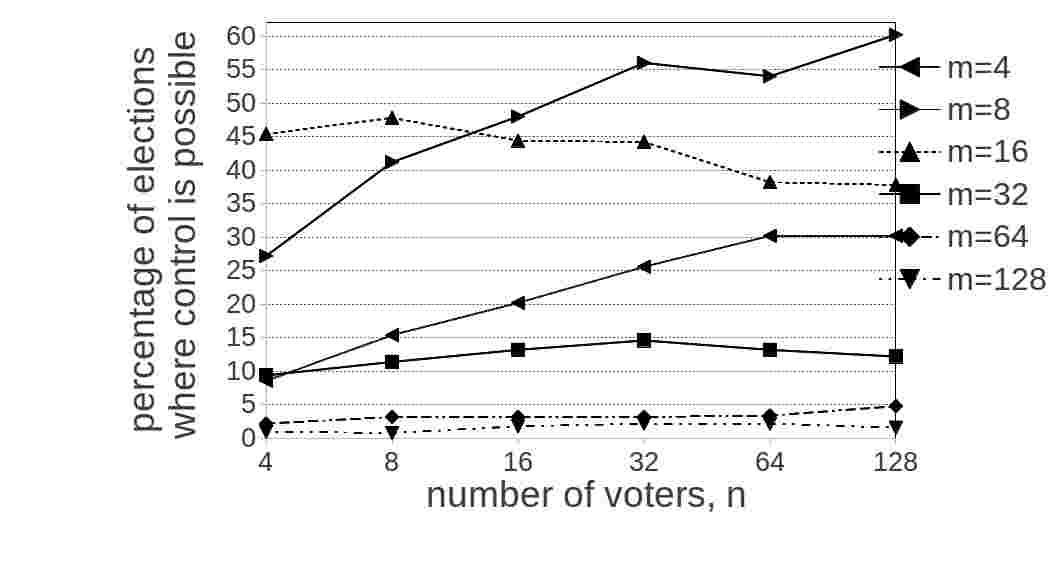}
		\caption{Results for Bucklin voting in the IC model for 
constructive control by partition of candidates in model TP. Number of candidates is fixed. }
\end{figure}

\newpage
\begin{figure}[ht]
\centering
	\includegraphics[scale=0.3]{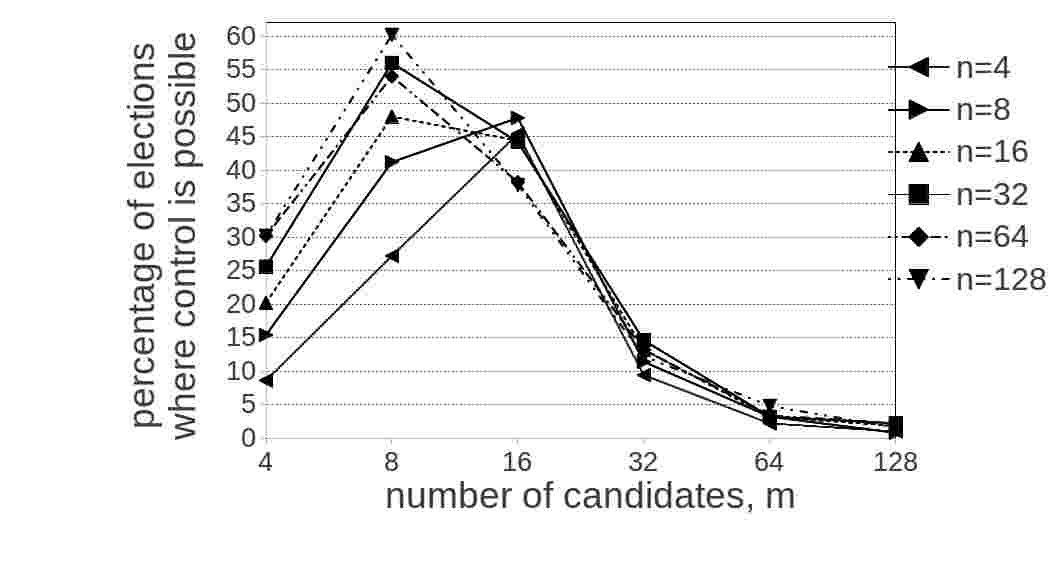}
		\caption{Results for Bucklin voting in the IC model for 
constructive control by partition of candidates in model TP. Number of voters is fixed. }
\end{figure}
%
\newpage
\begin{figure}[ht]
\centering
	\includegraphics[scale=0.3]{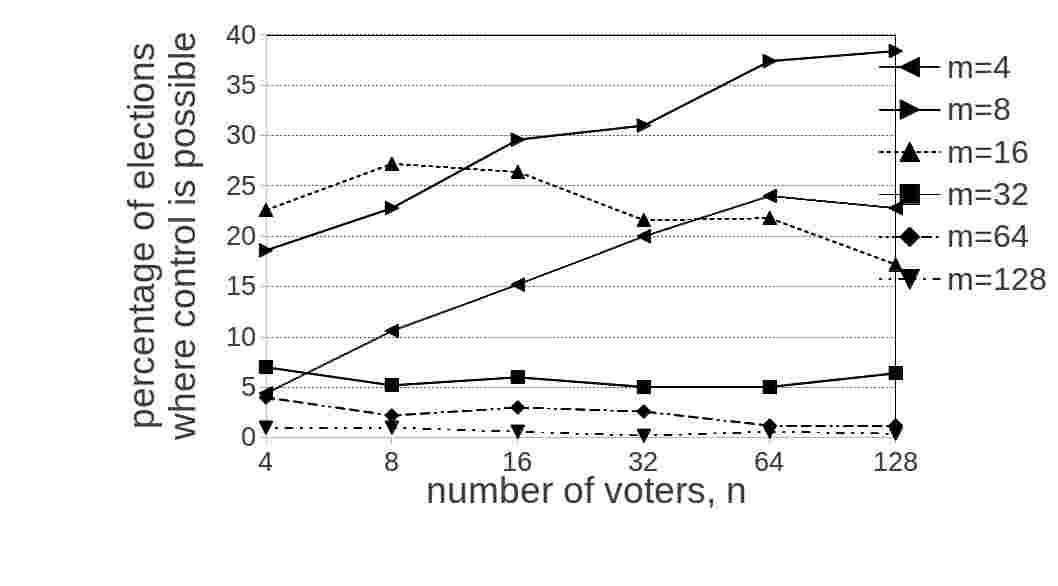}
	\caption{Results for Bucklin voting in the TM model for 
constructive control by partition of candidates in model TP. Number of candidates is fixed. }
\end{figure}

%
\newpage
\begin{figure}[ht]
\centering
	\includegraphics[scale=0.3]{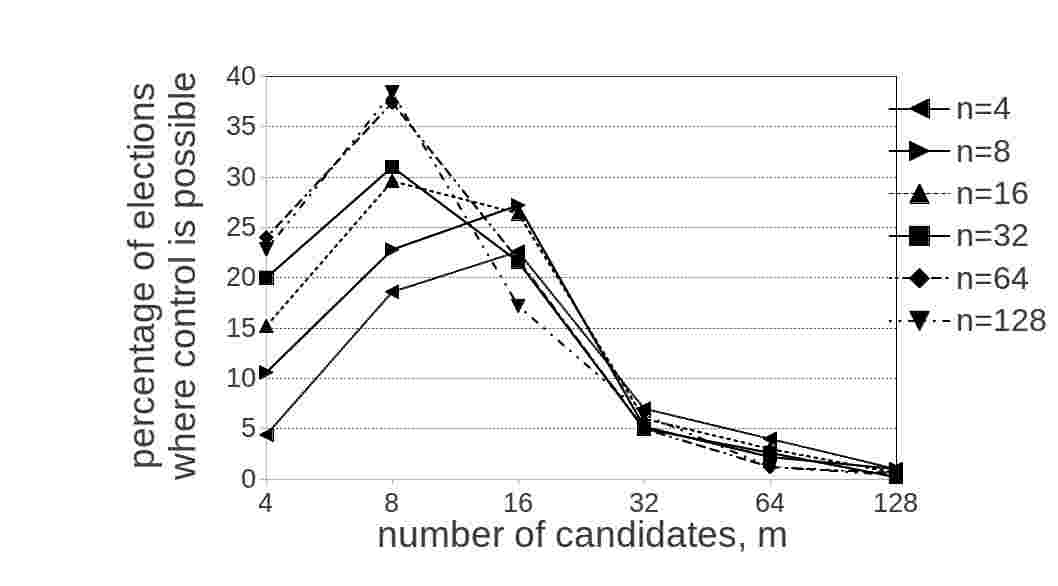}
	\caption{Results for Bucklin voting in the TM model for 
constructive control by partition of candidates in model TP. Number of voters is fixed. }
\end{figure}
%
\end{center}

\clearpage
\subsubsection{Computational Costs}
\begin{figure}[ht]
\centering
	\includegraphics[scale=0.3]{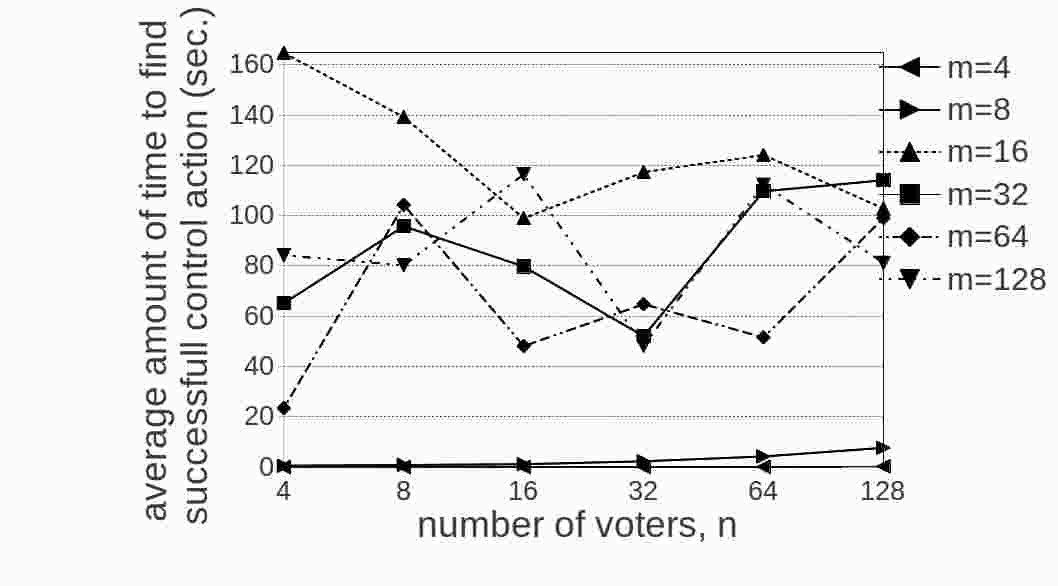}
	\caption{Average time the algorithm needs to find a successful control action for 
	constructive control by partition of candidates in model TP
	in Bucklin elections in the IC model. The maximum is $164,65$ seconds.}
\end{figure}
\begin{figure}[ht]
\centering
	\includegraphics[scale=0.3]{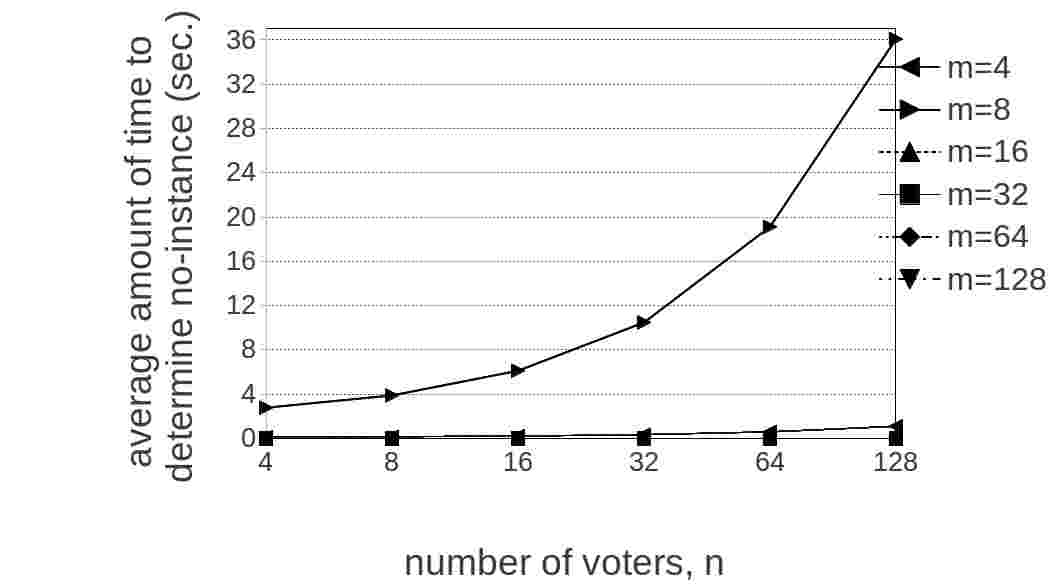}
	\caption{Average time the algorithm needs to determine no-instance of 
		constructive control by partition of candidates in model TP
	in Bucklin elections in the IC model. The maximum is $36,09$ seconds.}
\end{figure}
\begin{figure}[ht]
\centering
	\includegraphics[scale=0.3]{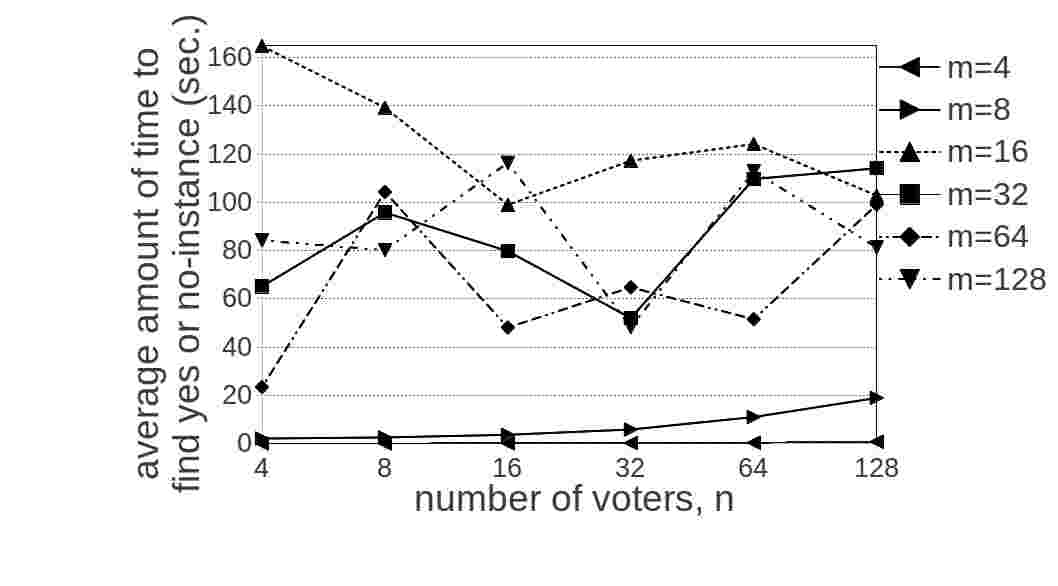}
	\caption{Average time the algorithm needs to give a definite output for 
	constructive control by partition of candidates in model TP
	in Bucklin elections in the IC model. The maximum is $164,65$ seconds.}
\end{figure}
\begin{figure}[ht]
\centering
	\includegraphics[scale=0.3]{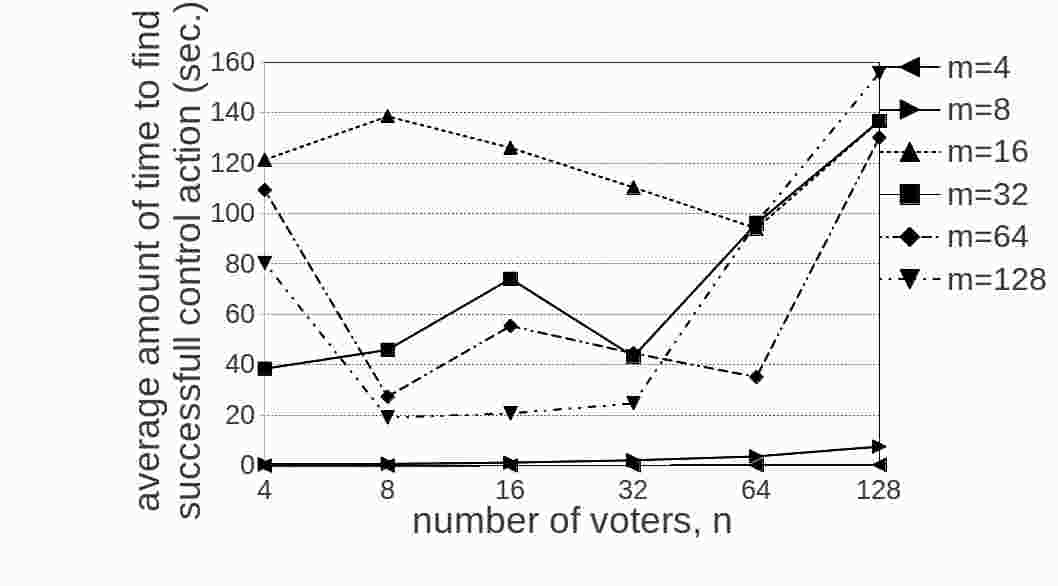}
	\caption{Average time the algorithm needs to find a successful control action for 
	constructive control by partition of candidates in model TP
	in Bucklin elections in the TM model. The maximum is $155,73$ seconds.}
\end{figure}
\begin{figure}[ht]
\centering
	\includegraphics[scale=0.3]{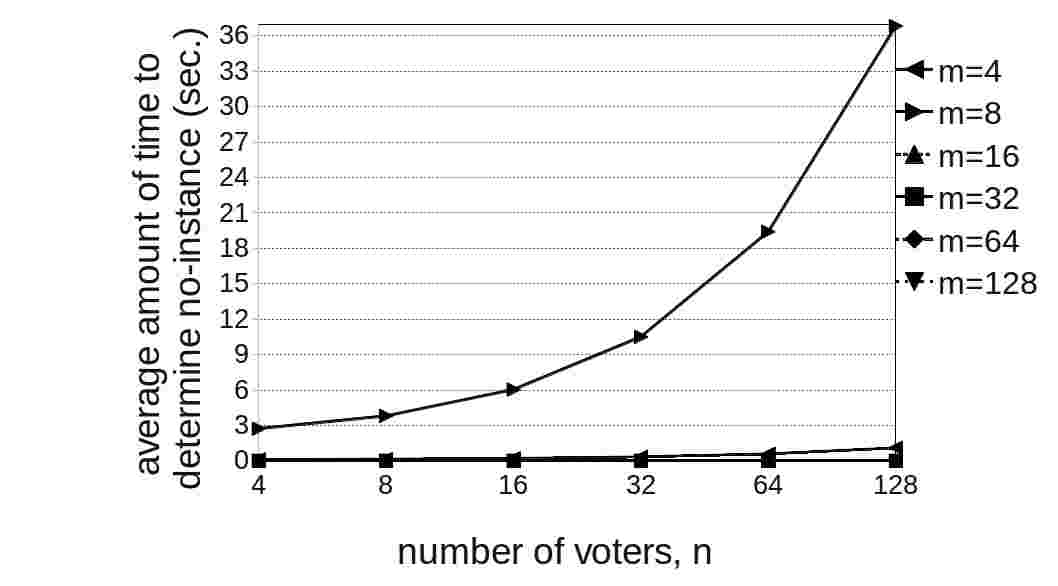}
	\caption{Average time the algorithm needs to determine no-instance of 
		constructive control by partition of candidates in model TP
	in Bucklin elections in the TM model. The maximum is $36,83$ seconds.}
\end{figure}
\begin{figure}[ht]
\centering
	\includegraphics[scale=0.3]{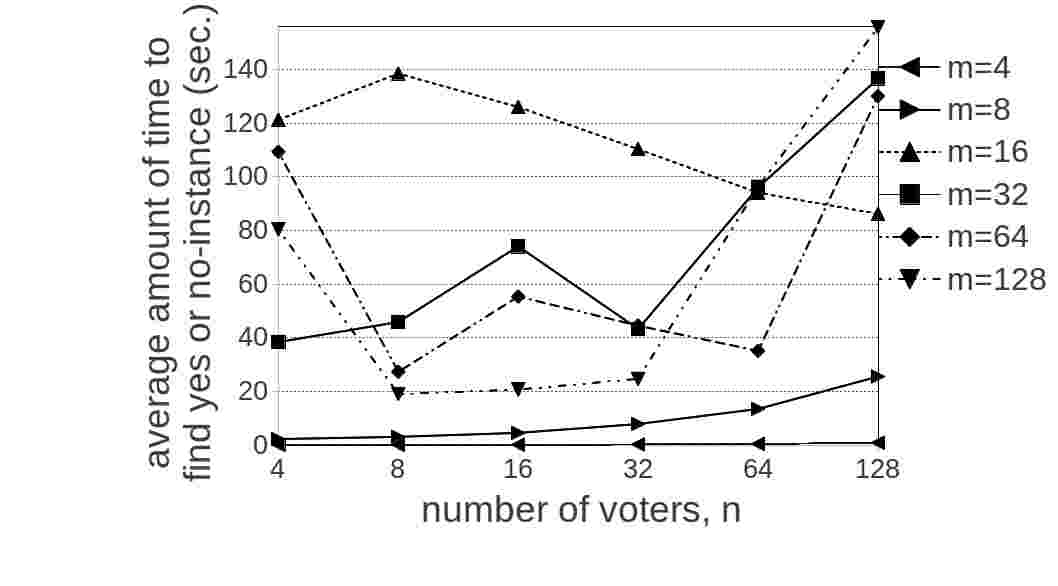}
	\caption{Average time the algorithm needs to give a definite output for 
	constructive control by partition of candidates in model TP
	in Bucklin elections in the TM model. The maximum is $155,73$ seconds.}
\end{figure}
\clearpage
\subsection{Destructive Control by Partition of Candidates in Model TP}
\begin{center}
\begin{figure}[ht]
\centering
	\includegraphics[scale=0.3]{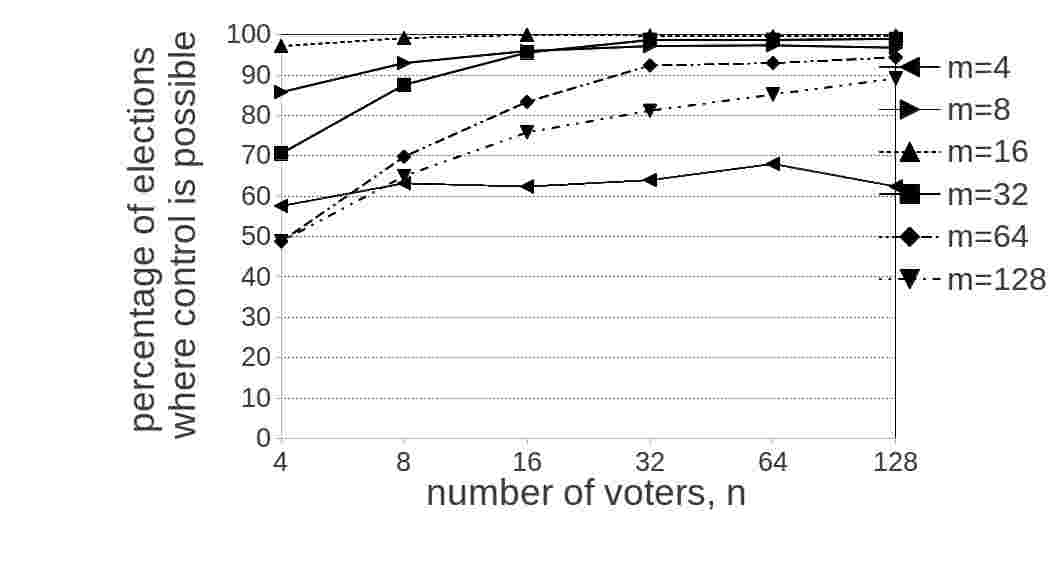}
		\caption{Results for Bucklin voting in the IC model for 
destructive control by partition of candidates in model TP. Number of candidates is fixed. }
\end{figure}

\newpage
\begin{figure}[ht]
\centering
	\includegraphics[scale=0.3]{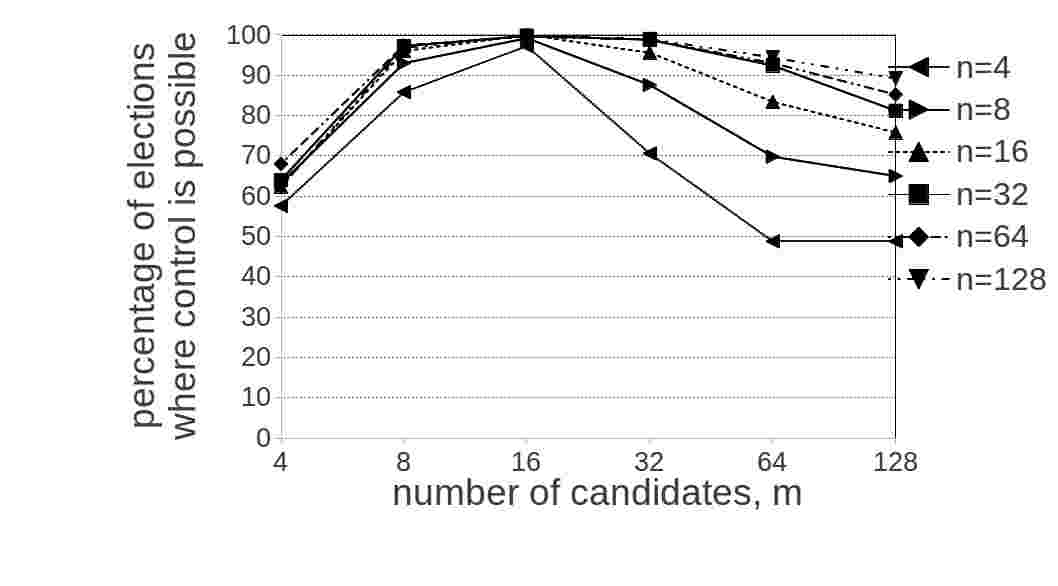}
		\caption{Results for Bucklin voting in the IC model for 
destructive control by partition of candidates in model TP. Number of voters is fixed. }
\end{figure}
%
\newpage
\begin{figure}[ht]
\centering
	\includegraphics[scale=0.3]{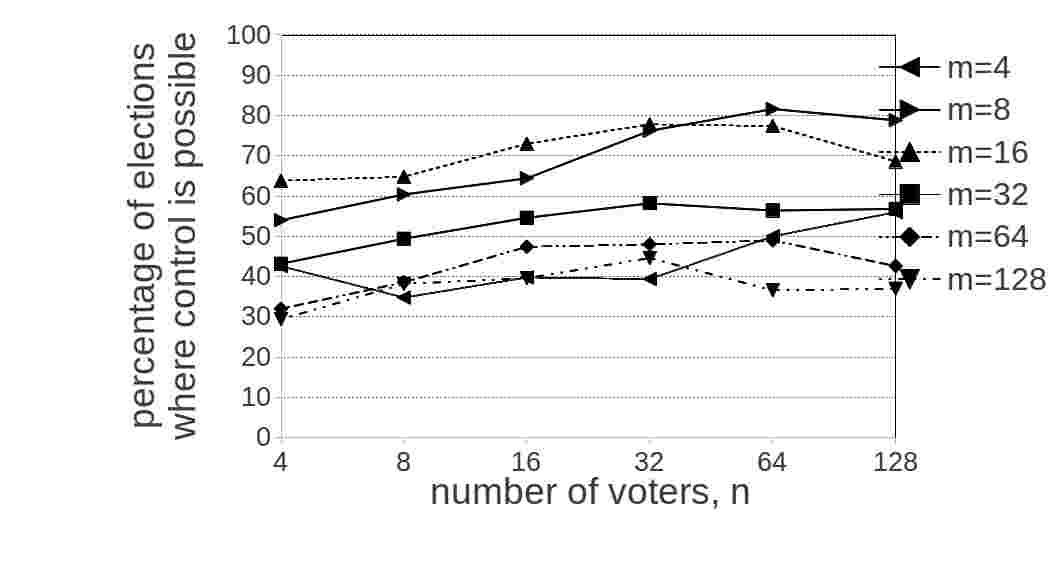}
	\caption{Results for Bucklin voting in the TM model for 
destructive control by partition of candidates in model TP. Number of candidates is fixed. }
\end{figure}
%
\newpage
\begin{figure}[ht]
\centering
	\includegraphics[scale=0.3]{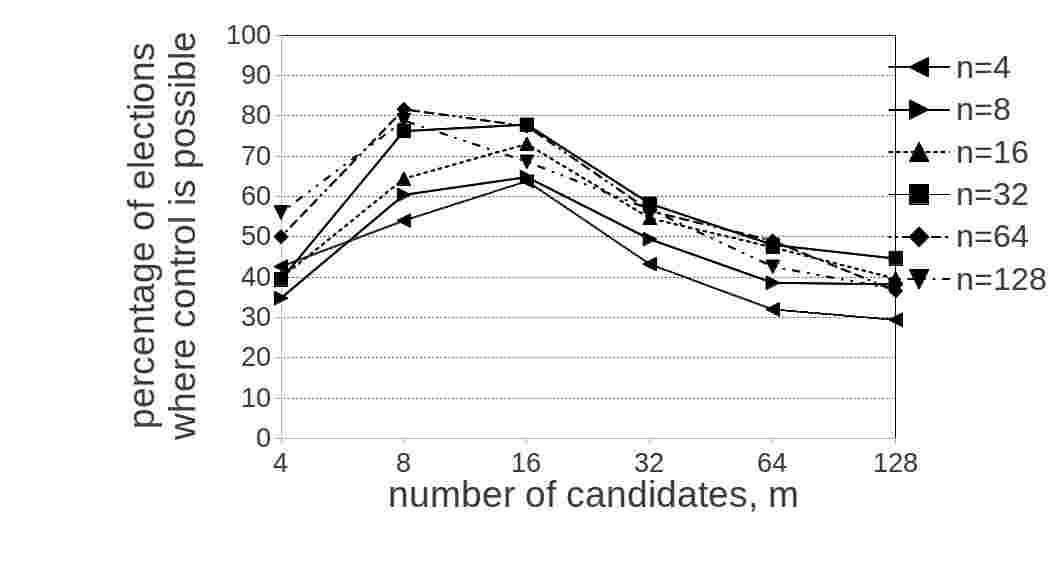}
	\caption{Results for Bucklin voting in the TM model for 
destructive control by partition of candidates in model TP. Number of voters is fixed. }
\end{figure}
%
\end{center}

\clearpage
\subsubsection{Computational Costs}
\begin{figure}[ht]
\centering
	\includegraphics[scale=0.3]{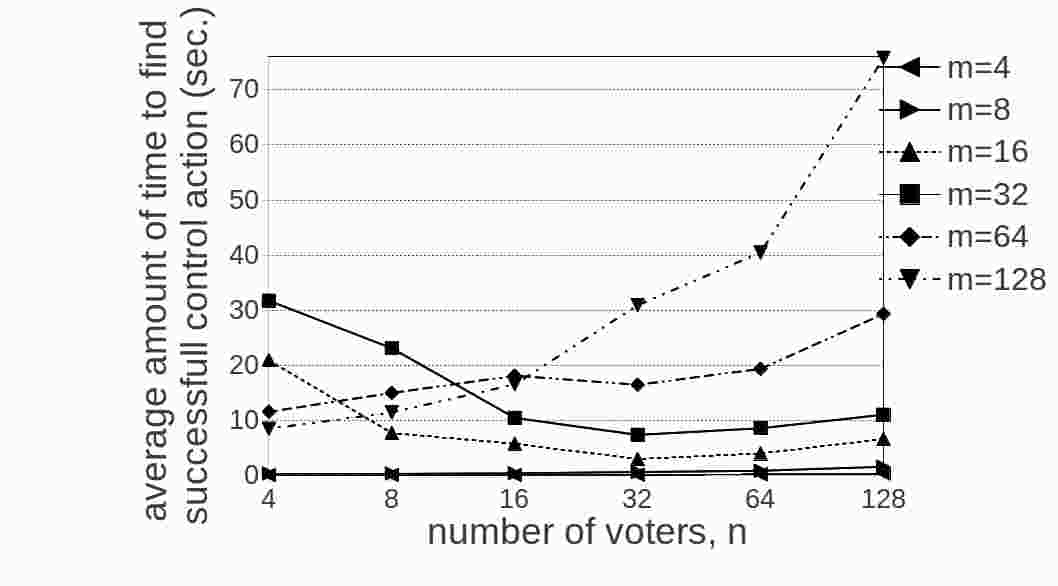}
	\caption{Average time the algorithm needs to find a successful control action for 
	destructive control by partition of candidates in model TP
	in Bucklin elections in the IC model. The maximum is $75,81$ seconds.}
\end{figure}
\begin{figure}[ht]
\centering
	\includegraphics[scale=0.3]{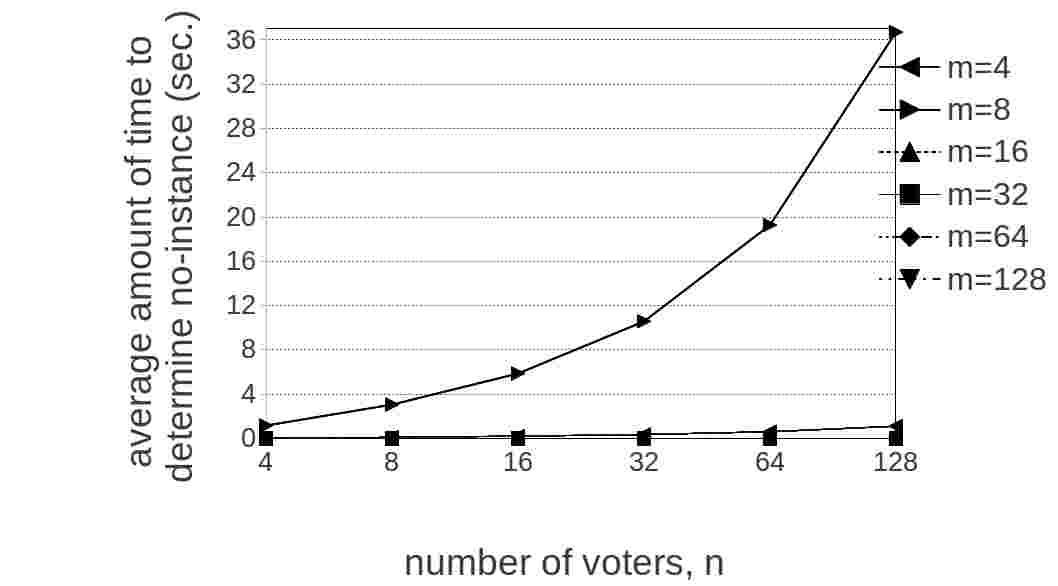}
	\caption{Average time the algorithm needs to determine no-instance of 
		destructive control by partition of candidates in model TP
	in Bucklin elections in the IC model. The maximum is $36,71$ seconds.}
\end{figure}
\begin{figure}[ht]
\centering
	\includegraphics[scale=0.3]{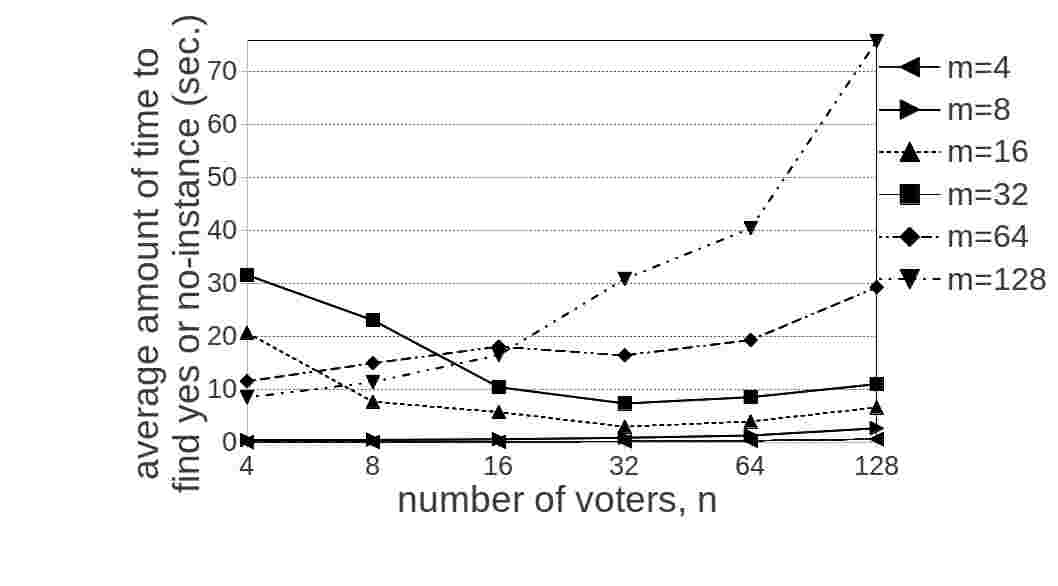}
	\caption{Average time the algorithm needs to give a definite output for 
	destructive control by partition of candidates in model TP
	in Bucklin elections in the IC model. The maximum is $75,81$ seconds.}
\end{figure}
\begin{figure}[ht]
\centering
	\includegraphics[scale=0.3]{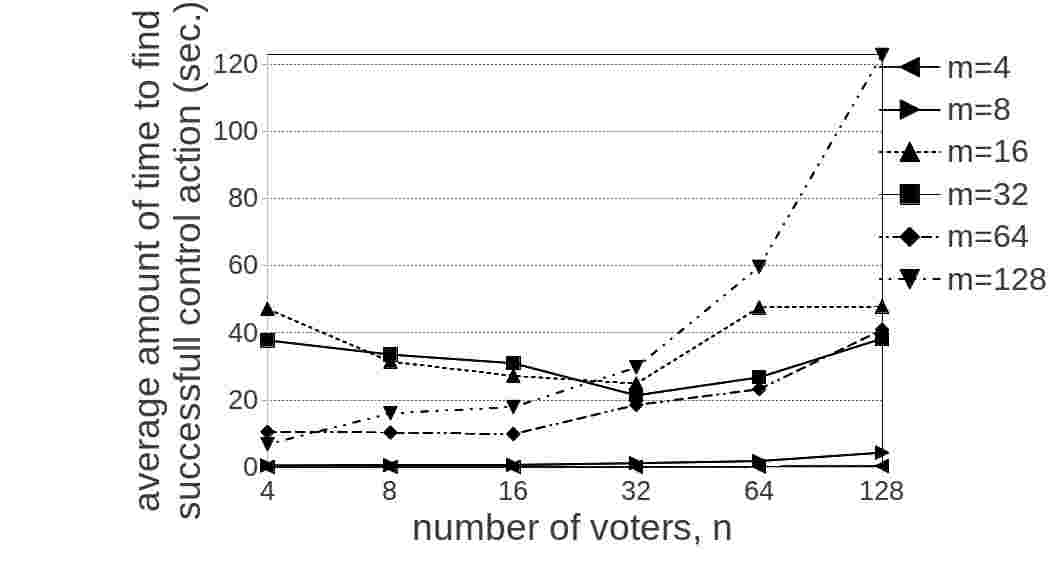}
	\caption{Average time the algorithm needs to find a successful control action for 
	destructive control by partition of candidates in model TP
	in Bucklin elections in the TM model. The maximum is $122,89$ seconds.}
\end{figure}
\begin{figure}[ht]
\centering
	\includegraphics[scale=0.3]{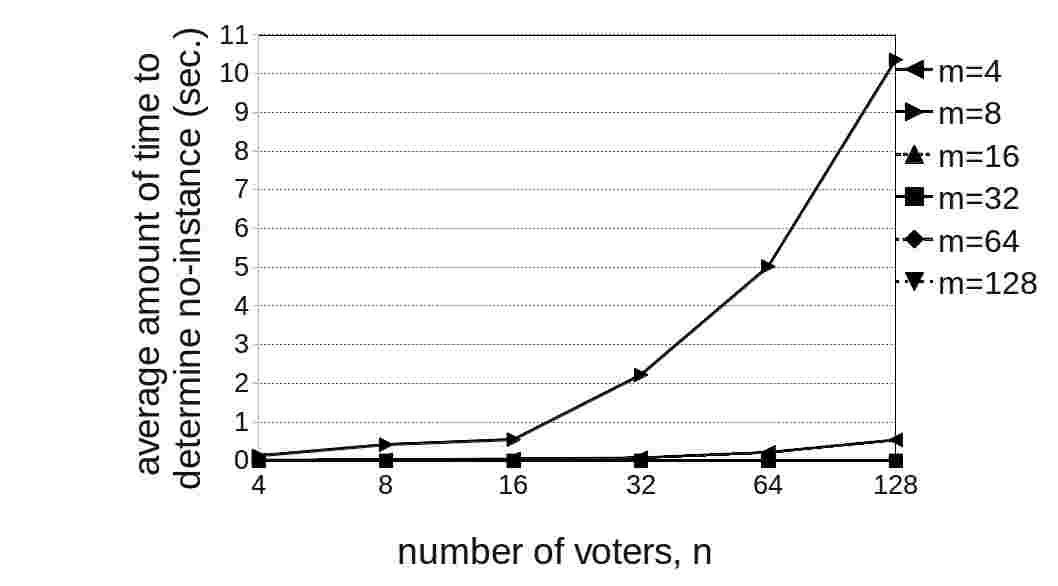}
	\caption{Average time the algorithm needs to determine no-instance of 
		destructive control by partition of candidates in model TP
	in Bucklin elections in the TM model. The maximum is $10,36$ seconds.}
\end{figure}
\begin{figure}[ht]
\centering
	\includegraphics[scale=0.3]{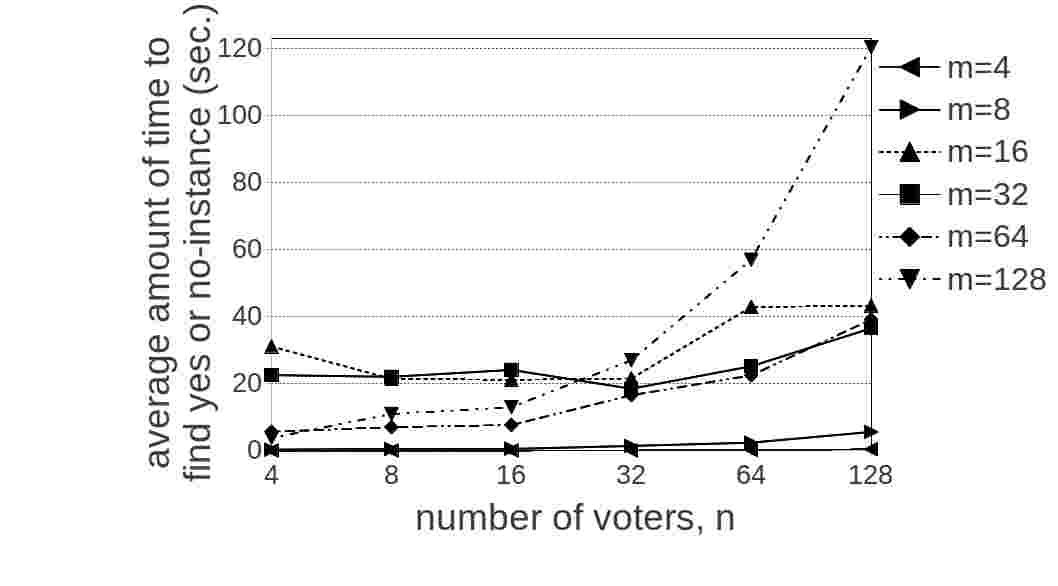}
	\caption{Average time the algorithm needs to give a definite output for 
	destructive control by partition of candidates in model TP
	in Bucklin elections in the TM model. The maximum is $120,28$ seconds.}
\end{figure}

\clearpage
\subsection{Constructive Control by Runoff Partition of Candidates in Model TE}
\begin{center}
\begin{figure}[ht]
\centering
	\includegraphics[scale=0.3]{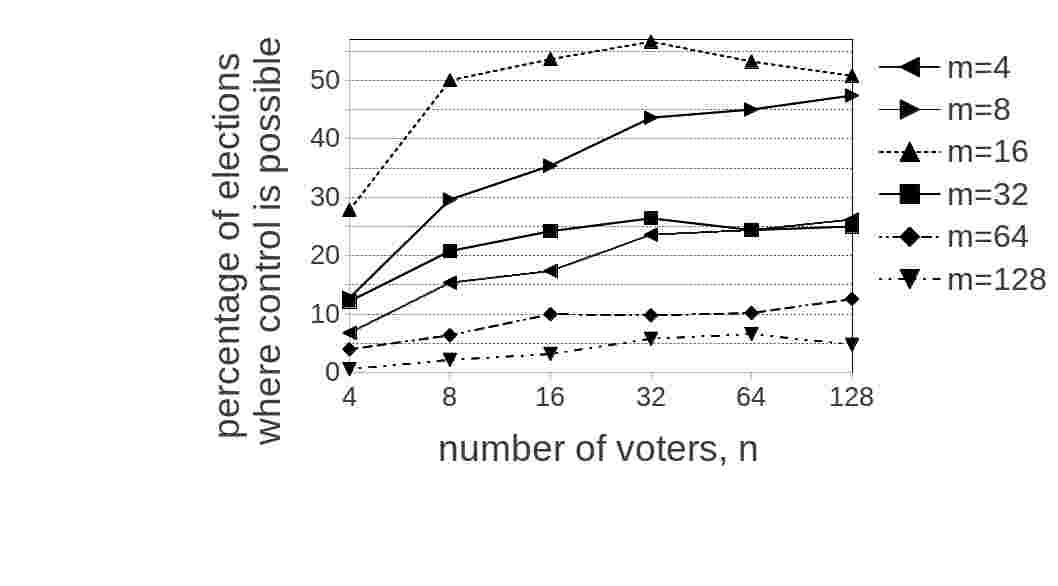}
		\caption{Results for Bucklin voting in the IC model for 
constructive control by runoff-partition  of candidates in model TE. Number of candidates is fixed. }
\end{figure}

\newpage
\begin{figure}[ht]
\centering
	\includegraphics[scale=0.3]{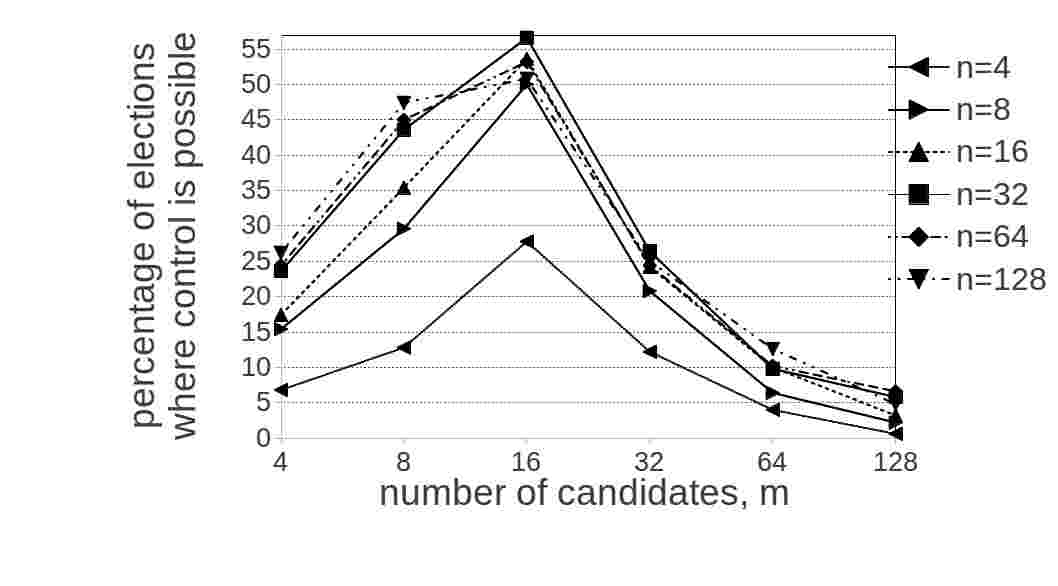}
		\caption{Results for Bucklin voting in the IC model for 
constructive control by runoff-partition  of candidates in model TE. Number of voters is fixed. }
\end{figure}
%
\newpage
\begin{figure}[ht]
\centering
	\includegraphics[scale=0.3]{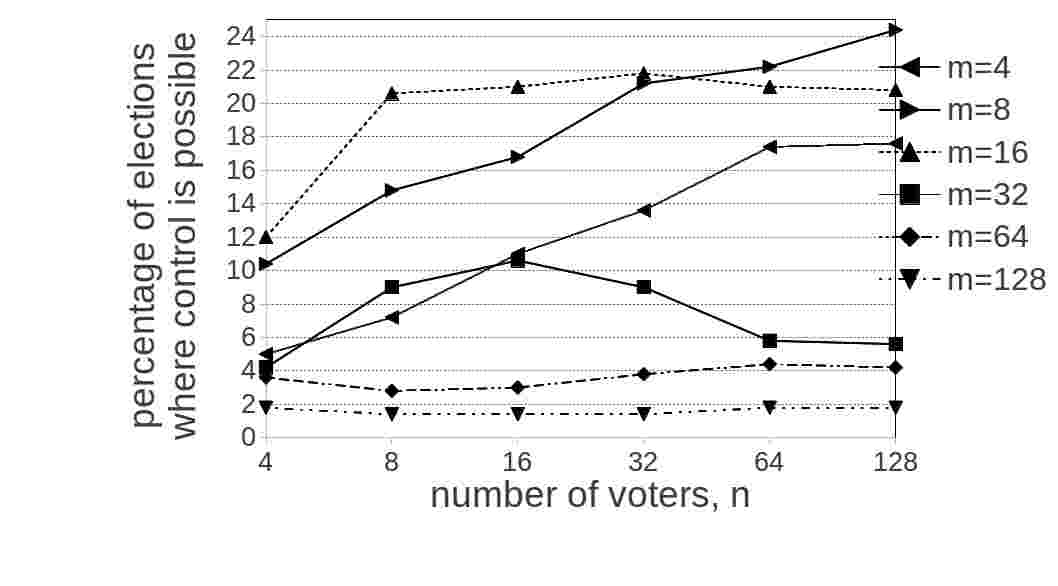}
	\caption{Results for Bucklin voting in the TM model for 
constructive control by runoff-partition  of candidates in model TE. Number of candidates is fixed. }
\end{figure}
%
\newpage
\begin{figure}[ht]
\centering
	\includegraphics[scale=0.3]{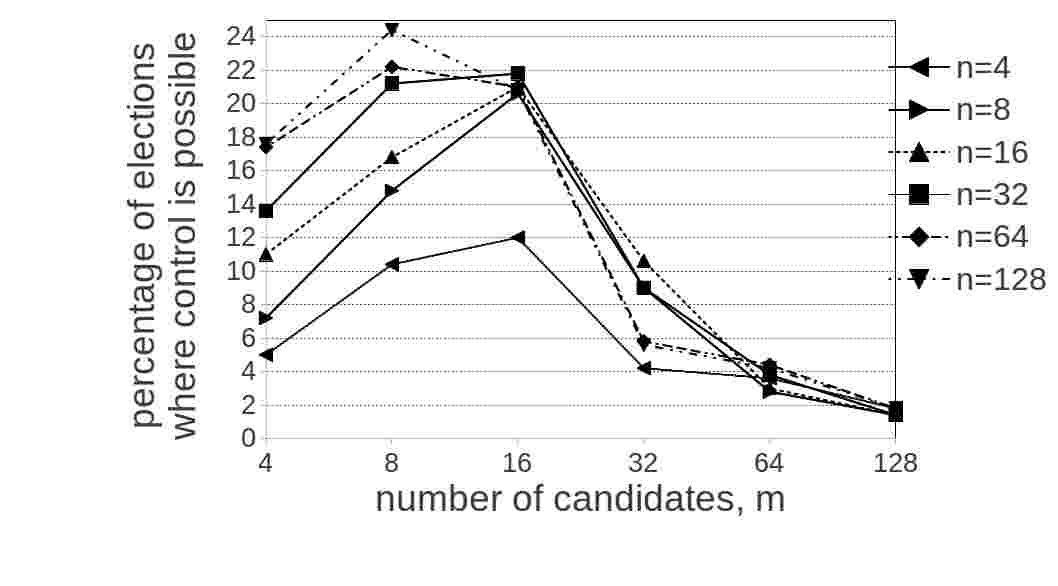}
	\caption{Results for Bucklin voting in the TM model for 
constructive control by runoff-partition  of candidates in model TE. Number of voters is fixed. }
\end{figure}
%
\end{center}

\clearpage
\subsubsection{Computational Costs}
\begin{figure}[ht]
\centering
	\includegraphics[scale=0.3]{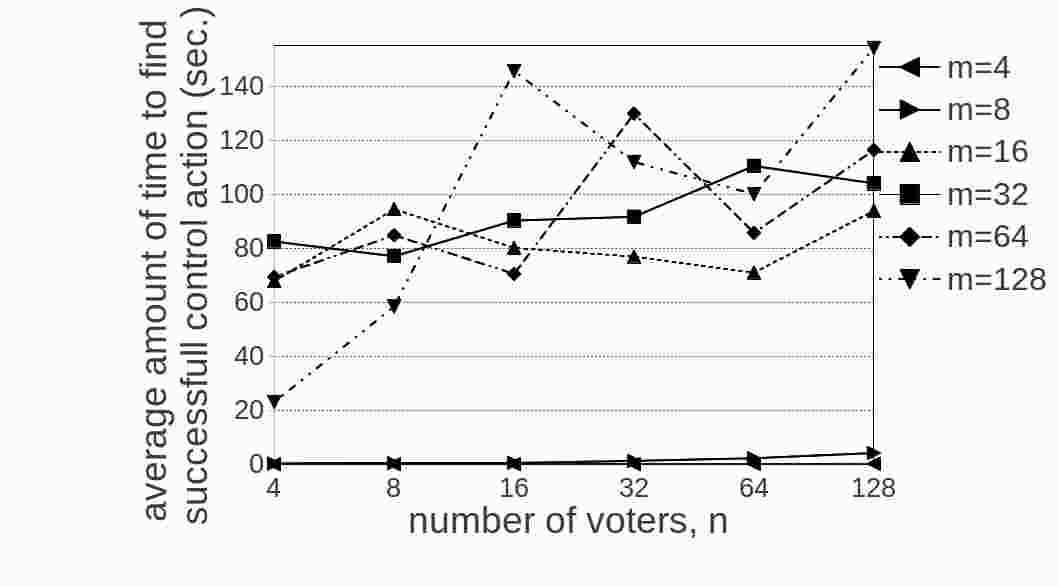}
	\caption{Average time the algorithm needs to find a successful control action for 
	constructive control by runoff-partition  of candidates in model TE
	in Bucklin elections in the IC model. The maximum is $154,23$ seconds.}
\end{figure}
\begin{figure}[ht]
\centering
	\includegraphics[scale=0.3]{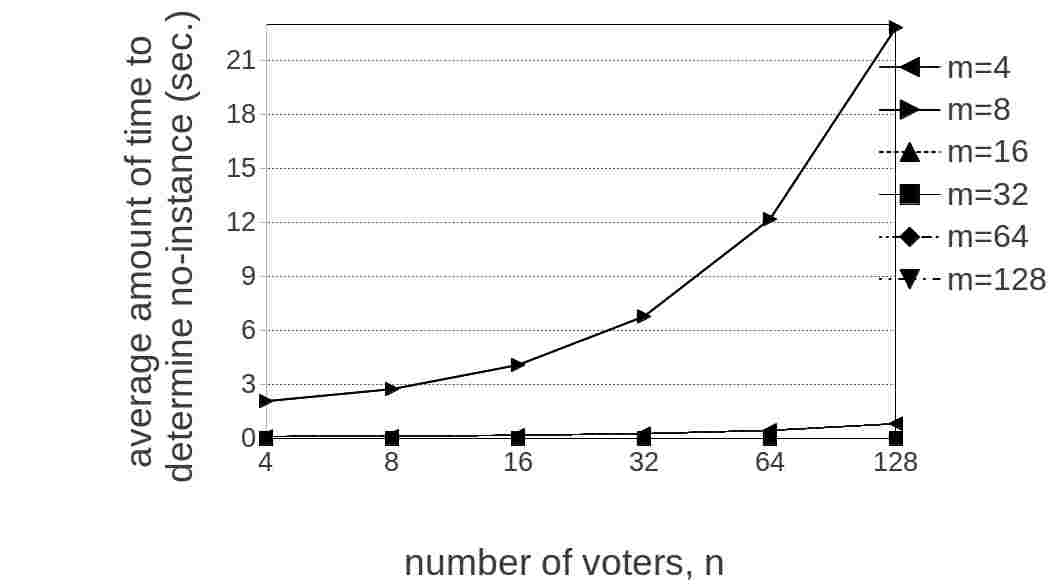}
	\caption{Average time the algorithm needs to determine no-instance of 
		constructive control by runoff-partition  of candidates in model TE
	in Bucklin elections in the IC model. The maximum is $22,8$ seconds.}
\end{figure}
\begin{figure}[ht]
\centering
	\includegraphics[scale=0.3]{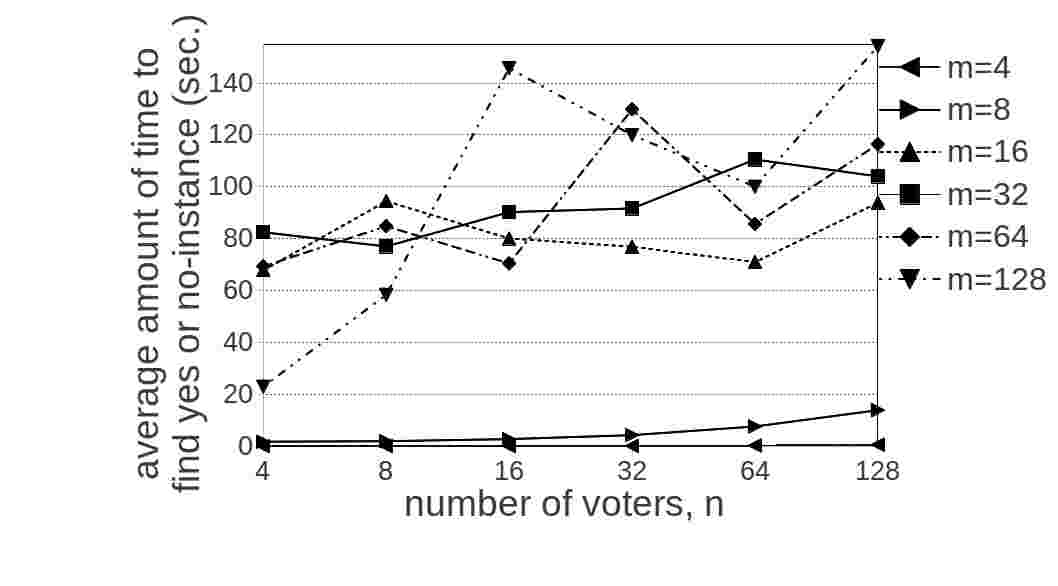}
	\caption{Average time the algorithm needs to give a definite output for 
	constructive control by runoff-partition  of candidates in model TE
	in Bucklin elections in the IC model. The maximum is $154,23$ seconds.}
\end{figure}
\begin{figure}[ht]
\centering
	\includegraphics[scale=0.3]{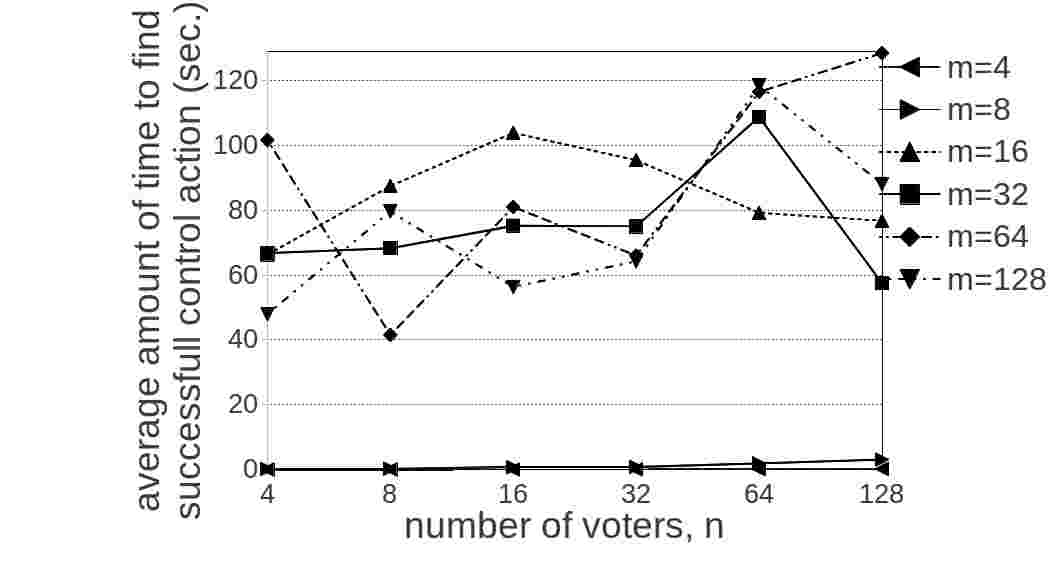}
	\caption{Average time the algorithm needs to find a successful control action for 
	constructive control by runoff-partition  of candidates in model TE
	in Bucklin elections in the TM model. The maximum is $128,41$ seconds.}
\end{figure}
\begin{figure}[ht]
\centering
	\includegraphics[scale=0.3]{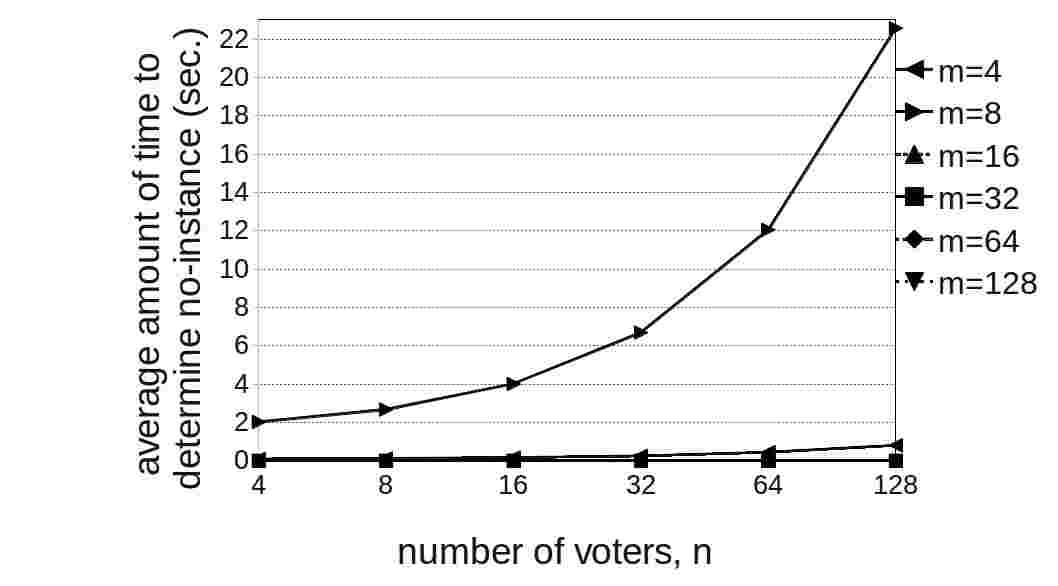}
	\caption{Average time the algorithm needs to determine no-instance of 
		constructive control by runoff-partition  of candidates in model TE
	in Bucklin elections in the TM model. The maximum is $22,56$ seconds.}
\end{figure}
\begin{figure}[ht]
\centering
	\includegraphics[scale=0.3]{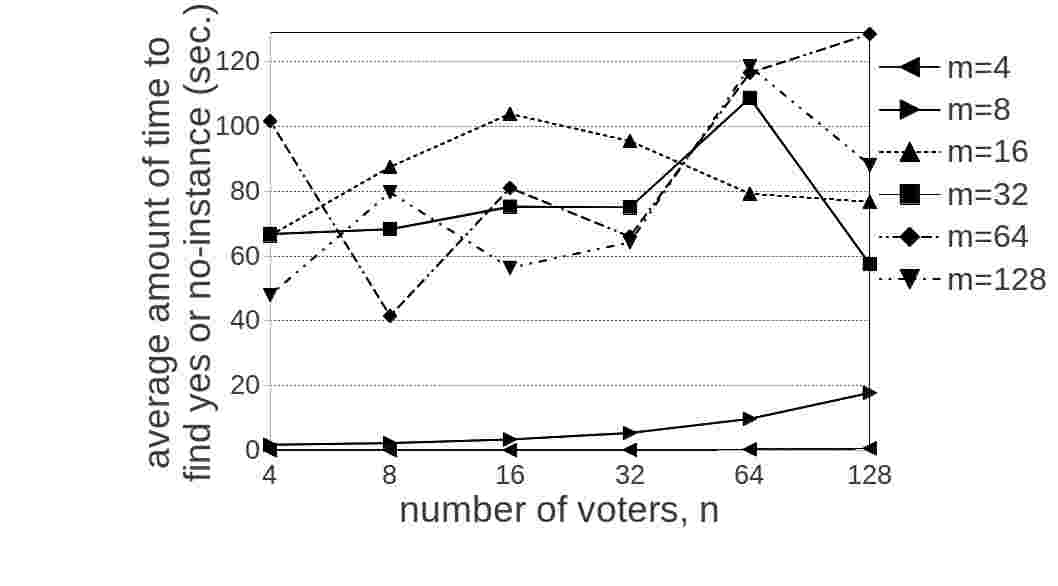}
	\caption{Average time the algorithm needs to give a definite output for 
	constructive control by runoff-partition  of candidates in model TE
	in Bucklin elections in the TM model. The maximum is $128,41$ seconds.}
\end{figure}

\clearpage
\subsection{Destructive Control by Runoff Partition of Candidates in Model TE}
\begin{center}
\begin{figure}[ht]
\centering
	\includegraphics[scale=0.3]{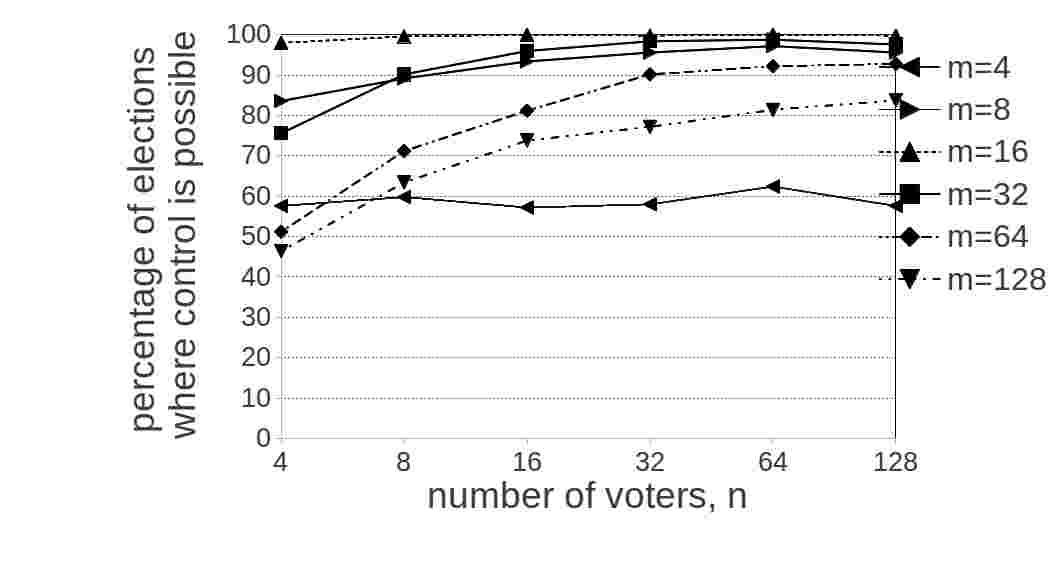}
		\caption{Results for Bucklin voting in the IC model for 
destructive control by runoff-partition  of candidates in model TE. Number of candidates is fixed. }
\end{figure}

\newpage
\begin{figure}[ht]
\centering
	\includegraphics[scale=0.3]{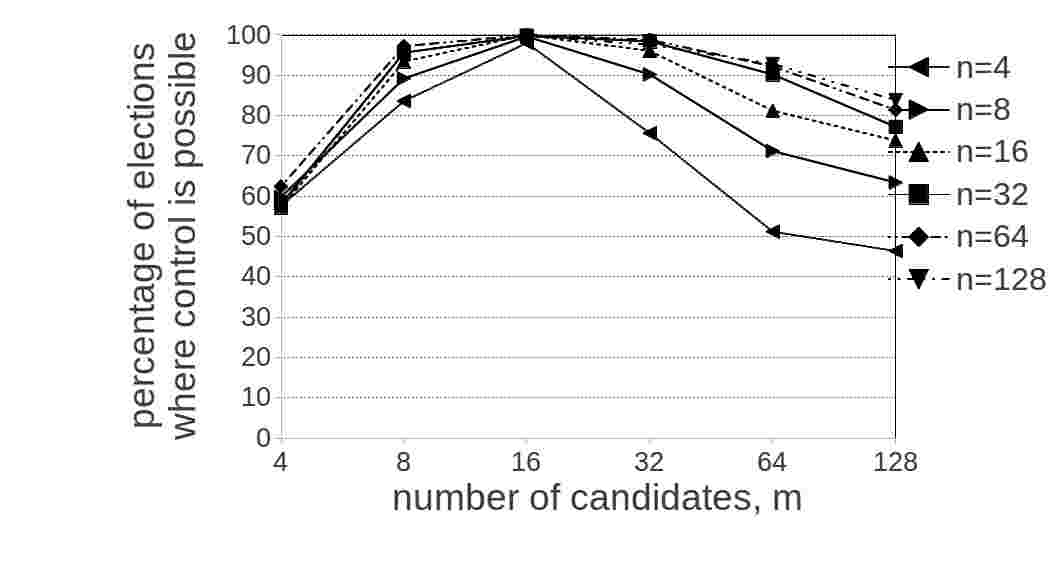}
		\caption{Results for Bucklin voting in the IC model for 
destructive control by runoff-partition  of candidates in model TE. Number of voters is fixed. }
\end{figure}
%
\newpage
\begin{figure}[ht]
\centering
	\includegraphics[scale=0.3]{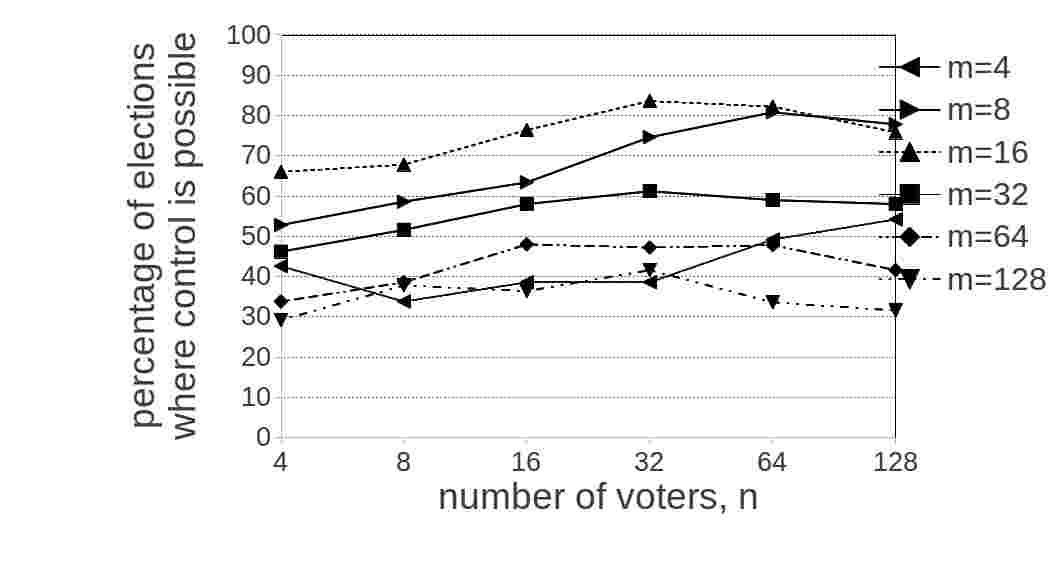}
	\caption{Results for Bucklin voting in the TM model for 
destructive control by runoff-partition  of candidates in model TE. Number of candidates is fixed. }
\end{figure}
%
\newpage
\begin{figure}[ht]
\centering
	\includegraphics[scale=0.3]{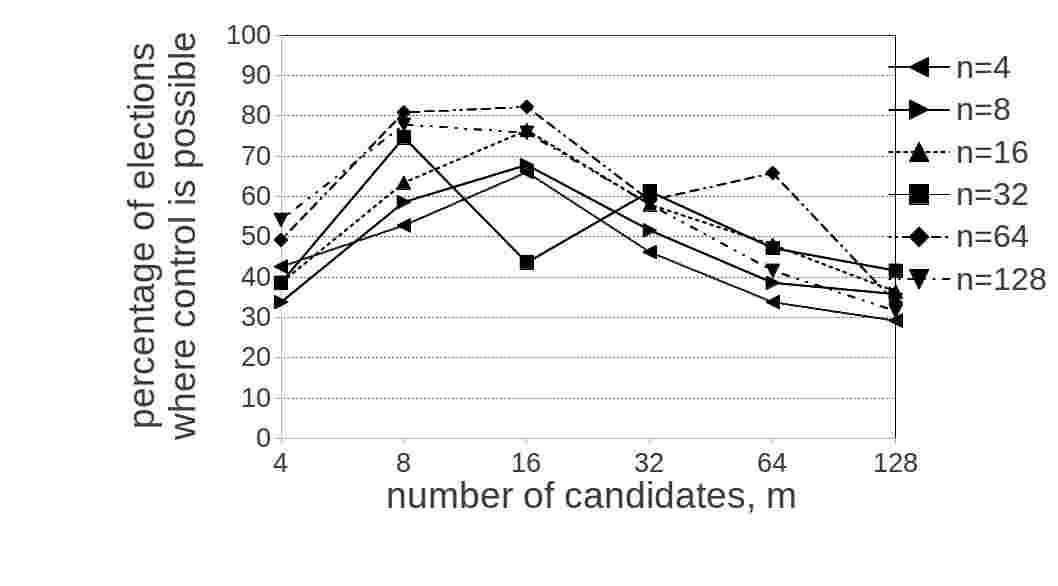}
	\caption{Results for Bucklin voting in the TM model for 
destructive control by runoff-partition  of candidates in model TE. Number of voters is fixed. }
\end{figure}
%
\end{center}

\clearpage
\subsubsection{Computational Costs}
\begin{figure}[ht]
\centering
	\includegraphics[scale=0.3]{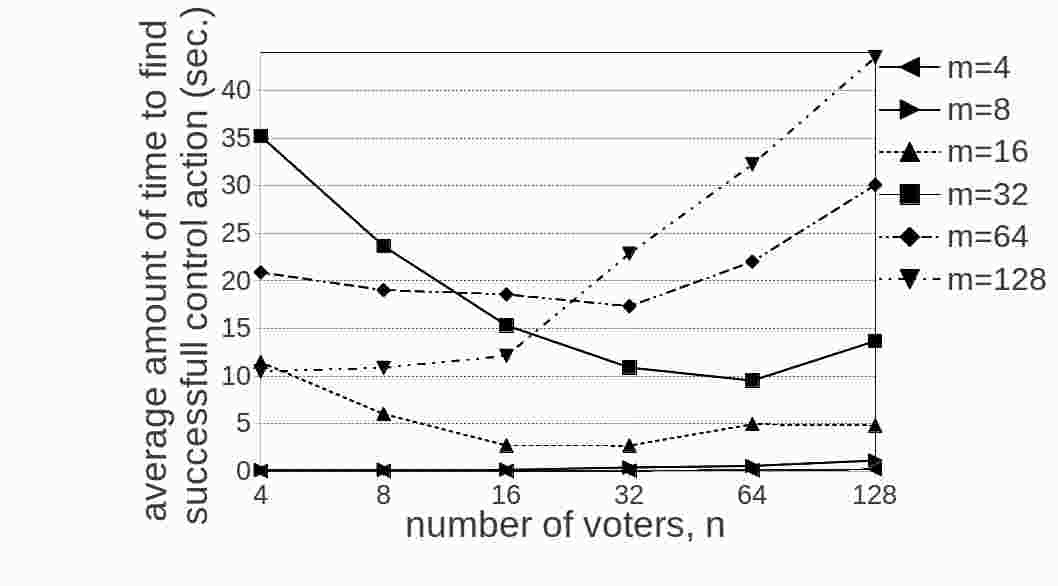}
	\caption{Average time the algorithm needs to find a successful control action for 
	destructive control by runoff-partition  of candidates in model TE
	in Bucklin elections in the IC model. The maximum is $43,52$ seconds.}
\end{figure}
\begin{figure}[ht]
\centering
	\includegraphics[scale=0.3]{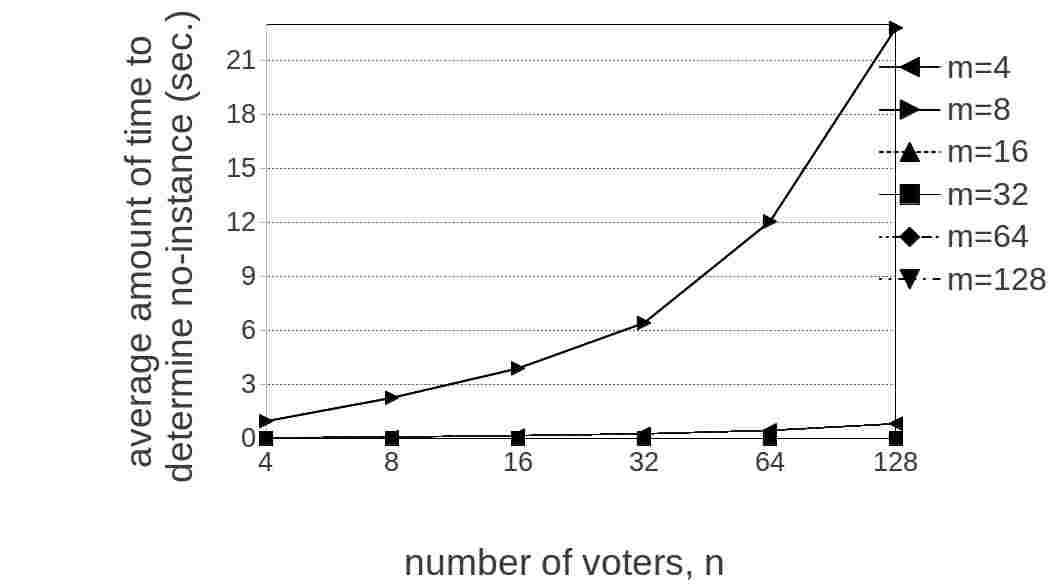}
	\caption{Average time the algorithm needs to determine no-instance of 
		destructive control by runoff-partition  of candidates in model TE
	in Bucklin elections in the IC model. The maximum is $22,78$ seconds.}
\end{figure}
\begin{figure}[ht]
\centering
	\includegraphics[scale=0.3]{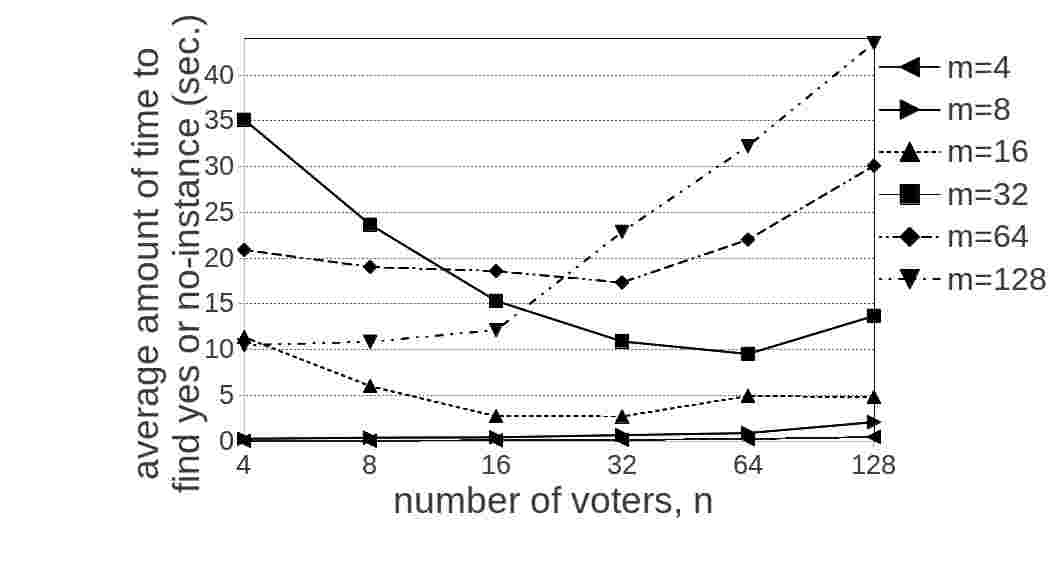}
	\caption{Average time the algorithm needs to give a definite output for 
	destructive control by runoff-partition  of candidates in model TE
	in Bucklin elections in the IC model. The maximum is $43,52$ seconds.}
\end{figure}
\begin{figure}[ht]
\centering
	\includegraphics[scale=0.3]{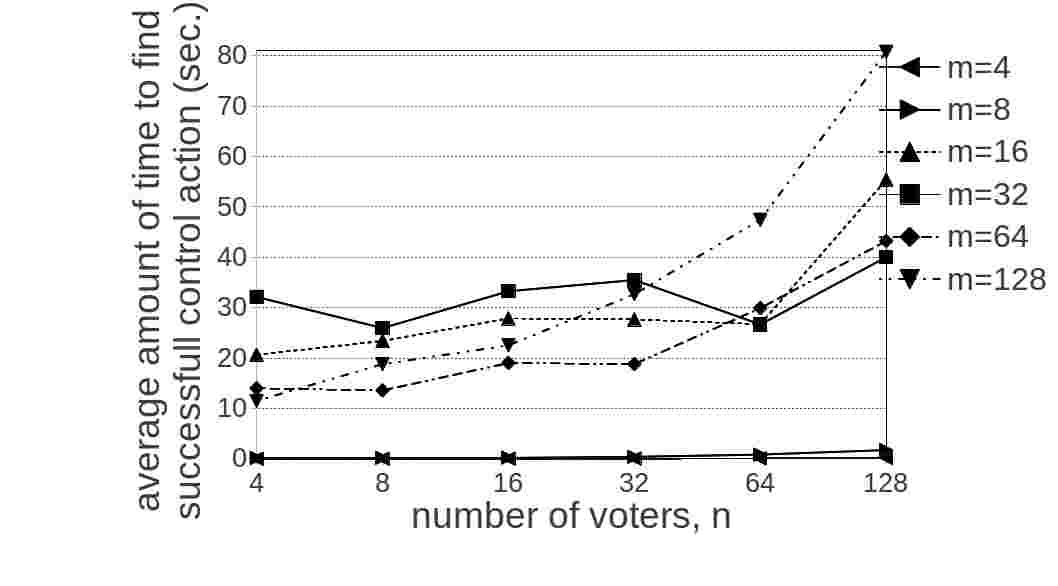}
	\caption{Average time the algorithm needs to find a successful control action for 
	destructive control by runoff-partition  of candidates in model TE
	in Bucklin elections in the TM model. The maximum is $80,81$ seconds.}
\end{figure}
\begin{figure}[ht]
\centering
	\includegraphics[scale=0.3]{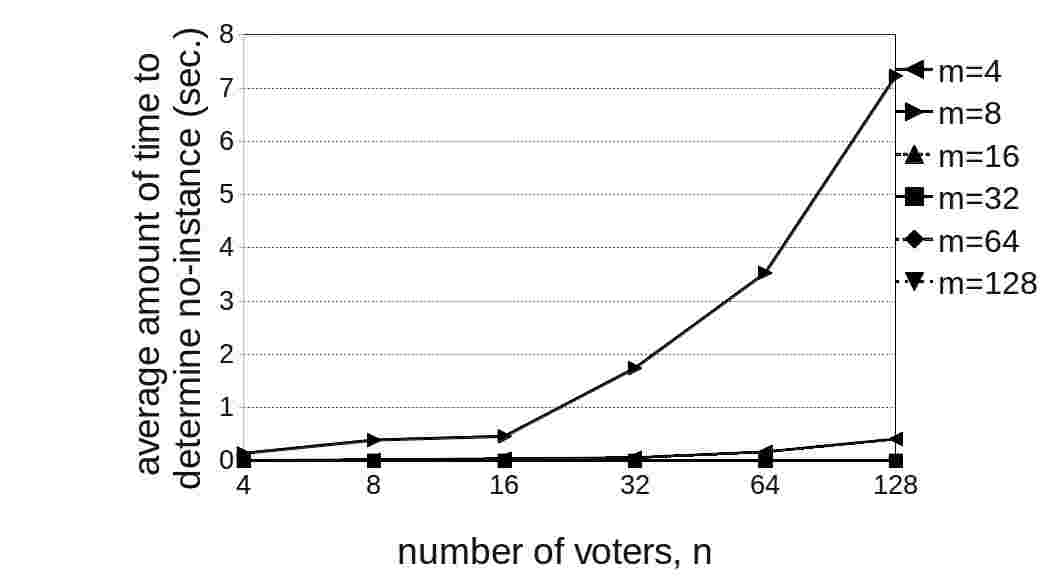}
	\caption{Average time the algorithm needs to determine no-instance of 
		destructive control by runoff-partition  of candidates in model TE
	in Bucklin elections in the TM model. The maximum is $7,23$ seconds.}
\end{figure}
\begin{figure}[ht]
\centering
	\includegraphics[scale=0.3]{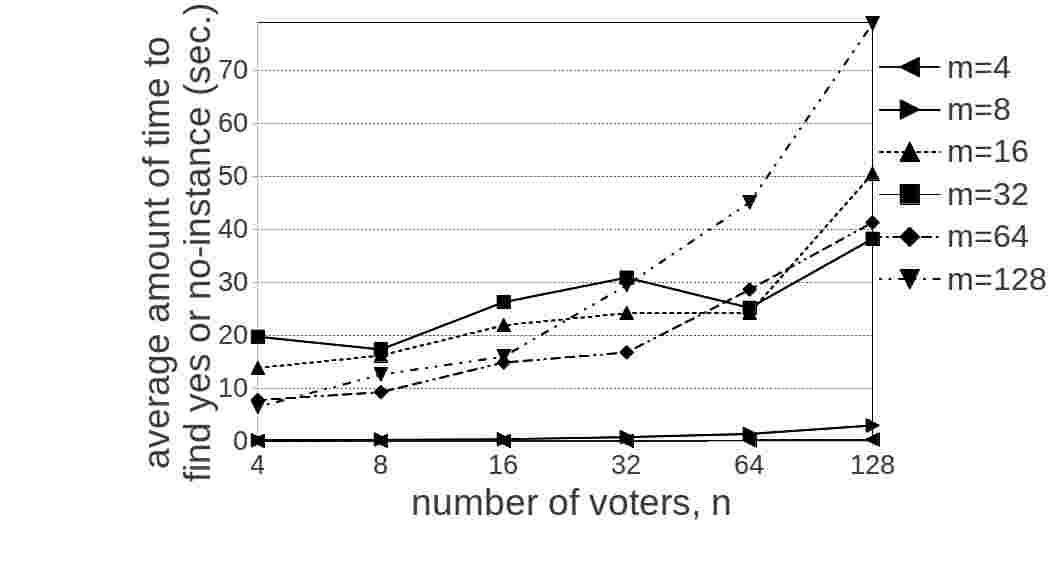}
	\caption{Average time the algorithm needs to give a definite output for 
	destructive control by runoff-partition  of candidates in model TE
	in Bucklin elections in the TM model. The maximum is $78,82$ seconds.}
\end{figure}

\clearpage
\subsection{Constructive Control by Runoff Partition of Candidates in Model TP}
\begin{center}
\begin{figure}[ht]
\centering
	\includegraphics[scale=0.3]{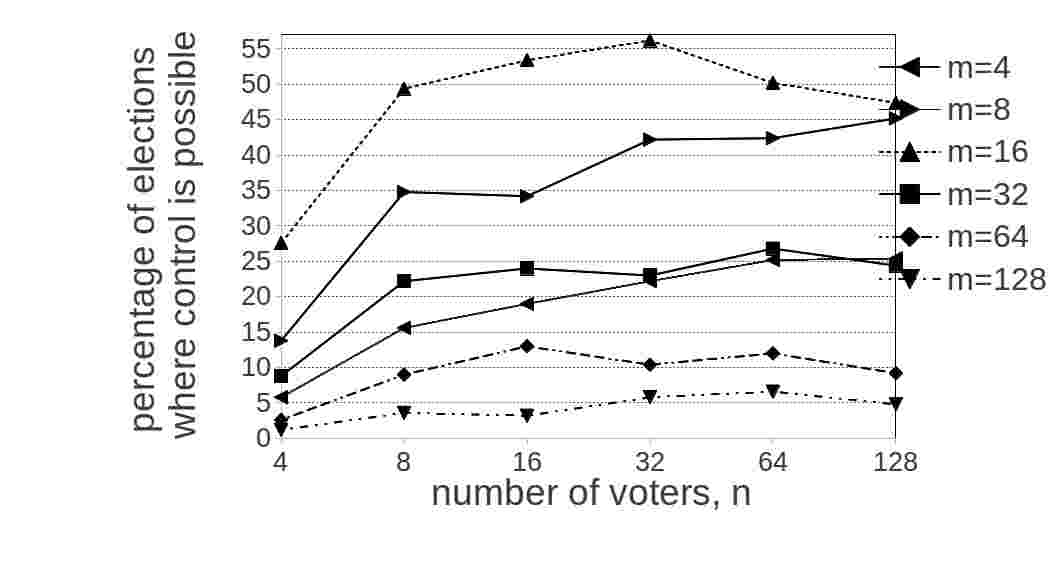}
		\caption{Results for Bucklin voting in the IC model for 
constructive control by runoff-partition  of candidates in model TP. Number of candidates is fixed. }
\end{figure}

\newpage
\begin{figure}[ht]
\centering
	\includegraphics[scale=0.3]{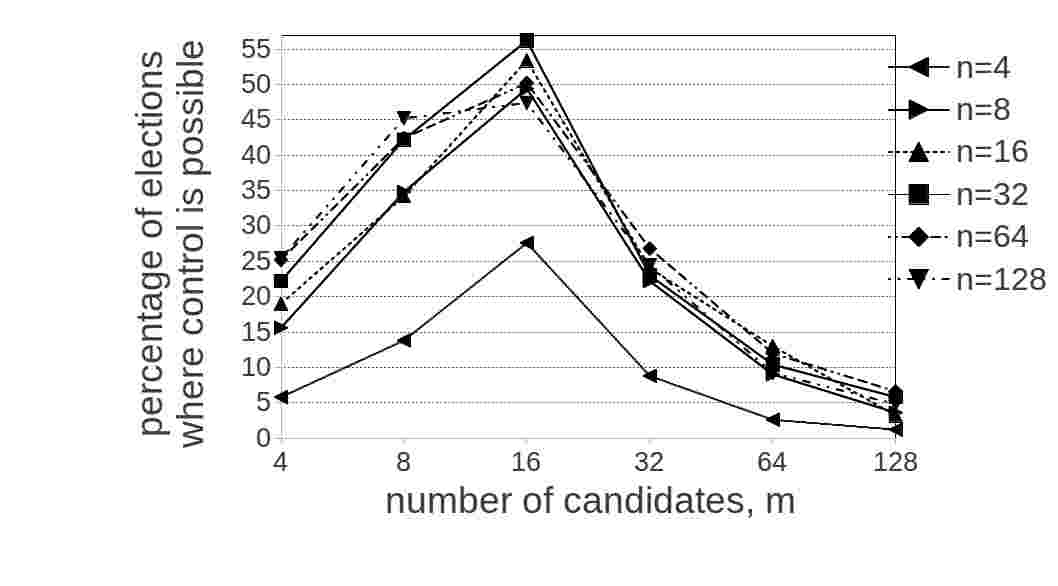}
		\caption{Results for Bucklin voting in the IC model for 
constructive control by runoff-partition  of candidates in model TP. Number of voters is fixed. }
\end{figure}
%
\newpage
\begin{figure}[ht]
\centering
	\includegraphics[scale=0.3]{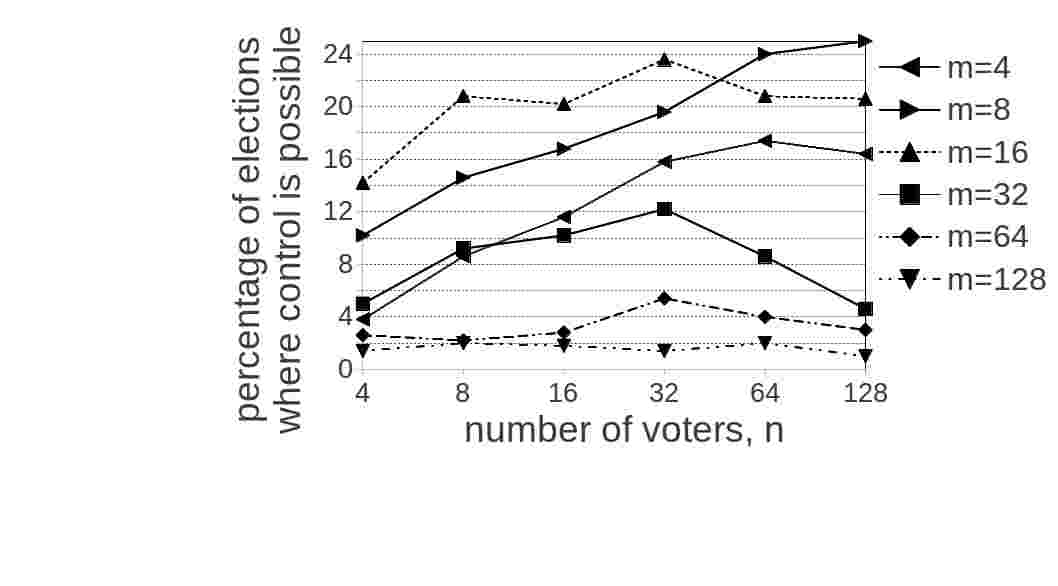}
	\caption{Results for Bucklin voting in the TM model for 
constructive control by runoff-partition  of candidates in model TP. Number of candidates is fixed. }
\end{figure}
%
\newpage
\begin{figure}[ht]
\centering
	\includegraphics[scale=0.3]{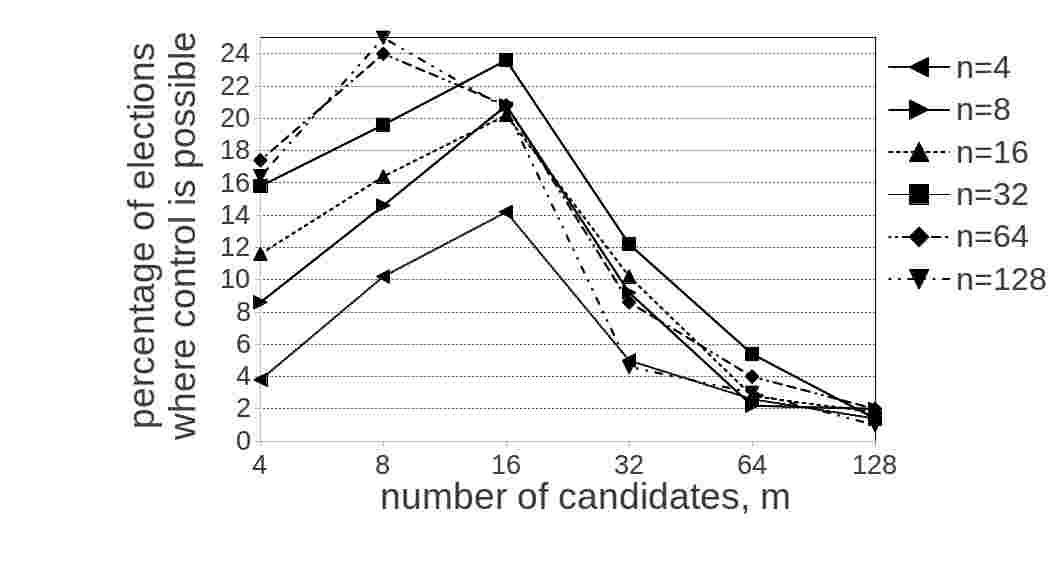}
	\caption{Results for Bucklin voting in the TM model for 
constructive control by runoff-partition  of candidates in model TP. Number of voters is fixed. }
\end{figure}
%
\end{center}

\clearpage
\subsubsection{Computational Costs}
\begin{figure}[ht]
\centering
	\includegraphics[scale=0.3]{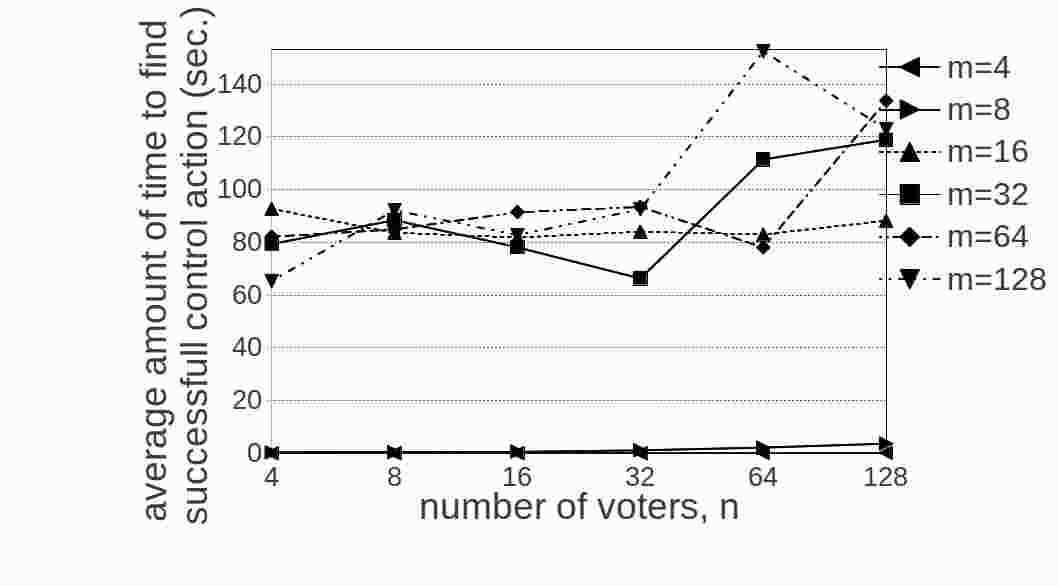}
	\caption{Average time the algorithm needs to find a successful control action for 
	constructive control by runoff-partition  of candidates in model TP
	in Bucklin elections in the IC model. The maximum is $152,45$ seconds.}
\end{figure}
\begin{figure}[ht]
\centering
	\includegraphics[scale=0.3]{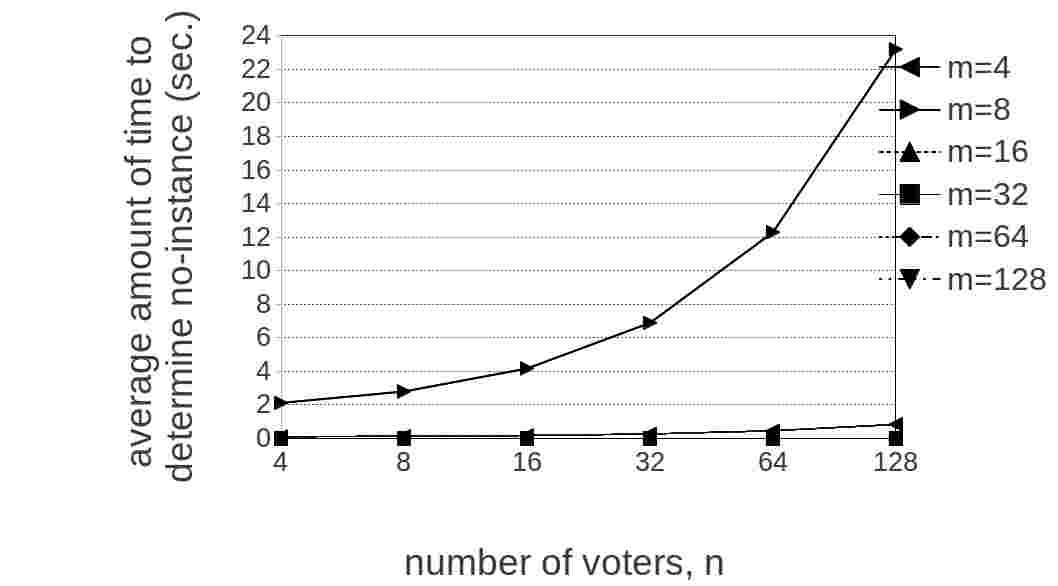}
	\caption{Average time the algorithm needs to determine no-instance of 
		constructive control by runoff-partition  of candidates in model TP
	in Bucklin elections in the IC model. The maximum is $23,19$ seconds.}
\end{figure}
\begin{figure}[ht]
\centering
	\includegraphics[scale=0.3]{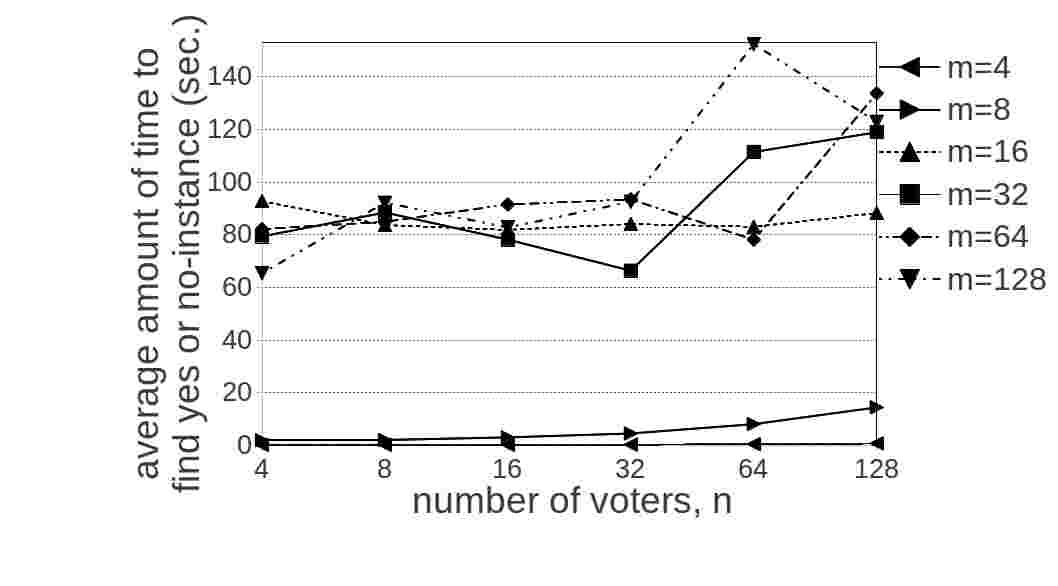}
	\caption{Average time the algorithm needs to give a definite output for 
	constructive control by runoff-partition  of candidates in model TP
	in Bucklin elections in the IC model. The maximum is $152,45$ seconds.}
\end{figure}
\begin{figure}[ht]
\centering
	\includegraphics[scale=0.3]{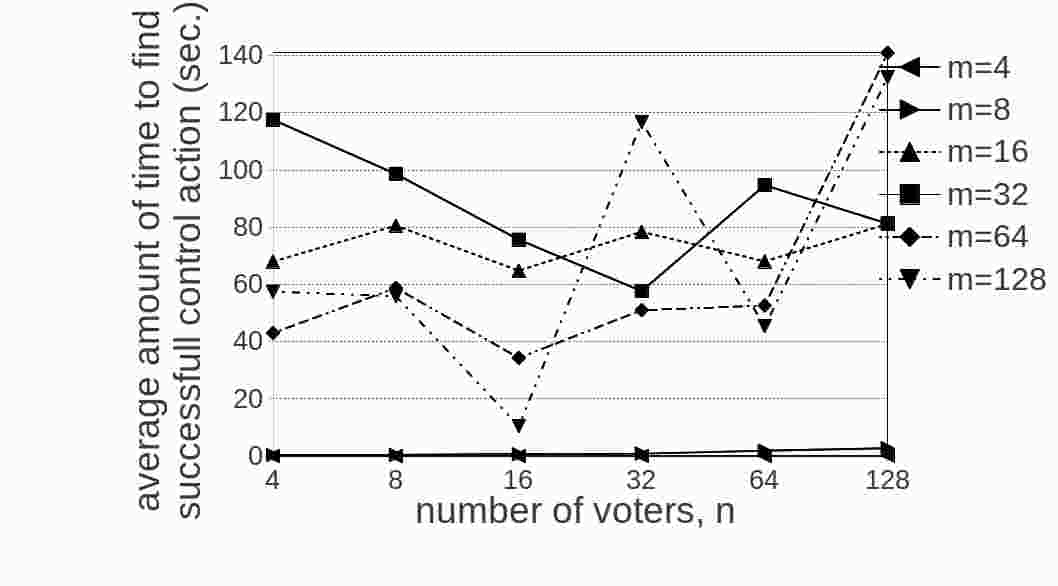}
	\caption{Average time the algorithm needs to find a successful control action for 
	constructive control by runoff-partition  of candidates in model TP
	in Bucklin elections in the TM model. The maximum is $140,93$ seconds.}
\end{figure}
\begin{figure}[ht]
\centering
	\includegraphics[scale=0.3]{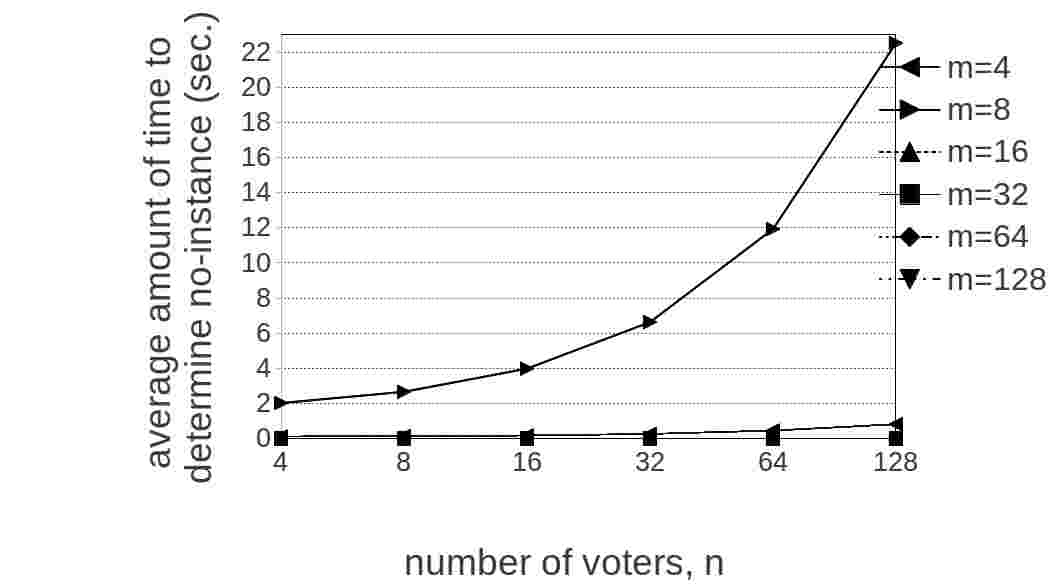}
	\caption{Average time the algorithm needs to determine no-instance of 
		constructive control by runoff-partition  of candidates in model TP
	in Bucklin elections in the TM model. The maximum is $22,53$ seconds.}
\end{figure}
\begin{figure}[ht]
\centering
	\includegraphics[scale=0.3]{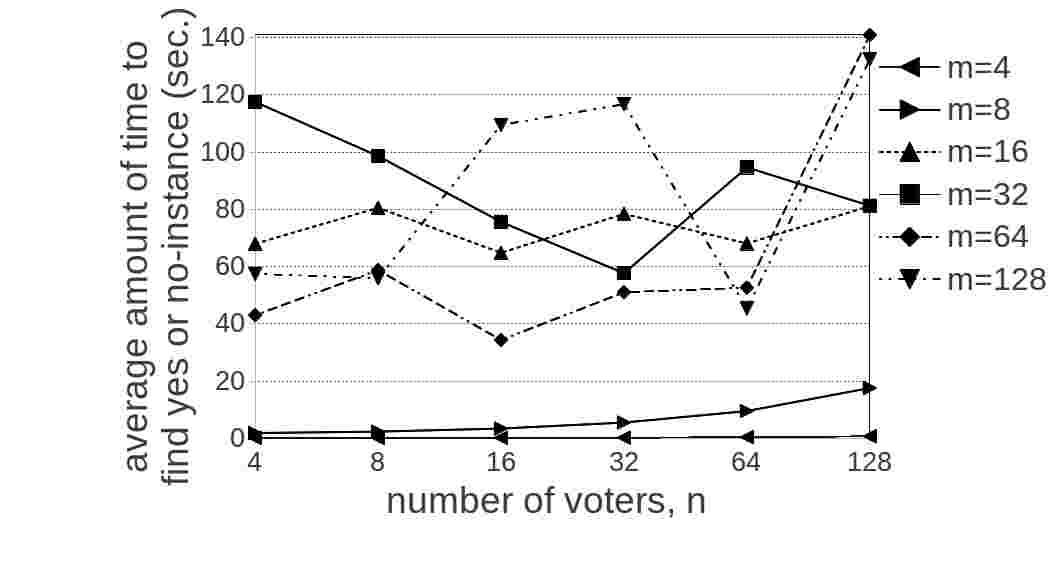}
	\caption{Average time the algorithm needs to give a definite output for 
	constructive control by runoff-partition  of candidates in model TP
	in Bucklin elections in the TM model. The maximum is $140,93$ seconds.}
\end{figure}

\clearpage
\subsection{Destructive Control by Runoff Partition of Candidates in Model TP}
\begin{center}
\begin{figure}[ht]
\centering
	\includegraphics[scale=0.3]{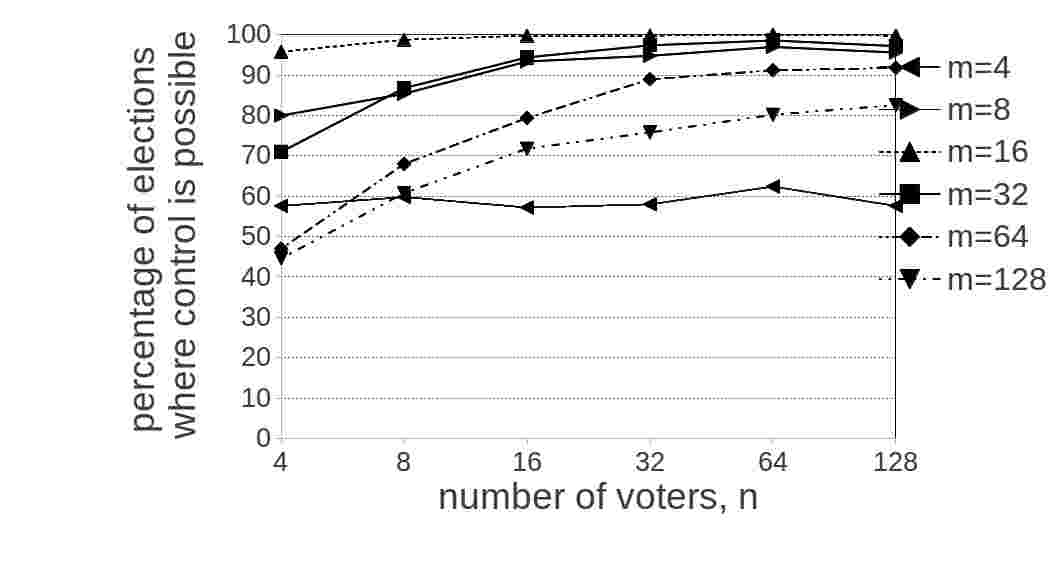}
		\caption{Results for Bucklin voting in the IC model for 
destructive control by runoff-partition  of candidates in model TP. Number of candidates is fixed. }
\end{figure}

\newpage
\begin{figure}[ht]
\centering
	\includegraphics[scale=0.3]{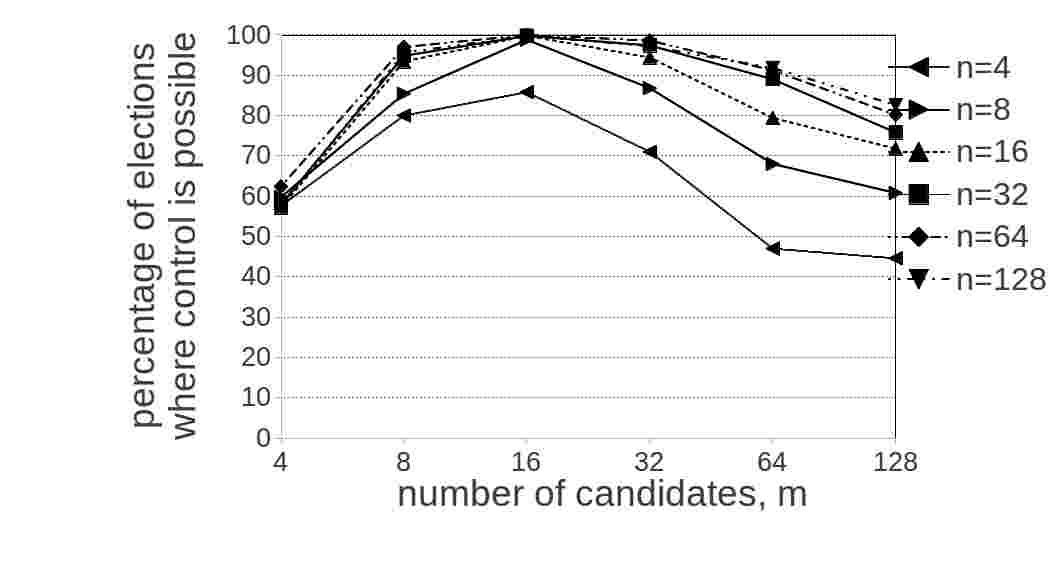}
		\caption{Results for Bucklin voting in the IC model for 
destructive control by runoff-partition  of candidates in model TP. Number of voters is fixed. }
\end{figure}
%
\newpage
\begin{figure}[ht]
\centering
	\includegraphics[scale=0.3]{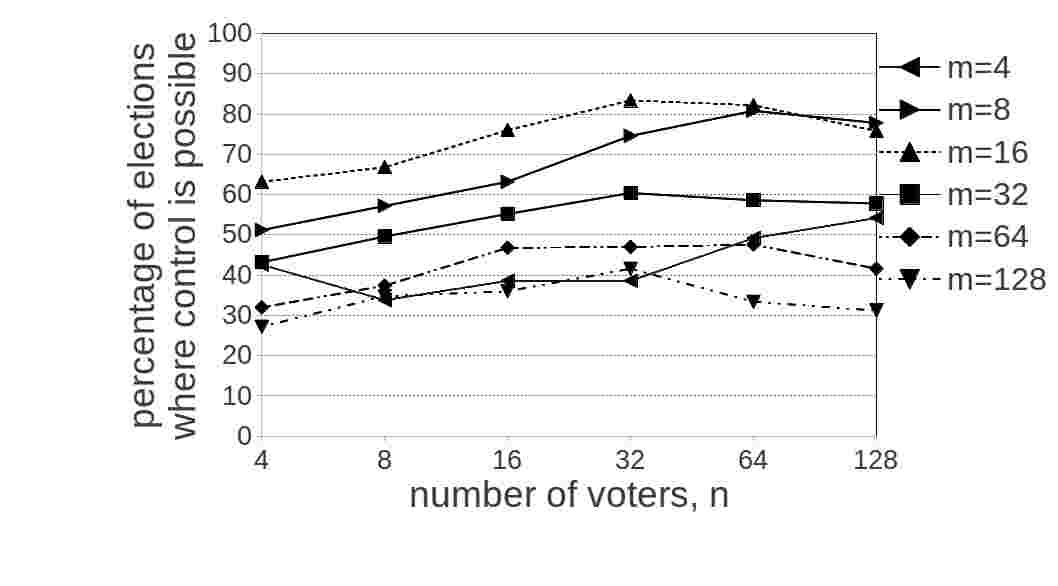}
	\caption{Results for Bucklin voting in the TM model for 
destructive control by runoff-partition  of candidates in model TP. Number of candidates is fixed. }
\end{figure}
%
\newpage
\begin{figure}[ht]
\centering
	\includegraphics[scale=0.3]{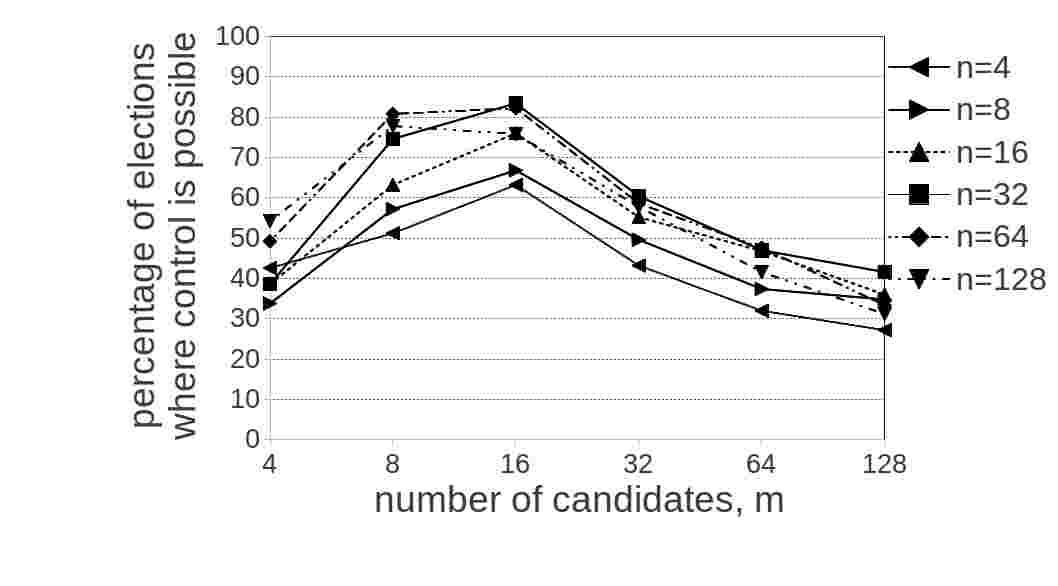}
	\caption{Results for Bucklin voting in the TM model for 
destructive control by runoff-partition  of candidates in model TP. Number of voters is fixed. }
\end{figure}
%
\end{center}

\clearpage
\subsubsection{Computational Costs}
\begin{figure}[ht]
\centering
	\includegraphics[scale=0.3]{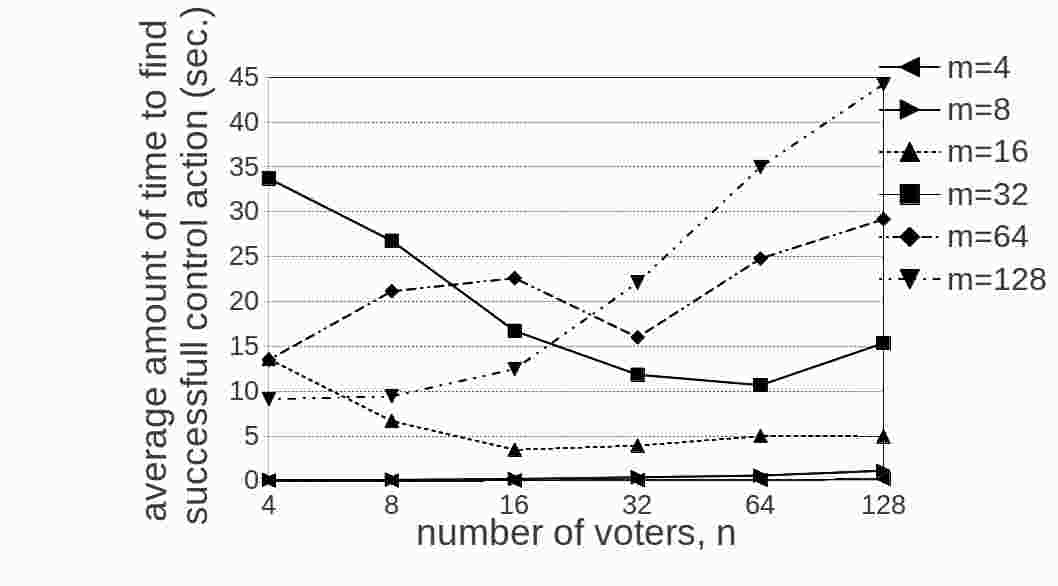}
	\caption{Average time the algorithm needs to find a successful control action for 
	destructive control by runoff-partition  of candidates in model TP
	in Bucklin elections in the IC model. The maximum is $44,23$ seconds.}
\end{figure}
\begin{figure}[ht]
\centering
	\includegraphics[scale=0.3]{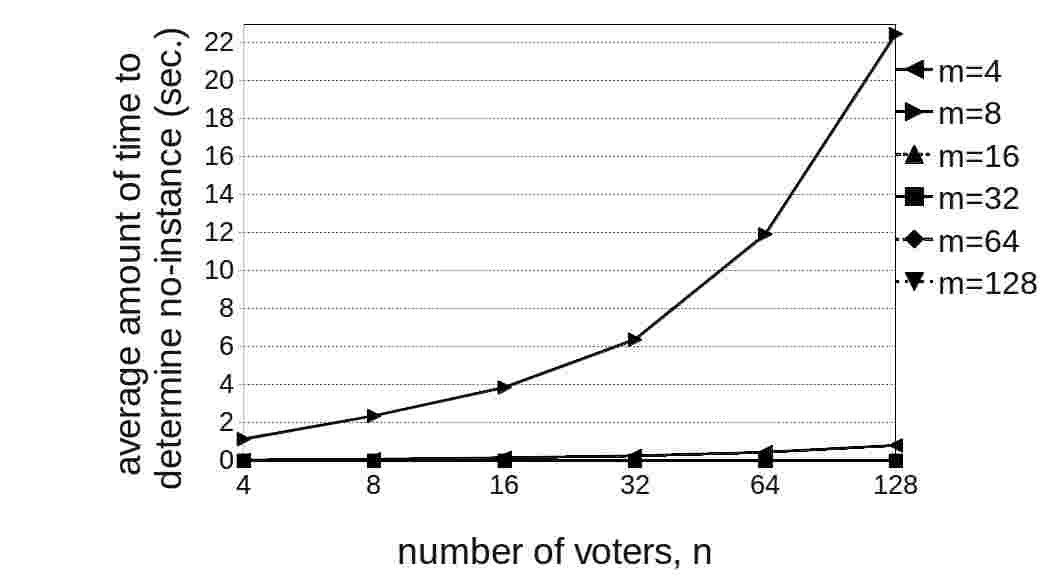}
	\caption{Average time the algorithm needs to determine no-instance of 
		destructive control by runoff-partition  of candidates in model TP
	in Bucklin elections in the IC model. The maximum is $22,47$ seconds.}
\end{figure}
\begin{figure}[ht]
\centering
	\includegraphics[scale=0.3]{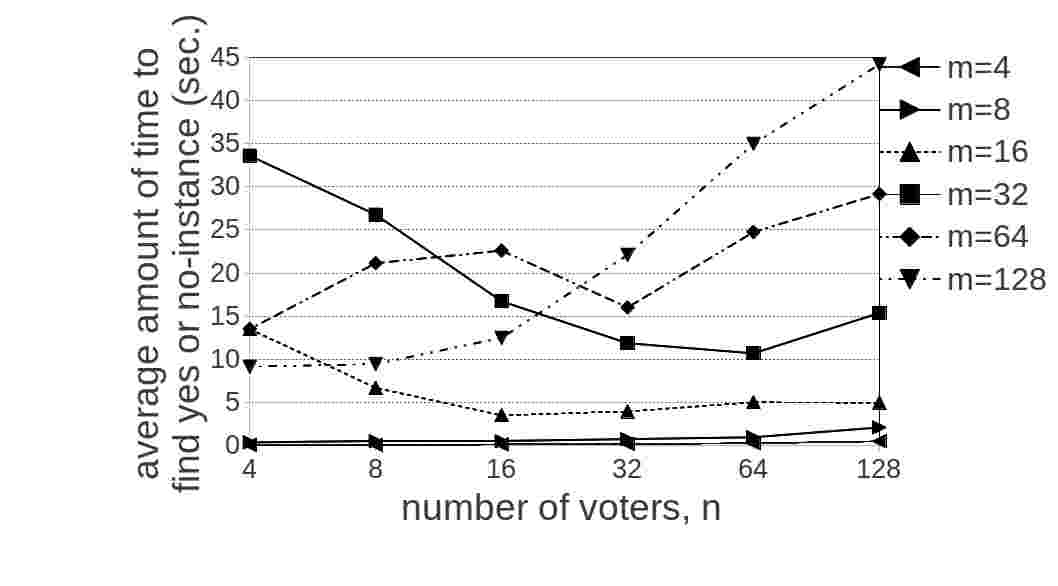}
	\caption{Average time the algorithm needs to give a definite output for 
	destructive control by runoff-partition  of candidates in model TP
	in Bucklin elections in the IC model. The maximum is $44,23$ seconds.}
\end{figure}
\begin{figure}[ht]
\centering
	\includegraphics[scale=0.3]{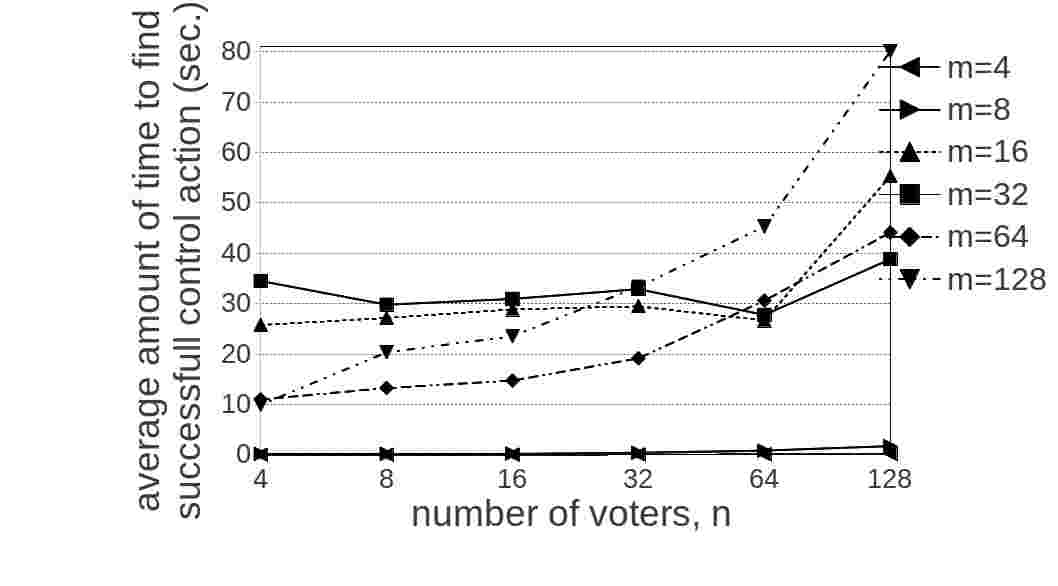}
	\caption{Average time the algorithm needs to find a successful control action for 
	destructive control by runoff-partition  of candidates in model TP
	in Bucklin elections in the TM model. The maximum is $80,1$ seconds.}
\end{figure}
\begin{figure}[ht]
\centering
	\includegraphics[scale=0.3]{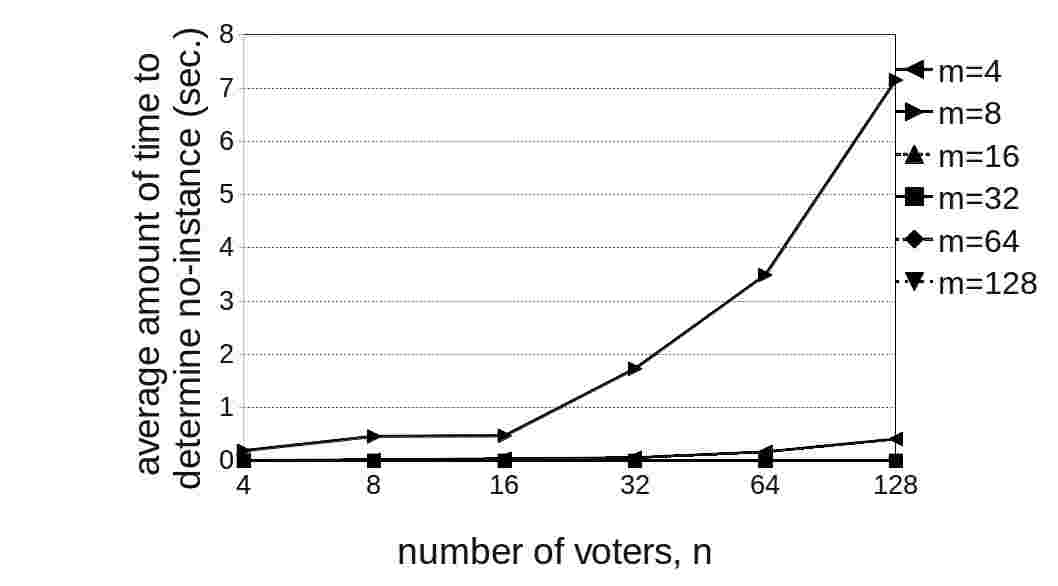}
	\caption{Average time the algorithm needs to determine no-instance of 
		destructive control by runoff-partition  of candidates in model TP
	in Bucklin elections in the TM model. The maximum is $7,15$ seconds.}
\end{figure}
\begin{figure}[ht]
\centering
	\includegraphics[scale=0.3]{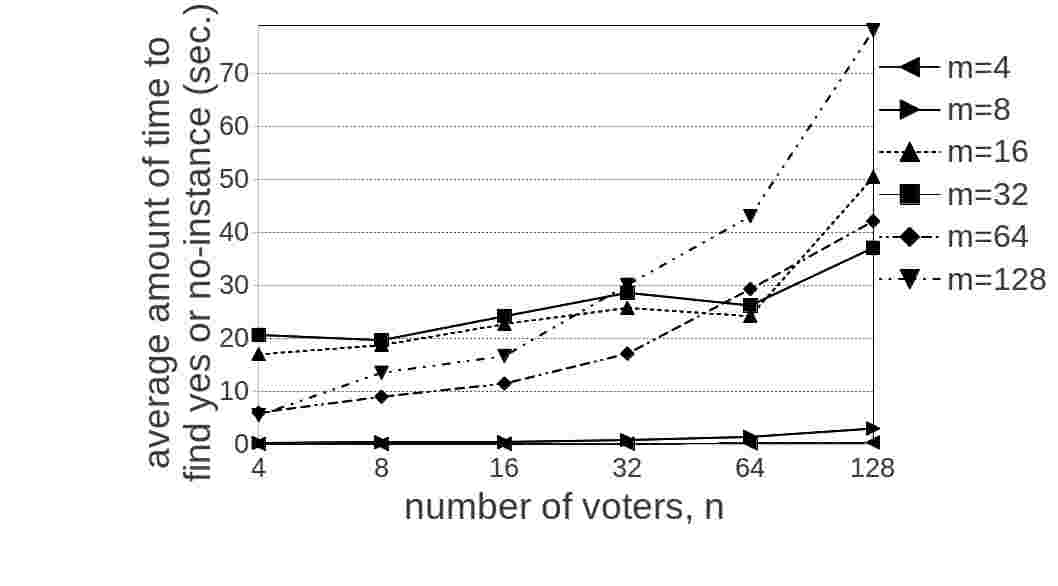}
	\caption{Average time the algorithm needs to give a definite output for 
	destructive control by runoff-partition  of candidates in model TP
	in Bucklin elections in the TM model. The maximum is $78,09$ seconds.}
\end{figure}

\clearpage
\subsection{Constructive Control by Adding Voters}
\begin{center}
\begin{figure}[ht]
\centering
	\includegraphics[scale=0.3]{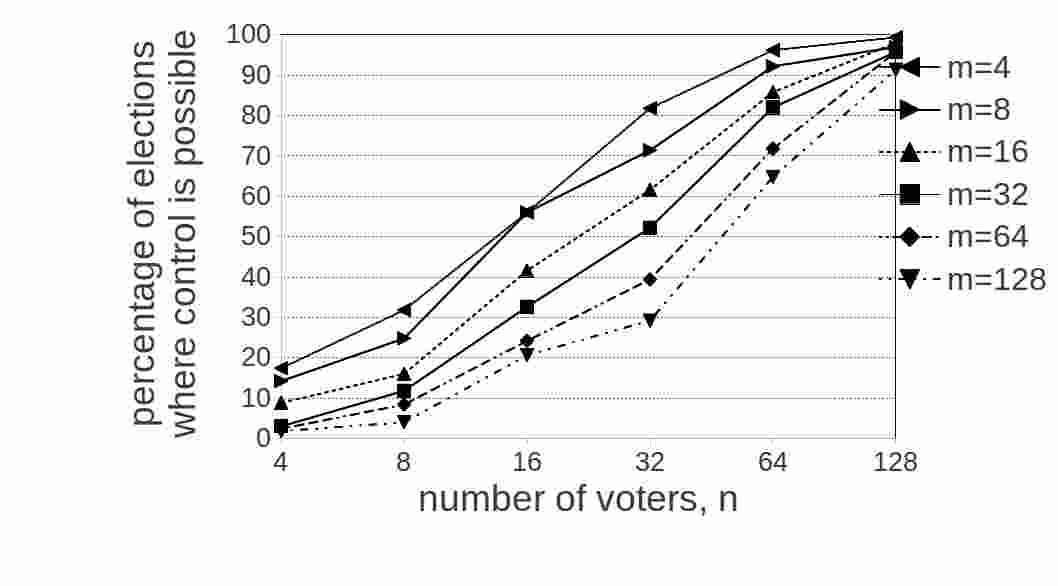}
	\caption{Results for Bucklin voting in the IC model for 
constructive control by adding voters. Number of candidates is fixed. }
\end{figure}


\newpage
\begin{figure}[ht]
\centering
	\includegraphics[scale=0.3]{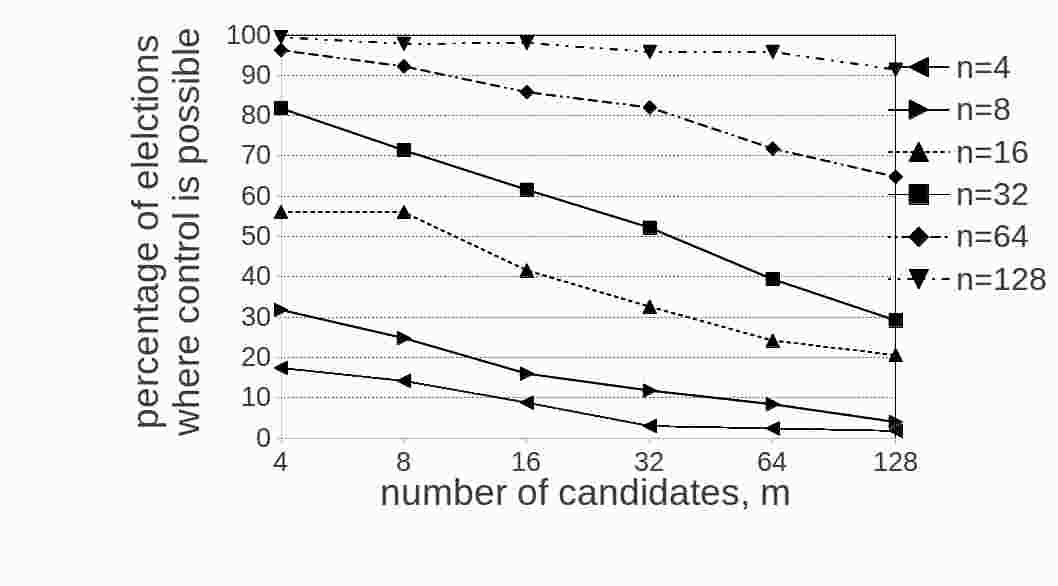}
	\caption{Results for Bucklin voting in the IC model for 
constructive control by adding voters. Number of voters is fixed. }
\end{figure}

%

\newpage
\begin{figure}[ht]
\centering
	\includegraphics[scale=0.3]{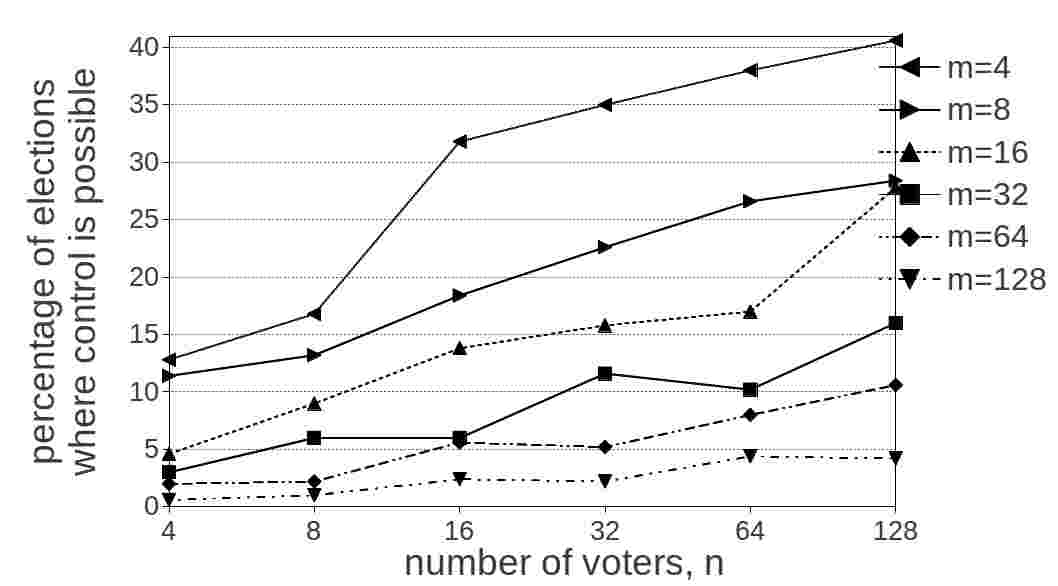}
	\caption{Results for Bucklin voting in the TM model for 
constructive control by adding voters. Number of candidates is fixed. }
\end{figure}

%

\newpage
\begin{figure}[ht]
\centering
	\includegraphics[scale=0.3]{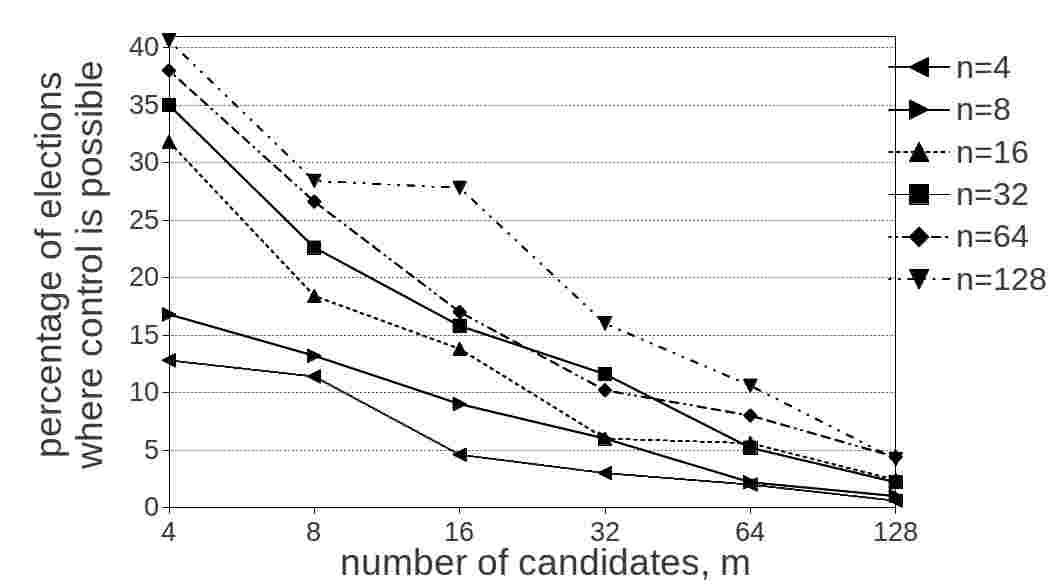}
	\caption{Results for Bucklin voting in the TM model for 
constructive control by adding voters. Number of voters is fixed.} 
\end{figure}

%
\end{center}

\clearpage
\subsubsection{Computational Costs}
\begin{figure}[ht]
\centering
	\includegraphics[scale=0.3]{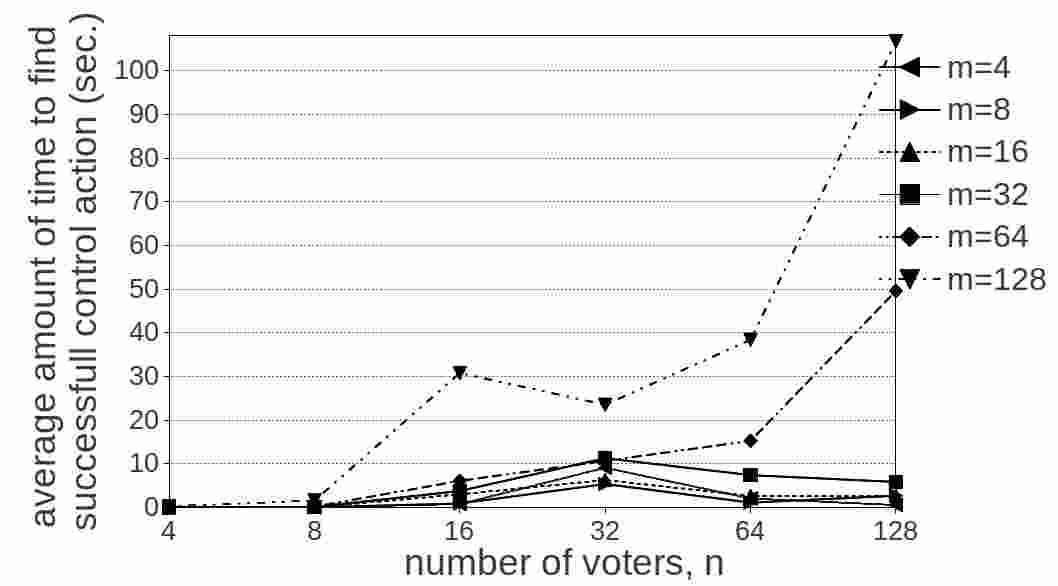}
	\caption{Average time the algorithm needs to find a successful control action for 
	constructive control by adding voters
	in Bucklin elections in the IC model. The maximum is $106,7$ seconds.}
\end{figure}

\begin{figure}[ht]
\centering
	\includegraphics[scale=0.3]{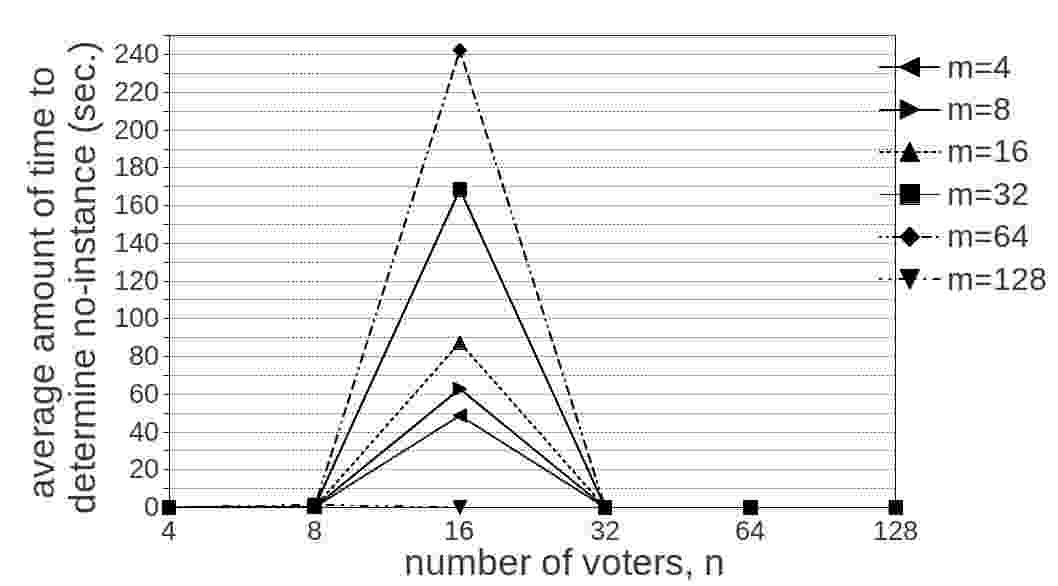}
	\caption{Average time the algorithm needs to determine no-instance of 
		constructive control by adding voters
	in Bucklin elections in the IC model. The maximum is $242,32$ seconds.}
\end{figure}

\begin{figure}[ht]
\centering
	\includegraphics[scale=0.3]{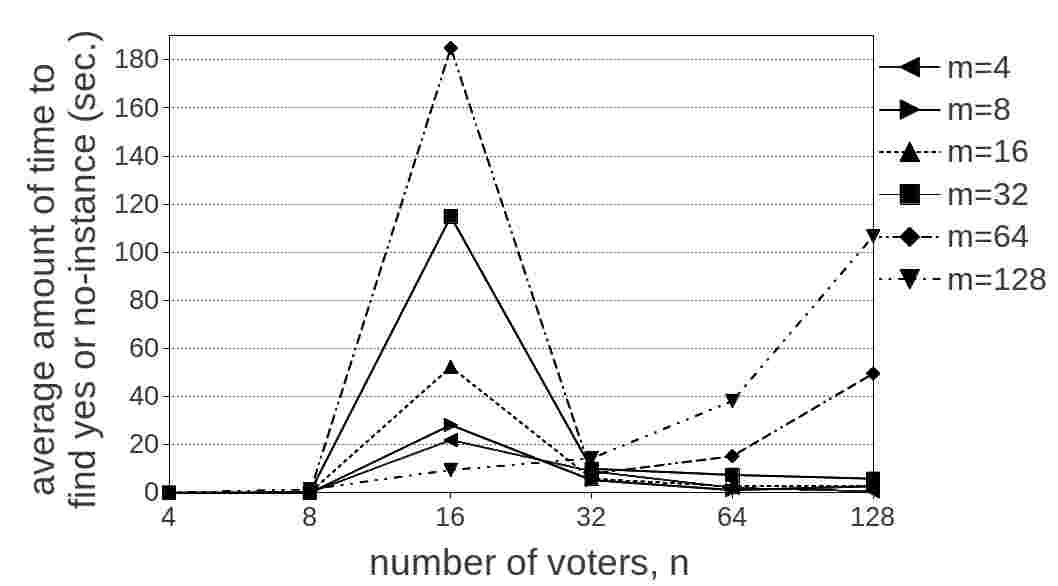}
	\caption{Average time the algorithm needs to give a definite output for 
	constructive control by adding voters
	in Bucklin elections in the IC model. The maximum is $184,92$ seconds.}
\end{figure}

\begin{figure}[ht]
\centering
	\includegraphics[scale=0.3]{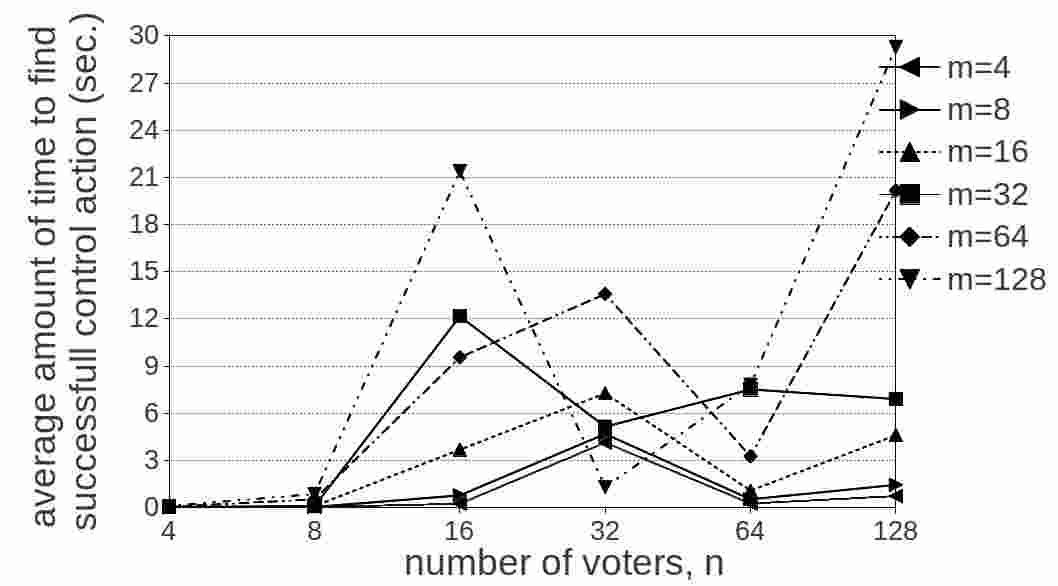}
	\caption{Average time the algorithm needs to find a successful control action for 
	constructive control by adding voters
	in Bucklin elections in the TM model. The maximum is $29,3$ seconds.}
\end{figure}

\begin{figure}[ht]
\centering
	\includegraphics[scale=0.3]{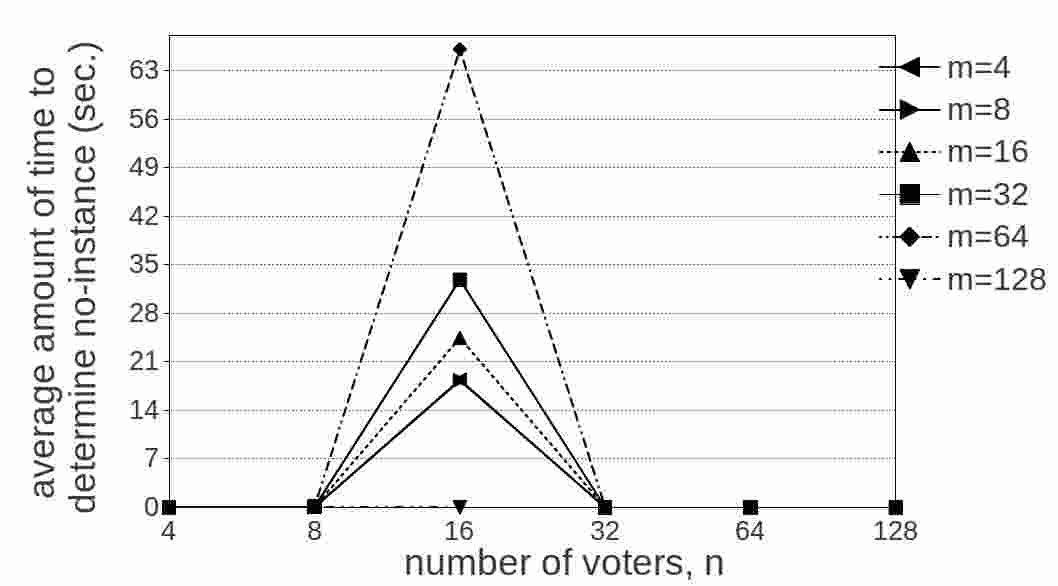}
	\caption{Average time the algorithm needs to determine no-instance of 
		constructive control by adding voters
	in Bucklin elections in the TM model. The maximum is $66,06$ seconds.}
\end{figure}
\begin{figure}[ht]
\centering
	\includegraphics[scale=0.3]{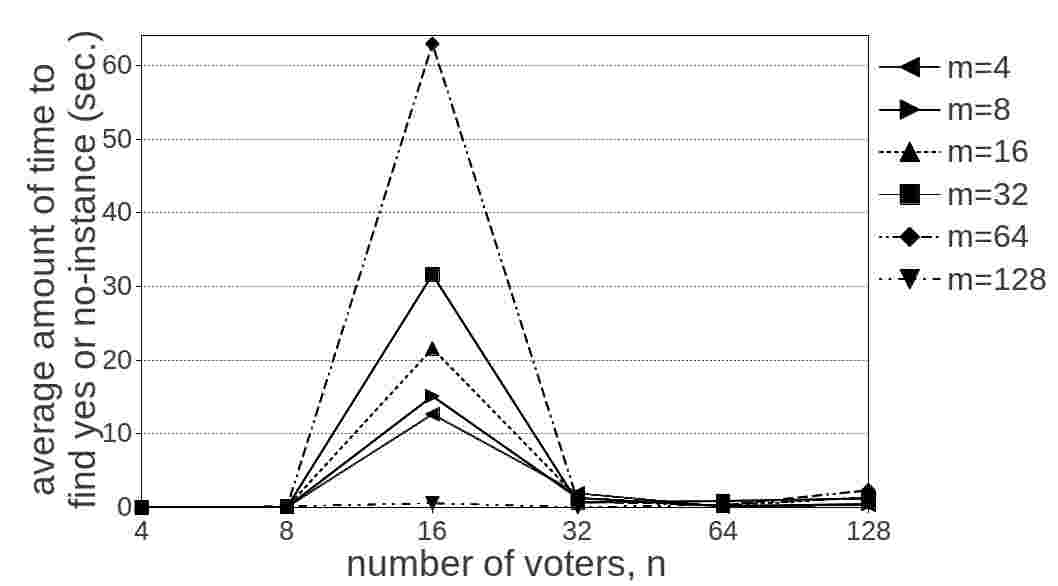}
	\caption{Average time the algorithm needs to give a definite output for 
	constructive control by adding voters
	in Bucklin elections in the TM model. The maximum is $62,9$ seconds.}
\end{figure}

\clearpage
\subsection{Constructive Control by Deleting Voters}
\begin{center}
\begin{figure}[ht]
\centering
	\includegraphics[scale=0.3]{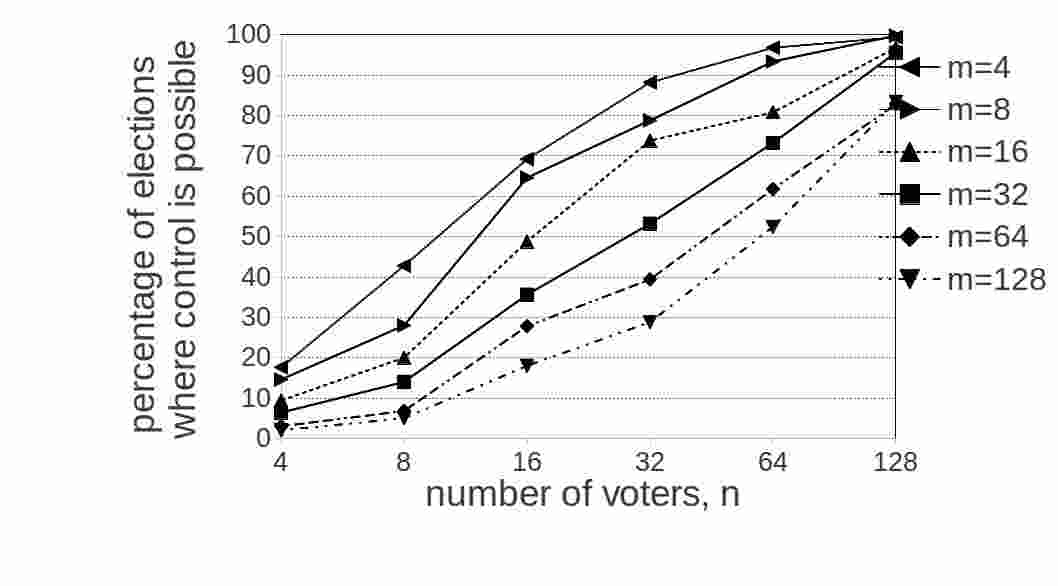}
		\caption{Results for Bucklin voting in the IC model for 
constructive control by deleting voters. Number of candidates is fixed. }
\end{figure}


\newpage
\begin{figure}[ht]
\centering
	\includegraphics[scale=0.3]{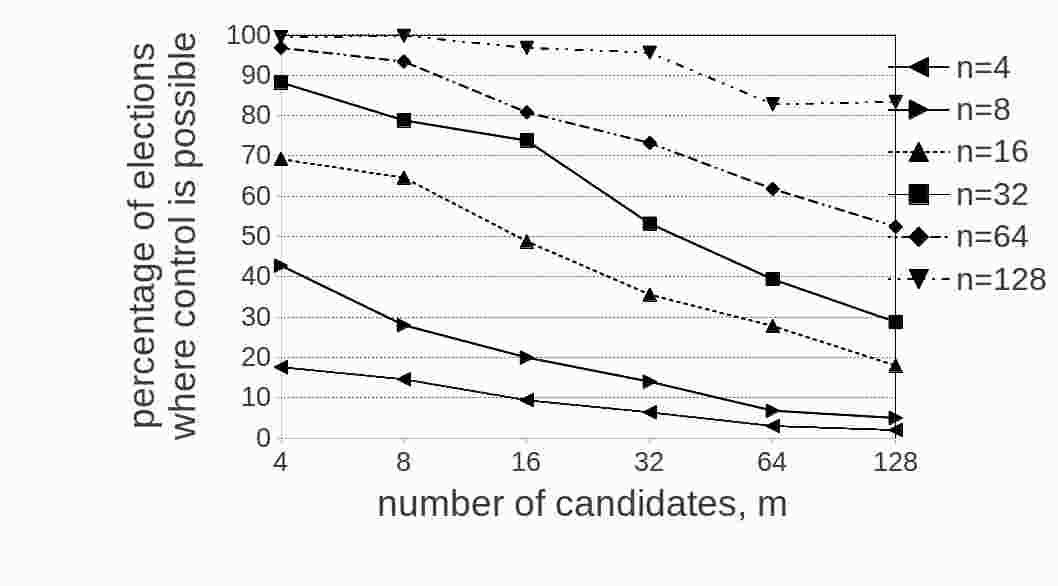}
		\caption{Results for Bucklin voting in the IC model for 
constructive control by deleting voters. Number of voters is fixed. }
\end{figure}

%

\newpage
\begin{figure}[ht]
\centering
	\includegraphics[scale=0.3]{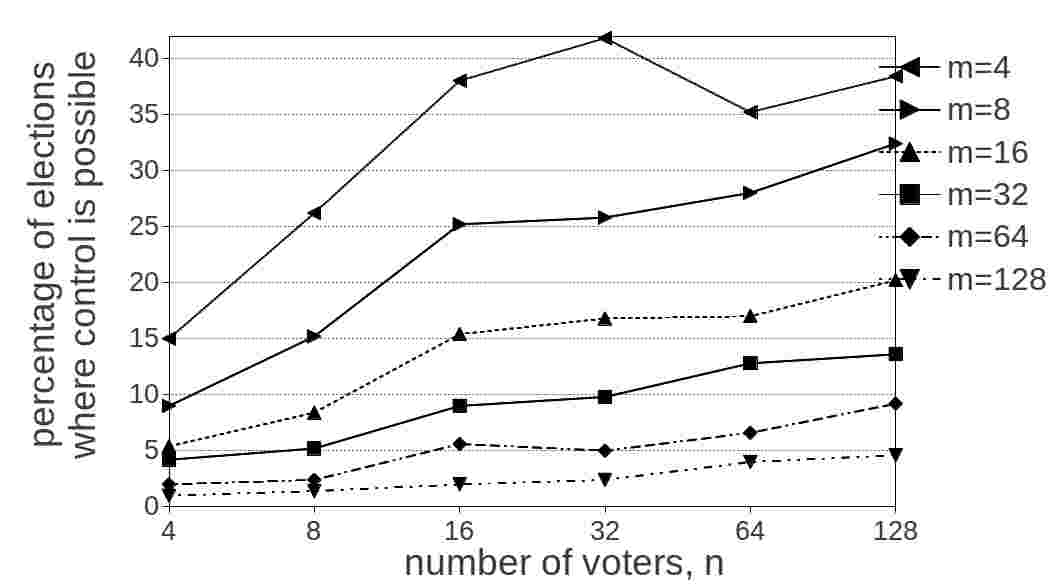}
	\caption{Results for Bucklin voting in the TM model for 
constructive control by deleting voters. Number of candidates is fixed. }
\end{figure}

%

\newpage
\begin{figure}[ht]
\centering
	\includegraphics[scale=0.3]{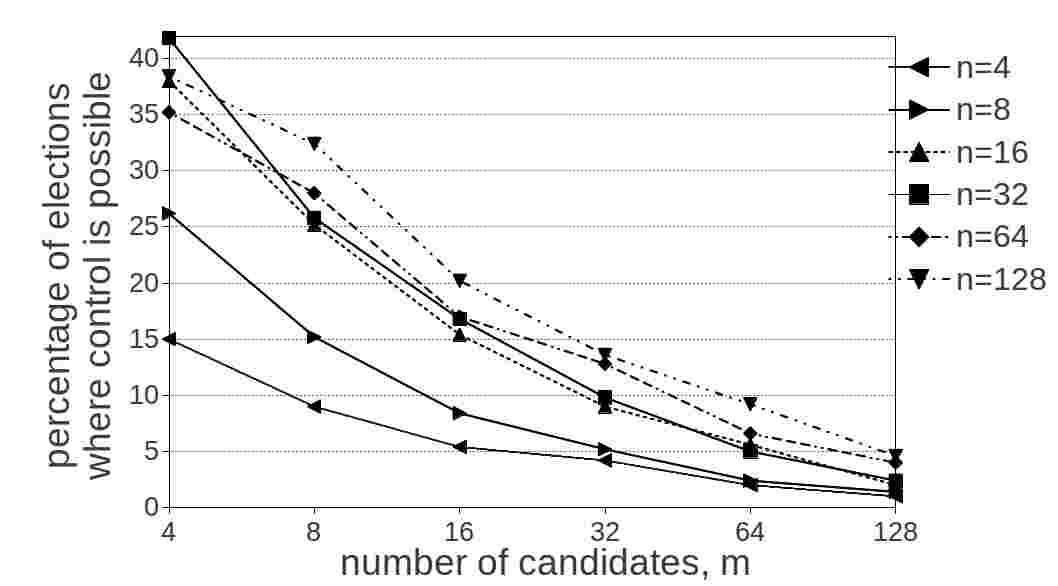}
	\caption{Results for Bucklin voting in the TM model for 
constructive control by deleting voters. Number of voters is fixed. }
\end{figure}

%
\end{center}

\clearpage
\subsubsection{Computational Costs}
\begin{figure}[ht]
\centering
	\includegraphics[scale=0.3]{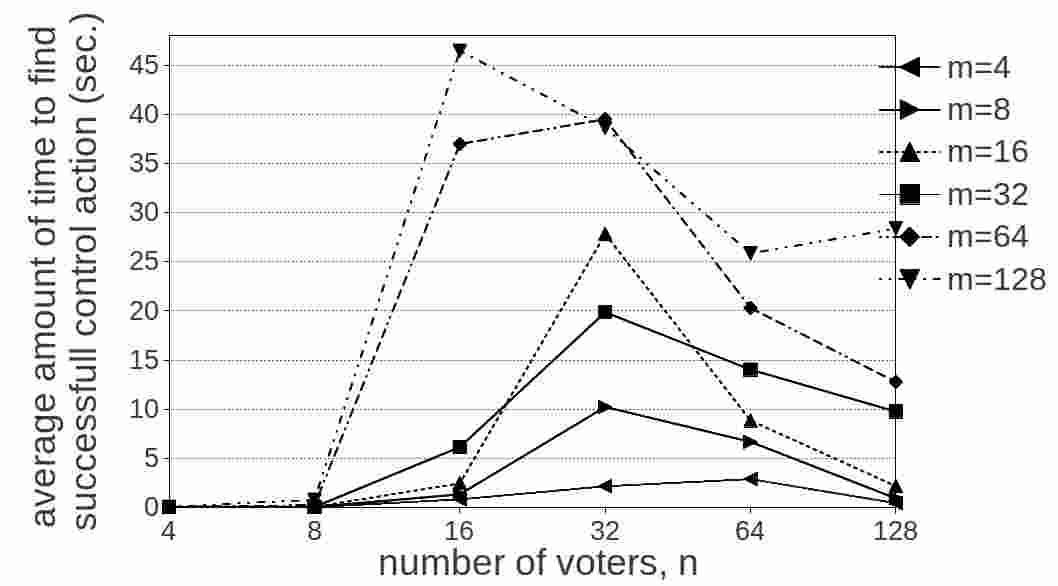}
	\caption{Average time the algorithm needs to find a successful control action for 
	constructive control by deleting voters
	in Bucklin elections in the IC model. The maximum is $46,5$ seconds.}
\end{figure}

\begin{figure}[ht]
\centering
	\includegraphics[scale=0.3]{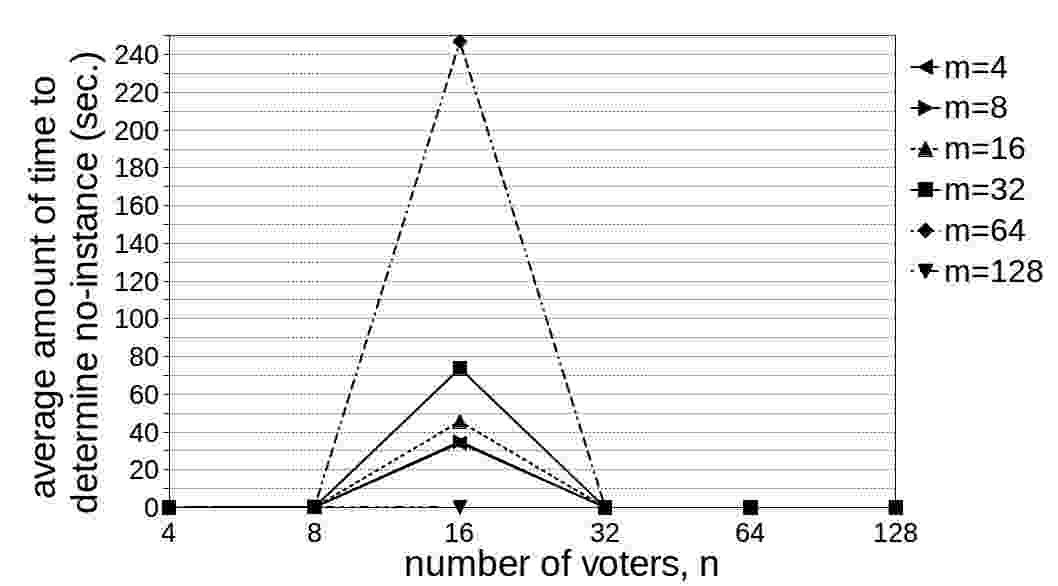}
	\caption{Average time the algorithm needs to determine no-instance of 
		constructive control by deleting voters
	in Bucklin elections in the IC model. The maximum is $247,02$ seconds.}
\end{figure}

\begin{figure}[ht]
\centering
	\includegraphics[scale=0.3]{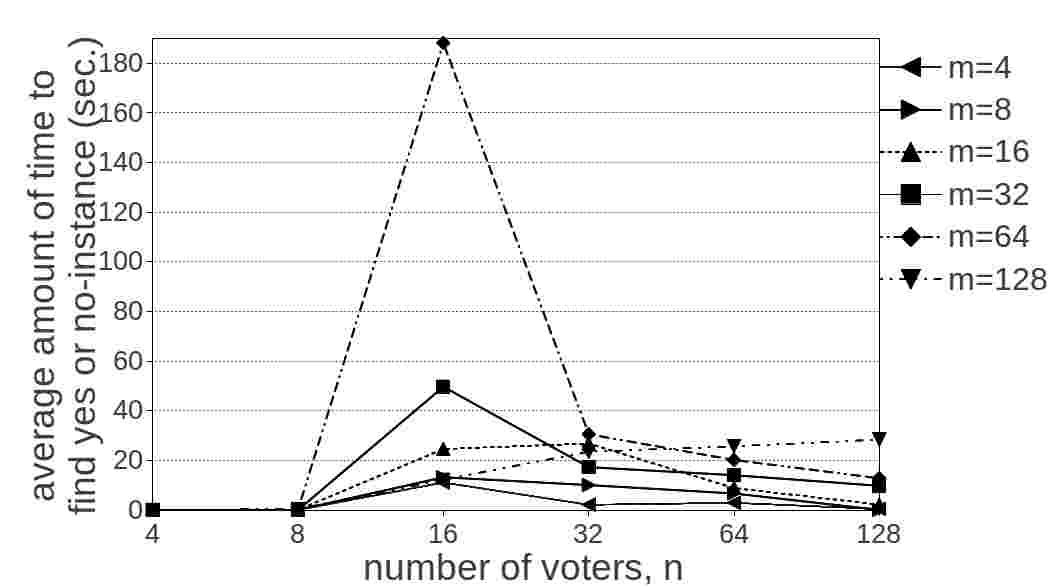}
	\caption{Average time the algorithm needs to give a definite output for 
	constructive control by deleting voters
	in Bucklin elections in the IC model. The maximum is $188,16$ seconds.}
\end{figure}

\begin{figure}[ht]
\centering
	\includegraphics[scale=0.3]{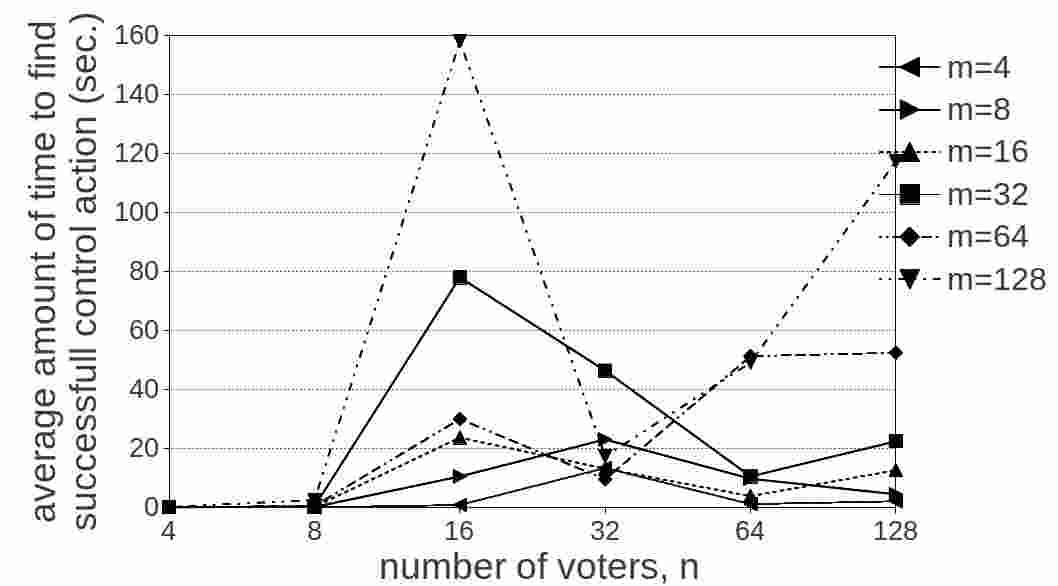}
	\caption{Average time the algorithm needs to find a successful control action for 
	constructive control by deleting voters
	in Bucklin elections in the TM model. The maximum is $158,2$ seconds.}
\end{figure}

\begin{figure}[ht]
\centering
	\includegraphics[scale=0.3]{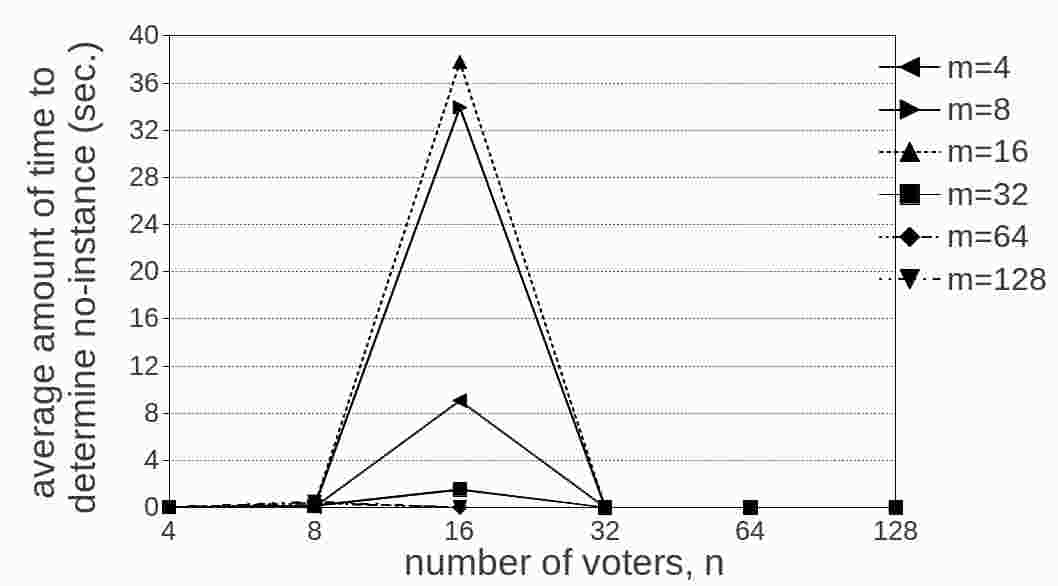}
	\caption{Average time the algorithm needs to determine no-instance of 
		constructive control by deleting voters
	in Bucklin elections in the TM model. The maximum is $37,76$ seconds.}
\end{figure}

\begin{figure}[ht]
\centering
	\includegraphics[scale=0.3]{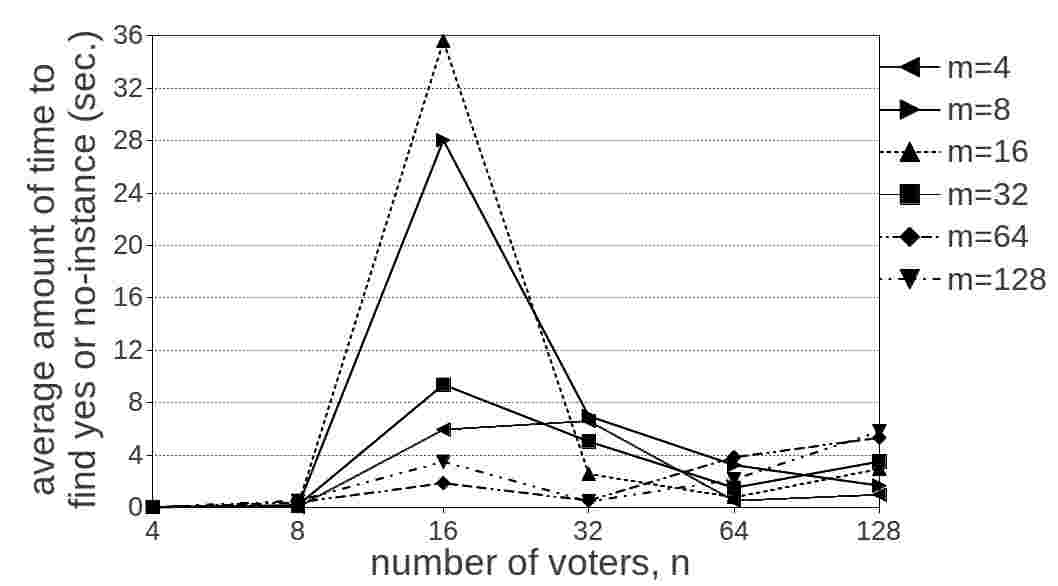}
	\caption{Average time the algorithm needs to give a definite output for 
	constructive control by deleting voters
	in Bucklin elections in the TM model. The maximum is $35,6$ seconds.}
\end{figure}

\clearpage
\subsection{Constructive Control by Partition of Voters in Model TE}
\begin{center}
\begin{figure}[ht]
\centering
	\includegraphics[scale=0.3]{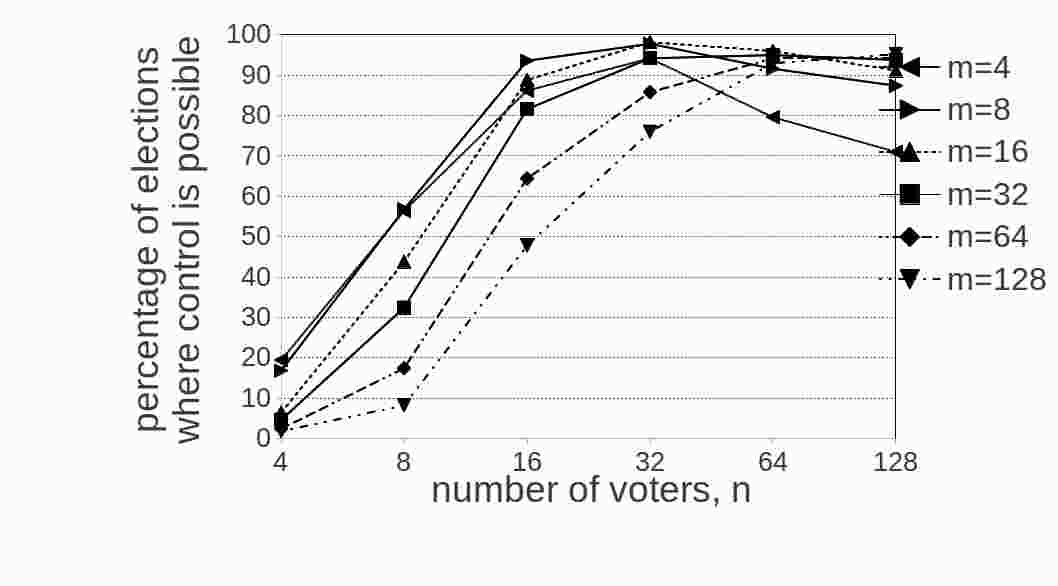}
	\caption{Results for Bucklin voting in the IC model for 
constructive control by partition of voters in model TE.  Number of candidates is fixed. }
\end{figure}

%
\newpage
\begin{figure}[ht]
\centering
	\includegraphics[scale=0.3]{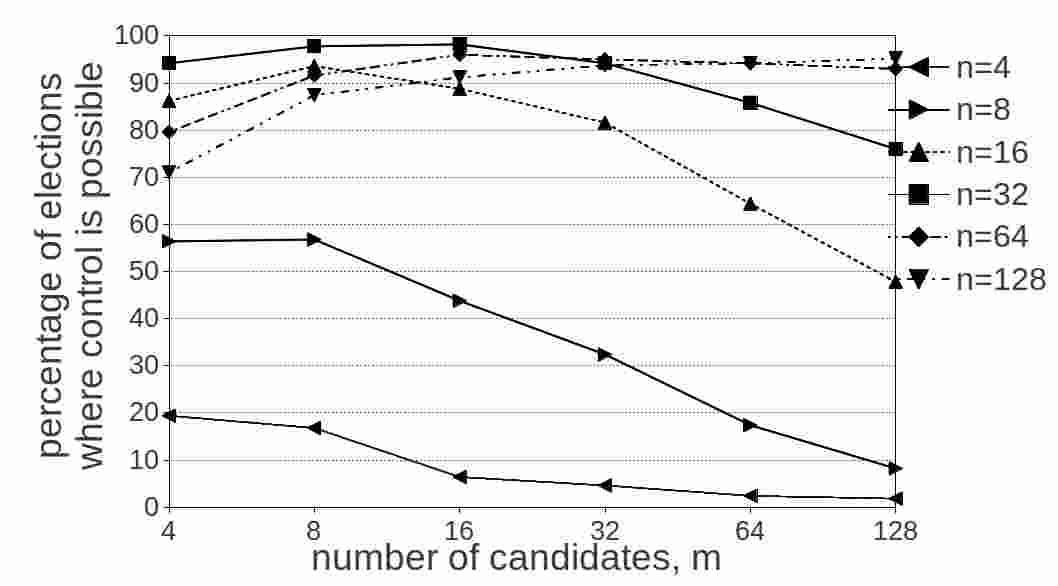}
	\caption{Results for Bucklin voting in the IC model for 
constructive control by partition of voters in model TE.  Number of voters is fixed. }
\end{figure}

%
%

\newpage
\begin{figure}[ht]
\centering
	\includegraphics[scale=0.3]{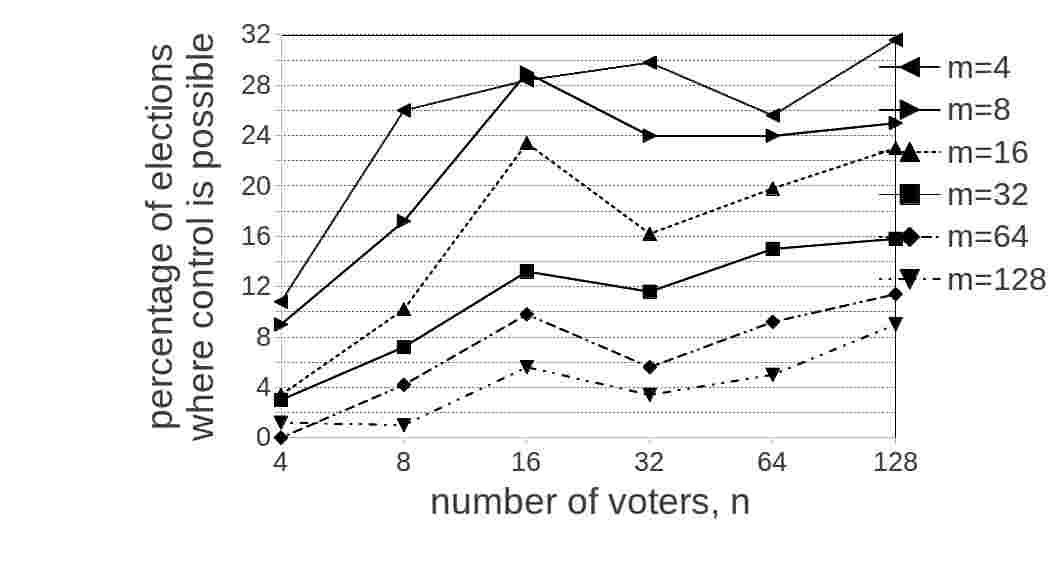}
	\caption{Results for Bucklin voting in the TM model for 
constructive control by partition of voters in model TE.  Number of candidates is fixed. }
\end{figure}

%
%

\newpage
\begin{figure}[ht]
\centering
	\includegraphics[scale=0.3]{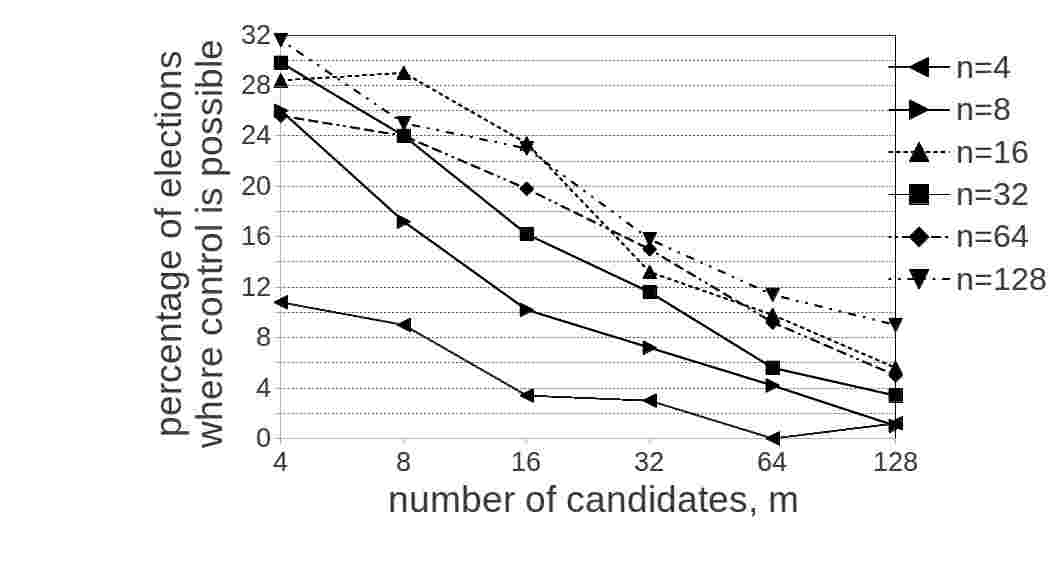}
	\caption{Results for Bucklin voting in the TM model for 
constructive control by partition of voters in model TE.  Number of voters is fixed. }
\end{figure}

%
\end{center}

\clearpage
\subsubsection{Computational Costs}
\begin{figure}[ht]
\centering
	\includegraphics[scale=0.3]{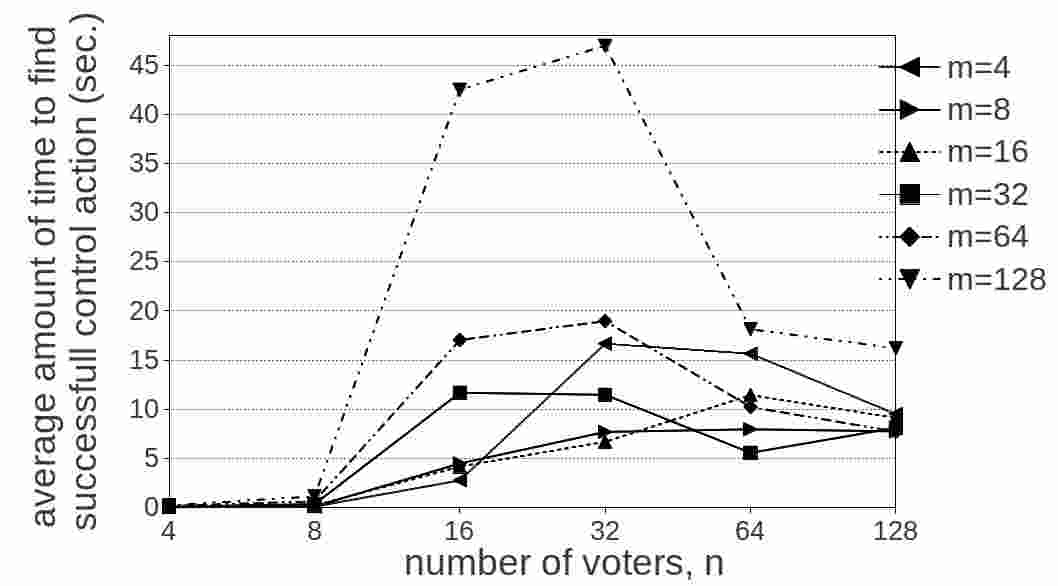}
	\caption{Average time the algorithm needs to find a successful control action for 
	constructive control by partition of voters in model TE
	in Bucklin elections in the IC model. The maximum is $47$ seconds.}
\end{figure}

\begin{figure}[ht]
\centering
	\includegraphics[scale=0.3]{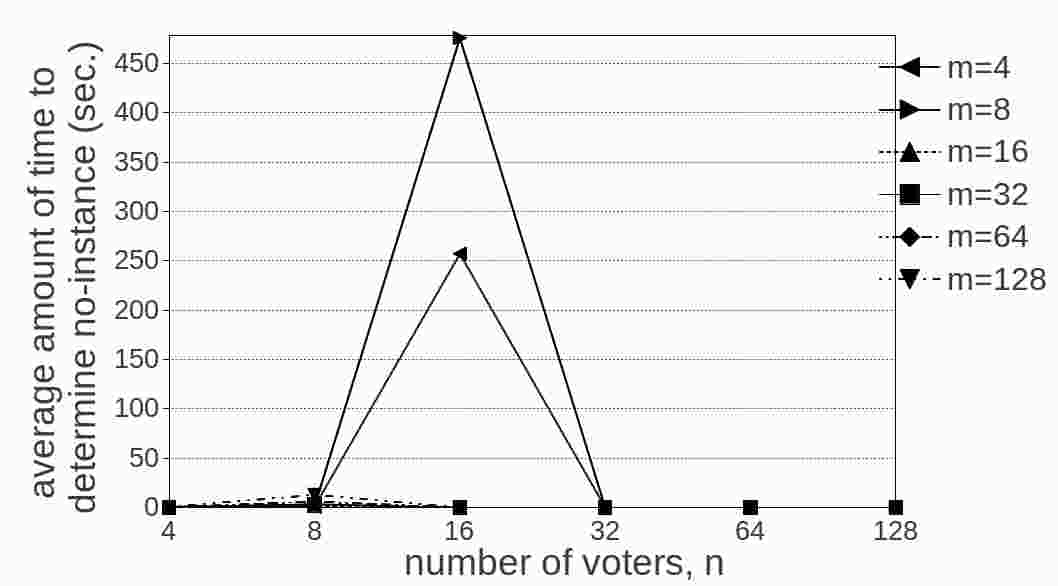}
	\caption{Average time the algorithm needs to determine no-instance of 
		constructive control by partition of voters in model TE
	in Bucklin elections in the IC model. The maximum is $475,79$ seconds.}
\end{figure}

\begin{figure}[ht]
\centering
	\includegraphics[scale=0.3]{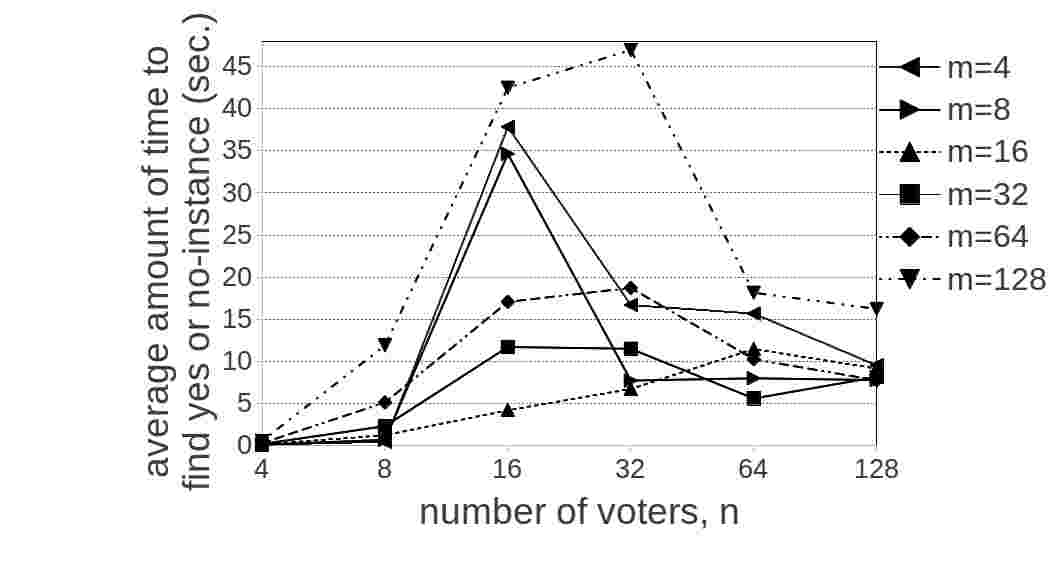}
	\caption{Average time the algorithm needs to give a definite output for 
	constructive control by partition of voters in model TE
	in Bucklin elections in the IC model. The maximum is $47$ seconds.}
\end{figure}

\begin{figure}[ht]
\centering
	\includegraphics[scale=0.3]{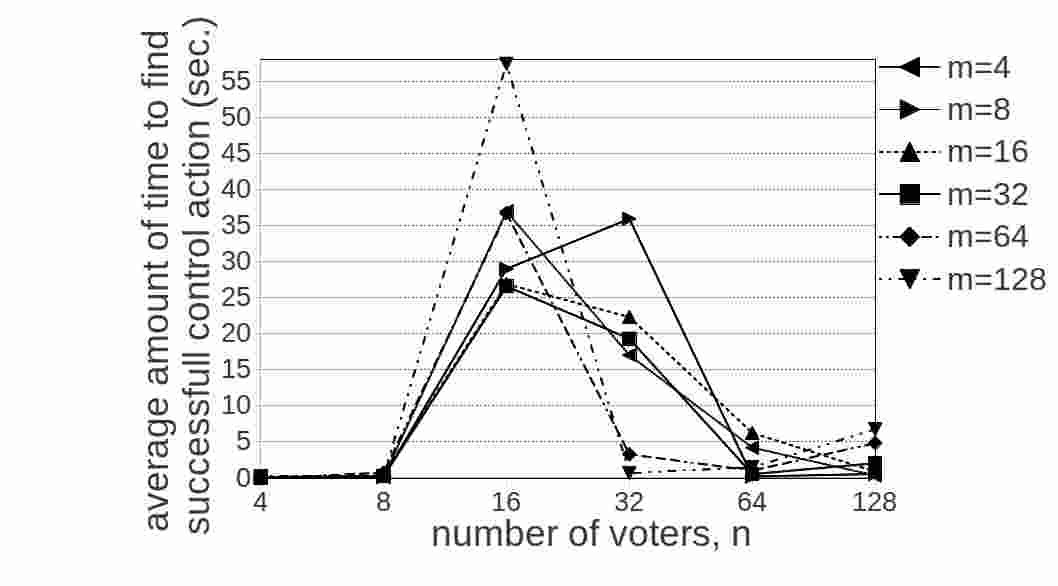}
	\caption{Average time the algorithm needs to find a successful control action for 
	constructive control by partition of voters in model TE
	in Bucklin elections in the TM model. The maximum is $57,39$ seconds.}
\end{figure}

\begin{figure}[ht]
\centering
	\includegraphics[scale=0.3]{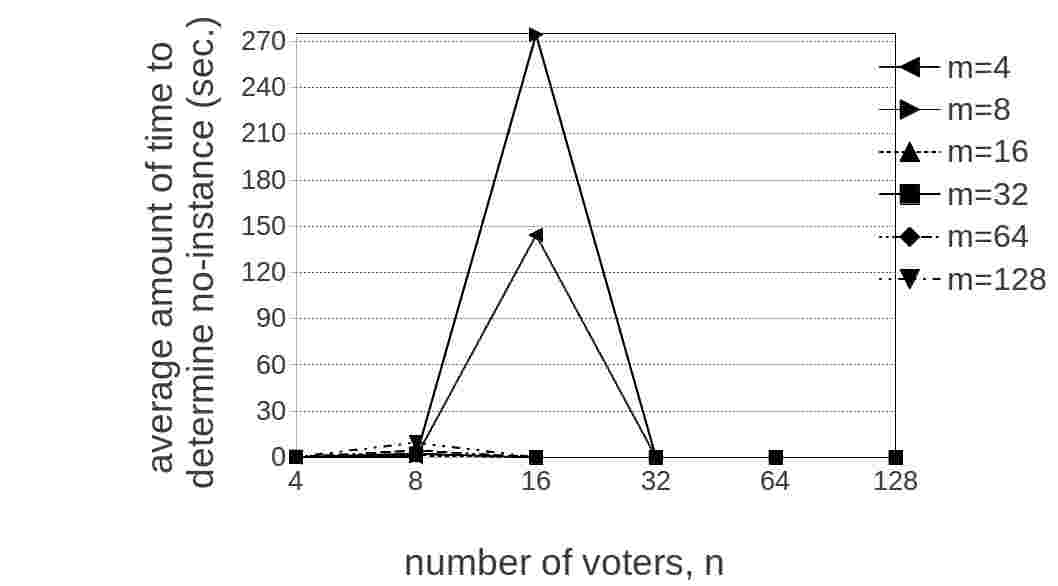}
	\caption{Average time the algorithm needs to determine no-instance of 
		constructive control by partition of voters in model TE
	in Bucklin elections in the TM model. The maximum is $274,43$ seconds.}
\end{figure}

\begin{figure}[ht]
\centering
	\includegraphics[scale=0.3]{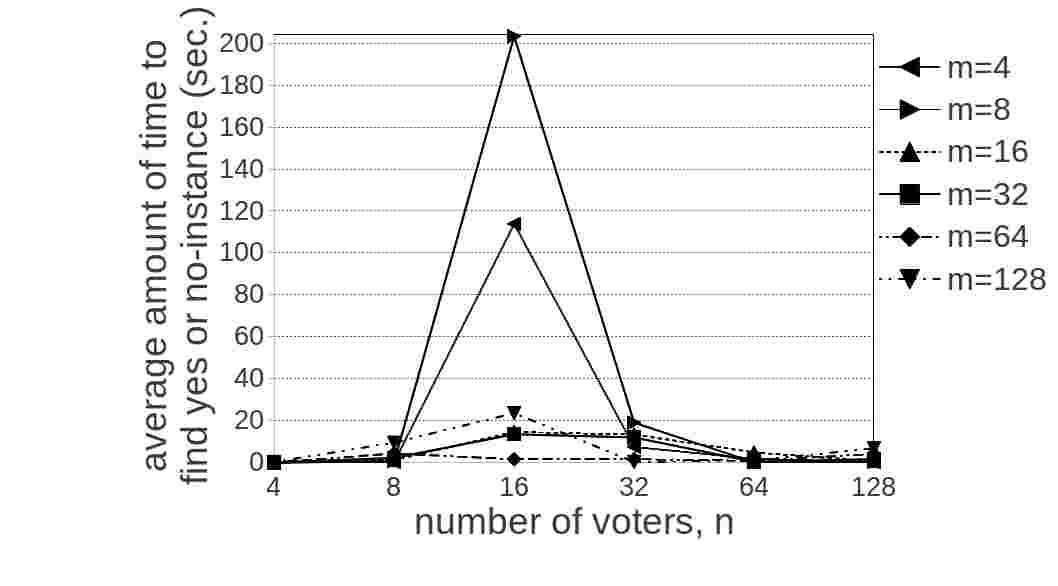}
	\caption{Average time the algorithm needs to give a definite output for 
	constructive control by partition of voters in model TE
	in Bucklin elections in the TM model. The maximum is $203,24$ seconds.}
\end{figure}

\clearpage
\subsection{Destructive Control by Partition of Voters in Model TE}
\begin{center}
\begin{figure}[ht]
\centering
	\includegraphics[scale=0.3]{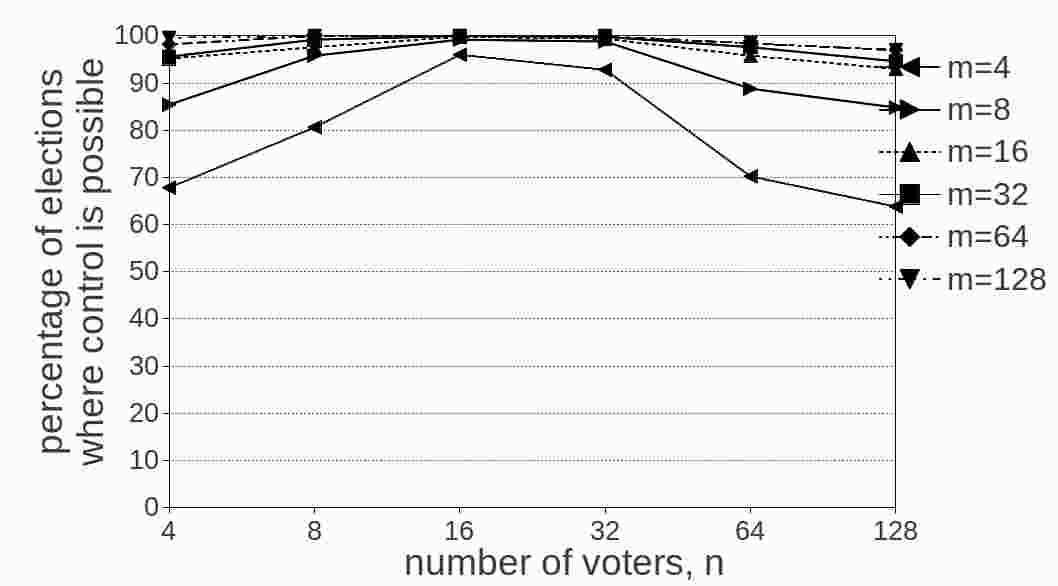}
	\caption{Results for Bucklin voting in the IC model for 
destructive control by partition of voters in model TE.  Number of candidates is fixed.}
\end{figure}

%
\newpage
\begin{figure}[ht]
\centering
	\includegraphics[scale=0.3]{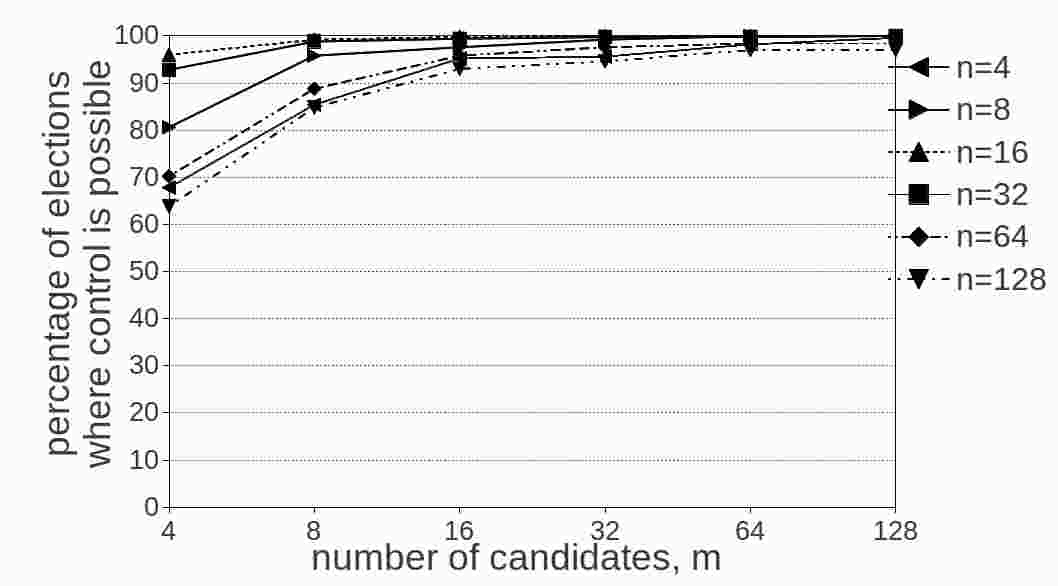}
	\caption{Results for Bucklin voting in the IC model for 
destructive control by partition of voters in model TE.  Number of voters is fixed.}
\end{figure}

%
%
\newpage
\begin{figure}[ht]
\centering
	\includegraphics[scale=0.3]{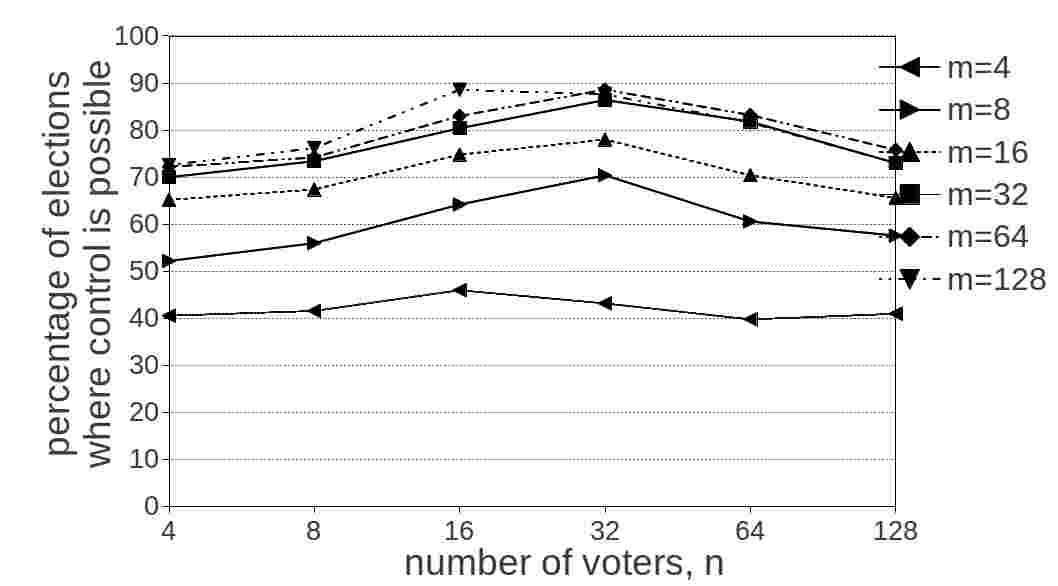}
	\caption{Results for Bucklin voting in the TM model for 
destructive control by partition of voters in model TE.  Number of candidates is fixed.}
\end{figure}

%
%

\newpage
\begin{figure}[ht]
\centering
	\includegraphics[scale=0.3]{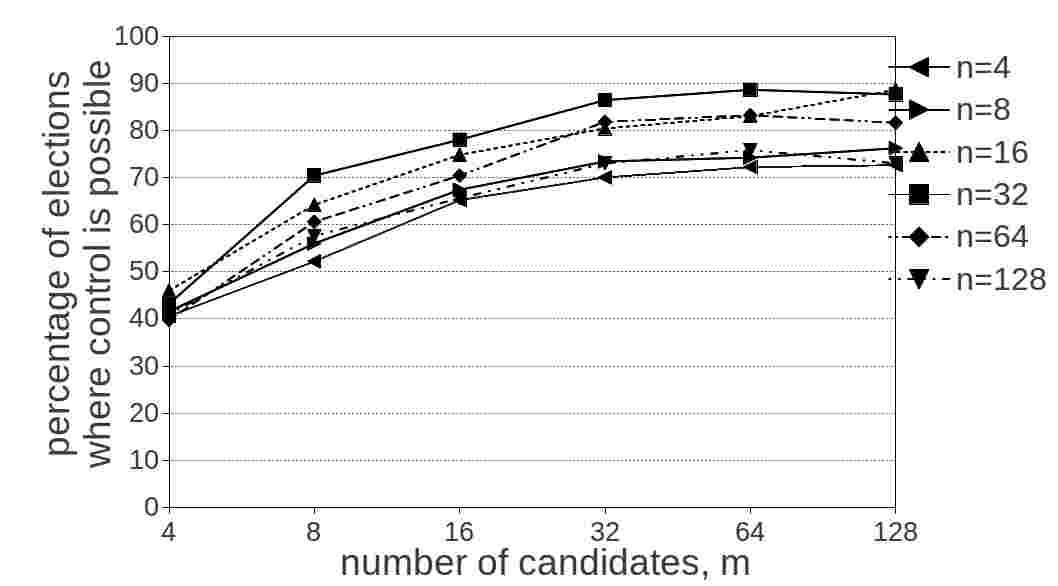}
	\caption{Results for Bucklin voting in the TM model for 
destructive control by partition of voters in model TE.  Number of voters is fixed.}
\end{figure}

%
\end{center}

\clearpage
\subsubsection{Computational Costs}
\begin{figure}[ht]
\centering
	\includegraphics[scale=0.3]{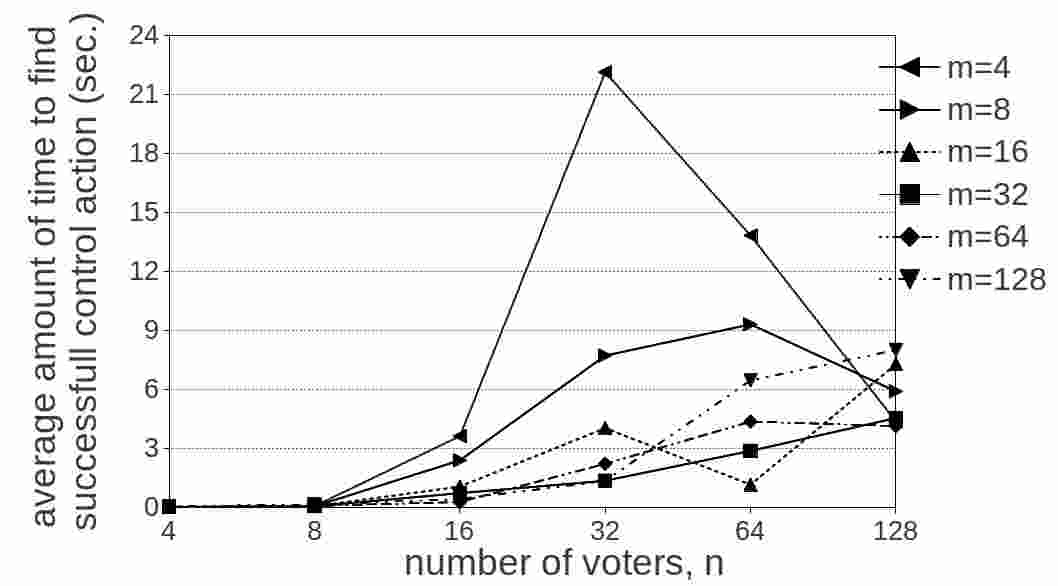}
	\caption{Average time the algorithm needs to find a successful control action for 
	destructive control by partition of voters in model TE
	in Bucklin elections in the IC model. The maximum is $22,15$ seconds.}
\end{figure}

\begin{figure}[ht]
\centering
	\includegraphics[scale=0.3]{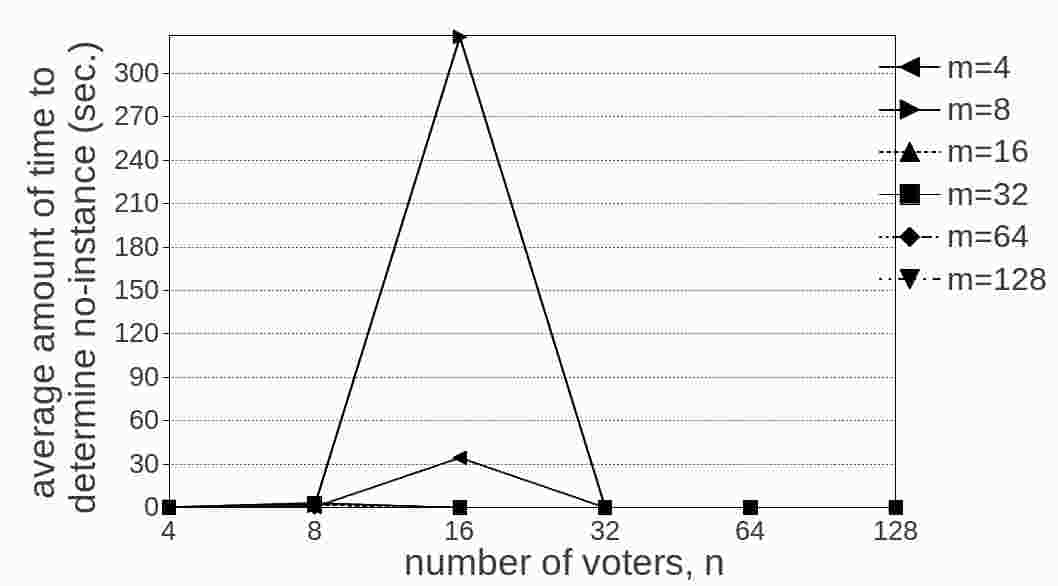}
	\caption{Average time the algorithm needs to determine no-instance of 
		destructive control by partition of voters in model TE
	in Bucklin elections in the IC model. The maximum is $325,26$ seconds.}
\end{figure}

\begin{figure}[ht]
\centering
	\includegraphics[scale=0.3]{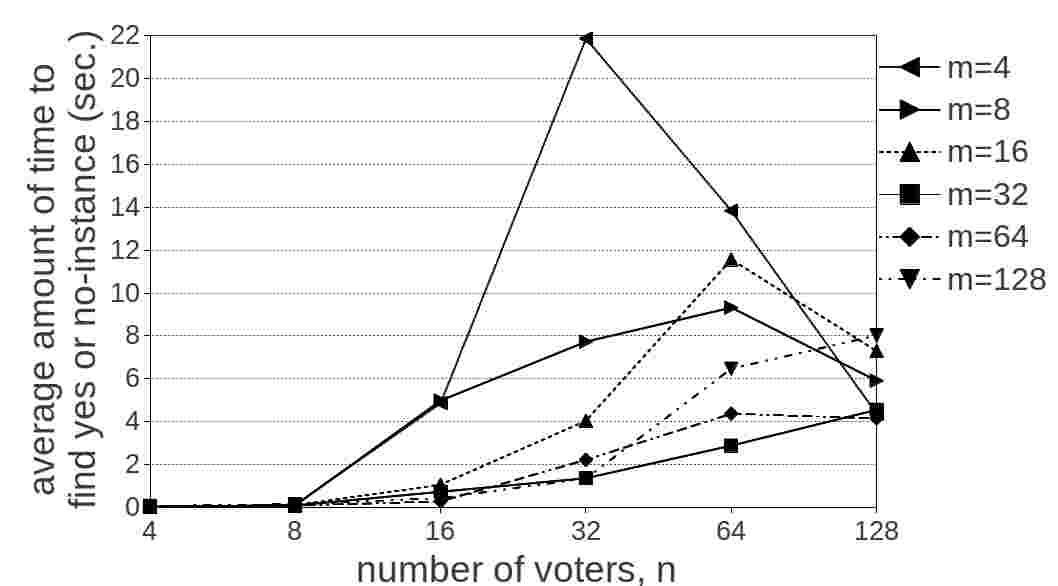}
	\caption{Average time the algorithm needs to give a definite output for 
	destructive control by partition of voters in model TE
	in Bucklin elections in the IC model. The maximum is $21,87$ seconds.}
\end{figure}

\begin{figure}[ht]
\centering
	\includegraphics[scale=0.3]{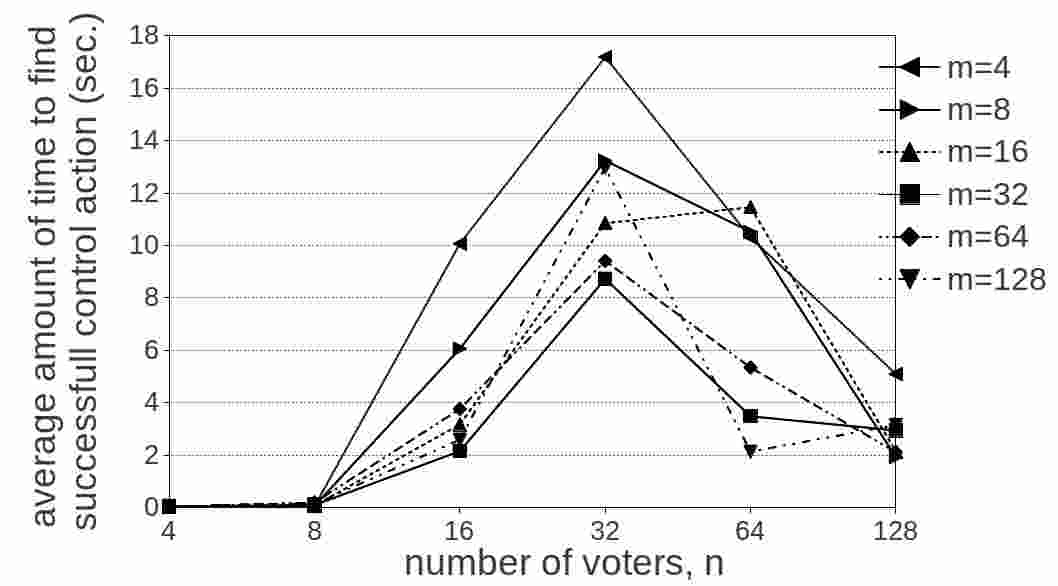}
	\caption{Average time the algorithm needs to find a successful control action for 
	destructive control by partition of voters in model TE
	in Bucklin elections in the TM model. The maximum is $17,19$ seconds.}
\end{figure}

\begin{figure}[ht]
\centering
	\includegraphics[scale=0.3]{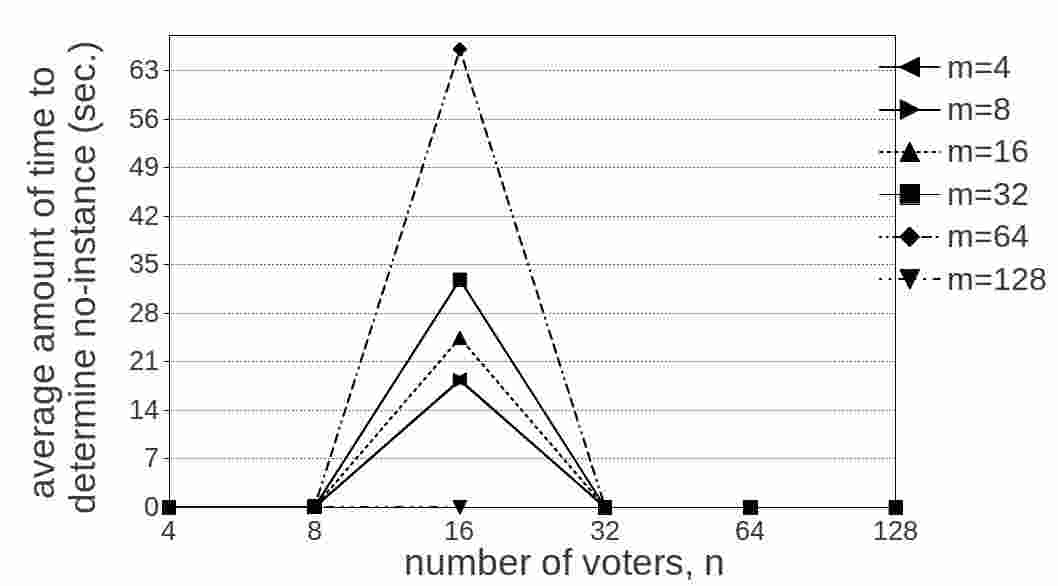}
	\caption{Average time the algorithm needs to determine no-instance of 
		destructive control by partition of voters in model TE
	in Bucklin elections in the TM model. The maximum is $57,94$ seconds.}
\end{figure}

\begin{figure}[ht]
\centering
	\includegraphics[scale=0.3]{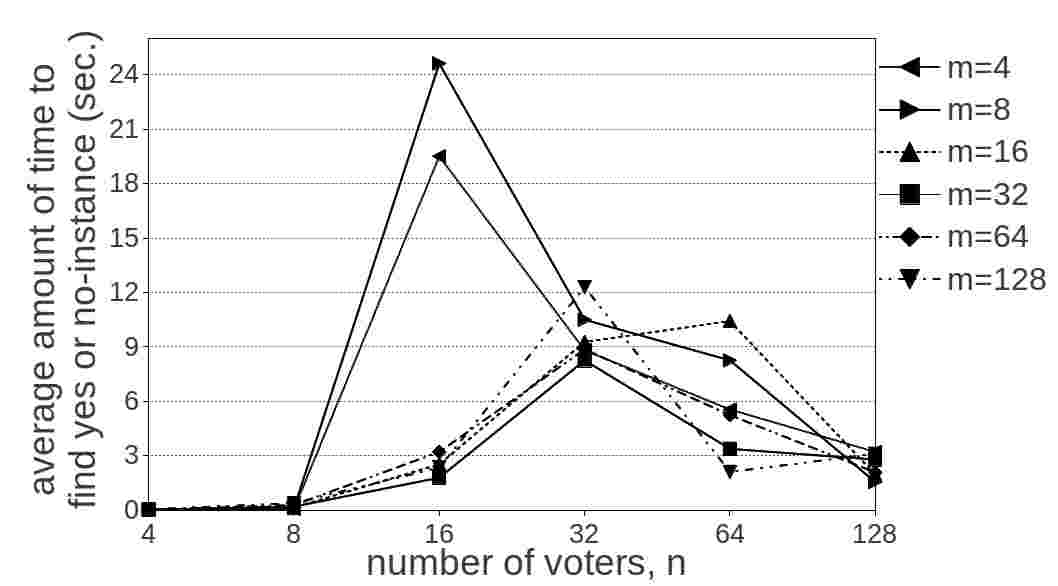}
	\caption{Average time the algorithm needs to give a definite output for 
	destructive control by partition of voters in model TE
	in Bucklin elections in the TM model. The maximum is $24,63$ seconds.}
\end{figure}

\clearpage
\subsection{Constructive Control by Partition of Voters in Model TP}
\begin{center}
\begin{figure}[ht]
\centering
	\includegraphics[scale=0.3]{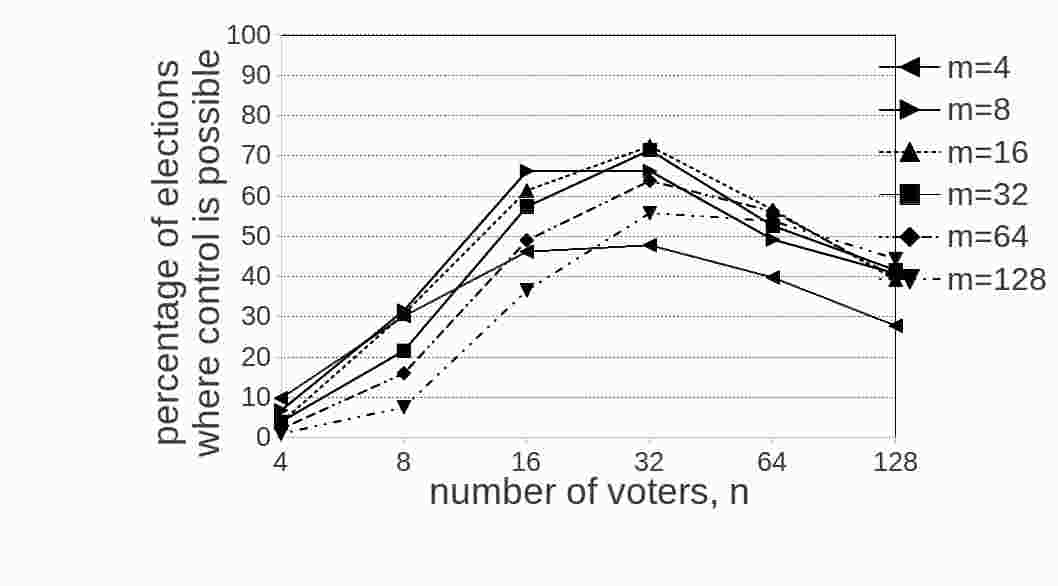}
	\caption{Results for Bucklin voting in the IC model for 
constructive control by partition of voters in model TP.  Number of candidates is fixed.}
\end{figure}

%
\newpage
\begin{figure}[ht]
\centering
	\includegraphics[scale=0.3]{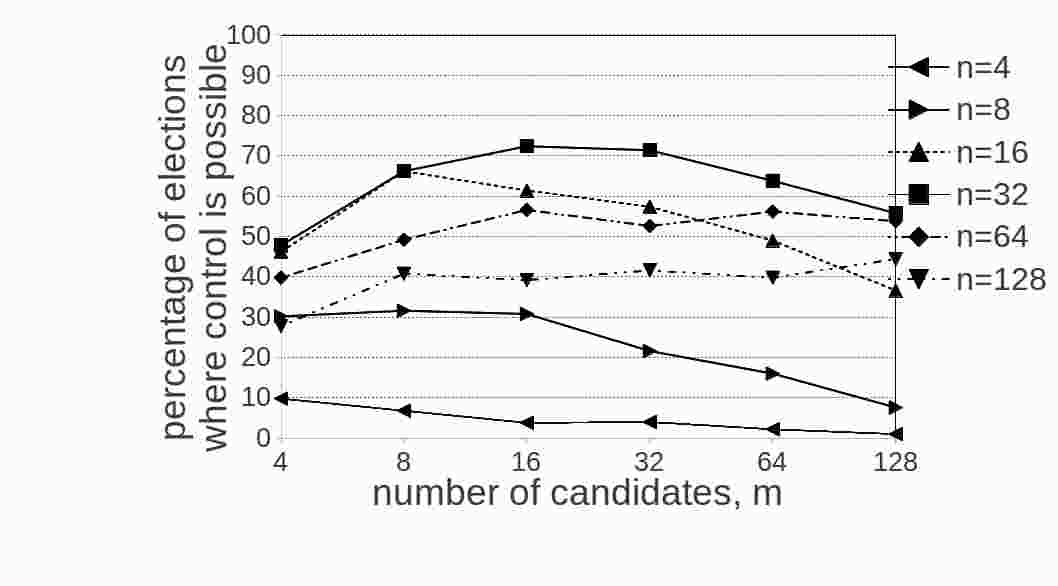}
	\caption{Results for Bucklin voting in the IC model for 
constructive control by partition of voters in model TP.  Number of voters is fixed.}
\end{figure}

%
%
\newpage
\begin{figure}[ht]
\centering
	\includegraphics[scale=0.3]{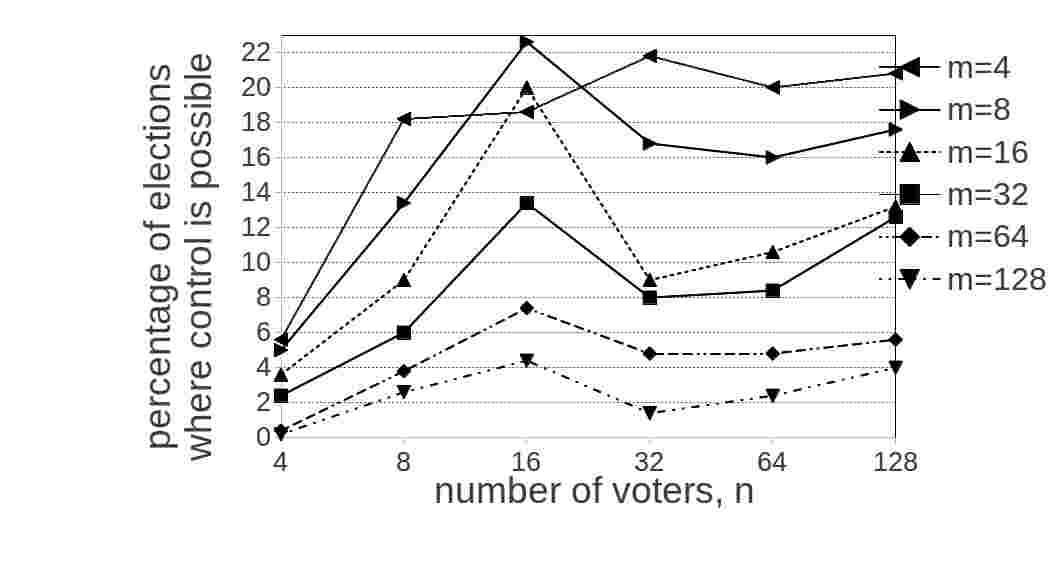}
	\caption{Results for Bucklin voting in the TM model for 
constructive control by partition of voters in model TP.  Number of candidates is fixed.}
\end{figure}

%
%
\newpage
\begin{figure}[ht]
\centering
	\includegraphics[scale=0.3]{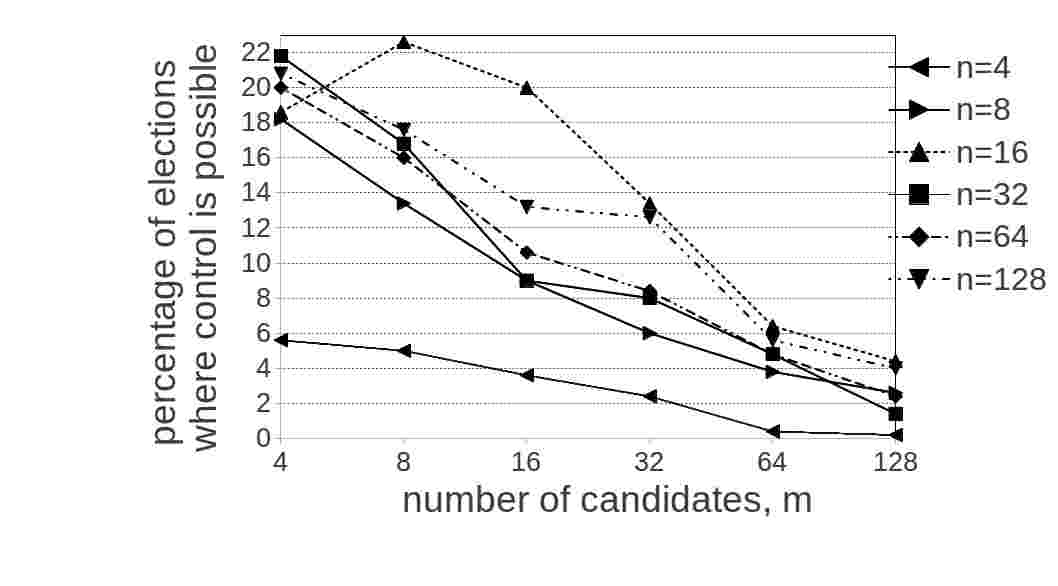}
	\caption{Results for Bucklin voting in the TM model for 
constructive control by partition of voters in model TP.  Number of voters is fixed.}
\end{figure}

%
	\end{center}

\clearpage
\subsubsection{Computational Costs}
\begin{figure}[ht]
\centering
	\includegraphics[scale=0.3]{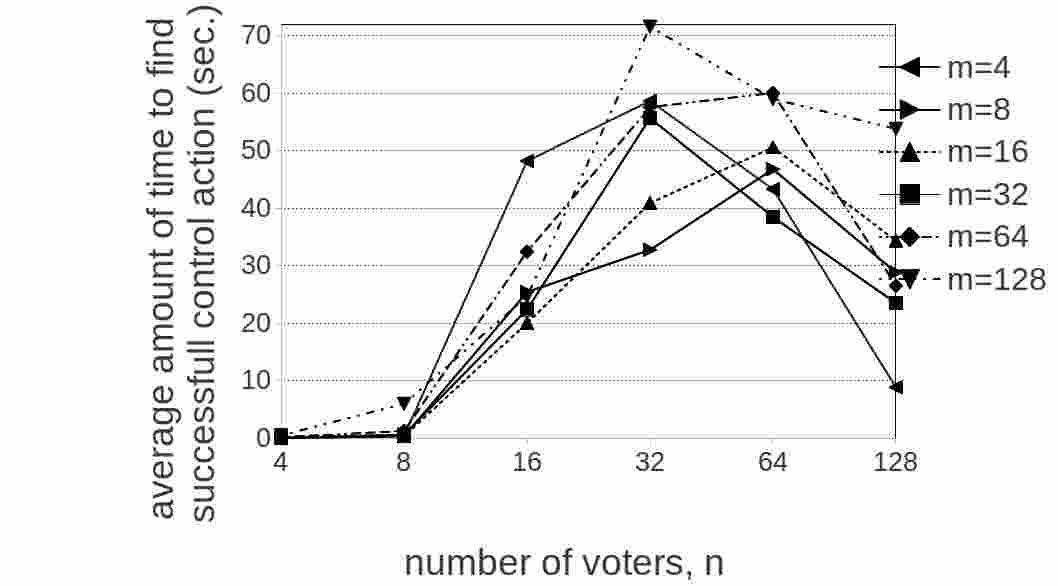}
	\caption{Average time the algorithm needs to find a successful control action for 
	constructive control by partition of voters in model TP
	in Bucklin elections in the IC model. The maximum is $71,61$ seconds.}
\end{figure}

\begin{figure}[ht]
\centering
	\includegraphics[scale=0.3]{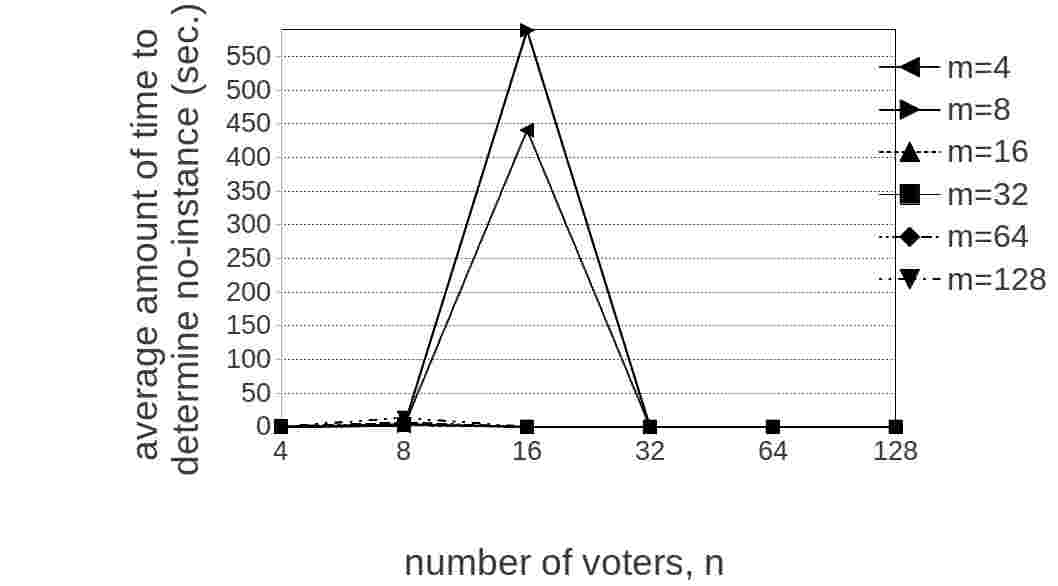}
	\caption{Average time the algorithm needs to determine no-instance of 
		constructive control by partition of voters in model TP
	in Bucklin elections in the IC model. The maximum is $588,49$ seconds.}
\end{figure}

\begin{figure}[ht]
\centering
	\includegraphics[scale=0.3]{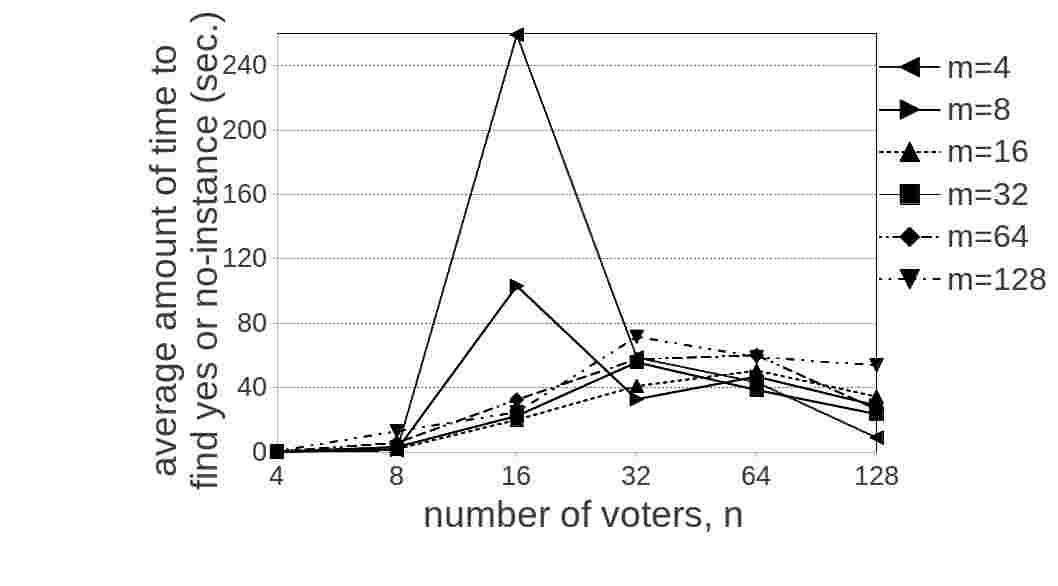}
	\caption{Average time the algorithm needs to give a definite output for 
	constructive control by partition of voters in model TP
	in Bucklin elections in the IC model. The maximum is $259,19$ seconds.}
\end{figure}

\begin{figure}[ht]
\centering
	\includegraphics[scale=0.3]{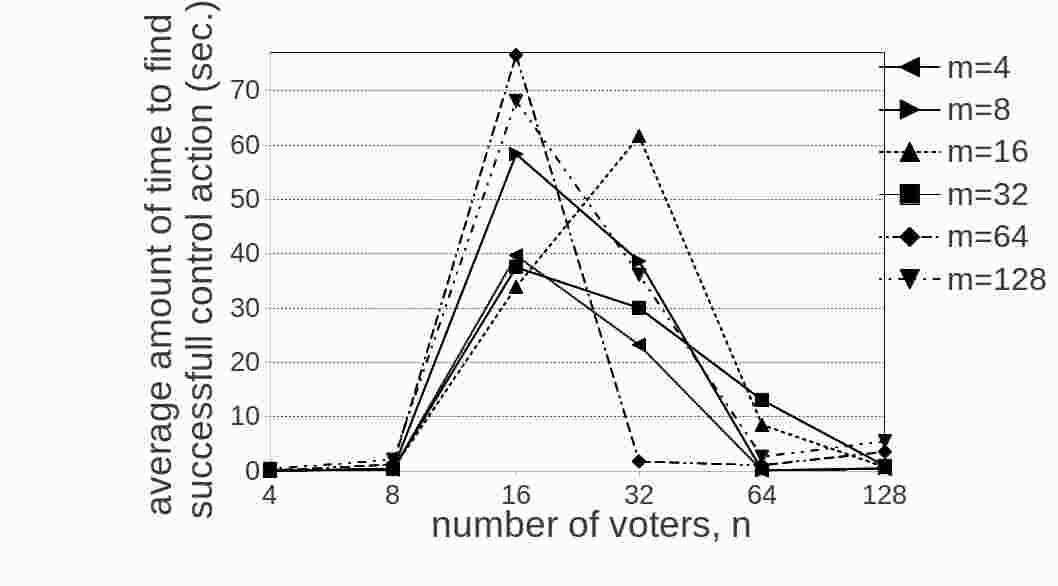}
	\caption{Average time the algorithm needs to find a successful control action for 
	constructive control by partition of voters in model TP
	in Bucklin elections in the TM model. The maximum is $76,56$ seconds.}
\end{figure}

\begin{figure}[ht]
\centering
	\includegraphics[scale=0.3]{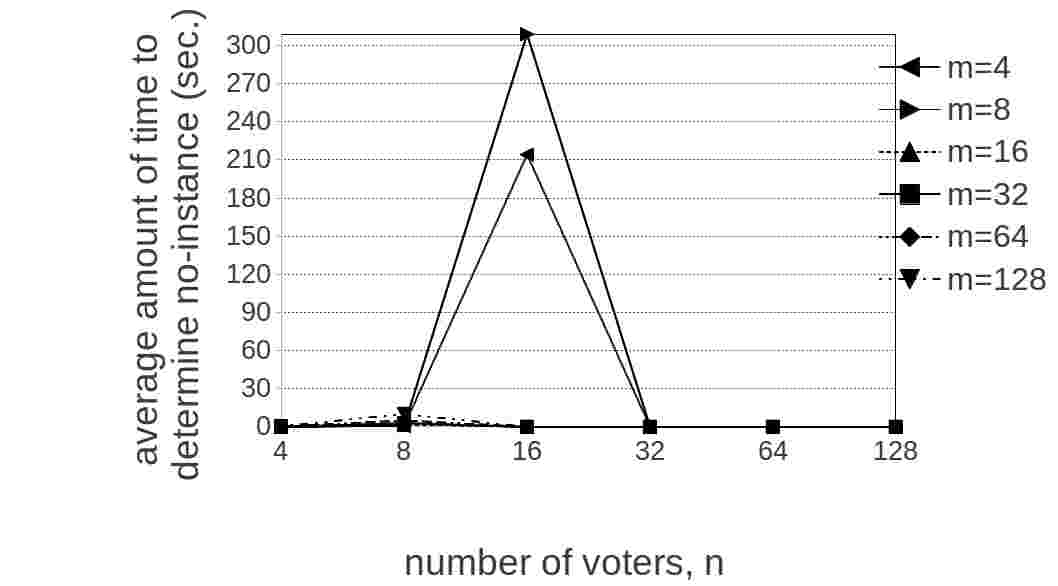}
	\caption{Average time the algorithm needs to determine no-instance of 
		constructive control by partition of voters in model TP
	in Bucklin elections in the TM model. The maximum is $308,87$ seconds.}
\end{figure}

\begin{figure}[ht]
\centering
	\includegraphics[scale=0.3]{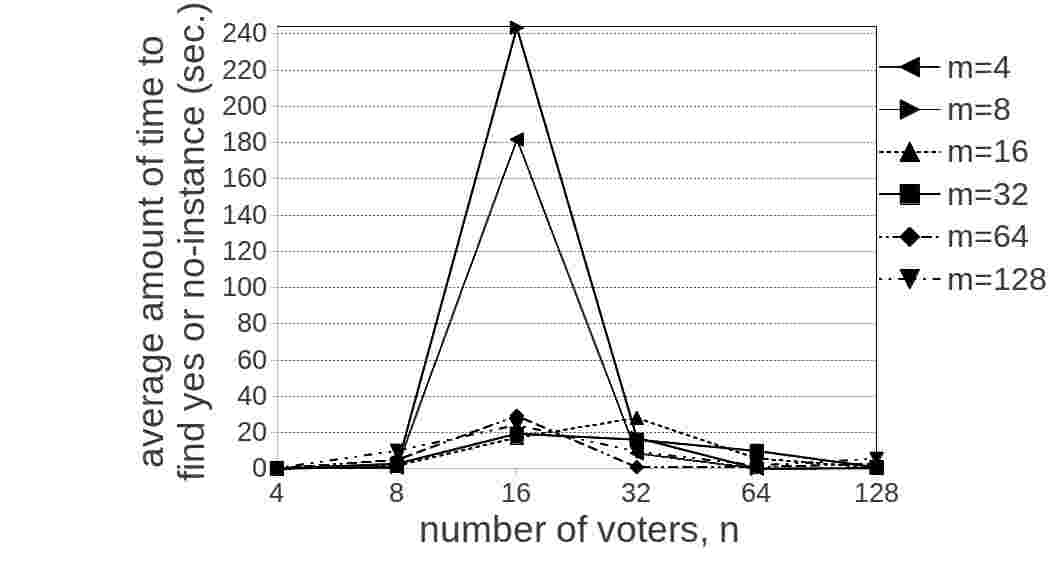}
	\caption{Average time the algorithm needs to give a definite output for 
	constructive control by partition of voters in model TP
	in Bucklin elections in the TM model. The maximum is $243,35$ seconds.}
\end{figure}

\clearpage
\subsection{Destructive Control by Partition of Voters in Model TP}
\begin{center}
\begin{figure}[ht]
\centering
	\includegraphics[scale=0.3]{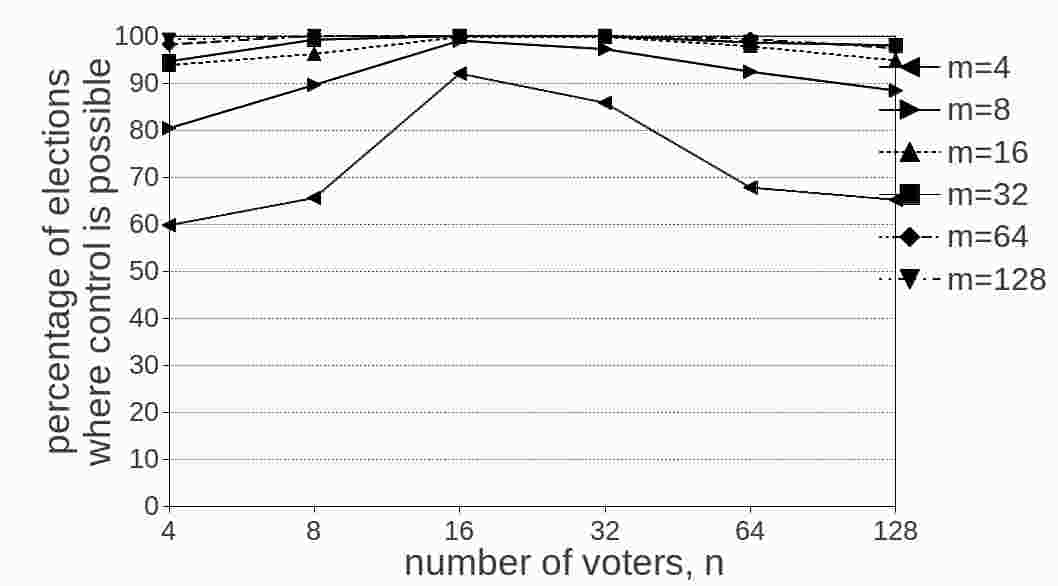}
	\caption{Results for Bucklin voting in the IC model for 
destructive control by partition of voters in model TP.  Number of candidates is fixed.}
\end{figure}

%
\newpage
\begin{figure}[ht]
\centering
	\includegraphics[scale=0.3]{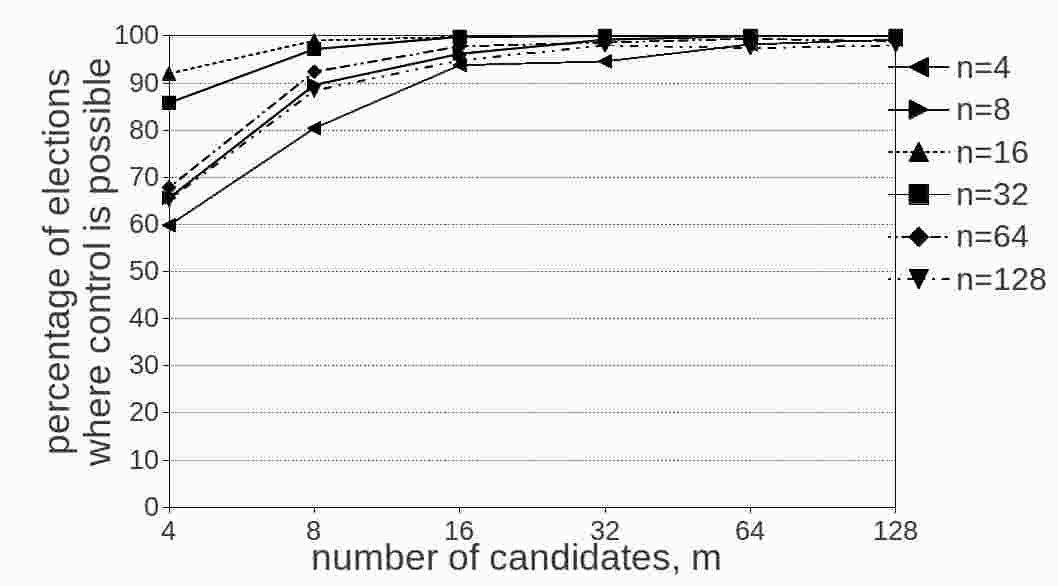}
	\caption{Results for Bucklin voting in the IC model for 
destructive control by partition of voters in model TP.  Number of voters is fixed.}
\end{figure}

%
%
\newpage
\begin{figure}[ht]
\centering
	\includegraphics[scale=0.3]{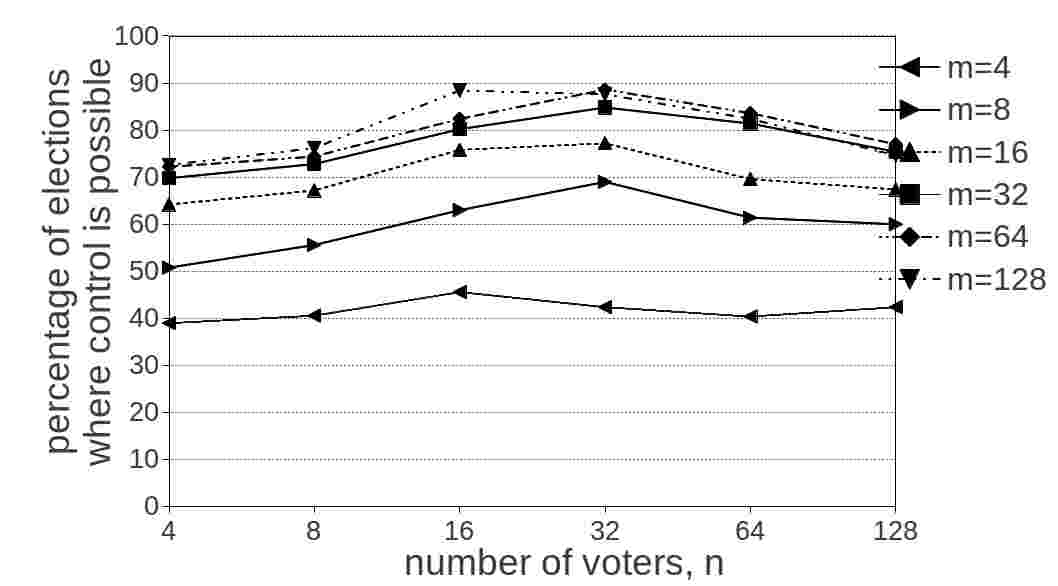}
	\caption{Results for Bucklin voting in the TM model for 
destructive control by partition of voters in model TP.  Number of candidates is fixed.}
\end{figure}

%
%
\newpage
\begin{figure}[ht]
\centering
	\includegraphics[scale=0.3]{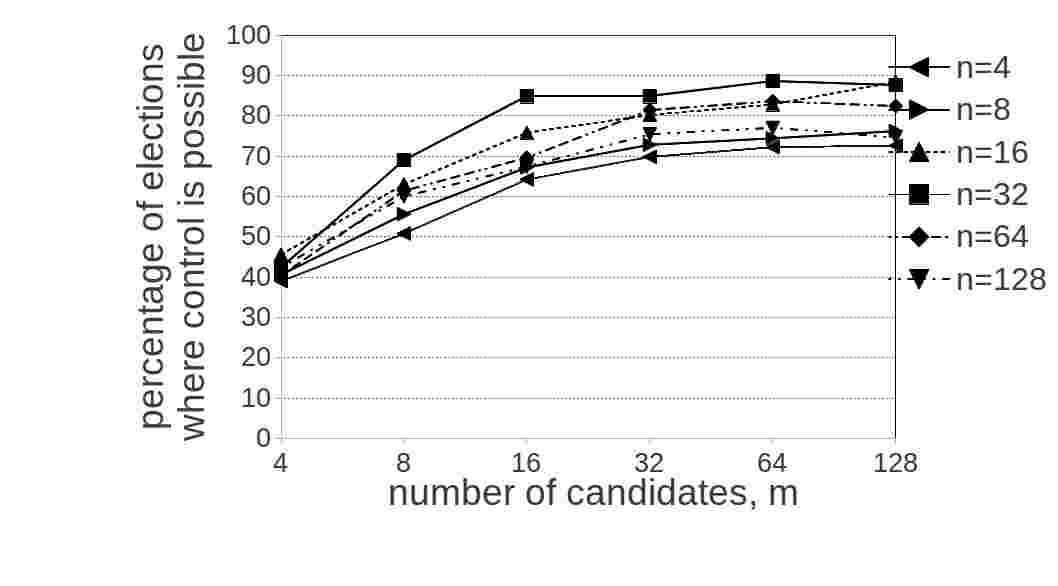}
	\caption{Results for Bucklin voting in the TM model for 
destructive control by partition of voters in model TP.  Number of voters is fixed.}
\end{figure}

%
\end{center}

\clearpage
\subsubsection{Computational Costs}
\begin{figure}[ht]
\centering
	\includegraphics[scale=0.3]{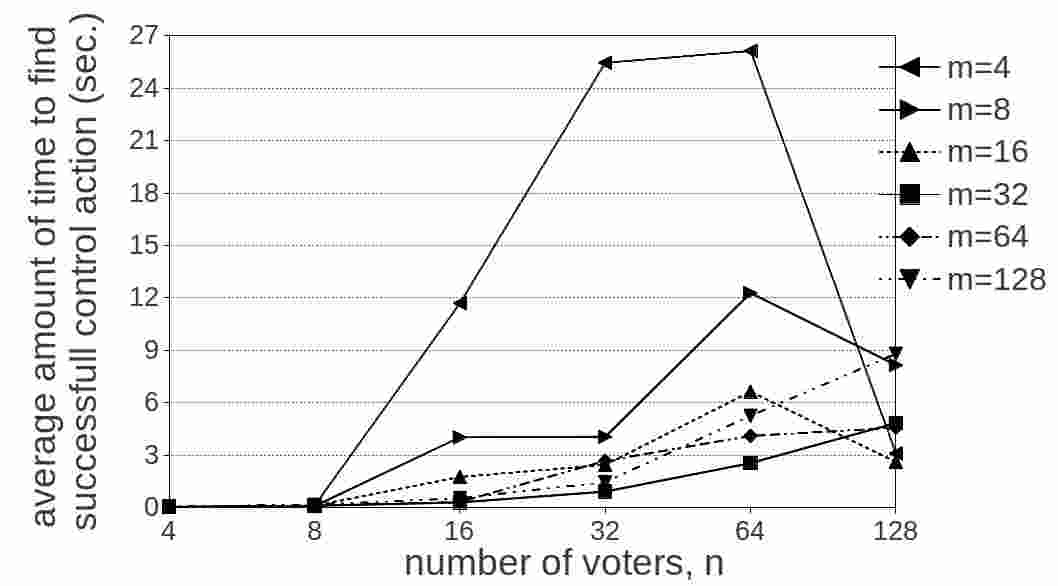}
	\caption{Average time the algorithm needs to find a successful control action for 
	destructive control by partition of voters in model TP
	in Bucklin elections in the IC model. The maximum is $26,14$ seconds.}
\end{figure}

\begin{figure}[ht]
\centering
	\includegraphics[scale=0.3]{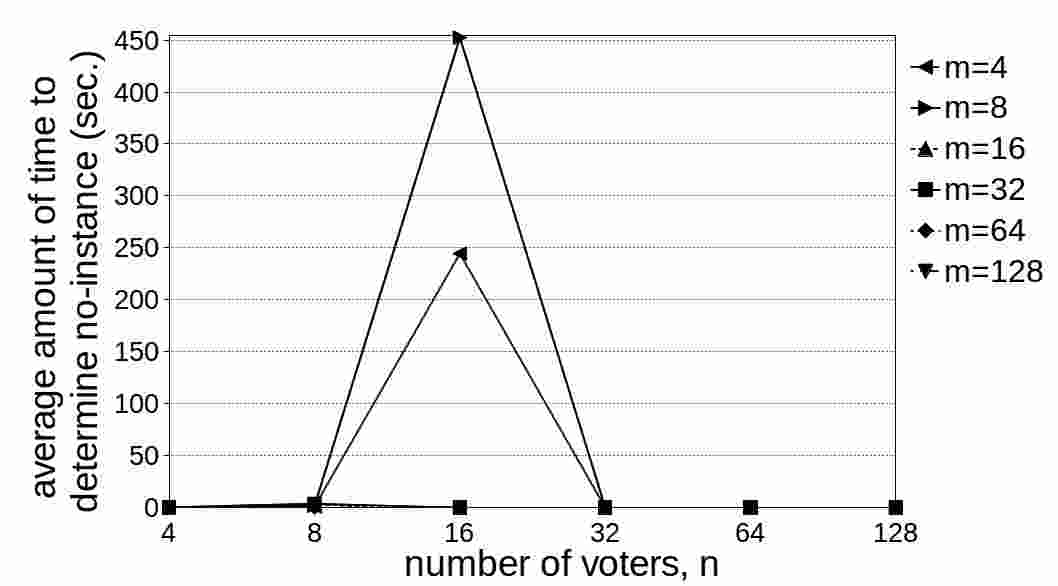}
	\caption{Average time the algorithm needs to determine no-instance of 
		destructive control by partition of voters in model TP
	in Bucklin elections in the IC model. The maximum is $452,32$ seconds.}
\end{figure}

\begin{figure}[ht]
\centering
	\includegraphics[scale=0.3]{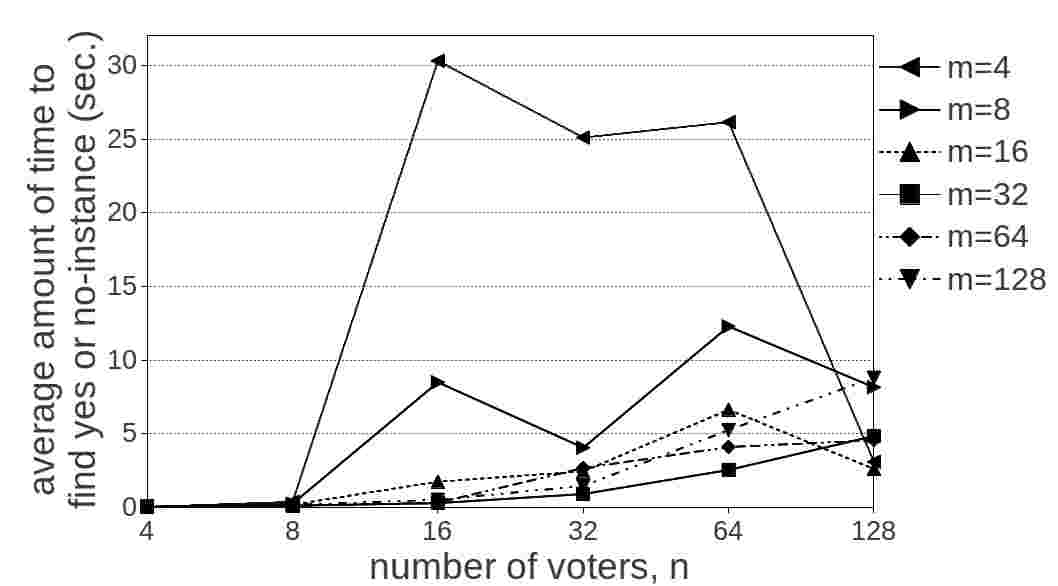}
	\caption{Average time the algorithm needs to give a definite output for 
	destructive control by partition of voters in model TP
	in Bucklin elections in the IC model. The maximum is $30,3$ seconds.}
\end{figure}

\begin{figure}[ht]
\centering
	\includegraphics[scale=0.3]{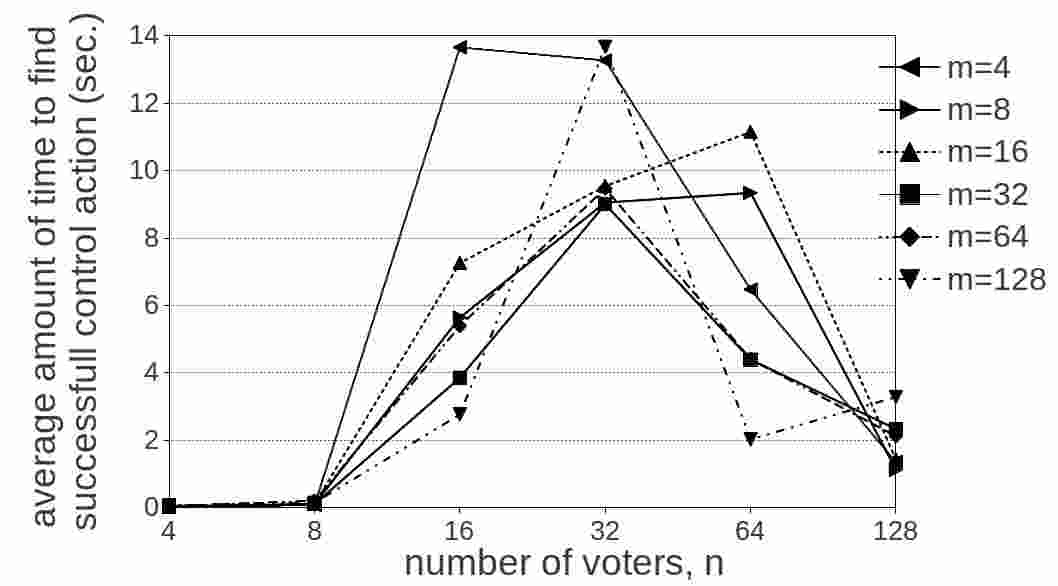}
	\caption{Average time the algorithm needs to find a successful control action for 
	destructive control by partition of voters in model TP
	in Bucklin elections in the TM model. The maximum is $13,68$ seconds.}
\end{figure}

\begin{figure}[ht]
\centering
	\includegraphics[scale=0.3]{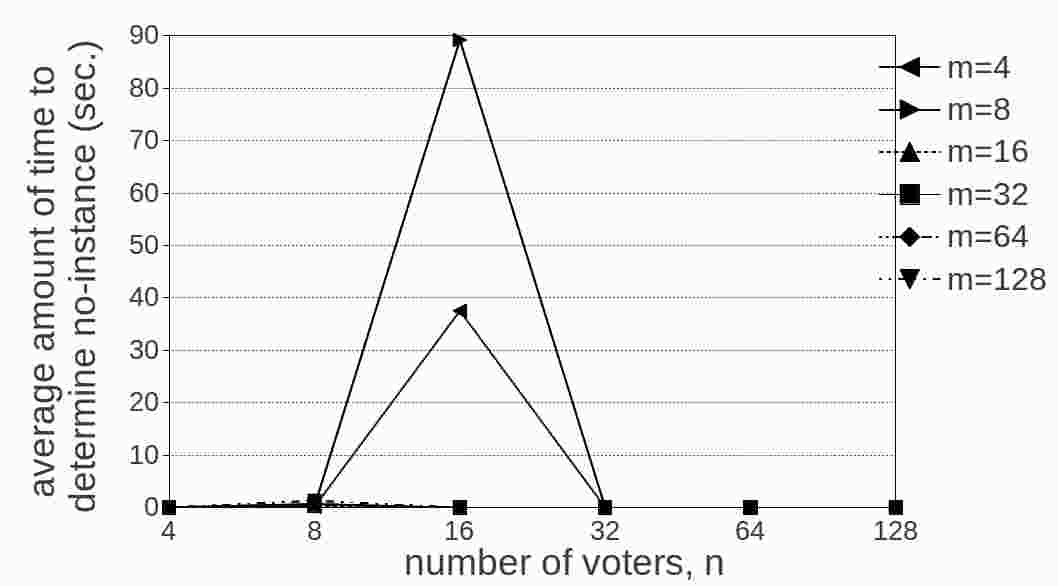}
	\caption{Average time the algorithm needs to determine no-instance of 
		destructive control by partition of voters in model TP
	in Bucklin elections in the TM model. The maximum is $89,22$ seconds.}
\end{figure}

\begin{figure}[ht]
\centering
	\includegraphics[scale=0.3]{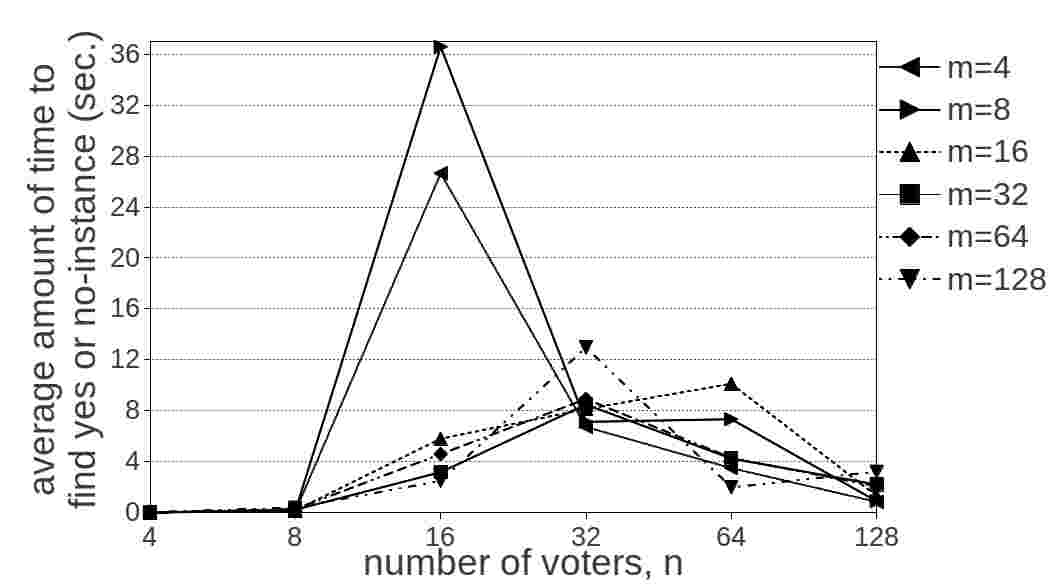}
	\caption{Average time the algorithm needs to give a definite output for 
	destructive control by partition of voters in model TP
	in Bucklin elections in the TM model. The maximum is $36,57$ seconds.}
\end{figure}

\clearpage
\section{Plurality Voting}

\subsection{Constructive Control by Adding Candidates}
\begin{center}
\begin{figure}[ht]
\centering
	\includegraphics[scale=0.3]{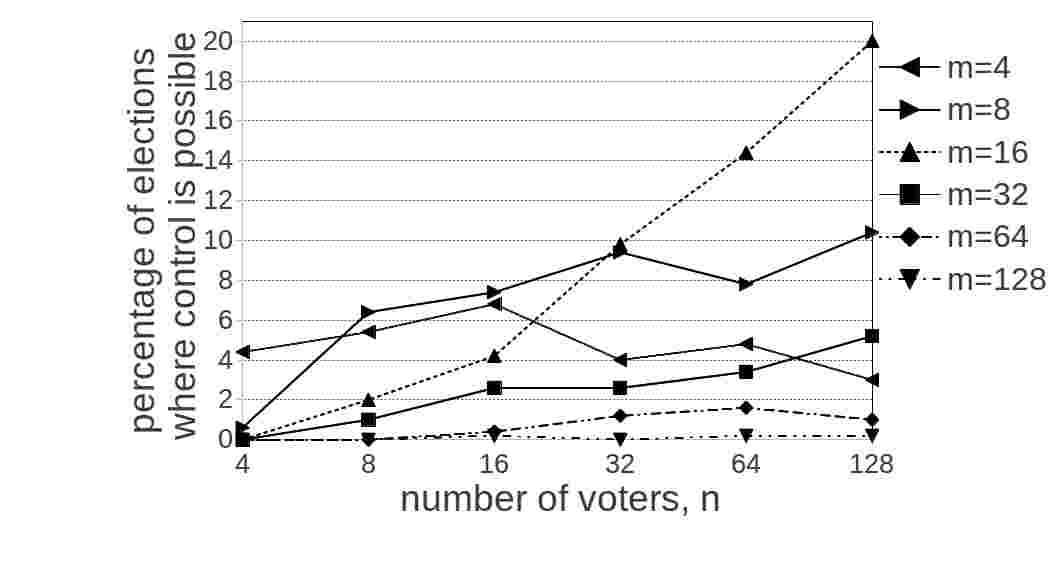}
		\caption{Results for plurality voting in the IC model for 
constructive control by adding candidates. Number of candidates is fixed. }
\end{figure}

\newpage
\begin{figure}[ht]
\centering
	\includegraphics[scale=0.3]{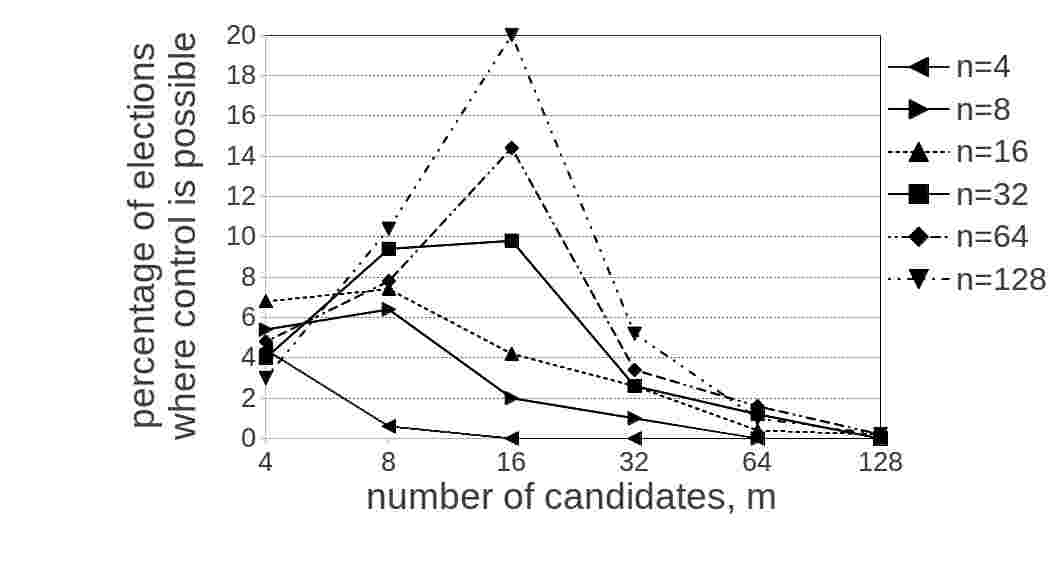}
		\caption{Results for plurality voting in the IC model for 
constructive control by adding candidates. Number of voters is fixed. }
\end{figure}
%

\newpage
\begin{figure}[ht]
\centering
	\includegraphics[scale=0.3]{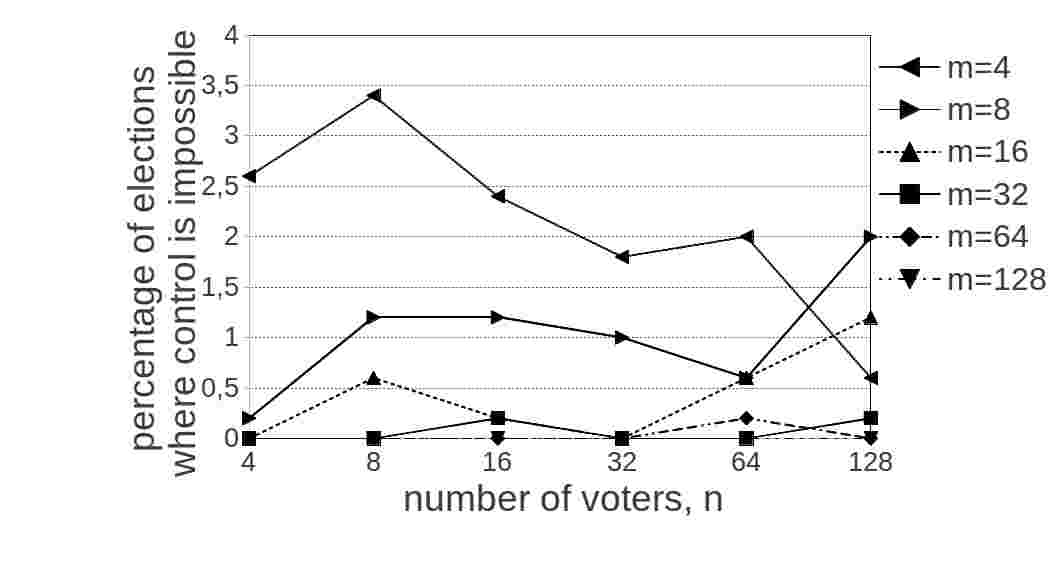}
		\caption{Results for plurality voting in the TM model for 
constructive control by adding candidates. Number of candidates is fixed. }
\end{figure}

%
\newpage
\begin{figure}[ht]
\centering
	\includegraphics[scale=0.3]{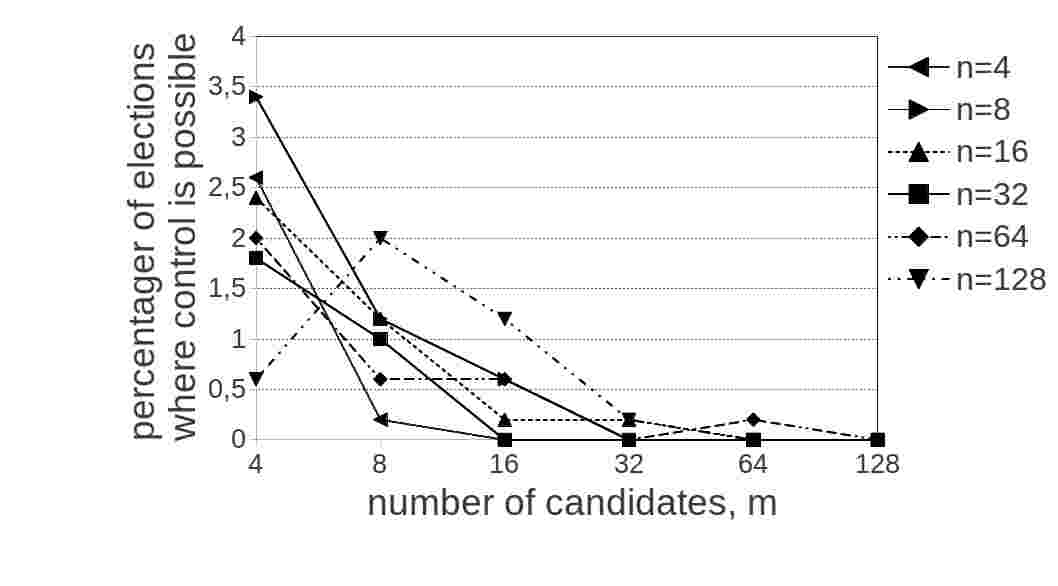}
		\caption{Results for plurality voting in the TM model for 
constructive control by adding candidates. Number of voters is fixed. }
\end{figure}
%
\end{center}

\clearpage
\subsubsection{Computational Costs}

\begin{figure}[ht]
\centering
	\includegraphics[scale=0.3]{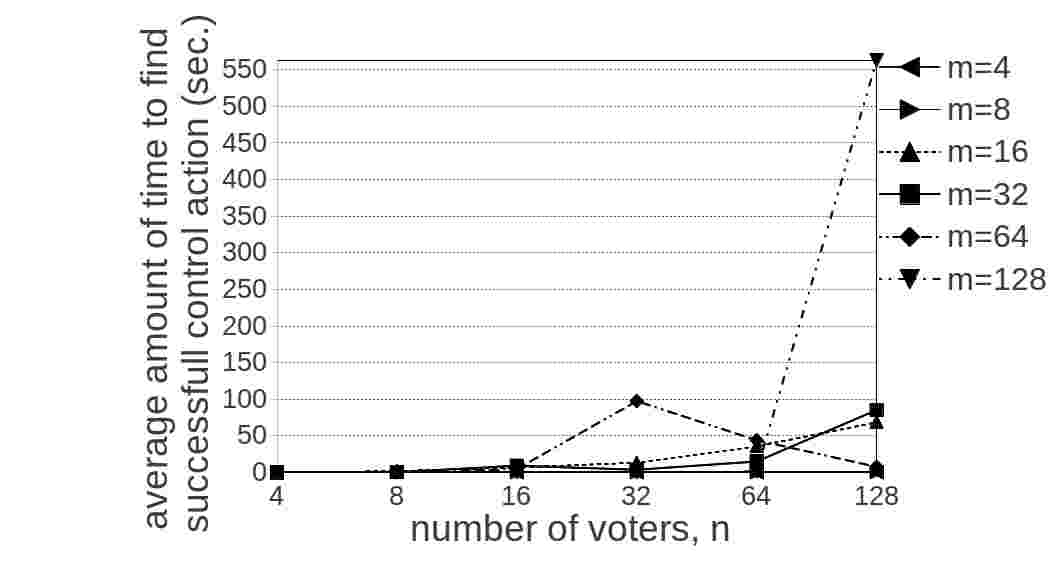}
	\caption{Average time the algorithm needs to find a successful control action for 
	constructive control by adding candidates
	in plurality elections in the IC model. The maximum is $562,76$ seconds.}
\end{figure}
\begin{figure}[ht]
\centering
	\includegraphics[scale=0.3]{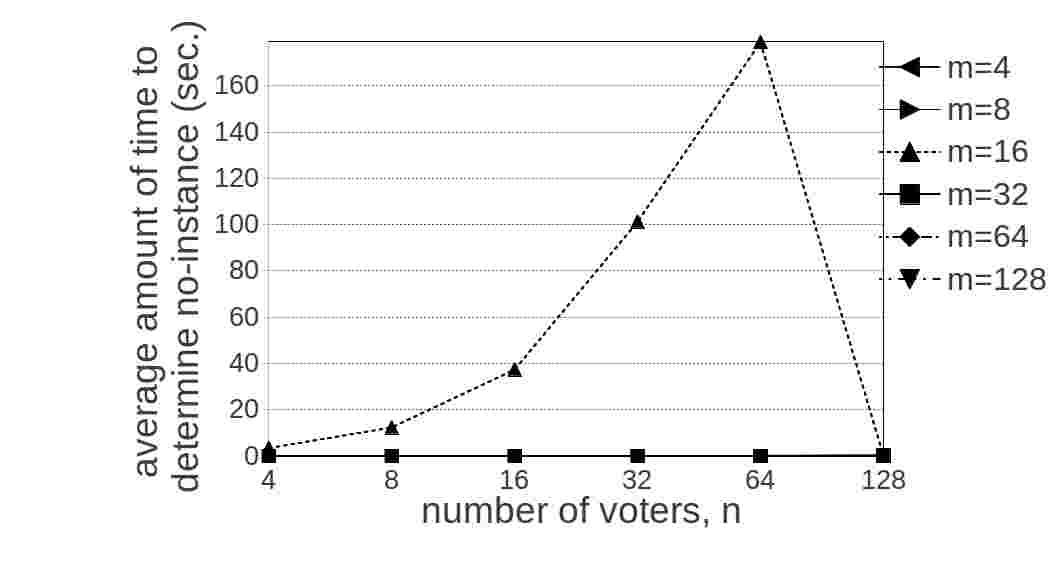}
	\caption{Average time the algorithm needs to determine no-instance of 
		constructive control by adding candidates
	in plurality elections in the IC model. The maximum is $178,74$ seconds.}
\end{figure}
\begin{figure}[ht]
\centering
	\includegraphics[scale=0.3]{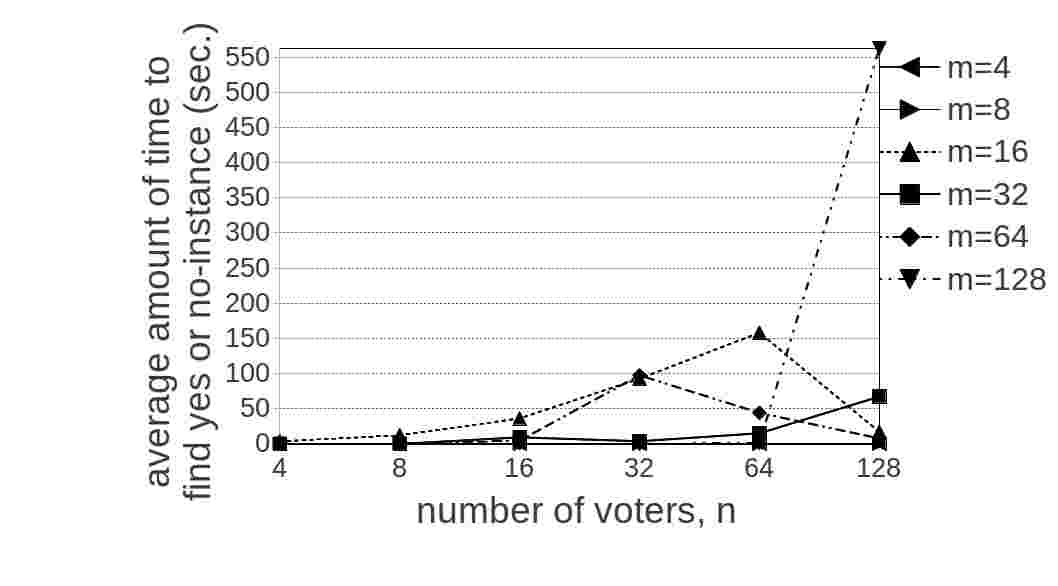}
	\caption{Average time the algorithm needs to give a definite output for 
	constructive control by adding candidates
	in plurality elections in the IC model. The maximum is $562,76$ seconds.}
\end{figure}
\begin{figure}[ht]
\centering
	\includegraphics[scale=0.3]{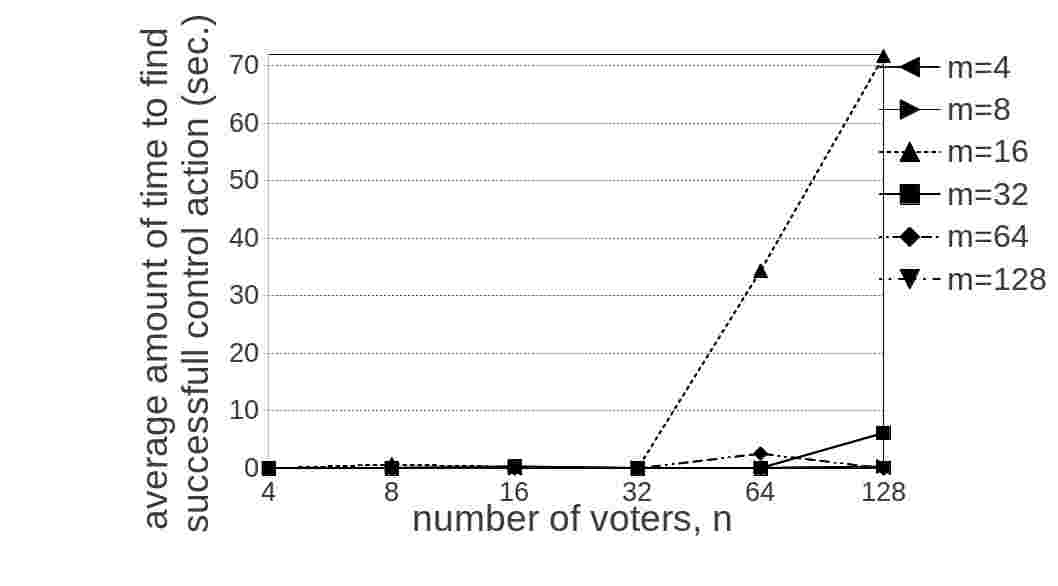}
	\caption{Average time the algorithm needs to find a successful control action for 
	constructive control by adding candidates
	in plurality elections in the TM model. The maximum is $71,66$ seconds.}
\end{figure}
\begin{figure}[ht]
\centering
	\includegraphics[scale=0.3]{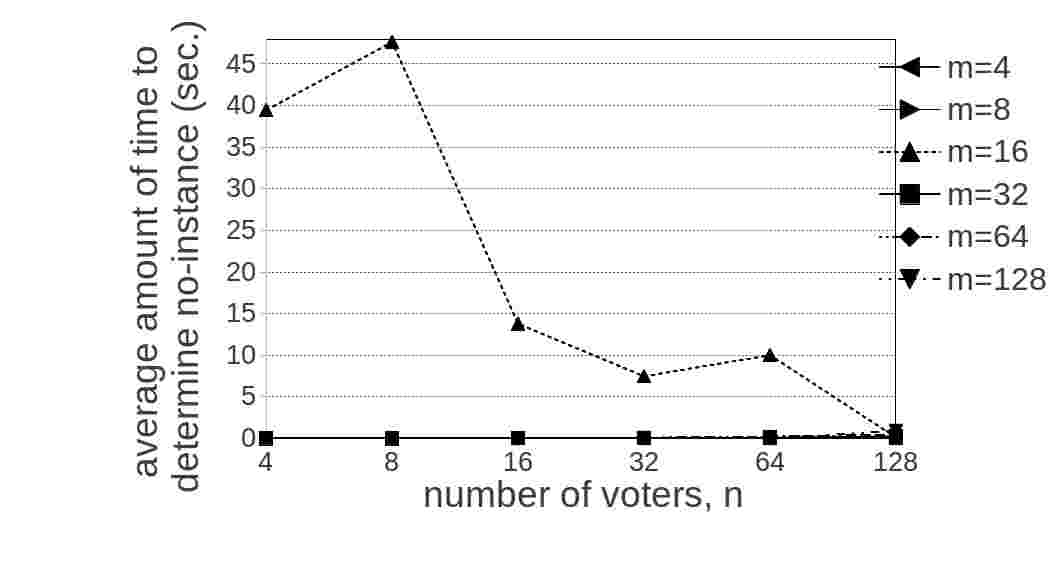}
	\caption{Average time the algorithm needs to determine no-instance of 
		constructive control by adding candidates
	in plurality elections in the TM model. The maximum is $47,64$ seconds.}
\end{figure}
\begin{figure}[ht]
\centering
	\includegraphics[scale=0.3]{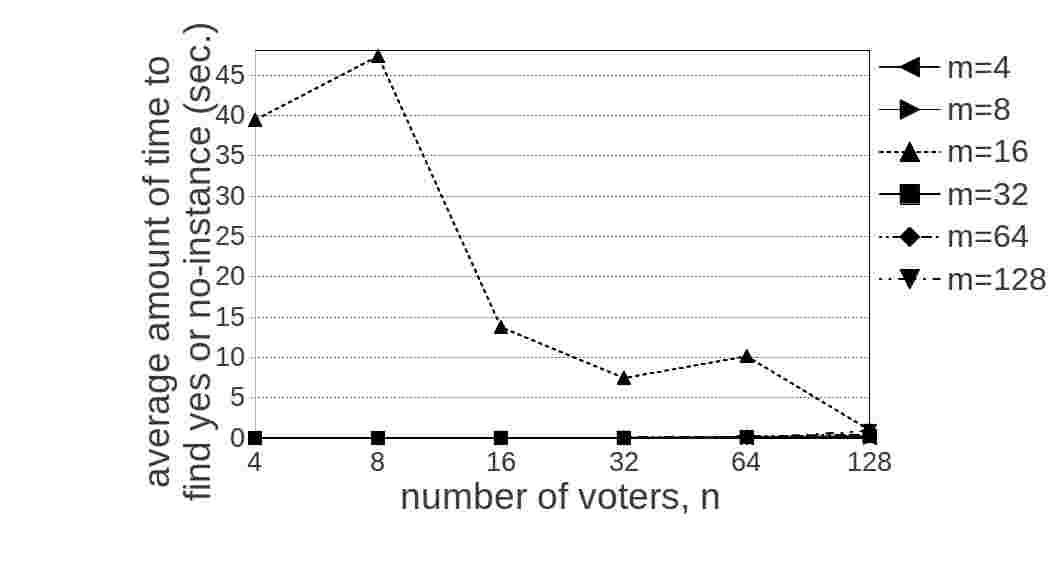}
	\caption{Average time the algorithm needs to give a definite output for 
	constructive control by adding candidates
	in plurality elections in the TM model. The maximum is $47,35$ seconds.}
\end{figure}

\clearpage
\subsection{Destructive Control by Adding Candidates}
\begin{center}
\begin{figure}[ht]
\centering
	\includegraphics[scale=0.3]{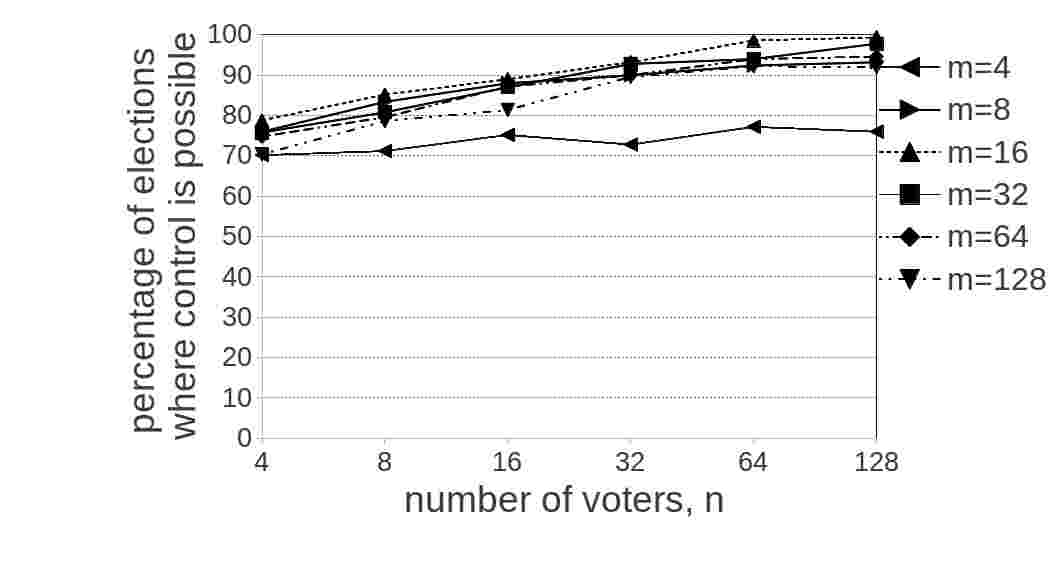}
		\caption{Results for plurality voting in the IC model for 
destructive control by adding candidates. Number of candidates is fixed. }
\end{figure}


\newpage
\begin{figure}[ht]
\centering
	\includegraphics[scale=0.3]{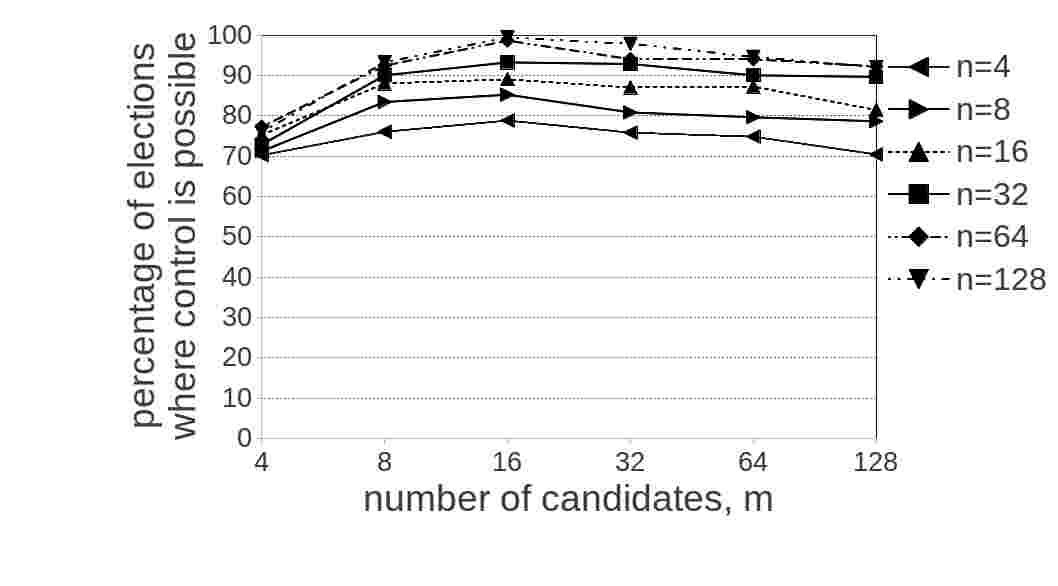}
		\caption{Results for plurality voting in the IC model for 
destructive control by adding candidates. Number of voters is fixed. }
\end{figure}

%

\newpage
\begin{figure}[ht]
\centering
	\includegraphics[scale=0.3]{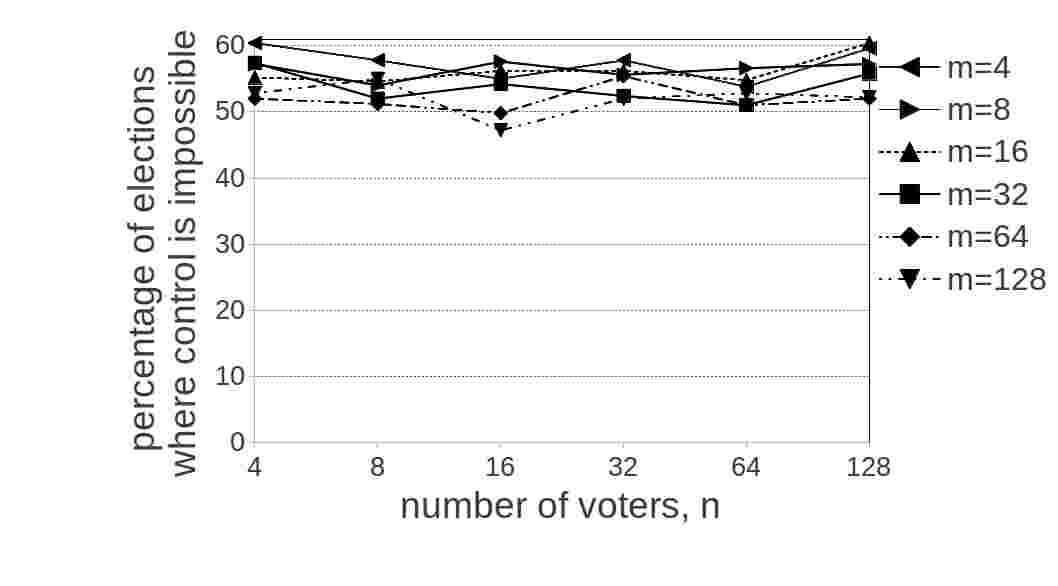}
		\caption{Results for plurality voting in the TM model for 
destructive control by adding candidates. Number of candidates is fixed. }
\end{figure}

%

\newpage
\begin{figure}[ht]
\centering
	\includegraphics[scale=0.3]{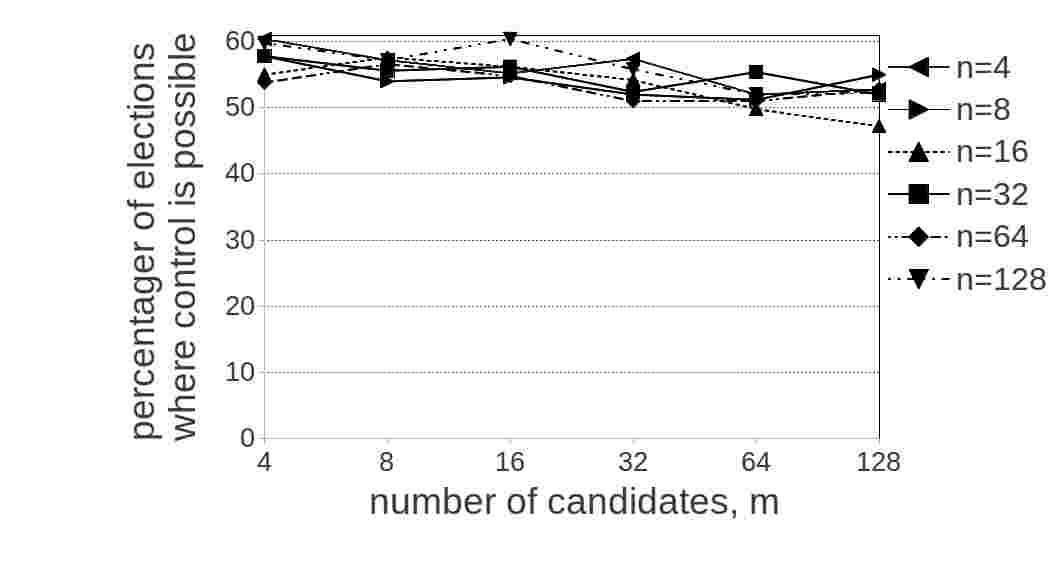}
		\caption{Results for plurality voting in the TM model for 
destructive control by adding candidates. Number of voters is fixed. }
\end{figure}

%
\end{center}

\clearpage
\subsubsection{Computational Costs}
\begin{figure}[ht]
\centering
	\includegraphics[scale=0.3]{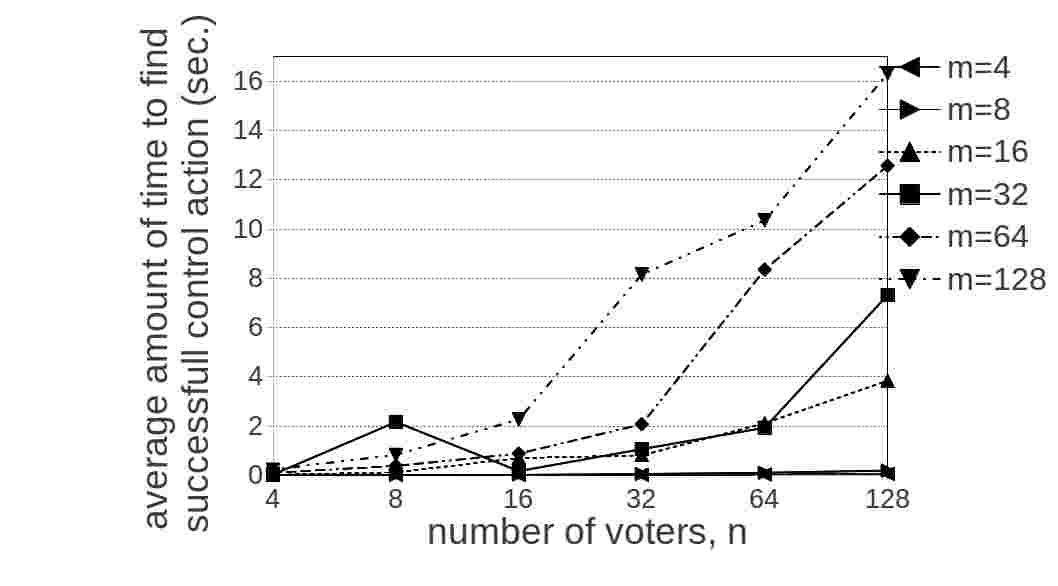}
	\caption{Average time the algorithm needs to find a successful control action for 
	destructive control by adding candidates
	in plurality elections in the IC model. The maximum is $16,32$ seconds.}
\end{figure}
\begin{figure}[ht]
\centering
	\includegraphics[scale=0.3]{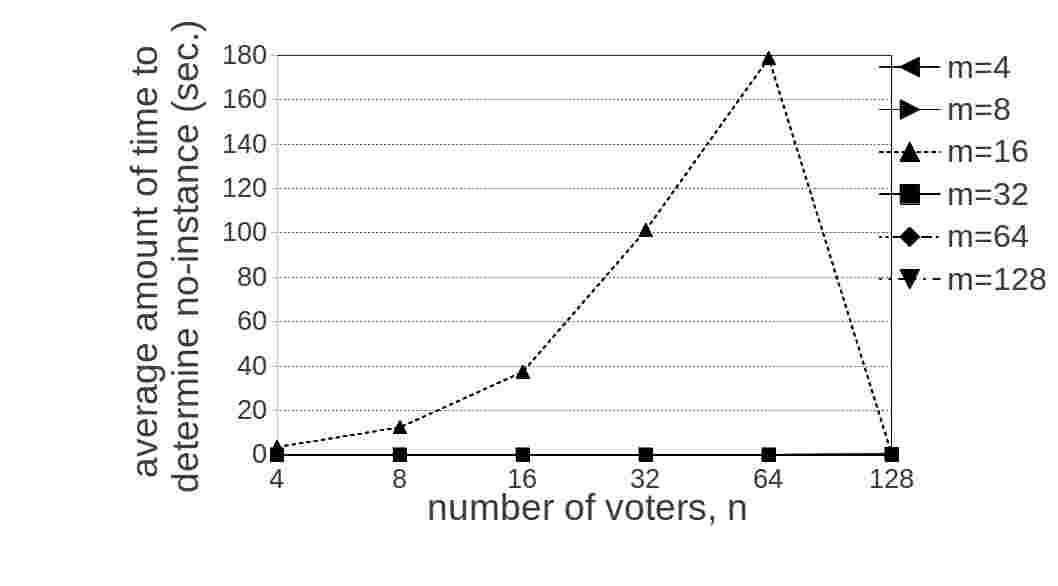}
	\caption{Average time the algorithm needs to determine no-instance of 
		destructive control by adding candidates
	in plurality elections in the IC model. The maximum is $356,7$ seconds.}
\end{figure}
\begin{figure}[ht]
\centering
	\includegraphics[scale=0.3]{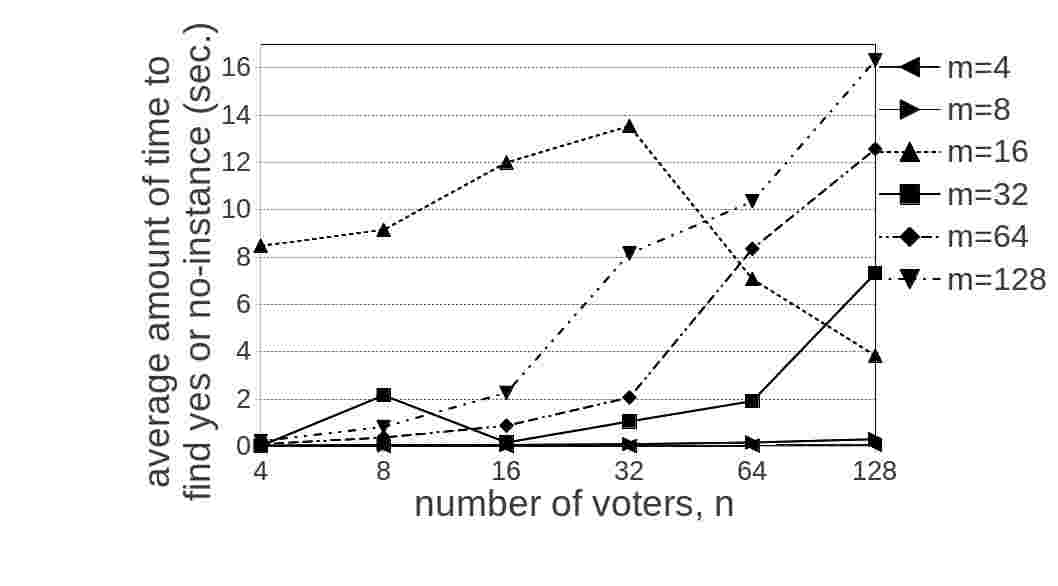}
	\caption{Average time the algorithm needs to give a definite output for 
	destructive control by adding candidates
	in plurality elections in the IC model. The maximum is $16,32$ seconds.}
\end{figure}
\begin{figure}[ht]
\centering
	\includegraphics[scale=0.3]{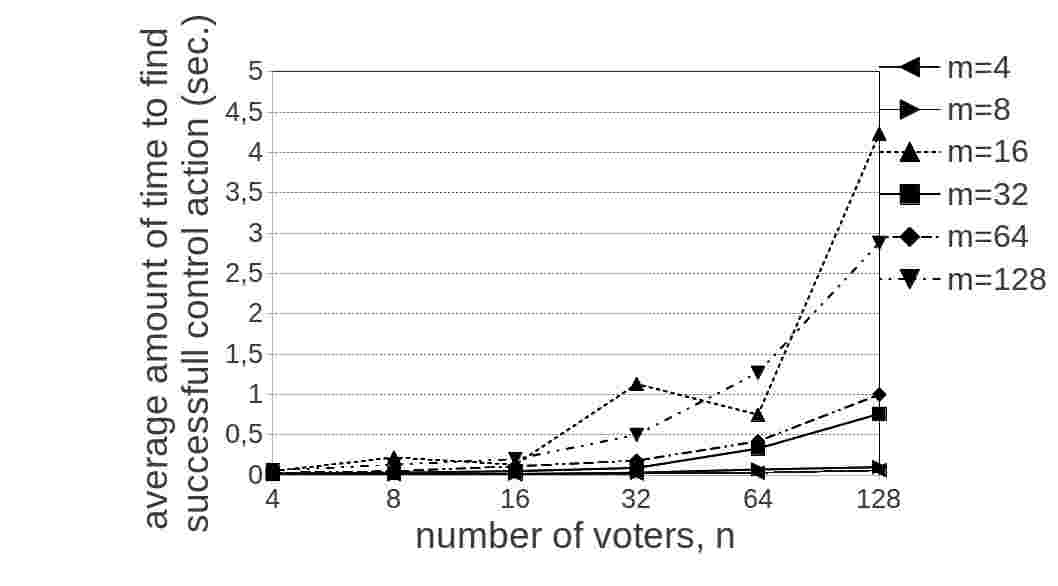}
	\caption{Average time the algorithm needs to find a successful control action for 
	destructive control by adding candidates
	in plurality elections in the TM model. The maximum is $4,23$ seconds.}
\end{figure}
\begin{figure}[ht]
\centering
	\includegraphics[scale=0.3]{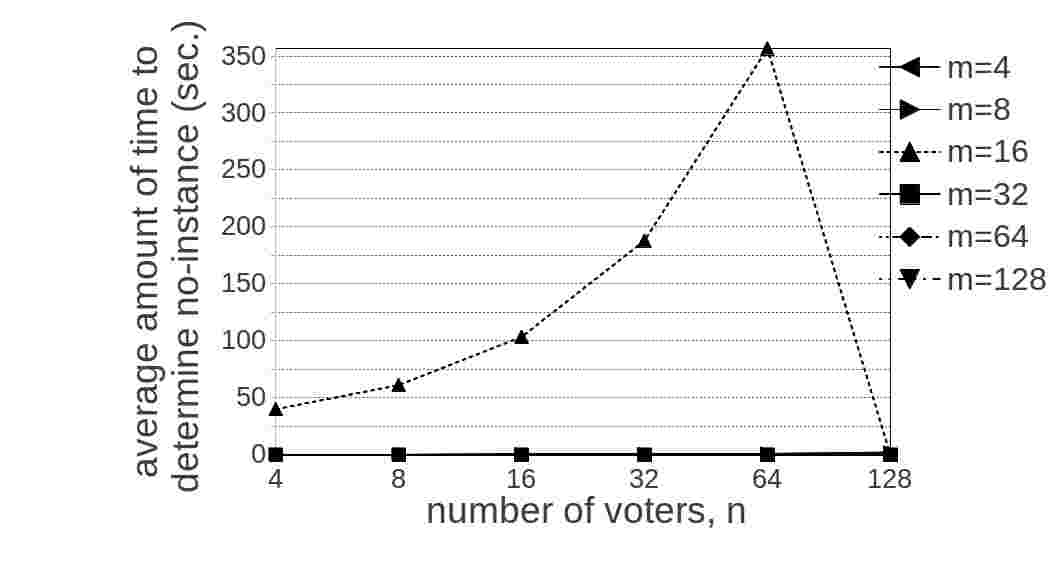}
	\caption{Average time the algorithm needs to determine no-instance of 
		destructive control by adding candidates
	in plurality elections in the TM model. The maximum is $356,7$ seconds.}
\end{figure}
\begin{figure}[ht]
\centering
	\includegraphics[scale=0.3]{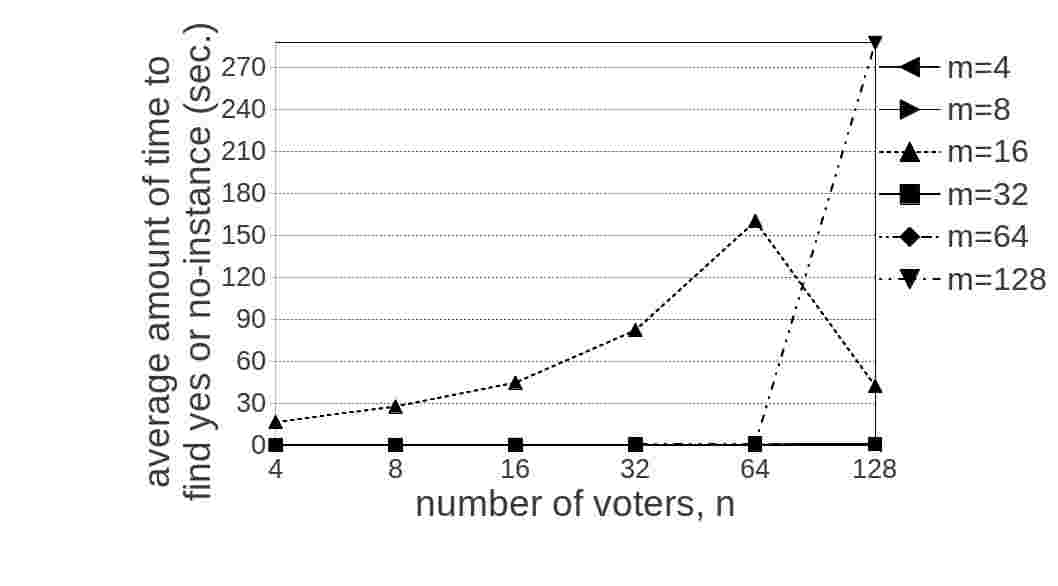}
	\caption{Average time the algorithm needs to give a definite output for 
		destructive control by adding candidates
	in plurality elections in the TM model. The maximum is $287,86$ seconds.}
\end{figure}

\clearpage
\subsection{Constructive Control by Deleting Candidates}
\begin{center}

\begin{figure}[ht]
\centering
	\includegraphics[scale=0.3]{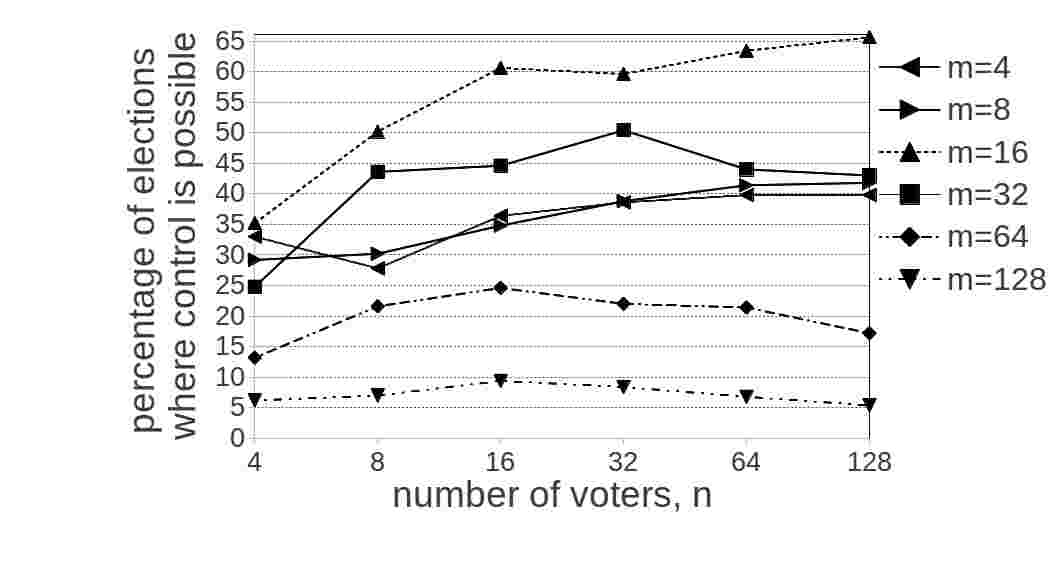}
		\caption{Results for plurality voting in the IC model for 
constructive control by deleting candidates. Number of candidates is fixed. }
\end{figure}


\newpage
\begin{figure}[ht]
 \centering
	\includegraphics[scale=0.3]{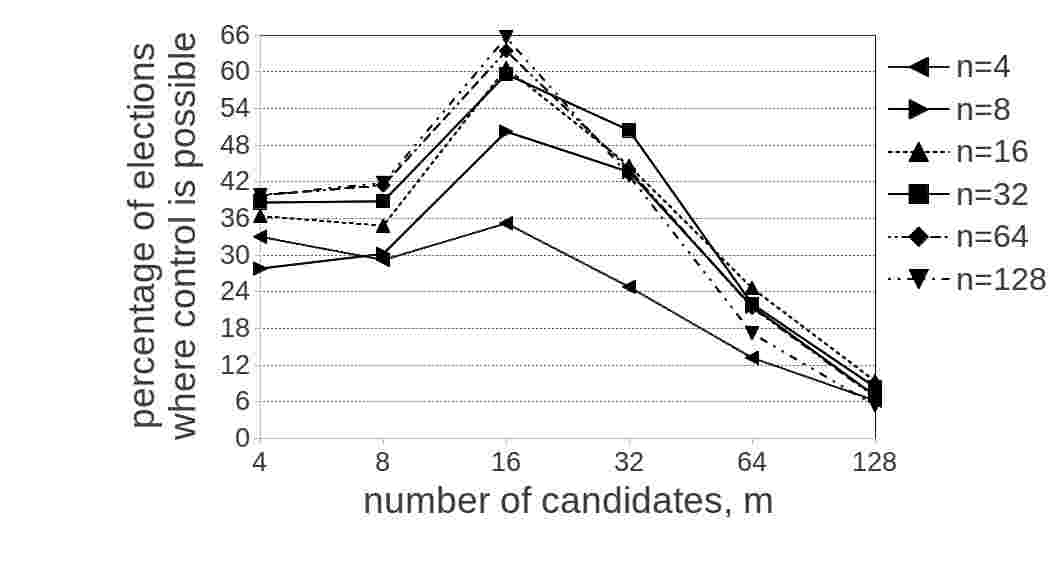}
		\caption{Results for plurality voting in the IC model for 
constructive control by deleting candidates. Number of voters is fixed. }
\end{figure}
%

\newpage
\begin{figure}[ht]
\centering
	\includegraphics[scale=0.3]{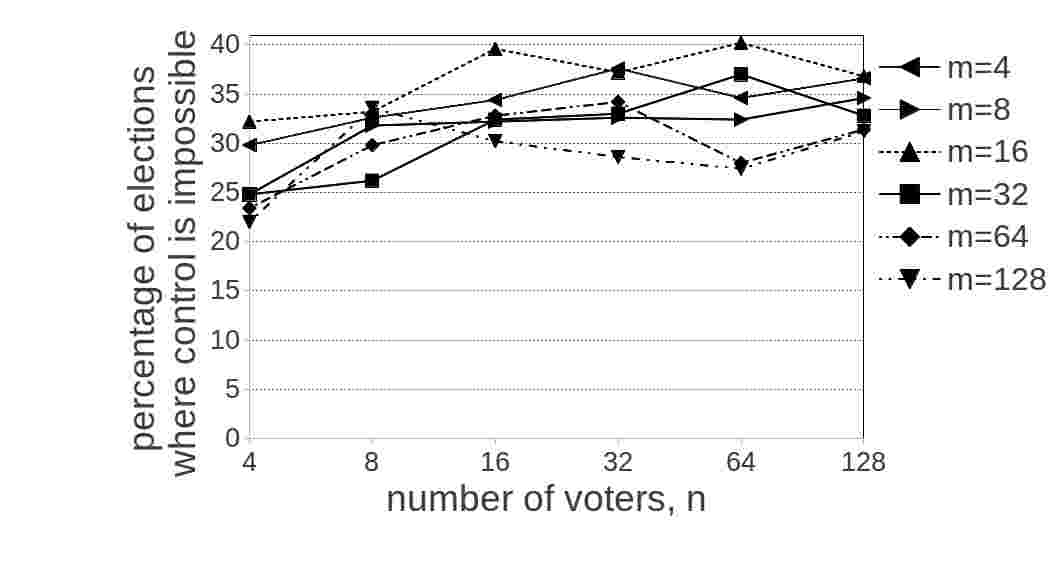}
		\caption{Results for plurality voting in the TM model for 
constructive control by deleting candidates. Number of candidates is fixed. }
\end{figure}

%

\newpage
\begin{figure}[ht]
\centering
	\includegraphics[scale=0.3]{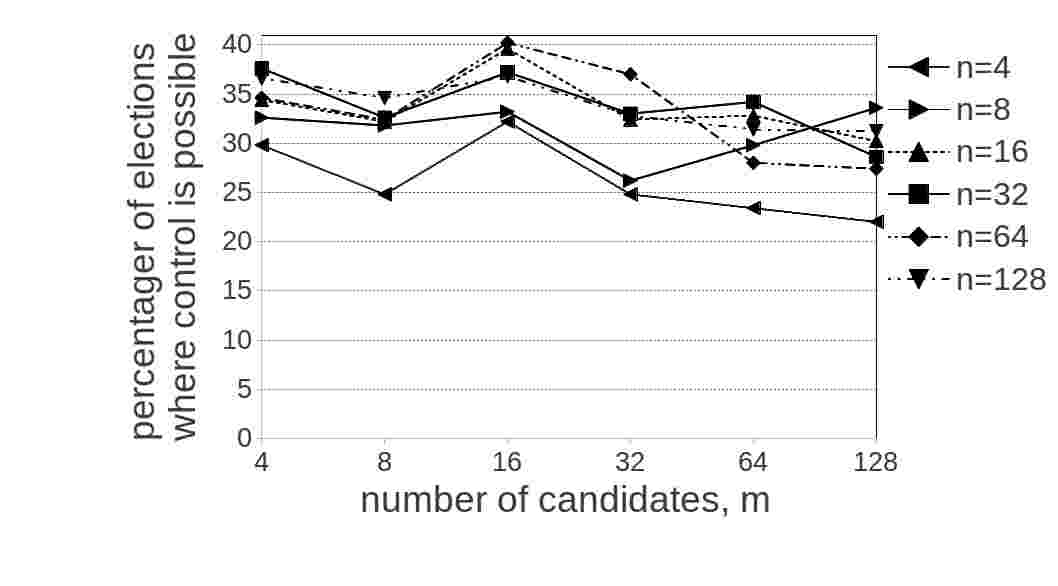}
		\caption{Results for plurality voting in the TM model for 
constructive control by deleting candidates. Number of voters is fixed. }
\end{figure}

%
 \end{center}

\clearpage
\subsubsection{Computational Costs}
\begin{figure}[ht]
\centering
	\includegraphics[scale=0.3]{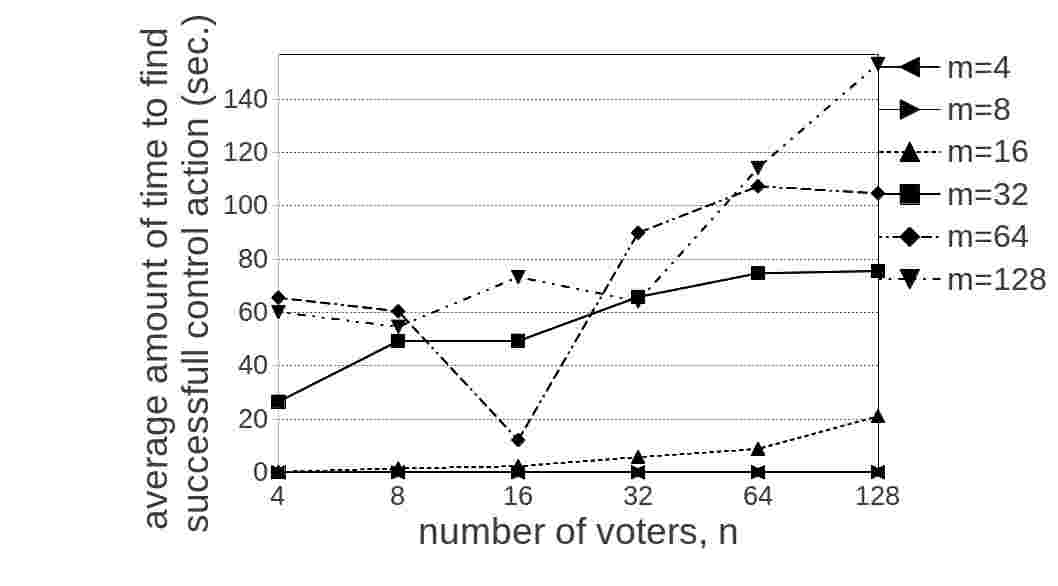}
	\caption{Average time the algorithm needs to find a successful control action for 
	constructive control by deleting candidates
	in plurality elections in the IC model. The maximum is $153,26$ seconds.}
\end{figure}
\begin{figure}[ht]
\centering
	\includegraphics[scale=0.3]{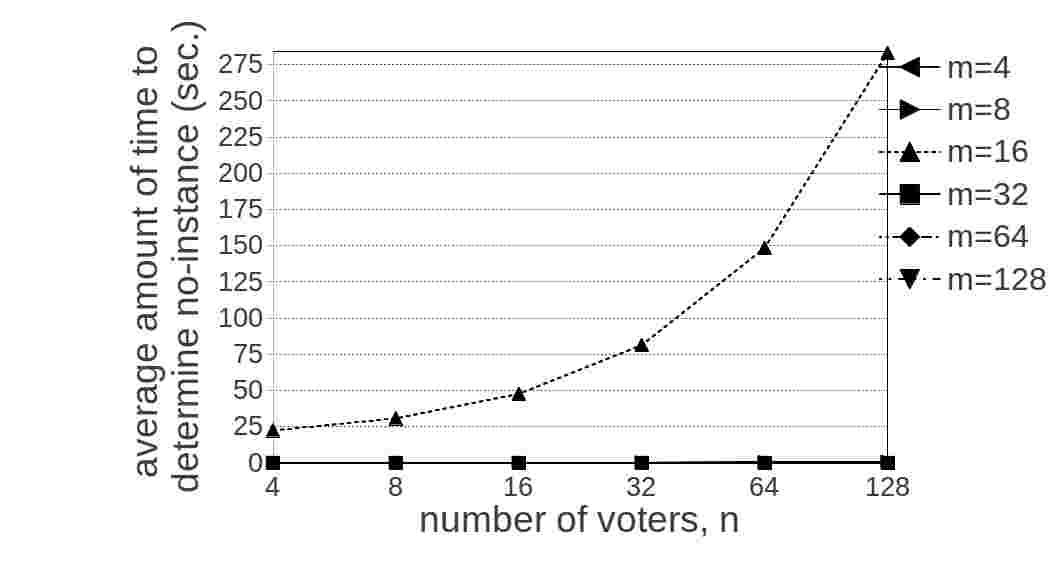}
	\caption{Average time the algorithm needs to determine no-instance of 
		constructive control by deleting candidates
	in plurality elections in the IC model. The maximum is $283,07$ seconds.}
\end{figure}
\begin{figure}[ht]
\centering
	\includegraphics[scale=0.3]{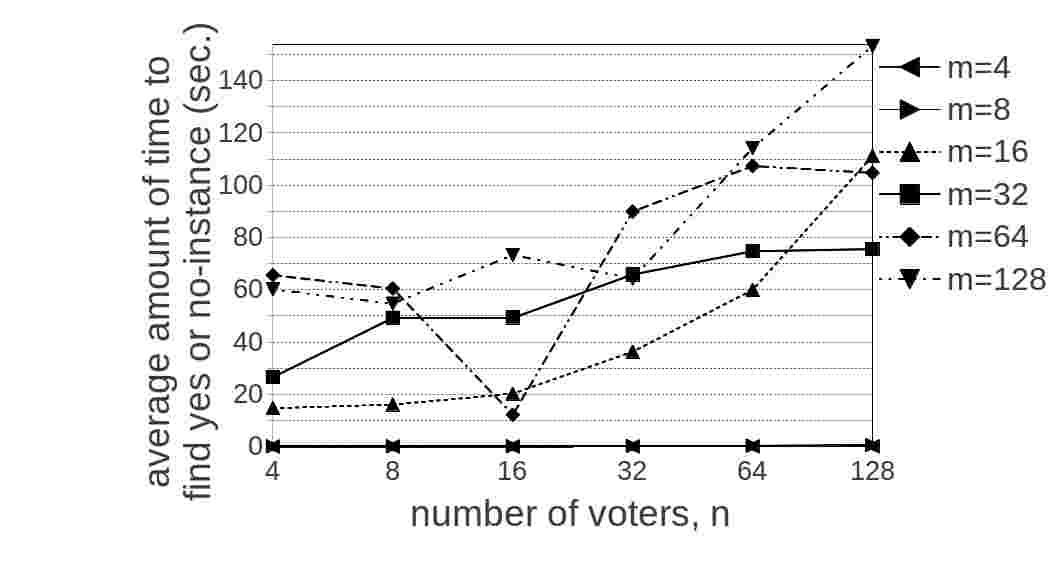}
	\caption{Average time the algorithm needs to give a definite output for 
	constructive control by deleting candidates
	in plurality elections in the IC model. The maximum is $153,26$ seconds.}
\end{figure}
\begin{figure}[ht]
\centering
	\includegraphics[scale=0.3]{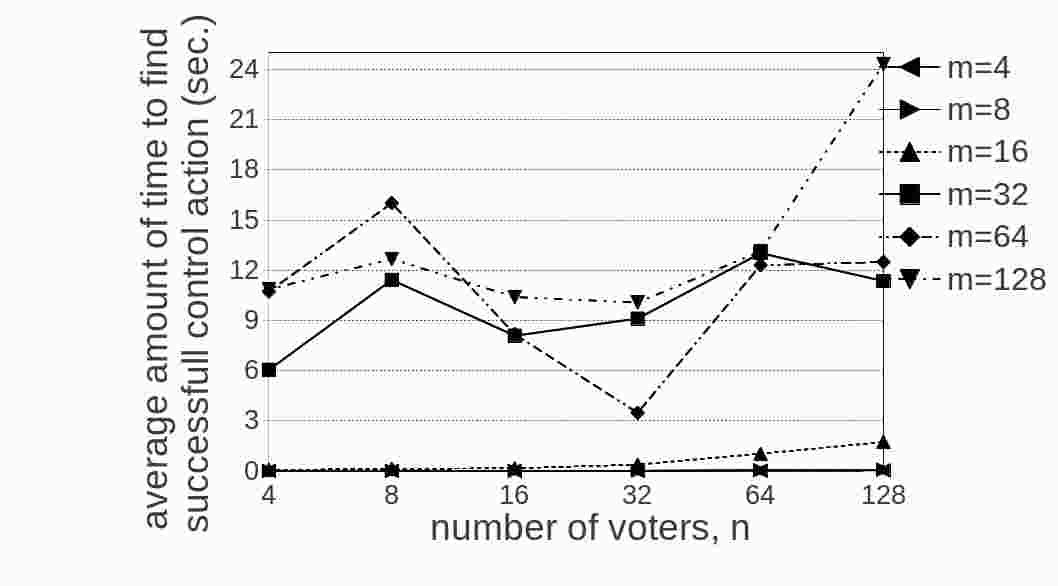}
	\caption{Average time the algorithm needs to find a successful control action for 
	constructive control by deleting candidates
	in plurality elections in the TM model. The maximum is $24,32$ seconds.}
\end{figure}
\begin{figure}[ht]
\centering
	\includegraphics[scale=0.3]{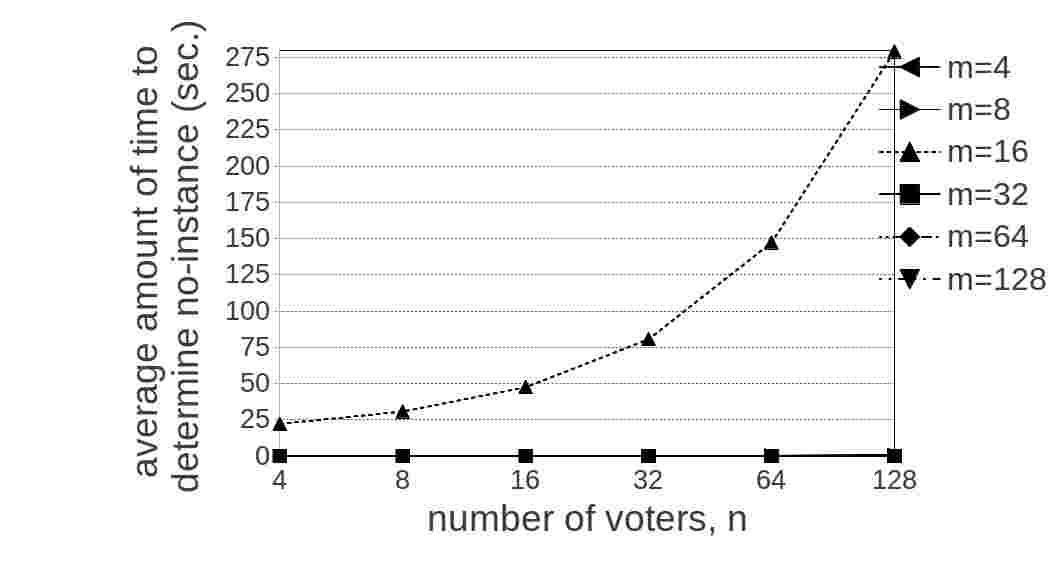}
	\caption{Average time the algorithm needs to determine no-instance of 
		constructive control by deleting candidates
	in plurality elections in the TM model. The maximum is $279,1$ seconds.}
\end{figure}
\begin{figure}[ht]
\centering
	\includegraphics[scale=0.3]{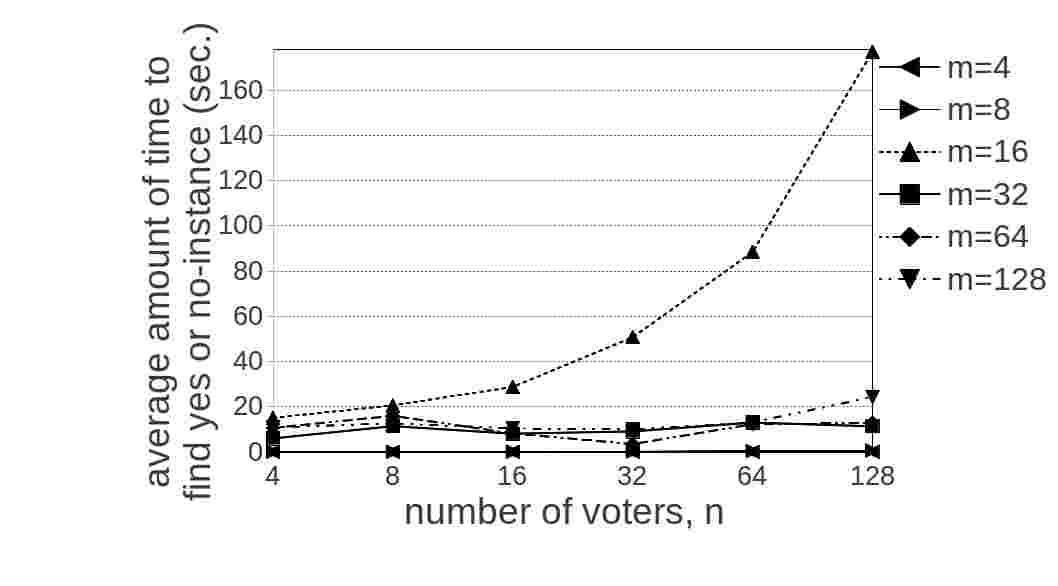}
	\caption{Average time the algorithm needs to give a definite output for 
	constructive control by deleting candidates
	in plurality elections in the TM model. The maximum is $177,03$ seconds.}
\end{figure}

\clearpage
\subsection{Destructive Control by Deleting Candidates}

\begin{center}
\begin{figure}[ht]
\centering
	\includegraphics[scale=0.3]{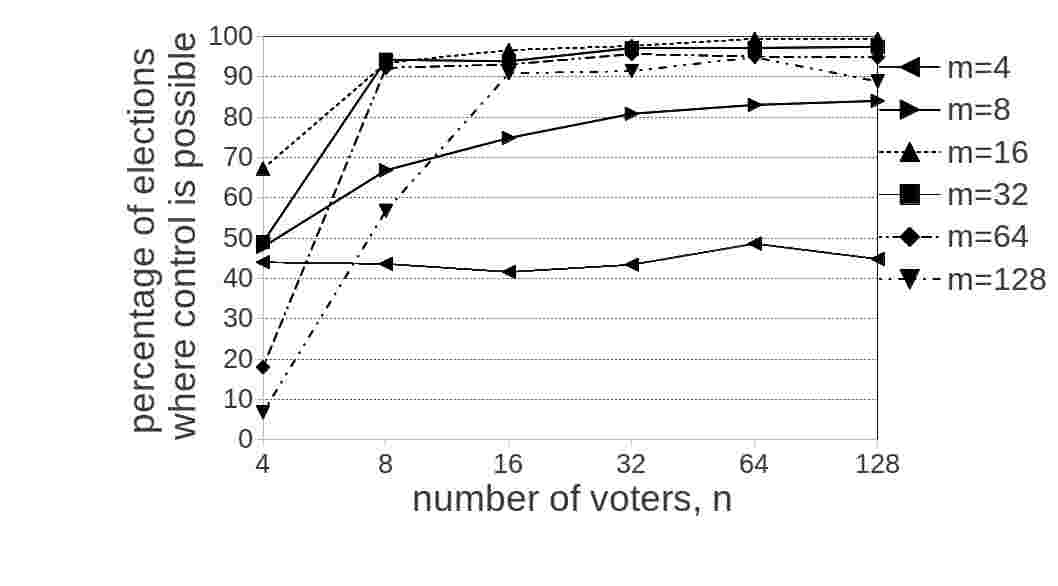}
		\caption{Results for plurality voting in the IC model for 
destructive control by deleting candidates. Number of candidates is fixed. }
\end{figure}


\newpage
\begin{figure}[ht]
\centering
	\includegraphics[scale=0.3]{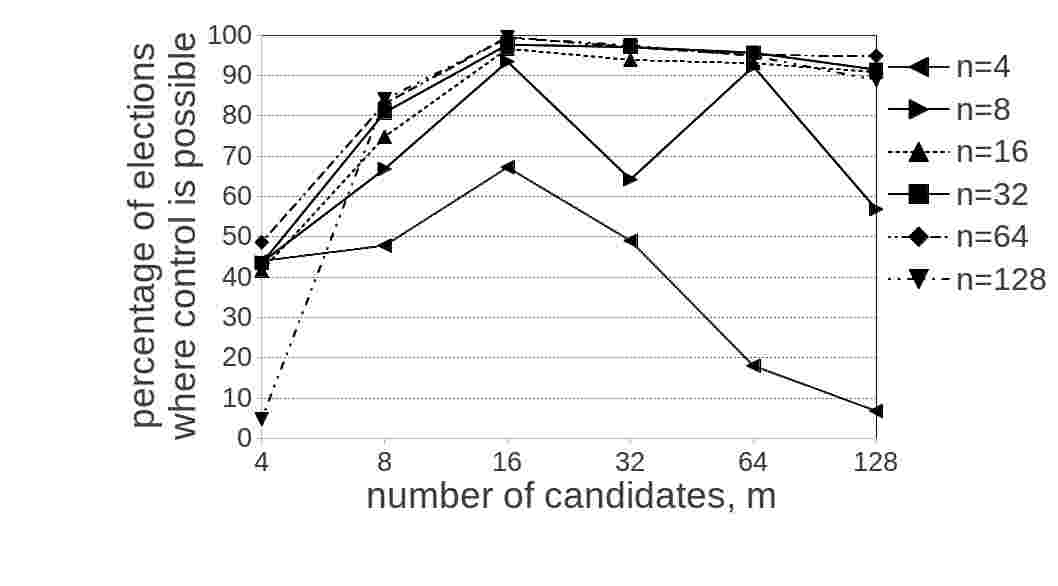}
		\caption{Results for plurality voting in the IC model for 
destructive control by deleting candidates. Number of voters is fixed. }
\end{figure}

%
\newpage
\begin{figure}[ht]
\centering
	\includegraphics[scale=0.3]{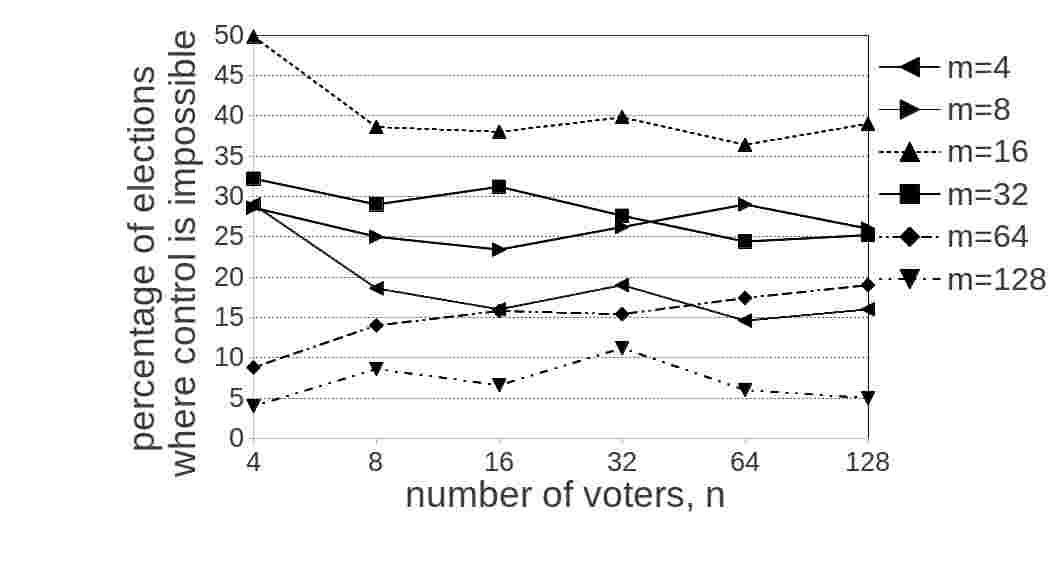}
		\caption{Results for plurality voting in the TM model for 
destructive control by deleting candidates. Number of candidates is fixed. }
\end{figure}

%
\newpage
\begin{figure}[ht]
\centering
	\includegraphics[scale=0.3]{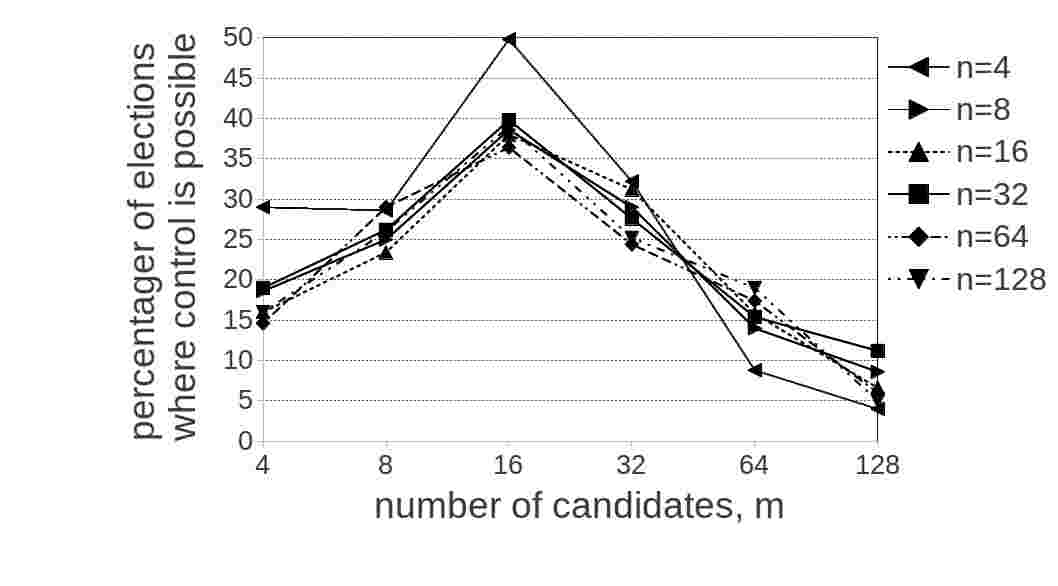}
		\caption{Results for plurality voting in the TM model for 
destructive control by deleting candidates. Number of voters is fixed. }
\end{figure}

%
\end{center}

\clearpage
\subsubsection{Computational Costs}
\begin{figure}[ht]
\centering
	\includegraphics[scale=0.3]{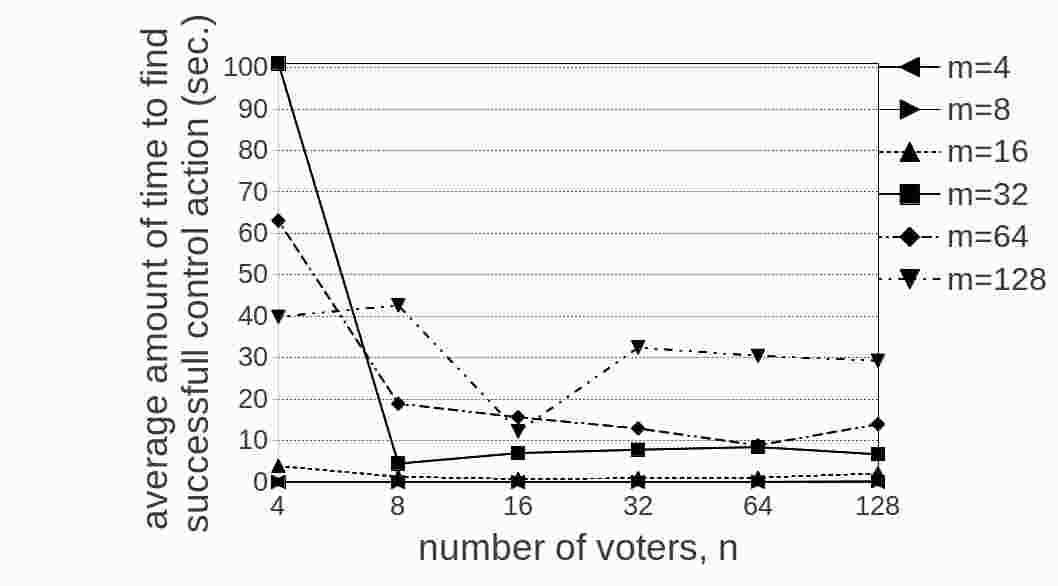}
	\caption{Average time the algorithm needs to find a successful control action for 
	destructive control by deleting candidates
	in plurality elections in the IC model. The maximum is $100,99$ seconds.}
\end{figure}
\begin{figure}[ht]
\centering
	\includegraphics[scale=0.3]{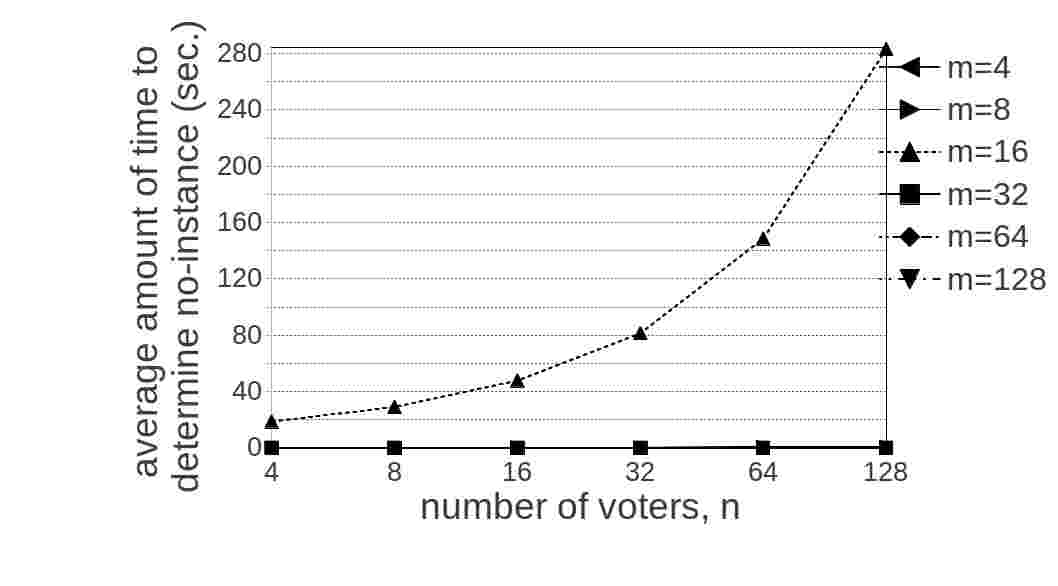}
	\caption{Average time the algorithm needs to determine no-instance of 
		destructive control by deleting candidates
	in plurality elections in the IC model. The maximum is $283,09$ seconds.}
\end{figure}
\begin{figure}[ht]
\centering
	\includegraphics[scale=0.3]{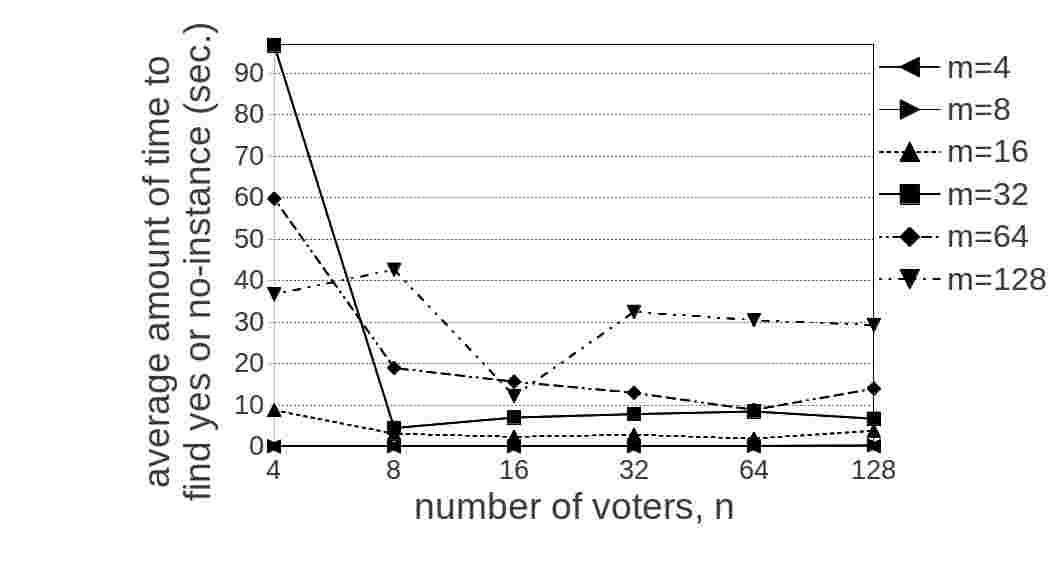}
	\caption{Average time the algorithm needs to give a definite output for 
	destructive control by deleting candidates
	in plurality elections in the IC model. The maximum is $96,65$ seconds.}
\end{figure}
\begin{figure}[ht]
\centering
	\includegraphics[scale=0.3]{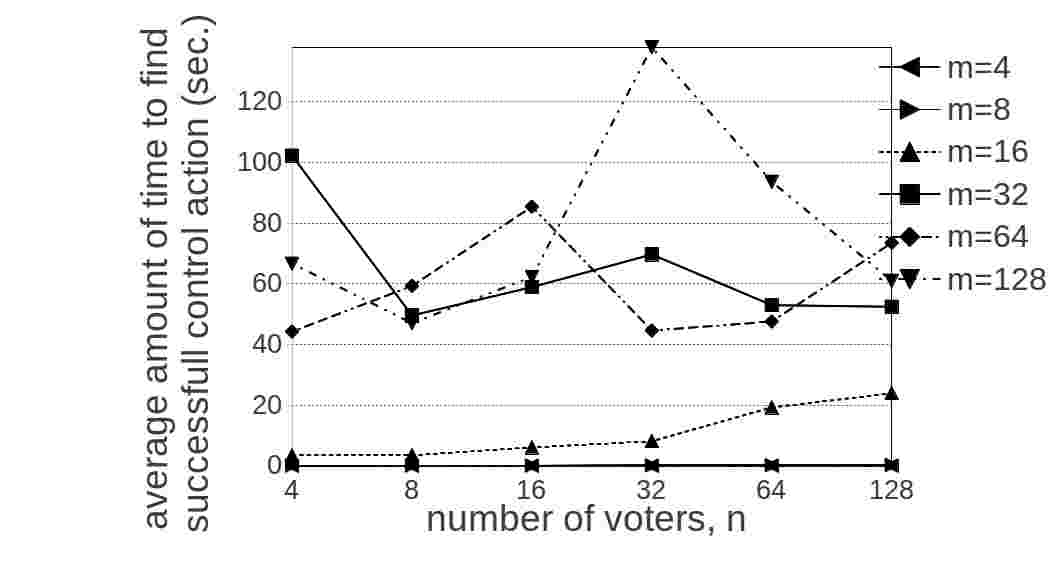}
	\caption{Average time the algorithm needs to find a successful control action for 
	destructive control by deleting candidates
	in plurality elections in the TM model. The maximum is $137,95$ seconds.}
\end{figure}
\begin{figure}[ht]
\centering
	\includegraphics[scale=0.3]{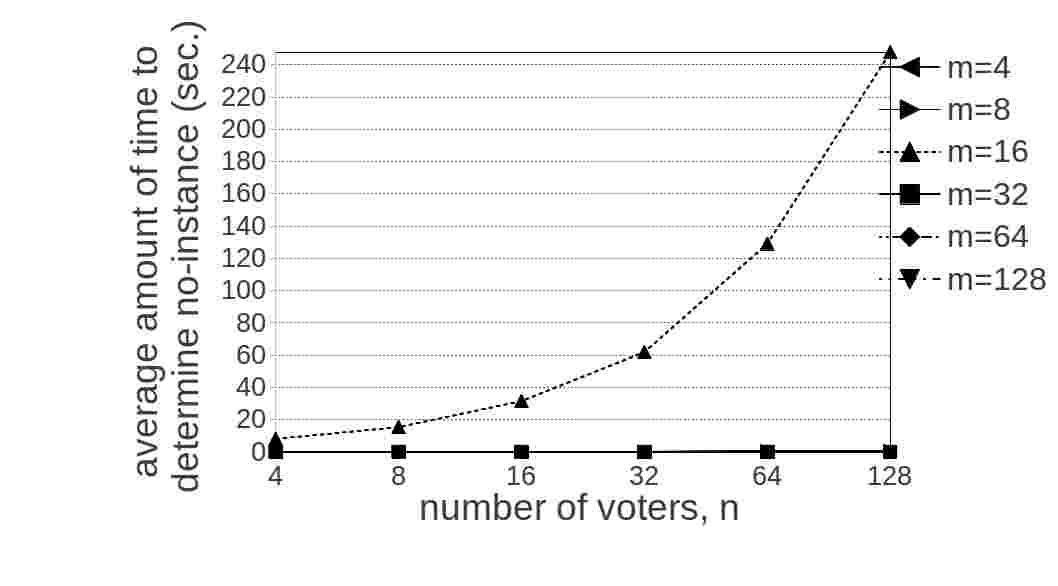}
	\caption{Average time the algorithm needs to determine no-instance of 
		destructive control by deleting candidates
	in plurality elections in the TM model. The maximum is $247,8$ seconds.}
\end{figure}
\begin{figure}[ht]
\centering
	\includegraphics[scale=0.3]{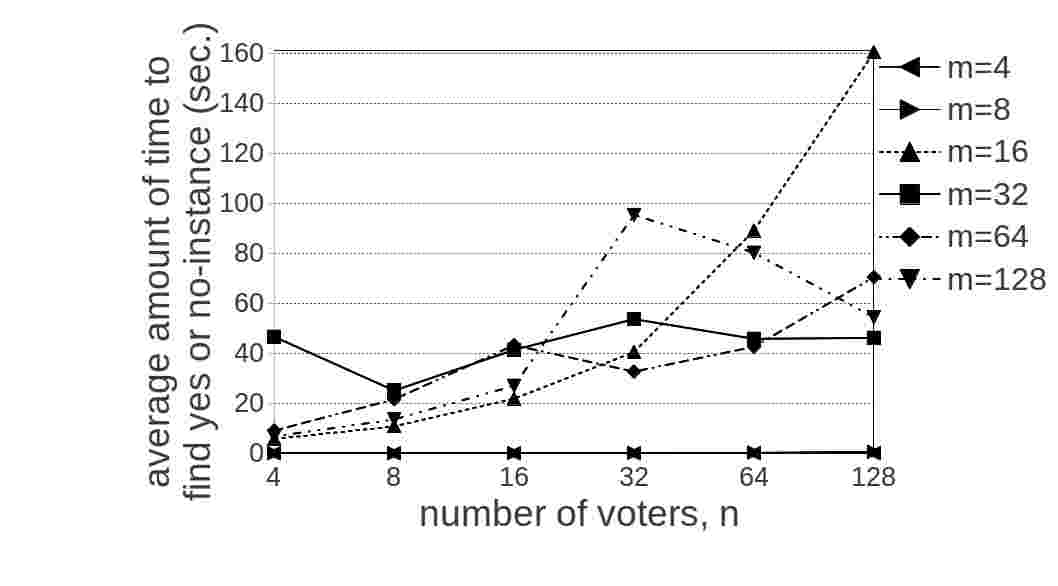}
	\caption{Average time the algorithm needs to give a definite output for 
	destructive control by deleting candidates
	in plurality elections in the TM model. The maximum is $160,46$ seconds.}
\end{figure}

\clearpage
\subsection{Constructive Control by Partition of Candidates in Model TE}

\begin{center}
\begin{figure}[ht]
\centering
	\includegraphics[scale=0.3]{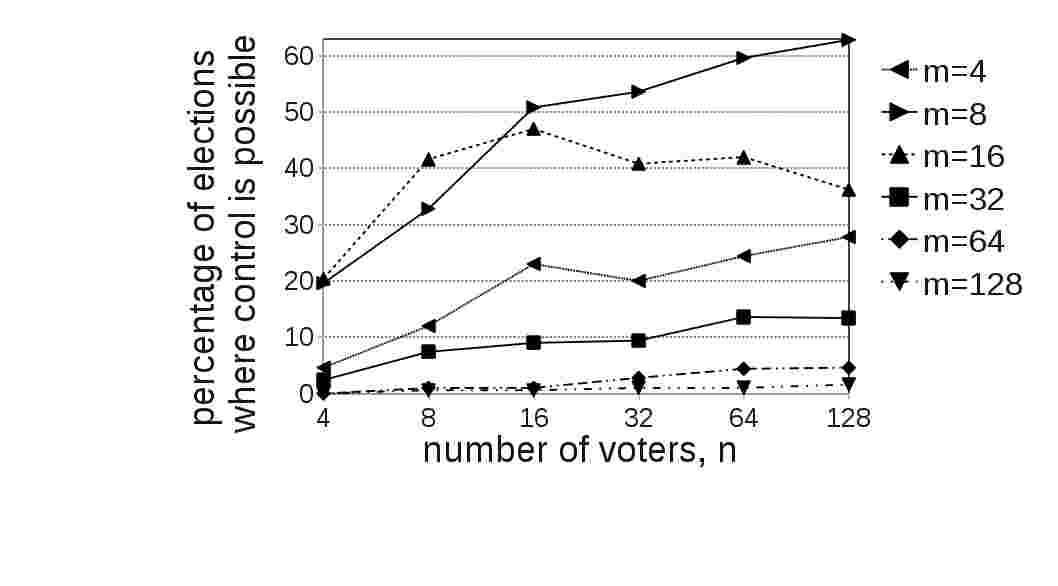}
		\caption{Results for plurality voting in the IC model for 
constructive control by partition of candidates in model TE. Number of voters is fixed. }
\end{figure}

\newpage
\begin{figure}[ht]
\centering
	\includegraphics[scale=0.3]{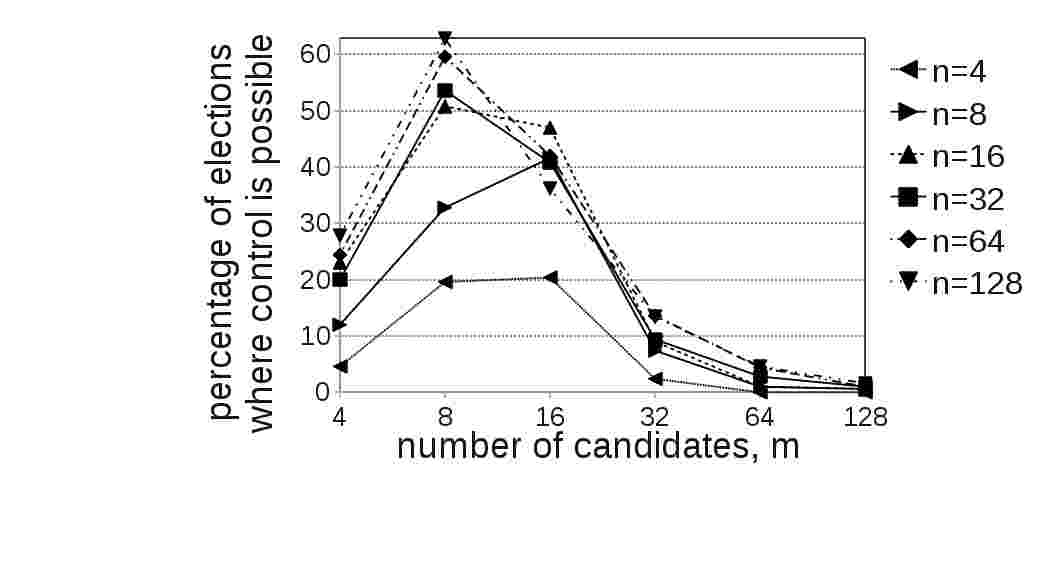}
		\caption{Results for plurality voting in the IC model for 
constructive control by partition of candidates in model TE. Number of voters is fixed. }
\end{figure}
%
\newpage
\begin{figure}[ht]
\centering
	\includegraphics[scale=0.3]{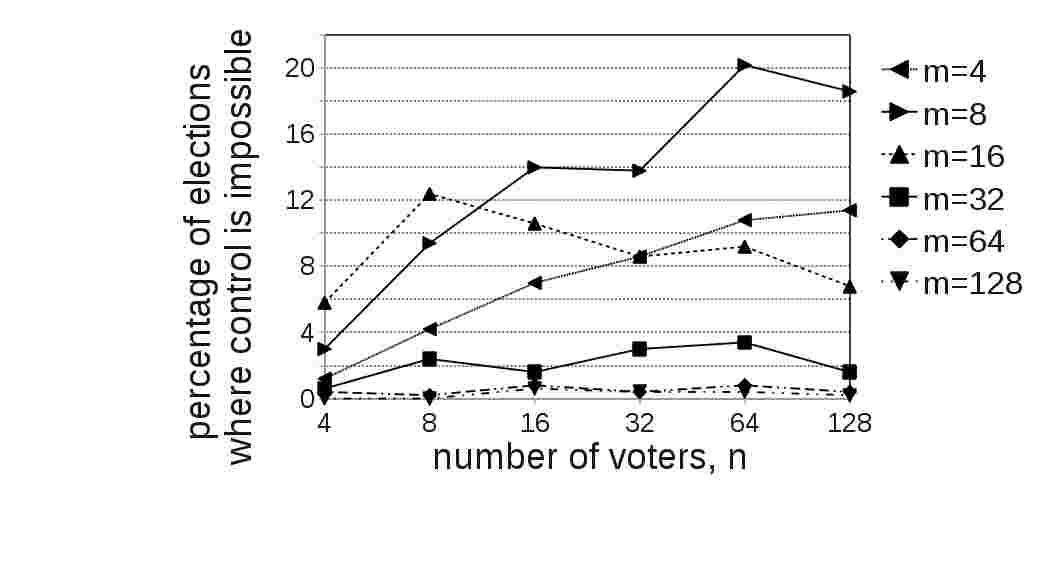}
	\caption{Results for plurality voting in the TM model for 
constructive control by partition of candidates in model TE. Number of candidates is fixed. }
\end{figure}
%
\newpage
\begin{figure}[ht]
\centering
	\includegraphics[scale=0.3]{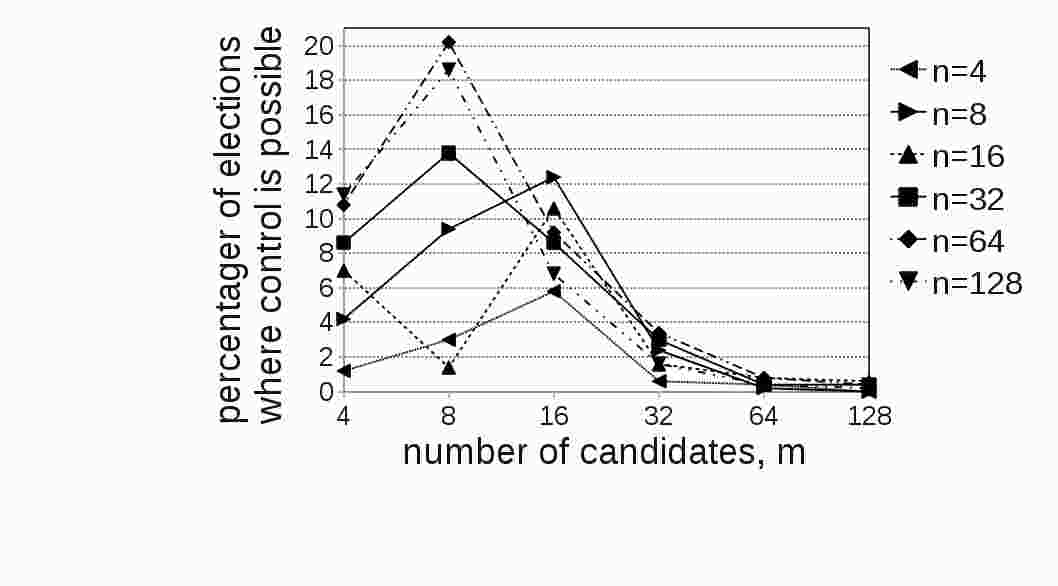}
	\caption{Results for plurality voting in the TM model for 
constructive control by partition of candidates in model TE. Number of voters is fixed. }
\end{figure}
%
\end{center}
\clearpage
\subsubsection{Computational Costs}
\begin{figure}[ht]
\centering
	\includegraphics[scale=0.3]{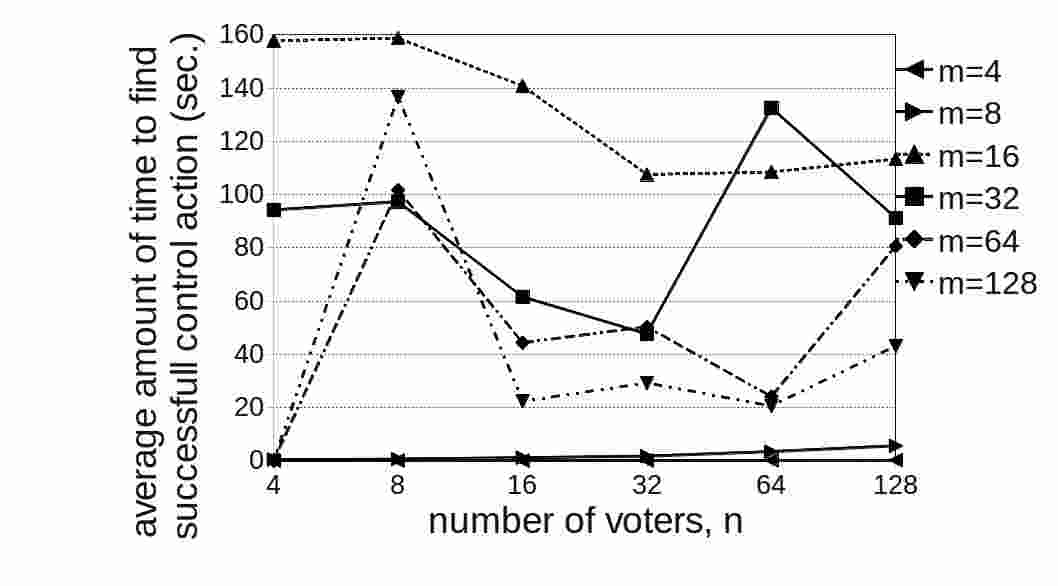}
	\caption{Average time the algorithm needs to find a successful control action for 
	constructive control by partition of candidates in model TE
	in plurality elections in the IC model. The maximum is $158,73$ seconds.}
\end{figure}
\begin{figure}[ht]
\centering
	\includegraphics[scale=0.3]{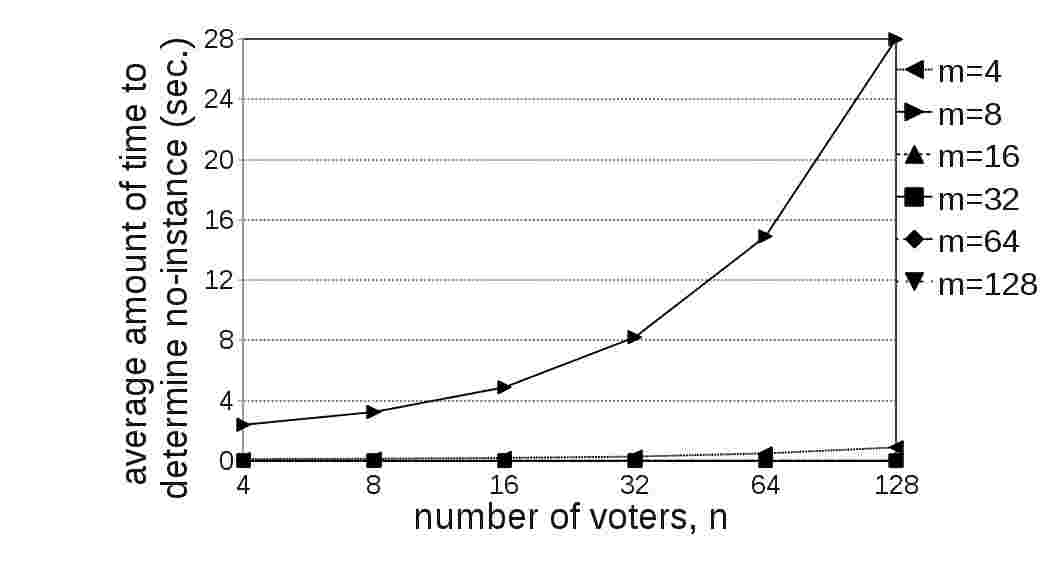}
	\caption{Average time the algorithm needs to determine no-instance of 
		constructive control by partition of candidates in model TE
	in plurality elections in the IC model. The maximum is $27,99$ seconds.}
\end{figure}
\begin{figure}[ht]
\centering
	\includegraphics[scale=0.3]{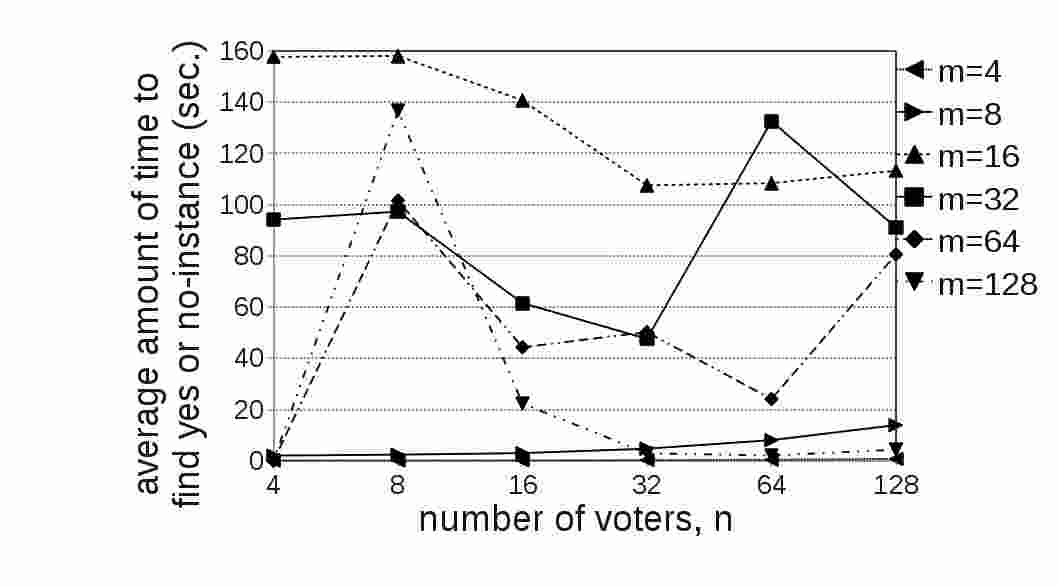}
	\caption{Average time the algorithm needs to give a definite output for 
	constructive control by partition of candidates in model TE
	in plurality elections in the IC model. The maximum is $158,11$ seconds.}
\end{figure}
\begin{figure}[ht]
\centering
	\includegraphics[scale=0.3]{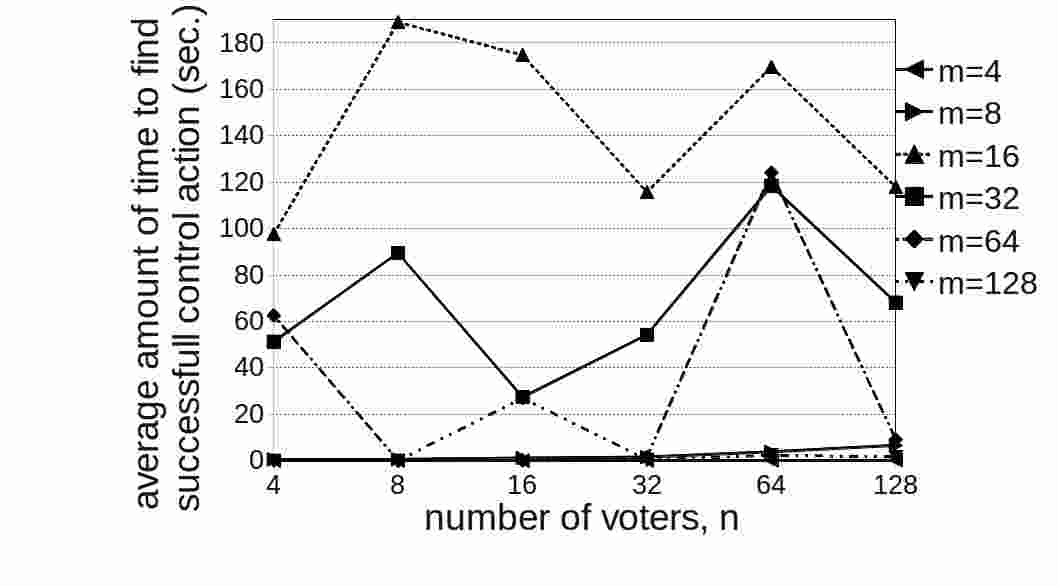}
	\caption{Average time the algorithm needs to find a successful control action for 
	constructive control by partition of candidates in model TE
	in plurality elections in the TM model. The maximum is $188,89$ seconds.}
\end{figure}
\begin{figure}[ht]
\centering
	\includegraphics[scale=0.3]{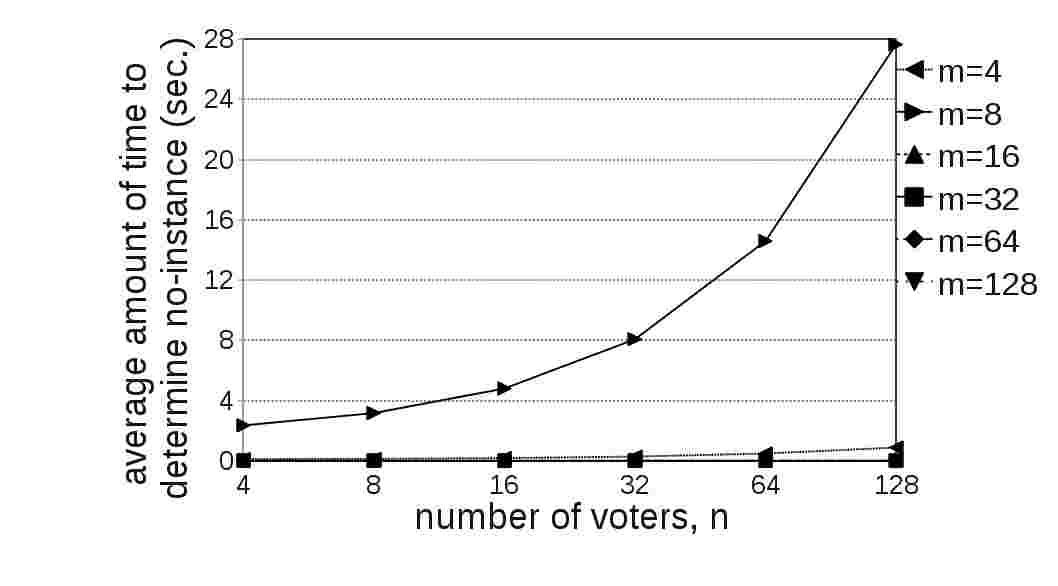}
	\caption{Average time the algorithm needs to determine no-instance of 
		constructive control by partition of candidates in model TE
	in plurality elections in the TM model. The maximum is $27,64$ seconds.}
\end{figure}
\begin{figure}[ht]
\centering
	\includegraphics[scale=0.3]{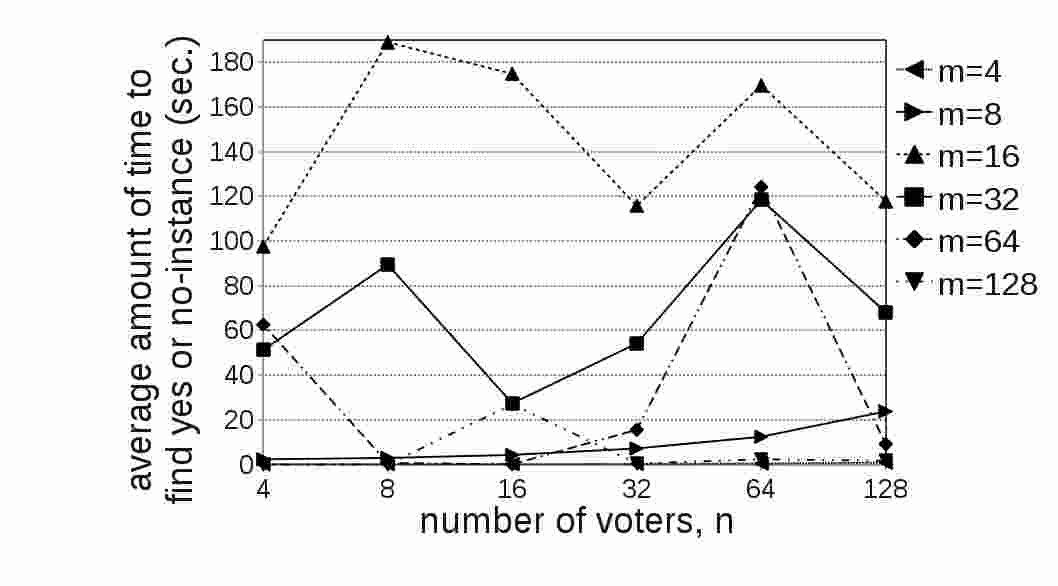}
	\caption{Average time the algorithm needs to give a definite output for 
	constructive control by partition of candidates in model TE
	in plurality elections in the TM model. The maximum is $188,89$ seconds.}
\end{figure}

\clearpage
\subsection{Destructive Control by Partition of Candidates in Model TE}
\begin{center}
\begin{figure}[ht]
\centering
	\includegraphics[scale=0.3]{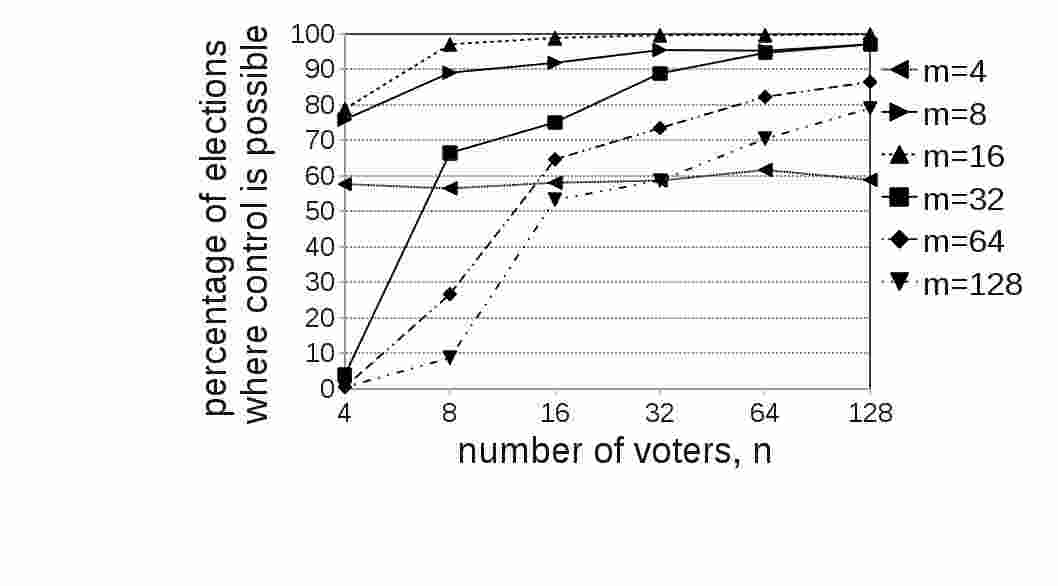}
		\caption{Results for plurality voting in the IC model for 
destructive control by partition of candidates in model TE. Number of candidates is fixed. }
\end{figure}


\newpage
\begin{figure}[ht]
\centering
	\includegraphics[scale=0.3]{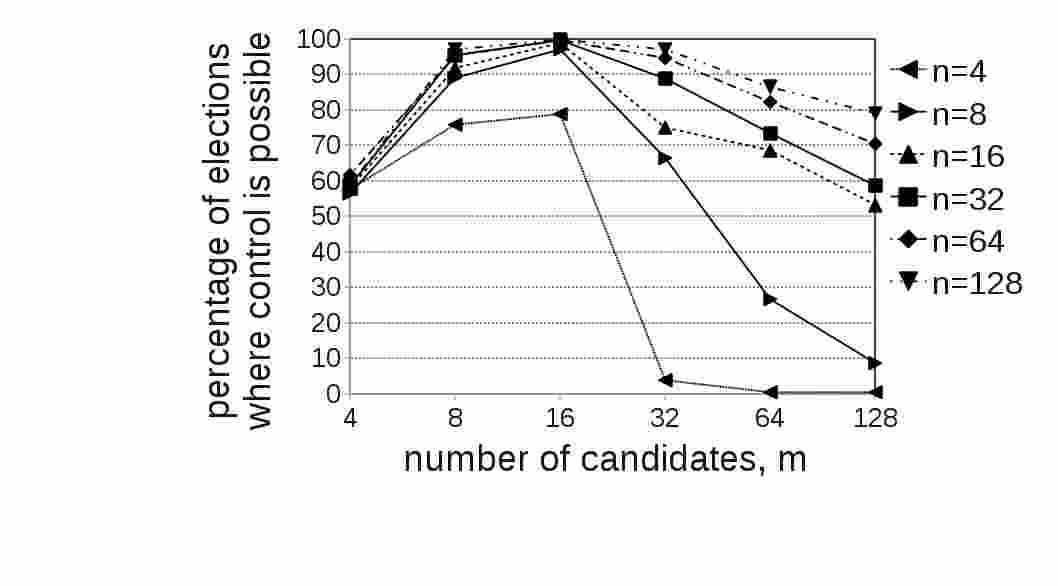}
		\caption{Results for plurality voting in the IC model for 
destructive control by partition of candidates in model TE. Number of voters is fixed. }
\end{figure}
%
\newpage
\begin{figure}[ht]
\centering
	\includegraphics[scale=0.3]{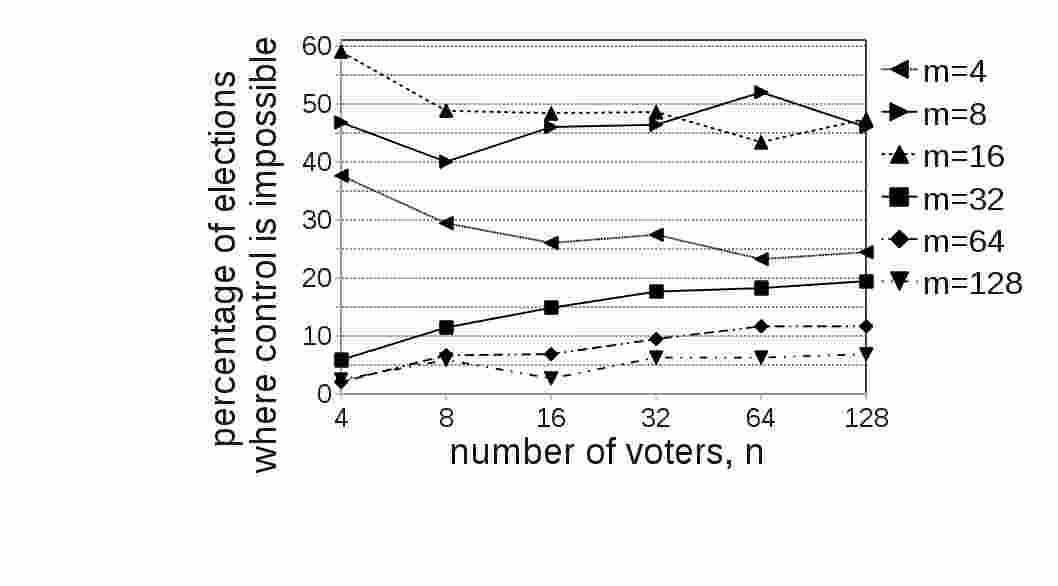}
	\caption{Results for plurality voting in the TM model for 
destructive control by partition of candidates in model TE. Number of candidates is fixed. }
\end{figure}
%
\newpage
\begin{figure}[ht]
\centering
	\includegraphics[scale=0.3]{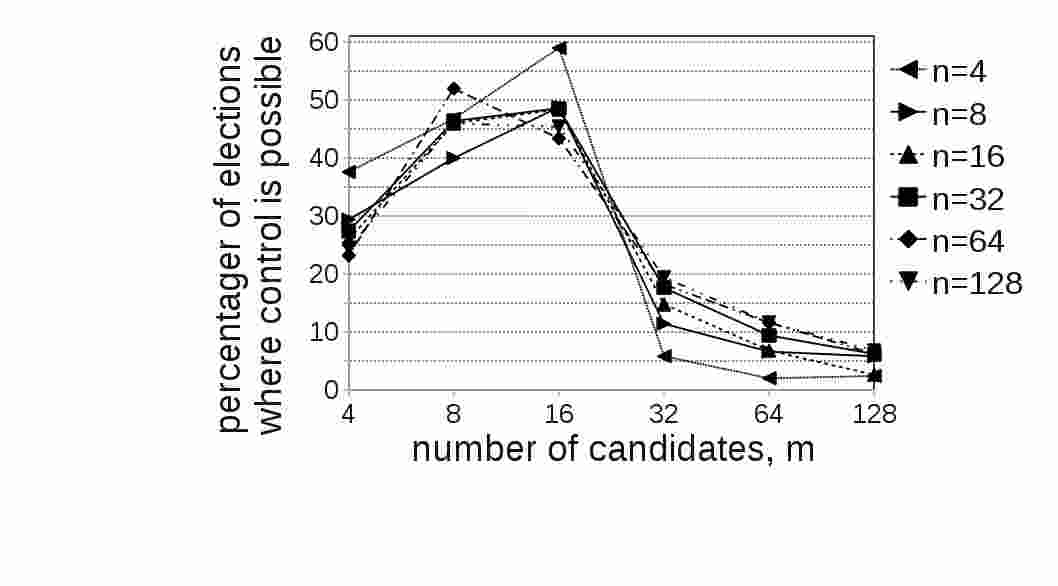}
	\caption{Results for plurality voting in the TM model for 
destructive control by partition of candidates in model TE. Number of voters is fixed. }
\end{figure}
%
\end{center}

\clearpage
\subsubsection{Computational Costs}
\begin{figure}[ht]
\centering
	\includegraphics[scale=0.3]{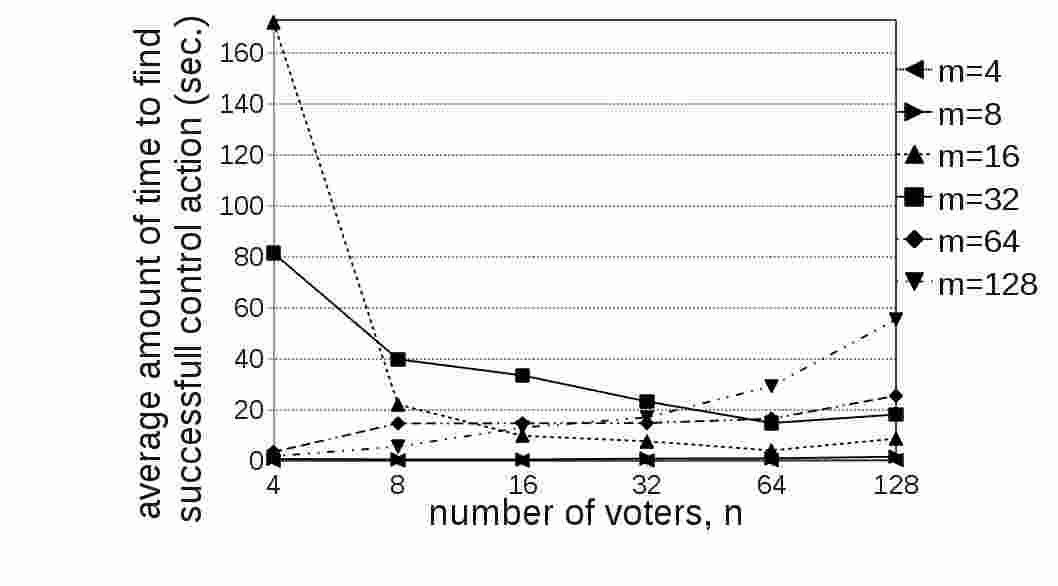}
	\caption{Average time the algorithm needs to find a successful control action for 
	destructive control by partition of candidates in model TE
	in plurality elections in the IC model. The maximum is $172,19$ seconds.}
\end{figure}
\begin{figure}[ht]
\centering
	\includegraphics[scale=0.3]{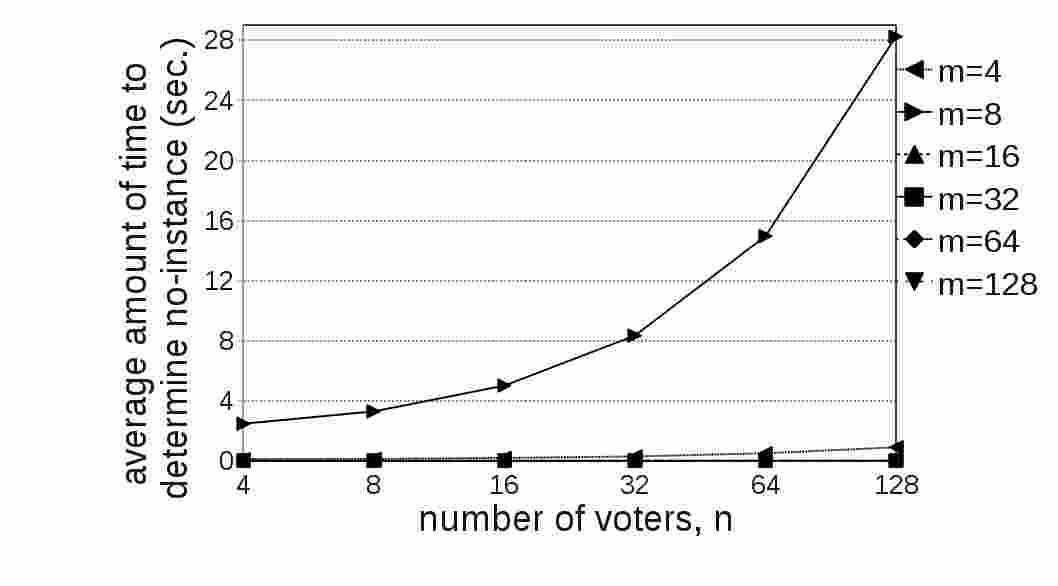}
	\caption{Average time the algorithm needs to determine no-instance of 
		destructive control by partition of candidates in model TE
	in plurality elections in the IC model. The maximum is $28,26$ seconds.}
\end{figure}
\begin{figure}[ht]
\centering
	\includegraphics[scale=0.3]{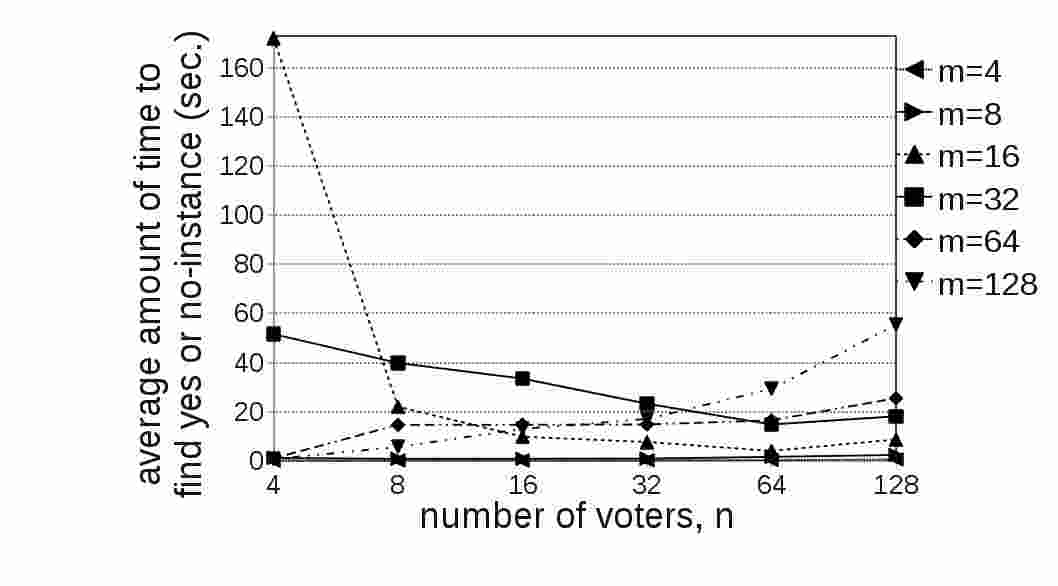}
	\caption{Average time the algorithm needs to give a definite output for 
	destructive control by partition of candidates in model TE
	in plurality elections in the IC model. The maximum is $172,19$ seconds.}
\end{figure}
\begin{figure}[ht]
\centering
	\includegraphics[scale=0.3]{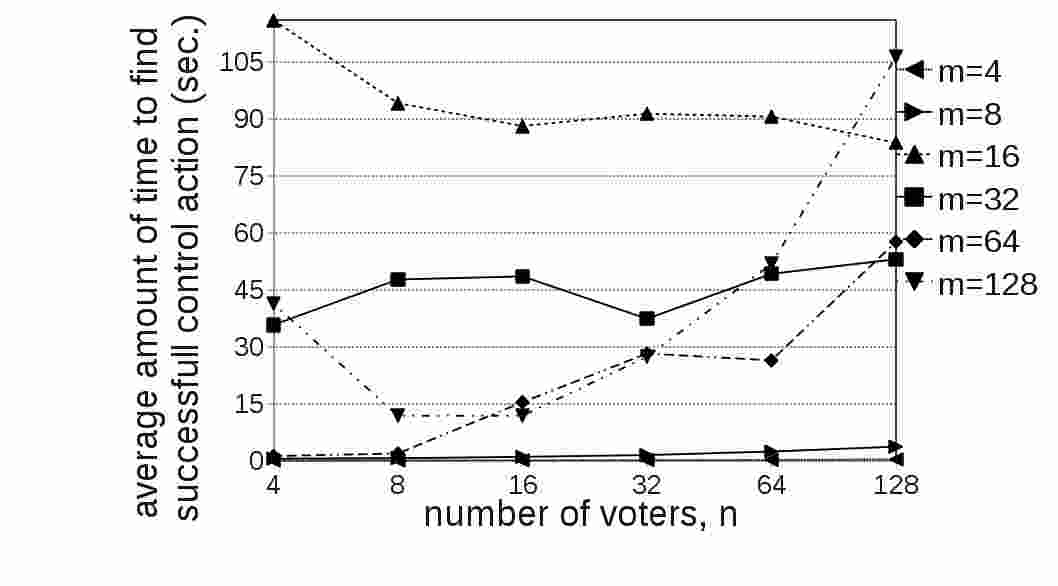}
	\caption{Average time the algorithm needs to find a successful control action for 
	destructive control by partition of candidates in model TE
	in plurality elections in the TM model. The maximum is $115,89$ seconds.}
\end{figure}
\begin{figure}[ht]
\centering
	\includegraphics[scale=0.3]{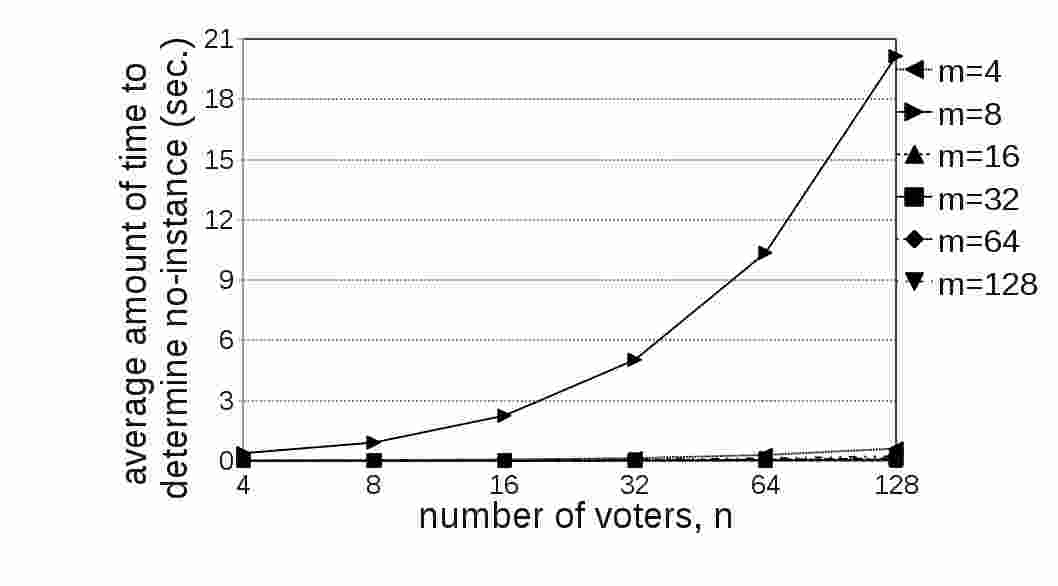}
	\caption{Average time the algorithm needs to determine no-instance of 
		destructive control by partition of candidates in model TE
	in plurality elections in the TM model. The maximum is $20,15$ seconds.}
\end{figure}
\begin{figure}[ht]
\centering
	\includegraphics[scale=0.3]{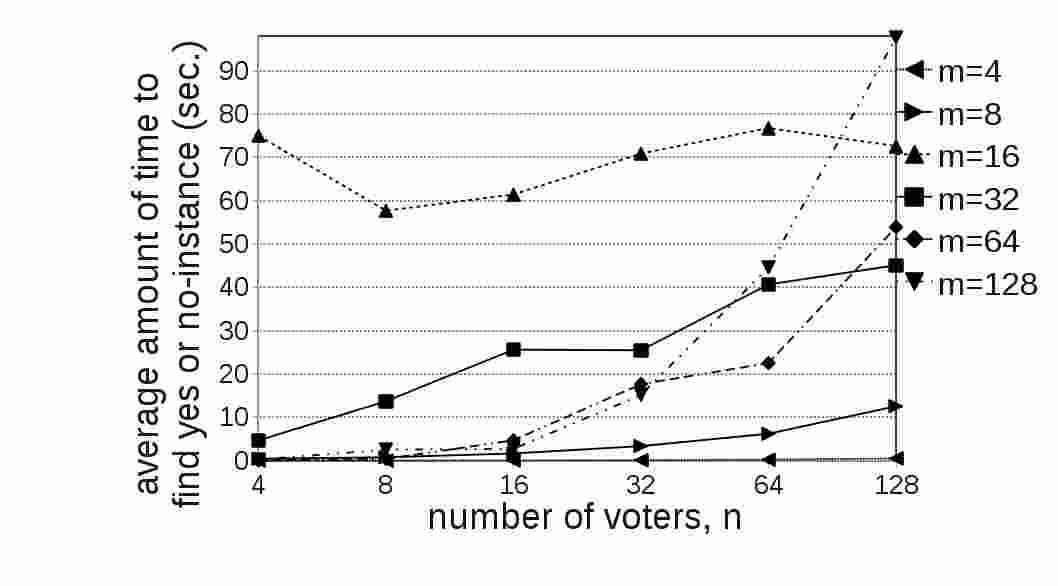}
	\caption{Average time the algorithm needs to give a definite output for 
	destructive control by partition of candidates in model TE
	in plurality elections in the TM model. The maximum is $97,65$ seconds.}
\end{figure}

\clearpage
\subsection{Constructive Control by Partition of Candidates in Model TP}
\begin{center}
\begin{figure}[ht]
\centering
	\includegraphics[scale=0.3]{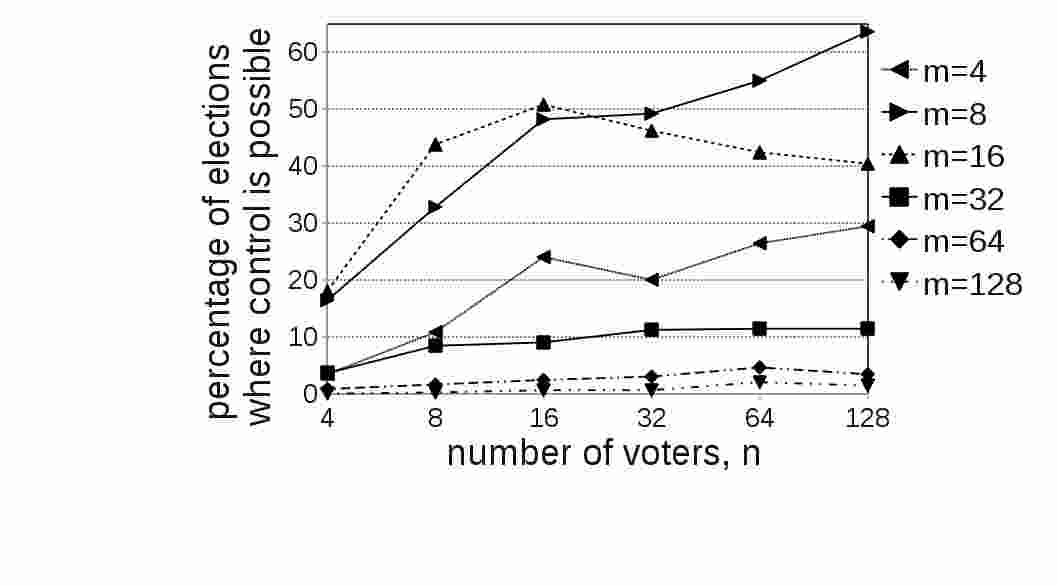}
		\caption{Results for plurality voting in the IC model for 
constructive control by partition of candidates in model TP. Number of candidates is fixed. }
\end{figure}

\newpage
\begin{figure}[ht]
\centering
	\includegraphics[scale=0.3]{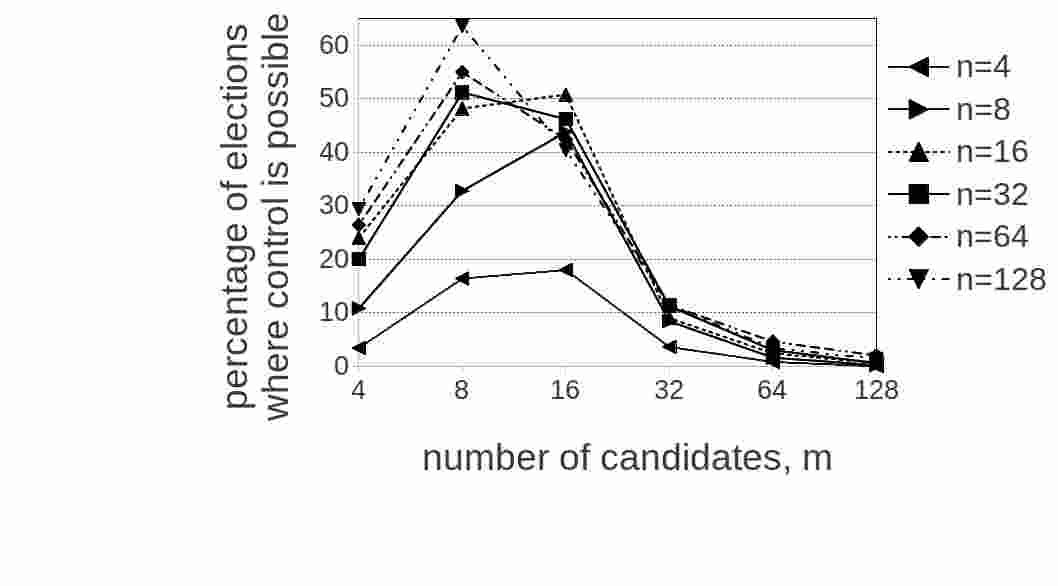}
		\caption{Results for plurality voting in the IC model for 
constructive control by partition of candidates in model TP. Number of voters is fixed. }
\end{figure}
%
\newpage
\begin{figure}[ht]
\centering
	\includegraphics[scale=0.3]{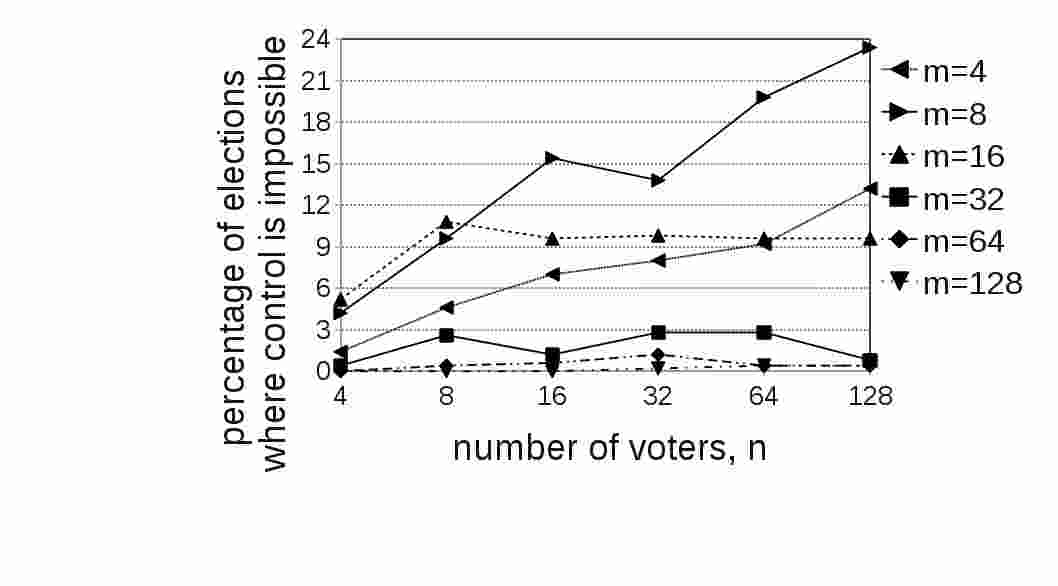}
	\caption{Results for plurality voting in the TM model for 
constructive control by partition of candidates in model TP. Number of candidates is fixed. }
\end{figure}
%
\newpage
\begin{figure}[ht]
\centering
	\includegraphics[scale=0.3]{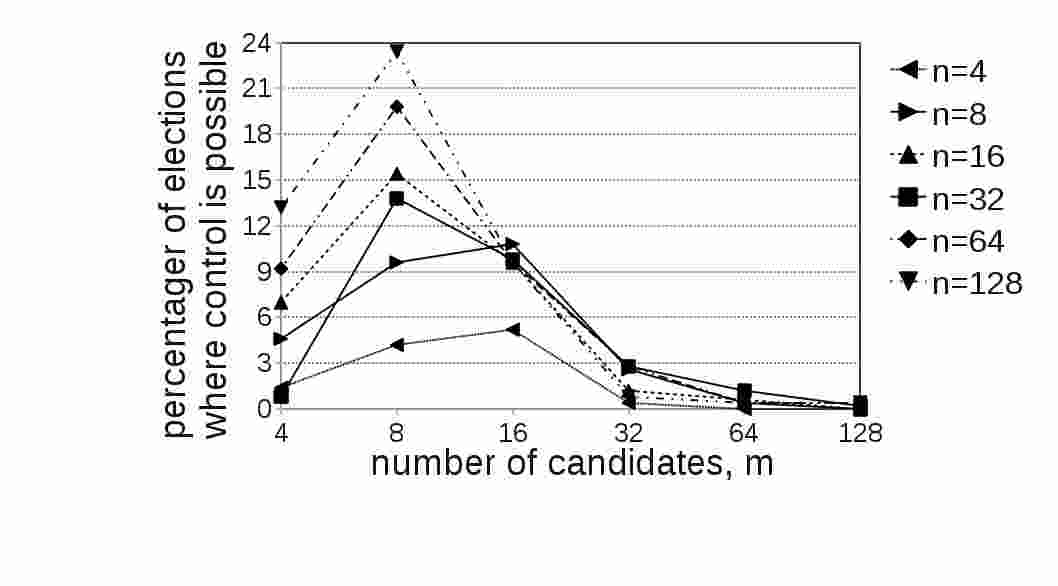}
	\caption{Results for plurality voting in the TM model for 
constructive control by partition of candidates in model TP. Number of voters is fixed. }
\end{figure}
%
\end{center}

\clearpage
\subsubsection{Computational Costs}
\begin{figure}[ht]
\centering
	\includegraphics[scale=0.3]{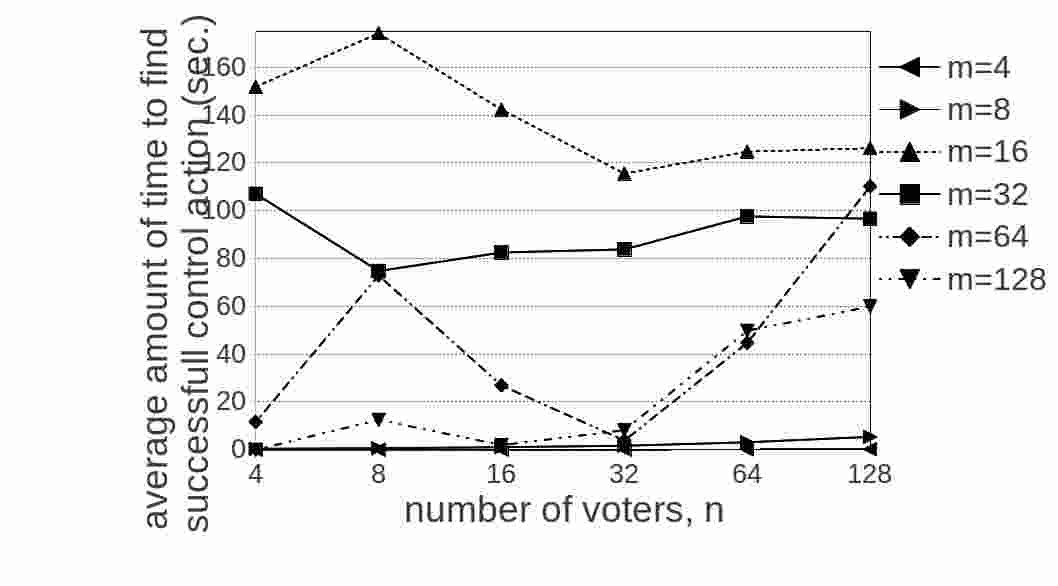}
	\caption{Average time the algorithm needs to find a successful control action for 
	constructive control by partition of candidates in model TP
	in plurality elections in the IC model. The maximum is $174,14$ seconds.}
\end{figure}
\begin{figure}[ht]
\centering
	\includegraphics[scale=0.3]{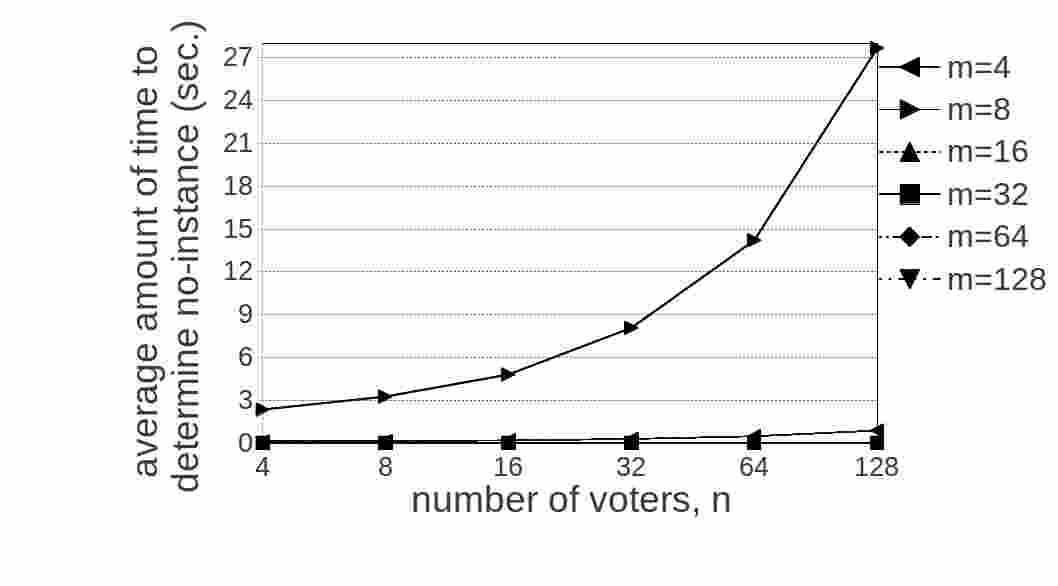}
	\caption{Average time the algorithm needs to determine no-instance of 
		constructive control by partition of candidates in model TP
	in plurality elections in the IC model. The maximum is $27,68$ seconds.}
\end{figure}
\begin{figure}[ht]
\centering
	\includegraphics[scale=0.3]{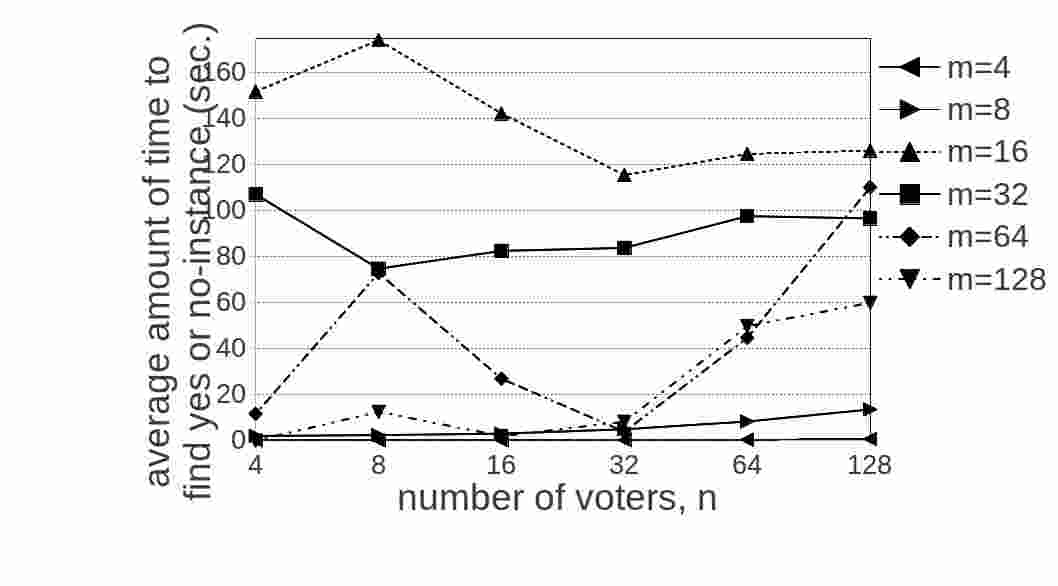}
	\caption{Average time the algorithm needs to give a definite output for 
	constructive control by partition of candidates in model TP
	in plurality elections in the IC model. The maximum is $174,14$ seconds.}
\end{figure}
\begin{figure}[ht]
\centering
	\includegraphics[scale=0.3]{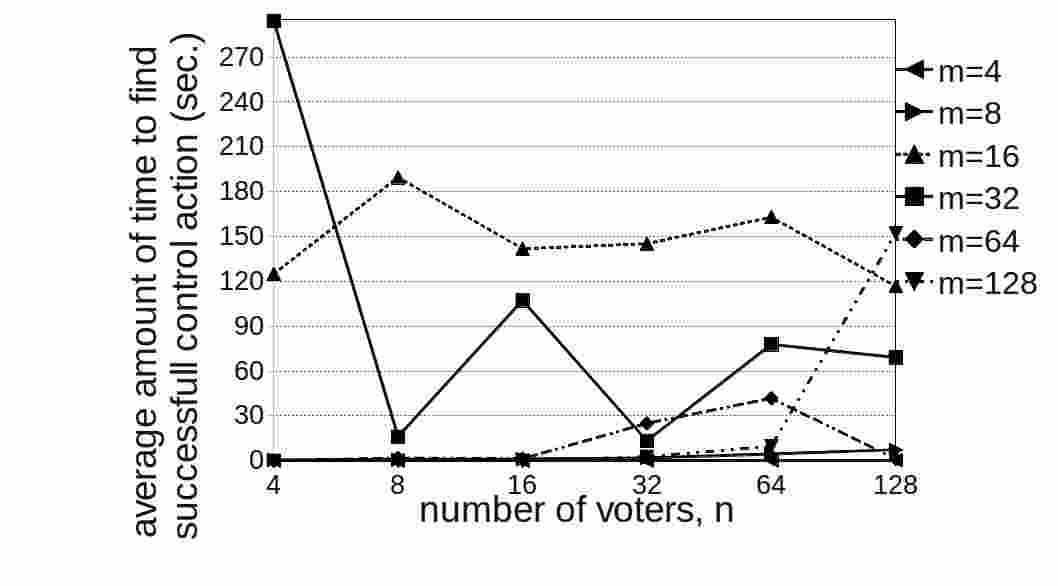}
	\caption{Average time the algorithm needs to find a successful control action for 
	constructive control by partition of candidates in model TP
	in plurality elections in the TM model. The maximum is $294,18$ seconds.}
\end{figure}
\begin{figure}[ht]
\centering
	\includegraphics[scale=0.3]{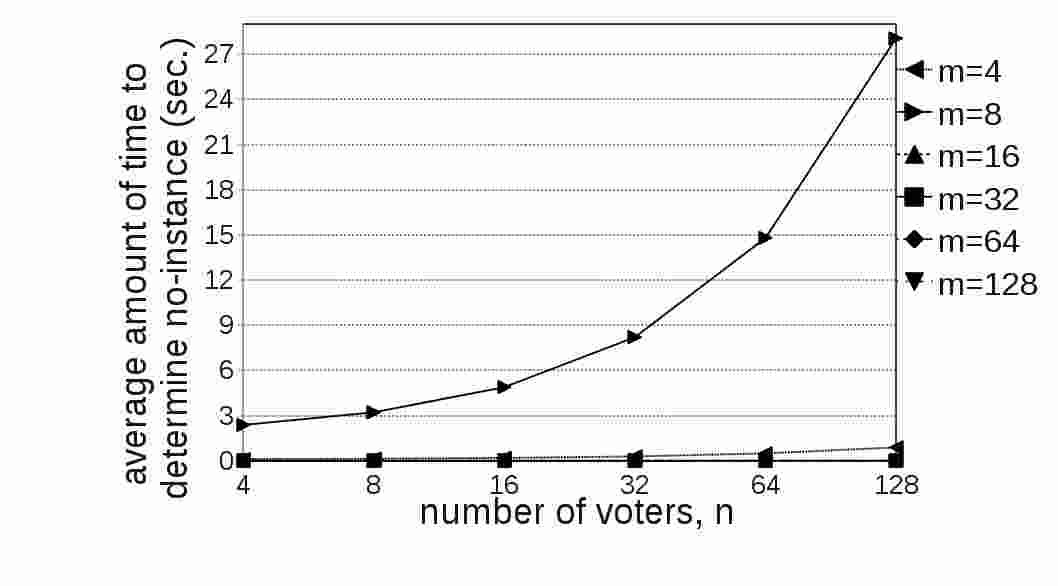}
	\caption{Average time the algorithm needs to determine no-instance of 
		constructive control by partition of candidates in model TP
	in plurality elections in the TM model. The maximum is $28,05$ seconds.}
\end{figure}
\begin{figure}[ht]
\centering
	\includegraphics[scale=0.3]{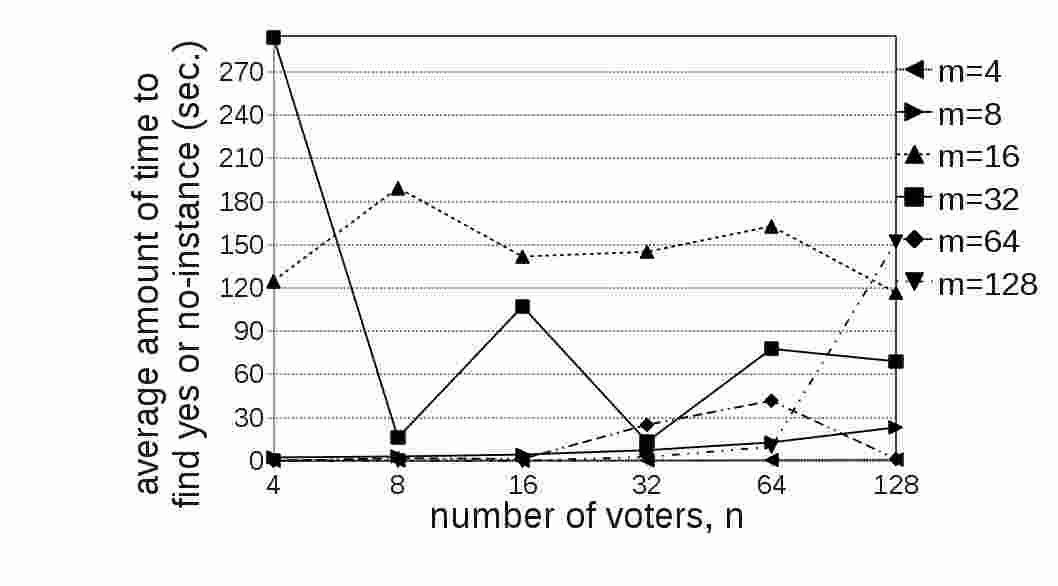}
	\caption{Average time the algorithm needs to give a definite output for 
	constructive control by partition of candidates in model TP
	in plurality elections in the TM model. The maximum is $294,18$ seconds.}
\end{figure}
\clearpage
\subsection{Destructive Control by Partition of Candidates in Model TP}
\begin{center}
\begin{figure}[ht]
\centering
	\includegraphics[scale=0.3]{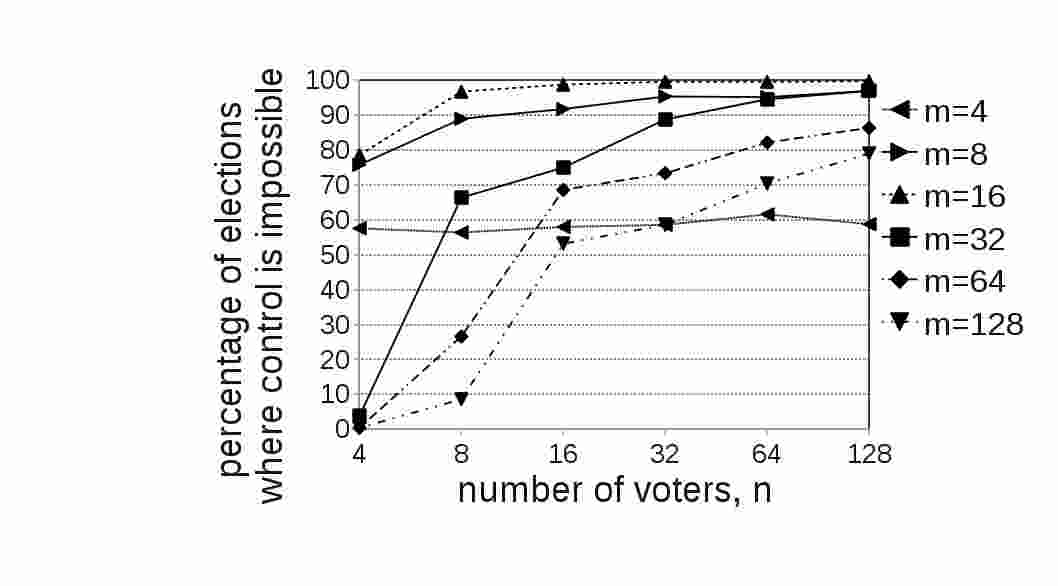}
		\caption{Results for plurality voting in the IC model for 
destructive control by partition of candidates in model TP. Number of candidates is fixed. }
\end{figure}

\newpage
\begin{figure}[ht]
\centering
	\includegraphics[scale=0.3]{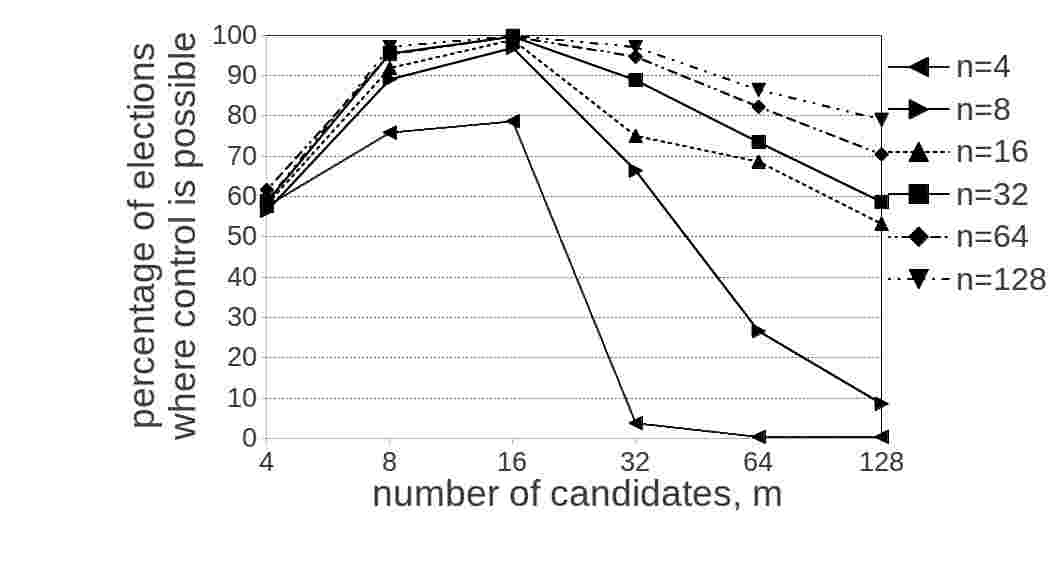}
		\caption{Results for plurality voting in the IC model for 
destructive control by partition of candidates in model TP. Number of voters is fixed. }
\end{figure}
%
\newpage
\begin{figure}[ht]
\centering
	\includegraphics[scale=0.3]{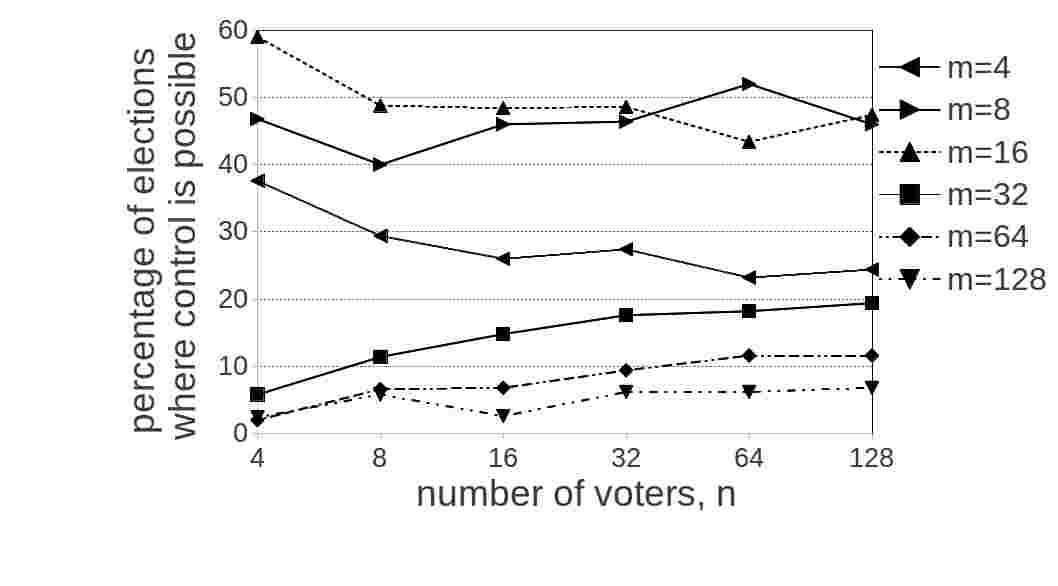}
	\caption{Results for plurality voting in the TM model for 
destructive control by partition of candidates in model TP. Number of candidates is fixed. }
\end{figure}
%
\newpage
\begin{figure}[ht]
\centering
	\includegraphics[scale=0.3]{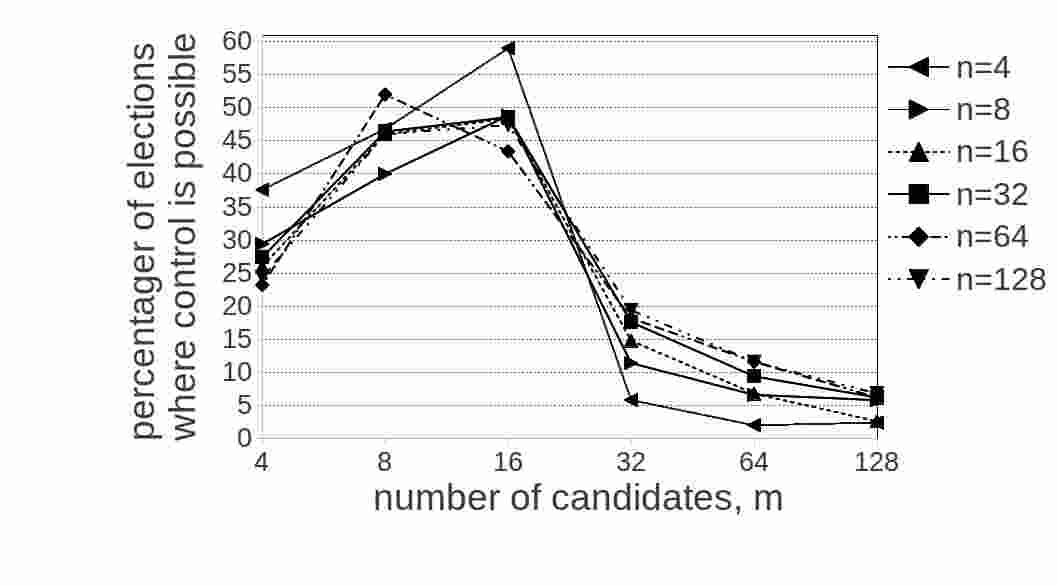}
	\caption{Results for plurality voting in the TM model for 
destructive control by partition of candidates in model TP. Number of voters is fixed. }
\end{figure}
%
\end{center}

\clearpage
\subsubsection{Computational Costs}
\begin{figure}[ht]
\centering
	\includegraphics[scale=0.3]{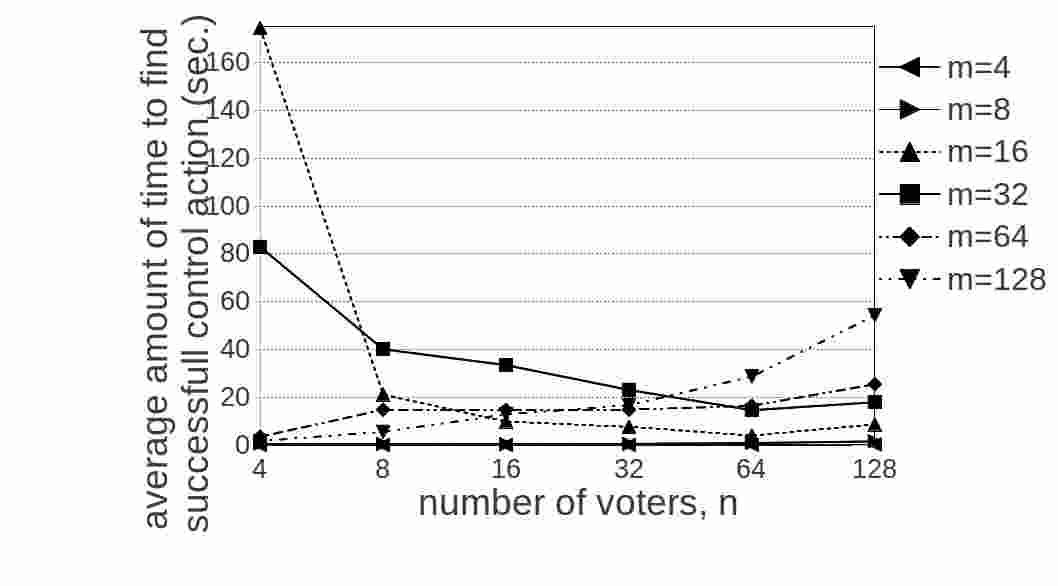}
	\caption{Average time the algorithm needs to find a successful control action for 
	destructive control by partition of candidates in model TP
	in plurality elections in the IC model. The maximum is $174,36$ seconds.}
\end{figure}
\begin{figure}[ht]
\centering
	\includegraphics[scale=0.3]{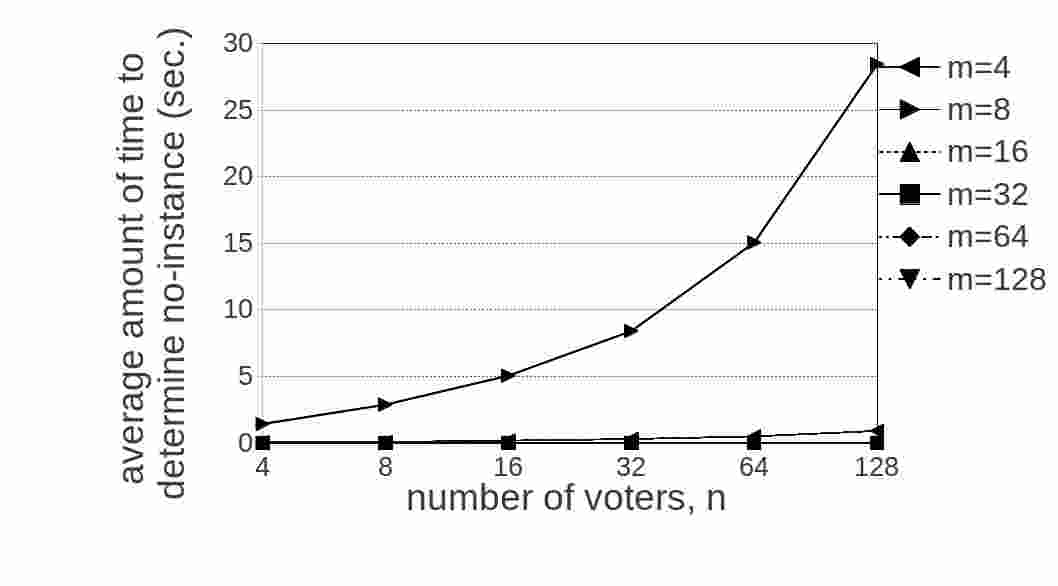}
	\caption{Average time the algorithm needs to determine no-instance of 
		destructive control by partition of candidates in model TP
	in plurality elections in the IC model. The maximum is $28,46$ seconds.}
\end{figure}
\begin{figure}[ht]
\centering
	\includegraphics[scale=0.3]{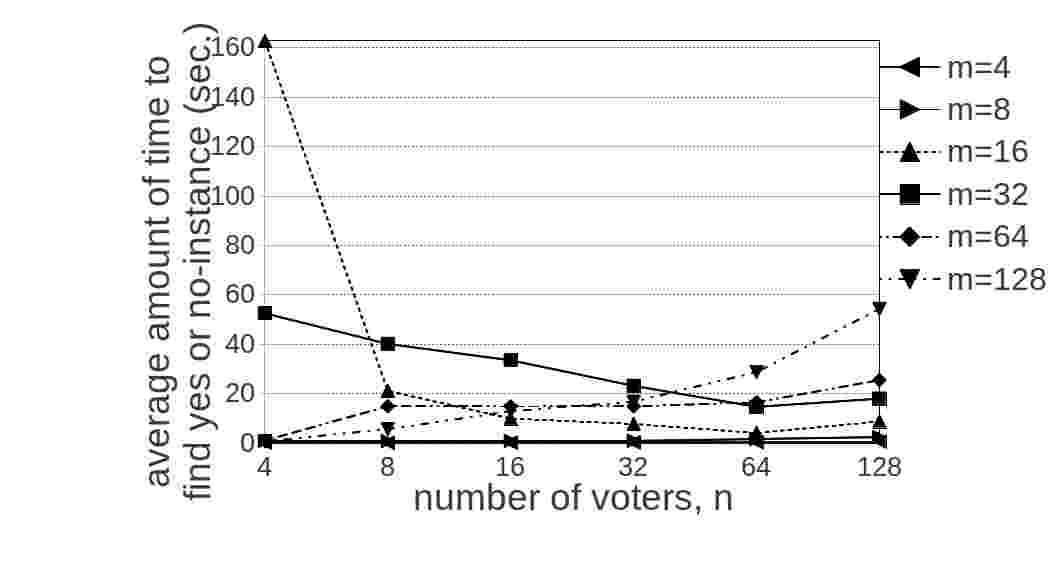}
	\caption{Average time the algorithm needs to give a definite output for 
	destructive control by partition of candidates in model TP
	in plurality elections in the IC model. The maximum is $162,76$ seconds.}
\end{figure}
\begin{figure}[ht]
\centering
	\includegraphics[scale=0.3]{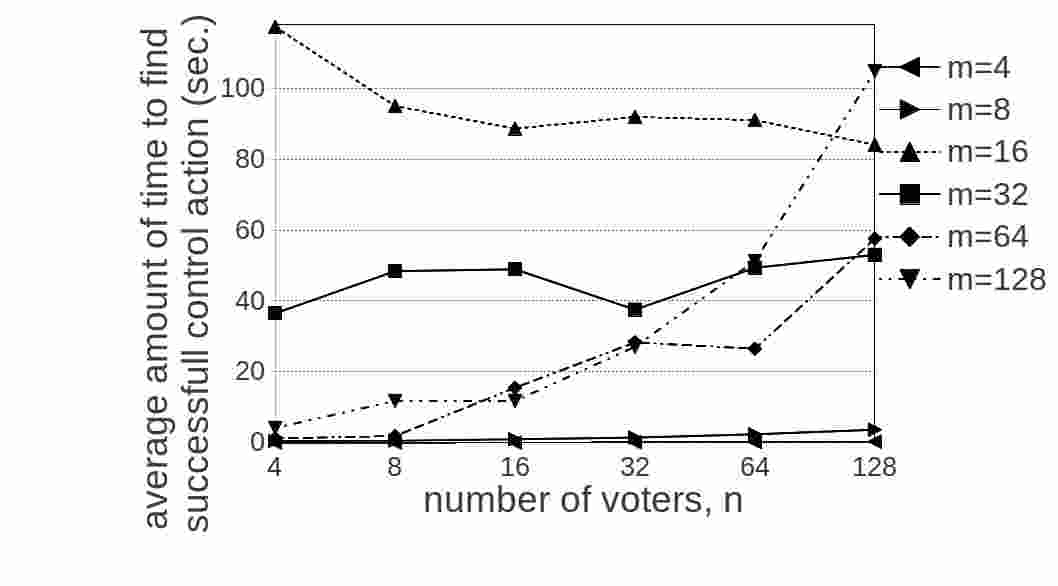}
	\caption{Average time the algorithm needs to find a successful control action for 
	destructive control by partition of candidates in model TP
	in plurality elections in the TM model. The maximum is $117,3$ seconds.}
\end{figure}

\begin{figure}[ht]
\centering
	\includegraphics[scale=0.3]{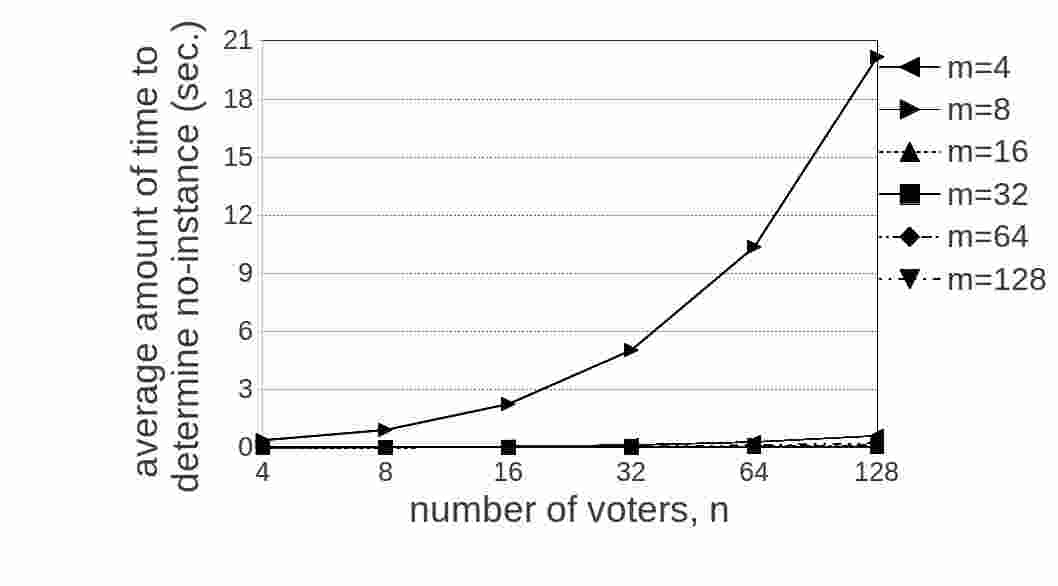}
	\caption{Average time the algorithm needs to determine no-instance of 
		destructive control by partition of candidates in model TP
	in plurality elections in the TM model. The maximum is $20,17$ seconds.}
\end{figure}
\begin{figure}[ht]
\centering
	\includegraphics[scale=0.3]{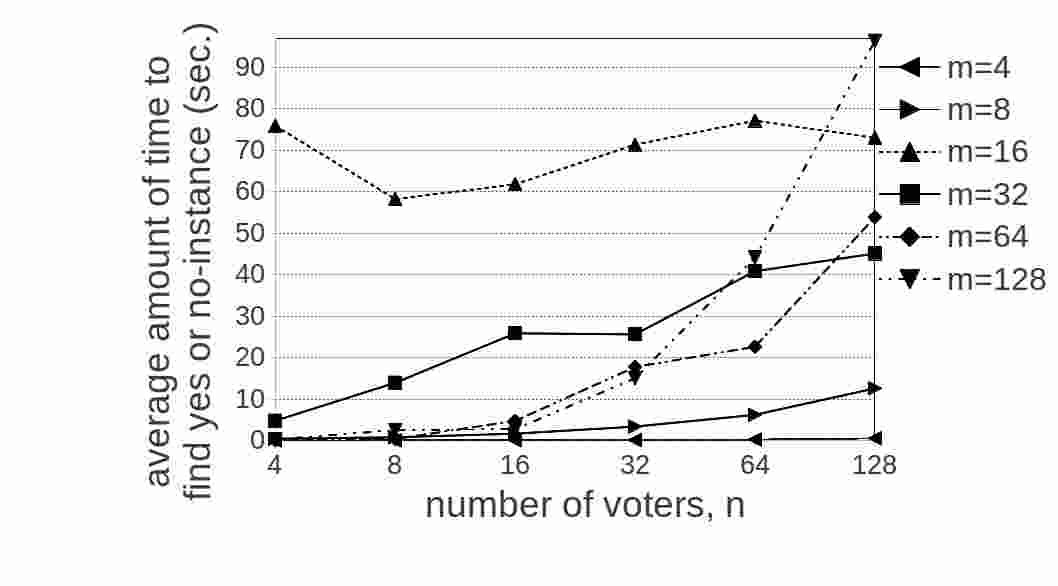}
	\caption{Average time the algorithm needs to give a definite output for 
	destructive control by partition of candidates in model TP
	in plurality elections in the TM model. The maximum is $96,35$ seconds.}
\end{figure}

\clearpage
\subsection{Constructive Control by Runoff Partition of Candidates in Model TE}
\begin{center}
\begin{figure}[ht]
\centering
	\includegraphics[scale=0.3]{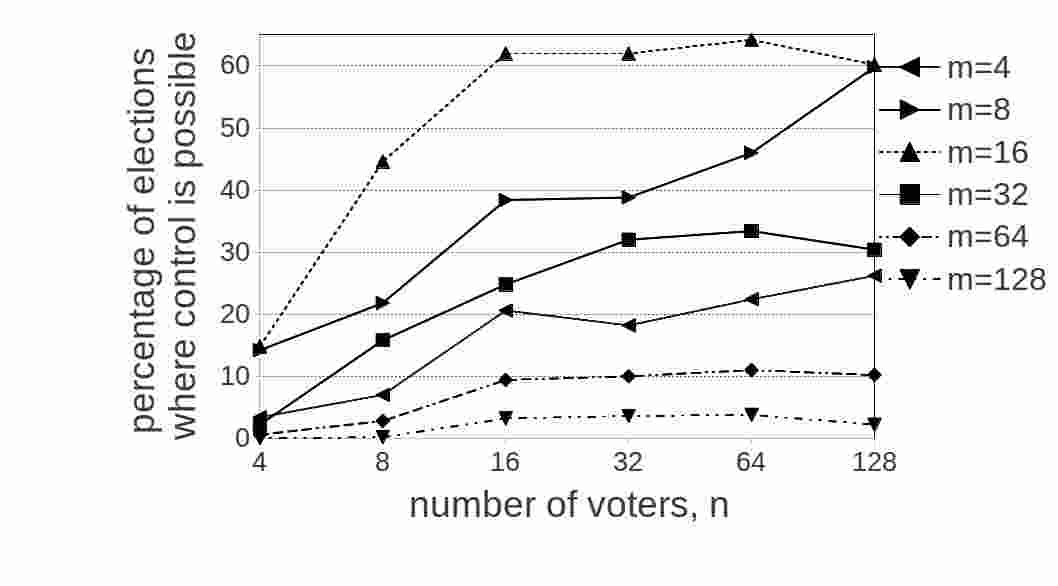}
		\caption{Results for plurality voting in the IC model for 
constructive control by runoff-partition  of candidates in model TE. Number of candidates is fixed. }
\end{figure}

\newpage
\begin{figure}[ht]
\centering
	\includegraphics[scale=0.3]{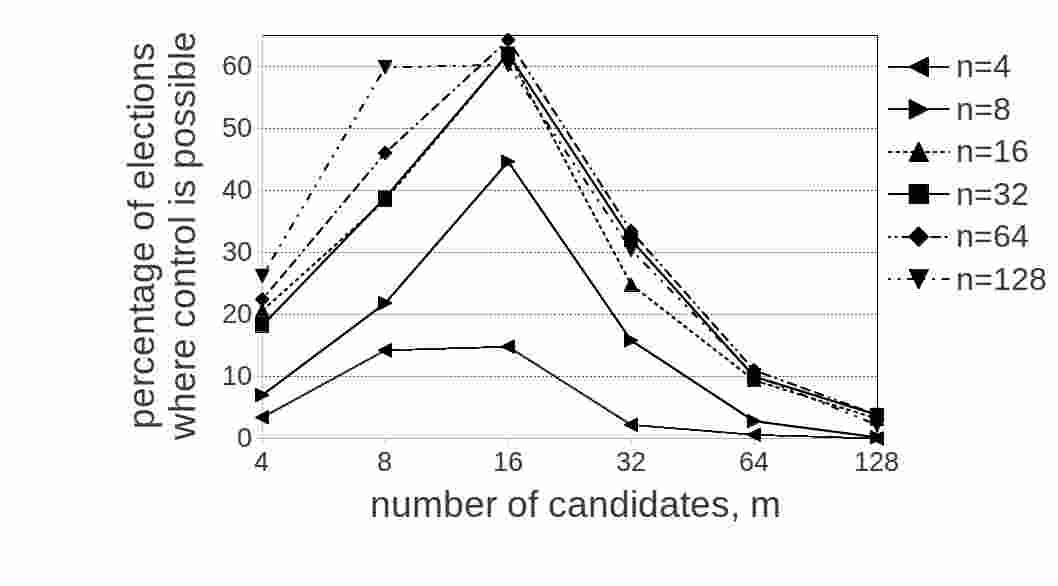}
		\caption{Results for plurality voting in the IC model for 
constructive control by runoff-partition  of candidates in model TE. Number of voters is fixed. }
\end{figure}
%
\newpage
\begin{figure}[ht]
\centering
	\includegraphics[scale=0.3]{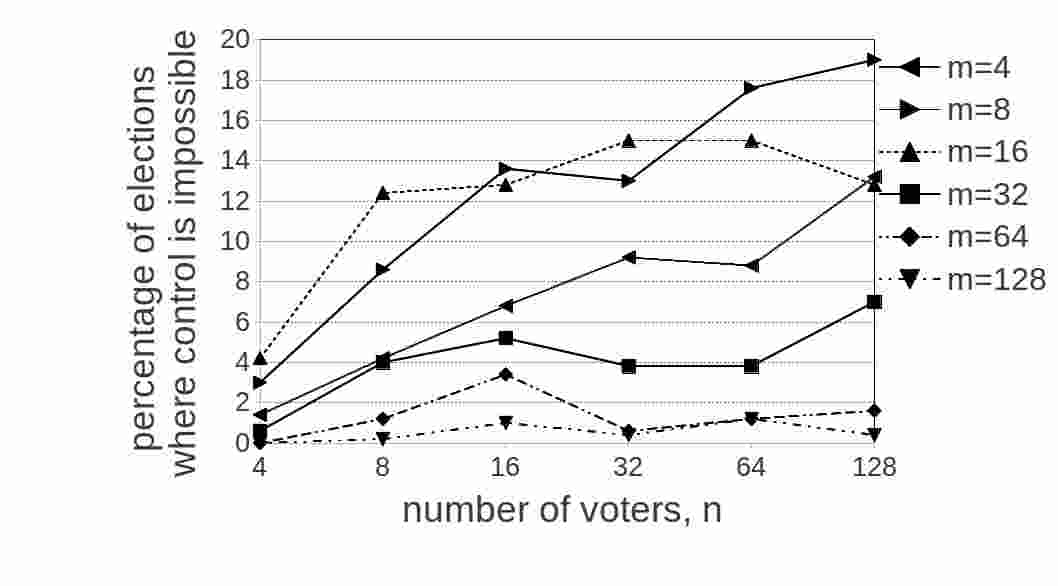}
	\caption{Results for plurality voting in the TM model for 
constructive control by runoff-partition  of candidates in model TE. Number of candidates is fixed. }
\end{figure}
%
\newpage
\begin{figure}[ht]
\centering
	\includegraphics[scale=0.3]{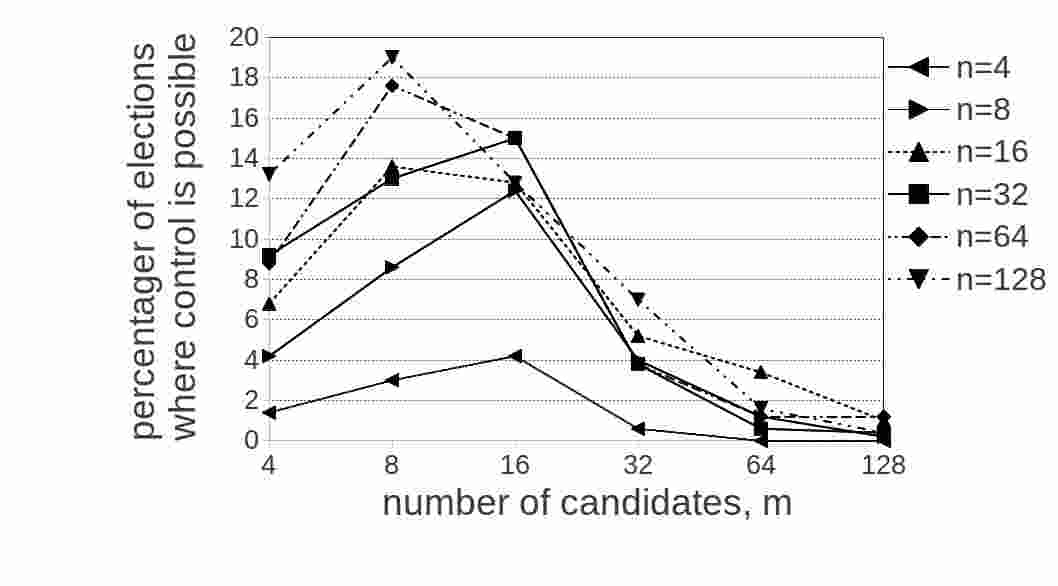}
	\caption{Results for plurality voting in the TM model for 
constructive control by runoff-partition  of candidates in model TE. Number of voters is fixed. }
\end{figure}
%
\end{center}

\clearpage
\subsubsection{Computational Costs}
\begin{figure}[ht]
\centering
	\includegraphics[scale=0.3]{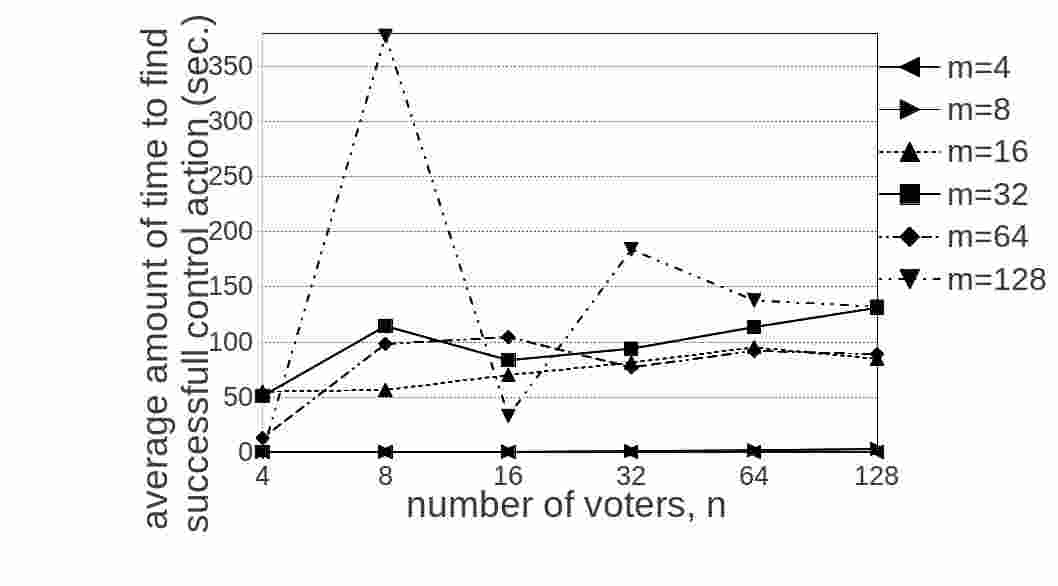}
	\caption{Average time the algorithm needs to find a successful control action for 
	constructive control by runoff-partition  of candidates in model TE
	in plurality elections in the IC model. The maximum is $378,02$ seconds.}
\end{figure}
\begin{figure}[ht]
\centering
	\includegraphics[scale=0.3]{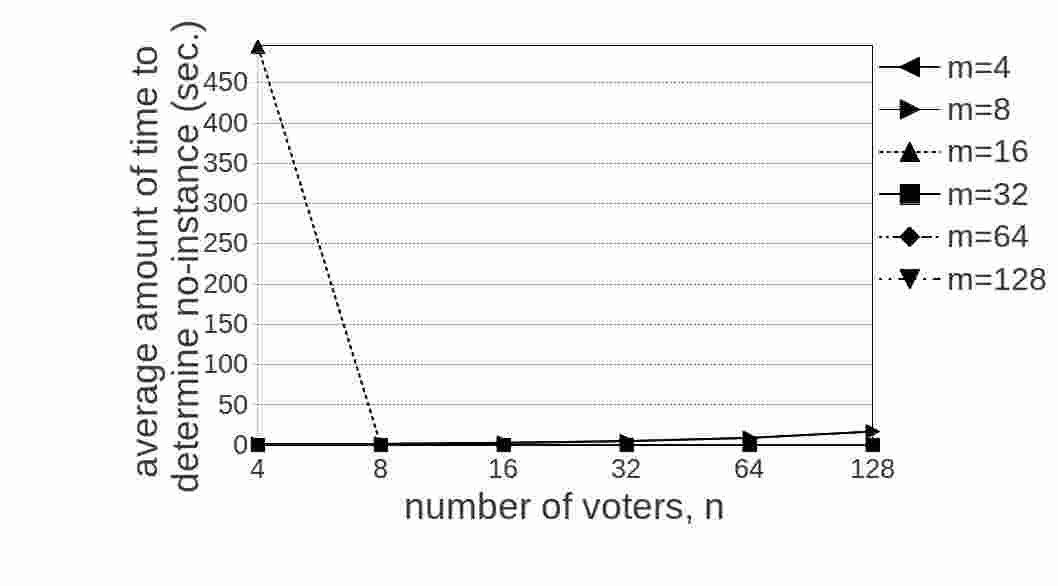}
	\caption{Average time the algorithm needs to determine no-instance of 
		constructive control by runoff-partition  of candidates in model TE
	in plurality elections in the IC model. The maximum is $495,26$ seconds.}
\end{figure}
\begin{figure}[ht]
\centering
	\includegraphics[scale=0.3]{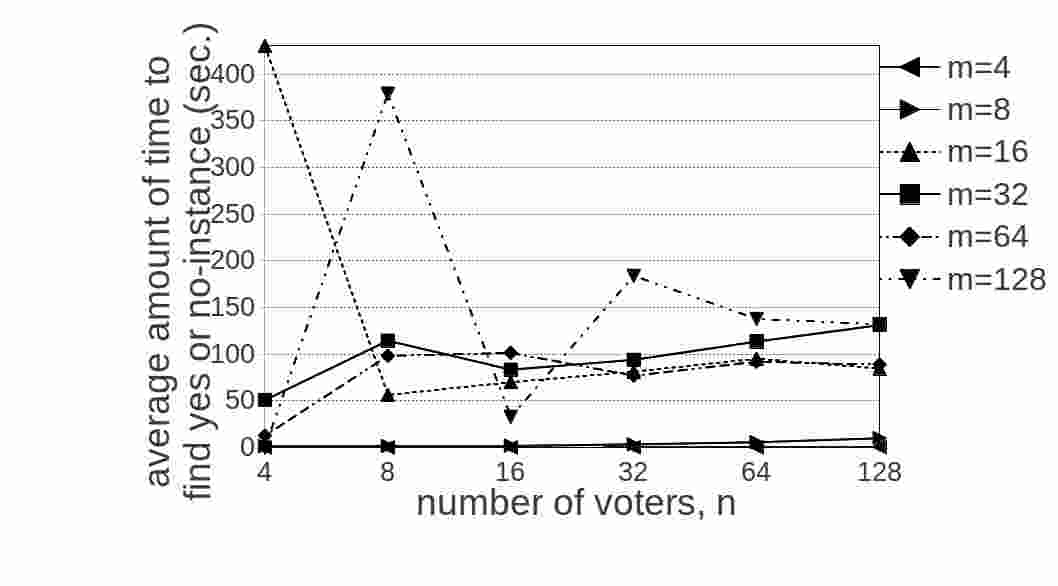}
	\caption{Average time the algorithm needs to give a definite output for 
	constructive control by runoff-partition  of candidates in model TE
	in plurality elections in the IC model. The maximum is $430,08$ seconds.}
\end{figure}
\begin{figure}[ht]
\centering
	\includegraphics[scale=0.3]{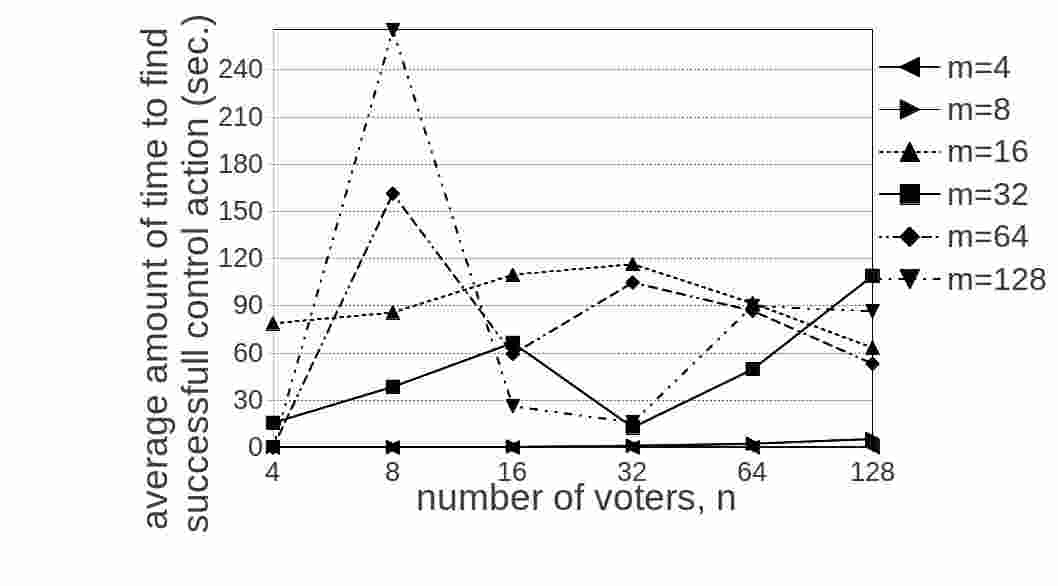}
	\caption{Average time the algorithm needs to find a successful control action for 
	constructive control by runoff-partition  of candidates in model TE
	in plurality elections in the TM model. The maximum is $265,22$ seconds.}
\end{figure}
\begin{figure}[ht]
\centering
	\includegraphics[scale=0.3]{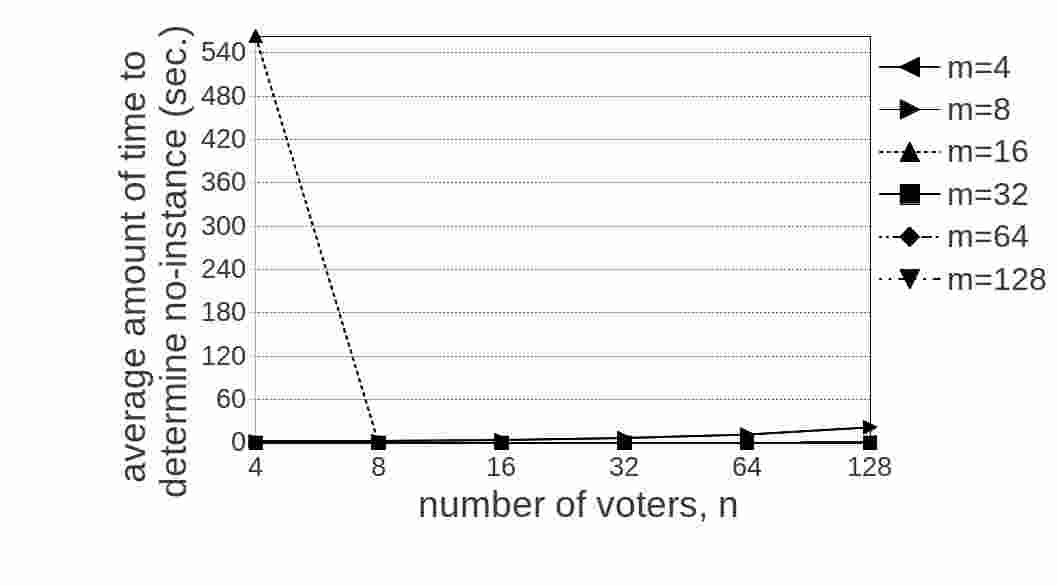}
	\caption{Average time the algorithm needs to determine no-instance of 
		constructive control by runoff-partition  of candidates in model TE
	in plurality elections in the TM model. The maximum is $562,86$ seconds.}
\end{figure}
\begin{figure}[ht]
\centering
	\includegraphics[scale=0.3]{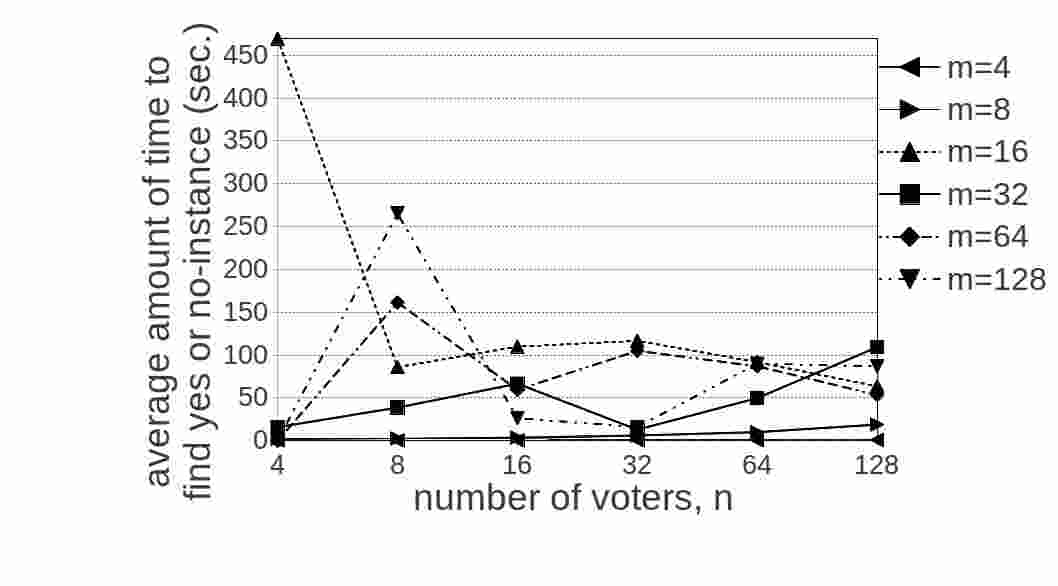}
	\caption{Average time the algorithm needs to give a definite output for 
	constructive control by runoff-partition  of candidates in model TE
	in plurality elections in the TM model. The maximum is $469,62$ seconds.}
\end{figure}

\clearpage
\subsection{Destructive Control by Runoff Partition of Candidates in Model TE}
\begin{center}
\begin{figure}[ht]
\centering
	\includegraphics[scale=0.3]{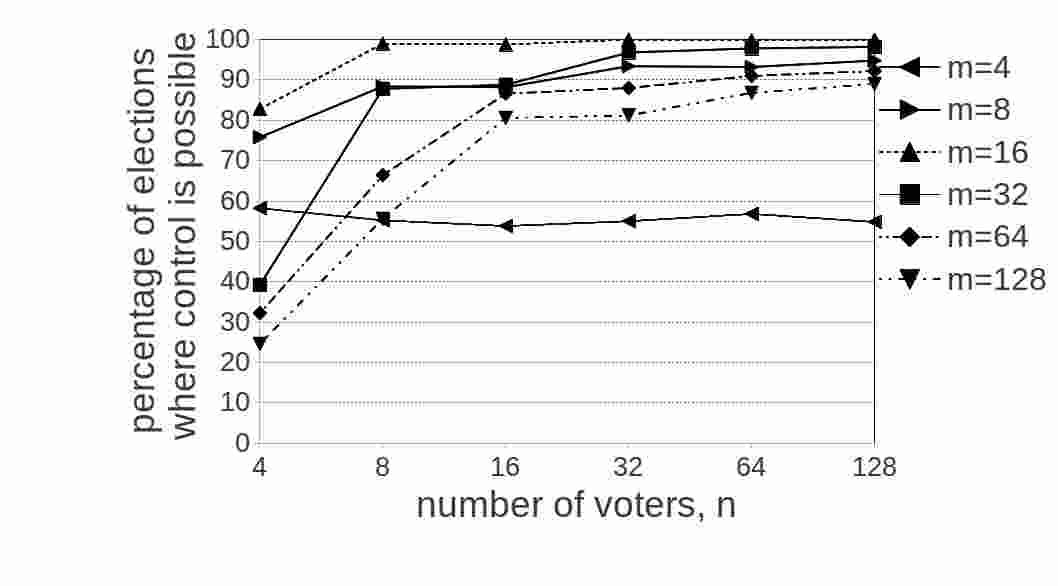}
		\caption{Results for plurality voting in the IC model for 
destructive control by runoff-partition  of candidates in model TE. Number of candidates is fixed. }
\end{figure}

\newpage
\begin{figure}[ht]
\centering
	\includegraphics[scale=0.3]{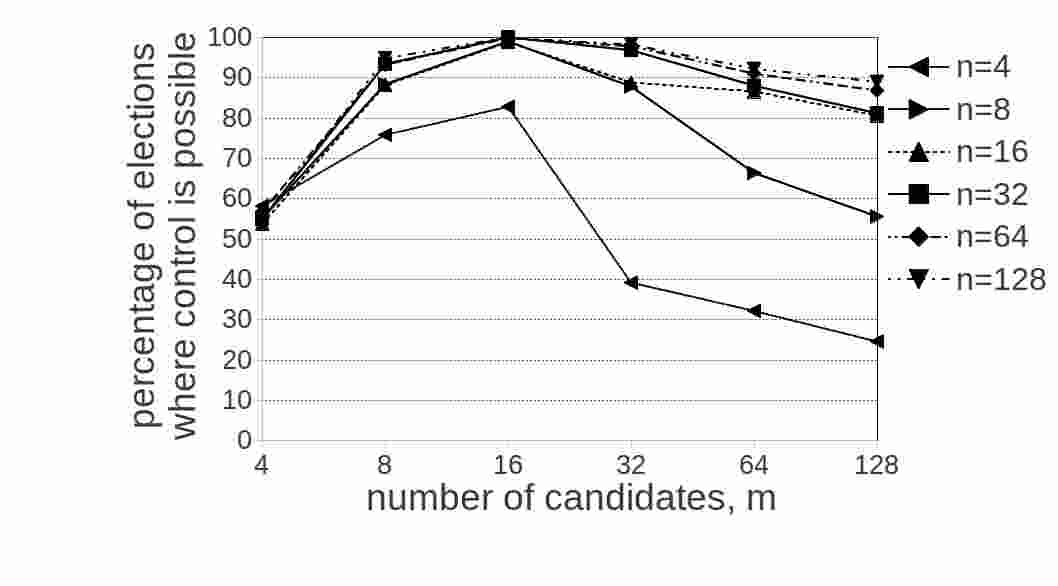}
		\caption{Results for plurality voting in the IC model for 
destructive control by runoff-partition  of candidates in model TE. Number of voters is fixed. }
\end{figure}
%
\newpage
\begin{figure}[ht]
\centering
	\includegraphics[scale=0.3]{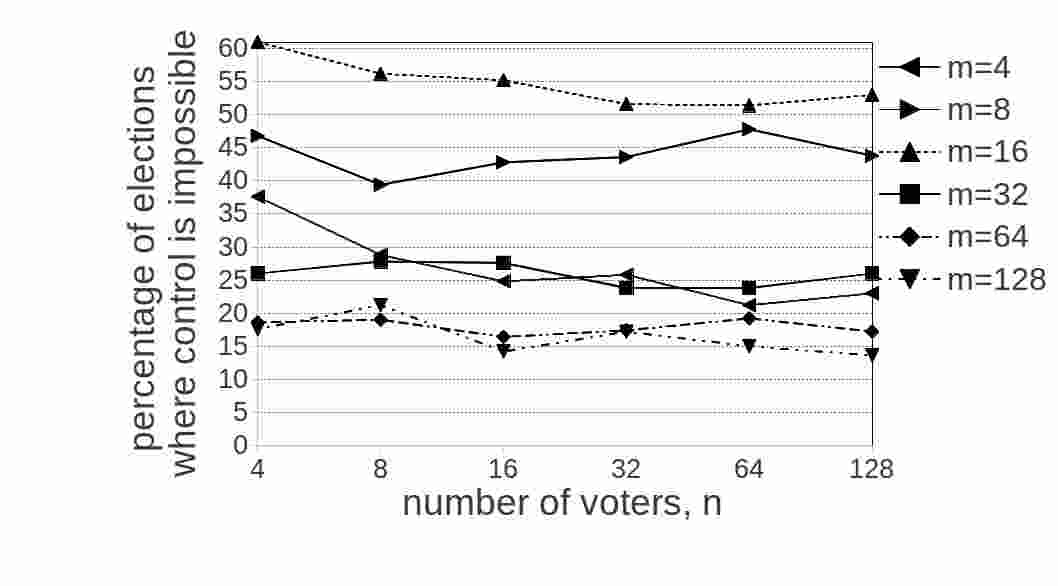}
	\caption{Results for plurality voting in the TM model for 
destructive control by runoff-partition  of candidates in model TE. Number of candidates is fixed. }
\end{figure}
%
\newpage
\begin{figure}[ht]
\centering
	\includegraphics[scale=0.3]{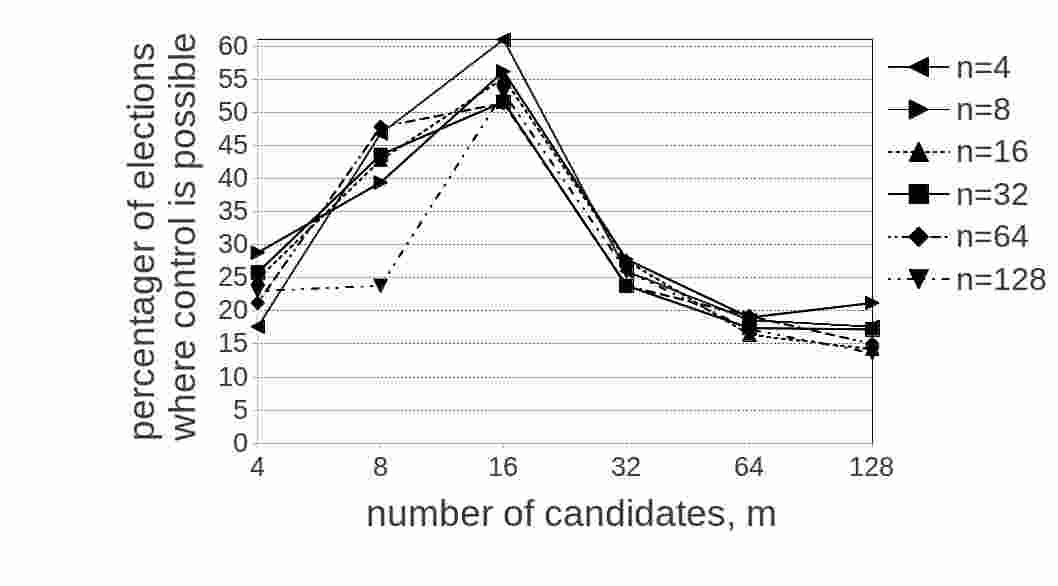}
	\caption{Results for plurality voting in the TM model for 
destructive control by runoff-partition  of candidates in model TE. Number of voters is fixed. }
\end{figure}
%
\end{center}

\clearpage
\subsubsection{Computational Costs}
\begin{figure}[ht]
\centering
	\includegraphics[scale=0.3]{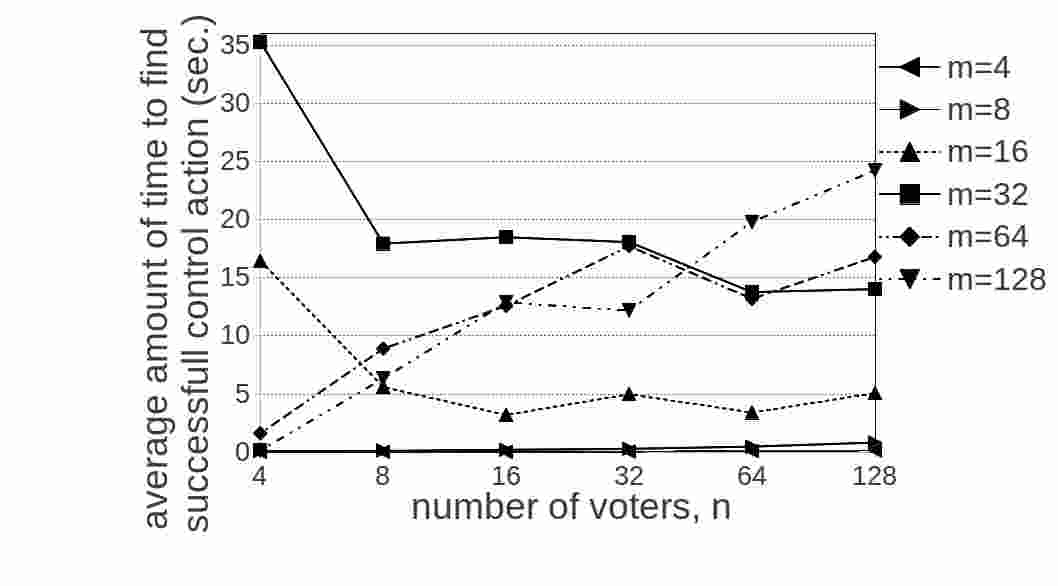}
	\caption{Average time the algorithm needs to find a successful control action for 
	destructive control by runoff-partition  of candidates in model TE
	in plurality elections in the IC model. The maximum is $35,3$ seconds.}
\end{figure}
\begin{figure}[ht]
\centering
	\includegraphics[scale=0.3]{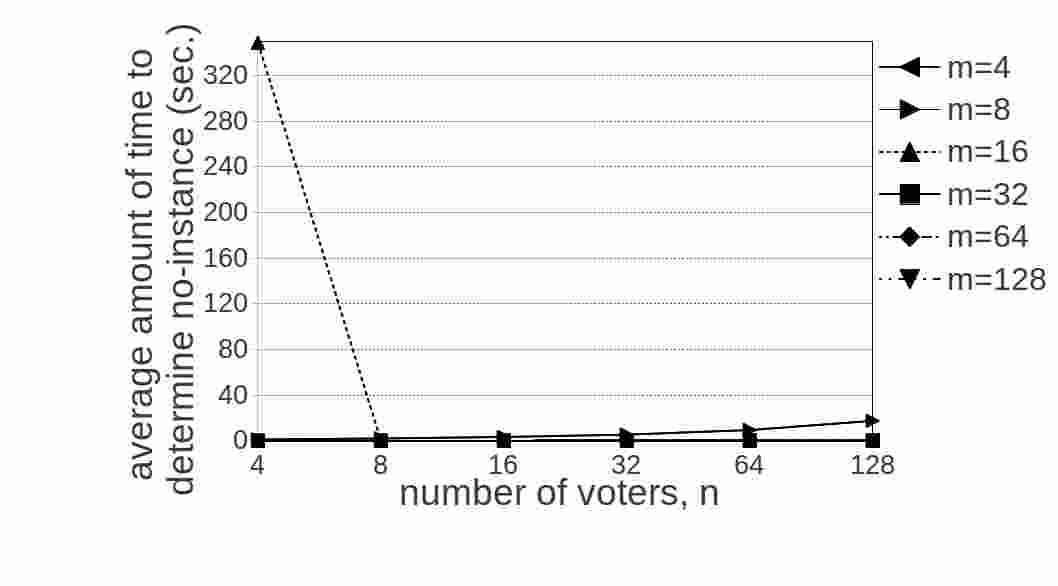}
	\caption{Average time the algorithm needs to determine no-instance of 
		destructive control by runoff-partition  of candidates in model TE
	in plurality elections in the IC model. The maximum is $348,65$ seconds.}
\end{figure}
\begin{figure}[ht]
\centering
	\includegraphics[scale=0.3]{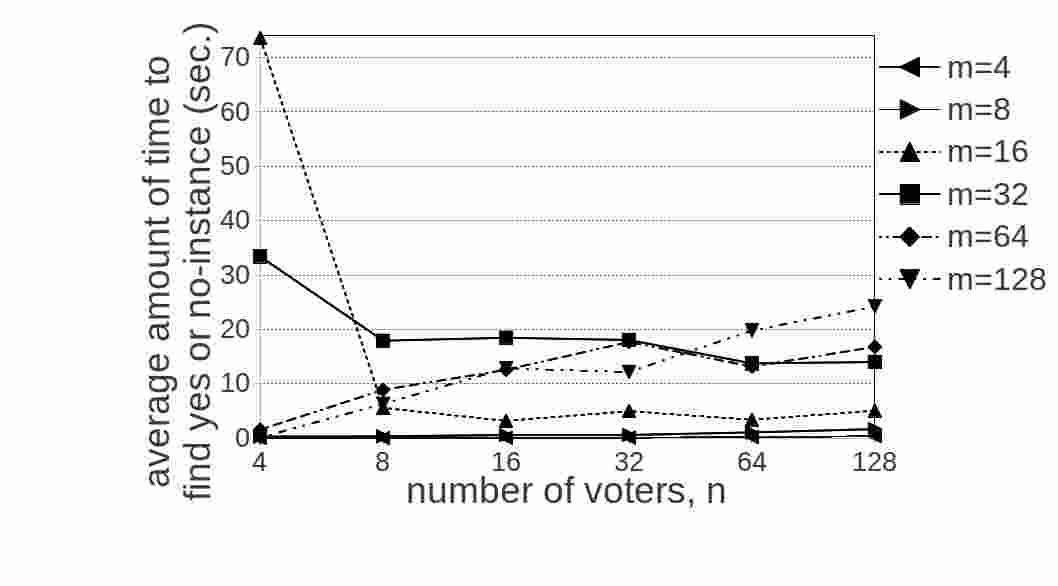}
	\caption{Average time the algorithm needs to give a definite output for 
	destructive control by runoff-partition  of candidates in model TE
	in plurality elections in the IC model. The maximum is $73,6$ seconds.}
\end{figure}
\begin{figure}[ht]
\centering
	\includegraphics[scale=0.3]{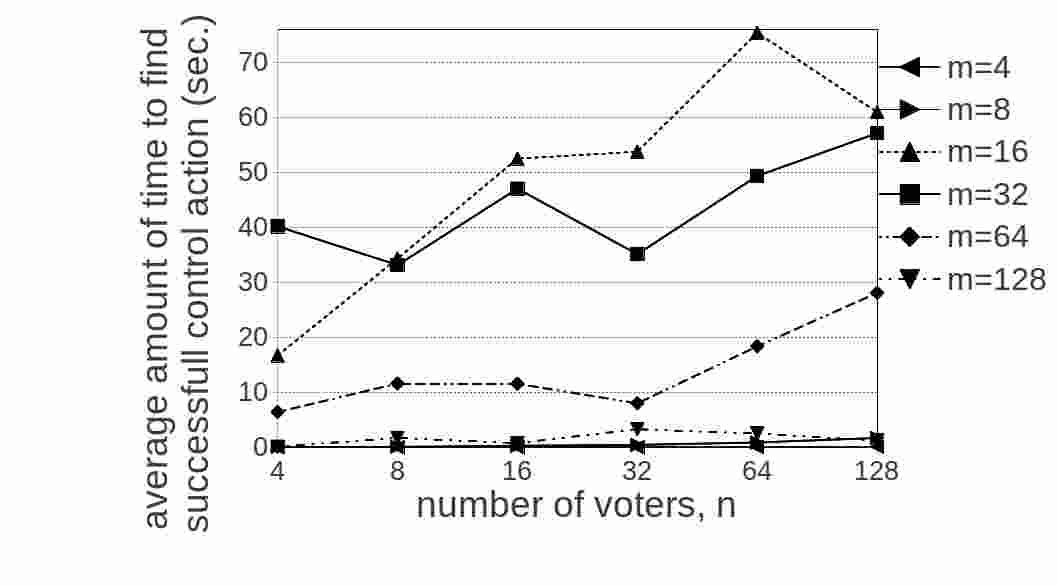}
	\caption{Average time the algorithm needs to find a successful control action for 
	destructive control by runoff-partition  of candidates in model TE
	in plurality elections in the TM model. The maximum is $75,31$ seconds.}
\end{figure}

\begin{figure}[ht]
\centering
	\includegraphics[scale=0.3]{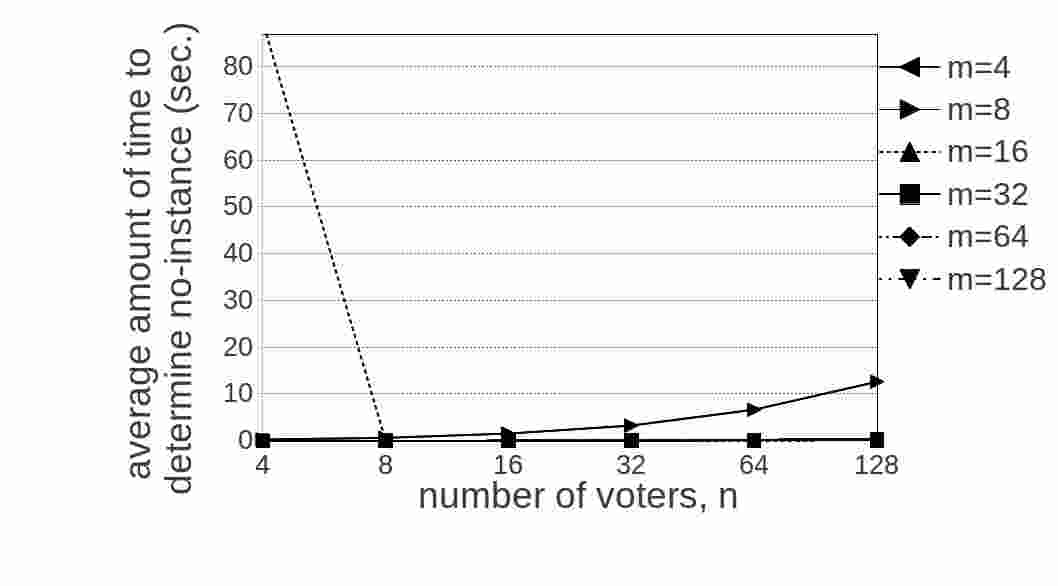}
	\caption{Average time the algorithm needs to determine no-instance of 
		destructive control by runoff-partition  of candidates in model TE
	in plurality elections in the TM model. The maximum is $89,69$ seconds.}
\end{figure}
\begin{figure}[ht]
\centering
	\includegraphics[scale=0.3]{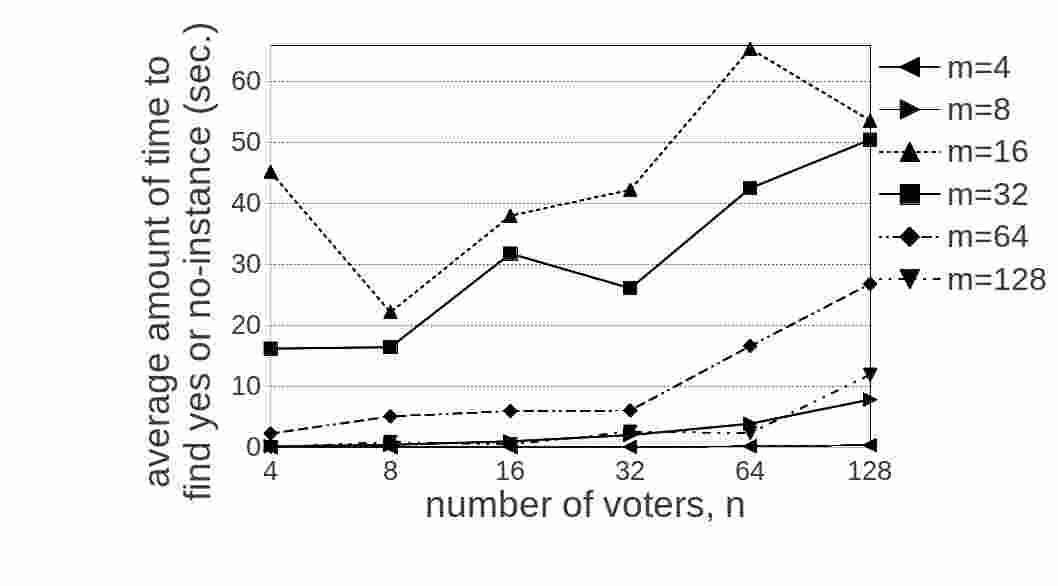}
	\caption{Average time the algorithm needs to give a definite output for 
	destructive control by runoff-partition  of candidates in model TE
	in plurality elections in the TM model. The maximum is $65,39$ seconds.}
\end{figure}

\clearpage
\subsection{Constructive Control by Runoff Partition of Candidates in Model TP}
\begin{center}
\begin{figure}[ht]
\centering
	\includegraphics[scale=0.3]{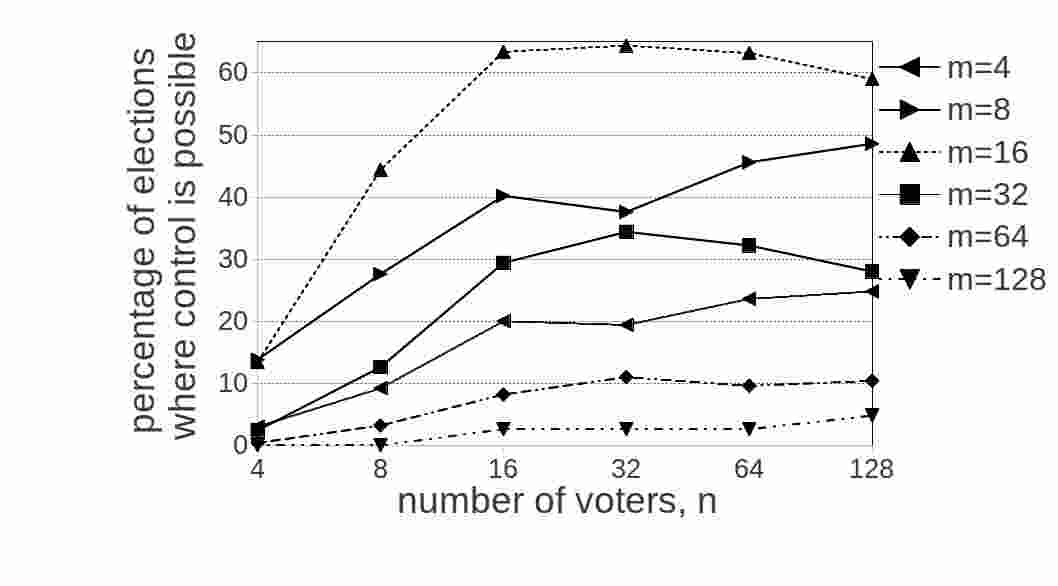}
		\caption{Results for plurality voting in the IC model for 
constructive control by runoff-partition  of candidates in model TP. Number of candidates is fixed. }
\end{figure}

\newpage
\begin{figure}[ht]
\centering
	\includegraphics[scale=0.3]{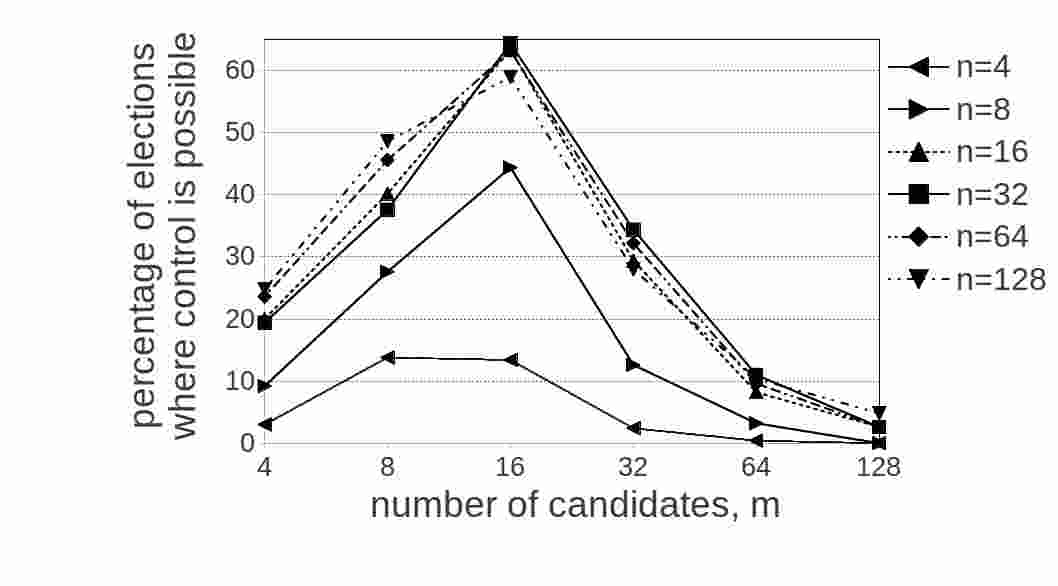}
		\caption{Results for plurality voting in the IC model for 
constructive control by runoff-partition  of candidates in model TP. Number of voters is fixed. }
\end{figure}
%
\newpage
\begin{figure}[ht]
\centering
	\includegraphics[scale=0.3]{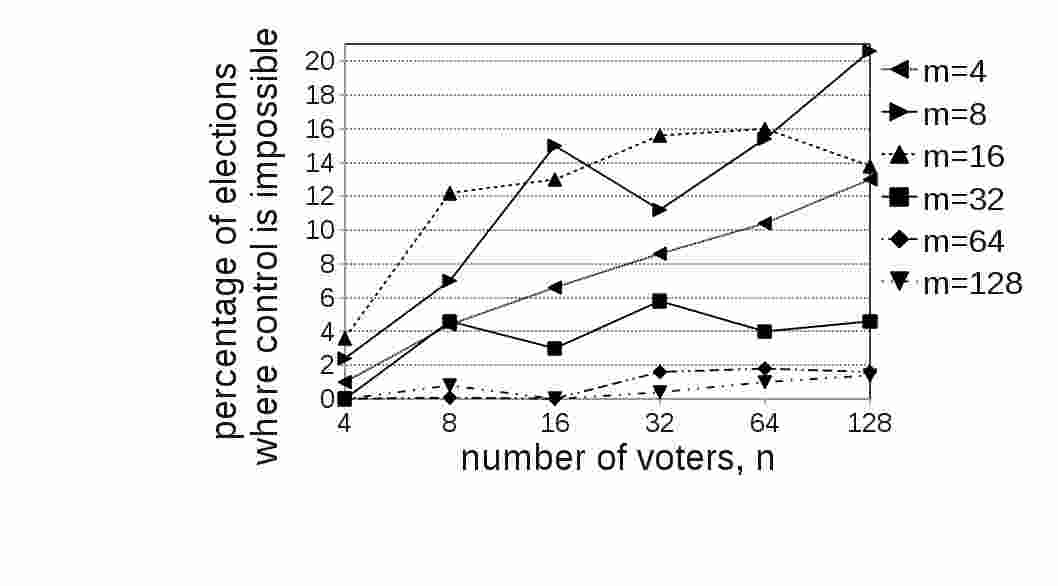}
	\caption{Results for plurality voting in the TM model for 
constructive control by runoff-partition  of candidates in model TP. Number of candidates is fixed. }
\end{figure}
%
\newpage
\begin{figure}[ht]
\centering
	\includegraphics[scale=0.3]{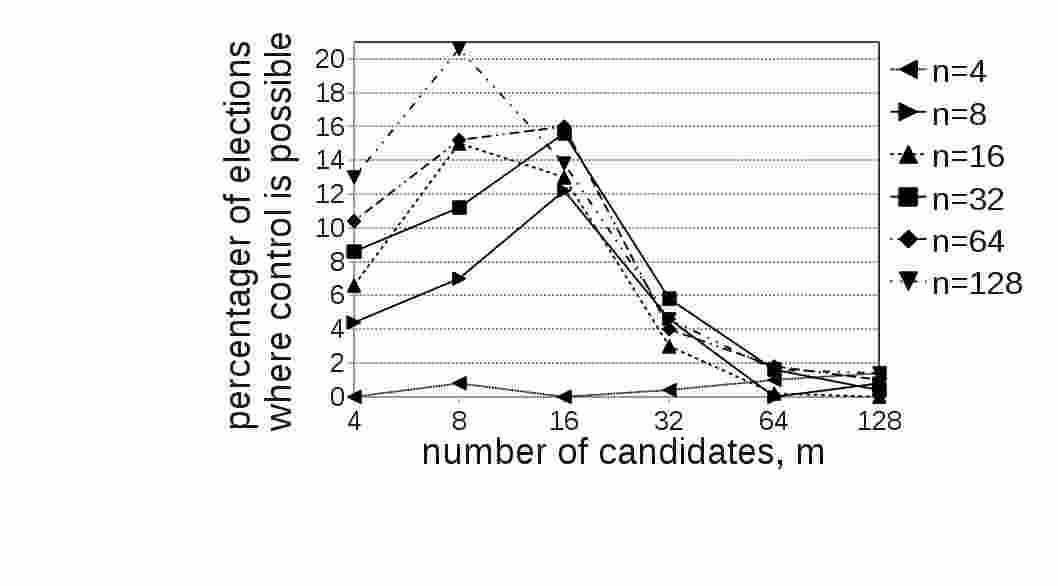}
	\caption{Results for plurality voting in the TM model for 
constructive control by runoff-partition  of candidates in model TP. Number of voters is fixed. }
\end{figure}
%
\end{center}

\clearpage
\subsubsection{Computational Costs}
\begin{figure}[ht]
\centering
	\includegraphics[scale=0.3]{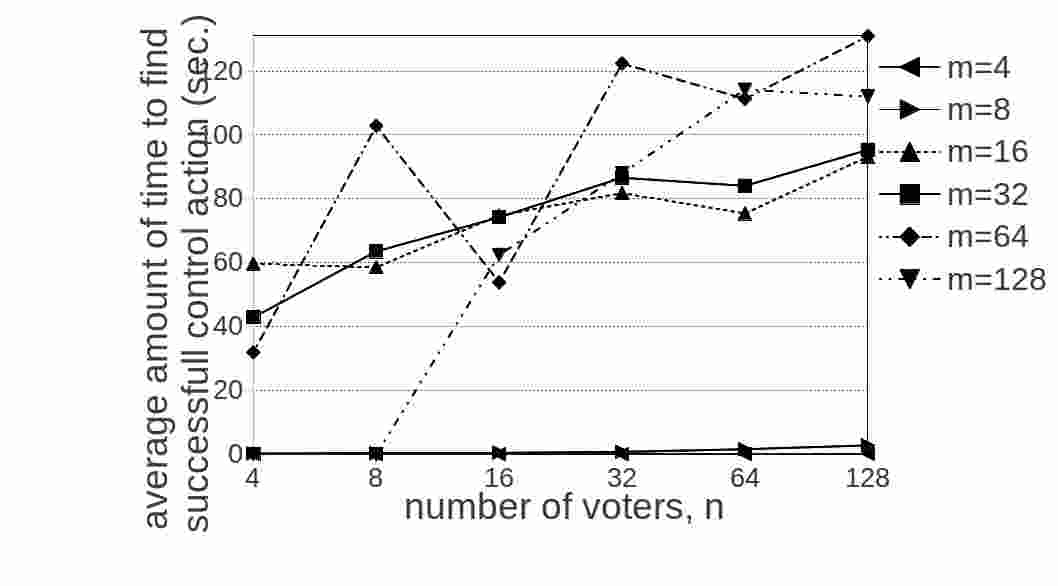}
	\caption{Average time the algorithm needs to find a successful control action for 
	constructive control by runoff-partition  of candidates in model TP
	in plurality elections in the IC model. The maximum is $130,91$ seconds.}
\end{figure}
\begin{figure}[ht]
\centering
	\includegraphics[scale=0.3]{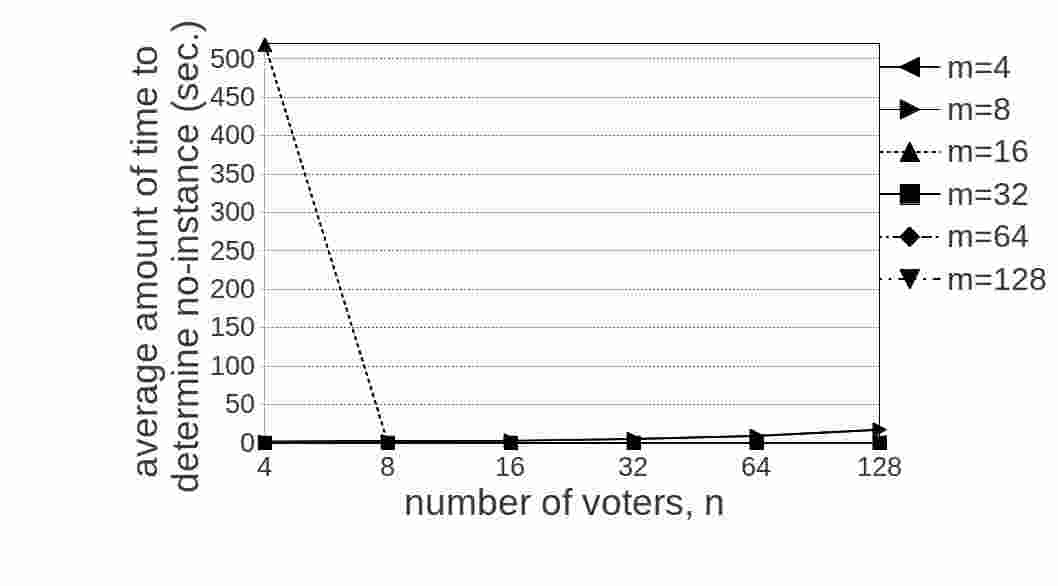}
	\caption{Average time the algorithm needs to determine no-instance of 
		constructive control by runoff-partition  of candidates in model TP
	in plurality elections in the IC model. The maximum is $518,67$ seconds.}
\end{figure}
\begin{figure}[ht]
\centering
	\includegraphics[scale=0.3]{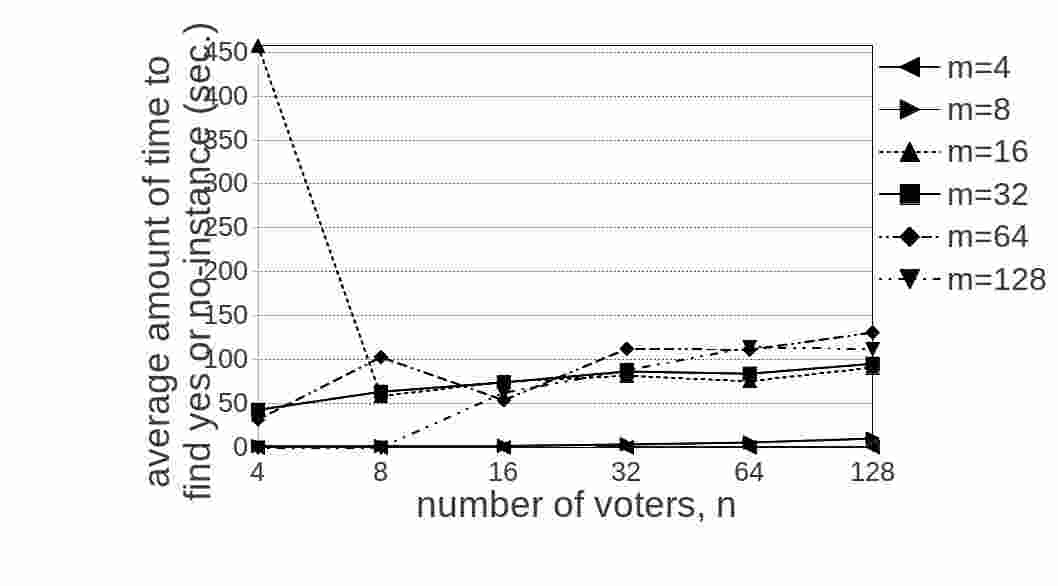}
	\caption{Average time the algorithm needs to give a definite output for 
	constructive control by runoff-partition  of candidates in model TP
	in plurality elections in the IC model. The maximum is $130,91$ seconds.}
\end{figure}
\begin{figure}[ht]
\centering
	\includegraphics[scale=0.3]{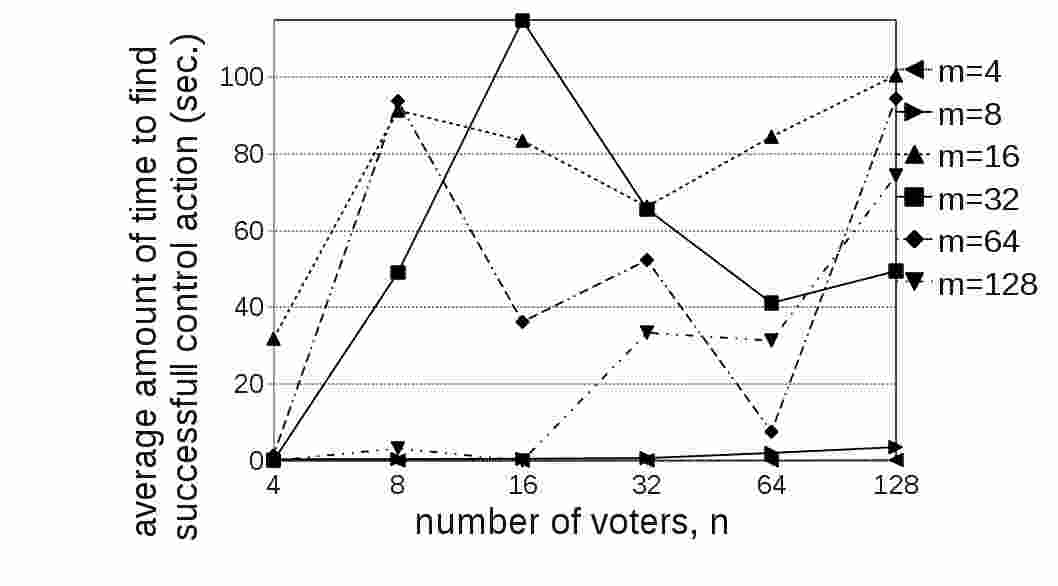}
	\caption{Average time the algorithm needs to find a successful control action for 
	constructive control by runoff-partition  of candidates in model TP
	in plurality elections in the TM model. The maximum is $114,84$ seconds.}
\end{figure}
\begin{figure}[ht]
\centering
	\includegraphics[scale=0.3]{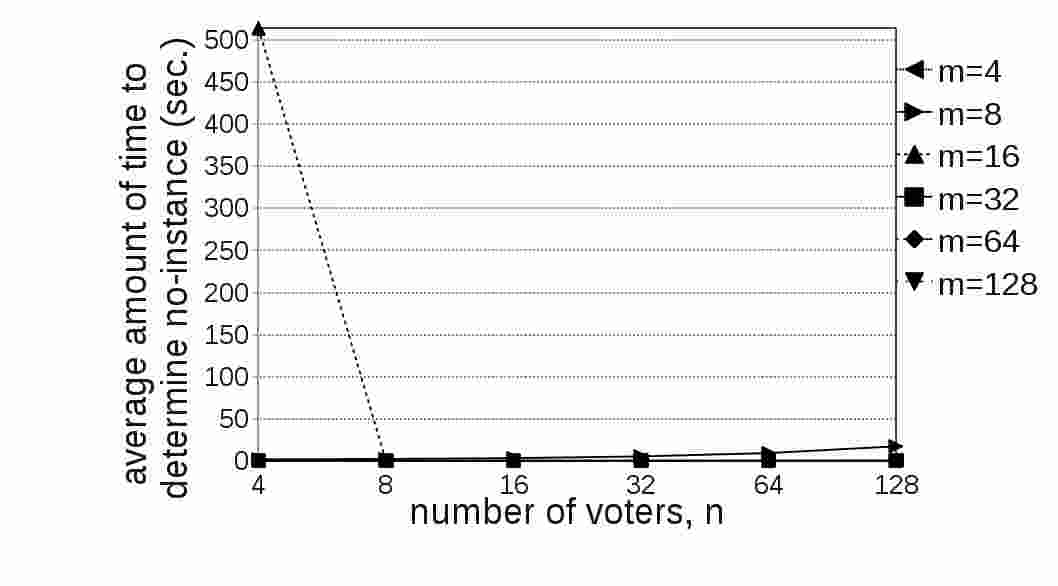}
	\caption{Average time the algorithm needs to determine no-instance of 
		constructive control by runoff-partition  of candidates in model TP
	in plurality elections in the TM model. The maximum is $514,44$ seconds.}
\end{figure}
\begin{figure}[ht]
\centering
	\includegraphics[scale=0.3]{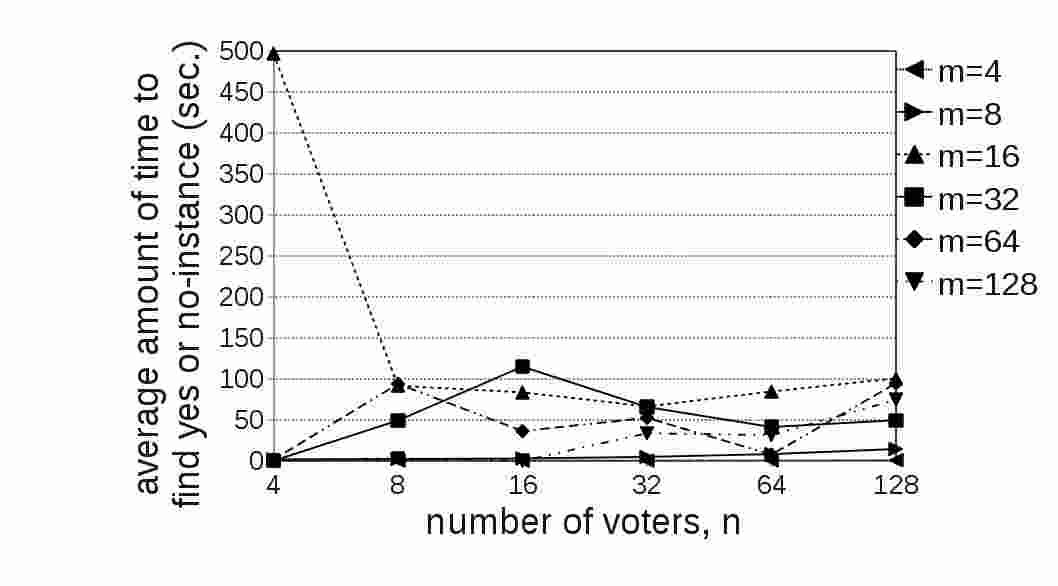}
	\caption{Average time the algorithm needs to give a definite output for 
	constructive control by runoff-partition  of candidates in model TP
	in plurality elections in the TM model. The maximum is $497,07$ seconds.}
\end{figure}

\clearpage
\subsection{Destructive Control by Runoff Partition of Candidates in Model TP}
\begin{center}
\begin{figure}[ht]
\centering
	\includegraphics[scale=0.3]{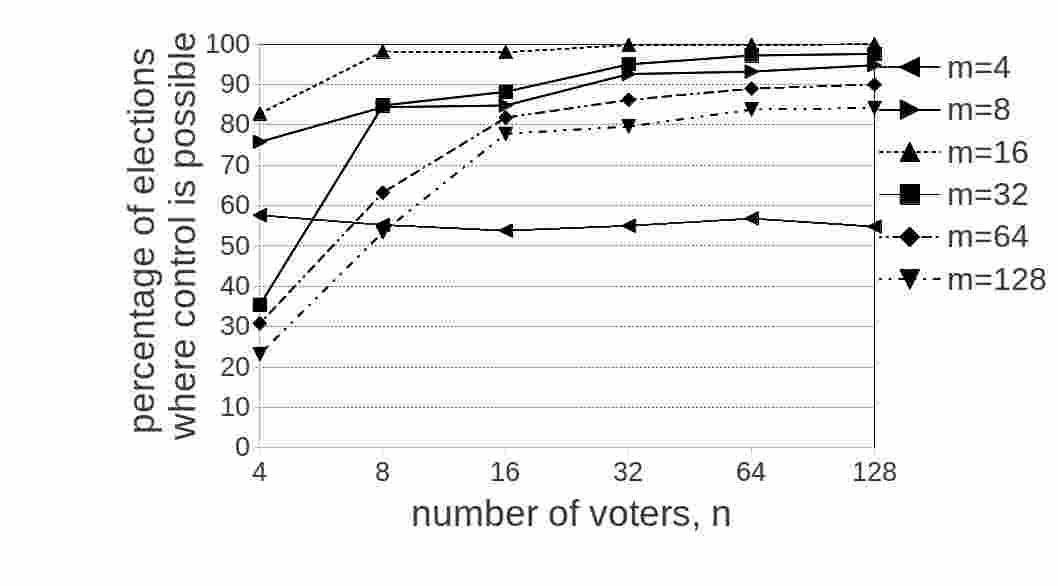}
		\caption{Results for plurality voting in the IC model for 
destructive control by runoff-partition  of candidates in model TP. Number of candidates is fixed. }
\end{figure}

\newpage
\begin{figure}[ht]
\centering
	\includegraphics[scale=0.3]{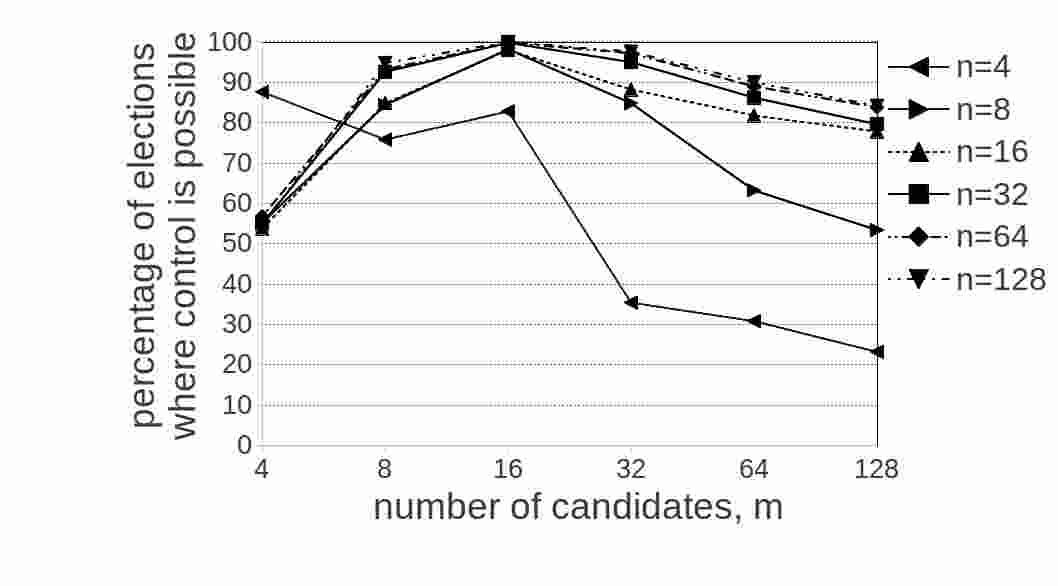}
		\caption{Results for plurality voting in the IC model for 
destructive control by runoff-partition  of candidates in model TP. Number of voters is fixed. }
\end{figure}
%
\newpage
\begin{figure}[ht]
\centering
	\includegraphics[scale=0.3]{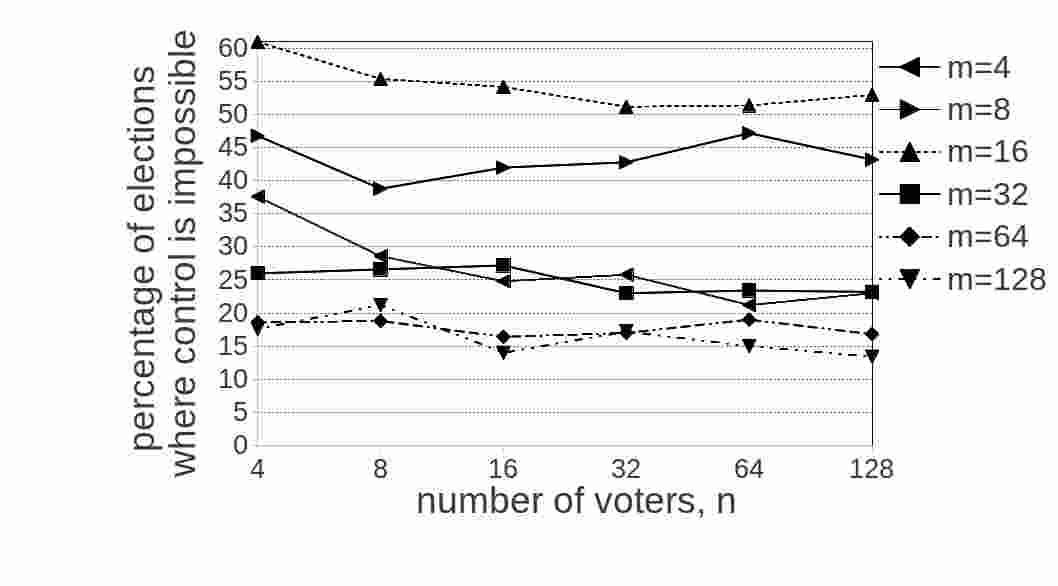}
	\caption{Results for plurality voting in the TM model for 
destructive control by runoff-partition  of candidates in model TP. Number of candidates is fixed. }
\end{figure}
%
\newpage
\begin{figure}[ht]
\centering
	\includegraphics[scale=0.3]{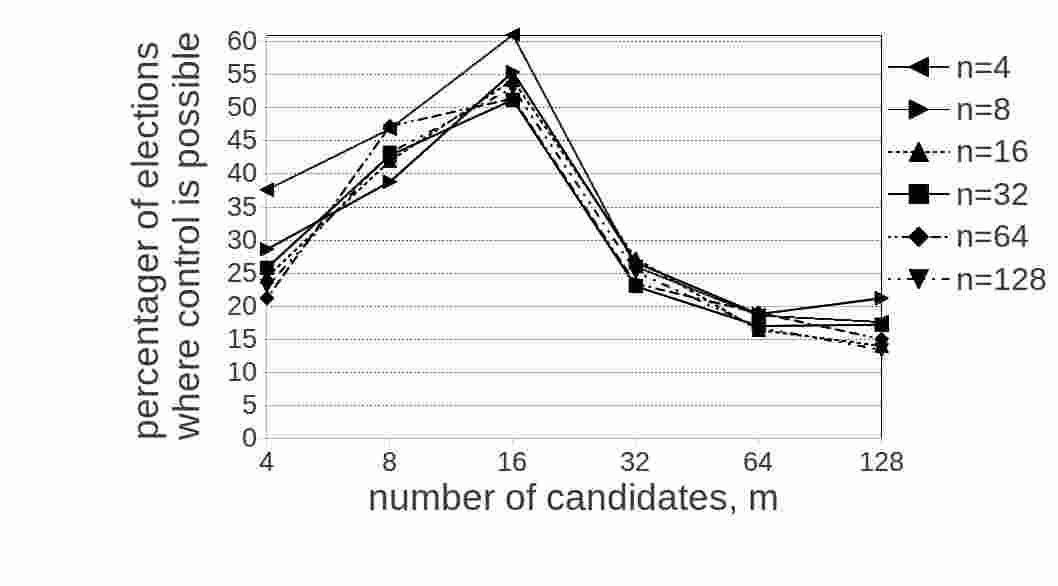}
	\caption{Results for plurality voting in the TM model for 
destructive control by runoff-partition  of candidates in model TP. Number of voters is fixed. }
\end{figure}

%
\end{center}

\clearpage
\subsubsection{Computational Costs}
\begin{figure}[ht]
\centering
	\includegraphics[scale=0.3]{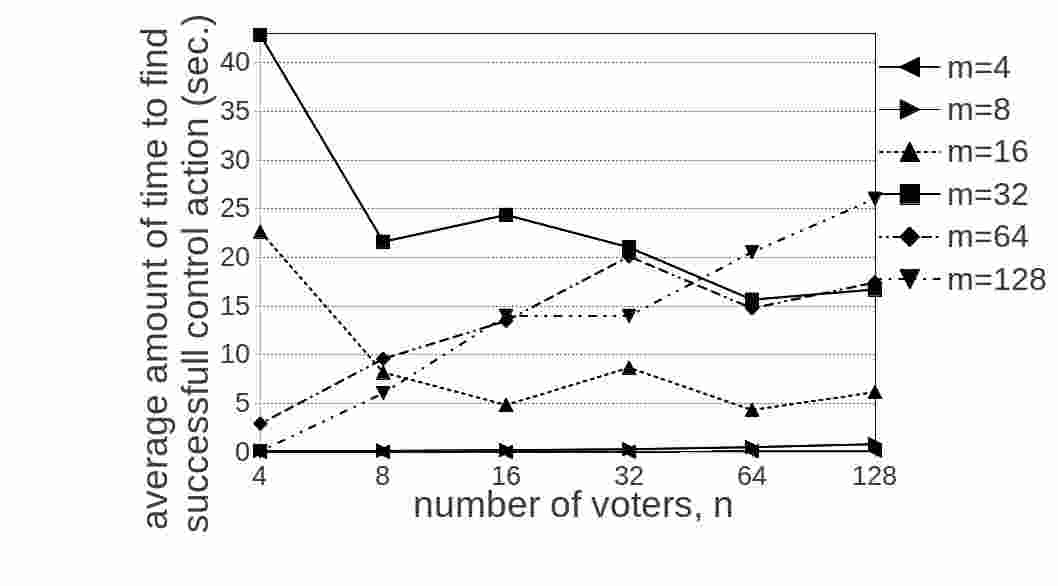}
	\caption{Average time the algorithm needs to find a successful control action for 
	destructive control by runoff-partition  of candidates in model TP
	in plurality elections in the IC model. The maximum is $42,85$ seconds.}
\end{figure}
\begin{figure}[ht]
\centering
	\includegraphics[scale=0.3]{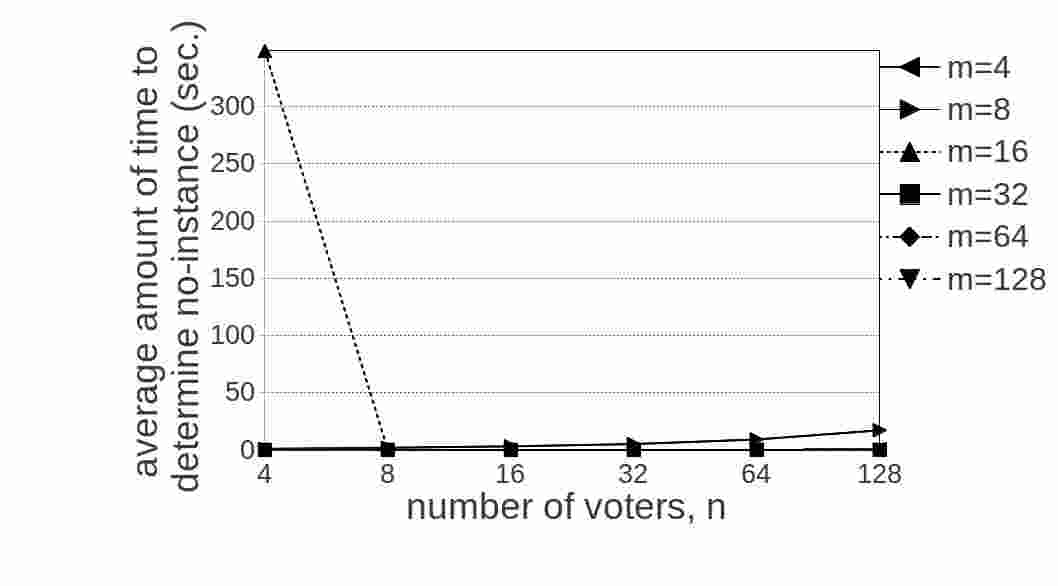}
	\caption{Average time the algorithm needs to determine no-instance of 
		destructive control by runoff-partition  of candidates in model TP
	in plurality elections in the IC model. The maximum is $348,43$ seconds.}
\end{figure}
\begin{figure}[ht]
\centering
	\includegraphics[scale=0.3]{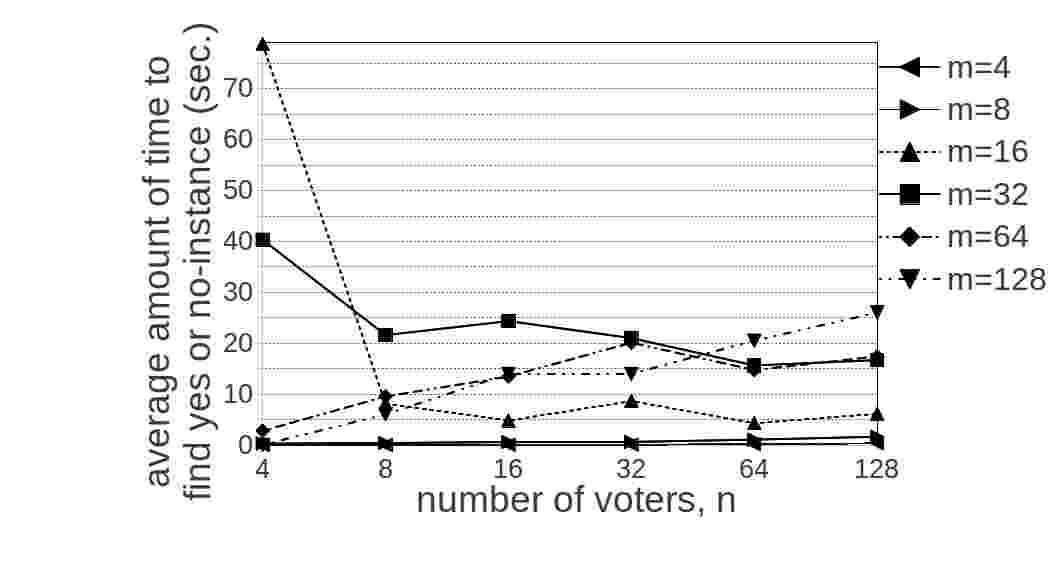}
	\caption{Average time the algorithm needs to give a definite output for 
	destructive control by runoff-partition  of candidates in model TP
	in plurality elections in the IC model. The maximum is $78,7$ seconds.}
\end{figure}
\begin{figure}[ht]
\centering
	\includegraphics[scale=0.3]{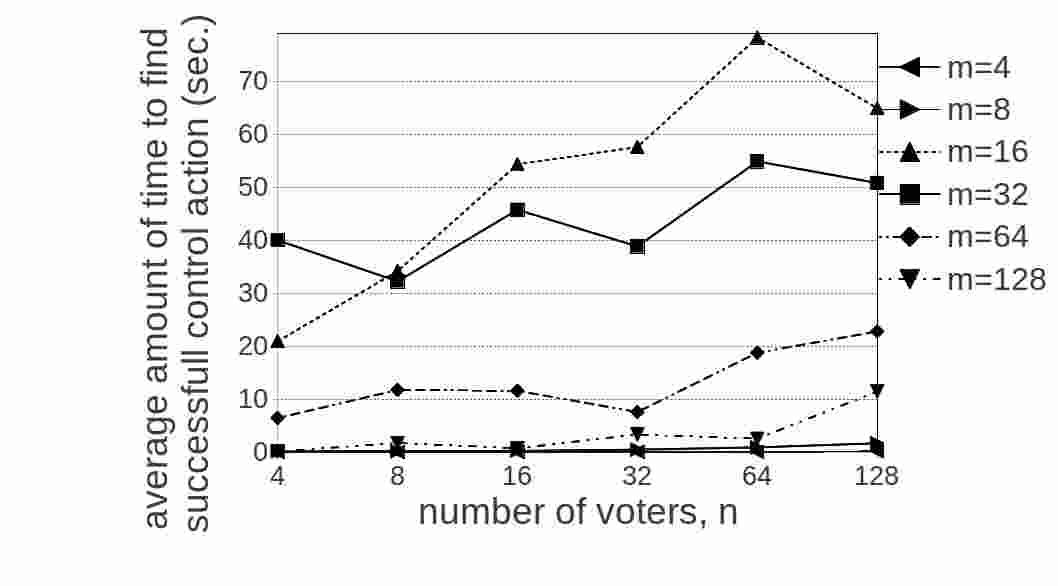}
	\caption{Average time the algorithm needs to find a successful control action for 
	destructive control by runoff-partition  of candidates in model TP
	in plurality elections in the TM model. The maximum is $78,26$ seconds.}
\end{figure}
\begin{figure}[ht]
\centering
	\includegraphics[scale=0.3]{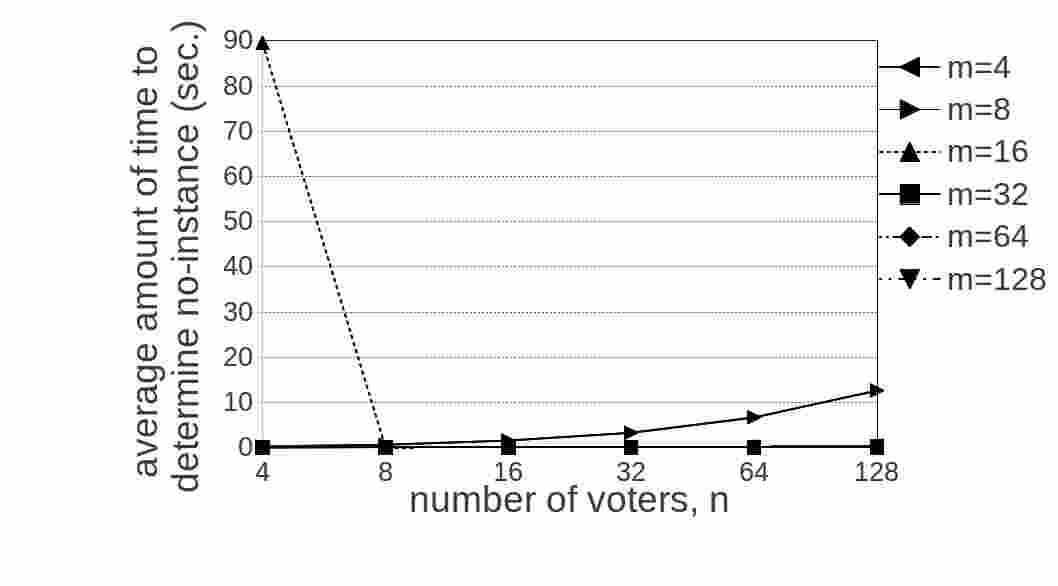}
	\caption{Average time the algorithm needs to determine no-instance of 
		destructive control by runoff-partition  of candidates in model TP
	in plurality elections in the TM model. The maximum is $89,53$ seconds.}
\end{figure}
\begin{figure}[ht]
\centering
	\includegraphics[scale=0.3]{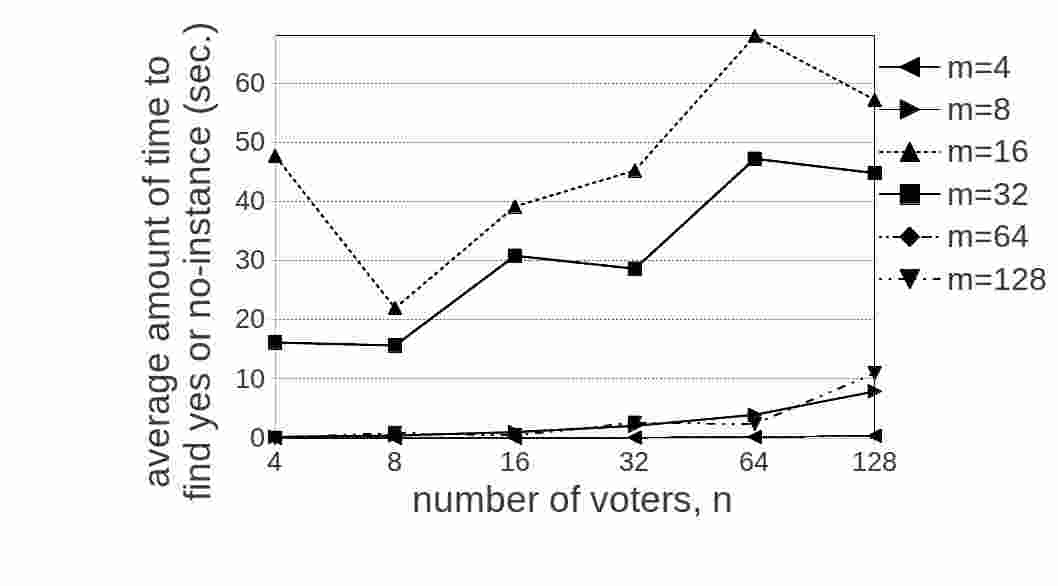}
	\caption{Average time the algorithm needs to give a definite output for 
	destructive control by runoff-partition  of candidates in model TP
	in plurality elections in the TM model. The maximum is $67,95$ seconds.}
\end{figure}

\clearpage
\subsection{Constructive Control by Partition of Voters in Model TP}
\begin{center}
\begin{figure}[ht]
\centering
	\includegraphics[scale=0.3]{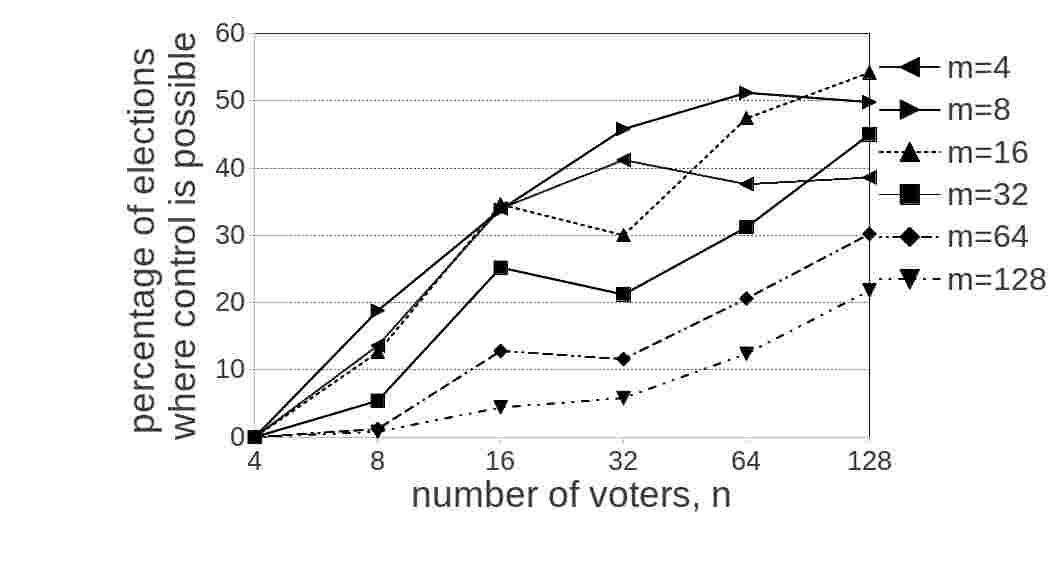}
	\caption{Results for plurality voting in the IC model for 
constructive control by partition of voters in model TP.  Number of candidates is fixed.}
\end{figure}

%

\newpage
\begin{figure}[ht]
\centering
	\includegraphics[scale=0.3]{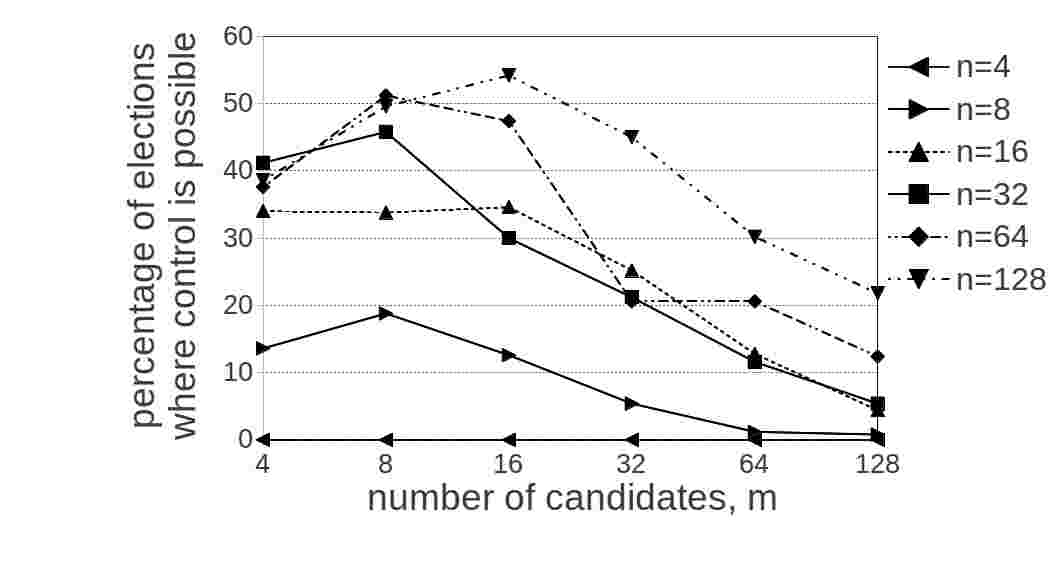}
	\caption{Results for plurality voting in the IC model for 
constructive control by partition of voters in model TP.  Number of voters is fixed.}
\end{figure}
%
%

\newpage
\begin{figure}[ht]
\centering
	\includegraphics[scale=0.3]{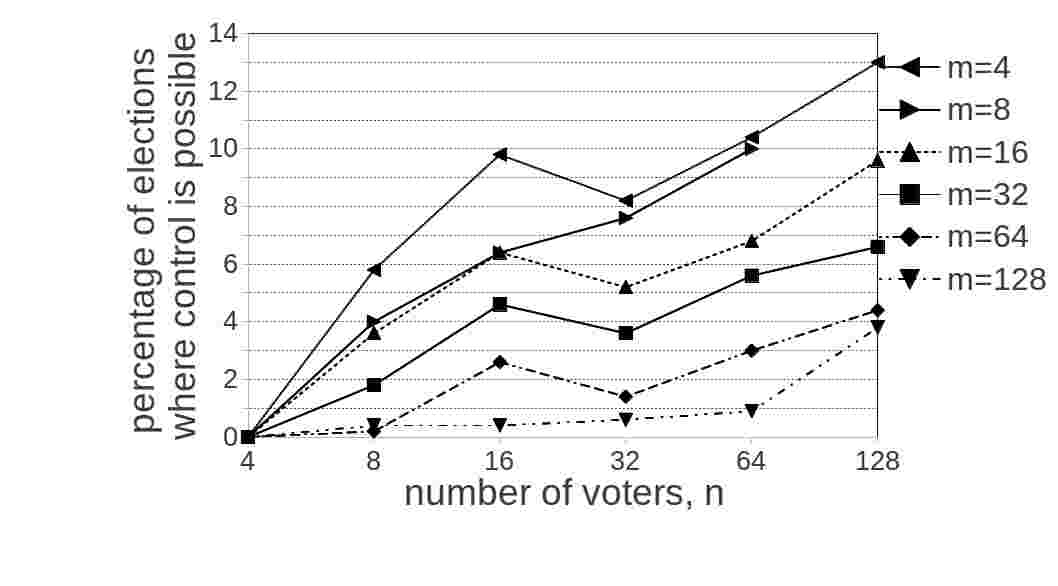}
	\caption{Results for plurality voting in the TM model for 
constructive control by partition of voters in model TP.  Number of candidates is fixed.}
\end{figure}

%
%

\newpage
\begin{figure}[ht]
\centering
	\includegraphics[scale=0.3]{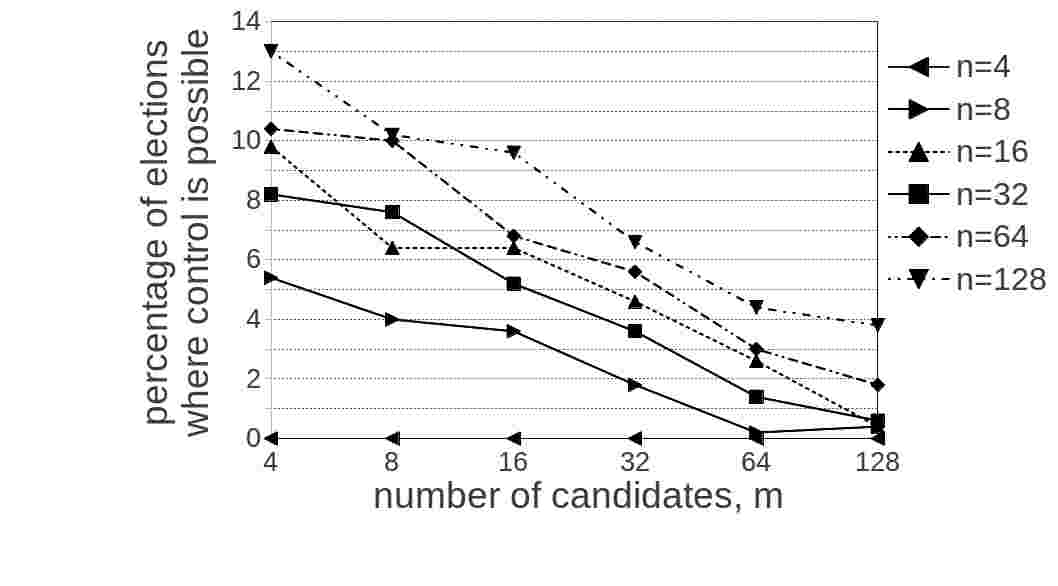}
	\caption{Results for plurality voting in the TM model for 
constructive control by partition of voters in model TP.  Number of voters is fixed.}
\end{figure}
%
\end{center}

\clearpage
\subsubsection{Computational Costs}
\begin{figure}[ht]
\centering
	\includegraphics[scale=0.3]{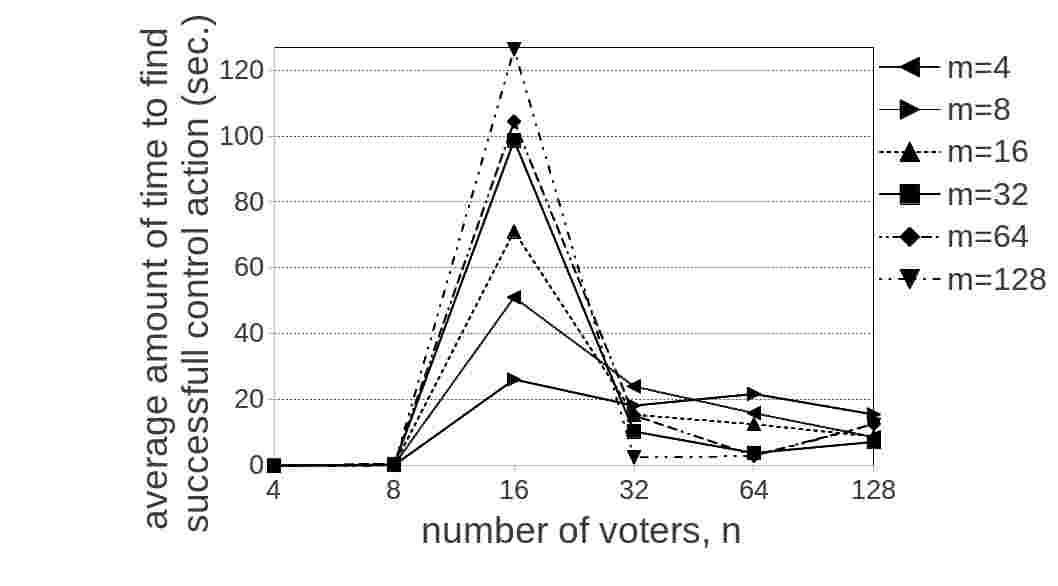}
	\caption{Average time the algorithm needs to find a successful control action for 
	constructive control by partition of voters in model TP
	in plurality elections in the IC model. The maximum is $126,26$ seconds.}
\end{figure}
\begin{figure}[ht]
\centering
	\includegraphics[scale=0.3]{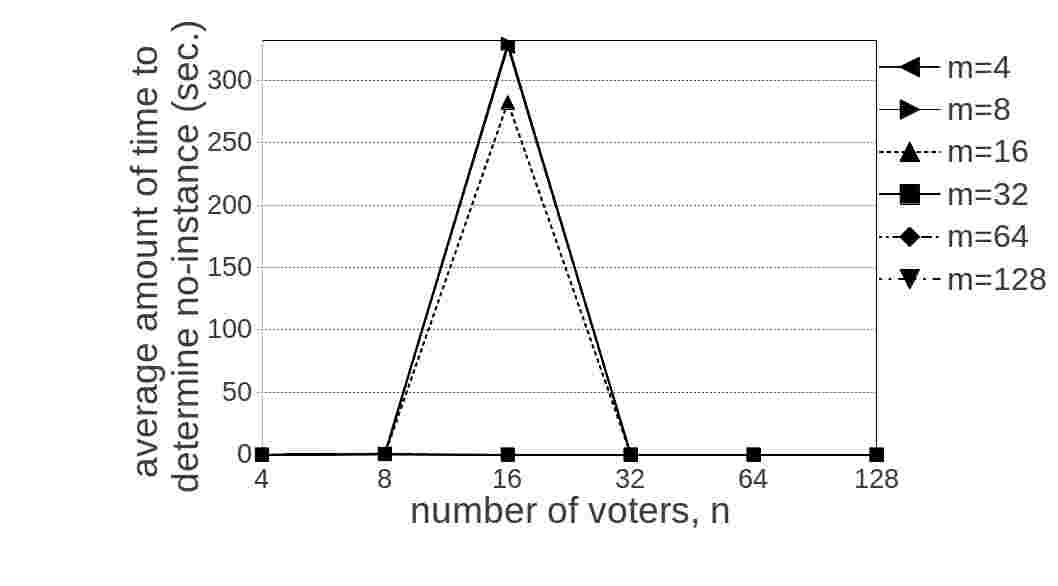}
	\caption{Average time the algorithm needs to determine no-instance of 
		constructive control by partition of voters in model TP
	in plurality elections in the IC model. The maximum is $329,07$ seconds.}
\end{figure}
\begin{figure}[ht]
\centering
	\includegraphics[scale=0.3]{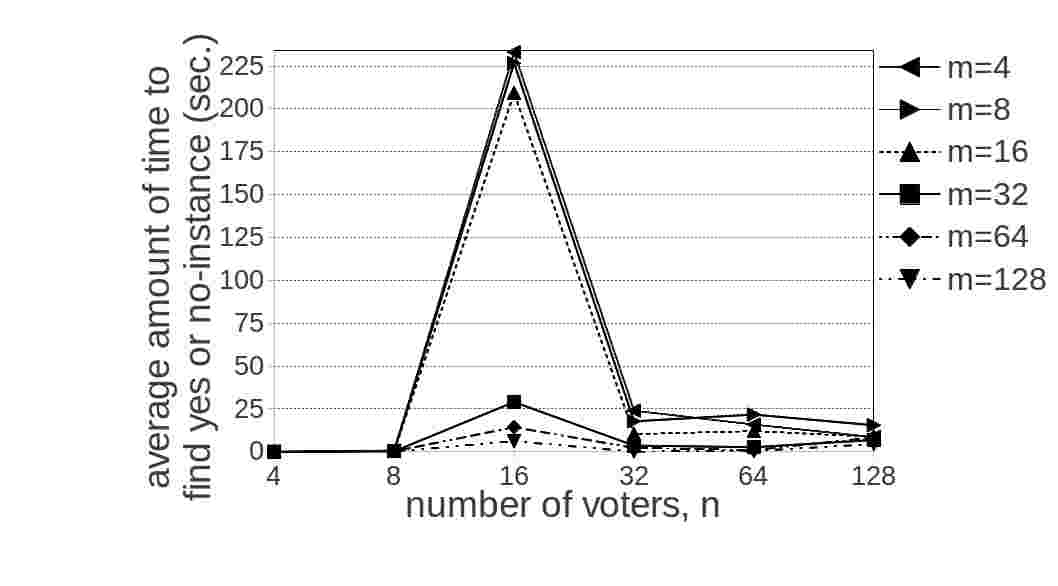}
	\caption{Average time the algorithm needs to give a definite output for 
	constructive control by partition of voters in model TP
	in plurality elections in the IC model. The maximum is $226,68$ seconds.}
\end{figure}
\begin{figure}[ht]
\centering
	\includegraphics[scale=0.3]{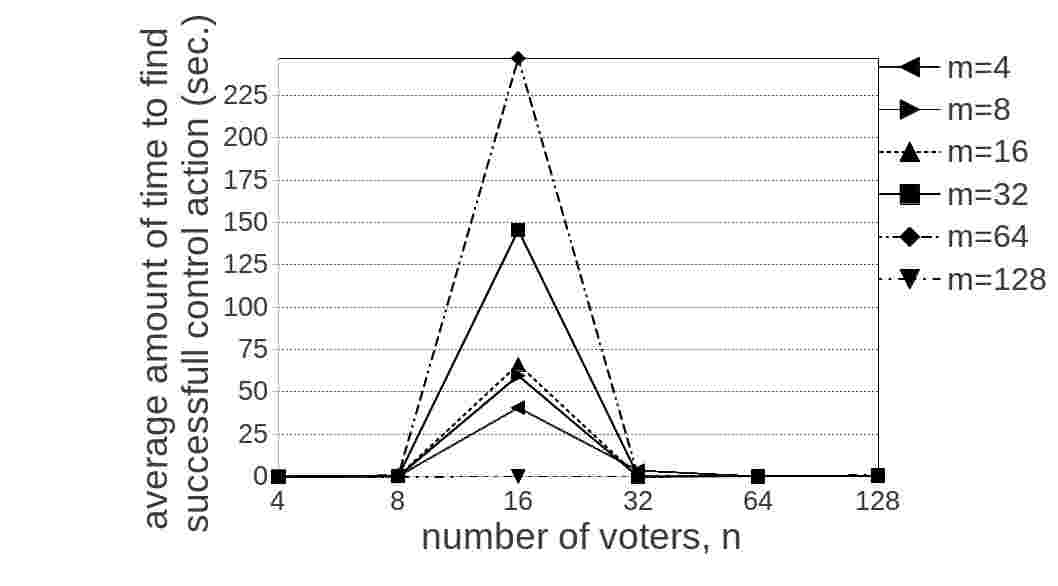}
	\caption{Average time the algorithm needs to find a successful control action for 
	constructive control by partition of voters in model TP
	in plurality elections in the TM model. The maximum is $246,85$ seconds.}
\end{figure}
\begin{figure}[ht]
\centering
	\includegraphics[scale=0.3]{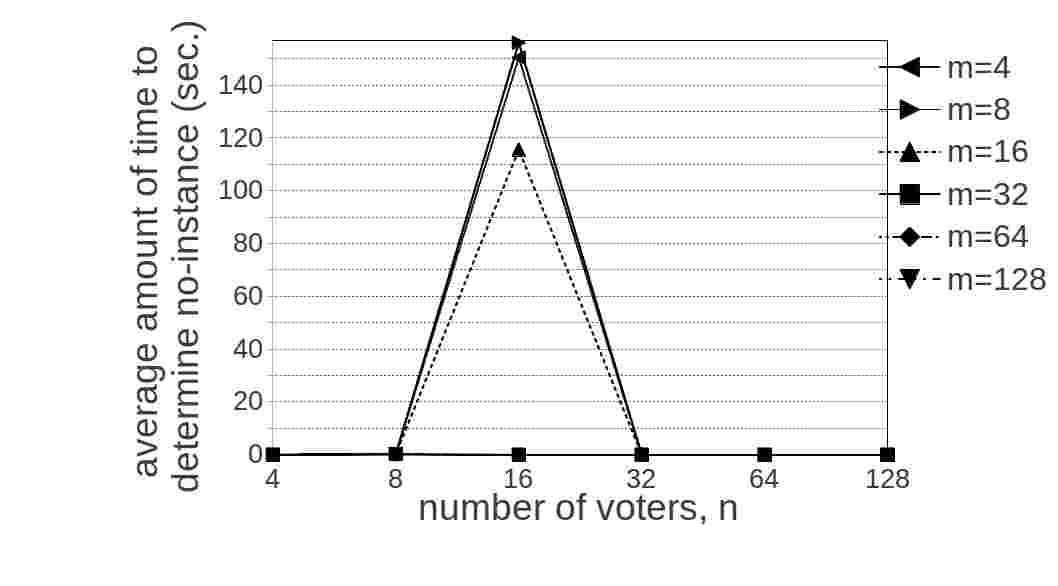}
	\caption{Average time the algorithm needs to determine no-instance of 
		constructive control by partition of voters in model TP
	in plurality elections in the TM model. The maximum is $156,07$ seconds.}
\end{figure}
\begin{figure}[ht]
\centering
	\includegraphics[scale=0.3]{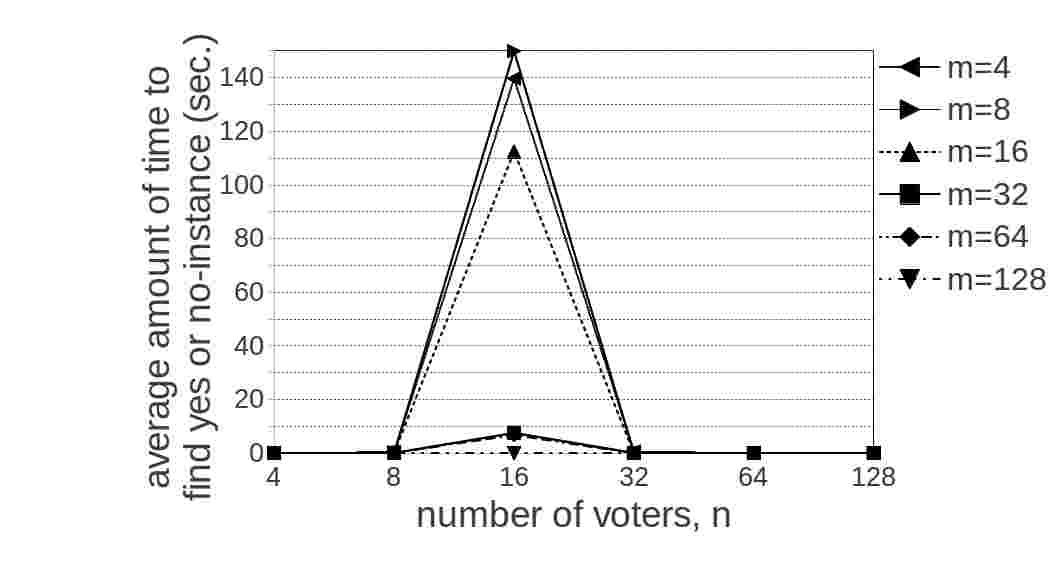}
	\caption{Average time the algorithm needs to give a definite output for 
	constructive control by partition of voters in model TP
	in plurality elections in the TM model. The maximum is $149,9$ seconds.}
\end{figure}

\clearpage
\subsection{Destructive Control by Partition of Voters in Model TP}
\begin{center}

\begin{figure}[ht]
\centering
	\includegraphics[scale=0.3]{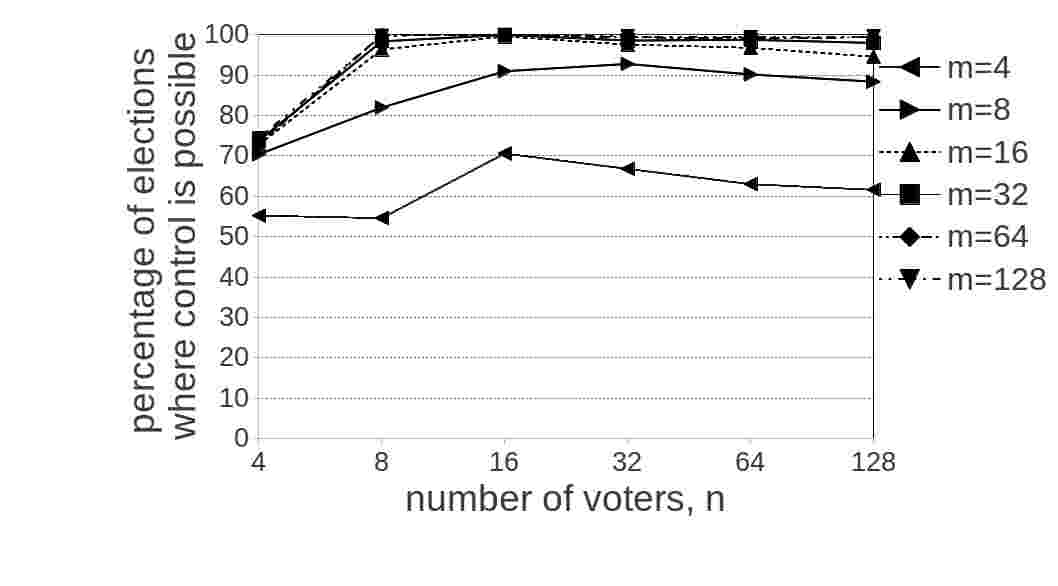}
	\caption{Results for plurality voting in the IC model for 
destructive control by partition of voters in model TP.  Number of candidates is fixed.}
\end{figure}

%

\newpage
\begin{figure}[ht]
\centering
	\includegraphics[scale=0.3]{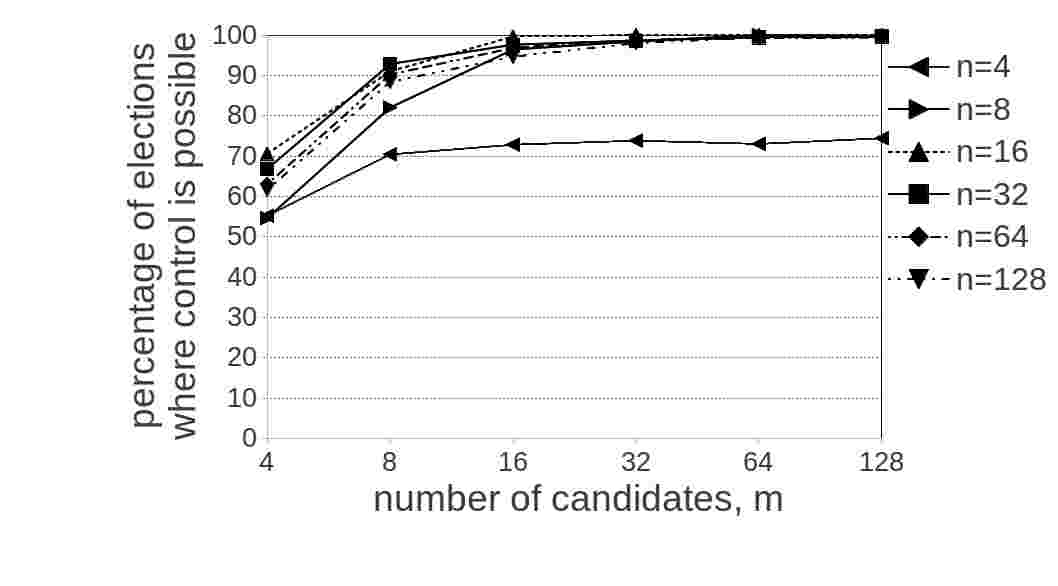}
	\caption{Results for plurality voting in the IC model for 
destructive control by partition of voters in model TP.  Number of voters is fixed.}
\end{figure}

%
%

\newpage
\begin{figure}[ht]
\centering
	\includegraphics[scale=0.3]{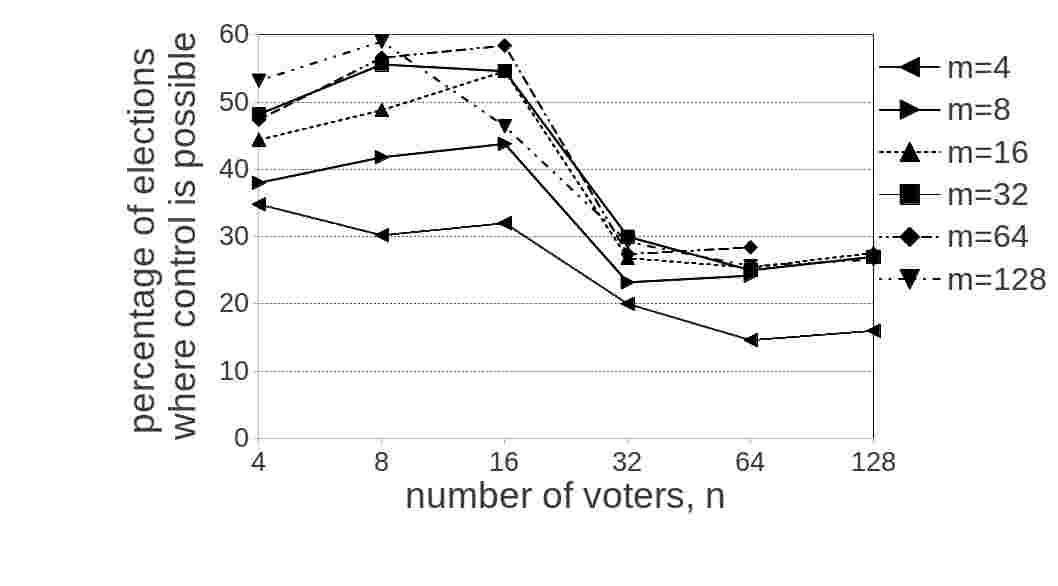}
	\caption{Results for plurality voting in the TM model for 
destructive control by partition of voters in model TP.  Number of candidates is fixed.}
\end{figure}

%
%

\newpage
\begin{figure}[ht]
\centering
	\includegraphics[scale=0.3]{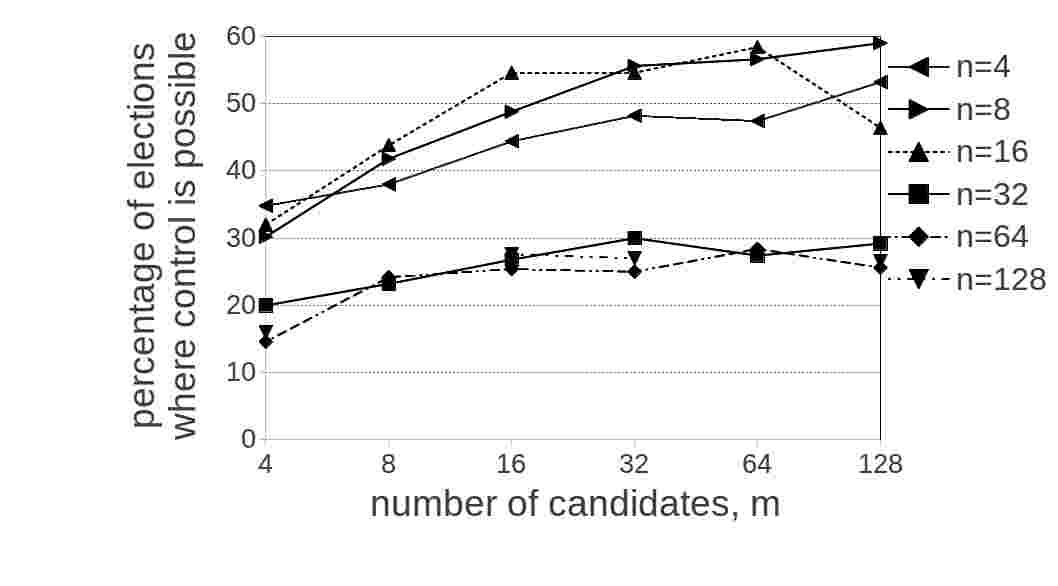}
	\caption{Results for plurality voting in the TM model for 
destructive control by partition of voters in model TP.  Number of voters is fixed.}
\end{figure}

%
\end{center}
\clearpage
\subsubsection{Computational Costs}
\begin{figure}[ht]
\centering
	\includegraphics[scale=0.3]{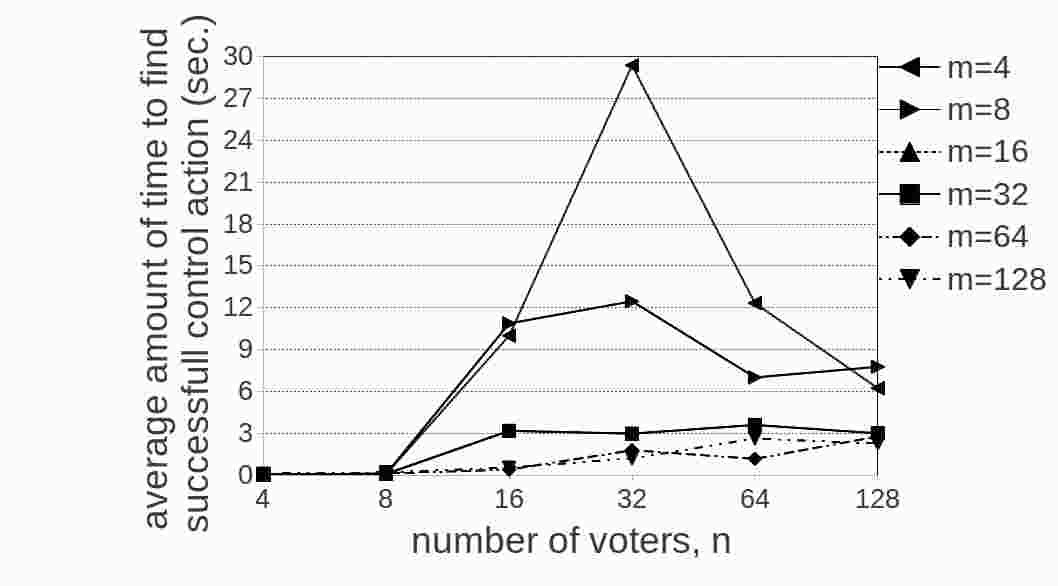}
	\caption{Average time the algorithm needs to find a successful control action for 
	destructive control by partition of voters in model TP
	in plurality elections in the IC model. The maximum is $29,37$ seconds.}
\end{figure}
\begin{figure}[ht]
\centering
	\includegraphics[scale=0.3]{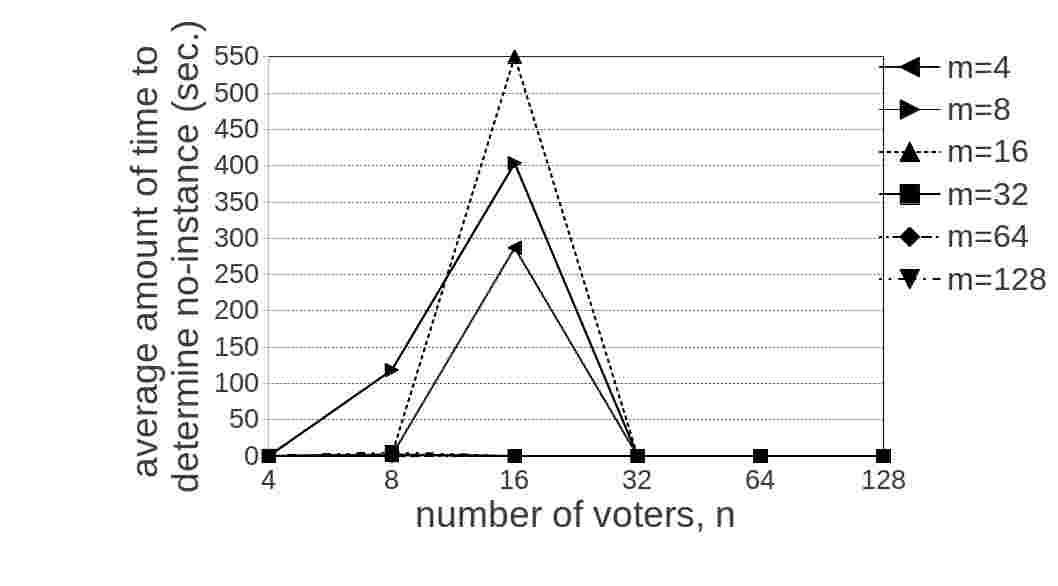}
	\caption{Average time the algorithm needs to determine no-instance of 
		destructive control by partition of voters in model TP
	in plurality elections in the IC model. The maximum is $550,39$ seconds.}
\end{figure}

\begin{figure}[ht]
\centering
	\includegraphics[scale=0.3]{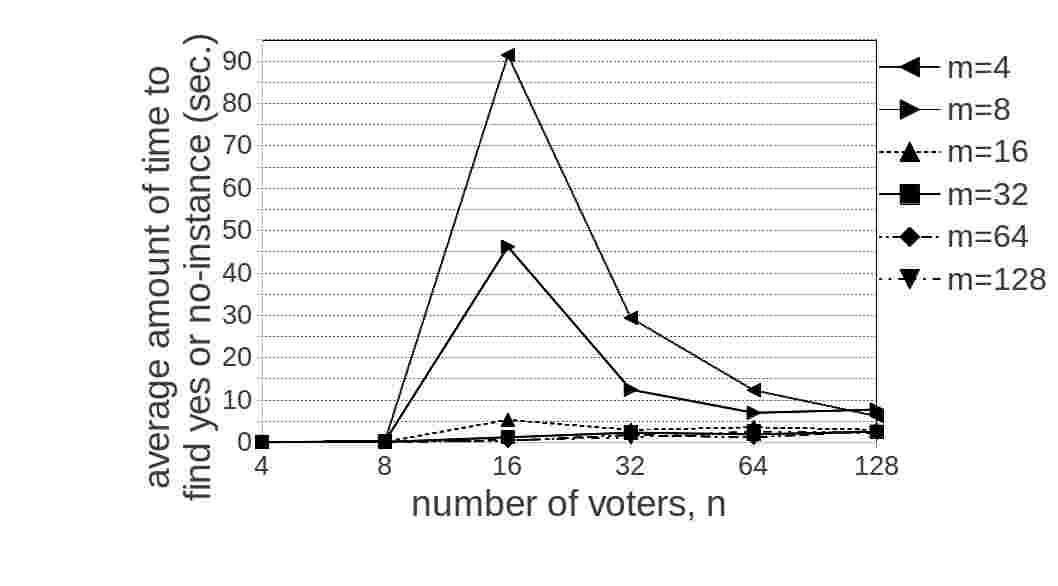}
	\caption{Average time the algorithm needs to give a definite output for 
	destructive control by partition of voters in model TP
	in plurality elections in the IC model. The maximum is $91,41$ seconds.}
\end{figure}
\begin{figure}[ht]
\centering
	\includegraphics[scale=0.3]{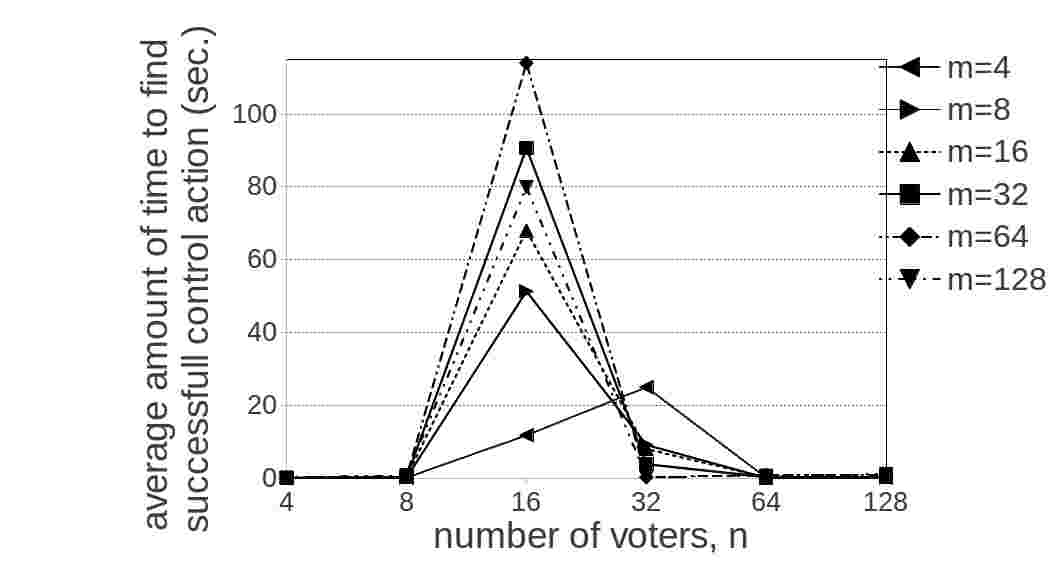}
	\caption{Average time the algorithm needs to find a successful control action for 
	destructive control by partition of voters in model TP
	in plurality elections in the TM model. The maximum is $114$ seconds.}
\end{figure}
\begin{figure}[ht]
\centering
	\includegraphics[scale=0.3]{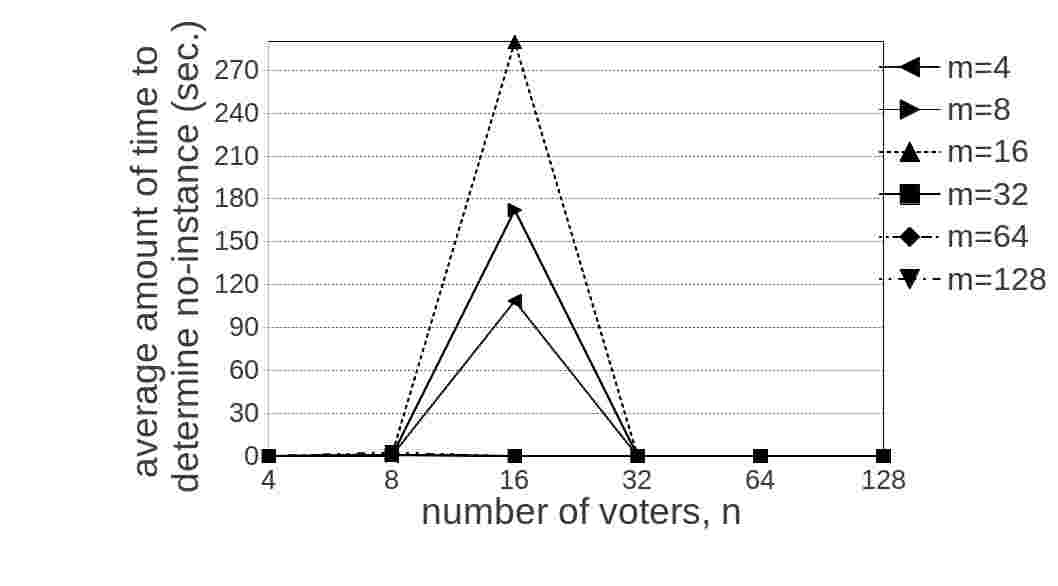}
	\caption{Average time the algorithm needs to determine no-instance of 
		destructive control by partition of voters in model TP
	in plurality elections in the TM model. The maximum is $289,7$ seconds.}
\end{figure}
\begin{figure}[ht]
\centering
	\includegraphics[scale=0.3]{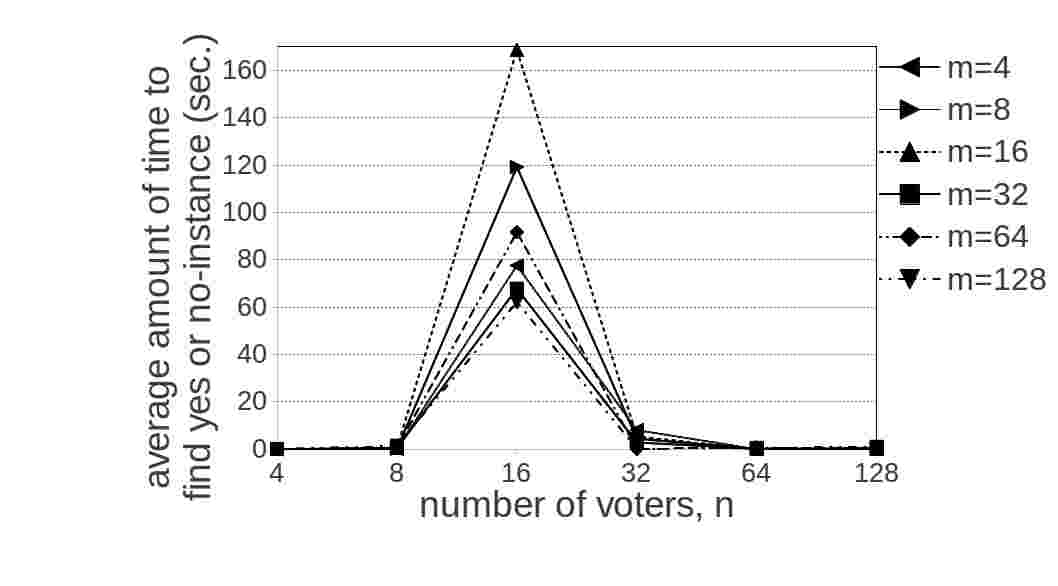}
	\caption{Average time the algorithm needs to give a definite output for 
	destructive control by partition of voters in model TP
	in plurality elections in the TM model. The maximum is $168,63$ seconds.}
\end{figure}